\magnification=\magstep1
\baselineskip12pt plus1pt
\hfuzz0.9pt
\vbadness6600

\input amstex
\input amssym.def
\input amssym
\input epsf

\newdimen\trpt
\trpt=1.095truept

\widowpenalty=10000
\clubpenalty=10000

\abovedisplayskip6pt plus2pt
\belowdisplayskip6pt plus2pt
\abovedisplayshortskip4pt plus2pt
\belowdisplayshortskip4pt plus2pt

\parindent15pt

 at10\trpt
\font\sc=cmcsc10
\font\fourmsy=msbm10 at14.4\trpt
\font\ninemsy=msbm9 at9\trpt
\def\sym#1{{\Bbb #1}}
\font\tytul=cmbx10 at20.74\trpt
\font\twbf=cmbx10 at17.28\trpt
\font\trz=cmbx10 at14.4\trpt
 at12\trpt
 at14.4\trpt
 at14.4\trpt
 at12\trpt
\font\ninerm=cmr9 at9\trpt
\font\nineit=cmti9 at9\trpt
\font\ninesy=cmsy9 at9\trpt
\font\ninei=cmmi9 at9\trpt
\font\sevenrm=cmr7 at7\trpt
\font\sevenit=cmti7 at7\trpt
\font\seveni=cmmi7 at7\trpt
\font\sevensy=cmsy7 at7\trpt
\font\fiverm=cmr5 at5\trpt
\font\fivei=cmmi5 at5\trpt
\font\fivesy=cmsy5 at5\trpt
\font\ninebf=cmbx9 at9\trpt
\font\nex=cmex10 at14.4\trpt
\font\ninex=cmex9 at9\trpt
\def\nine{%
\textfont0=\ninerm \scriptfont0=\sevenrm \scriptscriptfont0=\fiverm
\textfont1=\ninei \scriptfont1=\seveni \scriptscriptfont1=\fivei
\textfont2=\ninesy \scriptfont2=\sevensy \scriptscriptfont2=\fivesy
\textfont3=\ninex \scriptfont3=\ninex \scriptscriptfont3=\ninex
\def\rm{\fam0\ninerm}%
\textfont\itfam=\nineit
\def\it{\fam\itfam\nineit}%
\textfont\bffam=\ninebf
\def\bf{\fam\bffam\ninebf}%
\def\sym##1{\hbox{\ninemsy##1}}%
\normalbaselineskip=11\trpt
\setbox\strutbox=\hbox{\vrule height8\trpt depth3\trpt width0pt}%
\normalbaselines\rm}

\def\seven{%
\textfont0=\sevenrm \scriptfont0=\ninerm
\textfont1=\seveni \scriptfont1=\fivei
\textfont2=\sevensy \scriptfont2=\fivesy
\def\rm{\fam0\sevenrm}%
\textfont\itfam=\sevenit
\def\it{\fam\itfam\sevenit}%
\textfont\bffam=\sevenbf
\def\bf{\fam\bffam\sevenbf}%
\def\sym##1{\hbox{\sevenmsy##1}}%
\normalbaselineskip=9\trpt
\setbox\strutbox=\hbox{\vrule height6\trpt depth2\trpt width0pt}%
\normalbaselines\rm}

\font\fourrm=cmr10 at14.4\trpt
\font\fourit=cmti10 at14.4\trpt
\font\foursy=cmsy10 at14.4\trpt
\font\fouri=cmmi10 at14.4\trpt
\font\fourx=cmex10 at14.4\trpt
\font\fourbf=cmbx10 at14.4\trpt
\font\rorm=cmr10

\font\roi=cmmi10
\font\rosy=cmsy10
\def\four{%
\textfont0=\fourrm \scriptfont0=\rorm
\textfont1=\fouri \scriptfont1=\roi
\textfont2=\foursy \scriptfont2=\rosy
\textfont3=\fourx \scriptfont3=\fourx
\def\rm{\fam0\fourrm}%
\textfont\itfam=\fourit
\def\it{\fam\itfam\fourit}%
\textfont\bffam=\fourbf
\def\bf{\fam\bffam\fourbf}%
\def\sym##1{\hbox{\fourmsy##1}}%
\normalbaselineskip=17\trpt
\setbox\strutbox=\hbox{\vrule height10\trpt depth7\trpt width0pt}%
\normalbaselines\rm}
\let\got\frak

\def\fg{{\got g}}
\def\fe{{\got e}}
\def\fh{{\got h}}
\def\fm{{\got m}}
\def\gote{{\got e}}
\def\gf{{\got f}}

\def\cW{{\cal W}}

\hsize35.04truecc
\vsize48.18truecc

\def\centrowanie#1#2{{#1\let\\=\cr\tabskip14cc minus14cc
\halign to\hsize{\hfil##\hfil\cr
#2\cr}}}

\def\luz#1{\luzno#1?}
\def\luzno#1{\ifx#1?\let\next=\relax\b
\else \let\next=\luzno#1\a\fi\next}

\def\sp#1#2{\def\a{\kern#1pt}\def\b{\kern-#1pt}\luz{#2}}

\newbox\refbox
\newdimen\refwidth

\def\maxref=#1{\ninerm\setbox\refbox=\hbox{#1}\refwidth=\wd\refbox
\advance\refwidth by 6pt\advance\refwidth by.5em\rm}

\long\def\uref#1{%
\def\textindent##1{\indent\llap{##1\enspace\hskip6pt}\ignorespaces}% PLAIN
\vskip28pt plus2pt minus2pt
{\nine
\centerline{\ninebf References}
\vskip12pt plus2pt minus2pt
\parindent=\refwidth %%%%%%%%%%%%
\frenchspacing\rm
#1

}}

\footline={\hss\tenrm\folio\hss}

\def\eff{\operatorname{eff}\nolimits}
\def\ddiv{\operatorname{div}\nolimits}
\def\diag{\operatorname{diag}\nolimits}
\def\crt{\operatorname{crt}\nolimits}
\def\const{\operatorname{const}\nolimits}
\def\rank{\operatorname{rank}\nolimits}
\def\Int{\operatorname{int}\nolimits}
\def\gauge{\operatorname{gauge}\nolimits}
\def\Nabla{\mathop{\nabla}\limits}
\def\Ad{\operatorname{Ad}\nolimits}
\def\Tan{\operatorname{Tan}\nolimits}
\def\cot{\operatorname{cot}\nolimits}
\def\tot{\operatorname{tot}\nolimits}
\def\id{\operatorname{id}\nolimits}
\def\hor{\operatorname{hor}\nolimits}
\def\grad{\operatorname{grad}\nolimits}
\def\ver{\operatorname{ver}\nolimits}
\def\Span{\operatorname{Span}\nolimits}
\def\Spin{\operatorname{Spin}\nolimits}
\def\Sp{\operatorname{Sp}\nolimits}
\def\scal{\operatorname{scal}\nolimits}
\def\kin{\operatorname{kin}\nolimits}
\def\cat{\operatorname{cat}\nolimits}
\def\Var{\operatorname{Var}\nolimits}
\def\Ker{\operatorname{Ker}\nolimits}
\def\Tr{\operatorname{Tr}\nolimits}

\def\rF{\text{F}}
\def\rE{\text{E}}
\def\rA{\text{A}}
\def\rB{\text{B}}
\def\rC{\text{C}}
\def\rD{\text{D}}
\def\rE{\text{E}}
\def\rG{\text{G}}
\def\rH{\text{H}}
\def\rM{\text{M}}
\def\rN{\text{N}}
\def\rS{\text{S}}
\def\rT{\text{T}}
\def\rMC{\text{MC}}
\def\rAB{\text{AB}}
\def\rBA{\text{BA}}
\def\AD{\text{AD}}
\def\SL{\text{SL}}
\def\rCB{\text{CB}}
\def\rBC{\text{BC}}
\def\rBN{\text{BN}}
\def\rDB{\text{DB}}
\def\rDN{\text{DN}}
\def\rDM{\text{DM}}
\def\rYM{\text{YM}}
\def\rDE{\text{DE}}
\def\rBD{\text{BD}}
\def\rDC{\text{DC}}
\def\rAC{\text{AC}}
\def\rCA{\text{CA}}
\def\rCM{\text{CM}}
\def\rAM{\text{AM}}
\def\rBE{\text{BE}}
\def\rMA{\text{MA}}
\def\rMB{\text{MB}}
\def\rMN{\text{MN}}
\def\rNM{\text{NM}}
\def\rBM{\text{BM}}
\def\rAD{\text{AD}}
\def\SO{\text{SO}}
\def\SU{\text{SU}}
\def\GL{\text{GL}}
\def\gl{\text{gl}}
\def\Gl{\text{Gl}}

\def\nad#1#2{\mathbin{{\buildrel #2\over #1}}}

\def\Check#1{\mathop{#1}\limits^{\scriptscriptstyle \vee}}
\def\esssup{\operatorname{ess~sup}\limits}

\def\falkag{\mathrel{\lower6pt\hbox{${\scriptstyle\sim}$}\hskip-5pt g}}
\def\falkaL{\mathrel{\lower4pt\hbox{${\scriptstyle\sim}$}\hskip-6pt L}}
\def\falkaS{\mathrel{\lower4pt\hbox{${\scriptstyle\sim}$}\hskip-6pt S}}
\def\falkau{\mathrel{\lower4pt\hbox{${\scriptstyle\sim}$}\hskip-6pt u}}
\def\falkaT{\mathrel{\lower4pt\hbox{${\scriptstyle\sim}$}\hskip-6pt T}}
\def\falkaG{\mathrel{\lower4pt\hbox{${\scriptstyle\sim}$}\hskip-6pt G}}
\def\falkalam{\mathrel{\lower4pt\hbox{${\scriptstyle\sim}$}\hskip-6pt \lambda}}

\def\Te{\mathop{T}\limits}
\def\De{\mathop{D}\limits}

\def\lstrz#1{\mathop{\leftarrow}\limits_{\raise3pt\hbox{$\scriptstyle#1$}}}
\def\strz#1{\mathop{\to}\limits_{\raise3pt\hbox{$\scriptstyle#1$}}}

\def\tovarphi{\strz{\varphi}}
\def\tor{\strz{r_1}}
\def\toS{\strz{{\varSigma}_1}}
\def\leftor{\lstrz{r_2}}

\def\Wedge{\mathop{\wedge}\limits}
\def\Sum{\mathop{{\textstyle\sum}}\nolimits}

\def\dah#1{{\textstyle{\hat {\scriptstyle #1}}}}

\newdimen\krwys
\def\dkr#1{\leavevmode\setbox0\hbox{#1}\krwys=\dp0\advance\krwys by2pt
\rlap{\lower\krwys\hbox to\wd0{\hfil.\hfil}}#1}%%%%%% kropka pod litera

\nopagenumbers

\long\def\ab#1{\vskip18pt plus2pt minus2pt
\centerline{\tenbf Abstract}
\vskip12pt plus2pt minus2pt
#1}

\catcode`@=11
\def\eeeeqalign#1{\vcenter{\openup\jot\m@th\ialign{\strut\hfil$\displaystyle{
##}$&$\displaystyle{{}##}$\hfil&\hfil$\displaystyle{
##}$&$\displaystyle{{}##}$\hfil&\hfil$\displaystyle{
##}$&$\displaystyle{{}##}$\hfil&\hfil$\displaystyle{
##}$&$\displaystyle{{}##}$\hfil\crcr #1\crcr}}}

\def\eeeqalign#1{\vcenter{\openup\jot\m@th\ialign{\strut\hfil$\displaystyle{
##}$&$\displaystyle{{}##}$\hfil&\hfil$\displaystyle{
##}$&$\displaystyle{{}##}$\hfil&\hfil$\displaystyle{
##}$&$\displaystyle{{}##}$\hfil\crcr #1\crcr}}}

\def\eeqalign#1{\vcenter{\openup\jot\m@th\ialign{\strut\hfil$\displaystyle{
##}$&$\displaystyle{{}##}$\hfil&\hfil$\displaystyle{
##}$&$\displaystyle{{}##}$\hfil\crcr #1\crcr}}}

\def\lleqalignno#1{\displ@y \tabskip\centering
  \halign to\displaywidth{\hfil$\@lign\displaystyle{##}$\tabskip\z@skip
    &$\@lign\displaystyle{{}##}$\hfil&\hfil$\@lign\displaystyle{{}##}$%
    &$\@lign\displaystyle{{}##}$\hfil\tabskip\centering
    &\kern-\displaywidth\rlap{$\@lign##$}\tabskip\displaywidth\crcr
    #1\crcr}}

\def\displaylinesno#1{\vcenter {\openup \jot \m@th \ialign {\strut
\hfil $\displaystyle {##}$\hfil \crcr #1\crcr }}}

\def\ldisplaylinesno#1{\displ@y \tabskip\centering
\halign to\displaywidth{\hfill$\@lign\displaystyle{##}$\hfill\tabskip\centering
&\kern-\displaywidth\rlap{$\@lign##$}\tabskip\displaywidth\crcr #1\crcr }}
\catcode`@=12

\def\puww#1#2{\setbox0=\hbox{#1}\parindent=\wd0\advance\parindent by15pt
\def\textindent##1{\noindent\rlap{##1}\hskip\parindent\ignorespaces}
\item{#2}\parindent15pt}

\def\puw#1{\setbox0=\hbox{#1}\parindent=\wd0\advance\parindent by15pt
\def\textindent##1{\noindent\rlap{##1}\hskip\parindent\ignorespaces}
\item{#1}\parindent15pt}

\def\ov#1{#1\llap{$\overline{\phantom{\text{\rm#1}}}$}}
\def\ovv#1#2{#1\llap{$\overline{\phantom{\text{\rm#2}}}$}}

\def\un#1{\rlap{$\underline{\phantom{\text{\rm#1}}}$}#1}

\font\tim=cmsy10 at20.736truept
\font\timm=cmsy10 at24.8832truept
\def\bigtimes{\mathop{\mathchoice{\lower4pt\hbox{\timm\char'002}}{\lower3pt\hbox{\tim\char'002}\hskip-2pt}{}{}}}

\def\ii{\hglue-1pt$\hat{\hskip1pt\hbox{\i}}$}

\def\nt{\textfont1=\teni\textfont3=\nex}

\newdimen\szerd
\newdimen\szerg
\newdimen\roz
\newdimen\wx
\newdimen\wy

\def\mint_#1^#2{{\setbox0=\hbox{$\scriptstyle#1$}%
\setbox1=\hbox{$\scriptstyle#2$}%
\szerd=\wd0%
\szerg=\wd1%
\divide\szerd by2%
\divide\szerg by2%
\roz=\szerg%
\advance\roz by-\szerd%
\def\dol{\mathop{\vrule width\wd0 height-6pt depth6pt}\limits_{\box0}}%
\def\gora{\mathop{\vrule width\wd1 height10pt depth-10pt}\limits^{\box1}}%
\def\calka{\hbox{\nt$\int$}}%
\ifdim\szerg>10pt%
\ifdim\roz>4pt\wx=\szerg\advance\wx by-10pt%
\else\wx=\szerd\advance\wx by-6pt\fi%
\else%
\ifdim\szerd>6pt\wx=\szerd\advance\wx by-6pt\else\wx=0pt\fi\fi%
\ifdim\szerg>6pt%
\ifdim\roz>-4pt\wy=\szerg\advance\wy by-6pt%
\else\wy=\szerd\advance\wy by-10pt\fi%
\else%
\ifdim\szerd>10pt\wy=\szerd\advance\wy by-10pt\else\wy=0pt\fi\fi%
\rlap{\hskip\wx\hskip10pt\hskip-\szerg$\gora$}%
\rlap{\hskip\wx\hskip4pt\calka}%
\rlap{\hskip\wx\hskip6pt\hskip-\szerd$\dol$}%
\hskip\wx\hskip16pt\hskip\wy}}

\def\mmmint{\hskip4pt\hbox{\nt$\int$}\hskip4pt}
\def\mmint_#1{\mint_{#1}^{}}

\hyphenation{pa-ram-e-triza-tion}
\let\cal\Cal
\def\matr#1{\matrix #1 \endmatrix}
\def\pmatr#1{\pmatrix #1 \endpmatrix}
\def\ctg{\operatorname{ctg}}

\long\def\obraz#1{#1}
\long\def\bez#1{}
\immediate\openout0 aaa.str
{\advance\baselineskip by8pt plus.5pt minus.5pt

\null
\vskip3cc
\centerline{\twbf {M.}\ {W.}\ Kalinowski}
\vskip3cc
\centerline{\trz Pracownia Bioinformatyki IMDiK PAN}
\centerline{\trz Warszawa, ul. Dworkowa 3}
\centerline{\trz e-mail: markwkal@bioexploratorium.pl,}
\centerline{\trz mkalinowski@imdik.pan.pl}
%\centerline{\trz University of Economics}
%\centerline{\trz and Computer Science}
%\centerline{\trz Warsaw, Poland}
%\centerline{\trz Higher Vocational State School}
%\vskip.6cc
%\centerline{\trz in Che{\l}m, Poland}
\vskip.6cc
%\centerline{\trz }
\vskip6cc
\centerline{\tytul Scalar Fields}
\vskip6pt
\centerline{\tytul in The Nonsymmetric}
\vskip6pt
\centerline{\tytul Kaluza--Klein (Jordan--Thiry)}
\vskip6pt
\centerline{\tytul Theory}
\vfill
%{\twrm * Previous address: Department of Solid State Physics, University of \L \'od\'z, \L\'od\'z, Poland}

%{\twrm ** Partially supported by Badania W{\l}asne U{\L} Nr 505/485}
\eject

\vglue0.6\vsize
\font\ded=cmbxti10 at17.28\trpt
{\ded
\setbox0=\hbox{Professor Adam Bielecki}
\rightline{\vtop{\hsize\wd0
\centerline{To the memory}
\vskip6pt
\centerline{of my teacher}
\vskip6pt
\copy0}}}
\vfill\eject

\font\mota=cmbx12 at14.40\trpt
\font\motb=cmbxti10 at14.40\trpt
\null\vfill
\centerline{\ded Motto}
{\mota \centerline{W fizyce teoretycznej}
\centerline{potrzebujemy ka\D zdej matematyki,}
\centerline{a g\l\'ownie takiej,}
\centerline{kt\'orej jeszcze nie ma.}
\vskip5pt
\centerline{\hskip84\trpt \motb Adam Bielecki}
\vskip20pt
\centerline{(In theoretical physics}
\centerline{we need all the mathematics,}
\centerline{especially such a mathematics}
\centerline{which now does not exist.)}}
\vfill\eject

\null\vfill\eject

}

\def\leaderfill{\leaders\hbox to1em{\hss.\hss}\hfil}

\def\ilp{\rtimes}

%\tyt{{M.\ W.}\ Kalinowski}{Scalar fields in the nonsymmetric\\ Kaluza--Klein
%(Jordan--Thiry) theory}

\def\cfn{confinement}
\def\cn{connection}
\def\di{dimension}
\def\elm{electromagnetic}
\def\gr{gravitational}
\def\iv{invariant}
\def\so{solution}
\def\spt{space-time}
\def\s{symmetric}
\def\U{\text{U}}
\def\wrt{with respect to }
\def\KK{Kaluza--Klein}
\def\JT{Jordan--Thiry}
\def\nos{nonsymmetric}

$\hbox{}$
\null
\vglue3cc plus4pt minus4pt
\centerline{\bf Preface}
\centerline{\hfil}

In this book we construct the Nonsymmetric Jordan--Thiry Theory
unifying N.G.T. (Nonsymmetric Gravitation Theory), the Yang--Mills' field, the Higgs' fields and scalar
forces in a geometric manner. In this way we get masses from higher
dimensions. We discuss spontaneous symmetry breaking, the Higgs'
mechanism and a mass generation in the theory. The scalar field
${\varPsi}$ (as
in the classical Jordan--Thiry Theory) is connected to the effective
gravitational constant. This field is massive and has Yukawa-type
behaviour. We discuss the relation between ${\sym R}_{+}$ invariance and
${\bold U}(1)_{\rF}$
from G.U.T. (Grand Unified Theory) within Einstein $\lambda $-transformation, and fermion number
conservation. In this way we connect $\ov{W}_{\mu}$-field from N.G.T. with a gauge
field $\widetilde{A}^{\rF}_{\mu}$ from G.U.T. We derive the equation of motion for a test
particle from conservation laws in the hydrodynamic limit. We consider
a truncation procedure for a tower of massive $\rho _{k}$ (or ${\varPsi} _{k})$ scalar
fields using Friedrichs' theory and an approximation procedure for the
lagrangian involving Higgs' field up to the second order with respect
to $h_{\mu\nu} = g_{\mu\nu} - \eta _{\mu\nu}$ and $\zeta
k^{0}_{\tilde{a}\tilde{b}}$. The geodetic equations on the
Jordan--Thiry manifold are considered with an emphasis to terms
involving Higgs' field. We consider also field equations in
linear approximation and an influence of electromagnetic and
$\ov{W}_{\mu}$ fields on field equations.

We consider a dynamics of Higgs' field in the framework of cosmological
models involving the scalar field~$\varPsi$. The field~$\varPsi$ plays here a
r\^ole of a quintessence field. We consider phase transition in cosmological
models of the second and the first order. Due to a tower of massive scalar
fields~$\varPsi_k$ derived from the theory we are able to get a warp factor
known from some modern approaches. Using a warp factor we build a
time-machine in the theory.

We consider a mass of a quintessence particle, various properties of a
quintessence field. We calculate a speed of sound in a quintessence and
fluctuations of a quintessence caused by primordial metric fluctuations.

The \nos\ \KK\ (\JT) Theory has been developed here. It unifies the gauge invariance
principle with the coordinate invariance principle but in more than
four-dimensional space-time. In particular in the case of electromagnetic and
\gr\ interactions in 5-dimensional theory.

A general nonabelian Yang--Mills fields have been unified with gravity in
$(n+4)$-\di al space-time ($n$---a \di\ of gauge group).
The theory uses a non\s\ metric defined on a metrized (in a non\s\
way) principal fibre bundle over a \spt\ with a structural group $\U(1)$ in an
\elm\ case and in general case nonabelian semi-simple compact group~$G$. The
\cn\ on \spt\ and on a metrized principal fibre bundle is compatible with
this metric. This \cn\ is similar to a \cn\ from Einstein's Unified Field
Theory, however we use its higher \di al analogue. This \cn\ is
right-\iv\ \wrt an action of the group~$G$ (a~gauge group).

In the \elm\ case the metric and the \cn\ are bi-\iv\ \wrt the group~$\U(1)$.
The theory has been developed to
include a scalar field leading to an effective \gr\ constant and \spt\
dependent cosmological terms. It is possible to extend the theory to include
Higgs' fields and spontaneous symmetry breaking of a gauge group to get
massive vector boson fields. Some exact \so\ of the fields equations has been
found and a proposition  to solve a \cfn\ of color in QCD has
been posed.

The theory is fully relativistic and unifies \elm\ field, gauge fields,
Higgs' field and scalar forces with NGT (Non\s\ Gravitation Theory)
in a nontrivial way.
By `in a nontrivial way' we mean that we get from the theory
something more than NGT, ordinary Kaluza--Klein (\JT) Theory, classical
electrodynamics, Yang--Mills' field theory with Higgs' field and spontaneous
symmetry breaking. These new features are some kind of ``interference
effects'' between all of them.
This theory unifies two important approaches in higher-dimensional
philosophy: Kaluza--Klein principle and a \di al reduction principle.

The beautiful theories such as Kaluza--Klein theory (a~Kaluza miracle) and
its descendents should pass the following test if they are treated as real
unified theories. They should incorporate chiral fermions. Since the
fundamental scale in the theory is a Planck's mass, fermions should be
massless up to the moment of spontaneous symmetry breaking. Thus they should
be zero modes. In our approach they can obtain masses on a \di al reduction
scale. Thus they are zero modes in $(4+n_1)$-\di al case. In this way
$(n_1+4)$-\di al fermions are not chiral (according to very well known
Witten's argument on an index of a Dirac operator). Moreover, they are
not zero modes after a \di al reduction, i.e., in 4-\di al case. It means we
can get chiral fermions under some assumptions.

We expect some nonrelativistic effects leading
to nonnewtonian gravity.

The real motivation of this book is to find a geometric unifications of
fundamental interactions of Nature and to find applications in Modern
Cosmology including inflation, dark matter and dark energy using a paradigm
of physics, which unifies two fundamental concepts of invariance in physics:
coordinate invariance principle and gauge invariance principle. The first is
known in General Relativity and in vaiable alternative theories of gravitation. The
second is known in Electrodynamics and Yang--Mills field theory. Yang--Mills
field theory governs Standard Model, i.e.: Glashow--Salam--Weiberg (GSW) model of
electro-weak interactions and the theory of strong interactions (QCD) using
$\text{SU}(2)_L\times \U(1)$ and SU(3)$_c$ groups as gauge groups. It governs
also any Grand Unified Theorems.

The book is addressed to graduate students in theoretical physics,
mathematical physics, cosmology, particle physics and field theory. It is
also addressed to scientists working in these domains and to any reader who
is interested in these rapidly developing domains.

\vfill\eject
$\hbox{}$
\null
\vglue3cc plus4pt minus4pt
\centerline{\bf Contents}

\vglue24pt plus2pt

\setbox0=\hbox{112. }
%\setbox2=\hbox{1200. }
\setbox1=\hbox{ ~1100}
\newdimen\pp
\pp=\hsize\advance\pp by-\wd0\advance\pp by-\wd1
\def\str{\vrule height0.63\baselineskip depth0.27\baselineskip width0pt}
%\vbox{\halign to\hsize{\hbox to\wd0{\hfill#}&\hbox to\pp{~#}&\hfil#\cr
\def\lin#1* #2* #3\cr{\line{\hskip\wd0
\vbox{\hsize\pp \parindent0pt \parfillskip0pt
\textindent{#1 }\str#2\str}\hfil \vbox{\hsize\wd1 \line{\hfil#3}}}}
\lin * Introduction \leaderfill* 1\cr
\lin1.* Elements of Geometry\leaderfill * 7\cr
\lin2.* The Nonsymmetric Tensor on a Lie Group\leaderfill * 12\cr
\lin3.* The Nonsymmetric Metrization of the Bundle $\un{P}$\leaderfill * 21\cr
\lin4.* Preliminary Remarks\leaderfill * 30\cr
\lin5.* The Nonsymmetric Jordan--Thiry Theory over $V = E \times G/G_{0}$\leaderfill * 43\cr
\lin6.* Dimensional Reduction Procedure\leaderfill * 52\cr
\lin7.* Conformal Transformation of $g_{\mu\nu}$. Transformation of the Scalar
 Field $\rho $\leaderfill * 63\cr
\lin8.* Symmetry Breaking. Higgs Mechanism. Mass Generation. Cosmological\break
Constant\leaderfill * 66\cr
\lin9.* Variational Principle. Equations of Fields. Interpretation of the
Scalar\break Field~${\varPsi} $\leaderfill * 90\cr
\lin10.* Properties of the Scalar Field ${\varPsi} $. Cosmological Drifting of the
 Mass Scale \leaderfill * 106\cr
\lin11.* Fermion Number, ${\sym R}_{+}$ and ${\bold U}(1)_{\rF}$ Invariance\leaderfill * 110\cr
\lin12.* Fermion Number as the Second Gravitational Charge.
$\ov{W}_{\mu}$ and
 $\widetilde{A}^{\rF}_{\mu}$ Potentials\leaderfill * 115\cr
\lin13.* Equations of Motion for a Test Particle in the Nonsymmetric
 Jordan--Thiry Theory\leaderfill * 124\cr
\lin14.* On a Cosmological Origin of the Mass of the Scalar Field
${\varPsi}$ (or $\rho$). Cosmological Models with Quintessence and Inflation
\leaderfill * 128\cr
\lin15.* Geodetic Equations on the Manifold $\underline P$\leaderfill * 207\cr
\lin16.* A Tower of Scalar Fields in the Nonsymmetric-Nonabelian
 Jordan--Thiry Theory. The Warp Factor\leaderfill * 228\cr
\lin17.* The Approximation Procedure for the Lagrangian of the Higgs
Field in the Nonsymmetric Jordan--Thiry Theory\leaderfill * 269\cr
\lin18.* On some Cosmological Consequences of the Nonsymmetric Kaluza--Klein
(Jordan--Thiry) Theory\leaderfill * 287\cr
\lin19.* Properties of a Quintessence\leaderfill * 315\cr
\lin20.* Charge Confinement in the Nonsymmetric Kaluza--Klein
Theory\leaderfill * 363\cr
\lin21.* Gravito--Electromagnetic\ Waves Solutions in the Nonsymmetric
Kaluza--Klein Theory\leaderfill * 372\cr
\lin22.* An Influence of a Cosmological Constant on a Solution\ in the
Nonsymmetric Kaluza--Klein Theory\leaderfill * 377\cr
\lin * Conclusions\leaderfill * 386\cr
\lin * Appendix A\leaderfill * 391\cr
\lin * References\leaderfill * 401\cr
\lin * Acknowledgement\leaderfill * 412\cr
\lin * Abstract\leaderfill * 413\cr
\lin * A brief information of a text\leaderfill * 416\cr 

\vfill\eject

\footline={\hss\ninerm\folio\hss}

\pageno=1

\null
\vglue2cc plus4pt minus4pt
\centerline{\trz Introduction}
\vglue24pt plus2pt minus2pt
\write0{Introduction \the\pageno}

\def\KK{Kaluza--Klein}
\def\JT{Jordan--Thiry}

In this work we develop a unification of the Nonsymmetric
Gravitational Theory and gauge fields (Yang--Mills' fields) including
spontaneous symmetry breaking and the Higgs' mechanism with scalar
forces connected to the gravitational constant and cosmological terms
appearing as the so-called quintessence. The theory is
geometric and unifies tensor-scalar gravity with massive gauge theory
using a multidimensional manifold in a Jordan--Thiry manner (see Refs.
[1--70]). We use a nonsymmetric version of this theory (see [23], [24],
[25], [26], [27]). The general scheme is the following. We introduce
the principal fibre bundle over the base $V = E \times G/G_{0}$ with the
structural group $H$, where $E$ is a space-time, $G$ is a compact semisimple
Lie group, $G_{0}$ is its compact subgroup and $H$ is a semisimple compact
group. The manifold $M = G/G_{0}$ has an interpretation as a ``vacuum states
manifold" if $G$ is broken to $G_{0}$ (classical vacuum states). We define on
the space-time $E$, the nonsymmetric tensor $g_{\alpha\beta}$ from N.G.T., which is
equivalent to the existence of two geometrical objects
$$\eqalignno{\noalign{\vskip2pt}
\ov{g}&= g_{(\alpha \beta )}\overline{\theta }^\alpha \otimes \overline{\theta}^\beta \cr
\noalign{\vskip2pt}
\un{g}&= g_{[\alpha \beta ]}\overline{\theta }^\alpha \wedge \overline{\theta}^\beta
\cr\noalign{\vskip2pt}}
$$
the symmetric tensor $\ov{g}$ and the 2-form $\un{g}$. Simultaneously we introduce
on $E$ two connections from N.G.T. $\ov{W}_{\beta \gamma}^\alpha $ and
$\widetilde{\overline{\varGamma}}_{\beta \gamma}^\alpha $. On the homogeneous space $M$
we define the nonsymmetric metric tensor %\vadjust{\vskip2pt}
$$
g_{\tilde{a}\tilde{b}} = h^{0}_{\tilde{a}\tilde{b}}
+ \zeta k^{0}_{\tilde{a}\tilde{b}} %\vadjust{\vskip2pt}
$$
where $\zeta $ is the dimensionless constant, in a geometric way. Thus we
really have the nonsymmetric metric tensor on or $V = E \times
G/G_{0}$. %\vadjust{\vskip2pt}
$$
\gamma _{\rAB} =
\left(\vcenter{\offinterlineskip\tabskip0pt
\halign{\strut\tabskip5pt plus10pt minus10pt
\hfill#\hfill&\vrule height10pt#&\hfill#\hfill\tabskip0pt\cr
$\vrule height0pt depth6pt width0pt g_{\alpha \beta }$ && $0$\cr
\noalign{\hrule}
$\vrule height9pt depth4pt width0pt 0$ &&
$r^2g_{\tilde{a}\tilde{b}}$\cr}}\right) %\vadjust{\vskip2pt}
$$
$r$ is a parameter which characterizes the size of the manifold $M =
G/G_{0}$. Now on the principal bundle $\un{P}$ we define the connection $\omega $, which
is the 1-form with values in the Lie algebra of $H$.

After this we introduce the nonsymmetric metric on $\un{P}$
right-invariant with respect to the action of the group $H$, introducing
scalar field $\rho $ in a Jordan--Thiry manner (see Ref~[18]).
The only difference is that
now our base space has more dimensions than four. It is
$(n_{1}+4)$-dimensional, where $n_{1} = \dim (M) = \dim (G)-\dim (G_{0})$. In other
words, we combine the nonsymmetric tensor $\gamma _{\rAB}$ on $V$ with the
right-invariant nonsymmetric tensor on the group $H$ using the connection
$\omega $ and the scalar field $\rho $. We suppose that the factor $\rho $ depends on a
space-time point only. This condition can be abandoned and we consider
a more general case where $\rho =\rho (x,y)$, $x\in E$, $y\in M$ resulting in a tower of
massive scalar field $\rho _{k}, k=1,2\ldots .$ This is really the Jordan--Thiry
theory in the nonsymmetric version but with $(n_{1}+4)$-dimensional
``space-time". After this we act in the classical manner as in Refs.\
[1], [18], [22], [23], [24], [26], [27]. We introduce the linear connection
which is compatible with this nonsymmetric metric. This connection is
the multidimensional analogue of the connection
$\widetilde{\overline{\varGamma}}{}^{\alpha}_{\beta \gamma }$ on the space-time
$E$. Simultaneously we introduce the second connection $W$. The
connection $W$ is the multidimensional analogue of the $\ov{W}$-connection from
N.G.T. and Einstein's Unified Field Theory. It is the same as the
connection from Refs.\ [23], [24]. Now we calculate the Moffat--Ricci
curvature scalar $R(W)$ for the connection $W$ and we get the following
result. $R(W)$ is equal to the sum of the Moffat--Ricci curvature on the
space-time $E$ (the gravitational lagrangian in Moffat's theory of
gravitation), plus $(n_{1}+4)$-dimensional lagrangian for the Yang--Mills'
field from the Nonsymmetric Kaluza--Klein Theory plus the Moffat--Ricci
curvature scalar on the homogeneous space $G/G_{0}$ and the Moffat--Ricci
curvature scalar on the group $H$ plus the lagrangian for the scalar
field~$\rho $. The only difference is that our Yang--Mills' field is defined
on $(n_{1}+4)$-dimensional ``space-time" and the existence of the
Moffat--Ricci\vadjust{\vskip-.6pt}
curvature scalar of the connection on the homogeneous space
$G/G_{0}$.\vadjust{\vskip-.6pt}
All of these terms (including $R(\ov{W})$) are multiplied by some
factors depending on the scalar field~$\rho $.

This lagrangian depends on the point of $V = E \times G/G_{0}$ i.e.\ on the
point of the space-time $E$ and on the point %\vadjust{\vskip-1pt}
of $M = G/G_{0}$. The curvature
scalar on $G/G_{0}$ also depends on the point of $M$.

We now go to the group structure of our theory. We assume $G$
invariance of the connection $\omega $ on the principal fibre bundle $\un{P}$, the so
called Wang-condition (see Refs.\ [59], [71]--[76]). According to the
Wang-theorem and the Ref.~[72] the connection $\omega $ decomposes into the
connection $\widetilde{\omega }_{\rE}$ on the principal bundle $Q$ over space-time $E$ with
structural group $G$ and the multiplet of scalar fields ${\varPhi} $. Due to this
decomposition the multidimensional Yang--Mills' lagrangian decomposes
into: a 4-dimensional Yang--Mills' lagrangian with the gauge group $G$
from the Nonsymmetric Kaluza--Klein Theory, plus a polynomial of 4th
order with respect to the fields ${\varPhi} $, plus a term which is quadratic with
respect to the gauge derivative of ${\varPhi}$ (the gauge derivative with respect
to the connection $\widetilde{\omega }_{\rE}$ on a space-time $E$) plus a new term which is of 2nd
order in the ${\varPhi} $, and is linear with respect to the Yang--Mills' field
strength. After this we perform the dimensional reduction procedure
for the Moffat--Ricci scalar curvature on the manifold $\un{P}$. We average
$R(W)$ with respect to the homogeneous space $M = G/G_{0}$ as in Refs.~[73],
[74], [75]. In this way we get the lagrangian of our theory. It is the
sum of the Moffat--Ricci curvature scalar on $E$ (gravitational
lagrangian) plus a Yang--Mills' lagrangian with gauge group $G$ from the
Nonsymmetric Kaluza--Klein Theory (see [23]), plus a kinetic term for
the scalar field ${\varPhi} $, plus a potential $V({\varPhi} )$ which is of 4th order with
respect to ${\varPhi} $, plus ${\cal L}_{\Int}$ which describes a nonminimal interaction
between the scalar field ${\varPhi} $ and the Yang--Mills' field, plus cosmological
terms, plus lagrangian for scalar field $\rho $. All of these terms
(including $\ov{R}(\ov{W})$) are multiplied of course by some factors depending on
the scalar field $\rho $. We redefine tensor $g_{\mu \nu }$ and $\rho$ (as in [18], [19],
[50]) and pass from scalar field $\rho $ to ${\varPsi} $
$$ %\vadjust{\vskip-1pt}
\rho = e^{-{\varPsi} }
$$
After this we get lagrangian which is the sum of gravitational
lagrangian, Yang--Mills' lagrangian, Higgs' field lagrangian,
interaction term ${\cal L}_{\Int}$ and lagrangian for scalar field ${\varPsi} $ plus
cosmological terms. These terms depend now on the scalar field ${\varPsi} $. In
this way we have in our theory a multiplet of scalar fields $({\varPsi} ,{\varPhi} )$. As
in the Nonsymmetric-Nonabelian Kaluza--Klein Theory we get a
polarization tensor of the Yang--Mills' field induced by the
skewsymmetric part of the metric on the space-time and on the group $G$.
We get an additional term in the Yang--Mills' lagrangian induced by the
skewsymmetric part of the metric $g_{\alpha \beta }$ (see Ref.~[23]). We get also ${\cal L}_{\Int}$,
which is absent in the dimensional reduction procedure known up to now
(see [73]--[76], [77]--[80]). Simultaneously, our potential for the
scalar---Higgs' field really differs from the analogous potential from
Refs.\ [73]--[80]. Due to the skewsymmetric part of the metric on $G/G_{0}$
and on $H$ it has a more complicated structure. This structure offers
two kinds of critical points for the minimum of this potential: ${\varPhi} ^{0}_{\crt }$
and ${\varPhi} ^{1}_{\crt }$. The first is known in the classical, symmetric dimensional
reduction procedure (see [73], [80]) and corresponds to the trivial
Higgs' field (``pure gauge"). This is the ``true" vacuum state of the
theory.
The second, ${\varPhi} ^{1}_{\crt }$, corresponds to a more complex configuration.
This is only a local (no absolute) minimum of~$V$. It is a ``false"
vacuum. The Higgs' field is not a ``pure" gauge here. In the first case
the unbroken group is always $G_{0}$. In the second case, it is in general
different and strongly depends on the details of the theory: groups $G_{0}$,
$G$, $H$, tensors $\ell _{ab}$, $g_{\tilde{a}\tilde{b}}$ and the constants $\zeta$, $\xi $. It results in a
different spectrum of mass for intermediate bosons. However, the scale
of the mass is the same and it is fixed by a constant $r$ (``radius" of
the manifold $M = G/G_{0})$. In the first case $V({\varPhi} ^{0}_{\crt }) =0$, in the second
case it is, in general, not zero $V({\varPhi} ^{1}_{\crt }) \neq 0$. Thus, in the first case,
the cosmological constant is a sum of the scalar curvature on $H$ and
$G/G_{0}$, and in the second case, we should add the value $V({\varPhi} ^{1}_{\crt })$. We
proved that using the constant $\xi$ we are able in some cases to make the
cosmological constant as small as we want (it is almost zero,
from the observational data point of view). Here we can
perform the same procedure for the second term in the cosmological
constant using the constant $\zeta $. In the first case we are able to make
the cosmological constant sufficiently small but this is not possible
in general for the second case.

The transition from ``false" to ``true" vacuum occurs as a second
order phase transition (see [81]--[85]). We discuss this transition in
context of the first order phase transition in models of the Universe
(see [81]--[85]). In this work the interesting point is that there
exists an effective scale of masses, which depends on the scalar
field~${\varPsi} $.

Using Palatini variational principle we get an equation for fields
in our theory. We find a gravitational equation from N.G.T. with
Yang--Mills', Higgs' and scalar sources (for scalar field ${\varPsi} )$ with
cosmological terms. This gives us an interpretation of the scalar
field ${\varPsi} $ as an effective gravitational constant
$$
G_{\eff } = G_{N}e^{-(n+2){\varPsi} }
$$
We get an equation for this scalar field ${\varPsi} $. Simultaneously we get
equations for Yang--Mills' and Higgs' field. We also discuss the change
of the effective scale of mass, $m_{\eff }$ with a relation to the change of
the gravitational constant~$G_{\eff }$.

In the ``true" vacuum case we get that the scalar field ${\varPsi} $ is
massive and has Yukawa-type behaviour. In this way the weak
equivalence principle is satisfied. In the ``false" vacuum case the
situation is more complex. It seems that there are possible some
scalar forces with infinite range. Thus the two worlds constructed
over the ``true" vacuum and the ``false" vacuum seem to be completely
different: with different unbroken groups, different mass spectrum for
the broken gauge and Higgs' bosons, different cosmological constants
and with different behaviour for the scalar field ${\varPsi} $. The last point
means that in the ``false" vacuum case the weak equivalence principle
could be violated and the gravitational constant (Newton's constant)
would increase in distance between bodies.

We discuss in the work ${\sym R}_{+}$ and ${\bold U}(1)_{\rF}$ invariance in a continuation
of our consideration from Refs.\ [26], [27]. We develop the first
possibility from Section~8 of Ref.~[26]. In this way we decide that
${\bold U}(1)_{\rF}$ invariance from G.U.T. is a local invariance. Due to a
geometrical construction we are able to identify $\ov{W}_{\mu }$ from Moffat's
theory of gravitation with the four-potential $\widetilde{A}^{\rF}_{\mu }$ corresponding to the
${\bold U}(1)_{\rF}$ group (internal rotations connected to fermion charge). In this
way, the fermion number is conserved and plays the role of the second
gravitational charge. Due to the Higgs' mechanism
$\widetilde{A}^{\rF}_{\mu }$ is massive and
its strength, $\widetilde{H}^{\rF}_{\mu \nu }$, is of short range with Yukawa-type behaviour. This
has important consequences. The Lorentz-like force term (or
Coriolis-like force term) in the equation of motion for a test particle
is of short range with Yukawa-type behaviour. The range of this force
is smaller than the range of the weak interactions. Thus it is
negligible in the equation of motion for a test particle. We discuss
the possibility of the cosmological origin of the mass of the scalar
field ${\varPsi} $ and geodetic equations on $\un{P}$. We consider an infinite tower of
scalar fields ${\varPsi} _{k}(x)$ coming from the expansion of the field ${\varPsi} (x,y)$ on
the manifold $M=G/G_{0}$ %\vadjust{\vskip-.3pt}
into harmonics of the Beltrami--Laplace operator.
Due to Friedrichs' theory we can %\vadjust{\vskip-.3pt}
diagonalize an infinite matrix of
masses for ${\varPsi} _{k}$ transforming them into new fields
${\varPsi}'_k$. The truncation %\vadjust{\vskip-.3pt}
procedure means here to take a zero mass mode %\vadjust{\vskip-.3pt}
${\varPsi} _{0}$ and equal it to ${\varPsi} $ from
the preceding section.

We discuss here cosmological models involving field $\varPsi$ which plays a
r\^ole of a quintessence field. We find inflationary models of the Universe
and we discuss a dynamics of the Higgs field. Higgs' field dynamics undergoes
a second order phase transition which causes a phase transition in an
evolution of the Universe. This ends an inflationary epoch and changes an
evolution of the field~$\varPsi$. Afterwards we consider the
field~$\varPsi$ as a quintessence field building some cosmological models
with a quintessence and even with a $K$-essence. A~dynamics of a Higgs field
in several approximations gives us an amount of an inflation. We consider
also a fluctuation spectrum of a primordial fluctuations caused by a Higgs
field and a speed up factor of an evolution. We speculate on a future of the
Universe based on our simple model with a special behaviour of a
quintessence.

We consider a possibility to take seriously additional dimensions in our
theory in a framework similar to a Randall--Sundrum scenario. In our theory
the additional dimensions connecting to the manifold~$M$ (a~vacuum manifold)
could be considered in such a way. They are not directly observable because
the size of the manifold is very small. In order to see them it is necessary
to excite massive modes of the scalar field (a~tower of these fields). We
find an interesting toy model (a~5-dimensional model) which can describe a
possibility to travel with speed higher than the speed of light using the
fifth dimension. This dimension has nothing to do with the fifth dimension in
Kaluza--Klein theory. We do some analysis on an energy to excite a scalar
field~$\varPsi$ to get this special solution to the theory. We also discuss
some quantitative relations involving traveling signals in the model via the
fifth dimension. We consider the various possibilities to excite a warp
factor due to fluctuations of a tower of scalar fields finding a density of
an excitation energy. Eventually we find a zero energy (or almost zero)
condition for such an excitation.

We also develop the approximation %\vadjust{\vskip-.3pt}
 procedure for
the lagrangian involving the Higgs' field up to the
second %\vadjust{\vskip-.3pt}
order of
expansion with respect to %\vadjust{\vskip-2pt}
$h_{\mu \nu } = g_{\mu \nu } - \eta _{\mu
\nu }$ and $\zeta $. We find $V({\varPhi} )$,
${\cal L}\big(\Nabla^{\gauge}{\varPhi}\big)$,
${\cal L}_{\Int}(\widetilde{A},{\varPhi} )$ up to this order. %\vadjust{\vskip-1pt}

The book is organized as follows. In section~1 we introduce the
notation and the definitions of geometrical quantities which we use
throughout this part. In sec.~2 we define nonsymmetric tensor on a
group Lie. The third section deals with a nonsymmetric metrization of
the fibre bundle.We define the skewsymmetric tensor on $M=G/G_{0}$ and
examine its properties in the fourth section. Section 5 is devoted
to the Nonsymmetric Jordan--Thiry Theory over the manifold $V = E \times G/G_{0}$.
We introduce the connections $\omega ^{\rA}{}_{\rB}$ and
$W^{\rA}{}_{\rB}$ which are analogues of the
connections $\ov{W}^{\alpha }{}_{\beta }$ and $\overline{\omega}^{\alpha
}{}_{\beta }$ from N.G.T. We calculate Moffat--Ricci
curvature scalar for $W^{\rA}{}_{\rB}$ and the density for this scalar. In section
6 we perform the dimensional reduction procedure for $R(W)$ using
results from section 5. We find the integral of action in our theory
and the lagrangian. In section 7 we perform a conformal
transformation for the nonsymmetric tensor $g_{\mu \nu }$ and a transformation for
the scalar field $\rho $. We pass from $\rho $ to ${\varPsi} $. In section 8 we discuss
symmetry breaking in our theory with two kinds of critical points
corresponding to two minimas, to the ``true" and ``false" vacuums. We
deal also with the Higgs' mechanism, mass generation in the theory and
the scale of mass for intermediate bosons. We also discuss a problem
of cosmological terms and phase transition in the models of Universe
based on our theory. We consider a possibility of
nontrivial gauge configurations in our theory with topological
magnetic charges (monopoles). Section 9 is devoted to the Palatini
variational principle and field equations. We find the interpretation
of the scalar field ${\varPsi} $ as a gravitational ``constant". In section 10
we discuss the properties of the scalar field ${\varPsi} $. We prove that this
field is massive. We discuss also cosmological drifting of the
effective scale of masses for the intermediate bosons. In section 11
we discuss ${\sym R}_{+}$ and ${\bold U}(1)_{\rF}$ invariance. In section 12 we discuss the
geometrical construction which allows us to identify
$\ov{W}_{\mu }$ with the
four-potential $\widetilde{A}^{\rF}_{\mu }$ connected to ${\bold
U}(1)_{\rF}$ (fermion number subgroup of $G$).
In section 13 we derive equation of motion for a test particle from
conservation laws in the hydrodynamic limit. Section~14 is devoted
to the problem of a cosmological origin of the mass of a scalar field ${\varPsi}$
(or $\rho$). We consider here cosmological models with a quintessence and a
dynamics of Higgs' field with phase transitions.
Section 15 deals with geodetic equations on $\un{P}$ and geodetic equation
deviation. In
Sec.~16 we consider a tower of scalar fields coming from the factor $\rho $ depending
on a point of $M = G/G_{0}$. We consider here a toy model with a warp factor
and we find a zero energy condition to excite the factor using a tower of
scalar fields and $(n_1+4)$-dimensional equations for cosmological solutions.
We consider a possibility to use a warp factor in order to send an effective
superluminal signal and to create a time machine.
Section 17 is devoted to the approximation
procedure for the lagrangian involving Higgs' field and geodetic
equations on the manifold $\un{P}$. We consider also linear
approximation of full field equations and an influence of %\vadjust{\vskip-1pt}
$\ov
W_{\mu}$--field and electromagnetic field on field equations.

Section 18 is devoted to examination of cosmological consequences of our
theory. We calculate spectral functions of  primordial fluctuations and
masses of quintessence particles in both de Sitter phases of an evolution of
the Universe.
In Section~19 we consider properties of a quintessence. We calculate a mass
of a quintessence particle of order $10^{-5}\,$eV, a speed of sound for a
quintessence and various solutions of quintessence field equation.
We consider also primordial fluctuations of a quintessence field.

In Section 20 we consider a charge \cfn\ in the Nonsymmetric \KK\ Theory in an
\elm\ case, reminding some notion of this theory in a new setting. Section~21
is devoted to the gravito--\elm\ waves and in Section~22 we consider in
details an exact solution of field equation with cosmological constant in a
stationary and spherically symmetric case, which is a generalization of that
derived earlier.

In Appendix A we give some useful formulae in the Nonsymmetric \KK\ (\JT)
Theory.

\vfill\eject

\null
\vskip36pt plus4pt minus4pt
\centerline{\trz 1. Elements of Geometry}
\vskip24pt 
\write0{1. Elements of Geometry \the\pageno}
\def\Hor{\text{\rm Hor}}
\let\ol\overline

In this  section  we  introduce  notations  and  define  geometric
quantities used in the paper. We use a smooth principal fibre bundle $\un{P}$,
which includes in its definition the following list  of  differentiable
manifolds and smooth maps:

\vskip6pt plus2pt
a total (bundle) space $\un{P}$

a base space $E$; in our case it is a space  time  (or  $E\times G/G_{0}$  as  in
sec.~4)

a projection $\pi : \un P \rightarrow  E$

a map ${\varPhi} : \un P\times G \rightarrow  \un{P}$  defining the action of $G$ on $\un{P}$; if $a,b \in  G$   and
$\varepsilon  \in  G$  is the unit element then ${\varPhi} (a)\circ {\varPhi} (b) = {\varPhi} (ba)$  and ${\varPhi} (\varepsilon ) =\id$  and
${\varPhi} (a)p = {\varPhi} (p,a) = R_{a}p = pa$, moreover $\pi \circ {\varPhi} (a) = \pi$. 
$\omega $ is  a  1-form  of  a
connection on $\un{P}$ with values in the Lie algebra of the group $G$.

\vskip6pt plus2pt
Let ${\varPhi} '(a)$ be the tangent map to ${\varPhi} (a)$ whereas ${\varPhi} ^{*}(a)$ is contragradient to
${\varPhi} (a)$ at the point a. The form $\omega $ is a form of Ad-type~i.e.:
$$
{\varPhi} ^{*}(a)\omega  = \Ad_{a-1}\omega, \eqno(1.1)
$$
where $\Ad_{a} \in  \GL(\frak g)$ is the tangent map to the  internal  automorphism  of
the group $G$ (i.e.\ it is an adjoint representation of a group $G)$
$$
\text{\rm ad}_{a}(b) = aba^{-1}.
$$
Due to the form $\omega $ we can introduce the  distribution  field  of  linear
elements $H_{r}$, $r \in  \un{P}$, where $H_{r}\subset  T_{r}(\un{P})$   is  a  subspace  of  the  space
tangent to $\un{P}$ at a point $r$ and
$$
v \in  H_{r} \Leftrightarrow  \omega (v) = 0.\eqno(1.2)
$$
We have
$$
T_{r}(\un{P}) = V_{r}\oplus  H_{r},\eqno(1.3)
$$
where $H_{r}$ is called a subspace of horizontal vectors and $V_{r}$ of  vertical
vectors. For vertical vectors $v\in  V_{r}$ we have $\pi '(v) = 0$.

This means that $v$ is tangent to fibres. Let us define
$$
v = \hor (v) + \ver (v),\quad\ \hor(v) \in  H_{r},\ \ver(v) \in 
V_{r}.\eqno(1.4)
$$
It is well known that the distribution $H_{r}$ is equivalent to a choice  of
the connection~$\omega $. We can reproduce the connection form $\omega $ demanding that
$\pi'|_{H_{r}}:H_{r}\rightarrow  T_{\pi(r)}(E)$ is a vector space isomorphism $(\dim  H_{r}= \dim  E =4)$,
$H_{{\varPhi} (r,g)}= {\varPhi} '(g)H_{r}$, $(T_{\pi(r)}(E)$ is a tangent space to  space-time $E$  at  a
point $\pi (r))$. We use the operation ``hor" for forms, i.e.:
$$
(\hor \beta ) (X,Y) = \beta (\hor X,\hor Y),\eqno(1.5)
$$
where $X,Y \in  T_{r}(\un{P})$. The 2-form of curvature of the connection $\omega
$~is:
$$
{\varOmega}  = \hor \,d\omega .\eqno(1.6)
$$
It is also a form of Ad-type like $\omega $. ${\varOmega} $ obeys  the  structural  Cartan's
equation
$$
{\varOmega}  = d\omega  + \frac12 [\omega ,\omega ],\eqno(1.7)
$$
where 
$$
[\omega ,\omega ] (X,Y) = [\omega (X),\omega (Y)].
$$
Bianchi's identity for $\omega $ is:
$$
\hor \,d{\varOmega}  = 0.\eqno(1.8)
$$
For the principal fibre bundle we use the following  convenient  scheme
(Fig.~1A). The map $e: U \rightarrow  \un{P}$, $U \subset  E$
($U$ open),  so that $e\circ \pi = \id_{U}$  is
called a local section. From  the  physical  point  of  view  it  means
choosing the gauge. Thus
$$\eqalign{
e^{*}\omega  &= e^{*}(\omega ^{a}X_{a}) = A^{a}_{\mu }\overline{\theta
}^\mu X_{a} = A,\cr
e^{*}{\varOmega}  &= e^{*}({\varOmega} ^{a} X_{a}) =\frac12
F^{a}_{\mu \nu }\overline{\theta }^\mu \wedge \overline{\theta }^\nu 
X_{a}.\cr}\eqno(1.9)
$$
\obraz{
\midinsert \centerline{\epsfxsize=\hsize \epsfbox{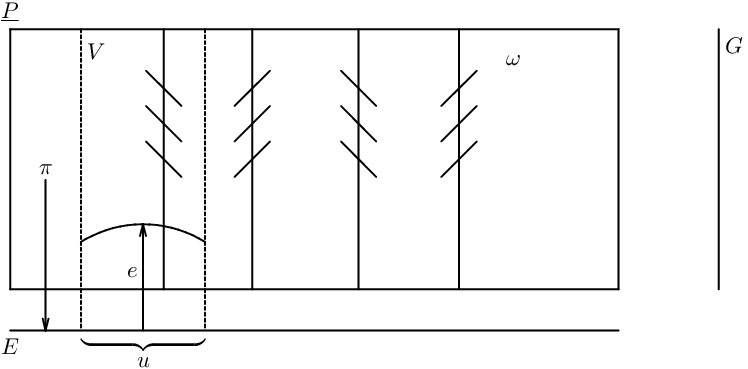}} 
\centerline{\nine Fig. 1A. The principal fibre bundle $\un{P}$} \endinsert

}Further we introduce a notation
$$
{\varOmega} ^{a} =\frac12 H^{a}_{\mu \nu } \theta ^{\mu }\wedge \theta
^{\nu },\eqno(1.10)
$$
where $\theta ^{\mu } = \pi ^{*}(\overline{\theta }^\mu )$ and
$$
F^{a}_{\mu \nu } = \partial _{\mu }A^{a}{}_{\nu } - \partial _{\nu
}A^{a}{}_{\mu } + C^{a}_{bc}A^{b}{}_{\mu }A^{c}{}_{\nu }
$$
$X_{a}$, $(a = 1,2\ldots \dim G = n)$ are generators of the Lie  algebra $\fg$ of  the
group~$G$ and
$$
[X_{a} ,X_{b}] = C^{c}{}_{ab}X_{c}.
$$

Analogically we can introduce a second local section $f:U\rightarrow P$,  and
corresponding to it $\ovv{A}{I}=f^{*}\omega $, $\ov{F}=f^{*}{\varOmega} $. For every $x\in U\subset E$ there is  an  element
$g  (x)\in  G$  such   that $f(x)=e(x)g  (x)=R_{g  (x)}e(x)={\varPhi}
(e(x),g  (x))$.   Due   to
Eq.~(1.1) and an analogical formula for ${\varOmega} $ one gets
$\ovv{A}{I} =\Ad_{g ^{-1}}A+g ^{-1}dg $  and
$\ov{F}=\Ad_{g ^{-1}}F$.  These  formulae  give  a  geometrical   meaning   of   gauge
transformation.

In this paper we use also a linear connection on manifolds $\un{P}$ and $E$
using the formalism of differential forms. So the basic quantity  is  a
1-form of a connection $\omega ^{\rA}{}_{\rB}$.This is an $R$-valued (coefficient) connection
form and it is referred to the principal fibre bundle of frames with $\un{P}$
or $E$ as a base. The 2-form of curvature is the following
$$
{\varOmega} ^{\rA}{}_{\rB} = d\omega ^{\rA}{}_{\rB} + \omega
^{\rA}{}_{\rC}\wedge \omega ^{\rC}{}_{\rB}\eqno(1.11)
$$
and the 2-form of torsion
$$
{\varTheta} ^{\rA} = D \theta ^{\rA},\eqno(1.12)
$$
where $\theta ^{\rA}$ are basic forms, $D$ means  the  exterior  covariant  derivative
with respect to $\omega ^{\rA}{}_{\rB}$. The following relations define  the  interrelation
between our symbols and generally used ones
$$\eqalignno{
\omega ^{\rA}{}_{\rB} &= {\varGamma} ^{\rA}{}_{\rBC}\theta ^{\rC},\cr
{\varTheta} ^{A} &=\frac12 Q^{\rA}{}_{\rBC}\theta ^{\rB}\wedge \theta
^{\rC},&(1.13)\cr
{\varOmega} ^{\rA}{}_{\rB} &= \frac12 R^{\rA}{}_{\text{\rm BCD}}\theta
^{\rC}\wedge \theta ^{\rD}.\cr}
$$
Where ${\varGamma} ^{\rA}{}_{\rBC}$ are coefficients of the connection (they do not have  to  be
symmetric in indices $B$ and $C$), $R^{\rA}{}_{\text{\rm BCD}}$ is a tensor of curvature and $Q^{\rA}{}_{\rBC}$
is a tensor of torsion. Covariant exterior differentiation with respect
to $\omega ^{\rA}{}_{\rB}$ is given by the formula
$$\eqalign{
D{\varXi} ^{\rA} &= d{\varXi} ^{\rA} + \omega ^{\rA}{}_{\rC}\wedge 
{\varXi} ^{\rC},\cr
D{\varSigma} ^{\rA}{}_{\rB} &= d{\varSigma} ^{\rA}{}_{\rB} + 
\omega ^{\rA}{}_{\rC}\wedge {\varSigma} ^{\rC}{}_{\rB} - 
\omega ^{\rC}{}_{\rB}\wedge {\varSigma}
^{\rA}{}_{\rC}.\cr}\eqno(1.14)
$$
The forms of curvature ${\varOmega} ^{\rA}{}_{\rB}$ and torsion ${\varTheta}
^{\rA}$ obey Bianchi's identities
$$\eqalign{
D{\varOmega} ^{\rA}{}_{\rB} &= 0,\cr
D{\varTheta} ^{\rA} &= {\varOmega} ^{\rA}{}_{\rB}\wedge \theta
^{\rB}.\cr}\eqno(1.15)
$$
In the paper we use also Einstein's $+$ and $-$  differentiations  for  the
nonsymmetric metric tensor $g_{\rAB}$.
$$
Dg _{\text{\rm A+B}-} = Dg _{\rAB} -g _{\text{\rm AD}} Q^{\rD}{}_{\text{\rm BC}}\theta
^{\rC},\eqno(1.16)
$$
where $D$ is the covariant exterior derivative with respect  to $\omega ^{\rA}{}_{\rB}$  and
$Q^{\rD}{}_{\rBC}$ is the tensor of torsion for $\omega ^{\rA}{}_{\rB}$.  In  a  homolonomic  system  of
coordinates we easily get:
$$
Dg _{\text{\rm A+B}-} = g _{\text{\rm A+B$-$;C}}\theta ^{\rC} = 
[g _{\rAB ,\rC}-g _{\rDB}{\varGamma} ^{\rD}{}_{\rAC}-g _{\AD}{\varGamma}
^{\rD}{}_{\rCB}]\theta ^{\rC}.\eqno(1.17)
$$
All quantities introduced in this section and their precise definitions
can be found in Refs.\ [51], [59], [60], [61].

Finally let us connect a general formalism of the principal  fibre
bundle with a formalism of a linear connection on $E$ or $\un P$.

Let $M$ be a $ m$-dimensional pseudo-Riemannian manifold with metric
$g$ of arbitrary signature. Let $T(M)$ be the tangent bundle and
$O(M,g)$  the
principal fibre bundle of  frames  (orthonormal  frames)  over  $M$.  The
structure group of $O(M,g)$ is the  group $\Gl (m,{\sym R})$  or  the  subgroup  of
$\Gl(m,{\sym R})$, $O(m-p,p)$ which leaves the  metric  invariant.  Let
${\varPi} $  be  the
projection of $O(M,g)$ onto $M$. Let $X$ be a tangent vector at a point $x$  in
$O(M,g)$. The canonical or soldering form $\widetilde{\theta }$ is  an $R^{m}$-valued  form  on
$O(M,g)$ whose $A$th component $\widetilde{\theta }^{\rA}$ at $x$ of $X$ is the
$A$th component  of ${\varPi} '(X)$
in the frame $x$. The connection form $\widetilde{\omega } = 
\omega ^{\rA}{}_{\rB}X^{\rB}{}_{\rA}$ is a 1-form  on $O(M,g)$
which takes its values in the Lie algebra $\text{\rm gl}(m,{\sym R})$  of
$\Gl(m,{\sym R})$  or  in
$o(m-p,p)$ of $O(m-p,p)$ and satisfies the structure equations
$$
d\widetilde{\omega } + \frac12 [\widetilde{\omega} ,\widetilde{\omega }] = 
\widetilde{{\varOmega} } = \widetilde{\Hor}\ d\widetilde{\omega
},\eqno(1.18)
$$
where $\widetilde{\Hor}$ is understood in the  sense  of 
$\widetilde{\omega }$   and 
$\widetilde{{\varOmega} } = \widetilde{{\varOmega} }^{\rA}{}_{\rB}
X^{\rB}{}_{\rA}$  is a
$\text{\rm gl}(m,{\sym R})
(o(m-p,p))$ valued 2-form of the curvature. We can write
Eq~(1.18)
using $R^{2m}$ valued forms and commutation relations  of  the  Lie  algebra
$\text{\rm gl}(m,{\sym R})(o(m,m-p))$
$$
\widetilde{{\varOmega} }^{\rA}{}_{\rB} = d\widetilde{\omega }^{\rA}{}_{\rB} 
+ \widetilde{\omega }^{\rA}{}_{\rC}\wedge \widetilde{\omega
}^{\rC}{}_{\rB}.\eqno(1.19)
$$
Taking any local section of $O(M,g)$ $e$, one can get forms of coefficients
of the connection, torsion, curvature, basic forms
$$\eqalign{
e^{*}\widetilde{\omega }^{\rA}{}_{\rB} &= \omega ^{\rA}{}_{\rB},\cr
e^{*}\widetilde{{\varOmega} }^{\rA}{}_{\rB} &= {\varOmega}
^{\rA}{}_{\rB},\cr
e^{*}\widetilde{\theta }^{\rA} &= \theta ^{\rA},\cr
e^{*}\widetilde{{\varTheta} }^{\rA} &= {\varTheta}^{\rA}.\cr}\eqno(1.20)
$$
\obraz{
\midinsert \centerline{\epsfxsize=\hsize \epsfbox{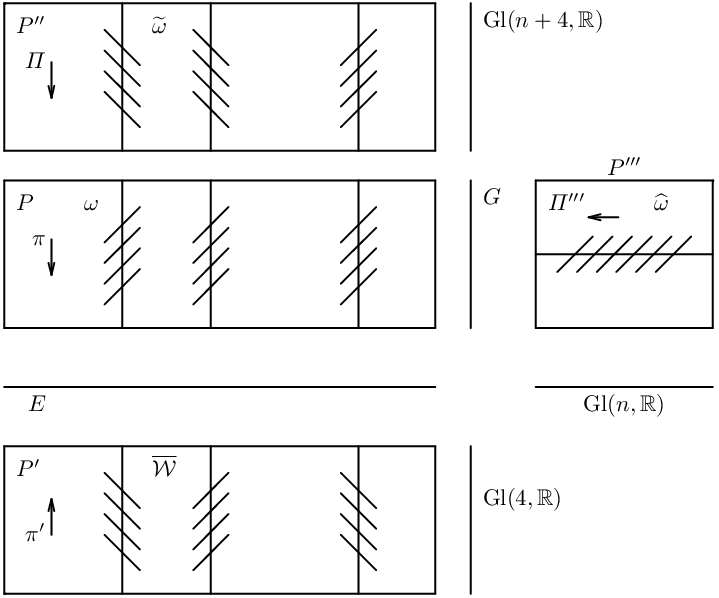}} 
\centerline{\nine Fig. 1B. Principal fibre bundles $\underline{P}$, $P'$, $P''$ and
$P'''$,}
\vskip-1pt
\centerline{\nine $P'''$ is principal fibre bundle of frames over $G$.}
\endinsert

}The forms of the right-hand side of the Eq.~(1.20)  are  the  forms
defined in Eq.~(1.11), (1.12), (1.13), (1.14) etc. We call this formalism
a linear (affine, metric, Riemannian, Einstein) connection on $M$.

In our theory it is necessary to consider at least four principal fibre
bundles. A principal fibre bundle $\un{P}$ over $E$ with, a structural group $G$ 
(a gauge group),  connection $\omega $  and  horizontality  operator  ``hor",  a
principal fibre bundle $P'$ of frames  over $(E,g )$  with  the  connection
$\widetilde{\omega }^{\alpha }{}_{\beta }X^{\beta }{}_{\alpha } =
\ol{\cal W}$, a  structural  group $\Gl(4,{\sym R})(O(1,3))$  an   operator   of
horizontality ``$\ol{\hor}$'', a principal fibre bundle $P$" of frames  over
$(\un{P},\gamma )$ 
(a   metrized   fibre   bundle $\un{P})$    with    a    structural    group
$\Gl(4+n,{\sym R})(O(n+3,1))$, a connection $\widetilde{\omega }^{\rA}{}_{\rB}
X^{\rB}{}_{\rA} = \widetilde{\omega }$ and   an   operator   of
horizontality ``$\ol{\hor}$'' and a principal fibre bundle of frames $P'''$ over $G$
with  a  projection ${\varPi}'''$,  operator  of  horizontality  
$(\hor)'''$, a
connection $\omega$ and the structural group $\Gl(n,{\sym R})$.  Moreover  in  order  to
simplify considerations we are using  in  the  work  the  formalism  of
linear  connection  coefficients  on  manifolds $(E,g)$, $(\un{P},\gamma )$  and   a
principal fibre bundle formalism for $\un{P}$ i.e.\ (a principal  fibre  bundle
over $E$ with the structural group $G$ a gauge group). In  the  meaning  of
the author of this work this is  a  way  to  make  the  formalism  more
natural and readable (see Fig.~1B).
\vfill\eject

\null
\vskip36pt plus4pt minus4pt
\centerline{\trz 2. The Nonsymmetric Tensor on a Lie Group}
\vskip24pt plus2pt minus2pt
\write0{2. The Nonsymmetric Tensor on a Lie Group \the\pageno}

Let $G$ be a Lie group and  let  us  define  on $G$ a  tensor  field
$h=h_{ab}v^{a}\otimes v^{b}$ and a field of a 2-form $k=k_{ab}v^{a}\wedge v^{b}$, where
$$
dv^{a} = -\frac12 C^{a}{}_{bc}v^{b}\wedge v^{c},\eqno(2.1)
$$
$v^{a}$ is a usual left-invariant frame on $G$, $C^{a}{}_{bc}$ are  structure  constants.
Let $X_{a}$ be generators of a Lie algebra $G-\fg$, $X_{a}$  are  left-invariant
vector fields on $G$ and they are dual to the forms $v^{a}$.
$$
[X_{a},X_{b}] = C^{c}{}_{ab}X_{c}.\eqno(2.2)
$$

Using $h$ and $k$ we construct a tensor field on $G$
$$
\ell _{ab} = h_{ab} + \mu k_{ab},\eqno(2.3)
$$
where $\mu $ is a real number. Let us remind that the left-invariant  vector
fields on $G$ are infinitesimal transformations of a right action of $G$ on
$G$. The symbol $\Ad_{G}(g )$ means a matrix of the adjoint representation of the
group $G$. Shortly we denote it $\Ad g $. $R$-means a right action of the  group
$G$ on $G$, $L$, a left action $(R(g ),L(g )$, $g \in G)$.

We are looking for the following $h$ and $k$
$$\eqalignno{
R^{*}(g )h &= h,&(2.4)\cr
R^{*}(g )k &= k,&(2.5)\cr}
$$
or in terms of the tensor $\ell _{ab}$
$$
R^{*}(g )\ell  = \ell .\eqno(2.6)
$$
The condition (2.5) can be rewritten
$$
(R^{*}(g )) k_{g _{1}}(X_{g _{1}},Y_{g _{1}}) = 
k_{g _1g }(X_{g _{1}}g ',Y_{g _{1}}g ') = k_{g _{1}}(X_{g _{1}},
Y_{g _{1}}),\eqno(2.5\text{\rm a})
$$
where $g ,g _{1} \in  G$.

Moreover $X$, $Y$ are left-invariant vector fields on $G$. Thus
$X_{g } = X_{\varepsilon }, = X$, $Y_{g } = Y_{\varepsilon } = Y$ and
$$
(R^{*}(g )) k_{g _{1}}(X,Y) = k_{g _1g }(Xg ',Yg ') = k_{g _{1}}
(X,Y),\eqno(2.5\text{\rm b})
$$
where $\varepsilon  \in  G$, is a unit element of $G$.

In order to find $h$ and $k$ satisfying (2.4) and (2.5)  we  define  a
linear connection on $G$ such that
$$
\widehat{\widetilde{\omega }}^{a}{}_{b} = - C^{a}{}_{bc}v^{c}.\eqno(2.7)
$$

Let the covariant differentiation with respect to $\widehat{\omega
}^{a}{}_{b}$ be $\widehat{\nabla }_{c}$ and an
exterior covariant differentiation $\widehat{D}$. 
It  is  easy  to  see  that  this
connection is flat
$$
\widehat{{\varOmega} }^{a}{}_{b} = d \widehat{\omega }^{a}{}_{b} + 
\widehat{\omega }^{a}{}_{c}\wedge \widehat{\omega }^{c}{}_{b} =
0\eqno(2.8)
$$
with non-zero torsion
$$
\widehat{{\varTheta} }^{a} = \widehat{D}v^{a} = dv^{a} + \widehat{\omega
}^{a}{}_{b}\wedge v^{c} =\frac12 C^{a}{}_{bc}v^{b}\wedge
v^{c}\eqno(2.8\text{\rm a})
$$
and with a tensor of torsion
$$
\widehat{Q}^{a}{}_{bc} = C^{a}{}_{bc}.\eqno(2.9)
$$

This connection is also metric. It means that  the  Killing--Cartan
tensor on the group $G$ is absolutely parallel with  respect  to
$\widehat{\omega }^{a}{}_{b}$. A
parallel transport according to this connection is a  right  action  of
the group $G$ on $G$.

One can easily find that (2.4), (2.5), (2.6) are equivalent to the
condition
$$
\widehat{\nabla }_{c}\ell _{ab} = 0.\eqno(2.10)
$$

Thus in order to find $h$ and $k$ we should solve  the  Eqs.~(2.10)  on
the group $G$. The tensor $\ell_{ab}$ satisfies an equation equivalent to
(2.10)
$$
X_{f}\ell _{cd} + \ell _{nd}C^{n}_{cf} + \ell _{cn}C^{n}_{df} = 0.
\eqno(2.11)
$$

It is easy to see that a bi-invariant  tensor $h$  on $G$  satisfies
(2.11) identically (for example a Killing--Cartan tensor).

Thus one gets for a tensor $k_{ab}$
$$
\widehat{\nabla }_{c}k_{ab} = X_{c}k_{ab} + k_{nb}C^{n}_{ac} + 
k_{an}C^{n}_{bc} = 0.\eqno(2.12)
$$

It is easy to see that if $k_{ab}$ satisfies (2.12) $b\cdot k_{ab}$  satisfies
this condition as well for $b=\const$.

In the case of an abelian group $k$ is bi-invariant on $G$.

The interesting case in our theory is a  semisimple  group  $G$.  In
this case $k_{ab}$ can not be bi-invariant. The only bi-invariant 2-form  on
the semisimple Lie group $G$ is a zero form. Moreover the  Eq.~(2.12)  has
always a solution on a  semisimple  group  and $k$  is  right-invariant.
Moreover we suppose that the symmetric part of $\ell $ is bi-invariant  (left
and right-invariant) and $k$ only right-invariant.

We can also define $k$ in a special way
$$
k(A,B) = h([A,B],V) ,\quad\  A=A^{a}X_{a},\ B=B^{a}X_{a},\eqno(2.13)
$$
where
$$
\widehat{\nabla }_{c}V_{d} = 0,\eqno(2.14)
$$
$V = V_{d}\otimes v^{d}$ is a covector field on $G$ (it is right-invariant) and $h$  is  a
Killing--Cartan tensor on $G$.

In order to be more familiar with the notion of a tensor $k$ we find
it for the group SO(3). In this  case  we  have  left-invariant  vector
fields
$$\eqalignno{
e_{1} &= \cos \psi \frac{\partial}{\partial\theta } - \sin \psi  
\biggl(\cos \theta \frac{\partial}{\partial \phi}  - \frac{1}{\sin
\theta }\frac{\partial }{\partial \phi}\biggr),\cr
\noalign{\vskip-1pt}
e_{2} &= \sin \psi \frac{\partial}{\partial\theta } + \cos \psi 
\biggl(\cot \theta \frac{\partial}{\partial\psi } - \frac{1}{\sin
\theta }\frac{\partial}{\partial\phi }\biggr),&(2.15)\cr
\noalign{\vskip-1pt}
e_{3} &= \frac{\partial}{\partial\psi },\cr
\noalign{\vskip-2pt}}
$$
such that 
$$ 
[e_{a},e_{b}] = - \varepsilon _{abc}e_{c} ;\quad\
a,b,c=1,2,3,\eqno(2.16)
$$
$\theta $, $\phi $, $\psi $  are Euler angles---usual parametrization of SO(3)
$$\eqalignno{
0 &\le  \theta  \le  \pi ,\cr
0 &\le  \psi  \le  2\pi ,&(2.17)\cr
0 &\le  \phi  \le  2\pi \cr}
$$
and $\varepsilon _{123} = 1$  and $\varepsilon _{abc}$ is a Levi--Civita 
symbol (see~[62])

In this case one can easily integrate (2.14) and finds
$$\eqalign{
V_{1}(\theta ,\phi ,\psi ) ={}& a(\cos \phi \cos \psi -\cos \theta
\sin \phi \sin \psi )+b\sin \psi \sin \theta\cr
&-c(\sin \phi \cos \psi +\cos \theta \cos \phi \sin \psi ),\cr
V_{2}(\theta ,\phi ,\psi ) ={}& a(\sin \psi \cos \phi +\cos \theta
\cos \psi \sin \phi )-b\cos \psi \sin \theta\cr
&+c(\cos \theta \cos \phi \cos \psi -\sin \phi \sin \psi ),\cr
V_{3}(\theta ,\phi ,\psi ) ={}& a\sin \phi \sin \theta +b\cos \theta
+c\sin \phi \sin \theta ,\quad\ a,b,c=\const.\cr}\eqno(2.18)
$$
In a simpler case $a = c = 0$, $b \neq  0$, one gets: 
$$\eqalignno{
V_{1} &= b\sin \theta \sin \psi ,\cr
V_{2} &= -b\sin \theta \cos \psi,&(2.19)\cr
V_{3} &= b\cos \theta ,\quad\ b=\const.\cr}
$$
For 
$$ 
k_{ab} = \varepsilon _{abc}V_{c}\eqno(2.20)
$$
we get 
$$\displaylines{
k_{ab}=\hfill\hbox{\tenrm (2.21)}\cr
\noalign{\vskip1pt}
{\seven \pmatrix &(a\sin \psi\sin \theta +b\cos \theta+ & -[a(\sin
\psi\cos \phi+\cos \theta \cos \psi\sin \phi)+\cr
0 & +c\sin \phi\sin \theta ) & -b\cos \psi\sin \theta \cr
& & +c(\cos \theta \cos \phi\cos \psi - \sin \phi\sin \psi)]\cr
\noalign{\vskip1pt}
\noalign{\vskip4pt}
-(a\sin \phi\sin \theta +b\cos \theta + & & [a(\alpha \phi \cos \psi
- \cos \theta \sin \phi \sin \psi)+\cr
+c\sin \phi\sin \theta ) & 0 & +b\sin \psi\sin \theta +\cr
& & -c(\sin \phi\cos \psi+\cos \theta \cos \phi\sin \psi)]\cr
a(\sin \psi\cos \phi+\cos \theta \cos \psi\sin \phi)+ & -[a(\cos
\phi\cos \psi - \cos \theta \sin \phi\sin \psi)+ & \cr
-b\cos \psi\sin \theta + & +b \sin \psi\sin \theta +& 0\cr
+c(\cos \theta \cos \phi\cos \psi - \sin \phi\sin \psi) & -c(\sin
\phi\cos \psi+\cos \theta \cos \phi\sin \psi)] &\cr \endpmatrix}\!,\cr}
$$
In a simpler case for $a = c = 0$, $b \neq  0$ one gets: 
$$
k_{ab} = \pmatrix 0 &,& b\cos \theta  &,& -b\sin \theta \cos \psi \cr
-b\cos \theta  &,& 0 &,& b\sin \theta \cos \psi \cr
b\sin \theta \cos \psi &,& -b\sin \theta \sin \psi &,&
0\cr \endpmatrix \eqno(2.22)
$$
Thus if we choose for $h$ a Killing--Cartan tensor on  SO(3)  (this  is  a
unique bi-invariant tensor on SO(3) {\it modulo}\/ constant factor)
$$
h_{ab} = -2\delta _{ab}\eqno(2.23)
$$
we easily get:
$$\displaylines{
\ell_{ab}=\hfill\cr
{\seven\pmatrix &\mu (a\sin \psi\sin \theta +\beta \cos \theta+ &
-\mu [\alpha (\sin\psi\cos \phi+\cos \theta \cos \psi\sin \phi)+\cr
-2 & +\gamma\sin \phi\sin \theta ) & -\beta \cos \psi\sin \theta+ \cr
& & +\gamma (\cos \theta \cos \phi\cos \psi - \sin \phi\sin \psi)]\cr
\noalign{\vskip4pt}
& & \mu [\alpha (\cos \phi \cos \psi- \cos \theta \sin \phi \sin \psi)+\cr
 & -2 & +\beta \sin \psi\sin \theta +\cr
& & -\gamma (\sin \phi\cos \psi+\cos \theta \cos \phi\sin \psi)]\cr
\noalign{\vskip4pt}
\mu (\sin \psi\cos \phi+\cos \theta \cos \psi\sin \phi)+ & -\mu
[\alpha (\cos\phi\cos \psi - \cos \theta \sin \phi\sin \psi)+ & \cr
-\beta \cos \psi\sin \theta + & +\beta  \sin \psi\sin \theta +&-2\cr
+\gamma (\cos \theta \cos \phi\cos \psi - \sin \phi\sin \psi) &
-\gamma (\sin \phi\cos \psi+\cos \theta \cos \phi\sin \psi)] &\cr\endpmatrix}\cr
\hfill \hbox{\tenrm (2.24)}\cr}
$$
where $\mu  = \eta \sqrt{a^2+b^2+c^2}$, $\eta ^{2} = 1$, $\alpha  =
{\displaystyle\frac{a}{\mu }}$, $\beta  = {\displaystyle\frac{b}{\mu }}$, $\gamma = {\displaystyle\frac{c}{\mu }}$.

In a simpler case i.e. for $a=c=0$, $b\neq 0$ one gets (absorbing $\beta $ by $\mu )$:
$$
\ell_{ab} = \pmatrix -2 &,& \mu \cos \theta  &,& -\mu \sin \theta \cos \psi \cr
-\mu \cos \theta  &,& -2 &,& \mu \sin \theta \cos \psi \cr
\mu \sin \theta \cos \psi &,& -\mu \sin \theta \sin \psi
&,&-2\cr \endpmatrix\eqno(2.24\text{\rm a})
$$
For an inverse tensor $\ell ^{ab}$ such that
$$ 
\ell ^{ab}\ell _{ac} = \ell ^{ba}\ell _{ca} = \delta ^{b}_{c}
\eqno(2.25)
$$
 we have 
$$ 
\ell ^{ab} = \frac{{\varDelta}^{ab}}{{\varDelta}}.\eqno(2.26)
$$
where  ${\varDelta}= \det (\ell _{ab}) = -2(4+\mu ^{2})$
${\varDelta}^{ab}$ is a factor matrix and 
$$\eqalignno{
{\varDelta}^{11}&= 4+\mu ^{2}\sin ^{2}\theta \sin ^{2}\psi ,\cr
{\varDelta}^{12}&= -(2\mu \cos \theta +\mu ^{2}\sin ^{2}\theta \sin \psi
\cos \psi ),\cr
{\varDelta}^{13}&= (\mu ^{2}\cos \theta \sin \theta \sin \psi  -2\mu
\sin \theta \cos \psi ),\cr
{\varDelta}^{21}&= (2\mu \cos \theta  - \mu ^{2}\sin ^{2}\theta \sin \psi
\cos \psi ),\cr
{\varDelta}^{22}&= (4+\mu ^{2}\sin ^{2}\theta \cos
^{2}\psi),&(2.27)\cr
{\varDelta}^{23}&= -(2\mu \sin \theta \sin \psi  + \mu ^{2}\cos \theta
\sin \theta \cos \psi ),\cr
{\varDelta}^{31}&= (\mu ^{2}\cos \theta \sin \theta \sin \psi  + 2\mu
\sin \theta \cos \psi ),\cr
{\varDelta}^{32}&= (2\mu \sin \theta \sin \psi  - \mu ^{2}\cos \theta
\sin \theta \cos \psi ),\cr
{\varDelta}^{33}&= (4+\mu ^{2}\cos ^{2}\theta ),\cr}
$$

In the case  of  SO(3) Eq.~(2.21)  is  the  most  general  tensor
satisfying (2.5) except a constant factor in front. Thus this  tensor
is unique for SO(3) {\it modulo}\/ a constant factor.

In the case of any SO($n$) one can find $k$ and $\ell $  similarly  using
Euler angles parametrization and so for classical groups SU($n$), Sp($2n$),
$G_{2}$, $F_{4}$, $E_{6}$, $E_{7}$, $E_{8}$. In the case of solvable and  nilpotent  groups  we  can
also try to find a bi-invariant skew-symmetric tensors.

Finally we  suggest  a  general  form  of  the  tensor $k_{ab}$  on  a
semi-simple group $G$  i.e.\  such  that  Eq.~(2.4)  is  satisfied.  The
solution of the Eq.~(2.10) and (2.12) are as follows
$$\eqalignno{
\ell _{ab}(e^{c}) &= \ell _{a'b'}(\varepsilon )(e^{\Ad'c})^{a'}_{a}
(e^{\Ad'c})^{b'}_{b}\cr
\noalign{\hbox{and}}
k_{ab}(e^{c}) &= k_{a'b'}(\varepsilon
)(e^{\Ad'c})^{a'}_{a}(e^{\Ad'c})^{b}_{b}.\cr}
$$
One writes
$$
k_{ab}(g ) = f_{a'b'}U^{a'}{}_{a}(g )U^{b'}{}_{b}(g ),\quad\ g \in
G,\eqno(2.28)
$$
where $U(g )=\Ad_{G}(g )$  is an adjoint representation of the group
$G$. It is easy to see that for (2.28) we have
$$\eqalignno{
\widehat{\nabla }_{c}k_{ab} &= 0,&(2.29)\cr
f_{ab} &= -f_{ba} =\const &(2.30)\cr}
$$
 and  it  is  defined  in  the  representation  space  of  the   adjoint
representation of the group $G$. In the case of the group SO(3) one has
$$\eqalignno{
f_{ab} &= \varepsilon _{abc}f_{c},&(2.31)\cr
k_{ab} &= \varepsilon _{abc}V_{c} &(2.31\text{\rm a})\cr}
$$
and
$$
V_{a} = f_{c'}U^{c'}{}_{a}(g ).\eqno(2.32)
$$
 If we choose $f_{c}=(0,0,b)$ we  get  Eq.~(2.19). Moreover  it  is  always
possible because an orthogonal (SO(3)) transformation can transform any
vector $f$ into $(0,0,\pm \| f\| )$, where $\| f\| $ is the length of $f$.

 The semisimple Lie group $G$ can  be  considered  a  Riemannian  manifold
equipped with a bi-invariant tensor $h$ (a Killing--Cartan tensor)  and  a
connection induced by this  tensor.  This  Riemannian  manifold  has  a
constant curvature. Such a manifold has a maximal group  of  isometries
$H$ of a dimension ${\displaystyle\frac12} n(n+1)$, 
$n=\dim G$ (see Ref~[59]). (The  isometry  is  here
understood in a sense of the metric measured along  geodetic  lines  in
a Riemannian geometry induced by a Killing--Cartan tensor.) This group  is
a Lie group. It is easy to see that for $G=\SO (3)$ we  have $H=\SO
(3)\ilp \SO(3)$ 
and  $\dim \SO (3)\ilp \SO (3)=6$,  $\dim \SO (3)=3$.  The   
group  SO(3)   leaves   the
Killing-Cartan tensor $h_{ab}$ invariant 
$$
h_{{a'b'}}A^{a'}{}_{a}A^{b'}{}_{b} = h_{ab},\eqno(2.33)
$$
where $A \in \SO (3)$. 

Moreover $f_{ab}$ has exactly 3 arbitrary parameters and  solutions  of
Eq.~(2.12) have the same freedom in arbitrary  constants.  This  suggests
that the tensor (2.28) could be in some sense unique {\it modulo}\/ an isometry
on SO(3) and a constant factor $b$. In this case  the  classification  of
$k_{ab}$ tensors on a SO(3)  could  be  reduced  to  the  classification  of
skewsymmetric tensors $f_{ab}$ with respect  to  the  action  of  the  group
SO(3). In general the situation is more complex, because SO($n$), $n=\dim G$
does not leave the commutator (Lie bracket) invariant.

Let us suppose that $G$ is compact. In this case we should find  all
inequivalent $f_{ab}$-tensors with respect to an  orthogonal  transformation
$A\in \SO (n)$. It means we should transform $f_{ab}$ to a canonical  form  via  an
orthogonal matrix i.e.
$$
(f_{ab}) = f \rightarrow  f'= (f'_{ab}) = A^{\rT}fA =
A^{-1}fA.\eqno(2.34)
$$
 For skewsymmetric matrices we have the following canonical forms,  the
so-called block-diagonal matrices
For $n = 2m$
$$
f = \left[\matrix \matrix 0&\xi^1\cr -\xi^1&0\cr\endmatrix && \text{\bf 0}\cr
&\ddots&\cr
\text{\bf 0} && \matrix 0&\xi^m\cr -\xi^m&0\cr \endmatrix \cr \endmatrix \right]\eqno(2.35)
$$
or for $n = 2m+1$
$$
f = \left[
\matrix
\matrix 0&\xi^1\cr -\xi^1&0\cr \endmatrix & &\text{\bf
0}\cr &\ddots&\cr
\text{\bf 0}&&
\matrix 0&\xi^m&0\cr -\xi^m&0&0\cr 0&0&0\cr\endmatrix\cr
\endmatrix\right],\eqno(2.36)
$$
 where $\xi ^{1},\xi ^{2}\ldots 
 \xi ^{m}$ are real numbers. In order to  find  them  we  should
solve a secular equation for $f$
$$\displaylines{
\hfill\det (\mu I_{n}-f) = \mu ^{n} + a_{1}(f)\mu ^{n-2} + a_{2}(f)\mu ^{n-2} + \ldots
\hfill\llap{(2.37)}\cr
(I_{n}=(\delta ^{i}_{j})_{i,j=1,2\ldots n}).\cr}
$$
 The coefficients $a_{1},a_{2} \ldots 
$  are invariant with respect to an action  of
the group O($n$) (SO($n$)) and they are functions of $\xi ^{1}, \ldots 
 \xi ^{m}$.  Thus  in
the case of a compact semisimple Lie group the skewsymmetric tensor $k_{ab}$
on $G$ is defined as follows
$$
k_{ab}(g ) = b\cdot
\widetilde{f}_{a'b'}U^{a'}{}_{a}(g )U^{b'}{}_{b}(g ),\eqno(2.38)
$$
 where $b$ is a constant real factor and  $(\widetilde{f}_{ab}) =
\widetilde{f}$ is given by 
$$
\widetilde{f}= A^{\rT} \left[\matr{\matr{0&1&\cr -1&0\cr}&&&\text{\bf 0}\cr
&\matr{0&\xi^1\cr -\xi^1&0\cr}&&&\cr
&&\ddots&\cr
\text{\bf 0}&&&\matr{0&\xi^{m-1}\cr
-\xi^{m-1}&0\cr}\cr}\right]A,\eqno(2.39)
$$
for $n=2m$, or
$$
\widetilde{f}= A^{\rT} \left[\matr{\matr{0&1&\cr -1&0\cr}&&&\text{\bf 0}\cr
&\matr{0&\xi^1\cr -\xi^1&0\cr}&&&\cr
&&\ddots&\cr
\text{\bf 0}&&&\matr{0&\xi^{m-1}\cr -\xi^{m-1}&0\cr}\cr}\right]A,\eqno(2.39\text{\rm a})
$$
for $n=2m+1$.

 Supposing that $h_{ab} = \diag(\lambda _{1},\lambda _{2},\ldots 
,\lambda _{n})$ where $n=2m$ or $n=2m+1$ one gets
$$
\ell _{ab}(\varepsilon ) = A^{\rT}
\left[\matr{\matr{\lambda _1&\zeta^1\cr \zeta^1&\lambda _2\cr}
&&&\text{\bf 0}\cr
&\matr{\lambda _3&\xi^2\cr -\xi^2&\lambda _4\cr}& & \cr
&&\ddots&&&\cr
\text{\bf 0}&&&\matr{\lambda _{2m-1}&\xi^m\cr -\zeta^m&\lambda
_{2m}\cr}\cr}\right]A=A^{\rT}\widetilde{\ell}(\varepsilon)A\eqno(2.39*)
$$
 for $n = 2m$ or
$$
\ell _{ab}(\varepsilon ) = A^{\rT}
\left[\matr{\matr{\lambda _1&\zeta^1\cr \zeta^1&\lambda _2\cr}&&&\text{\bf 0}\cr
&\matr{\lambda _3&\xi^2\cr -\xi^2&\lambda _4\cr}&&\cr
&&\ddots&\cr
\text{\bf 0}&&\matr{\lambda _{2m-1}&\xi^m\cr -\zeta^m&\lambda
_{2m}\cr}&\cr\noalign{\vskip2pt}
&&&\lambda
_{2m+1}\cr}\right]A=A^{\rT}\widetilde{\ell}(\varepsilon)A\eqno(2.39\text{\rm a}*)
$$
for $n = 2m+1$.

Moreover if $G$ is compact, we have $\lambda _{i}= \lambda $, $i = 1,2\ldots 
n$ and $\lambda <0$.  This  is
because  any bi-invariant  symmetric  tensor  is  proportional  to  the
Killing--Cartan tensor. In particular the Tr-tensor  commonly  used  in
Yang--Mills' theory is proportional to $h_{ab}$. Thus $h_{ab} = \lambda (Tr)_{ab}= \lambda \delta
_{ab}$, $\lambda  <0$. (For a particular normalization of generators
$\Tr(\{X_{a},X_{b}\})=2\delta _{ab})\}$.
Let us remark that, in general if $k_{ab}(\varepsilon )$  and $h_{ab}$  commute  (in  this
moment I do not suppose that $G$ is compact) we have $\ell _{ab}(\varepsilon ) =
(A^{-1}\widetilde{\ell}(\varepsilon )A)_{ab}$ where $A \in  \Gl(n,{\sym R})$ 
and $\ell _{ab}(g ) =
U^{a'}{}_{a}(g )U^{b'}{}_{b}(g )(A^{-1}\widetilde{\ell}(\varepsilon
)A)_{a'b'}$.

One can say, of course, that $k_{ab}$, tensors are defined with more
arbitrarility than bi-invariant, symmetric tensors. This is because $k$ is
only right-invariant.

Let us notice that\vadjust{\vskip-6pt}
$$
f_{ab} = k_{ab}(\varepsilon )\vadjust{\vskip-6pt}\eqno(2.40)
$$
$(\varepsilon $ is a unit element of $G)$  and\vadjust{\vskip-6pt}
$$
R_{g },k_{ab}(g ) = k_{ab}(g g ') =
k_{cd}(g )U^{c}{}_{a}(g ')U^{d}{}_{b}(g '),\vadjust{\vskip-6pt}\eqno(2.41)
$$
 where  $g ,g ' \in G$.

In the case of $G=\SO (3)$ $k_{ab}$ is unique  up  to  a  isometry  of  the
Riemannian manifold with the bi-invariant tensor as a metric tensor and
a constant factor. This suggests that the $k_{ab}$ tensor given in the  form
$(2.13\div 14)$ and (2.31--32) is an analogue of the Killing--Cartan tensor
for $k_{ab}$ (skewsymmetric). Moreover the vector $f$ can be transformed by an
orthogonal ($O(n)$) transformation into
$$
\mathop{(0,0,\ldots,\pm \|f\|)}\limits_{n\ \text{\rm times}}.\eqno(2.42)
$$
 Thus one gets
$$
k_{ab}(g ) = b\cdot
C^{c}{}_{ab}f^{0}{}_{c'}U^{c'}{}_{c}(g ),\eqno(2.43)
$$
 where $b$ is a constant factor and
$$
(f^{0}_{c'}) = f^{0} = \mathop{\underline{(0,0,\ldots,1)}}\limits_{n\ \text{\rm
times}}.\eqno(2.44)
$$
 Thus we can write $k$ in a more compact form
$$
k(A,B)(g ) = b\cdot h([A,B],\Ad_{g }f^{0}),\eqno(2.45)
$$
where $A=A^{a}X_{a}$, $B=B^{a}X_{a}$

 Using a bi-invariancy of the Killing--Cartan tensor one can write
$$
k(A,B)(g ) = b\cdot h(\Ad_{g ^{-1}}[A,B],f^{0}).\eqno(2.45\text{\rm a})
$$
 Moreover if there is $\widetilde{g }\in G$ such that
$\widetilde{g }^{2}=g $ we get
$$
k(A,B)(g ^{2}) = b\cdot
h(\Ad_{\tilde{g }^{-1}}[A,B],\Ad_{\tilde{g }} f^{0}).\eqno(2.46)
$$

 We find an  interpretation  of  a  factor $b$  for $k$  given  by  formulae
(2.45--46).

One gets
$$
k_{ab}k^{ab} = h^{aa'}h^{bb'}k_{ab}k_{a'b'} = b^{2}\| \Ad_{g }f^{O}\| ^{2} 
= b^{2}.\eqno(2.47)
$$
 Thus we have
$$
b = \pm \sqrt{k_{ab}k^{ab}}. \eqno(2.48)
$$
 Finally let us notice that we can repeat  the  considerations  changing
right(left)-invariant to left(right)-invariant in all  the  places.  In
this case we can consider left-invariant 2-form $k$  and  left-invariant
nonsymmetric tensor on a Lie group~$G$.

In general we can simply say that our right invariant tensor on a group $G$
can be constructed as follows. We define a skew-symmetric tensor at a unit
element of a group and afterwards propagate it on~$G$ via group translations
(if the group is simply connected). If it is not simply connected, we define
it for a universal covering group or only for simply connected part
containing a unit element.
\vfill\eject

\null
\vskip36pt plus4pt minus4pt
\centerline{{\four\bf 3. The Nonsymmetric Metrization of the Bundle $\un P$}}
\vskip24pt plus2pt minus2pt
\write0{3. The Nonsymmetric Metrization of the Bundle $\underline{P}$ \the\pageno}

Let us introduce the principal fibre bundle $\un{P}$ over the space-time
$E$ with the structural group $G$ and with the projection $\pi $. Let us suppose
that $(E,g )$ is a manifold with a nonsymmetric metric tensor of the
signature $(-,-,-,+)$
$$
g _{\mu \nu } = g _{(\mu \nu )} + g _{[\mu \nu ]}\eqno(3.1)
$$
Let us introduce a natural frame on $\un{P}$
$$
\theta ^{\rA} = (\pi ^{*}(\overline{\theta }^\alpha ),\ \theta ^{a}=\lambda \omega
^{a}),\quad\ \lambda = \const.\eqno(3.2)
$$
It is convenient to introduce the following notations. Capital Latin
indices $\rA,\rB,\rC$ run $1,2,3,\ldots 
n+4$, $n=\dim G$. Lower Greek indices $\alpha ,\beta ,\gamma ,\delta =
1,2,3,4$ and lower Latin cases $\text a,\text b,\text c,\text d = 5,6,\ldots 
n+4$. The symbol ``$-$" over
$\theta ^{\alpha }$ and over other quantities indicates that these quantities are
defined on $E$.

It is easy to see that the existence of the nonsymmetric metric on
$E$ is equivalent to introducing two independent geometrical quantities
on $E$.
$$\eqalignno{
\ov{g } &= g _{\alpha \beta }\overline{\theta }^\alpha \otimes \overline{\theta
}^{\beta } = g _{(\alpha \beta )}\overline{\theta }^\alpha \otimes
\overline{\theta }^{\beta },&(3.3)\cr
\un{g } &= g _{\alpha \beta }\overline{\theta }^{\alpha }\wedge
\overline{\theta }^{\beta } = g _{[\alpha \beta ]}\overline{\theta }^{\alpha }\wedge
\overline{\theta }^{\beta },&(3.4)\cr}
$$
i.e. the symmetric metric tensor $\ov{g }$ on $E$ and 2-form
$\un{g }$. On the group $G$ we
can introduce a bi-invariant symmetric tensor called the Killing--Cartan
tensor.
$$
h(A,B) = \Tr (\Ad'_{\rA}\circ \Ad'_{\rB}),\eqno(3.5)
$$
where $\Ad_{\rA}(C) = [A,C]$ (it is tangent to Ad i.e.\ it is an
``infinitesimal" Ad-transforma\-tion). It is easy to see that
$$
h(A,B) = h_{ab}A^{a}\cdot B^{b},\eqno(3.6)
$$
where
$$
h_{ab} = C^{c}{}_{ad}C^{d}{}_{bc} ,\quad h_{ab} = h_{ba} ,\quad A = A^{a}X_{a}
,\quad B = B^{a}X_{a}.
$$
This tensor is distinguished by the group structure, but there are of
course other bi-invariant tensors on G. Normally it is supposed that $G$
is semisimple. It means that $\det (h_{ab}) \neq 0$. In this construction we use
$\ell _{(ab)}= h_{ab}$ (the bi-invariant tensor on $G)$ in order to get a proper
limit (i.e.\ the Nonabelian Kaluza--Klein Theory) for $\mu =0$.

For a natural 2-form $k$ on $G$, or a natural skewsymmetric right-invariant
tensor we choose $k$ described in sec.~2, $k$ is zero for ${\bold U}(1)$. Let us
turn to the nonsymmetric natural metrization of $\un{P}$. Let us suppose that:
$$\eqalignno{
\overline{\gamma }(X,Y) &=\ov{g }(\pi 'X,\pi 'Y) + \lambda ^{2}\rho ^{2}h(\omega (X),\omega
(Y)),&(3.7)\cr
\underline{\gamma }(X,Y) &=\un{g }(\pi 'X,\pi 'Y) + \mu \lambda ^{2}\rho ^{2}
k(\omega (X),\omega (Y)),&(3.8)\cr}
$$
$\mu $ = const and is dimensionless, $X, Y \in \Tan(\un{P})$ and $\rho = \rho (x)$ is a
scalar field defined on $E$. The first formula (2.9) was introduced by
A.~Trautman (in the case with $\rho = 1)$ for the symmetric natural
metrization of $\un{P}$ and it was used to construct the Kaluza--Klein Theory
for ${\bold U}(1)$ and nonabelian generalizations of this theory. It is easy to
see that
$$\eqalignno{
\overline{\gamma }&= \pi ^{*}\ov{g }\oplus \rho ^{2}h_{ab}\theta ^{a}\otimes \theta
^{b},&(3.9)\cr
\underline{\gamma }&= \pi ^{*}\un{g }+ \mu \rho ^{2}k_{ab}\theta ^{a}\wedge \theta
^{b},&(3.10)\cr}
$$
or
$$\eqalignno{
\gamma _{(AB)}&= \left(\vcenter{\offinterlineskip\tabskip0pt
\halign{\strut\tabskip5pt plus10pt minus10pt
\hfill#\hfill&\vrule height10pt#&\hfill#\hfill\tabskip0pt\cr
$\vrule height0pt depth6pt width0pt g _{(\alpha \beta)}$ && $0$\cr 
\noalign{\hrule}
$\vrule height9pt depth4pt width0pt 0$ && $\rho^2h_{ab}$\cr}}\right),&(3.11)\cr
\gamma _{[AB]}&=\left(\vcenter{\offinterlineskip\tabskip0pt
\halign{\strut\tabskip5pt plus10pt minus10pt
\hfill#\hfill&\vrule height10pt#&\hfill#\hfill\tabskip0pt\cr
$\vrule height0pt depth6pt width0pt g_{[\alpha \beta] }$ && $0$\cr 
\noalign{\hrule}
$\vrule height9pt depth4pt width0pt 0$ && $\mu \rho^2k_{ab}$\cr}}\right).&(3.12)\cr}
$$
For
$$
\gamma _{\rAB} = \gamma _{(\rAB)} + \gamma _{[\rAB]}
$$
one easily gets
$$
\gamma _{\rAB} =\left(\vcenter{\offinterlineskip\tabskip0pt
\halign{\strut\tabskip5pt plus10pt minus10pt
\hfill#\hfill&\vrule height10pt#&\hfill#\hfill\tabskip0pt\cr
$\vrule height0pt depth6pt width0pt g _{\alpha \beta }$ && $0$\cr 
\noalign{\hrule}
$\vrule height9pt depth4pt width0pt 0$ && $\rho^2\ell_{ab}$\cr}}\right),\eqno(3.13)
$$
where $\ell _{ab} = h_{ab} + \mu k_{ab}$. Tensor $\gamma _{AB}$ has this simple form in the natural
frame on $\un{P}, \theta ^{A}$. This frame is unholonomical, because:
$$
d\theta ^{a} =\frac{\lambda }{2}\biggl(H^{a}{}_{\mu \nu }\theta ^{\mu }
\wedge \theta ^{\nu } -\frac{1}{\lambda ^2}C^{a}{}_{bc}\theta ^{b}\wedge 
\theta ^{c}\biggr) \neq 0\eqno(3.14)
$$
$\gamma $ is invariant with respect to the right action of the group on 
$\un{P}$. In the case with $k_{ab} = 0$ we have
$$
\gamma _{\rAB}=\left(\vcenter{\offinterlineskip\tabskip0pt
\halign{\strut\tabskip5pt plus10pt minus10pt
\hfill#\hfill&\vrule height10pt#&\hfill#\hfill\tabskip0pt\cr
$\vrule height0pt depth6pt width0pt g _{\alpha \beta }$ && $0$\cr 
\noalign{\hrule}
$\vrule height9pt depth4pt width0pt 0$ && $\rho^2h_{ab}$\cr}}\right).\eqno(3.15)
$$
For the electromagnetic case $(G = {\bold U}(1))$ one easily finds
$$
\gamma _{\rAB}=\left(\vcenter{\offinterlineskip\tabskip0pt
\halign{\strut\tabskip5pt plus10pt minus10pt
\hfill#\hfill&\vrule height10pt#&\hfill#\hfill\tabskip0pt\cr
$\vrule height0pt depth6pt width0pt g _{\alpha \beta }$ && $0$\cr 
\noalign{\hrule}
$\vrule height9pt depth4pt width0pt 0$ && $-\rho^2$\cr}}\right).\eqno(3.16)
$$
Now let us take a section $e: E \rightarrow \un{P}$ and attach to it a frame $\upsilon ^{a}$,
$a = 5,6\ldots 
n+4$, selecting $X^{\mu } = \const$ on a fibre in such a way that $e$ is
given by the condition $e^{*}\upsilon ^{a} = 0$ and the fundamental fields $\zeta _{a}$ such
that $\upsilon ^{a}(\zeta _{b})=\delta ^{a}{}_{b}$ satisfy $[\zeta _{b},
\zeta _{a}]=\frac{1}{\lambda }C^{c}{}_{ab}\zeta _{c}$.

Thus we have
$$
\omega =\frac{1}{\lambda }\upsilon ^{a}X_{a} + \pi ^{*}(A^{a}{}_{\mu }
\overline{\theta }^\mu )X_{a},
$$
where
$$
e^{*}\omega = A = A^{a}{}_{\mu }\overline{\theta }^{\mu }X_{a}.
$$
In this frame the tensor $\gamma $ takes the form
$$
\gamma _{\rAB} = \left(\vcenter{\offinterlineskip\tabskip0pt
\halign{\strut\tabskip5pt plus10pt minus10pt
\hfill#\hfill&\vrule height10pt#&\hfill#\hfill\tabskip0pt\cr
$\vrule height0pt depth6pt width0pt g _{\alpha \beta } +\lambda ^2\rho^2\ell_{ab}A^a{}_\alpha
A^b{}_\beta$ && $\lambda \rho^2\ell_{cb}A^c{}_\alpha $\cr 
\noalign{\hrule}
$\vrule height9pt depth4pt width0pt \lambda \rho^2\ell_{cb}A^c{}_\beta $ &&
$\rho^2\ell_{ab}$\cr}}\right),\eqno(3.17)
$$
where
$$
\ell _{ab} = h_{ab} + \mu k_{ab}.
$$
This frame is also unholonomic. One easily finds
$$
d\upsilon ^{a} =\frac{-1}{2\lambda } C^{a}{}_{bc}\upsilon ^{b}\wedge \upsilon
^{c}.\eqno(3.18)
$$
The nonsymmetric theory of gravitation uses the nonsymmetric metric
$g _{\mu \nu }$ such that
$$
g _{\mu \nu }g ^{\beta \nu } = g _{\nu \mu }g ^{\nu \beta } = 
\delta ^{\beta }_{\mu },\eqno(3.19)
$$
where the order of indices is important. If $G$ is semisimple and $k_{ab} = 0$
$$
\ell _{ab} = h_{ab} ,\quad\ \det (h_{ab}) \neq 0
$$
and
$$
h_{ab}h^{bc} = \delta ^{b}_{a}.\eqno(3.20)
$$
Thus one easily finds in this case:
$$
\gamma _{\rAC}\gamma ^{\rBC} = \gamma _{\rCA}\gamma ^{\rCB} = \delta
^{\rB}{}_{\rA}.\eqno(3.21)
$$
where the order of indices is important. We have the same for the
electromagnetic case $(G = {\bold U}(1))$. In general if $\det (\ell _{ab}) \neq 0$ then
$$
\ell _{ab}\ell ^{ac} = \ell _{ba}\ell ^{ca} = \delta
^{c}_{b},\eqno(3.22)
$$
where the order of indices is important. From (3.22) we have (3.21)
for the general nonsymmetric metric $\gamma $.

It is easy to see that
$$\eqalign{{\varPhi} '(g )\overline{\gamma }&= \overline{\gamma },\cr
{\varPhi} '(g )\underline{\gamma }&= \underline{\gamma }\cr}\eqno(3.23)
$$
and $\gamma _{\rAB}$ is invariant tensor with respect to the right-action of the
group $G$ on $\un{P}$ (this means it is covariant and right-invariant).

In the case of any abelian group the condition (3.23) is
stronger and we get that $\gamma _{\rAB}$ is bi-invariant. Thus in the case of
$G={\bold U}(1)$ (electromagnetic case)
$$
\xi _{5}\overline{\gamma }= 0 = \xi _{5}\underline{\gamma },\eqno(3.24)
$$
where $\xi _{\rA}$ is a dual base
$$
\theta ^{\rA}(\xi _{\rB}) = \delta ^{\rA}{}_{\rB},\eqno(3.25)
$$
$A,B = 1,2,3,4,5$
$$
\xi _{\rA} = (\xi _{\alpha },\xi _{5}).\eqno(3.26)
$$

Let us come back to the connection $\widehat{\omega }^{a}{}_{b}$ defined on the group
$G$. For
a typical fibre is diffeomorphic to $G$ we can define $\widehat{\omega
}^{a}{}_{b}$ on every fibre
$F_{x}\simeq G$, $x \in E$. Due to a local trivialization of the bundle $P$ we can
define $\widehat{\omega }^{a}{}_{b}$ on every set $U\times G$, where 
$U \subset E$ and is open. Thus we get a
linear connection on $P$ such that
$$
\widehat{\omega }^{\rA}_{\rB} = \left(\vcenter{\offinterlineskip\tabskip0pt
\halign{\strut\tabskip5pt plus10pt minus10pt
\hfill#\hfill&\vrule height10pt#&\hfill#\hfill\tabskip0pt\cr
$\vrule height0pt depth6pt width0pt 0$ && $0$\cr 
\noalign{\hrule}
$\vrule height9pt depth4pt width0pt 0$ && $-\frac{1}{\lambda }c^a{}_{bc}\theta
^c$\cr}}\right)\eqno(3.27)
$$
defined in a frame $\theta ^{\rA} = (\pi ^{*}(\overline{\theta }^{\alpha }),\theta ^{a})$ 
where $\overline{\theta }^{\alpha }$ is a frame on $E$ and $\theta ^{a}$
is a horizontal lift base.

This connection can be examined in a systematic way. Let us
introduce a metric on $\un P$ in a following way
$$
p = \pi ^{*}\eta \oplus h_{ab}\theta ^{a}\otimes \theta ^{b},
\eqno(3.28)
$$
where $\eta =\eta _{\mu \nu }\overline{\theta }^{\mu }\otimes \overline{\theta
}^{\nu }$ is a Minkowski tensor and $h_{ab}$ is a Killing--Cartan
tensor on $G$. One gets
$$
p_{\rAB} = \left(\vcenter{\offinterlineskip\tabskip0pt
\halign{\strut\tabskip5pt plus10pt minus10pt
\hfill#\hfill&\vrule height10pt#&\hfill#\hfill\tabskip0pt\cr
$\vrule height0pt depth6pt width0pt \eta_{\alpha \beta }$ && $0$\cr 
\noalign{\hrule}
$\vrule height9pt depth4pt width0pt 0$ && $h_{ab}$\cr}}\right)\ \hbox{and}\ p^{\rAB}=
\left(\vcenter{\offinterlineskip\tabskip0pt
\halign{\strut\tabskip5pt plus10pt minus10pt
\hfill#\hfill&\vrule height10pt#&\hfill#\hfill\tabskip0pt\cr
$\vrule height0pt depth6pt width0pt \eta^{\alpha \beta }$ && $0$\cr 
\noalign{\hrule}
$\vrule height9pt depth4pt width0pt 0$ && $h^{ab}$\cr}}\right).\eqno(3.29)
$$
The connection $\widehat{\omega }^{\rA}{}_{\rB}$ can be defined as
$$
\widehat{\omega }^{\rA}{}_{\rB} = \left(\vcenter{\offinterlineskip\tabskip0pt
\halign{\strut\tabskip5pt plus10pt minus10pt
\hfill#\hfill&\vrule height10pt#&\hfill#\hfill\tabskip0pt\cr
$\vrule height0pt depth6pt width0pt \pi^*\widehat{\overline{\omega}}^\alpha {}_\beta $ && $0$\cr 
\noalign{\hrule}
$\vrule height9pt depth4pt width0pt 0$ && $\phi_x^0\widehat{\omega}^a{}_b$\cr}}\right),\eqno(3.27\text{\rm a})
$$
where $\widehat{\overline{\omega}}^\alpha {}_\beta $ is a trivial connection on the Minkowski space, $\widehat{\omega
}^{a}{}_{b}$ is the
connection defined in sec.~2 and $\phi _{x}$ is a diffeomorphism
$\phi _{x}:F_{x}\rightarrow G$, $x\in U$.

It is easy to check that
$$
\widehat{D}p_{\rAB} = 0 = \widehat{D}p^{\rAB},\eqno(3.30)
$$
where $\widehat{D}$ is an exterior covariant differential with respect to $\widehat{\omega
}^{\rA}{}_{\rB}$. One
can easily calculate the torsion for $\widehat{\omega }^{\rA}{}_{\rB}$
$$\eqalignno{
\widehat{Q}^{a}{}_{\mu \nu } &= \lambda H^{a}{}_{\mu \nu },&(3.31\text{\rm
a})\cr
\widehat{Q}^{a}{}_{bc} &= \frac{1}{\lambda }C^{a}{}_{bc}&(3.31\text{\rm
b})\cr}
$$
and the curvature tensor
$$
\widehat{R}^{a}{}_{b\mu \nu } = \lambda X_{b}H^{a}{}_{\mu \nu
}\eqno(3.32)
$$
(the remaining torsion and curvature components are zero).

The connection $\widehat{\omega }^{\rA}{}_{\rB}$ is neither flat nor torsionless. Moreover it is
still metric as a connection $\widehat{\omega }^{a}{}_{b}$ from sec.~2.

The covariant differentiation with respect to this connection is
connected to the right action of the group $G$ on $\un P$. Thus the condition
of the right-invariance of the $p$-form ${\varXi}
^{\rA_{1}\ldots \rA_{l}}{}_{\rB_{1}\ldots \rB_{m}}$ on $\un P$ is
equivalent to
$$
\widehat{\nabla }_{a}{\varXi} ^{\rA_{1}\ldots \rA_{l}}{}_{\rB_{1}\ldots
\rB_{m}}=0,\eqno(3.33)
$$
where $\widehat{\nabla }_{k}$ is a covariant derivative with respect to $\widehat{\omega
}^{\rA}{}_{\rB}$ in vertical
directions on $\un P$. This means right-invariance of ${\varXi} $. This can be written
$$
\widehat{\nabla }_{\ver (X)}{\varXi} = 0,\eqno(3.33\text{\rm a})
$$
ver is understood in the sense of $\omega $.
$$
{\varPhi} '(g ){\varXi} = {\varXi} ,\eqno(3.34)
$$
where $g \in G$ and ${\varXi} = ({\varXi} )^{\rA_{1}\ldots
\rA_{l}\rB_{1}\ldots \rB_{m}}) = (p^{\rB_{1}\rB'_1}p^{\rB_{2}\rB'_2}\ldots 
p^{\rB_{m}\rB'_m}{\varXi} ^{\rA_{1}\ldots \rA_{l}}{}_{\rB_{1}\ldots
\rB_{m}})$.

\noindent For a connection on a bundle $\un P$, $\omega $, its curvature ${\varOmega} $ one gets
$$
\widehat{\nabla }_{k}\omega = \widehat{\nabla }_{k}{\varOmega} =
0.\eqno(3.34*)
$$
Thus we can rewrite Eq.~(3.23)
$$
\widehat{\nabla }_{a}\underline{\gamma } = \widehat{\nabla }_{a}\overline{\gamma }=
0.\eqno(3.35)
$$
This means that
$$\eqalignno{
\widehat{\nabla }_{a} \gamma _{\rAB}&= 0,&(3.36)\cr
\noalign{\hbox{or}}
\widehat{\nabla }_{\ver (X)}\gamma &= 0.&(3.36\text{\rm a})\cr}
$$
For every linear connection $\omega ^{\rA}{}_{\rB}$ defined on $\un P$ compatible in some sense
with $\gamma _{\rAB}$ we get
$$
{\varPhi} ^{*}(g )\omega _{\rAB} =\Ad_{g }\omega
_{\rAB},\eqno(3.37)
$$
which means that $\omega _{\rAB}$ is right-invariant with respect to the
right-action of the group $G$ on $P$. The same we say for a 2-form of
torsion and 2-form of curvature derived for $\omega ^{\rA}{}_{\rB}$ i.e.
$$
\widehat{\nabla }_{a}{\varOmega} ^{\rA}{}_{\rB} = \widehat{\nabla }_{a}
{\varTheta} ^{\rA} = 0.\eqno(3.38)
$$
The curvature scalar is invariant with respect to the right action of
the group $G$ on $P$.
$$
0 = \widehat{\nabla }_{a}R = X_{a}R = \zeta _{a}R.\eqno(3.39)
$$
The condition (3.37) is the same as in the classical Kaluza--Klein
(Jordan--Thiry) theory in a nonabelian case. A parallel transport with
respect to the connection $\widehat{\omega}^{\rA}{}_{\rB}$ means simply a right action of the group
$G$ on $\un P$.

Our subject of investigations consists in looking for a generalization
of the geometry from Einstein Unified Field Theory (the so called
Einstein--Kaufmann Theory (see [4,5], [61])) defined on $P$ i.e.\ for a
connection $\omega ^{\rA}{}_{\rB}$ such that
$$
D\gamma _{\rAB} = \gamma _{\rAD}Q^{\rD}{}_{\rBE}\theta
^{\rE},\eqno(3.40)
$$
where $D$ is an exterior, covariant differential with respect to the
connection $\omega ^{\rA}{}_{\rB}$ and $Q^{\rD}{}_{\rBE}$ is a tensor of torsion for $\omega
^{A}{}_{B}$. We suppose
that this connection is right-invariant with respect the right action
of the group $G$.

We can write Eqs.~(3.37--39) for a torsion, curvature and a scalar of
curvature for $\omega ^{\rA}{}_{\rB}$. In this way we consider
Einstein--Kaufmann $G$-structure
on the bundle of linear frames over the manifold $\un{P}$
(i.e.\ a right $G$-structure).

We can repeat all the considerations changing right(left)-invariant
into 
left(right)-invariant in all places.

In this section we define $\omega ^{\rA}{}_{\rB}$ as a collection of 1-forms defined on the
manifold $\un{P}$ (a gauge bundle manifold) and we choose for $\omega
^{\rA}{}_{\rB}$ a lift
horizontal frame (connected to the connection $\omega $ on the gauge bundle).

The collection of 1-forms $\omega ^{\rA}{}_{\rB}$ becomes a linear connection on $\un{P}$ iff it
satisfies the following transformation properties
$$
\omega^{,\rA'}{}_{\rB'} = {\varSigma} ^{-1\rA'}{}_{\rA} (p)\omega
^{\rA}{}_{\rB} {\varSigma}^{-1\rB}{}_{\rB'}(p) -
{\varSigma}^{-1\rA'}{}_{\rA}(p)\,d{\varSigma}^{\rA}{}_{\rB'}(p),
\eqno(3.41)
$$
where
$$
\sum (p) \in \GL(n + 4,{\sym R}),\quad\ p \in U_{p} \subset P
$$
and
$$
\theta ^{C}={\varSigma}^{\rC}{}_{\rC'}(p)\theta ^{,C'}\eqno(3.42)
$$
is a simultaneous transformation property of a frame. Having $\omega
^{\rA}{}_{\rB}$ with 
(3.41--42) transformation properties we can lift it on a principal fibre
bundle of frames over $\un{P}$ with the structural group $GL(n+4,{\sym R})$ getting a
1-form of connection $\widetilde{\omega}$
$$
\widetilde{\omega}_{p} = \Ad_{\GL(n+4,R)}(g^{-1}_{p})[{\varPi} ^{*}
(\omega ^{\rA}{}_{\rB}X^{\rB}{}_{\rA})-g^{-1}_{p}\,dg_{p}],\eqno(3.43)
$$
where ${\varPi} $ is a projection defined on this principal fibre bundle of
frames and
$$\displaylines{
g_{p}:z\in {\varPi} ^{-1}(U_{p}) \rightarrow g_{p}(z) = (pr_{\GL(n+4,R)}
{\varPsi} _{p}(z))^{-1} \in \Gl(n+4,{\sym R}),\cr
p \in U_{p} \subset \un{P},\cr}
$$
pr means a projection on $\Gl(n+4,{\sym R})$ in a local trivialization of the
bundle $P''$, ${\varPsi} $ is an action of $\GL(n+4,{\sym R})$ on principal fibre bundle of
frames over~$\un{P}$, ${\varPsi} \rightarrow \Gl(n+4,{\sym R})\times P''
\rightarrow P'' $, ${\varPsi} _{p}$ is defined for $\Gl(n+4,{\sym
R})\times U_{p}$.
In this way we have an action of $\GL(n+4,{\sym R})$ on the bundle and for
$\widetilde{\omega}$.
$$
{\varPsi} ^{*}(g)\widetilde{\omega}=
\Ad_{\GL(n+4,R)}(g^{-1})\widetilde{\omega},\eqno(3.44)
$$
$X^{\rA}{}_{\rB}$ are generators of the Lie algebra $\gl(n+4,{\sym R})$ of
$\GL(n+4,{\sym R})$ and $g \in \Gl(n+4,{\sym R})$.
For a soldering forms $\widetilde{\theta }^{\rA}$ one gets
$$
\widetilde{\theta }^{\rA} = g_{p}{\varPi} ^{*}(\theta
^{\rA}).\eqno(3.45)
$$
Taking any two sections of the principal fibre bundle of $P''$ frames $E$
and $F$ such that
$$\eqalignno{
E^{*}\widetilde{\omega}&= \omega
^{,\rA'}{}_{\rB},X^{\rB'}{}_{\rA'},&\cr
F^{*}\widetilde{\omega}&= \omega
^{\rA}{}_{\rB}X^{\rB}{}_{\rA},&(3.46)\cr
E^{*}\widetilde{\theta }^{\rA} &= \theta ^{,\rA},\cr
F^{*}\widetilde{\theta }^{\rA} &= \theta ^{\rA},&(3.47)\cr}
$$
one gets transformation properties (3.41) and (3.42). In such a way
that
$$
E(p) = F(p) {\varSigma}(p).\eqno(3.48)
$$
Eq.~(3.40) can be rewritten in a more compact form
$$
\nabla \gamma = S,\eqno(3.49)
$$
where 
$$\displaylines{
S(X,Y,Z) = [\Tr(\gamma \otimes Q)](X,Y,Z) = \Sum_{\rA}\gamma (X,e_{\rA})
\theta ^{\rA}(Q(Y,Z)),\cr
Q(Y,Z) = -Q(Z,Y),\cr}
$$
is a torsion of the connection $\widetilde{\omega}$, $X$, $Y$, $Z$ are contravariant vector
fields and $\theta ^{\rA}$, $e_{\rB}$
$$
\theta ^{\rA}(e_{\rB}) = \delta ^{\rA}{}_{\rB}\ \hbox{are dual
bases}.
$$
Or in a different shape
$$
\nabla _{Z}\gamma (X,Y) = S(X,Y,Z),\eqno(3.50)
$$
$\nabla $ is a covariant derivative with respect to the connection
$\widetilde{\omega}$ on the
fibre bundle of frames.

Moreover now we consider $\gamma $, $Q$, $X$, $Y$, $Z$ etc.\ as geometrical objects
living on appropriate associate fibre bundles to the fibre bundle of
frames. The condition (3.50) gives us Einstein--Kaufmann connection
$\widetilde{\omega}$ on
the principal fibre bundle of frames over $P$. For
$\widetilde{\omega}$ is right-invariant
with respect to the action of group $G$ on this bundle of frames (lifted
to this bundle from $\un{P}$, the condition (3.50) is also right-invariant.
Let us consider the right-invariance of the Einstein connection on
$(\un{P},\gamma )$ from a different point of view.

Let ${\varPhi} :G\times \un{P}\rightarrow \un{P}$ be a right action of the group $G$ on $\un{P}$ and let ${\varPhi}
^{*}(g )$
be a contragradient map to ${\varPhi}'(g )$, a tangent map to ${\varPhi} $ at 
$g \in $G. Let
$\sum :G\rightarrow \GL(n+4),{\sym R})$ be a homomorphism of groups and let us consider a
connection $\widetilde{\omega }$ on a fibre bundle of frames over $P$ with a structural group
$\GL(n+4,{\sym R})$ compatible in Einstein-Kaufmann sense 
with the nonsymmetric
tensor~$\gamma$. The connection is right-invariant with respect to the action
of the group $G$ on $P$ if one has
$\widehat{{\varPhi}}^{*}(g )\widetilde{\omega } = \Ad_{\GL(n+4,R)}
(\sum (g ^{-1})\widetilde{\omega }$ ($\widehat{{\varPhi}}$ is an
action of $G$ on $P$ lifted to this bundle) or for any local section $E$ of
the bundle of frames over $\un{P}$
$$
{\varPhi} ^{*}(g ){\varGamma} = \Ad_{\GL(n+4,R)}
\Bigl(\Sum (g ^{-1})\Bigr){\varGamma} + \Sum ^{-1}(g )
d\Sum (g ),\eqno(3.51)
$$
where ${\varGamma} = E^{*}\widetilde{\omega }$
$$
{\varGamma} = {\varGamma} ^{\rA}{}_{\rBC}\theta
^{\rC}X^{\rB}{}_{\rA}\ \hbox{and}
$$
Thus one gets
$$\eqalignno{
{\varPhi} ^{*}(g ){\varGamma} ^{\rA'}{}_{\rB'\rC'}, ={}& \Sum
^{\rA'}{}_{\rA}(g^{-1}){\varGamma}^{\rA}{}_{\rBC'}(g^{-1})\Sum^{\rB}{}_{\rB'}(
g^{-1})\Sum ^{ \rC}{}_{\rC'}(g^{-1}) +&\cr
&+ \Sum ^{-1\rA'}{}_{\rA}(g )d_{\rC}\Sum^{\rA}{}_{\rB'}(g )
\Sum ^{\rC}{}_{\rC'}(g ),&(3.52)\cr}
$$
where $d_{\rC}$ is a vector field dual to $\theta ^{\rC}$. The reper transforms
$$
{\varPhi} ^{*}(g )\theta ^{\rC} = \Sum ^{\rC}{}_{\rC'}(g^{-1})\theta
^{\rC'}.\eqno(3.52')
$$

Let us take the lift horizontal basis. In this case one gets
$$
\Sum ^{\rA}{}_{\rB}(g ) = \left(\vcenter{\offinterlineskip\tabskip0pt
\halign{\strut\tabskip5pt plus10pt minus10pt
\hfill#\hfill&\vrule height10pt#&\hfill#\hfill\tabskip0pt\cr
$\vrule height0pt depth6pt width0pt \delta^{\alpha}{}_{\beta }$ && $0$\cr 
\noalign{\hrule}
$\vrule height9pt depth4pt width0pt 0$ && $\ov{U}^a{}_b(g )$\cr}}\right),\quad
\Sum (g)=\id_E\otimes \ov{U}(g ),\eqno(3.53)
$$
where $U^{a}{}_{b}(g ) = (\Ad_{\rG}(g ^{-1}))^{a}{}_{b}$ is a matrix of the adjoint representation of
$G$. Thus we have
$$\eqalignno{
{\varPhi} ^{*}(g )\theta ^{\alpha } &= \theta ^{\alpha },\cr
{\varPhi} ^{*}(g )\theta ^{a} &= \ov{U}^{a}{}_{a'}(g )\theta^{a'}&(3.54)\cr
{\varPhi} ^{*}(g ){\varGamma} ^{\alpha }{}_{\beta \gamma } &= 
{\varGamma} ^{\alpha }_{\beta \gamma },\cr
{\varPhi} ^{*}(g ){\varGamma} ^{a'}{}_{\beta '\gamma '} &=
\ov{U}^{a'}{}_{a}(g ^{-1}){\varGamma} ^{a}{}_{\beta '\gamma
'},&(3.55)\cr
{\varPhi} ^{*}(g ){\varGamma} ^{a'}{}_{\beta 'c'} &= \ov U^{a'}{}_{a}
(g ^{-1}){\varGamma} ^{a}{}_{\beta 'c}U^{c}{}_{c'}(g ),\cr
{\varPhi} ^{*}(g ){\varGamma} ^{a'}{}_{b'\gamma } &= \ov U^{a'}{}_{a}
(g ^{-1}){\varGamma} ^{a}{}_{b\gamma }U^{b}{}_{b'}(g ^{-1}),\cr
{\varPhi} ^{*}(g ){\varGamma} ^{\alpha '}{}_{\beta 'a'}&= 
{\varGamma} ^{\alpha }{}_{\beta 'a}\ov U^{a}{}_{a'}(g ^{-1}).\cr}
$$
For $\omega ^{a}{}_{b} = \widetilde{{\varGamma}}^{a}{}_{bc}\theta ^{c}$ one gets
$$
{\varPhi} ^{*}(g )\omega ^{a'}{}_{b'} = \ov U^{a'}{}_{a}(g ^{-1})
\omega ^{a}{}_{b}\ov U^{b}{}_{b'}(g ^{-1}).\eqno(3.56)
$$
In this way $\widetilde{{\varGamma} }^{a}{}_{bc}$ has tensorial properties in the lift horizontal basis
(Ad-type).

Moreover Eq.~(3.56) has a natural interpretation a right-invariancy of
the connection $\widetilde{\omega }^{a}{}_{b}$ with respect to the right action of the group $G$ on
$G$. The second equation of Eq.~(3.54) means Ad-property of the connection
on the principal fibre bundle $\un{P}$ (a gauge bundle).

Eq.~(3.56) can be rewritten in a more familiar form
$$
R^{*}(g )\widetilde{{\varGamma} } = \Ad\Bigl(\widetilde{\Sum}
(g ^{-1})\Bigr)\widetilde{{\varGamma} } + \widetilde{\Sum}
(g )d\widetilde{\Sum}(g ),\eqno(3.57)
$$
where $\widetilde{{\varGamma} } = \widetilde{\omega
}^{a}{}_{b}Y^{b}{}_{a}$ and $Y^{b}{}_{a}$ are generators of the Lie algebra 
$g \ell (n,R)$ of
the group $\GL(n,{\sym R})$ and $\Ad_{\GL(n,R)}$ is an adjoint representation of
$\GL(n,{\sym R})$
and $R$ is a right action of $G$ on $G$.
$$
\widetilde{\Sum}:G\rightarrow \GL(n,{\sym R})\eqno(3.58)
$$
is a smooth homomorphism of groups such that
$$
\widetilde{\Sum}^{a}{}_{b}(g ) = (\Ad_{G}(g ^{-1}))^{a}{}_{b} = 
\ov U^{a}{}_{b}(g ).\eqno(3.59)
$$
In this way we are coming to the notion of the fibre bundle of frames
over a group $G$ and to the right invariant connection defined on this
bundle. Eq.~(3.57) can be rewritten:
$$
\widehat{R}^{*}(g ) \widehat{\omega} = \widehat{\omega},\eqno(3.60)
$$
where $\widehat{\omega}$ is a connection on the principal fibre bundle of frames over $G$
with the structural group $\GL(n,{\sym R})$ and
$$
\widetilde{{\varGamma} } = f^{*}\widehat{\omega},\eqno(3.61)
$$
$f$ is a local section of this bundle. $\widehat{R}$ is an action of $G$ on $G$ lifted
to the bundle of frames.

Let us notice that our considerations are valid for any connection
defined on a fibre bundle of frames over $\un{P}$, not only for
Einstein--Kaufmann one. The same we can say for a connection on a fibre
bundle of frames over $G$. The above considerations justify some
$\Ad_{G}$-properties of $\omega ^{\rA}{}_{\rB}$ defined on manifold
$\un P$ (a gauge bundle) and $\widetilde{\omega }^{a}{}_{b}$
defined on $G$ (a group manifold). They are treated there as some 1-forms
defined on $P$ or $G$ according to our conventions. From Eq.~(3.51) one
easily gets transformation laws for the curvature
$$
\widehat{{\varPhi}}^{*}(g )\widetilde{{\varOmega} } = 
\widetilde{{\varOmega} }\eqno(3.62)
$$
and from (3.60)
$$
\widehat{R}^{*}(g )\widetilde{{\varOmega} } = \widetilde{{\varOmega}
}.\eqno(3.63)
$$
For the curvature scalars we get
$$\eqalignno{
\widehat{{\varPhi}}^{*}(g )R(\widetilde{\omega }) &= R(\widetilde{\omega
}),&(3.64)\cr
\widehat{R}^{*}(g )R(\widetilde{\omega }) &= R(\widetilde{\omega
}).&(3.65)\cr}
$$
In this work we consider a non-symmetric metrization of the
principal fibre bundle defined over a base manifold $V=E\times
G/G_0$, where $E$ is a space-time and $M=G/G_0$ is a certain
homogeneous space.
\vfill\eject

\null
\vskip36pt plus4pt minus4pt
\centerline{{\trz 4. Preliminary Remarks}}
\vskip24pt plus2pt minus2pt
\write0{4. Preliminary Remarks \the\pageno}

Let us introduce the principal fibre bundle $P$ over the $V = E \times M$
with the structural group $H$ and with the projection $\pi $, where $M = G/G_{0}$
is a homogeneous space, $G$ is a compact Lie group (semisimple), $G_{0} \subset G$ is
a semisimple compact subgroup of $G$ and let $\omega = \omega ^{a}X_{a}$ is a connection
defined on $P$ (Fig.~2A). Notice that now we have in the place of $E$,
$E\times G/G_{0}$ and in
the place of $G$, $H$. Let us suppose that $(V,\gamma )$ is a manifold with a
nonsymmetric metric tensor
$$
\gamma _{\rAB} = \gamma _{(\rAB)} + \gamma _{[\rAB]}.
\eqno(4.1)
$$
\obraz{
\midinsert \centerline{\epsfxsize\hsize \epsfbox{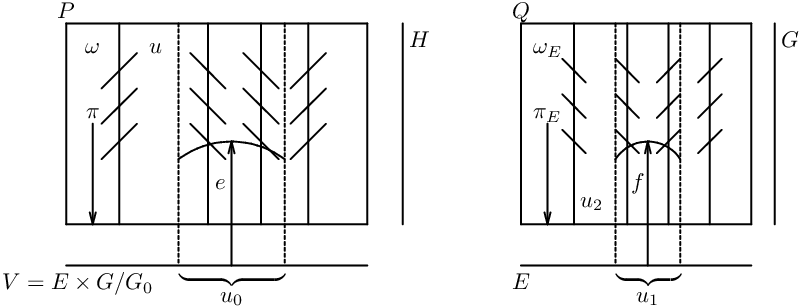}}
\centerline{\nine Fig. 2A. Principal fibre bundles $P$ and $Q$}
\endinsert

}The signature of the tensor $\gamma$ is $(+ - - -,\underbrace{- - - \ldots -}_{n_1})$.
Let us introduce a natural 
frame on $\un{P}$
$$
\theta ^{\tilde{\rA}} =(\pi ^{*}(\theta ^{\rA}),\ \theta ^{a} = \lambda \omega
^{a}),\quad\ \lambda =\const.\eqno(4.2)
$$
It is convenient to introduce the following notations. Capital Latin
indices with tilde $\widetilde{A}$, $\widetilde{B}$, $\widetilde{C}$ run $1,2,3,\ldots 
,m+4$, $m = \dim H + \dim M = n + \dim M = n + n_{1}$, $n_{1} = \dim
M$, $n = \dim H$. Lower Greek indices $\alpha , \beta , \gamma ,
\delta = 1,2, 3, 4$ and lower Latin cases $a, b, c, d = n_{1}+5, n_{1}+6,
\ldots ,m+4$. Capital
Latin indices $A, B, C = 1, 2, 3,\ldots ,n_{1}+4$,
lower Latin cases with tilde $\widetilde a,\widetilde b,\widetilde c$ run $5,6,\ldots,n_1+4$. 
The symbol ``$-$" over 
$\theta ^{\rA}$ and 
over other quantities indicates that these quantities are defined on
$V$.
We have of course $\dim H = n$, $\dim M = \dim (G/G_{0}) = n_{1}$, $\dim G 
= n_{2}$, $\dim G_{0} =n_{2}-n_{1}$, and $m = n_{1}+n$. It is easy to see that the existence of the
nonsymmetric metric on $V$ is equivalent to introducing the two
independent geometrical quantities on $E$.
$$\eqalignno{
\overline{\gamma }&= \gamma _{\rAB}\overline{\theta }^{\rA} \otimes \overline{\theta}^{\rB} = 
\gamma _{(\rAB)}\overline{\theta }^{\rA} \otimes \overline{\theta}^{\rB},&(4.3)\cr
\underline{\gamma }&= \gamma _{\rAB}\overline{\theta }^{\rA}\wedge \overline{\theta
}^{\rB} = \gamma _{[\rAB]}\overline{\theta }^{\rA}\wedge \overline{\theta
}^{\rB},&(4.4)\cr}
$$
i.e. the symmetric metric tensor $\overline{\gamma }$ on $V$ and the 2-form
$\underline{\gamma }$. On the group
$H$ we can introduce a bi-invariant symmetric tensor called the
Killing--Cartan tensor.
$$
h(A,B) = h_{ab} A^{a} \cdot B^{b},\eqno(4.5)
$$
where
$$
h_{ab} = h_{ba},\quad A = A^{a}X_{a},\quad B = B^{a}X_{a},
$$
$X_{a}$---are generators of a Lie algebra of $H,(\fh)$.

It is supposed that $H$ is semisimple; it means that $\det (h_{ab})\neq
0$. Let us
introduce a skewsymmetric right-invariant tensor on $H$
$$
k(A,B) = k_{bc}A^{b} \cdot B^{c} ,\quad\ k_{bc} = -k_{cb}.\eqno(4.6)
$$
Let us turn to the nonsymmetric metrization of $\un{P}$ .Let us suppose that
$$\eqalignno{
\overline{\varkappa }(X,Y) &= \overline{\gamma }(\pi ' X,\pi ' Y) + 
\rho ^{2}\lambda ^{2}h(\omega (X),\omega (Y)),&(4.7)\cr
\underline{\varkappa }(X,Y) &=\underline{\gamma }(\pi ' X,\pi ' Y) + 
\rho ^{2}\xi \lambda ^{2}k(\omega (X),\omega (Y)),&(4.8)\cr}
$$
$\xi = \const$ and is dimensionless, $X, Y \in \Tan(\un{P})$ and 
$\rho = \rho (x)$ is a scalar
field on $E$. It is easy to see that:
$$\eqalignno{
\overline{\varkappa }&= \pi ^{*}\overline{\gamma }\oplus \rho ^{2} h_{ab} \theta ^{a} \otimes \theta
^{b},&(4.9)\cr
\underline{\varkappa }&= \pi ^{*}\underline{\gamma }\oplus \rho ^{2} \xi k_{ab} \theta
^{a}\wedge \theta ^{b},&(4.10)\cr}
$$
For
$$
\varkappa _{\tilde{\rA}\tilde{\rB}}= \varkappa _{(\tilde{\rA}\tilde{\rB})} + 
\varkappa _{[\tilde{\rA}\tilde{\rB}]}
$$
one easily gets
$$
\varkappa _{\tilde{\rA}\tilde{\rB}} =
\left(\vcenter{\offinterlineskip\tabskip0pt
\halign{\strut\tabskip5pt plus10pt minus10pt
\hfill#\hfill&\vrule height10pt#&\hfill#\hfill\tabskip0pt\cr
$\vrule height0pt depth6pt width0pt \gamma _{\rAB}$ && $0$\cr 
\noalign{\hrule}
$\vrule height9pt depth4pt width0pt 0$ &&
$\rho^2\ell_{ab}$\cr}}\right),\eqno(4.11)
$$
where $\ell _{ab} = h_{ab} + \xi k_{ab}$ and $\det (\ell _{ab})\neq 0$.
One easily writes:
$$
\varkappa _{\tilde{\rA}\tilde{\rC}}\varkappa ^{\tilde{\rB}\tilde{\rC}} = 
\varkappa _{\tilde{\rC}\tilde{\rA}}\varkappa ^{\tilde{\rC}\tilde{\rB}} = 
\delta ^{\tilde{\rB}}{}_{\tilde{\rA}},\eqno(4.12)
$$
where the order of indices is important. The signature of the
tensor $\varkappa$ is\hfill\break $(+ - - -, \underbrace{- - - -\ldots
-}_{n_1},\underbrace{- - - \ldots -}_{n})$. If $\det (\ell _{ab})\neq 0$ then
$$
\ell _{ab}\ell ^{ac} = \ell _{ba}\ell ^{ca} = \delta ^{c}_{b},\eqno(4.13)
$$
where the order of indices is important.
$$
[X_{a},X_{b}] = C^{c}_{ab}X_{c}\eqno(4.13\text{\rm a})
$$
are commutation relation for the Lie algebra of $H,(\fh)$.

Let us pass to the nonsymmetric metrization of the homogeneous
space $G/G_{0}$. Let $M$ be a homogeneous space for $G$ (with left action of
group $G$) (see [86], [87]). Let us suppose that the Lie algebra of $G$, $\fg$
has the following reductive decomposition
$$
\fg = \fg_{0}\dot{+}\fm ,\eqno(4.14)
$$
where $\fg_{0}$ is a Lie algebra of $G_{0}$ (a subalgebra of 
$\fg$) and $\fm$ (the
complement to the subalgebra $\fg_{0})$ is Ad $G_{0}$ invariant,
$\dot{+}$ means a direct
sum. Such a decomposition might not be unique in general, but we shall
assume that one has been chosen. Sometimes one assumes a stronger
condition for $\fm$, the so called symmetry requirement
$$
[\fm,\fm] \subset \fg_{0}.
$$
Let us introduce the following notation for generation of $\fg$.
$$
Y_{i} \in \fg,\quad Y_{\tilde{i}} \in \fm,\quad Y_{\dah{a}} \in
\fg_{0}.
$$
This is a decomposition of a basics of $\fg$ according to (4.14).

If $G$ is a connected Lie group and $G_{0}$ is a closed subgroup, $G/G_{0}$ is the
space of cosets $\{gG_{0}\}, \varphi : G \to G/G_{0}$ is defined by $\varphi (g)
= \{gG_{0}\}$, $g\in G$. Let
us suppose that $G$ is a compact Lie group and define a field of 2-form
$k^{0}(x)$, $x\in M$ on $G/G_{0}$. Let us notice that $G \tovarphi M = G/G_{0}$ gives us a
principal fibre bundle over $M$ with a projection $\varphi $ and a
structural group $G_{0}$. We use the complement $\fm$ in order to introduce a
coordinate system on $G/G_{0}$. Let $g \in G$ and let ``$\exp $" be an exponential
mapping $\exp :\fg\to G$. Thus
$$
(x_{5},x_{6},\ldots ,x_{n_{1}+4})\to \varphi \Bigl(g \exp 
\sum_{\tilde a} x_{\tilde a}Y_{\tilde a}\Bigr)\eqno(*)
$$
is a diffeomorphism of a neighbourhood of zero in $\fm$ on a neighbourhood
of $\varphi (g)$ in $G/G_{0}$. An inverse mapping is a local coordinate system in a
neighbourhood of $\varphi (g) \in G/G_{0}$ on $M = G/G_{0}$

The mapping $\varphi :G\to G/G_{0}$ has a tangent mapping $\varphi '$ which maps 
$\fg$ in a
tangent space to $M = G/G_{0}$ at a point $o = \{\epsilon G_{0}\}$. A kernel of $\varphi '$ is equal
to $\fg_{0}$. The operator of left action $\tau(g) : hG_{0}\to ghG_{0}$ satisfies $\varphi oL(g) =
\tau(g)o\varphi $, $h \in G$.

For $\varphi oR_{g_{0}}= \varphi $, $\Ad_{G}(g)Y = R'_{g^{-1}}oL'_{g}(Y)$ we get
$\varphi 'o \Ad_{g}(g_{0})Y = \tau'(g_{0})_{0}o\varphi '(Y)$, $Y \in\fg$.

Thus $\fm\simeq \fg/\fg_{0} \simeq \Tan_{0}(G/G_{0})$ and using the left action of the group $G$
on $M$ we can get vector fields at any point $x \in M$ from $\Tan_{0}(M) \simeq
\fm$. Thus
let $Y_{\tilde a} \in \fm$. Due to an isomorphism of $\fm$ and $\Tan_{0}(M)$ we get
$$
\widetilde{Y}_{\tilde a}(x) = L'_{g}(o),\quad\ x = go,\ g \in G,
$$
$Y(x)$ is left invariant on $M$. If $k(g)$ is a field of 2-forms on $G$ and
being left-invariant we define a field of 2-form $k^{0}(x)$ on $M$ in the
following way: Let us restrict it to the subspace $\fm$ (a complement)
$$
k' (g) = k(g)|_{\fm},
$$
i.e $\Wedge_{g\in G} k(g)|_{\fm}(x,y) = o$ if $x$ or $y$ do not belong to $\fm$
and consider a 2-form $k^{0}(x)$.

The form $k(g)$ is left-invariant on $G$. The form $k(g)|_{\fm}$ is $G_{0}$
Ad'-invariant, for $\fm$ is $\Ad'_{G_0}$ invariant. Let us consider a local section
$\sigma _{1} : U_{1}\to G$, $U_{1}\subset M$ of the bundle $G$. We get
$$
k^{\sigma _1}(x) = \sigma ^{*}_{1}k(\sigma _{1}(x))|_{\fm},\quad\
x\in U_1.\eqno(4.15)
$$
Let us take a different local section $\sigma _{2} : U_{2}\to G$, $U_{2} \subset M$
$$
k^{\sigma _{2}}(x) = \sigma ^{*}_{2}k(\sigma _{2}(x))|_{\fm}.\eqno(4.16)
$$
One has
$$
\sigma _{2}(x) = p(x)\sigma _{1}(x),\eqno(4.17)
$$
where $p(x) \in G_{0}$
$$
x \in U_{1} \cap U_{2}.
$$
Thus one obtains
$$
k^{\sigma _{2}}(x) = R^{*}_{p(x)}k^{\sigma _{1}}(x) = k^{\sigma
_{1}}(x).\eqno(4.18)
$$
Thus we get
$$
k^{0}(x) = k^{\sigma }(x),\eqno(4.19)
$$
for any local section of $G$. Let us examine a transformation properties
of $k^{0}$ with respect to the left action of the group $G$ on $M$. There is a
natural action of $G$ on left cosets, such that $G$ acts with an 
isotropy subgroup $G_{0}$. One gets
$$
L^{*}_{g}k^{0}(x) = k^{0}(x).\eqno(4.21)
$$
In this way $k^{0}(x)$ is left invariant with respect to the action of the
group $G$ on $M = G/G_{0}$ .

This fact allows us to define $k$ being left-invariant on $G$ in such a
way that $k = \varphi ^{*}k^{0}$ and $k^{0}$ is left-invariant 2-form on $M = G/G_{0}$.
This gives us a construction of a left-invariant 2-form on $G$ if we have
a left-invariant 2-form on $M = G/G_{0}$.

In the same way we can define a tensor field $h^{0}(x)$ on
$G/G_{0}$, $x\in G/G_{0}$
using tensor field $h$ on $G$. Moreover if we suppose that $h$ is a
biinvariant (Ad-invariant) metric on $G$ (a Killing--Cartan tensor) we
have a simpler construction.

The complement $\fm$ is a tangent space to the point $\{\varepsilon G_{0}\}$ of $M, \varepsilon $ is a
unit element of $G$. We restrict $h$ to the space $\fm$ only. Thus we have
$h^{0}(\{\varepsilon G_{0}\})$ at the one point of $M$. Now we propagate $h^{0}(\{fG_{0}\})$ using a
left action of the group $G$.
$$
h^{0}(\{fG_{0}\}) = (L^{-1}_{f})^{*}(h^{0}(\{\varepsilon G_{0}\}),\eqno(4.22)
$$
$h^{0}(\{\varepsilon G_{0}\})$ is of course Ad $G_{0}$ invariant tensor defined on
$\fm$ and
$$
L^{*}_{f}h^{0} = h^{0},
$$
where $f\in G$.

Let us find $k^{0}(x)$ in the case of $\SO(3)/\SO(2)\simeq S^{2}$, where SO(2) subgroup
is defined by the exponentiation of $e_{3}$, i.e. rotations around an axis $z$
and $\fm\cong \Span\{ e_{1}, e_{2}\}$. One finds
$$
k(g)|_{\fm} = k' (g) = 2k_{12} v^{1}\wedge v^{2} ,
$$
where $v^{1}$ and $v^{2}$ are dual with respect to $e_{1}$ and $e_{2}$ being
right-invariant vector fields on SO(3). Thus
$$\eqalignno{
k' _{\tilde{a}\tilde{b}}(\vartheta ,\phi ,\psi ) &= 
\pmatr{0&\cos \vartheta \cr -\cos \vartheta &0\cr},&(4.23)\cr
k^{0}_{\tilde{a}\tilde{b}}(\vartheta ,\phi ) &= 
\pmatr{0&\cos \vartheta \cr -\cos \vartheta &0\cr}.&(4.23\text{\rm c})\cr}
$$
We can also try to find and classify such tensors for all homogeneous
spaces. This will be done elsewhere. The classification of homogeneous
spaces for which $G$ and $G_{0}$ are nilpotent or solvable will be done
elsewhere as well. In this case the same construction as for $h^{0}$ is
applicable for $k^{0}$ (coming from bi-invariant from defined on a nilpotent
or a solvable Lie algebra).

Let $f^{i}{}_{jk}$ are the structure constants of the Lie algebra $\fg$, i.e.
$$
[Y_{j}, Y_{k}] = f^{i}{}_{jk} Y_{i},\eqno(4.24)
$$
$Y_{j}$ are generators of Lie algebra $\fg$. Let us construct a natural
left-invariant frame on $M = G/G_{0}$. To do this we choose the local
section $\sigma : U \rightarrow G/G_{0}$, of the natural bundle 
$G \rightarrow G/G_{0}$ where $U\subset M = G/G_{0}$ (see Fig.~2B).

\obraz{
\midinsert \centerline{\epsfxsize\hsize \epsfbox{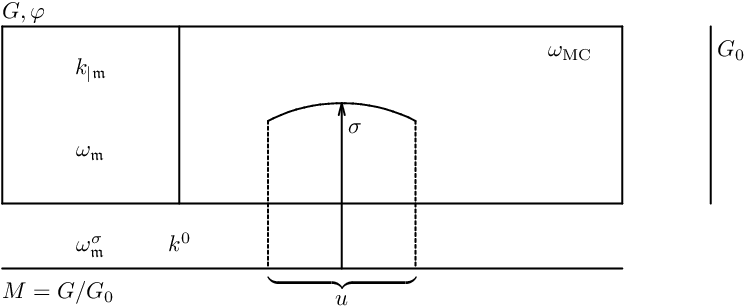}}
\centerline{\nine Fig. 2B. The principal fibre bundle $G$ over $G/G_0=M$}
\vskip-1pt
\centerline{\nine and some details on the constructions of the form $k^0(x)$
on $M$}
\endinsert

}The local section $\sigma $ can be considered an introduction of coordinate
system on $M$ i.e $(*)$. Let $\sigma ^{*}$ means a pull-back of $\sigma $ and let us define
$$
\omega ^{\sigma }_{\rMC} = \sigma ^{*}\omega _{\rMC},\eqno(4.25)
$$
where $\omega _{\rMC}$ is a left-invariant Maurer--Cartan form. Using the
decomposition (4.14) we have
$$
\omega ^{\sigma }_{\rMC} = \omega ^{\sigma }_{0} + \omega ^{\sigma
}_{\fm} = \theta ^{\dah{i}}Y_{\dah{i}} + \overline{\theta }^{\tilde
a}Y_{\tilde a}.\eqno(4.26)
$$
It is easy to see that $\theta ^{\tilde a}$ is the natural left-invariant frame on $M$ and
we have
$$\eqalignno{
h^{0} &= h^{0}_{\tilde{a}\tilde{b}}\overline{\theta }^{\tilde a} \otimes 
\overline{\theta }^{\tilde b},&(4.27)\cr
k^{0} &= k^{0}_{\tilde{a}\tilde{b}}\overline{\theta }^{\tilde a}\wedge
\overline{\theta }^{\tilde b}.&(4.28)\cr}
$$
According to our notation we have $\widetilde{a}, \widetilde{b}= 5, 6,\ldots 
,n_{1}+4$. Thus we have a nonsymmetric metric on $M$
$$
\gamma _{\tilde{a}\tilde{b}} = r^{2} (h^{0}_{\tilde{a}\tilde{b}}+\zeta 
k^{0}_{\tilde{a}\tilde{b}}) = r^{2} g_{\tilde{a}\tilde{b}},\eqno(4.29)
$$
where $\zeta $ is a real dimensionless parameter and $r$ has 
 length-dimension. 
$r$ characterizes the size of the manifold $M$ and plays the role of a
``radius". The tensor field $g_{\tilde{a}\tilde{b}}\overline{\theta
}^{\tilde a} \otimes \overline{\theta }^{\tilde b} = (h^{0}{}_{\tilde{a}\tilde{b}}+
\zeta k^{0}_{\tilde{a}\tilde{b}})\overline{\theta }^{\tilde a} \otimes
\overline{\theta }^{\tilde b}$ is
left-invariant on $M= G/G_{0}$. For $k^{0}$ is left-invariant with respect to the
action of the group $G$ on $M$ one gets
$$
Y_{\ell } k^{0} = 0\eqno(4.30)
$$
and in particular
$$\eqalign{
Y_{\tilde{\ell}} k^{0} &= 0,\cr
Y_{\tilde{\ell}} \in \fm &= \fg/\fg_{0}.\cr}\eqno(4.31)
$$
Let $\widehat{\widetilde{D}}$ be an exterior covariant derivative with respect to the
Riemannian connection on $M$ induced by the tensor field $h^{0}$.

If
$$
\widehat{\widetilde{D}} k^{0} = dk^{0} = 0\eqno(4.32)
$$
the form $k^{0}$ is closed and it can define a symplectic 
structure on $M$ (of course if $k^{0}$ is nondegenerate and $\dim M = 2k$, i.e. it
is even). In this case we have
$$
\widehat{\widetilde{\nabla}}_{[\tilde a} k_{\tilde b \tilde c]}= 0.\eqno(4.33)
$$
In this case we can define an almost complex structure compatible with
the Riemannian and the symplectic structure on $M$ i.e
$$
J: T(M)\to T(M),\quad\ J^{2}= -\id,\eqno(4.34)
$$
such that
$$
h^{0}(JX,Y) = k^{0}(X,Y),\eqno(4.35)
$$
for any $X,Y \in T(M)$.

If those three structure are compatible we have:
$$
\widehat{\widetilde{\nabla}}_{z}J = 0 ,\quad\ z \in T(M).\eqno(4.36)
$$
The almost complex structure $J$ can be integrated into a complex
structure 
(a unique one) if the torsion (torsion of $J$, not of a
nonriemannian connection) vanishes i.e.
$$
S(X,Y) = [X,Y] + J[JX,Y] + J[X,JY] - [JX,JY] = 0 ,\quad\ X,Y \in 
T(M).\eqno(4.37)
$$
In this case we get K\"ahlerian structure on the manifold $M$. If $\dim M
=2$, this can be always achieved.

Thus in the case of $\dim G/G_{0} = 2$ we can always integrate the
left-invariant symmetric tensor $g_{\tilde{a}\tilde{b}}$ introduced in this section if
$\widehat{\widetilde{\nabla}}_{\tilde c}k^{0}_{\tilde{a}\tilde{b}}= 0$ into 
K\"ahlerian geometry.

In particular we get in this way for $M = S^{2} = \SO(3)/\SO(2)$ a K\"ahlerian
geometry which is holomorphically equivalent to the standard one.
Holomorphically means here also conformally.

In the case of K\"ahlerian structure, one easily gets that
$$
\widehat{\widetilde{D}}g_{\tilde{a}\tilde{b}} = 0.\eqno(4.38)
$$
Eq.~(4.38) has important consequences.

The K\"ahlerian structure on $M = G/G_{0}$ ($G$-invariant) induces a
left-invariant 2-form on $M$, $k^{0}$. Moreover $k^{0}$ induces a skew-symmetric
form on $G$. This has some important consequences, because we can induce
on an $G$ a left-invariant nonsymmetric tensor coming from K\"ahlerian
structure on $G/G_{0}$.

Thus we are able to write down the nonsymmetric metric on $V = E \times M =
E \times G/G_{0}$.
$$
\gamma _{\rAB} =
\left(\vcenter{\offinterlineskip\tabskip0pt
\halign{\strut\tabskip5pt plus10pt minus10pt
\hfill#\hfill&\vrule height10pt#&\hfill#\hfill\tabskip0pt\cr
$\vrule height0pt depth6pt width0pt g_{\alpha \beta }$ && $0$\cr 
\noalign{\hrule}
$\vrule height9pt depth4pt width0pt 0$ &&
$r^2g_{\tilde{a}\tilde{b}}$\cr}}\right),\eqno(4.39)
$$
where
$$\displaylines{
\eqalign{g_{\alpha \beta } &= g_{(\alpha \beta )} + g_{[\alpha \beta]},\cr
g_{\tilde{a}\tilde{b}} &= h^{0}_{\tilde{a}\tilde{b}} + \zeta 
k^{0}_{\tilde{a}\tilde{b}},\cr}\cr
\noalign{\vskip6pt}
k^{0}_{\tilde{a}\tilde{b}} = -k^{0}_{\tilde{a}\tilde{b}} ,\quad\ 
h^{0}_{\tilde{a}\tilde{b}} = h^{0}_{\tilde{a}\tilde{b}},\cr}
$$
$\alpha , \beta = 1, 2, 3, 4$. $\widetilde{a}, \widetilde{b}= 5, 6,\ldots 
,n_{1}+4 = \dim M+4 = \dim G-\dim G_{0}+4$. The
frame $\overline{\theta }^{\tilde a}$ is unholonomic one:
$$
d\overline{\theta }^{\tilde a} = {1\over 2} \varkappa ^{\tilde
a}{}_{\tilde{b}\tilde{c}}\overline{\theta }^{\tilde b}\wedge \overline{\theta
}^{\tilde c},\eqno(4.40)
$$
where $\varkappa ^{\tilde a}{}_{\tilde{b}\tilde{c}}$ are coefficients of nonholonomicity and they depend on the
point of the manifold $M = G/G_{0}$ (they are not constant in general).
They depend on the section $\sigma $ and on the constants $f^{\tilde
a}{}_{\tilde{b}\tilde{c}}$.

Let us pass to the group structure of our theory. We have there three
groups $H$, $G$ and $G_{0}$ which have up to now been disconnected. Let us
suppose that there exists a homomorphism $\mu $ between $G_{0}$ and $H$
$$
\mu : G_{0} \rightarrow H,\eqno(4.41)
$$
such that the centralizor of $\mu (G_{0})$ in $H$, $C^{\mu }$ is isomorphic to 
$G$. $C^{\mu }$,
the centralizor of $\mu (G_{0})$ in $H$ is the set of all elements of $H$ which
commute with elements of $\mu (G_{0})$ (it is a subgroup of $H$, of course).
$$
C^{\mu } = G,\eqno(4.42)
$$
This means that $H$ has the following structure:
$$
\mu (G_{0}) \otimes G \subset H.\eqno(4.43)
$$
If we suppose that $\mu $ is an isomorphism between $G_{0}$ and $\mu (G_{0})$ we get:
$$
G_{0} \otimes G \subset H.\eqno(4.44)
$$
Let us denote $\mu ' $ as the tangent map to $\mu $ at a unit element. Thus $\mu ' $ is
the differential of $\mu $ acting on the Lie algebra elements. Let us
suppose that the connection $\omega $ on the fibre bundle $P$ is invariant under
group action of $G$ on the manifold $V = E\times G/G_{0}$. According to 
Ref.~[23]
this means the following. Let $e$ be a local section of $P$, $e: U \subset 
V\to {P}$
and $A = e^{*}\omega $. Then for every $g\in G$ there exists a gauge transformation $\rho _{g}$
such that:
$$
f^{*}(g)A =\Ad_{\rho _{g}^{-1}}A + \rho ^{-1}_{g}\,d\rho _{g},\eqno(4.45)
$$
$f^{*}$ means a pull-back of the action $f$ of the group $G$ on manifold 
$V$.
According to Refs.\ [71], [72], [73] we are able to write a general
form for such an $\omega $. Following Ref.~[72] we have:
$$
\omega = \widetilde{\omega }_{E} + \mu ' \circ \omega ^{\sigma }_{0} + 
{\varPhi} \circ \omega ^{\sigma }_{\fm},\eqno(4.46)
$$
where $\omega ^{\sigma }_{0} + \omega ^{\sigma }_{\fm} = \omega
^{\sigma }_{\rMC}$ are components of the pull-back of the
Maurer--Cartan form from the decomposition (4.14), $\widetilde{\omega
}_{\rE}$ is the
connection defined on the fibre bundle $Q$ over space-time $E$ with
structural group $C^{\mu }$ and the projection $\pi _{\rE}$. Moreover
$C^{\mu }=G$, and $\widetilde{\omega }_{E}$ is a
1-form with values in the Lie algebra $\fg$. This connection describes the
ordinary Yang--Mills' field with gauge group $G = C^{\mu }$ on the space-time
$E$.
${\varPhi} $ is a function on $E$ with values in the space $\widetilde{S}$ of linear maps:
$$
{\varPhi} : \fm\rightarrow \fh\eqno(4.47)
$$
satisfying
$$
{\varPhi} ([X_{0},X]) = [\mu ' X_{0} , {\varPhi} (X)] ,\quad\ X_{0} \in 
\fg_{0}\eqno(4.48)
$$
thus
$$\eeqalign{
\widetilde{\omega }_{\rE} &= \widetilde{\omega }^{i}_{\rE} Y_{i} ,\quad\ &Y_{i}
&\in \fg,\cr
\omega ^{\sigma }_{0} &= \theta ^{\dah{i}} Y_{\dah{i}} ,\quad\
&Y_{\dah{i}} &\in \fg_{0},\cr
\omega ^{\sigma }_{\fm} &=\overline{\theta }^{\tilde a} Y_{\tilde a}
,\quad\ &Y_{\tilde a} &\in \fm.\cr}
$$

Let us write condition (4.46) in the base of left-invariant form
$\theta ^{\dah{i}}$ and $\overline{\theta }^{\tilde a}$, which span respectively dual spaces to
$\fg_{0}$ and $\fm$.

It is easy to see that
$$
{\varPhi} \circ \omega ^{\sigma }_{\fm} = {\varPhi} ^{a}_{\tilde
a}(x)\overline{\theta }^{\tilde a} X_{a} ,\quad\ X_{a}\in \fh\eqno(4.49)
$$
and
$$
\mu ' = \mu ^{a}_{\dah{i}} \theta ^{\dah{i}} X_{a},\eqno(4.50)
$$
where ${\varPhi} ^{a}_{\tilde a}(x)$ is the scalar field on $E$ and $\mu
^{a}_{\dah{i}}$ is a constant matrix. From
(4.48) one gets
$$
{\varPhi} ^{c}_{\tilde b}(x) f^{\tilde b}_{\dah{i}\tilde a}= \mu
^{a}_{\dah{i}} {\varPhi} ^{b}_{\tilde a}(x) C^{c}{}_{ab},\eqno(4.51)
$$
where $f^{\tilde b}_{\dah{i}\tilde a}$ are structure constants of the Lie algebra 
$\fg$ and $C^{c}{}_{ab}$ are
structure constants of the Lie algebra $\fh$. (4.51) is a constraint on
the scalar field ${\varPhi} ^{a}_{\tilde a}(x)$. For a curvature of 
$\omega $ one gets: 
$$\eqalignno{
{\varOmega} ={}& {1\over 2} H_{\rAB}\theta ^{\rA}\wedge \theta ^{\rB} 
X_{c}=\cr
\noalign{\vskip4pt}
={}& {1\over 2} \widetilde{H}^{i}{}_{\mu \nu } \theta ^{\mu }\wedge
\theta ^{\nu } \alpha ^{c}_{i} X_{c} + \mathop{\nabla_{\mu}}\limits^{\gauge} 
{\varPhi} ^{c}_{\tilde a} \theta ^{\mu }\wedge \theta ^{\tilde a} X_{c}
+\cr
\noalign{\vskip4pt}
&+ {1\over 2} C^{c}{}_{ab} {\varPhi} ^{a}_{\tilde a} {\varPhi}
^{b}_{\tilde b} \theta ^{\tilde a}\wedge \theta ^{\tilde b} X_{c} - 
{1\over 2} \mu ^{c}_{\dah{i}} f^{\dah{i}}_{\tilde a\tilde b} \theta
^{\tilde a}\wedge \theta ^{\tilde b} X_{c} 
- {1\over 2} {\varPhi} ^{c}_{\tilde d} f^{\tilde d}_{\tilde a\tilde
b}\theta ^{\tilde a}\wedge \theta ^{\tilde b} X_{c}.&(4.52)\cr}
$$
Thus we have: 
$$\displaylines{
\hfill H^{c}_{\mu \nu } = \alpha ^{c}_{i} \widetilde{H}^{i}_{\mu \nu
},\hfill\llap{(4.53)}\cr
\hfill H^{c}_{\tilde a\tilde b} = \mathop{\nabla_{\mu}}\limits^{\gauge}
{\varPhi}^{c}_{\tilde a} = -H^{c}_{\tilde a\mu
},\hfill\llap{(4.54)}\cr
\hfill H^{c}_{\tilde a\tilde b} = C^{c}_{ab} {\varPhi} ^{a}_{\tilde
a} {\varPhi} ^{b}_{\tilde b} - \mu ^{c}_{\dah{i}} f^{\dah{i}}_{\tilde a\tilde b}
- {\varPhi} ^{c}_{\tilde d} f^{\tilde d}_{\tilde a\tilde
b},\hfill\llap{(4.55)}\cr}
$$
where $\mathop{\nabla_{\mu}}\limits^{\gauge}$ means gauge derivative with respect to the connection
$\widetilde{\omega }_{E}$
defined on the bundle $Q(E,G)$, 
$$ 
Y_{i} = \alpha ^{c}_{i} X_{c}\eqno(4.56)
$$
and $\widetilde{H}^{i}_{\mu \nu }$ is the curvature of the connection
$\widetilde{\omega }_{\rE}$ in the base $\{Y_{i}\}$,
generators of the Lie algebra of the Lie group $G$, $\fg$, $\alpha ^{c}_{i}$ is the matrix
which connects $\{Y_{i}\}$ with $\{X_{c}\}$. Now we would like to remind that
indices $a$, $b$, $c$ refer to the Lie algebra $\fh$, $\widetilde{a}$,
$\widetilde{b}$, $\widetilde{c}$ to the space $\fm$
(tangent space to $M)$, $\widehat i$, $\widehat j$, $\widehat k$ to the Lie 
algebra $\fg_{0}$ and $i$, $j$, $k$ to the
Lie algebra of the group $G$, $\fg$. The matrix $\alpha ^{c}_{i}$ establishes a direct
relation between generators of the Lie algebra of the subgroup of the
group $H$ isomorphic to the group $G$.

Finally let us consider a different way to introduce a $G$-invariant
(left-invariant) skewsymmetric form on $M = G/G_{0}$. Let us suppose that $G$
is a connected compact simple Lie group with center $\{\varepsilon \}$ and $G_{0}$ has
nondiscrete center and is a maximal connected proper subgroup of 
$G$.
Really we need suppose that a center of $G_{0}$, $Z_{G_{0}}$ is isomorphic to
$U(1)$ (a circle group). In this case (see the first position of 
Ref.~[86] p.~382) there is an Hermitian structure (really K\"ahlerian) invariant with
respect to the action of the group~$G$.

Let us denote the unique $G$-invariant almost complex structure on $M$ by
$J^{2} = -1$. $J$ is integrable to complex structure and let us denote by
$h^{0}$ a
$G$-invariant Riemannian structure on $M$, which is {\bf Hermitian} 
$h^{0}(JX,JY) =h^{0}(X,Y)$, $X,Y \in \Tan (M)$ and
$\widehat{\widetilde{\nabla}}_{x}J = 0$,
$\widehat{\widetilde{\nabla}}$ is covariant derivative induced by
a Riemannian connection on $M$.

This allows us to introduce a skewsymmetric $G$-invariant 2-form on
$M$. One has 
$$ 
k^{0}(X,Y)\mathbin{\buildrel \text{\rm df}\over =} h^{0}(JX,Y).\eqno(4.57)
$$
One easily gets that 
$$\eqalignno{
k^{0}(X,Y) &= - k^{0}(Y,X),&(4.58)\cr
\widehat{\widetilde{\nabla}}_{z}k^{0} &= 0&(4.59)\cr}
$$
and $k^{0}$ is $G$-invariant.

Thus one can write
$$
\widetilde{g}(X,Y) = h^{0}(X,Y) + \zeta k^{0}(X,Y) = h^{0}(X,Y) + 
\zeta h^{0}(JX,Y).\eqno(4.60)
$$
Let us notice that we get a stronger result, for a symmetric space then
for a group $G$. The form $k^{0}$ is defined uniquely for any $G$-invariant
metric. We can define the metric $h^{0}$ using a bi-invariant tensor on
$G$ (a
Killing-Cartan tensor), supposing the reductive decomposition of $\fg$ or
even the symmetry requirement. For the group $G$ is compact any
bi-invariant tensor on $G$ is proportional to Killing--Cartan tensor. In
this way any nonsymmetric metric on $M = G/G_{0}$ is proportional to
(4.60). To be safe we can use in the place 
of~$h^{0}$, ${1\over 2}(h^{0}(X,Y) +
h^{0}(JX,JY))$.

Thus we get a left-invariant $(G$-invariant) nonsymmetric tensor on
$G$, such that
$$
\widehat{\widetilde{\nabla}}_{x}g= 0.\eqno(4.61)
$$
In particular let us notice that $S^{2}$ is an Hermitian symmetric space
(and K\"ahlerian) with SO(3) invariant scalar product and SO(3) invariant
symplectic form.

In the first position of Ref.~[86] p.~518 S.Helgason quoted all the
K\"ahlerian structures on the symmetric spaces of compact type:
$$\displaylines{
\text{\rm SU}\,(p + q)/S(U_{p}\times U_{q}) ,\ \dim M = 2pq,\quad\ p,q =
1,2\ldots,\cr
\SO(2n)/U(n) ,\ \dim M = n(n-1),\quad\ n = 1,2\ldots ,\cr
\hfill \SO(p +2)/\SO(p)\times \SO(2) ,\ \dim M = 2p,\quad\ p = 2,3\ldots ,\hfill\llap{(4.62)}\cr
\Sp(n)/U(n) ,\ \dim M = n(n + 1),\quad\ n = 1,2\ldots ,\cr
(\fe_{6(-78)}, so(10) + R),\ \dim M = 32,\cr
(\fe_{7(-133)} , \fe_{6} + R),\ \dim M = 54.\cr}
$$
On every of these manifold of even dimension (real) and compact there
is a K\"ahlerian structure. Thus there is a symmetric metric and a
skewsymmetric 2-form (nondegenerate) inducing a nonsymmetric metric
which is covariantly constant with respect to the Riemannian connection
(induced by a symmetric metric tensor) and left-invariant with respect
to the action of the group $G$ on $M$. In this way a torsion of the
connection compatible with nonsymmetric tensor on $M$ vanishes. This has
some interesting consequences, which we use in further sections. The
skewsymmetric $G$-invariant 2-form on $M$ induces in a natural way a
skewsymmetric left-$G$-invariant form on $G$, which together with the
Killing--Cartan form on $G$ form a nonsymmetric metric on $G$. Now, however
the covariant derivative of the skewsymmetric tensor $G$ with respect to
the Riemannian connection on $G$ (induced by a Killing--Cartan tensor)
does not vanish in general.

In this way the torsion of the connection compatible with the
nonsymmetric tensor on $G$ does not vanish. This is different as for
the Hermitian homogeneous space with left-invariant 2-form. In 
sec.~5 we find a general shape for such a connection which can be
applied for our case i.e for the skewsymmetric form obtained from the
left-invariant 2-form on $M$.

Let us calculate how many nonsymmetric (Einstein--Kaufmann)
left-invariant structure we get in this way from~(4.62).
One gets
$$
E_{6} - 1,\quad\ E_{7} - 1,\vadjust{\vskip-15pt}\eqno(4.63)
$$
$$\eqalignno{\vadjust{\vskip-15pt}
\Sp(n) - 1 ,\quad\ &n = 1,2\ldots ,\cr
\SU(n) - \biggl[{n+1\over 2}\biggr] ,\quad\ &n = 2,3\ldots ,\cr
\SO(2n + 1) - 1 ,\quad\ &n = 1,2\ldots ,\cr
\SO(2n) - 2 ,\quad\ &n = 1,2\ldots ,\cr}
$$
Thus except $SU(n)$ we have really 1 or 2 inequivalent Einstein--Kaufmann
connections on $G$ obtained from K\"ahlerian structure on a homogeneous
space (being a symmetric space).

Moreover for lower dimensions we have some coincidences, i.e. 
$$\eqalignno{
\SU(2)/S(U_{1}\times U_{1}) &\simeq \Sp(1)/U(1),\cr
\SO(5)/\SO(3)\times \SO(2) &\simeq \Sp(2)/U(2),&(4.64)\cr
\SU(4)/S(U(2)\times U(2)) &\simeq \SO(6)/U(3),\cr
\SO(8)/\SO(4)\times \SO(4) &\simeq \SO(8)/U(4).\cr}
$$
Notice the absence in (4.63) $G_{2}$, $F_{4}$ and $E_{8}$. Moreover we can introduce
on those groups left-invariant skewsymmetric forms. Let us notice that
in (4.62) we have the symmetric space $\SU(4)/S(U_{3}\times U_{1})$ which can have
some physical applications. The same we can say for $\SO(10)/U(5)$ and for
the last two symmetric spaces in (4.62).

We can introduce almost complex (almost K\"ahlerian) structures on
homogeneous spaces which are not symmetric spaces. They have the
following shape: $G/K$, where $K$ is a fixed point set of an automorphism
of $G$ of order 3. If $\dim (G/K) \ge 6$ these manifolds are not complex
(K\"ahlerian). For example we have the following possibilities: 
\vskip6pt
\puww{1.}{1.} $G = \SU(n)/Z_{n}$,

\puww{3.}{} $K = S(U(r_{1})\times U(r_{2})\times U(r_{3}))/Z_{n}$,

\puww{3.}{} $0 \le r_{1} \le r_{2} \le r_{3} $, $r_{1} + r_{2} + r_{3} = n$,
$$ 
0 < r_{2}\eqno(4.65)
$$

\puww{3.}{2.} $G =\SO(2n + 1)$,

\puww{3.}{} $ K = U(r)\times \SO(2n - 2r + 1)$,

\puww{3.}{} $ 1 \le r \le n$.

\puw{3.} $G = \Sp(n)/Z_{2}$,

\puww{3.}{} $K =\{U(r)\times \Sp(n-r)\}/Z_{2} $, $1 \le r \le n$.

\puww{3.}{4.} $G =\SO(2n)/Z_{2} $, $n\ge 3$,

\puww{3.}{} $K =\{U(r)\times \SO(2n-2r)\}/Z_{2} $, $1 \le r \le n$.

\vskip6pt
Except the above example we have the following
\obraz{$$\eeqalign{
&G = G2,\quad\ &K &= U(2),\cr
&G = F4,\quad\ &K &= (\Spin (7)\times T^{1})/Z_{2},\cr
&G = F4,\quad\ &K &= (\Sp(3)\times T^{1})/Z_{2},\cr
&G = E6/Z_{3} ,\quad\ &K &= (\SO(10)\times \SO(2))/Z_{2},\cr
&G = E6/Z_{2} ,\quad\ &K &= (S(U(5)\times U(1))\times \SU(2))/Z_{2},\cr
&G = E6/Z_{2} ,\quad\ &K &= (\SU(6)/Z_{3}\times T^{1})/Z_{2},\cr
&G = E6/Z_{2} ,\quad\ &K &= [(\SO(8)\times \SO(2))\times \SO(2)]/Z_{2},\cr
&G = E7/Z_{2} ,\quad\ &K &= E6\times T^{1}/Z_{3},\cr
&G = E7/Z_{2} ,\quad\ &K &= (\SU(2)\times (\SO(10)\times \SO(2)))/Z_{2},\cr
&G = E7/Z_{2} ,\quad\ &K &= S(U(7)\times U(1))/Z_{1},\cr
&G = E8,\quad\ &K &= \SO(14)\times \SO(2),\cr
}\eqno(4.66)
$$
$$\eeqalign{
&G = E8,\quad\ &K &= (E7\times T^{1})/Z_{2},\cr
&G = G2,\quad\ &K &= \SU(3),\cr
&G = F4,\quad\ &K &= (\SU(3)\times \SU(3))/Z_{3},\cr
&G = E6/Z_{3} ,\quad\ &K &= (\SU(3)\times \SU(3)\times 
\SU(3))/(Z_{2}\times Z_{2}),\cr
&G = E7/Z_{3} ,\quad\ &K &= (\SU(3)\times (\SU(6)/Z_{2}))/Z_{3},\cr
&G = E8,\quad\ &K &= (\SU(3)\times E6)/Z_{3},\cr
&G = E8,\quad\ &K &= \SU(9)/Z_{3},\cr
&G = \Spin(8) ,\quad\ &K &= \SU(3)/Z_{3},\cr
&G =\Spin(8) ,\quad\ &K &= G2.\cr}\eqno(4.66~\hbox{\sevenrm cont.})
$$
}\bez{$$\eeqalign{
&G = G2,\quad\ &K &= U(2),\cr
&G = F4,\quad\ &K &= (\Spin (7)\times T^{1})/Z_{2},\cr
&G = F4,\quad\ &K &= (\Sp(3)\times T^{1})/Z_{2},\cr
&G = E6/Z_{3} ,\quad\ &K &= (\SO(10)\times \SO(2))/Z_{2},\cr
&G = E6/Z_{2} ,\quad\ &K &= (S(U(5)\times U(1))\times \SU(2))/Z_{2},\cr
&G = E6/Z_{2} ,\quad\ &K &= (\SU(6)/Z_{3}\times T^{1})/Z_{2},\cr
&G = E6/Z_{2} ,\quad\ &K &= [(\SO(8)\times \SO(2))\times \SO(2)]/Z_{2},\cr
&G = E7/Z_{2} ,\quad\ &K &= E6\times T^{1}/Z_{3},\cr
&G = E7/Z_{2} ,\quad\ &K &= (\SU(2)\times (\SO(10)\times \SO(2)))/Z_{2},\cr
&G = E7/Z_{2} ,\quad\ &K &= S(U(7)\times U(1))/Z_{1},\cr
&G = E8,\quad\ &K &= \SO(14)\times \SO(2),\cr
&G = E8,\quad\ &K &= (E7\times T^{1})/Z_{2},\cr
&G = G2,\quad\ &K &= \SU(3),\cr
&G = F4,\quad\ &K &= (\SU(3)\times \SU(3))/Z_{3},\cr
&G = E6/Z_{3} ,\quad\ &K &= (\SU(3)\times \SU(3)\times 
\SU(3))/(Z_{2}\times Z_{2}),\cr
&G = E7/Z_{3} ,\quad\ &K &= (\SU(3)\times (\SU(6)/Z_{2}))/Z_{3},\cr
&G = E8,\quad\ &K &= (\SU(3)\times E6)/Z_{3},\cr
&G = E8,\quad\ &K &= \SU(9)/Z_{3},\cr
&G = \Spin(8) ,\quad\ &K &= \SU(3)/Z_{3},\cr
&G =\Spin(8) ,\quad\ &K &= G2.\cr}\eqno(4.66)
$$
}One can find all the details concerning a construction of an almost complex
structure~on the above homogeneous spaces in the fourth
position of Ref.~[86]. Due to the almost complex structure on $G/K$ we can
construct a left-invariant skewsymmetric form on~$G$. Let us remind to
the reader that in this case $G/K$ is not a symmetric space.

In this way we can introduce some nonsymmetric metrics on:
$$\displaylines{
\SO(2n) - n , n \ge 3,\cr
\SO(8) - 6,\cr
\Sp(n) - n,\cr
\SO(2n + 1) - n,\cr}
$$
$\SU(n) - a$ number of partitions of $n$ in such a way that $n = r_{1} + r_{2} + r_{3}$,
$r_{1} \le r_{2} \le r_{3} $, $r_{2} > 0 $, $G2 - 2 $, $F4 - 3
$, $E6 - 5 $, $E7 - 4 $, $E8 - 4$
different Einstein--Kaufmann structures coming from almost K\"ahlerian
structures on (4.65--66).

All details concerning constructions presented here can be found
in Refs. [7], [8], [9], [10], [16], [21]--[27], [51], [52], [64], [70],
[71]--[77].

\vfill\eject

\null
\vskip36pt plus4pt minus4pt
\centerline{{\four\bf 5. The Nonsymmetric Jordan--Thiry Theory}}
\vskip3pt
\centerline{{\four\bf over $V=E\times G/G_{0}$}} 
\vskip24pt plus2pt minus2pt
\write0{5. The Nonsymmetric Jordan--Thiry Theory
over $V=E\times G/G_{0}$ \the\pageno}

Let $P$ be the principal fibre bundle with the structural group $H$,
over $V = E \times G/G_{0}$ with a projection $\pi $ and let us define on this bundle
a connection $\omega$ (Fig.~2A). Let us suppose that $H$ is semisimple. On the
base $V = E \times G/G_{0}$ we define a nonsymmetric metric tensor such that:
$$\eqalign{
\gamma _{\rAB} &= \gamma _{(\rAB)} + \gamma _{[\rAB]},\cr
\gamma _{\rAB} \gamma ^{\rCB} &= \gamma _{\rBA} \gamma ^{\rBC} =
\delta^{\rC}_{\rA},\cr}\eqno(5.1)
$$
where the order of the indices is important. We also define on $V$ the
connection $\overline{\omega}^{\rA}{}_{\rB}$,
$$
\overline{\omega }^{\rA}{}_{\rB} =
\overline{\varGamma}^{\rA}{}_{\rBC}\overline{\theta }^{\rC}\eqno(5.2)
$$
such that:
$$
\ov{D}\gamma _{\text{\rm A+B}-} = \ov{D} \gamma _{\rAB} - \gamma
_{\rAD}\ov{Q}^{\rD}{}_{\rBC}(\overline{\varGamma}) \theta ^{\rC} =
0,\eqno(5.3)
$$
where $\ov{D}$ is the exterior covariant derivative with respect to 
$\overline{\omega }^{\rA}{}_{\rB}$ and
$\ov{Q}^{\rA}{}_{\rBC}(\overline{\varGamma})$ is the torsion of
$\overline{\omega }^{\rA}{}_{\rB}$. Using (4.39) one easily finds that the
connection (5.2) has the following shape:
$$
\overline{\omega }^{\rA}{}_{\rB} = \left(\vcenter{\offinterlineskip\tabskip0pt
\halign{\strut\tabskip5pt plus10pt minus10pt
\hfill#\hfill&#&\hfill#\hfill\tabskip0pt\cr
$\vrule height0pt depth6pt width0pt \widetilde{\overline{\omega }}^\alpha{}_\beta $ &\vrule height10pt& $0$\cr 
\noalign{\hrule}
$\vrule height12pt depth4pt width0pt 0$ &\vrule height14pt& $\widehat{\overline{\omega
}}^{\tilde a}{}_{\tilde b}$\cr}}\right),\eqno(5.4)
$$
where $\widetilde{\overline{\omega }}^\alpha {}_\beta$ is the connection on the space-time $E$ and 
$\widehat{\overline{\omega }}^{\tilde a}{}_{\tilde b}$ is the
connection on the manifold $M = G/G_{0}$ with the following properties:
$$\displaylines{
\widetilde{{\ov{D}}} g_{\alpha +\beta -} = \widetilde{{\ov{D}}}
g_{\alpha \beta } - g_{\alpha \delta } \widetilde{\ov{Q}}^\delta
{}_{\beta \gamma }(\widetilde{{\overline{\varGamma}}}) \theta ^{\gamma } =
0,\cr
\hfill \widetilde{\ov{Q}}^{\alpha }{}_{\beta \alpha
}(\widetilde{\overline{\varGamma}}) = 0,\hfill\llap{(5.5)}\cr
\hfill \widehat{{\ov{D}}} g_{\tilde a+\tilde b-} = \widehat{{\ov{D}}}
g_{\tilde a\tilde b}- g_{\tilde a\tilde d}\widehat{{\ov{Q}}}^{\tilde
d}{}_{\tilde b\tilde c}(\widehat{{\overline{\varGamma}}}) \overline{\theta
}^{\tilde c} = 0,\hfill\llap{(5.6)}\cr}
$$
$\ov{D}$ is the exterior covariant derivative with respect to the connection
$\overline{\omega }^\alpha {}_\beta $ on $E$, $\ov{Q}^{\alpha }{}_{\beta
\gamma }(\overline{\omega})$ is the tensor of torsion for 
$\overline{\omega }^{\alpha }{}_{\beta }$. $\widehat{\ov{D}}$ means the
exterior covariant derivative with respect to the connection 
$\widehat{\overline{\omega }}^{\tilde a}{}_{\tilde b}$ and
$\widehat{\ov{Q}}^{\tilde a}{}_{\tilde b\tilde
c}(\widehat{\overline{\varGamma}})$ is the tensor of torsion for 
$\widehat{\overline{\omega }}^{\tilde a}{}_{\tilde b}$. The second condition of
Eq.~(5.5) is not necessary to define Einstein connection on $E$. Moreover
we suppose it to be in line with some classical results. On the
space-time $E$ we also define the second affine connection
$\ov{W}^{\alpha }{}_{\beta }$ such that:
$$
\ov{W}^{\alpha }{}_{\beta } = \overline{\omega }^{\alpha }{}_{\beta } - 
{2\over 3} \delta ^{\alpha }{}_{\beta }\ov{W},\eqno(5.7)
$$
where
$$
\ov{W}= \ov{W}_{\gamma } \overline{\theta }^\gamma = 
{1\over 2}(\ov{W}^{\sigma }{}_{\gamma \sigma } - \ov{W}^{\sigma
}{}_{\sigma \gamma })\overline{\theta }^\gamma .
$$
Thus on the space-time $E$ we have all the geometrical quantities from
N.G.T.: two connections $\overline{\omega }^{\alpha }{}_{\beta }$ and $\ov{W}^{\alpha }{}_{\beta }$ and the nonsymmetric metric
$g_{\alpha \beta}$ (see [18], [34], [47], [65], [66]). Now let us turn to the nonsymmetric
metrization of the bundle $P$. According to section 4 we have:
$$
\varkappa _{\tilde{\rA}\tilde{\rB}} = 
\left(\vcenter{\offinterlineskip\tabskip0pt
\halign{\strut\tabskip5pt plus10pt minus10pt
\hfill#\hfill&\vrule height10pt#&\hfill#\hfill\tabskip0pt\cr
$\vrule height0pt depth6pt width0pt \gamma _{\rAB}$ && $0$\cr 
\noalign{\hrule}
$\vrule height9pt depth4pt width0pt 0$ &&
$\rho^2\ell_{ab}$\cr}}\right),\eqno(5.8)
$$
where $\rho = \rho (x,y)$ is a scalar field on $E \times M$ and
$$\displaylines{
\gamma _{\rAB} =
\left(\vcenter{\offinterlineskip\tabskip0pt
\halign{\strut\tabskip5pt plus10pt minus10pt
\hfill#\hfill&\vrule height10pt#&\hfill#\hfill\tabskip0pt\cr
$\vrule height0pt depth6pt width0pt g_{\alpha \beta }$ && $0$\cr 
\noalign{\hrule}
$\vrule height9pt depth4pt width0pt 0$ &&
$r^2g_{\tilde{a}\tilde{b}}$\cr}}\right)\quad\
\hbox{and}\cr
\ell _{ab} = h_{ab} + \xi k_{ab} ,\quad\ \det (\ell _{ab}) \neq 
0\cr}
$$
or
$$\eqalign{
\varkappa _{(\tilde{\rA}\tilde{\rB})} \theta ^{\tilde{\rA}} \otimes 
\theta ^{\tilde{\rB}} &= \gamma _{(\rAB)} \theta ^{\rA} \otimes \theta
^{\rB} + \rho ^{2} h_{ab} \theta ^{a} \otimes \theta ^{b},\cr
\varkappa _{[\tilde{\rA}\tilde{\rB}]}\theta ^{\tilde{\rA}}\wedge \theta
^{\tilde{\rB}} &= \gamma _{[\rAB]} \theta ^{\rA}\wedge \theta ^{\rB} + 
\rho ^{2} \xi k_{ab} \theta ^{a}\wedge \theta ^{b},\cr}\eqno(5.9)
$$
where $\theta ^{a} = \lambda \omega ^{a}$. From the Kaluza--Klein Theory and
Jordan--Thiry we know
that $\lambda $ is proportional to $\sqrt{G_{N}}$ (see [1], [2], [7],
[8], [9], [10], [18]), $\lambda \sim \sqrt{G_{N}}$. We work with such a system of units that 
$\lambda = 2$.

\vskip4pt plus2pt
Now we define on $P$, a connection $\omega^{\tilde{\rA}}{}_{\tilde{\rB}}$ right-invariant
(Ad-covariant) with respect to the action of group $H$ on $P$ such that:
$$
D\gamma _{\tilde{\rA}+\tilde{\rB}-} = D\gamma _{\tilde{\rA}\tilde{\rB}} - 
\gamma _{\tilde{\rA}\tilde{\rD}}
Q^{\tilde{\rD}}{}_{\tilde{\rB}\tilde{\rC}}
({\varGamma} ) \theta ^{\tilde{\rC}} = 0,\eqno(5.10)
$$
where $\omega^{\tilde{\rA}}{}_{\tilde{\rB}} =
{\varGamma}^{\tilde{\rA}}{}_{\tilde{\rB}\tilde{\rC}} \theta
^{\tilde{\rC}}$. $D$ is the exterior covariant derivative with
respect to the connection $\omega^{\tilde{\rA}}{}_{\tilde{\rB}}$ 
and $Q^{\tilde{\rA}}{}_{\tilde{\rB}\tilde{\rC}}({\varGamma})$ is the tensor of torsion for
the connection $\omega^{\tilde{\rA}}{}_{\tilde{\rB}}$. After some
calculations one finds:
$$\displaylines{
\omega^{\tilde{\rA}}{}_{\tilde{\rB}}=\hfill\cr
\hfill
\left(\vcenter{\nine\offinterlineskip\tabskip0pt
\halign{\strut\tabskip5pt plus10pt minus10pt
\hfill#\hfill&\vrule height10pt#&\hfill#\hfill\tabskip0pt\cr
$\vrule height0pt depth6pt width0pt \pi^*(\overline{\omega
}^{\rA}{}_{\rB})-\rho^2\ell_{db}\gamma ^{\rMA}L^d{}_{\rMB}\theta ^b$ &&
$L^a{}_{\rBC}\theta ^C-{\textstyle{1\over \rho}}\gamma
_{\rBD}\widetilde{\gamma}^{(\rDC)}\rho_{,C}\theta ^a $\cr 
\noalign{\hrule}
$\vrule height9pt depth4pt width0pt \rho^2\ell_{bd}\gamma
^{\rAB}(2H^d{}_{\rCB}-L^d{}_{\rCB})\theta ^C-\rho\widetilde{\gamma
}^{(\rAB)}\rho_{,\rB}\ell_{bc}\theta ^C$ &&
${\textstyle{1\over \rho}}\gamma
_{\rBD}\widetilde{\gamma}^{(\rDC)}\rho_{,C}\delta ^a_b\theta
^{\rB}+\widetilde{\omega }^a{}_b$\cr}}\right),\quad (5.11)\cr}
$$
where $\widetilde{\gamma }^{(\rAB)}$ is the inverse tensor for $\gamma
_{(\rAB)}$
$$\eqalign{
\gamma _{(\rAB)} \widetilde{\gamma }^{(\rAC)} &= \delta
^{\rC}{}_{\rB},\cr
L^{d}{}_{\rMB} &= -L^{d}{}_{\rBM},\cr}\eqno(5.12)
$$
is an Ad-type tensor (Ad-covariant) on $P$ such that
$$\displaylines{
\hfill\ell _{dc} \gamma _{\rMB} \gamma ^{\rCM} L^{d}{}_{\rCA} + 
\ell _{cd} \gamma _{\rAM} \gamma ^{\rMC} L^{d}{}_{\rBC} = 
2\ell _{cd} \gamma _{\rAM} \gamma ^{\rMC}
H^{d}{}_{\rBC},\hfill\llap{(5.13)}\cr
\widetilde{\omega }^{a}{}_{b} = \widetilde{{\varGamma} }^{a}{}_{bc} 
\theta ^{c},\cr
\hfill \ell _{db} \widetilde{{\varGamma} }^{d}{}_{ac} + 
\ell _{ad}\widetilde{{\varGamma} }^{d}{}_{cb} = -\ell
_{db}C^{d}{}_{ac},\hfill\llap{(5.14)}\cr
\widetilde{{\varGamma} }^{d}{}_{ac} = -\widetilde{{\varGamma}
}^{d}{}_{ca} ,\quad\ \widetilde{{\varGamma} }^{d}{}_{ad} = 0.\cr}
$$
We define on $P$ a second connection:
$$
W^{\tilde{\rA}}{}_{\tilde{\rB}} = \omega ^{\tilde{\rA}}{}_{\tilde{\rB}} - 
{4\over 3(m+2)} \delta ^{\tilde{\rA}}{}_{\tilde{\rB}}\ov{W}.\eqno(5.15)
$$
Thus we have on $P$ all $(m+4)$-dimensional analogues of geometrical
quantities from N.G.T., i.e.:
$$
W^{\tilde{\rA}}{}_{\tilde{\rB}} , \omega
^{\tilde{\rA}}{}_{\tilde{\rB}}\quad\hbox{and}\quad \varkappa
_{\tilde{\rA}\tilde{\rB}} .
$$
Formulae (5.11)--(5.13) are similar to the analogous formulae from
Refs.~[23], [24] (see also [18]). The one difference is that now we have $E \times M$ in
place of $E$. The connection (5.15) is analogous to the connection $W$
from Refs.~[19], [23] and from Part~4. The form $\ov{W}$ is horizontal one,
hor $\ov{W} = \ov{W}$ (in the sense of the connection $\omega $ on the bundle~$P$).

Let us calculate the Moffat-Ricci curvature scalar for the
connection $W^{\tilde{\rA}}{}_{\tilde{\rB}}$.
$$
R(W) = \varkappa ^{\tilde{\rA}\tilde{\rB}} 
\biggl(R^{\tilde{\rC}}{}_{\tilde{\rA}\tilde{\rB}\tilde{\rC}}(W)+
{1\over
2}R^{\tilde{\rC}}{}_{\tilde{\rC}\tilde{\rA}\tilde{\rB}}(W)\biggr),\eqno(5.16)
$$
where $R^{\tilde{\rA}}{}_{\tilde{\rB}\tilde{\rC}\tilde{\rD}}(W)$ is the curvature tensor for the connection
$W^{\tilde{\rA}}{}_{\tilde{\rB}}$.
Using results from Refs.\ [19], [23], [24] one gets:
$$\eqalignno{
R(W) ={}&\ov{R}(\ov{W}) + {1\over r^{2}}
R(\widehat{\overline{\varGamma}}) + {1\over \lambda ^{2} \rho ^{2}} 
\widetilde{R}(\widetilde{{\varGamma} })+&\cr
&- {\lambda ^{2} \rho ^{2}\over 4} \ell _{ab}
(2H^{a}H^{b}-L^{a\rMN}H^{b}{}_{MN})
- {2\lambda ^{2}\widetilde{\rM}\over \rho ^{2}} \widetilde{\gamma
}^{(\rBN)}\rho _{,\rB} \rho _{,\rN} + \un P(\rho ),\qquad\ &(5.17)\cr\noalign{\vskip-3pt}}
$$
where:
$$\eqalignno{\noalign{\vskip4pt}
\un P(\rho ) ={}& {\lambda ^{2}n\over 8\rho } \gamma _{\rDB} 
\widetilde{\gamma }^{(\rDC)} \rho _{,\rC} \gamma
^{\rMN}\ov{Q}^{\rB}{}_{\rMN}(\overline{\omega }) +\cr
\noalign{\vskip4pt}
&+ {\lambda ^{2}n\over 4\rho ^{2}}\overline{\nabla}_{\rA} (\rho
\widetilde{\gamma}^{(\rAB)}\rho _{,\rB}) + {\lambda ^{2}\over 4} 
n\gamma ^{\rBC}\overline{\nabla}_{\rC} \biggl({1\over \rho }\gamma _{\rBD}
\widetilde{\gamma }^{(\rDE)}\rho _{,\rE}\biggr) +\cr
\noalign{\vskip4pt}
&+ {\lambda ^{2}\rho \over 8} \gamma ^{\rMN} \biggl\{\overline{\nabla}_{\rM} 
\biggl({1\over \rho }\gamma _{\rDN}\widetilde{\gamma }^{(\rDC)}\rho
_{,\rC}\biggr)
- \overline{\nabla}_{\rN} \biggl({1\over \rho }\gamma _{\rDM}\widetilde{\gamma
}^{(\rDC)}\rho _{,\rC}\biggr)\biggr\},&(5.18)\cr
&\qquad\ \widetilde{\rM} = \ell ^{[dc]} \ell _{[dc]} - n(n-1),&(5.19)\cr}
$$
$\ov{Q}^{\rB}{}_{\rMN}(\overline{\omega })$ means the torsion of the connection 
$\overline{\omega }^{\rA}{}_{\rB}$ on $V = E \times M = E \times G/G_{0}$
and $\overline{\nabla}_{\rA}$ is the covariant derivative with respect to this connection.
$\ov R(\ov W)$ is the Moffat--Ricci curvature scalar on the space-time $E$ for the
connection $\ov{W}^{\alpha }{}_{\beta }$.
$R(\widehat{\overline{\varGamma}})$ is the Moffat--Ricci curvature scalar for the
connection $\widehat{\overline{\omega }}^{\alpha }{}_{\beta }$ on the homogeneous space $M =
G/G_{0}$, $\widetilde{R}(\widetilde{{\varGamma} })$ is the
Moffat--Ricci curvature scalar for the connection $\widetilde{\omega
}^{a}{}_{b}$. Notice that in
the curvature scalar we pass from $\lambda =2$ to the arbitrary value of this
constant.
$$\eqalignno{
H^{a} ={}& \gamma ^{[\rAB]} H^{a}{}_{[\rAB]} = g^{[\alpha \beta ]} 
H^{a}{}_{\alpha \beta } + {1\over r^{2}} g^{[\tilde a\tilde b]} 
H^{a}{}_{\tilde a\tilde b},&(5.20)\cr
L^{aMN} ={}& \gamma ^{\rAM} \gamma ^{\rBN} L^{a}{}_{\rAB} = 
\delta ^{\rM}{}_{\mu } \delta ^{\rN}{}_{\gamma } g^{\alpha \mu } 
g^{\beta \gamma } L^{a}{}_{\alpha \beta } +\cr
&+ {1\over r^{2}} (g^{\alpha \mu } g^{\tilde b\tilde n}
L^{a}{}_{\alpha \tilde{b}} + g^{\tilde a\tilde n} g^{\beta \gamma } 
L^{a}{}_{\tilde a\beta }) \delta ^{\rM}{}_{\mu } \delta
^{\rN}{}_{\tilde n} +\cr
&+ {1\over r^{4}} g^{\tilde a\tilde m} g^{\tilde b\tilde n}
L^{a}{}_{\tilde a\tilde b} \delta ^{\rM}{}_{\tilde m} \delta
^{\rN}{}_{\tilde n}.&(5.21)\cr}
$$
Let us consider the condition (5.13). One gets:
$$\eqalignno{
\ell _{dc} g_{\mu \beta } g^{\gamma \mu } L^{d}{}_{\gamma \alpha } + 
\ell _{cd} g_{\alpha \mu } g^{\mu \gamma } L^{d}{}_{\beta \gamma } &= 
2\ell _{cd} g_{\alpha \mu } g^{\mu \gamma } H^{d}{}_{\beta \gamma
},&(5.22)\cr
\ell _{dc} g_{\tilde m\tilde b} g^{\tilde c\tilde m} L^{d}{}_{\tilde
c\tilde a} + \ell _{cd} g_{\tilde a\tilde m} g^{\tilde m\tilde c} 
L^{d}{}_{\tilde b\tilde c} &= 2\ell _{cd} g_{\tilde a\tilde m} 
g^{\tilde m\tilde c} H^{d}{}_{\tilde b\tilde c},&(5.23)\cr
\ell _{dc} g_{\mu \beta } g^{\gamma \mu } L^{d}{}_{\gamma \tilde a} + 
\ell _{cd} g_{\tilde a\tilde m} g^{\tilde m\tilde c} L^{d}{}_{\beta \tilde
c} &= 2\ell _{cd} g_{\tilde a\tilde m} g^{\tilde m\tilde c}
H^{d}{}_{\beta \tilde c}.&(5.24)\cr}
$$
One finds that:
$$\eqalignno{
-\ell _{ab} L^{a\rMN} H^{b}{}_{\rMN} ={}& -\ell _{ab}
\biggl(g^{\alpha \mu } g^{\beta \nu } L^{a}{}_{\alpha \beta } 
H^{b}{}_{\mu \nu }+\cr
&+ {2\over r^{2}} g^{\alpha \mu } g^{\tilde b\tilde n}
L^{a}{}_{\alpha \tilde b} H^{b}{}_{\mu \tilde n}
+ {1\over r^{4}} g^{\tilde a\tilde m} g^{\tilde b\tilde n} L^{a}{}_{\tilde
a\tilde b} H^{b}{}_{\tilde m\tilde n}\biggr)=\cr
={}&-\ell _{ab} \biggl(L^{a\mu \nu }H^{b}{}_{\mu \nu }+{2\over r^{2}}
g^{\tilde b\tilde n}L^{\alpha \mu }{}_{\tilde b}H^{b}{}_{\mu \tilde
n}+\cr
&+ {1\over r^{4}}g^{\tilde a\tilde m}g^{\tilde b\tilde n}L^{a}{}_{\tilde
a\tilde b}H^{b}{}_{\tilde m\tilde n}\biggr),&(5.25)\cr}
$$
where: 
$$\eqalignno{
L^{\alpha \mu \nu } &= g^{\alpha \mu } g^{\beta \nu } L^{a}{}_{\alpha
\beta },\cr
L^{\alpha \mu }{}_{\tilde b} &= g^{\alpha \mu } L^{a}{}_{\alpha \tilde
b}.\cr}
$$
For $\ell _{ab} H^{a} H^{b} = h_{ab} H^{a} H^{b}$, we have the following:
$$
h_{ab} H^{a} H^{b} = h_{ab} H^{a}{}_{0} H^{b}{}_{0} + {2\over r^{2}} 
h_{ab} H^{a}{}_{0} H^{b}{}_{1} + {1\over r^{4}} h_{ab} H^{a}{}_{1}
H^{b}{}_{1},\eqno(5.26)
$$
where 
$$ 
H^{a}{}_{0} = g^{[\alpha \beta ]} H^{a}{}_{\alpha \beta }\eqno(5.27)
$$
and
$$
H^{a}{}_{1} = g^{[\tilde a\tilde b]} H^{a}{}_{\tilde a\tilde b}.\eqno(5.28)
$$
And finally we get for $R(W)$:
$$\eqalignno{
R(W) ={}& \ov{R} (\ov{W}) + {1\over r^{2}}
R(\widehat{\overline{\varGamma}}) + {4\over \lambda ^{2} r^{2}} 
\widetilde{R}(\widetilde{{\varGamma} }) 
- {\lambda ^{2} \rho ^{2}\over 4} \ell _{ab} (2H^{a}{}_{0}H^{b}{}_{0}-
L^{a\mu \nu }H^{b}{}_{\mu \nu }) +\cr
\noalign{\vskip6pt}
&- {\lambda ^{2} \rho ^{2}\over 4r^{2}} \ell _{ab} 
(4H^{(a}_{0}H^{b)}_{1}-2g^{\tilde b\tilde n}L^{a\mu }_{\tilde b}
H^{b}{}_{\mu \tilde n}) +\cr
\noalign{\vskip6pt}
&- {\lambda ^{2} \rho ^{2}\over 4r^{4}} \ell _{ab} 
(2H^{a}{}_{1}H^{b}{}_{1}-g^{\tilde a\tilde m}g^{\tilde b\tilde
n}L^{a}{}_{\tilde a\tilde b}H^{b}{}_{\tilde m\tilde n} +\cr
\noalign{\vskip6pt}
&- {\widetilde{M}\over \rho ^{2}} g_{\gamma \delta }
\widetilde{g}^{(\delta \nu )} \rho _{,\nu } \widetilde{g}^{(\gamma
\beta )} \rho _{,\beta } + \un{P}(\rho ),&(5.29)\cr}
$$
where $\widetilde{g}^{(\delta \nu )}$ is the inverse tensor of 
$g_{(\alpha \beta )}$ such that:
$$
\widetilde{g}^{(\delta \nu )} g_{(\delta \mu )} = \delta ^{\nu
}{}_{\mu }.\eqno(5.30)
$$
Let us calculate the density for the Moffat--Ricci curvature scalar for
the connection $W^{\tilde{\rA}}{}_{\tilde{\rB}}$. We have:
$$
\sqrt{|\varkappa|} R(W) = \sqrt{-g} r^{n_{1}} \sqrt{|\widetilde{g}|}
\sqrt{|\ell|} \rho ^{n} R(W),\eqno(5.31)
$$
where
$$
g = \det (g_{\alpha \beta }) ,\quad \widetilde{g} = \det (g_{\tilde a\tilde
b}) ,\quad \ell = \det (\ell _{ab})\eqno(5.32)
$$
and
$$
\varkappa = \det (\varkappa _{\tilde{\rA}\tilde{\rB}}) = g \cdot 
r^{n_{1}} \cdot \widetilde{g} \cdot \rho ^{n} \ell .\eqno(5.33)
$$
After some calculation one gets:\vadjust{\eject}
$$\eqalignno{
\sqrt{|\varkappa|} R(W) ={}& \sqrt{-g}r^{n_{1}}\sqrt{|\widetilde{g}|}
\ell \biggl\{\rho ^{n}\ov{R}(\ov{W}) + {\widetilde{R}(\widetilde{{\varGamma} })
\over \lambda ^{2}\rho ^{2-n}} + {\rho ^{n}\over r^{2}}
R(\widehat{\overline{\varGamma}}) +\cr
&+ {\lambda ^{2}\over 4} \rho ^{n+2} \ell _{ab} (2H^{a}{}_{0}H^{b}{}_{0}-
L^{a\mu \nu }H^{b}{}_{\mu \nu }) +\cr
&+ {\lambda ^{2}\over 4r^{2}} \rho ^{n+2} \ell _{ab} 
(4H^{(a}_{0}H^{b)}_{1}-2g^{\tilde b\tilde n}L^{a\mu }{}_{\tilde b}
H^{b}{}_{\mu \tilde n}) +\cr
&+ {\lambda ^{2}\over 4r^{4}} \rho ^{n+2} \ell _{ab} 
(2H^{a}_{1}H^{b}_{1}-g^{\tilde a\tilde m}g^{\tilde b\tilde n}
L^{a}{}_{\tilde a\tilde b}H^{b}{}_{\tilde m\tilde n}) +\cr
&+ {\lambda ^{2}\over 4} \rho ^{n-2} (\ov{M}\widetilde{g}^{(\gamma
\mu )} \rho _{,\gamma } \rho _{,\mu } +\cr
&+ n^{2}g^{[\mu \nu ]} g_{\delta \mu } \widetilde{g}^{(\delta \gamma )} 
\rho _{,\nu } \rho _{,\gamma })\biggr\}+ \partial
_{\rM}K^{\rM},&(5.34)\cr}
$$
where
$$
\ov{\rM}= (\ell ^{[dc]}\ell _{[dc]} -3n(n-1))\eqno(5.35)
$$
and
$$
K^{\rM} = {n\over 2} \rho ^{n-1} \sqrt{|\varkappa|}
(5\widetilde{\gamma }^{(\rMC)}-\gamma ^{\rNM}\gamma _{\rDN}
\widetilde{\gamma }^{(\rDC)})\cdot \rho _{,\rC}.\eqno(5.35')
$$
If we define an integral of action: 
$$ 
S \sim \mmint_U \sqrt{|\varkappa|} R(W) d^{m+4}x,\eqno(5.36)
$$
(where $d^{m+4}x = d^{4}x\, d\mu _{\rH}(h)\cdot dm(y)$, $d\mu _{\rH}(h)$ is a biinvariant measure on a
group $H$ and $dm(y)$ is a measure on $M$ induced by a biinvariant measure on
a group $G$, $x\in E$, $h\in H$, $y\in M)$, and the variation principle for $R(W)$ then the
full divergence $\partial _{\rM}K^{\rM}$ does not play any role. It could play a certain
role in topological problems.

One can write
$$ 
K^{\mu } = {n\over 2} \rho ^{n-1}\sqrt{g\widetilde{g}} 
(5\tilde{g}^{(\mu \gamma )}-g^{\nu \mu }g_{\delta \nu
}\widetilde{g}^{(\delta \gamma )}) \rho _{,\gamma }
$$
and 
$$ 
K^{\tilde m} = {n\over 2r^{2}} \rho ^{n-1}\sqrt{g\widetilde{g}}
(5g^{(\tilde m\tilde c)} - g^{\tilde n\tilde m} g_{\tilde d\tilde n} 
\widetilde{g}^{(\tilde d\tilde c)} ) \rho _{,\tilde c}.
$$
Thus we really only have to deal with $B(W)$:
$$
B(W) =\sqrt{\varkappa} R(W) - \partial _{\rM}K^{\rM}.\eqno(5.37)
$$
In the calculations we use results from Refs.~[23], [24],
[26], [27] and we suppose that $\rho = \rho (x)$ is really the scalar field on
the manifold $E$ (space-time). The latest means that: 
$$ 
\rho _{,\tilde a} = 0,\eqno(5.38)
$$
for every $\widetilde a= 5, 6, 7,\ldots ,n_{1}+4$.

Moreover it is interesting to consider a more general case with $\rho $
depending on $x\in E$ and $y\in M$ i.e. 
$$ 
\rho = \rho (x,y).\eqno(5.39)
$$
In this case we expand $\rho $ into a complete set of real functions defined
on $M$. 
$$ 
\rho = \sum^{\infty }_{k=0}\rho _{k}(x) \chi _{k}(y)\eqno(5.40)
$$
and consider such a $\rho $ that: 
$$ 
\|\widetilde{\rho }(x)\|^{2} = \sum^{\infty }_{k=0}\rho ^{2}_{k}(x) <
\infty.\eqno(5.41)
$$
The condition (5.41) means that $\widetilde{\rho }(x) \in \ell ^{2}$ 
(a Hilbert space) $x\in E$ i.e. 
$\ell ^{2}$--valued function defined on
$E$. If functions $\chi _{k}(y)$ defined on $M$ form
an orthogonal basis in $L^{2}(M, dm, \sqrt{|\widetilde{g}|})$ i.e. 
$$ 
\|\chi _{k}\|^{2} = {1\over V_{2}}\mmint_{\rM}\sqrt{|\widetilde{g}|}
\chi ^{2}_{k}(y)\, dm(y) < \infty\eqno(5.42)
$$
and
$$ 
(\chi _{k}, \chi _{l}) = {1\over
V_{2}}\mmint_{\rM}\sqrt{|\widetilde{g}|}\chi _{k}(y) \chi _{l}(y)\, dm(y)
= \|\chi _{k}\|^{2} \delta _{kl}\quad\ k,l = 0,1,2\ldots \eqno(5.43)
$$
and for every real $f(y)$ such that
$$\displaylines{
\hfill \|f\|^{2} = {1\over V_{2}}\mmint_{\rM}\sqrt{|\widetilde{g}|}
f^{2}(y)\, dm(y) < \infty,\hfill\llap{(5.44)}\cr
\hfill f(y) = \sum^{\infty }_{k=0}f_{k} \chi
_{k}(y),\hfill\llap{(5.45)}\cr
\hfill\sum^{\infty }_{k=0}f^{2}_{k} < \infty,\hfill\llap{(5.46)}\cr
\hfill f_{k} = {1\over V_{2}}\mmint_{\rM}\sqrt{|\widetilde{g}|}
f(y) \chi _{k}(y)\, dm(y).\hfill\llap{(5.47)}\cr}
$$
The convergence of (5.40), (5.45) is understood in the sense of $L^{2}$
norm. Thus in this more general case we have to do with a tower of
scalar fields $\rho _{k}(x)$, $k=0,1,2,3\ldots $

We can arrange a basis $\chi _{k}(y)$ in such way that:
$$
\chi _{0}(y) = 1.\eqno(5.48)
$$
It is easy to see that this function is normalized to one:
$$
\|\chi _{0}\|= 1.\eqno(5.49)
$$
Using a Schmidt orthogonalization procedure one gets remaining $\chi
_{k}$, $k=1,2\ldots $ from generalized harmonics on $M$ such that:
$$
(\chi _{k}, \chi _{l}) = \delta _{kl}.\eqno(5.50)
$$
In this way the condition (5.38) means a truncation condition for $\rho $
i.e.
$$
\rho (x) = \rho _{0}(x) + \sum^{\infty }_{k=1}\rho _{k}(x) \chi _{k}(y)
\eqno(5.51)
$$
and we consider only the first term $\rho _{0}$
$$
\rho (x) = \rho _{0}(x).\eqno(5.52)
$$
The general treatment of $\rho $ is given in section~16.

The generalized harmonics on $(M,h^{0})$ are eigenfunctions of the usual
Beltrami--Laplace operator on $(M,h^{0})$ (which is an elliptic operator for
$M$ is compact and without boundaries).
$$
{\varDelta} \eta _{k} = a_{k} \eta _{k},\eqno(5.53)
$$
where $\eta _{k}$ is an eigenfunction of ${\varDelta} $ and $a_{k}$ is an eigenvalue, $k=0,1,2\ldots $
$$
u{\varDelta} = {1\over 2} (d*d*+*d*d)\eqno(5.54)
$$
is the Beltrami--Laplace operator on $M$, $*$---means a Hodge's star and $d$
is an exterior derivative on $M$, $u$ means a volume form on $M$ ($n_{1}$-form).
Sometimes we define $\delta = *d*$. One can write ${\varDelta} $ in a more convenient form:
$$
{\varDelta} f = \ddiv(\grad(f)) = - {1\over \sqrt{|h^0|}} 
(h^{0\tilde a\tilde b}\sqrt{|h^0|}
f_{,\tilde b})_{,\tilde a},\eqno(5.54\text{\rm a})
$$
where $h^{0} = \det (h^{0}_{\tilde a\tilde b})$.

It is easy to see that $a_{k} = 0$ corresponds to a constant function (a
manifold $M$ is compact).

Functions $\eta _{k}$ form a basis in a different Hilbert space of $L^{2}$-type i.e.
$\eta _{k} \in L^{2}(M,dm)$. Moreover $L^{2}(M,dm)$ and 
$L^2(M,dm,\sqrt{|\widetilde{g}|})$ are unitary
equivalent since measures $d\mu _{1}=dm$ and $d\mu
_{2}=\sqrt{|\widetilde{g}|}\,dm$ are such that $\mu _{2}\ll\mu _{1}$
and $\mu _{1}\ll\mu _{2}$. The isomorphism
$$\displaylines{
{\bold U} : L^{2}(M,dm,\sqrt{|\widetilde{g}|}) \rightarrow 
L^{2}(M,dm),\cr
f_{2} = {\bold U}(f_{1}) =|\widetilde{g}|^{1/4} f_{1}\cr}
$$
establishes this equivalence.

The truncation procedure can be obtained in a different way taking
$$
\rho (x,y) = \rho _{0}(x) \rho _{1}(y) ,\quad\ x\in E ,\ y\in M.
$$
In this way after averaging over the manifold $M$ we get basically the
same shape of the lagrangian with exactly the same factors depending on
the scalar field $\rho _{0}(x)$.

It is worth to notice that the formulae (5.21)--(5.23) do not change
under the general conformal transformation---the redefinition of the
tensor $g_{\mu \nu }$. This is a general property of these formulae i.e.
$$
g_{\mu \nu } \rightarrow C\cdot g_{\mu \nu },\eqno(5.55)
$$
where $C = C(x)$ is the conformal factor.

$R(W)$ is invariant with respect to the right action
of the group $H$ on $\un{P}$. Thus an integration over the group $H$ is trivial.
Let us consider the following two $\Ad_{H}$-type 2-forms with values in the
Lie algebra of $H,(\fh)$. 
$$\eqalignno{
\ov{L}&={1\over 2}L^{d}{}_{\rMB}\theta ^{\rM}\wedge\theta
^{\rB}X_{d}&(5.56)\cr
\noalign{\vskip-2pt}
\noalign{\hbox{and}}
\noalign{\vskip-2pt}
\ov{Q}&={1\over 2}Q^{d}{}_{\rMB}\theta ^{\rM}\wedge\theta
^{\rB}X_{d}.&(5.57)\cr
\noalign{\vskip-2pt}
\noalign{\hbox{One gets}}
\noalign{\vskip-2pt}
\ov{L}&= {\varOmega} - {1\over 2}\ov{Q},&(5.58)\cr}
$$
where ${\varOmega} $ is the curvature of the connection $\omega $ on
$P$ (over $E \times G/G_{0})$ and
$Q^{d}{}_{\rMB}$ is torsion in additional gauge dimensions. This generalizes our
considerations in Nonabelian Kaluza-Klein Theory to higher dimensional
space-time.

\vfill\eject

S
\null
\vskip36pt plus4pt minus4pt
\centerline{{\trz 6. Dimensional Reduction Procedure}}
\vskip24pt plus2pt minus2pt
\write0{6. Dimensional Reduction Procedure \the\pageno}

Let us consider equation (5.21), (5.23) and (5.24) {\it modulo}\/ equation
(4.37), (4.38) and (4.39). One finds:
$$
\ell _{ij} g_{\mu \beta } g^{\gamma \mu } \widetilde{L}^{i}{}_{\gamma
\alpha } + \ell _{ji} g_{\alpha \mu } g^{\mu \gamma }
\widetilde{L}^{i}{}_{\beta \gamma } = 2 \ell _{ji} g_{\alpha \mu } 
g^{\mu \gamma } \widetilde{H}^{i}{}_{\beta \gamma },\eqno(6.1)
$$
where $\ell _{ij} = \ell _{cd} \alpha ^{c}{}_{i} \alpha ^{d}{}_{j}$ is a 
right-invariant nonsymmetric metric on the group $G$ and
$$
L^{c}{}_{\mu \nu } = \alpha ^{c}{}_{i} \widetilde{L}^{i}{}_{\mu \nu
},\eqno(6.2)
$$
$\widetilde{L}^{i}{}_{\mu \nu }$ plays the role of the second tensor of strength (an induction
tensor) for the Yang--Mills' field with the gauge group $G$.
$\widetilde{H}^{i}{}_{\mu \nu }$ is of
course the first tensor of strength of this field. The polarization
tensor is defined as usual (see~[18], [19], [23], [24], [26], [27]).
$$
\widetilde{L}^{i}{}_{\mu \nu } = \widetilde{H}^{i}{}_{\mu \nu } - 4\pi 
\widetilde{M}^{i}{}_{\mu \nu }.\eqno(6.3)
$$
We introduce two $\Ad_{G}$-type 2-forms with values in the Lie algebra
$\fg$ (of $G)$
$$\eqalignno{
\widetilde{L} &={1\over 2}\widetilde{L}^{i}{}_{\mu \nu }\theta ^{\mu }\wedge 
\theta ^{\nu }Y_{i},\cr
\widetilde{M} &={1\over 2}\widetilde{M}^{i}{}_{\mu \nu }\theta ^{\nu
}\wedge\theta ^{\nu }Y_{i}\cr}
$$
and one easily writes
$$
\widetilde{L} = \widetilde{{\varOmega} }_{\rE} - 4\pi \widetilde{M} = 
\widetilde{{\varOmega} }_{\rE} -{1\over 2}Q,
$$
where $\widetilde{Q} ={1\over 2}\widetilde{Q}^{i}{}_{\mu \nu }
\theta ^{\mu }\wedge\theta ^{\nu }Y_{i}$.

In this way we get a geometrical interpretation of a Yang--Mills'
induction tensor in terms of the curvature and torsion in additional
dimensions (see~Ref.~[18]).

From (5.24) and (4.38) one gets:
$$
\ell _{dc} g_{\mu \beta } g^{\gamma \mu } L^{d}{}_{\gamma \tilde a} + 
\ell _{cd} g_{\tilde a\tilde m} g^{\tilde m\tilde c} L^{d}{}_{\beta \tilde
c} = 2\ell _{cd} g_{\tilde a\tilde m} g^{\tilde m\tilde c}
\Nabla^{\gauge}{}_{\beta }({\varPhi} ^{d}{}_{\tilde c}).\eqno(6.4)
$$
From (5.22) and (4.39) one gets:
$$
\ell _{dc} g_{\tilde m\tilde b}g^{\tilde c\tilde m} L^{d}{}_{\tilde c\tilde a} 
+ \ell _{cd} g_{\tilde a\tilde m} g^{\tilde m\tilde c} L^{d}{}_{\tilde
b\tilde c} 
= 2 \ell _{cd} g_{\tilde a\tilde m} g^{\tilde m\tilde c} 
(C^{d}_{ab}{\varPhi} ^{a}_{\tilde b}{\varPhi} ^{b}_{\tilde b}-
\mu ^{d}_{\dah{i}}f^{\dah{i}}_{\tilde b\tilde c}-{\varPhi} ^{d}_{\tilde
d}f^{\tilde d}_{\tilde b\tilde c}),\eqno(6.5)
$$

One can rewrite the formula (6.5) in the following shape
$$\eqalignno{
&\ell _{dc} g_{\tilde m\tilde b} g^{\tilde c\tilde m} L^{d}{}_{\tilde c\tilde a}
(1) + \ell _{cd} g_{\tilde a\tilde m} g^{\tilde m\tilde c} L^{d}{}_{\tilde
b\tilde c}(1)= 2 \ell _{cd} g_{\tilde a\tilde m} g^{\tilde m\tilde c} C^{d}_{ab} 
{\varPhi} ^{a}{}_{\tilde b}{\varPhi} ^{b}{}_{\tilde c},&(6.5\text{\rm a})\cr
&\ell _{dc} g_{\tilde m\tilde b} g^{\tilde c\tilde m} L^{d}_{\tilde
c\tilde a}(2) + \ell _{cd} g_{\tilde a\tilde m} g^{\tilde m\tilde c} 
L^{d}{}_{\tilde b\tilde c}(2)
= 2 \ell _{cd} g_{\tilde a\tilde m}g^{\tilde m\tilde c} \mu
^{d}{}_{\dah{i}}f^{\dah{i}}{}_{\tilde b\tilde c},&(6.5\text{\rm b})\cr
&\ell _{dc} g_{\tilde m\tilde b} g^{\tilde c\tilde m} L^{d}{}_{\tilde
c\tilde a}(3) + \ell _{cd} g_{\tilde a\tilde m} g^{\tilde m\tilde c} 
L^{d}{}_{\tilde b\tilde c}(3)
= 2 \ell _{cd} g_{\tilde a\tilde m} g^{\tilde m\tilde c} 
f^{\tilde d}{}_{\tilde b\tilde c}{\varPhi} ^{d}{}_{\tilde d}&(6.5\text{\rm c})\cr}
$$
and
$$
L^{d}{}_{\tilde a\tilde b} = L^{d}{}_{\tilde a\tilde b}(1) - L^{d}{}_{\tilde a\tilde
b}(2) - L^{d}{}_{\tilde a\tilde b}(3).\eqno(6.5*)
$$
Moreover we can write $L(i)$, $i = 1,2,3$ in the following forms
$$\eqalignno{
L^{d}{}_{\tilde a\tilde b}(1) &= C^{d}{}_{be}{\varPsi} ^{b}{}_{\tilde a}
{\varPsi} ^{e}{}_{\tilde b},&(6.5\text{\rm a}*)\cr
L^{d}{}_{\tilde a\tilde b}(2) &= \mu ^{d}{}_{\dah{i}}\cdot m^{\dah{i}}{}_{\tilde b\tilde c},&(6.5\text{\rm b}*)\cr
L^{d}{}_{\tilde a\tilde b}(3) &= r^{\tilde d}{}_{\tilde a\tilde b}
{\varPhi} ^{d}{}_{\tilde d},&(6.5\text{\rm c}*)\cr}
$$
where ${\varPsi} ^{b}{}_{\tilde a}(x)$ is a multiple of scalar fields transforming with respect to
groups $G$ and $H$ as ${\varPhi} ^{b}{}_{\tilde a}(x)$, $m^{\dah{i}}{}_{\tilde b\tilde c}$ are left-invariant skewsymmetric 2-forms on
$M$ with values in the Lie algebra $\fg_{0}$, $r^{\tilde d}{}_{\tilde a\tilde
b}$ is a left-invariant tensor field on $M$.
One gets
$$\eqalignno{
\ell _{dc} g_{\tilde m\tilde b} g^{\tilde c\tilde m} {\varPsi}
^{b}{}_{\tilde c}{\varPsi} ^{e}{}_{\tilde a} + \ell _{cd} g_{\tilde a\tilde
m} g^{\tilde m\tilde c} {\varPsi} ^{b}{}_{\tilde b}{\varPsi} ^{e}{}_{\tilde
c} &= 2 \ell _{cd} g_{\tilde a \tilde m } g^{\tilde m \tilde c } 
{\varPhi} ^{b}{}_{\tilde b }{\varPhi} ^{e}{}_{\tilde c
},&(6.5\text{\rm a}{*}{*})\cr
\ell _{dc} g_{\tilde m \tilde b } g^{\tilde c \tilde m }
m^{\dah{i}}{}_{\tilde c \tilde a } + \ell _{cd} g_{\tilde a \tilde m } 
g^{\tilde m \tilde c } m^{\dah{i}}{}_{\tilde b \tilde c }
&= 2 \ell _{cd} g_{\tilde a \tilde m } g^{\tilde m \tilde c } 
f^{\dah{i}}{}_{\tilde b \tilde c },&(6.5\text{\rm b}{*}{*})\cr
\ell _{dc} g_{\tilde m \tilde b } g^{\tilde c \tilde m }
r^{\tilde d }{}_{\tilde c \tilde a } + \ell _{cd} g_{\tilde a \tilde
m } g^{\tilde m \tilde c } r^{\tilde d }{}_{\tilde b \tilde c }
&= 2 \ell _{cd} g_{\tilde a \tilde m } g^{\tilde m \tilde c } 
f^{\tilde d }{}_{\tilde a \tilde b }&(6.5\text{\rm c}{*}{*})\cr}
$$
and
$$
L^{d}{}_{\tilde a \tilde b } = C^{d}{}_{be}{\varPsi} ^{b}{}_{\tilde
a }{\varPsi} ^{e}{}_{\tilde b } - \mu ^{d}{}_{\dah{i}}\cdot
m^{\dah{i}}{}_{\tilde a\tilde b } + r^{\tilde d }{}_{\tilde a \tilde b
}{\varPhi}^{d}{}_{\tilde d }.\eqno(6.5{*}{*}{*})
$$
It seems that sometimes the last formula will be easier to apply in
order to find an explicit form of the Higgs' potential.

Now we have the following formula for $B(W)$:
$$\eqalignno{
B(W) ={}& \sqrt{-g|\widetilde{g}||\ell|}
\biggl\{r^{n_{1}}\rho ^{n}\ov{R}(\ov{W}) + {\rho ^{n-2}\over \lambda ^{2}} 
r^{n_{1}} \widetilde{R}(\widetilde{{\varGamma} })+\phantom{\}}\cr
\noalign{\vskip2pt}
&+ r^{n_{1}-2}\rho^n \widehat{\ov{R}}(\widehat{\overline{\varGamma}})
- r^{n_{1}}\rho ^{n+2} {\lambda ^{2}\over 4} \widetilde{\ell }_{ij} 
(2\widetilde{H}^{i}\widetilde{H}^{j}-\widetilde{L}^{i\mu\nu
}\widetilde{H}^{j}{}_{\mu\nu }) +\cr
\noalign{\vskip2pt}
&+ {\lambda ^{2}\rho ^{n+2}r^{n_{1}-2}\over 2} \ell _{ab} g^{\tilde b
\tilde n} L^{a\mu}{}_{\tilde b } \Nabla^{\gauge}{}_\mu {\varPhi}
^{b}{}_{\tilde n} +\cr
\noalign{\vskip2pt}
&- {\lambda ^{2}\rho ^{n+2}r^{n_{1}-4}\over 4} \ell _{ab}
[2g^{[\tilde a \tilde b ]} (C^{a}_{cd}{\varPhi} ^{c}{}_{\tilde a }
{\varPhi} ^{d}{}_{\tilde b }-\mu ^{a}{}_{\dah{i}}f^{\dah{i}}{}_{\tilde
a \tilde b }-{\varPhi} ^{a}{}_{\tilde d }f^{\tilde d }{}_{\tilde a
\tilde b} ) \times \cr
\noalign{\vskip2pt}
&\times g^{[\tilde n\tilde m ]} (C^{b}_{ef}{\varPhi} ^{e}{}_{\tilde
n}{\varPhi} ^{f}{}_{\tilde m }-\mu ^{b}{}_{\dah{i}}f^{\dah{j}}_{\tilde
n\tilde m }-{\varPhi} ^{b}{}_{\tilde e}f^{\tilde e}{}_{\tilde
n\tilde m }) +&(6.6)\cr}
$$
$$
\eqalignno{
&- g^{\tilde a \tilde m } g^{\tilde b \tilde n} L^{a}{}_{\tilde a
\tilde b }\cdot (C^{b}_{cd}{\varPhi} ^{c}{}_{\tilde m }{\varPhi}
^{d}{}_{\tilde n}-\mu ^{b}{}_{\dah{i}}f^{\dah{i}}{}_{\tilde m \tilde n}-
{\varPhi} ^{b}{}_{\tilde e}f^{\tilde e}{}_{\tilde m \tilde n})]+\cr
\noalign{\vskip2pt}
&- {\lambda ^{2}\rho ^{n+2}r^{n_{1}-2}\over 2} h_{ab} \alpha ^{a}{}_{i} 
\widetilde{H}^{i} g^{[\tilde a \tilde b ]} (C^{b}_{cd}{\varPhi}
^{c}{}_{\tilde a }{\varPhi} ^{d}{}_{\tilde b }-\mu ^{b}{}_{\dah{i}}
f^{\dah{i}}{}_{\tilde a \tilde b }-{\varPhi} ^{b}{}_{\tilde d }
f^{\tilde d }{}_{\tilde a \tilde b }) +\cr
\noalign{\vskip2pt}
&-\phantom{\{} r^{n_{1}} \rho ^{n-2} (\ov{M} \widetilde{g}^{(\gamma \mu)} 
\rho _{,\gamma } \rho _{,\mu} + n^{2}g^{[\mu\nu ]} g_{\delta \mu} 
\widetilde{g}^{(\delta \gamma )} \rho _{,\nu } \rho _{,\gamma }
)\biggr\}.&(6.6~\hbox{\sevenrm cont.})\cr}
$$
Where\vadjust{\vskip-5pt}
$$
\widetilde{L}^{i\mu\nu } = g^{\alpha \mu} g^{\beta \nu }
\widetilde{L}^{i}{}_{\alpha \beta }\vadjust{\vskip-5pt}\eqno(6.7)
$$
and $\widetilde{L}^{i}{}_{\alpha \beta }$ obeys (6.1).

One can define the following two $\Ad_{H}$-type 2-forms with value in the Lie
algebra of $H,(\fh)$\vadjust{\vskip-4pt}
$$
\Check{L} = {1\over 2} L^{d}{}_{\tilde a \tilde b }\theta ^{\tilde
a }\wedge \theta ^{\tilde b }X_{d},\quad\
\Check{Q} = {1\over 2}Q^{d}{}_{\tilde a \tilde b }\theta ^{\tilde a
}\wedge\theta ^{\tilde b }X_{d},\vadjust{\vskip-4pt}
$$
in such a way that\vadjust{\vskip-4pt}
$$
\Check{L}=\Check{\Omega}-{1\over 2}\Check{Q},\vadjust{\vskip-4pt}
$$
where $\Check{\Omega} = {\displaystyle{1\over 2}}H^{d}{}_{\tilde a \tilde b }\theta
^{\tilde a }\wedge \theta ^{\tilde b }X_{d}$. These forms are defined on
$P$ (over $G/G_{0})$. In
this way we get similar formulae for the Higgs' field and for
Yang--Mills' field.

Let us define the following quantity $M^{d}{}_{\tilde a \tilde b}$ such that
$$
L^{d}{}_{\tilde a \tilde b } = H^{d}{}_{\tilde a \tilde b } - 4\pi
M^{d}{}_{\tilde a \tilde b }.\eqno(6.8)
$$
This quantity is an analogue of the polarization tensor for the Higgs'
field.

In this way we have
$$
\Check{Q}= 8\pi \Check{M}= 4\pi M^{d}{}_{\tilde a \tilde b }
\theta ^{\tilde a }\wedge \theta ^{\tilde b }X_{d}.\eqno(6.8\text{\rm a})
$$
The form $\Check{M}$ is an $\Ad_{\rH}$-type 2-form.
$$
\widetilde{H}^{i} = g^{[\alpha \beta ]} \widetilde{H}^{i}{}_{\alpha
\beta },\eqno(6.9)
$$
$L^{a}{}_{\alpha \beta }$ obeys (6.5) and scalar field 
${\varPhi} ^{a}{}_{\tilde b}$ satisfies the following
constraints:
$$
{\varPhi} ^{c}{}_{\tilde b } f^{\tilde b }{}_{\dah{i}\tilde a } = 
\mu ^{a}{}_{\dah{i}} {\varPhi} ^{b}{}_{\tilde a } C^{c}_{ab}.\eqno(6.10)
$$
Let us define the integral of action for $B(W)$
$$
S = - {1\over V_{1} V_{2}
r^{n_{1}}}\mmint_{{\bold U}}((B(W)\,d^{n}x)d^{n_{1}}x) d^{4}x,\eqno(6.11)
$$
where ${\bold U} = V \times M \times H$, $V \subset E$.
$$
V_{1} = \mmint_{H} \sqrt{-\ell}\,d^{n}x,\eqno(6.12)
$$
where $d^{n}x = d\mu _{\rH}(h)$ is a biinvariant measure on a group $H$ and
$$
V_{2} = \mmint_{M} \sqrt{|\widetilde{g}|}\,d^{n_{1}}x,\eqno(6.13)
$$
$d^{n_{1}}x = dm(y)$ is a measure on $M$ which is quasi-invariant with respect to
the action of $G$ on $M$ and $d^{4}x$ means integration over space-time
coordinates. After some calculations one gets:
$$
S = - \mmint_{V}\sqrt{-g} B(\ov{W}, g, \widetilde{A}, \rho ,
{\varPhi})\, d^{4}x,\quad\ V \subset E,\eqno(6.14)
$$
where
$$\eqalignno{
B(\ov{W}, g, \widetilde{A}, \rho , {\varPhi}) ={}& \rho ^{n}\cdot 
\ov{R}(\ov{W}) + {\lambda^2\over 4}\biggl[8\pi 
\rho ^{n+2} {\cal L}_{\rYM}(\widetilde{A}) 
+ \rho ^{n+2} {2\over r^{2}} {\cal L}_{\kin} (\Nabla^{\gauge} {\varPhi}
)+\cr
& + \rho ^{n+2} {1\over r^{4}} V({\varPhi} )
- \rho ^{n+2} {4\over r^{2}} {\cal L}_{\Int}({\varPhi}
,\widetilde{A}) + \rho ^{n-2} {\cal L}_{\scal}(\rho )\biggr]+ \lambda
_{c},\qquad\ &(6.15)\cr}
$$
$\ov{R}(\ov{W})$ is the Moffat--Ricci curvature scalar on the space-time $E$ and plays
the role of the gravitational lagrangian
$$
{\cal L}_{\rYM}(\widetilde{A}) = - {1\over 8\pi } \ell _{ij} 
(2\widetilde{H}^{i}\widetilde{H}^{j}-\widetilde{L}^{i\mu\nu }
\widetilde{H}^{j}{}_{\mu\nu })\eqno(6.16)
$$
is the lagrangian for the Yang--Mills' field with the gauge group $G$ in
the Nonsymmetric-Nonabelian Kaluza--Klein Theory
(see Appendix A for further details). $\widetilde{H}^{i}$ is defined in
(6.9) and $\widetilde{L}^{i\mu\nu }$ by (6.1).
$$\eqalignno{
{\cal L}_{\kin}(\Nabla^{\gauge} {\varPhi} ) ={}& {1\over
V_{2}}\mmint_{M}\sqrt{|\widetilde{g}|} d^{n_{1}}x (\ell
_{ab}g^{\tilde b \tilde n }L^{a\mu}{}_{\tilde b
}\Nabla^{\gauge}{}_\mu {\varPhi} ^{b}{}_{\tilde n }) =\cr
={}& \ell _{ab} g^{\alpha \mu} {1\over V_{2}}\mmint_{M}\sqrt{|\widetilde{g}|}
d^{n_{1}}x (g^{\tilde b \tilde n }L^{a}{}_{\alpha \tilde b
}\Nabla^{\gauge}{}_\mu {\varPhi} ^{b}{}_{\tilde n })&(6.17)\cr}
$$
is the kinetic part of the lagrangian for the scalar field ${\varPhi}
^{b}{}_{\tilde n }$. It is
easy to see that ${\cal L}_{\kin}$ is a quadratic form with respect to 
$\Nabla^{\gauge}{\varPhi}$ (gauge
derivative with respect to the connection $\widetilde{\omega }_{\rE})$ and is invariant with
respect to the action of the groups $H$ and $G$.
$$\eqalignno{
V({\varPhi} ) ={}& {\ell _{ab}\over V_{2}}\mmint_{M}\sqrt{|\widetilde{g}|}
d^{n_{1}}x[2g^{[\tilde m \tilde n ]}(C^{a}_{cd} {\varPhi}
^{c}{}_{\tilde m } {\varPhi} ^{d}{}_{\tilde n } - \mu ^{a}_{\dah{i}} 
f^{\dah{i}}{}_{\tilde m \tilde n }
- {\varPhi} ^{a}{}_{\tilde e} f^{\tilde e}{}_{\tilde m \tilde n })
g^{[\tilde a \tilde b ]}\cdot\cr
&\cdot (C^{b}_{ef}{\varPhi} ^{e}{}_{\tilde
a }{\varPhi} ^{f}{}_{\tilde b }-\mu ^{b}{}_{\dah{j}}f^{\dah{j}}{}_{\tilde a \tilde b }-{\varPhi} ^{b}{}_{\tilde a }f^{\tilde d
}{}_{\tilde a \tilde b }) +\cr
&- g^{\tilde a \tilde m } g^{\tilde b \tilde n } L^{a}{}_{\tilde a
\tilde b } (C^{b}_{cd}{\varPhi} ^{c}{}_{\tilde m }{\varPhi}
^{d}{}_{\tilde n }-\mu ^{b}_{\dah{i}}f^{\dah{i}}{}_{\tilde m \tilde n }-
{\varPhi} ^{b}{}_{\tilde e }f^{\tilde e }{}_{\tilde m \tilde n})]&(6.18)\cr}
$$
is the self-interacting term for the field ${\varPhi} $. It is invariant with
respect to the action of the groups $H$ and $G$. This term is a polynomial
of 4th order in ${\varPhi}$'s (a Higgs' field potential term).
$$
{\cal L}_{\Int}({\varPhi} ,\widetilde{A}) = h_{ab} \mu ^{a}_{i} 
\widetilde{H}^{i}\un{g}^{[\tilde a \tilde b ]} \cdot (C^{b}_{cd}
{\varPhi} ^{c}{}_{\tilde a }{\varPhi} ^{d}{}_{\tilde b }-
\mu ^{b}{}_{\dah{i}}f^{\dah{i}}{}_{\tilde a \tilde b }-{\varPhi}
^{b}{}_{\tilde d }f^{\tilde d }{}_{\tilde a \tilde b })\eqno(6.19)
$$
is the term describing nonminimal coupling between the scalar field ${\varPhi} $
and the Yang--Mills' field where
$$\displaylines{
\hfill \un{g}^{[\tilde a \tilde b ]} = {1\over V_{2}}\mmint_{M}\sqrt{|\widetilde{g}|}
d^{n_{1}}x g^{[\tilde a \tilde b ]},\hfill\llap{(6.20)}\cr
\hfill
\lambda _{c} = {\rho ^{n}\over \lambda ^{2}} \widetilde{R}
(\widetilde{{\varGamma} }) + {\rho ^{n}\over r^{2}V_{2}}\mmint_{M}\sqrt{|\widetilde{g}|}
\widehat{\ov{R}}(\widehat{\overline{\varGamma}})\, d^{n_{1}}x
= {\rho ^{n}\over \lambda ^{2}} \widetilde{R}(\widetilde{{\varGamma} }) + 
{\rho ^{n}\over r^{2}} \widetilde{\un P}.\hfill\llap{(6.21)}\cr}
$$
This term is also invariant with respect to the action of groups $H$ and
$G$. One can write (6.17) and (6.18) in a different form.
$$\eqalignno{
{\cal L}_{\kin}(\Nabla^{\gauge}{\varPhi} ) ={}& \ell _{ab} 
(g^{\beta \nu }L^{a\mu}{}_{\tilde b }\Nabla^{\gauge}{}_{\mu}{\varPhi}
^{b}{}_{\tilde n })_{av}\cr
={}& \ell _{ab} g^{\alpha \mu} (g^{\tilde b \tilde n }L^{a}{}_{\alpha
\tilde b }\Nabla^{\gauge}{}_\mu {\varPhi} ^{b}{}_{\tilde n
})_{av},&(6.22)\cr
V({\varPhi} ) ={}& 2 \ell _{ab}([g^{([\tilde m \tilde n ][\tilde a
\tilde b ])} (C^{a}_{cd}{\varPhi} ^{c}{}_{\tilde m }{\varPhi}
^{d}{}_{\tilde n }-\mu ^{a}{}_{\dah{i}}f^{\dah{i}}{}_{\tilde m \tilde n }-
{\varPhi} ^{a}{}_{\tilde e }f^{\tilde e }{}_{\tilde m \tilde n }) \times\cr
&\times (C^{b}_{ef}{\varPhi} ^{e}{}_{\tilde a }{\varPhi}
^{f}{}_{\tilde b }-\mu ^{b}_{\dah{j}}f^{\dah{j}}{}_{\tilde a \tilde b }-
{\varPhi} ^{b}{}_{\tilde d }f^{\tilde d }{}_{\tilde a \tilde b })+\cr
&+ \un{g}^{\tilde a \tilde m \tilde b \tilde n } L^{a}{}_{\tilde a
\tilde b } (C^{b}_{cd}{\varPhi} ^{c}{}_{\tilde m }{\varPhi}
^{d}{}_{\tilde n }-\mu ^{b}{}_{\dah{i}}f^{\dah{i}}{}_{\tilde m \tilde n }-
{\varPhi} ^{b}{}_{\tilde e }f^{\tilde e }{}_{\tilde m \tilde n
})])_{av},&(6.23)\cr}
$$
where
$$\displaylines{
(\ldots )_{av} = {1\over V_{2}} \mmint_{M}\sqrt{|\widetilde{g}|}\, dx^{n_{1}}
(\ldots ),\cr
\hfill\un{g}^{\tilde a \tilde m \tilde b \tilde n } = g^{\tilde a \tilde m } 
g^{\tilde b \tilde n },\hfill\llap{(6.24)}\cr
\un{g}^{([\tilde m \tilde n ][\tilde a \tilde b ])} = g^{[\tilde m
\tilde n ]} \cdot g^{[\tilde a \tilde b ]}\cr}
$$
(see Appendix A for further details).
The connection $\widehat{{\varGamma}}^{\tilde a }{}_{\tilde b
\tilde c }$ on $M$ satisfies the following compatibility
conditions:
$$
g_{\tilde d \tilde b }\widehat{\overline{\varGamma}}^{\tilde d
}{}_{\tilde a \tilde c } + g_{\tilde a \tilde d }
\widehat{\overline{\varGamma}}^{\tilde d }{}_{\tilde c \tilde b } = 
g_{\tilde a \tilde d } \varkappa ^{\tilde d }{}_{\tilde b \tilde c } 
+ g _{\tilde a \tilde b ,\tilde c },\eqno(6.25)
$$
where $\varkappa ^{\tilde d }{}_{\tilde b \tilde c }$ are nonholonomicity coefficients,
``$,$'' means the action of a
vector field defined on $M$ and induced by a left invariant vector field
on $G$. For $g_{\tilde a \tilde b ,\tilde c }$ one easily finds
$$
g_{\tilde a \tilde b ,\tilde c }(x) = \zeta k^{0}{}_{\tilde a \tilde
b ,\tilde c } + h^{0}{}_{\tilde a \tilde b ,\tilde c }.\eqno(6.26)
$$

For ${\cal L}_{\scal}(\rho )$ we have the following:
$$
{\cal L}_{\scal}(\rho ) = (\ov{M}\widetilde{g}^{(\gamma \mu)} 
\rho _{,\gamma } \rho _{,\mu} + n^{2}g^{[\mu\nu ]} g_{\delta\mu} 
\widetilde{g}^{(\delta \gamma )} \rho _{,\nu } \rho _{,\gamma }),\eqno(6.27)
$$
where
$$
\ov{M} = (\ell _{[dc]}\ell ^{[dc]}-3n(n-1)) \ge 0.\eqno(6.27\text{\rm a})
$$
According to section 3 $B(W,\widetilde{A},{\varPhi} ,{\varPsi} )$ is invariant with respect to the right
action of the group $G$ on the bundle $Q(E, G)$. Thus we do not see any
additional dimensions. They can be easily dropped due to the
integration over the group $G$.

Let us consider $M = S^{2} \cong \SO(3)/\SO(2)$ (see sec.~4 and in particular
Eq.~(4.23)). One gets for $h^{0} = h^{0}{}_{\tilde a \tilde b
}\widetilde{V}^{\tilde a }\otimes \widetilde{V}^{\tilde b }$
$$
h^{0}{}_{\tilde a \tilde b } = \pmatr{1&0\cr 0&1\cr},\eqno(6.28)
$$
where
$$
\widetilde{V}^{1} = d\theta ,\quad\ \widetilde{V}^{2} = \sin
\theta\, d\varphi \eqno(6.29)
$$
and
$$
d\widetilde{V}^{1} = 0 ,\quad\ d\widetilde{V}^{2} = \cot \theta
\widetilde{V}^{1}\wedge \widetilde{V}^{2}.\eqno(6.30)
$$
On $S^{2} = \SO(3)/\SO(2)$ we introduce a coordinate system in a known way, i.e.
$$
(\theta ,\varphi ) =\widetilde{\varphi}(g\exp (\theta e_{1}+\varphi
e_{2})),\quad\ g \in \SO(3) ,\ \SO(3) \strz{\tilde\varphi} S^{2}\eqno(*)
$$
and getting it in a neighbourhood of $\pmatr{\theta _{0},\varphi _{0}}$. In particular for 
$(0,0)$.
In this way we have a local trivialization of the bundle $G $, $(\theta ,\varphi ,\psi )$.
Moreover we have a local section of $G $, $\sigma $: in a neighbourhood of 
$\widetilde{\varphi}(g)$ due to $(*)$. The forms $\widetilde{V}_{1}$,
$\widetilde{V}_{2}$ are pullbacks of $V_{1}$, $V_{2}$.
$$
\widetilde{V}_{i} = \sigma ^{*}V_{i} ,\quad\ i = 1,2.
$$
They are dual to left-invariant vector fields $\widetilde{e}_{1}$,
$\widetilde{e}_{2}$ on $M$, $\widetilde{e}_{i}(\theta ,\varphi ) =
L'_{g^{-1}}e_{i}(0,0) $, $i = 1,2$. We have
$$
\pmatr{\widetilde{e}_{1}(\theta ,\varphi )\cr
\widetilde{e}_{2}(\theta ,\varphi )\cr} = \pmatr{{\displaystyle{\partial\over
\partial\theta }}&\cr\noalign{\vskip3pt}
{\displaystyle{1\over \sin \theta }}&{\displaystyle{\partial\over \partial
\varphi}}\cr}\quad \hbox{and}\quad
[\widetilde{e}_{1},\widetilde{e}_{2}] = - \cot \theta
\widetilde{e}_{2}.
$$
Thus we have
$$
g_{\tilde a \tilde b } = \pmatr{1&\zeta \cos \theta \cr
-\zeta \cos \theta &1\cr}.\eqno(6.31)
$$
Substituting Eqs.~(6.30--31) into (6.25) one finds
$$
\widehat{\overline{\varGamma}}^{1}{}_{11} =
\widehat{\overline{\varGamma}}^{2}{}_{22} = 0\eqno(6.32)
$$
and
$$\eqalignno{
\widehat{\overline{\varGamma}}^{1}{}_{22}
+\widehat{\overline{\varGamma}}^{2}{}_{11} - \zeta \cos \theta
\widehat{\overline{\varGamma}}^{2}{}_{21} &= \zeta \cos \theta ,\cr
\widehat{\overline{\varGamma}}^{2}{}_{11}
+\widehat{\overline{\varGamma}}^{1}{}_{21} + \zeta \cos \theta
\widehat{\overline{\varGamma}}^{2}{}_{21} &= -3\zeta \sin \theta ,\cr
\widehat{\overline{\varGamma}}^{1}{}_{12}
+\widehat{\overline{\varGamma}}^{1}{}_{21} - \zeta \cos \theta
\widehat{\overline{\varGamma}}^{2}{}_{12} + \zeta \cos \theta
\widehat{\overline{\varGamma}}^{2}{}_{21} &= 2\zeta \sin \theta ,&(6.33)\cr
\widehat{\overline{\varGamma}}^{2}{}_{12} + 
\widehat{\overline{\varGamma}}^{1}{}_{22} + \zeta \cos \theta
\widehat{\overline{\varGamma}}^{1}{}_{12} &= 0,\cr
\widehat{\overline{\varGamma}}^{1}{}_{22}
+\widehat{\overline{\varGamma}}^{2}{}_{21} + \zeta \cos \theta
\widehat{\overline{\varGamma}}^{2}{}_{21} &= 2\cot \theta ,\cr
\widehat{\overline{\varGamma}}^{2}{}_{21}
+\widehat{\overline{\varGamma}}^{2}{}_{12} + \zeta \cos \theta
\widehat{\overline{\varGamma}}^{1}{}_{21} - \zeta \cos \theta
\widehat{\overline{\varGamma}}^{1}{}_{12} &= -2\cot \theta .\cr}
$$
The existence of the connection $\widehat{\overline{\varGamma}}$ (a nonsymmetric) is guaranteed by the
theorem from the second position of Ref.~[3]. In order to apply it let us
transform $h^{0}$ and $g$ to holonomic system of coordinates.

One gets
$$\displaylines{
\hfill g_{\tilde{a}'\tilde{b}'}= \pmatr{1&\zeta \sin \theta \cos
\theta \cr
\noalign{\vskip4pt}
-\zeta\sin \theta \cos \theta &\sin ^{2}\theta
\cr},\hfill\llap{\hbox{(6.34a)}}\cr
\noalign{\vskip4pt}
\hfill h^{0}{}_{\tilde a '\tilde b '}= \pmatr{1&0\cr 0&\sin ^{2}\theta
\cr},\hfill\llap{\hbox{(6.34b)}}\cr
\noalign{\vskip4pt}
\hfill g = \det (g_{\tilde a '\tilde b '}) = \sin ^{2}\theta 
(1+\zeta ^{2}\cos ^{2}\theta ) \neq 0,\hfill\llap{\hbox{(6.35a)}}\cr
\noalign{\vskip4pt}
\hfill h^{0} = \det (h^{0}{}_{\tilde a '\tilde b'}) = 4\sin ^{2}\theta \neq 
0,\hfill\llap{\hbox{(6.35b)}}\cr
\noalign{\vskip4pt}
\hfill \sigma = {(2g_{[12]})^2\over g} = {4\cos ^2\theta \over
(1+\zeta^2\cos ^2\theta )},\hfill\llap{\hbox{(6.35c)}}\cr
\noalign{\vskip4pt}
\hfill W = \biggl({\partial \sigma \over \partial \theta},
{\partial \sigma \over \partial \varphi }\biggr) = (\sigma_{,\theta}
,0)\neq 0,\hfill\llap{\hbox{(6.35d)}}\cr}
$$
Thus all the conditions are satisfied except $\theta = 0$, $\pi $, 
$0\le \varphi \le 2\pi $ and because
of this the nonsymmetric connection $\widehat{\overline{\omega}}$ exist 
(except singular line $\theta =0 $, $\pi$, $0\le \varphi \le 2\pi$, where it is symmetric
i.e.\ Riemannian one). The system of linear
equations (6.33) can be solved because its matrix is invertible. The
determinant of this matrix is equal to
$$
2(2\zeta ^{2}\cos ^{2}\theta -\zeta ^{3}\cos ^{3}\theta +\zeta \cos
\theta-1)
$$
and nonzero.

Let us write the remaining connection coefficients. One gets
$$\displaylines{
\widehat{\overline{\varGamma}}^{1}{}_{22} =\hfill\cr
{(3\zeta^2\cos ^2\theta -10\zeta^4\cos ^4\theta -6\zeta ^4\cos
^3\theta \sin \theta -4\zeta ^3{\cos ^4\theta \over \sin \theta
}-3\zeta ^2\cos \theta \sin \theta -2\cot\theta )\over 
2(\zeta ^3\cos ^3\theta -2\zeta \cos ^2\theta -\zeta\cos\theta+1)},\cr
\noalign{\vskip4pt}
\widehat{\overline{\varGamma}}^{2}{}_{11} =
{5\zeta \sin \theta (\zeta ^2\cos ^2\theta -2\zeta ^2\cos \theta \sin
\theta +1)\over 2(2\zeta ^2\cos ^2\theta -\xi^3\cos \theta +\zeta
\cos \theta -1)},\cr
\noalign{\vskip4pt}
\widehat{\overline{\varGamma}}^{2}{}_{21} = 
{2\zeta \cos \theta (\zeta ^2\cos \theta \sin \theta +2\cat \theta
-\zeta \cos \theta -2\zeta \sin \theta )\over
(2\zeta ^2\cos ^2\theta -\zeta ^3\cos ^3\theta +\zeta \cos \theta
-1)},\cr
\noalign{\vskip4pt}
\widehat{\overline{\varGamma}}^{1}{}_{21} =
{\textstyle{(2\zeta ^3\cos ^3\theta +4\zeta \cos \theta \cot \theta +2\zeta
^3\cos ^2\theta \sin \theta +3\zeta \sin \theta -2\zeta ^2\cos \theta
\sin \theta -\zeta \cos \theta -4\zeta ^2\cos \theta \cot \theta)\over
2(\zeta ^3\cos ^3\theta -2\zeta \cos ^2\theta -\zeta\cos\theta+1)}},\cr
\noalign{\vskip4pt}
\widehat{\overline{\varGamma}}^{1}{}_{12} =\hfill\cr
{(4\zeta ^3\cos \theta \sin \theta -5\zeta \sin \theta +\zeta ^2\cos
^2\theta +3\zeta ^2\cos \theta \sin \theta +2\zeta \cos \theta \cot
\theta -\zeta \cos \theta -2\cot\theta )\over
(2\zeta ^2\cos ^2\theta -\zeta^3 \cos ^3\theta -\zeta\cos\theta+1)},\cr
\noalign{\vskip4pt}
\widehat{\overline{\varGamma}}^{2}{}_{12} \!=\!
{\textstyle{(7\zeta ^4\cos ^3\theta \sin \theta +12\zeta ^2\cos ^2\theta
\cot\theta -5\zeta ^2\cos \theta \sin \theta -4\cot\theta +\zeta
^4\cos ^4\theta -4\zeta ^4\cos ^4\theta \cot\theta -\zeta ^2\cos
^2\theta \over
2(2\zeta ^2\cos ^2\theta -\zeta^3 \cos ^3\theta +\zeta\cos\theta-1)}},\cr
\hfill(6.36)\cr}
$$
Eqs.~(6.32--3) and (6.36) give us an example of Einstein--Kaufmann
connection on a homogeneous space $- S^{2} \cong \SO(3)/\SO(2)$. Let us calculate
the Moffat--Ricci scalar of the curvature for the connection 
$\widehat{\overline{\varGamma}}^{\tilde a }{}_{\tilde b \tilde c }$,
$(\widehat{\overline{\omega}}^{\tilde a }{}_{\tilde b })$. In order to do this let us find
$g ^{\tilde a \tilde b }$. One finds
$$
g ^{\tilde a \tilde b } = {1\over (1+\zeta^2\cos ^2\theta )}
\pmatr{1&-\zeta \cos \theta \cr \zeta \cos \theta &1\cr}.\eqno(6.37)
$$
Moreover it is easier to calculate the scalar of curvature using
results from the second position of Ref.~[4]. In this paper one has found
the general solution of compatibility conditions for Einstein--Kaufmann
connection in 2-dimensional case. In particular
$$
\widehat{\overline{\omega}}^{\tilde n }{}_{\tilde b } = 
\widetilde{\omega }^{\tilde n }{}_{\tilde b } + {1\over \ov{g}} 
h^{\tilde n \tilde a }\widetilde{\nabla }_{\tilde a }k^{0}{}_{\tilde
b \tilde m }V^{\tilde m } - {2\over \ov{g}}
[(\widetilde{\nabla}_{\alpha } k^{0}_{(\tilde b }{}^{\tilde n })
k^{0}{}_{\tilde m )}{}^{\tilde a }]V^{\tilde m },\eqno(6.38)
$$
where $\widetilde{\omega }^{\tilde n }{}_{\tilde b }$ is the Riemannian connection for
$h^{0}{}_{\tilde n \tilde b } = g _{(\tilde n \tilde b )}$.
$$
\ov{g }={\det(g _{\tilde a \tilde b })\over \det(h^o{}_{\tilde a
\tilde b })}
$$
is a scalar, and $\widetilde{\nabla }$ means a covariant derivative with respect to $\widetilde{\omega
}^{\tilde a }{}_{\tilde b }$.

One gets
$$\eqalignno{
\widehat{\ov{R}}(\widehat{\overline{\varGamma}})={}&
g^{\tilde b \tilde m }\widetilde{R}_{\tilde b \tilde m
}(\widetilde{{\varGamma}})+{1\over 4}\zeta g^{\tilde b \tilde m }
\biggl(\widetilde{\nabla}_{\tilde b }\biggl({1\over \ov{g}}
\nabla_{\tilde a }k^{0\tilde a }{}_{\tilde m }\biggr)
-\widetilde{\nabla }_{\tilde m }\biggl({1\over \ov{g}}
\tilde{\nabla }_{\tilde a }k^{0\tilde a }{}_{\tilde b
}\biggr)\biggr)+\cr
\noalign{\vskip4pt}
&- \zeta ^{2}g^{\tilde b \tilde m }\biggl(\widetilde{\nabla}_{\tilde
q }\biggl({1\over \ov{g}}(\widetilde{\nabla}k^{0}_{(\tilde b
}{}^{\tilde q}) k^{0}_{\tilde m )}\biggr)
- \widetilde{\nabla }_{(\tilde m }
\biggl({1\over \ov{g}}(\widetilde{\nabla}_{\tilde a }k^{0}{}_{\tilde b)}{}^{\tilde q})k^{0}{}_{\tilde q}{}^{\tilde a
}\biggr)\biggr),&(6.39)\cr}
$$
where $\widetilde{R}_{\tilde b \tilde m }(\widetilde{{\varGamma} })$ 
is the Ricci tensor for the Riemannian connection $\widetilde{{\varGamma} }
(\widetilde{\omega }^{\tilde a }{}_{\tilde b })$.
One gets after some algebra
$$\eqalignno{
\widehat{\ov{R}}(\widehat{\overline{\varGamma}})={}& R(\theta ,\zeta)
= {2\over (1+\zeta ^2\cos ^2\theta )} + {\zeta ^2\cos ^2\theta
(1+9\zeta ^2-8\zeta ^4+\zeta ^2(16\zeta ^2-1)\cos ^2\theta )\over
2(1+\zeta ^2\cos ^2\theta )^3}+\cr
\noalign{\vskip4pt}
&- {4\zeta ^2\sin \theta (1-4\cos \theta -4\zeta ^2\cos ^3\theta
)\over 2(1+\zeta ^2\cos ^2\theta )^3}.&(6.39\text{\rm a})\cr}
$$
In a general case i.e.\ for $\dim M\ge 3$ we should transform Eq.~(6.25) 
to the holonomic system of coordinates and to use recurrence relations for the
connection in order to find it (see Ref.~[3] for more details).

In this case we can find a general solution (Ref.~[3]).
$$
\overline{\omega }^{\tilde n }{}_{\tilde z } = \widetilde{\omega }^{\tilde
n }{}_{\tilde z } + u^{\tilde n }{}_{\tilde z }\eqno(6.40)
$$
(see Appendix A for further details),
where $\widetilde{\omega }^{\tilde n }{}_{\tilde z }$ is a Riemannian connection for
$h^{0}{}_{\tilde a \tilde b }$ and
$$\displaylines{
\hfill u^{\tilde n }{}_{\tilde z } = u^{\tilde n }{}_{\tilde z \tilde m }\theta
^{\tilde m },\hfill\llap{(6.41)}\cr
u^{\tilde n }{}_{\tilde z \tilde m } = {1\over 2}(L^{\tilde n
}{}_{\tilde z \tilde m } - 2\zeta ^{2}k^{0}{}_{[\tilde m }{}^{\tilde
a }L_{\tilde z ]\tilde a \tilde b }k^{0\tilde n \tilde b }) + 
\zeta h^{0\tilde n \tilde e }\{L^{\tilde a }{}_{\tilde e (\tilde z }
k^{0}{}_{\tilde m )\tilde a }+\hfill\cr
\hfill + \zeta k^{0}{}_{\tilde c}{}^{\tilde b }(k^{0}{}_{(\tilde m}{}^{\tilde c}
L_{\tilde z )\tilde a \tilde b }k^{0}{}_{\tilde e }{}^{\tilde a } - 
L_{\tilde e \tilde a \tilde b }k^{0}_{(\tilde z }k^{0}{}_{\tilde m
)}{}^{\tilde c })\},\quad\ (6.42)\cr}
$$
for $L_{\tilde a \tilde b \tilde c }$ one has
$$
L_{\tilde a \tilde b \tilde c }= -\zeta (\widetilde{\nabla}_{\tilde a
}k^{0}{}_{\tilde b \tilde c } + \widetilde{\nabla}_{\tilde b
}k^{0}{}_{\tilde a \tilde b } -
\widetilde{\nabla}_{c}k^{0}{}_{\tilde a \tilde b }),\eqno(6.43)
$$
$\widetilde{\nabla}$ means a covariant derivative with respect to the connection
$\widetilde{\omega }^{\tilde a }{}_{\tilde b }$.

For a scalar of curvature one gets:
$$
\widehat{\ov{R}}(\widehat{\overline{\varGamma}}) =
g^{\tilde b \tilde d }\widetilde{R}_{\tilde b \tilde d
}+g^{\tilde b \tilde d }(\widetilde{\nabla}_{\tilde a
}u^{\tilde a }{}_{\tilde b \tilde d }-\widetilde{\nabla}_{\tilde a
}u^{\tilde a }{}_{\tilde b \tilde a })
+{1\over 2}g^{\tilde b \tilde d }(\widetilde{\nabla}_{\tilde b
}u^{\tilde a }{}_{\tilde a \tilde d }-\widetilde{\nabla}_{\tilde a
}u^{\tilde a }{}_{\tilde a \tilde b }).\eqno(6.44)
$$
Let us notice that if $k^{0}$ is induced by the K\"ahlerian structure on
$M = G/G_{0}$ (see sec.~4) we have
$$
L_{\tilde a \tilde b \tilde c }= 0\eqno(6.45)
$$
and
$$
\widehat{\overline{\omega }}^{\tilde n }{}_{\tilde z } = \widetilde{\omega
}^{\tilde n }_{\tilde z }.\eqno(6.46)
$$
In this case one simply gets
$$
\widehat{\ov{R}}(\widehat{\overline{\varGamma}})
= g^{\tilde b \tilde d }\widetilde{R}_{\tilde b \tilde
d }(\omega ).\eqno(6.47)
$$
Moreover the Riemannian Ricci tensor can be easily calculated in terms
of structure constants of $G$ and $G_{0}$. One gets
$$
\widetilde{R}_{\tilde b \tilde c }(\omega ) = - {1\over 4}
h_{\tilde b \tilde c } - {3\over 4}
h_{\dah{e}\dah{f}}f^{\tilde{e}\tilde{a}}{}_{\tilde c}f^{\tilde{f}}
{}_{\tilde b \tilde a }.\eqno(6.48)
$$
Thus we have finally from (6.47)
$$
\widehat{\ov{R}}(\widehat{\overline{\varGamma}}) = -{1\over 4}
g^{\tilde b \tilde c }h_{\tilde b \tilde c } - 
{3\over 4} h_{\dah{e}\dah{f}}g^{\tilde b \tilde c
}f^{\dah{e}\dah{a}}{}_{\tilde c }f^{\dah{f}}{}_{\tilde b \tilde a },\eqno(6.49)
$$
i.e. in the case of $\widehat{\overline{\omega }}^{\tilde n }{}_{\tilde z } 
= \widetilde{\omega }^{\tilde n }{}_{\tilde z }$. In general one gets
Eq.~(6.44) where the first term is given by Eq.~(6.49).

Let us reconsider $S^{2}$ in a different context i.e introducing on $S^{2}$
Riemannian and symplectic structures coming from ${\sym C}$ via stereographic
projection.

One gets
$$
\xi = {x\over 1+|z|^2},\quad\ \eta = {y\over 1+|z|^2},\quad\ 
\chi = {|z|^2\over 1+|z|^2},\eqno(6.50)
$$
where $(\xi ,\eta ,\chi ) \in S^{2}$ and $z = x + iy \in {\sym C}$
and
$$
x = {\xi\over 1-\chi},\quad\ y = {\eta\over 1-\chi}.\eqno(6.50\text{\rm a})
$$
Using spherical coordinates on $S^{2}$ one easily obtains
$$
z = x + iy = \ctg {\theta \over 2}e^{-i\varphi},\eqno(6.51)
$$
where $- {\displaystyle{\pi\over 2}} < \theta <{\displaystyle{\pi\over
2}}$, $0\le \varphi <2\pi $ and
$$\eqalignno{
\xi &= \sin \theta \cos \varphi ,\cr
\eta &= \sin \theta \sin \varphi,&(6.51\text{\rm a})\cr
\chi &= \cos \theta.\cr}
$$
On the complex plane ${\sym C}$ we have a K\"ahlerian structure induced by a
Hermitian metric tensor
$$
p = dz\,d\ov{z}= (dx^{2} + dy^{2}).\eqno(6.52)
$$
The symplectic form is given by
$$
\ov{k} = {1\over 2i} dz \wedge d\ov{z}= dx \wedge dy.\eqno(6.53)
$$
After some calculation one gets using Eq.~(6.51)
$$
p = {1\over 4\sin ^4{\displaystyle{\theta \over 2}}} (d\theta ^{2}+\sin ^{2}\theta\,
d\varphi^{2})\eqno(6.52\text{\rm a})
$$
and
$$
\ov{k} = {1\over 4\sin ^4{\displaystyle{\theta \over 2}}} \biggl(-{1\over 2}\sin \theta\,
d\theta \wedge d\varphi^{2}\biggr).\eqno(6.53\text{\rm a})
$$
The tensor
$$
ds^{2} = d\theta ^{2} + \sin ^{2}\theta \,d\varphi ^{2}\eqno(6.54)
$$
is a Riemannian metric tensor on $S^{2}$ induced by the euclidean metric in
$E^{3}$. It is conformally equivalent to the flat geometry on ${\sym
C}$ (locally). The symplectic form
$$
-{1\over 2}\sin \theta\, d\theta \wedge d\varphi \eqno(6.55)
$$
corresponds to the symplectic form on ${\sym C}$ with the same conformal factor
$(4\sin ^{4}{\theta \over 2})^{-1}$. The metric $ds^{2} = d\theta ^{2} + 
\sin ^{2}\theta\, d\varphi ^{2}$ and the 2-form
$k^{0} = \sin \theta \, d\theta \wedge d\varphi $ from Riemannian and symplectic structure on
$S^{2}$ (really
K\"ahlerian). Thus we have on $S^{2} a$ nonsymmetric tensor
$$
g_{\tilde a \tilde b } = h^0_{\tilde a \tilde b } + \zeta
k^{0}{}_{\tilde a \tilde b },\eqno(6.56)
$$
such that
$$
\widetilde{\nabla}_{\tilde a }k^{0}{}_{\tilde b \tilde c } = 0,\eqno(6.57)
$$
$\widetilde{\nabla}$ means covariant derivative with respect to Riemannian metric induced
by (6.54) on $S^{2}$
$$\eqalignno{
g_{\tilde a \tilde b } &= \pmatr{1&\zeta\sin \theta \cr
-\zeta\sin \theta &\sin ^{2}\theta \cr},&(6.58)\cr
\noalign{\vskip4pt}
g^{\tilde a \tilde b } &= {1\over (1+\zeta ^2)\sin^2 \theta }
\pmatr{\sin ^{2}\theta &-\zeta\sin \theta \cr \zeta\sin \theta &
1\cr}.&(6.59)\cr}
$$
One gets 
$$
\widehat{\ov{R}}(\widehat{\overline{\varGamma}})= 
g^{\tilde a \hat{\tilde
b}}\widetilde{R}(\widetilde{\widehat{\omega }}) = 
{4\over (1+\zeta ^2)}.\eqno(6.60)
$$
The form $k^{0}$ introduced here can be lifted to SO(3) to get a
left-invariant 2-form on SO(3).

\vfill\eject

\null
\vskip36pt plus4pt minus4pt
\centerline{{\four\bf 7. Conformal Transformation of $g_{\mu\nu}$.}}
\vskip3pt
\centerline{{\four\bf Transformation of the Scalar Field $\rho $}}
\vskip24pt plus2pt minus2pt
\write0{7. Conformal Transformation of $g_{\mu\nu}$.
Transformation of the Scalar Field $\rho $ \the\pageno}

In section 6 we obtained the lagrangian density
$B(\ov{W}, g, \widetilde{A}, \rho , {\varPhi})$ such that:
$$\eqalignno{
\noalign{\vskip4pt}
B(\ov{W}, g, \widetilde{A}, \rho , {\varPhi}) ={}& \sqrt{-g}=\biggl\{\rho ^{n}
\cdot \ov{R}(\ov{W}) 
+ {\lambda ^{2}\over 4}\biggl[8\pi \rho ^{n+2} {\cal L}_{\rYM}
(\widetilde{A}) +\phantom{\}}\cr
\noalign{\vskip5pt}
&+ \rho ^{n+2} {2\over r^{2}} {\cal L}_{\kin}
(\Nabla^{\gauge} {\varPhi} ) 
- \rho ^{n+2} {1\over r^{4}} V({\varPhi} ) - \rho ^{n+2} {4\over r^{2}} 
{\cal L}_{\Int}({\varPhi} , \widetilde{A})\biggr]+\cr
\noalign{\vskip5pt}
&- \phantom{\{}{\lambda ^{2}\over 4} \rho ^{n-2} {\cal L}_{\scal}(\rho ) + 
{4\over \lambda ^{2}\rho ^{2-n}} \widetilde{R}(\widetilde{{\varGamma} }) 
+ \rho ^{n} {1\over r^{2}} \widetilde{\un{P}}\biggr\},&(7.1)\cr
\noalign{\vskip4pt}}
$$
where ${\cal L}_{\rYM}$, ${\cal L}_{\kin}$, $V$, ${\cal L}_{\Int}$, 
${\cal L}_{\scal}$ and $\widetilde{\un{P}}$ are defined in section 6 by
formulae (6.16), (6.17), (6.18), (6.19) and (6.21). This lagrangian
density is a generalization of the lagrangian from Bergmann's paper
(see Ref.~[50]). P.~G.~Bergmann considers the lagrangian for the
tensor-scalar theory of gravitation including Jordan--Thiry theory and
Brans--Dicke theory. Our lagrangian is more general for two reasons.
We have nonsymmetric metric tensor $g_{\mu\nu }$ as a metric 
$(\ov{R}(\ov{W})$ is the
lagrangian of gravitational field from the Nonsymmetric Theory of
Gravitation). The lagrangian (7.1) possesses also apart from lagrangian
for gauge field (in the Bergmann's paper it is an electromagnetic
field) lagrangian for Higgs' field coupled to Yang--Mills' field. We
also get two terms which play the role of the cosmological terms. In
the Bergmann's paper there are four arbitrary functions of scalar field
$\rho $, $f_{1}$, $f_{2}$, $f_{3}$, $f_{4}$. Here we have more such functions:
$$\eeqalign{
f_{1}(\rho ) &= \rho ^{n} ,\quad\ &f_{2}(\rho ) &= 8\pi \rho
^{n+2},\cr
\noalign{\vskip4pt}
f_{2'}(\rho ) &= - {2\over r^{2}} \rho ^{n+2} ,\quad\ &f_{2''}(\rho ) &= {1\over r^{4}} \rho
^{n+2},\cr
\noalign{\vskip4pt}
f_{2'''}(\rho ) &= {4\over r^{2}} \rho ^{n+2} ,\quad\ &f_{3}(\rho ) &= \rho
^{n+2},\cr
\noalign{\vskip4pt}
f_{4'}(\rho ) &= {4\over \lambda ^{2}\rho ^{2-n}} ,\quad\ &f_{4''}(\rho ) &= {1\over r^{2}} \rho
^{n-2}.\cr}\eqno(7.2)
$$
\looseness-2Now we proceed with the conformal transformation for the metric
$g_{\mu\nu }$ and
the transformation of the scalar field $\rho $. This is only the
redefinition of $g_{\mu\nu }$ and $\rho $.
$$\displaylines{\hfill
\rho = e^{-{\varPsi}},\hfill\llap{(7.3)}\cr
\noalign{\vskip4pt}
\hfill g_{\mu\nu } \rightarrow e^{n{\varPsi} }\cdot g_{\mu\nu } = 
{1\over \rho ^{n}} g_{\mu\nu }.\hfill\llap{(7.4)}\cr}
$$
This procedure comes of course from Ref.~[50]. The only difference is
that $g_{\mu\nu }$ is now nonsymmetric. After transformations (7.3) and (7.4) we
get the following:
$$\displaylines{
B(\ov{W}, g, \widetilde{A}, {\varPsi} , {\varPhi})=
\sqrt{-g}\biggl\{\ov{R}(\ov{W}) + {\lambda ^{2}\over 4}\biggl(8\pi 
e^{-(n+2){\varPsi} } {\cal L}_{\rYM}(\widetilde{A})
+ {2e^{-2{\varPsi} }\over r^{2}} {\cal L}_{\kin}
(\Nabla^{\gauge}{\varPhi})+\hfill\cr
\noalign{\vskip8pt}
\hfill - {e^{(n-2){\varPsi} }\over r^{4}} V({\varPhi} )
- {4e^{(n-2){\varPsi} }\over r^{2}} {\cal L}_{\Int}({\varPhi} , 
\widetilde{A})\biggr)- {\lambda ^{2}\over 4} {\cal L}_{\scal}({\varPsi} )
+ {1\over \lambda ^{2}} e^{(n+2){\varPsi} } \widetilde{R}(
\widetilde{{\varGamma} }) + {e^{n{\varPsi} }\over r^{2}}
\widetilde{\un{P}}\biggr\},\quad (7.5)\cr}
$$
where
$$\displaylines{
\hfill {\cal L}_{\scal}({\varPsi} ) = (\ov{M}\widetilde{g}^{(\gamma
\nu )} + n^{2}g^{[\mu \nu ]} g_{\delta \mu } \widetilde{g}^{(\delta
\gamma )}){\varPsi} _{,\nu } {\varPsi} _{,\gamma
},\hfill\llap{(7.6)}\cr
\ov{M}= (\ell ^{[dc]}\ell _{[dc]}-3n(n-1)).\cr}
$$
It is easy to see that the scalar field ${\varPsi} $ is chargeless (it has no
colour charges). However, it couples the gauge (Yang--Mills') field and
the Higgs' field due to the terms:
$$\displaylines{
\hfill 8\pi e^{-(n+2){\varPsi} } {\cal
L}_{\rYM}(\widetilde{A}),\hfill\llap{\hbox{(7.7a)}}\cr
\noalign{\vskip4pt}
\hfill + {2e^{-2{\varPsi} }\over r^{2}} {\cal L}_{\kin}
(\Nabla^{\gauge} {\varPhi} ),\hfill\llap{\hbox{(7.7b)}}\cr
\hfill - {e^{(n-2){\varPsi} }\over r^{4}} V({\varPhi}
),\hfill\llap{\hbox{(7.7c)}}\cr
\noalign{\vskip4pt}
\hfill - {4e^{(n-2){\varPsi} }\over r^{2}} {\cal L}_{\Int}({\varPhi} ,
\widetilde{A}),\hfill\llap{\hbox{(7.7d)}}\cr}
$$
It also couples the cosmological terms:
$$\displaylines{
\hfill {4\over \lambda ^{2}} e^{(n+2){\varPsi} } 
\widetilde{R}(\widetilde{{\varGamma}
}),\hfill\llap{\hbox{(7.8a)}}\cr
\hfill {e^{n{\varPsi} }\over r^{2}} \widetilde{\un{P}}.\hfill\llap{\hbox{(7.8b)}}\cr}
$$
These six terms (7.7a--d) and (7.8a--b) suggest that the scalar field is
massive. This is different than in Brans--Dicke theory, where the
scalar field couples the trace of the energy-momentum tensor for
matter. Our results from this section generalize results from Refs.
[23], [24].

We can consider a more general case for the lagrangian in the
theory i.e.
$$
(R(W)+\beta )\sqrt{|\varkappa|}.\eqno(7.9)
$$
In this case we get an additional cosmological term $\beta \rho ^{n}$ or in terms of
the field ${\varPsi} $ and after a redefinition of the nonsymmetric metric
$g_{\mu \nu }$
$$
\Bigl(\beta \!\!\mmmint\! dm(y)\sqrt{|\widetilde{g}|}\Bigr)e^{n{\varPsi}}.\eqno(7.10)
$$
This term can be added to remaining cosmological terms in the theory.
Moreover we do not consider this term anymore because it is not in the
real spirit of the Kaluza--Klein (Jordan--Thiry) Theory.

\vfill\eject

\null
\vskip36pt plus4pt minus4pt
\centerline{{\four\bf 8. Symmetry Breaking. Higgs' Mechanism.}}

\vskip3pt
\centerline{{\four\bf Mass Generation. Cosmological Constant}}
\vskip24pt plus4pt minus4pt
\write0{8. Symmetry Breaking. Higgs' Mechanism.
 Mass Generation. Cosmological Constant \the\pageno}

Before passing to symmetry breaking in our theory we do some
cosmetic manipulations, connecting constants. The connection $\omega $ on the
fibre bundle $P$ has no correct physical dimensions. It is necessary to
pass in all formulae from $\omega $ to 
$\alpha _{S}{\displaystyle{1\over\sqrt{\hbar
c}}} \omega $
$$
\omega \rightarrow \alpha _{S} {1\over\sqrt{\hbar c}} \omega ,\eqno(8.1)
$$
where $\hbar$ is the Planck's constant, $c$ is the velocity of light in the
vacuum and $\alpha _{S}$ is dimensionless coupling constant for the 
Yang--Mills'
field if this field couples matter. For example in the electromagnetic
case $\alpha _{S} = {\displaystyle{1\over \sqrt{137}}}$. We use $\alpha _{g} = \alpha ^{2}_{S}
= {\displaystyle{g^{2}\over \hbar c}}$ where $g$ is a coupling constant
for a gauge field. The redefinition of $\omega$ is equivalent to a usual
treatment in a local
section $e:V\supset U\to P$, $e^*\omega={\displaystyle{g\over \hbar c}}A$. 
Now our quantities have correct physical dimensions.

Using (8.1) one easily writes the integral of action (6.11):
$$\eqalignno{
S ={}& - {1\over r^{2}}\mmint_{{\bold U}}\sqrt{-g}\,d^{4}x\biggl[
\ov{R}(\ov{W}) + {8\pi \lambda ^{2}\alpha ^{2}_{s}\over 4c\hbar} \cdot 
\biggl(e^{-(n+2){\varPsi} } \cdot {\cal L}_{\rYM} +\cr
\noalign{\vskip4pt}
&+ {e^{-2{\varPsi}}\over 4\pi r^2} {\cal L}_{\kin}-
{e^{(n-2){\varPsi}}\over 8\pi r^2}V({\varPhi} ) -
{e^{(n-2){\varPsi}}\over 2\pi r^2} {\cal L}_{\Int} ({\varPhi}
,\widetilde{A})
-{\cal L}_{\scal}({\varPsi})\biggr)+\lambda _c\biggr].&(8.2)\cr}
$$
Thus we get the integral of action for the matter described by the
Yang--Mills' field and scalar field coupled to gravity. If we want to
be in line with the ordinary coupling between gravity and matter we
should put:
$$
{8\pi \lambda ^{2} \alpha ^{2}_{S}\over 4c\hbar} = {8\pi G_{\rN}\over
c^{4}}.\eqno(8.3)
$$
One gets:
$$
\lambda = {2\over \alpha _{\rS}} \ell _{pl} = {2\over \sqrt{\alpha
_g}} \ell _{pl},\eqno(8.4)
$$
where $\ell _{pl}$ is the Planck's length $\ell _{pl} =
\sqrt{{\displaystyle{G_{\rN}\hbar\over c^{3}}}} \simeq 10^{-33}$~cm. In this case
we have:
$$
\lambda _{c} = \biggl({e^{(n+2){\varPsi} } \alpha ^{2}_{S}\over \ell ^{2}_{pl}}
\widetilde{R}(\widetilde{{\varGamma} })+
{e^{n{\varPsi} }\over r^{2}}\widetilde{\un{P}}\biggr).\eqno(8.5)
$$
For $V({\varPhi} )$ one gets:
$$\eqalignno{
V({\varPhi} ) ={}& 2 \ell _{ab} \biggl(\un{g}^{([\tilde m \tilde n
],[\tilde a \tilde b ])} 
\times \biggl(\alpha _{S}{1.\over\sqrt{\hbar c}}C^{a}_{cd}{\varPhi}
^{c}{}_{\tilde m }{\varPhi} ^{d}{}_{\tilde n }-{1\over \alpha _{S}}
\sqrt{\hbox{\dkr{$\hbar$}} c}\mu ^{a}{}_{\dah{i}}f^{\dah{i}}{}_{\tilde m \tilde n }
-{\varPhi} ^{a}{}_{\tilde e }f^{\tilde e }{}_{\tilde m \tilde n }\biggr) \times\cr
\noalign{\vskip4pt}
&\times \biggl(\alpha _{S}{1.\over\sqrt{\hbar c}}C^{b}_{ef}{\varPhi}
^{e}{}_{\tilde a }{\varPhi} ^{f}{}_{\tilde b }-{1\over \alpha _{S}}
\sqrt{\hbox{\dkr{$\hbar$}} c}\mu ^{b}{}_{\dah{j}}f^{\dah{j}}{}_{\tilde a \tilde b }
-{\varPhi} ^{b}{}_{\tilde d }f^{\tilde d }{}_{\tilde a \tilde
b}\biggr)\biggr)_{av}+\cr
\noalign{\vskip4pt}
&- {\ell _{ab}\over V_{2}} \mmint_{\rM}\sqrt{|\widetilde{g}|}\,d^{n_{1}}x
\biggl[g^{\tilde a \tilde n } g^{\tilde b \tilde m } L^{a}{}_{\tilde a
\tilde b }\cdot\cr
\noalign{\vskip4pt}
&\times \biggl(\alpha _{S}{1.\over \sqrt{\hbar c}}C^{b}_{cd}
{\varPhi} ^{c}{}_{\tilde m }{\varPhi} ^{d}{}_{\tilde n }-
{1\over \alpha _{S}}\sqrt{\hbox{\dkr{$\hbar$}} c}\mu ^{b}{}_{\dah{i}}f^{hat
i}{}_{\tilde m \tilde n }-{\varPhi} ^{b}{}_{\tilde e }f^{\tilde e
}{}_{\tilde m \tilde n }\biggr)\biggr],&(8.6)\cr}
$$
where
$$
\un{g}^{([\tilde m \tilde n ],[\tilde a \tilde b ])} = 
g^{[\tilde m \tilde n ]} g^{[\tilde a \tilde b ]}\eqno(8.7)
$$
(see Appendix A for further details).
Let us pass to spontaneous symmetry breaking and the Higgs' mechanism
in our theory. In order to do this we look for the critical points
(the minima) of the potential $V({\varPhi} )$. The scalar factor before $V({\varPhi} )$ is
positive and has no influence on these considerations. However, our
field ${\varPhi} $ satisfies the constraints
$$
{\varPhi} ^{c}{}_{\tilde b } f^{\tilde b }{}_{\dah{i}\tilde a } - 
\mu ^{a}{}_{\dah{i}} {\varPhi} ^{b}{}_{\tilde a } C^{c}_{ab} = 0.\eqno(8.8)
$$
Thus we must look for the critical points of
$$
V' = V + \psi ^{\dah{i}\tilde d }{}_{c} ({\varPhi} ^{c}{}_{\tilde b }
f^{\tilde b }{}_{\dah{i}\tilde d }-\mu ^{a}{}_{\dah{i}}
{\varPhi} ^{b}{}_{\tilde d }C^{c}_{ab}),\eqno(8.9)
$$
where $\psi ^{\dah{i}\tilde d }{}_{c}$ is a Lagrange multiplier. Let us calculate 
${\delta V'\over \delta {\varPhi} }$. One finds 
$$\eqalignno{
{\delta V' \over \delta {\varPhi} ^{w}{}_{\tilde v }} ={}& \ell
_{ab}\biggl(\biggl\{\biggl[ 4\un{g}^{([\tilde m \tilde n ],[\tilde a
\tilde b ])} 
\biggl(\alpha _{S}{1\over \sqrt{\hbar c}}C^{a}_{cd}
{\varPhi} ^{c}{}_{\tilde m }{\varPhi} ^{d}{}_{\tilde n }-
{1\over \alpha _{S}}\sqrt{\hbar c}\mu ^{a}{}_{\dah{i}}
f^{\dah{i}}{}_{\tilde m \tilde n }-{\varPhi} ^{a}{}_{\tilde c }
f^{\tilde c }_{\tilde m \tilde n }\biggr)\biggr] +\cr
\noalign{\vskip4pt}
&-\un{g}^{\tilde n \tilde a ,\tilde m \tilde b } \cdot 
L^{a}{}_{\tilde n \tilde m } -\un{g}^{\tilde m \tilde n ,\tilde r
\tilde p } \cdot \biggl( {\delta L^{a}{}_{\tilde m \tilde n }\over 
\delta H^{b}{}_{\tilde a \tilde b }}\biggr)\times \cr
\noalign{\vskip4pt}
&\times \biggl(\alpha _{S}{1\over \sqrt{\hbar c}}C^{d}_{ce}
{\varPhi} ^{c}{}_{\tilde r }{\varPhi} ^{e}{}_{\tilde p }-
{1\over \alpha _{S}}\sqrt{\hbar c}\mu ^{a}{}_{\dah{i}}
f^{\dah{i}}{}_{\tilde r \tilde p }-{\varPhi} ^{d}{}_{\tilde c }
f^{\tilde c }_{\tilde r \tilde p }\biggr)\times\cr 
\noalign{\vskip4pt}
&\times \biggl(\alpha _{S}{1\over \sqrt{\hbar c}}C^{d}_{ce}
{\varPhi} ^{c}{}_{\tilde r }\delta ^{f}{}_{w}\delta ^{\tilde v
}{}_{\tilde b }-\delta ^{b}{}_{w}\delta ^{\tilde d }{}_{\tilde v }
f^{\tilde d }{}_{\tilde a \tilde b }\biggr)\biggr\}\biggr)_{av}+\cr 
\noalign{\vskip4pt}
&+ \psi ^{\dah{i}\tilde d }{}_{c} (\delta ^{c}{}_{w}
f^{\tilde v }{}_{\dah{i}\tilde d }-\mu ^{a}{}_{\dah{i}}C^{c}{}_{aw}
\delta ^{\tilde v }{}_{\tilde d }),&(8.10)\cr}
$$
where
$$
H^{b}{}_{\alpha \beta } = \biggl(\alpha _{S}{1\over \sqrt{\hbar
c}}C^{d}_{cd} {\varPhi} ^{c}{}_{\tilde a }{\varPhi} ^{d}_{\tilde b } - 
{1\over \alpha _{S}}\sqrt{\hbar c}\mu ^{b}{}_{\dah{i}} 
f^{\dah{i}}{}_{\tilde a \tilde b } - {\varPhi} ^{b}{}_{\tilde c } 
f^{\tilde c }{}_{\tilde a \tilde b }\biggr)\eqno(8.11)
$$
and ${\displaystyle{\delta L^{a}{}_{\tilde m \tilde n }\over 
\delta H^{b}{}_{\tilde a \tilde b }}}$ satisfies the following equation: 
$$\displaylines{
\hfill\ell _{dc} g_{\tilde m \tilde b } g^{\tilde c \tilde m } 
{\delta L^{d}{}_{\tilde e \tilde a }\over \delta H^{w}{}_{\tilde p
\tilde q }} + \ell _{cd} g_{\tilde a \tilde m } g^{\tilde m \tilde c } 
{\delta L^{d}{}_{\tilde b \tilde e }\over \delta H^{w}{}_{\tilde p
\tilde q }}
= 2\ell _{cd} g_{\tilde a \tilde m } g^{\tilde m \tilde c } 
\delta ^{d}{}_{w} \delta ^{\tilde p}{}_{\tilde b } \delta ^{\tilde
q}_{\tilde c }.\hfill\llap{(8.12)}\cr}
$$
It is easy to see that, if
$$
H^{a}{}_{\tilde m \tilde n } = 0,\eqno(8.13)
$$
then
$$
{\delta V' \over \delta {\varPhi} } = 0,\eqno(8.14)
$$
if (8.8) is satisfied. This was noticed in Refs.\ [73], [78] and it is
known in the symmetric theory. $H^{a}{}_{\tilde m \tilde n }$ is the part of the curvature of $\omega $
over the manifold $M$. Thus this means that ${\varPhi} ^{0}_{\crt}$ , satisfying (8.13), is
a ``pure gauge". If potential $V({\varPhi} )$ is positively defined, then we have
the absolute minimum of~$V$.
$$
V({\varPhi} ^{0}_{\crt}) = 0.\eqno(8.15)
$$
But apart from this solution there are some others. Potential $V$ (see
(8.6)) is the result of the subtraction of two positively defined
terms. This means that there exist some local minima of $V$ and the
shape of the potential is shown on Fig.~3. Values of $V$ at these local
minima are not zero. If the potential is not positively defined, the
critical point ${\varPhi} ^{0}_{\crt}$ is not a point of the minimum and the shape of the
potential is as on Fig.~4. In this case we have no minimum. All of
these properties depend on values of the constants $\xi $ and $\zeta $ and on
groups $G$, $G_{0}$, $H$. The case with no minimum is not interesting from the
physical point of view. Radiative corrections may change the situation
as shown on Fig.~5 (see [81], [88]). Let us examine the situation
described in Fig.~3. At the point of local minimum we have:
$$
V({\varPhi} ^{1}_{\crt}) \neq 0\eqno(8.16)
$$
and
$$
H^{a}{}_{\tilde m \tilde n }({\varPhi} ^{1}_{\crt}) \neq 0.\eqno(8.17)
$$
\obraz{
\midinsert \centerline{\epsfxsize\hsize \epsfbox{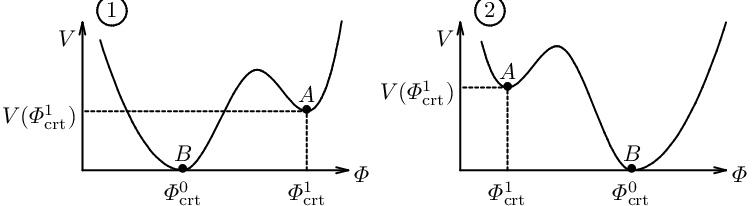}}
\centerline{\nine Fig. 3. Two shapes of the potential $V({\varPhi})$ 
in the case that there exist two minima,}
\vskip-1pt
\centerline{\nine absolute and local (the ``true'' and
``false'' vacua)}
\endinsert

}This means that ${\varPhi} ^{1}_{\crt}$ is not a ``pure gauge" and the gauge configuration
connected to ${\varPhi} ^{1}_{\crt}$ is not trivial. This indicates that the local
minimum is not a vacuum state. It is a ``false vacuum" in
contradistinction to ``true vacuum" for the absolute minimum 
${\varPhi} ^{0}_{\crt}$. It
is possible that the ``tunnel effect" from point $A$ to $B$ (see Fig.~3)
results in a second or first order phase transition (see [81--85]). Such
a possibility was considered by Weinberg and a second (local) minimum
occurred due to radiative corrections. Here we have this from the very
beginning on the classical level.

\obraz{
\midinsert \centerline{\epsfxsize\hsize \epsfbox{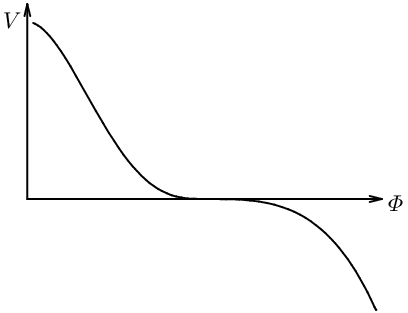}}
\centerline{\nine Fig. 4. The shape of potential $V({\varPhi})$ in the case
without minima}
\endinsert

}Now we answer the question of what is symmetry breaking in our
theory if we choose one of the critical values of ${\varPhi}
^{0}_{\crt}$ (We choose one
of the degenerated vacuum states and the spontaneous breaking of
symmetry takes place). In Ref.~[73] it was shown that if
$H^{a}{}_{\tilde m \tilde n } = 0$ and
(8.8) is satisfied then the symmetry is reduced to $G_{0}$. In the case of
the second minimum (local minimum---false vacuum), the unbroken
symmetry will be in general different. Let us called it $G'_0$ and its Lie
algebra $\fg'_0$. This will be the symmetry which preserves 
${\varPhi} ^{1}_{\crt}$ and the
constraint (8.8) imposed on ${\varPhi} ^{1}_{\crt}$. It is easy to see that the Lie
algebra of this unbroken group preserves Eq.~(8.10) and constraint (8.8)
under Ad-action. Thus it is necessary to find all symmetries of these
equations. The other possibility is to find symmetries of the
hypersurface, defined by $V({\varPhi} ) = \const$. and the constraints (8.8). This
strongly depends on the details concerning $\xi $, $\zeta $, $G$,
$G_{0}$, $H$ in
contradistinction to the absolute minimum 
${\varPhi} ^{0}_{\crt}$---true vacuum. Maybe it
would be possible to find a similar geometrical interpretation as in
the case of ${\varPhi} ^{0}_{\crt}$. But in general this group will be completely
different.

\obraz{
\midinsert \centerline{\epsfxsize\hsize \epsfbox{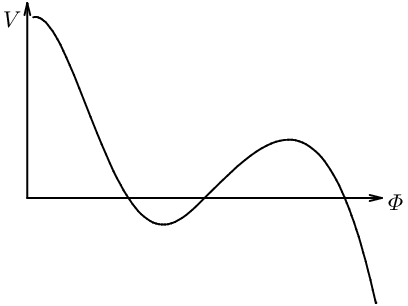}}
\centerline{\nine Fig. 5. The possible influence of radiative corrections}
\vskip-1pt
\centerline{\nine on the
shape of the potential $V({\varPhi})$ in the case without minima}
\endinsert

}For the symmetry group of $V$ is larger than $G$ (it is $H)$ we expect some
scalars which remains massless after the symmetry breaking in both
cases (i.e.\ $k=0$, $k=1$, ``true" and ``false" vacuum cases). They became
massive only through radiative corrections. They are often referred as
the pseudo-Goldstone bosons.

Let us consider a problem of the Higgs' multiplet ${\varPhi}
^{c}{}_{\tilde a }$ on $E$. One
can find a representation space (N) of ${\varPhi} ^{c}{}_{\tilde a }$ in the following way (see
Ref.~[80]). Let
$$
\Ad_{\rG} \rightarrow \sum^{}_{i}\oplus \un{n}_{i} \oplus \Ad_{\rG_{0}}
\eqno(8.18)
$$
be the decomposition of the adjoint representation of $G$, where $\un{n}_{i}$ are
irreducible representations of $G_{0}$ and let us consider the branching
rule of $\Ad_{\rH}$
$$
\Ad_{H} \rightarrow \sum^{}_{j}\oplus (\un{n}'_j\otimes
\un{m}_{j}),\eqno(8.19)
$$
where $\un{n}'_j$ are irreducible representations of $G_{0}$ and $\un{m}_{j}$ are irreducible
representations of $G$. The latest formula comes from the known fact
that
$$
G_{0} \otimes G \subset H.\eqno(8.20)
$$
Thus for each pair $(\un{n}_{i},\un{n}'_j)$ where $\un{n}_{i}$ and 
$\un{n}'_j$ are identical irreducible
representations of $G_{0}$ there is an $\un{m}_{j}$ multiplet of Higgs' field on 
$E$.
In this way we can decompose ${\varPhi} $ into a sum:
$$
{\varPhi} =\sum^{}_{(\underline{n}_{i},\underline{n}'_j)}\oplus {\varPhi}
^{(\underline{n}_i\underline{n}'_j)}_{\underline{m}_{j}},\quad \hbox{or}\quad 
N = \sum^{}_{(\underline{n}_{i},\underline{n}'_j)}\oplus \un{m}_{j}.
$$
Thus the multiplet of Higgs' field is quite complicated in
contradistinction to the usual case where the Higgs' field belongs to
the adjoint representation of a chosen group. Moreover in our case we
have a smaller number of parameters in the theory. The theory (it
means a symmetry breaking) is established by a coupling constant $\alpha_{S}$, 
a radius $r$, parameters coming from the nonsymmetricity of the theory $\xi$, 
$\zeta$ a homomorphism $\mu $, an embedding of $G$ in $H (\alpha ^{c}_{i})$ 
and an embedding of $G_{0}$ in
$G$ i.e.\ the manifold $M$.

In the classical approach we need much more parameters in order to get
a desired symmetry breaking pattern.

Let us notice that in the branching rule for $\Ad_{H}$ we have a term
$\Ad G_{0} \otimes {\sym I}$ and in the decomposition for Ad $G$, the Ad $G_{0}$. Thus the space
$N$ contains the representation space for ${\sym I}$.

Let us pass to the integral of action (8.2) in the two vacuum
cases: ${\varPhi} ^{0}_{\crt}$ and ${\varPhi} ^{1}_{\crt}$. Let us expand the Higgs' field ${\varPhi} ^{\tilde a}_{b}$ in the
neighbourhood of $({\varPhi} ^{k}_{\crt})^{a}{}_{\tilde b } $, $k =0,1$.
$$
{\varPhi} ^{a}{}_{\tilde b } = ({\varPhi}
^{k}_{\crt})^{a}{}_{\tilde b } + (\phi ^{k})^{a}{}_{\tilde b }\eqno(8.21)
$$
and apply this formula for $e^{*}\mathop{\nabla_{\mu}}\limits^{\gauge}{\varPhi}
^{b}{}_{\tilde a }$ and $V({\varPhi} )$. We get
$$\displaylines{
\hfill e^{*} \mathop{\nabla_{\mu}}\limits^{\gauge} {\varPhi} ^{b}{}_{\tilde a } = e^{*} 
\mathop{\nabla_{\mu}}\limits^{\gauge}(\phi ^{k})^{b}{}_{\tilde a }
+ \alpha _{S} {1\over \sqrt{\hbar c}}(({\varPhi}
^{k}_{\crt})^{a}{}_{\tilde a } C^{b}_{ac} \alpha ^{c}_{j}
\widetilde{A}^{j}_{\mu} +
({\varPhi} ^{k}_{\crt})^{a}{}_{\tilde b } f^{\tilde b }{}_{\tilde a j} 
\widetilde{A}^{j}{}_{\mu}),\hfill(8.22)\cr
\hfill V({\varPhi} ) = V({\varPhi} ^{k}_{\crt}) + \widetilde{V}^{k}
(\phi ^{k}) ,\quad\ k = 0, 1,\hfill\llap{(8.23)}\cr}
$$
where $V({\varPhi} ^{k}_{\crt})$ is the value of $V$ for the critical value of ${\varPhi} $ and $\widetilde{V}^{k}$ is
the polynomial of fourth order in $\phi ^{k}$. If we put (8.19) to (6.4) and to
(6.5) we can solve these equations with respect to $L^{a}{}_{\tilde m
\tilde n }$ and $L^{b}{}_{\mu\alpha }$ in a
certain order of approximation (there are some procedures in N.G.T.
(see Refs.\ [79], [80] and sec.~17). Thus we can expand the kinetic
term ${\cal L}_{\kin}(\mathop{\nabla_{\mu}}\limits^{\gauge}{\varPhi})$ with respect to the decomposition (8.19) and we get a
mass-matrix for vector bosons $\widetilde{A}^{j}_{\mu}$ which strongly depends on ${\varPhi} ^{k}_{\crt}$.
$$
- e^{-2{\varPsi} } N^{\mu\nu } M^{2}_{ij}({\varPhi} ^{k}_{\crt}) 
\widetilde{A}^{i}_{\mu} \widetilde{A}^{j}_{\nu }.\eqno(8.24)
$$

Let us consider the expression
$$
({\varPhi} ^{k}_{\crt})^{m}{}_{\tilde n } C^{b}_{ms} \alpha ^{s}_{i} + 
({\varPhi} ^{k}_{\crt})^{b}{}_{\tilde m } f^{\tilde m }{}_{\tilde n i}\eqno(*)
$$
in order to find its interpretation. One easily notices that it equals
to
$$
([\Ad'_{\rH}(Y_i)+\Ad'_{\rG}(Y_{i})]\cdot [{\varPhi}
^{k}_{\crt}])^{b}{}_{\tilde m }
$$
($\Ad_{\rH}$ and $\Ad_{\rG}$ mean the adjoint representation of the group $H$ and $G$
respectively.)

Thus if $k=0$ $(*)$ equals zero for $Y_{i}\in \fg_{0}$ and if $k=1$ $(*)$ equals zero for
$Y_{i}\in \fg'_0$. The latest statements come from the invariancy of the vacuum
state with respect to the action of the group $G_{0}$ for $k=0$
($G'_0$ for $k=1)$.
Generators of $\fg_{0}$ $(\fg'_0)$ should annihilate vacuum state. Thus the matrix
elements $M^{2}_{ij}({\varPhi} ^{k}_{\crt})$ are zero for $i$, $j$ corresponding to
$\fg_0$ $(\fg'_0)$.

From the invariancy of the potential $V$ with respect to the action of
the group $G$ one gets
$$
{\partial^2V\over \partial{\varPhi}^b{}_{\tilde n }\partial
{\varPhi}^d{}_{\tilde d }}\bigg|_{{\varPhi} ={\varPhi} ^{k}_{\crt}} 
(T^{b}{}_{\tilde n }{}^c{}_{\tilde c })_{i}[{\varPhi}
^{k}_{\crt}]^{\tilde c }{}_{c} = 0,\eqno({*}{*})
$$
where
$$
(T^{b}{}_{\tilde n }{}^c{}_{\tilde c })_{i}[{\varPhi}
^{k}_{\crt}]^{\tilde c }{}_{c} = [{\varPhi}
^{k}_{\crt}]^{m}{}_{\tilde n }C^{b}_{ms}\alpha ^{s}_{i} + 
[{\varPhi} ^{k}_{\crt}]^{b}{}_{\tilde m }f^{\tilde m }{}_{\tilde n
i}.
$$
Matrix $N^{\mu\nu }$ depends on the tensor $g_{\alpha \beta }$ and for 
$g_{\alpha \beta } = g_{\beta \alpha }$ we have $N^{\mu\nu } =
g^{\mu\nu }$. In the zero$^{\text{\rm th}}$ order of approximation for 
$g_{\alpha \beta }$ we get:
$$
- e^{-2{\varPsi} } \eta ^{\mu\nu } M^{2}_{ij}({\varPhi} ^{k}_{\crt} )
\widetilde{A}^{i}{}_{\mu} \widetilde{A}^{j}{}_{\nu },\eqno(8.24\text{\rm a})
$$
where $\eta ^{\mu\nu }$ is the Minkowski tensor. Eigenvalues of $M^{2}_{ij}$
$({\varPhi} ^{k}_{\crt}$ are the
squares of the masses of the gauge bosons. The secular equation
$\det (M^{2}-m^{2}J)=0$ gives us a mass spectrum of massive vector bosons.

Thus there is an orthogonal matrix $(A^{i}{}_{j})=A$ such that 
$A^{T}=A^{-1}$ and
$$
A^{-1}M^{2}({\varPhi} ^{k}_{\crt})A = 
\left[\matr{m^{2}_{1}({\varPhi} ^{k}_{\crt})& \ldots &0\cr
\vdots&\ddots&\vdots\cr
0&\ldots&m^{2}_{\ell _{k}}({\varPhi}_{\crt})\cr}\right],\quad\ 
k=0,1.\eqno(8.25)
$$
$\ell _{0}=n_{1}$, $\ell _{1}=\dim G- \dim G'_0$.

In this way we transform the broken vector fields into massive vector
fields
$$
\widetilde{B}^{i'}{}_{\mu}=\sum^{l_{k}}_{j=1}A^{i'}{}_{j} 
\widetilde{A}^{j}{}_{\mu},\eqno(8.26)
$$
such that
$$
\eta^{\mu\nu } \sum^{l_{k}}_{j=1}m^{2}_{j} ({\varPhi}^{k}_{\crt}) 
\widetilde{B}^{j}{}_{\mu} \widetilde{B}^{j}_{\nu }= \eta ^{\mu\nu } 
M^{2}_{ij} ({\varPhi} ^{k}_{\crt}) \widetilde{A}^{i}_{\mu} 
\widetilde{A}^{j}_{\nu }\eqno(8.27)
$$
(see Appendix A for further details).
The matrix $A$ depends of course, on the critical point of the Higgs'
potential.

Moreover we should remember the formula (4.51) which is a constraint
for Higgs field. In this way we should differentiate with respect to
independent components of ${\varPhi} $. In all the formulae concerning Higgs
field we suppose (removing spurious components) that we have to do
with independent components. Solving the equation (4.51) we get
independent components of Higgs field. Let the components be 
$\widehat{{\varPhi} }^{a_{1}}{}_{\tilde a _{1}}$ and
$$
{\varPhi} ^{a}{}_{\tilde a } = {\varPhi} ^{a}{}_{\tilde a }
(\widehat{{\varPhi}} ^{a_{1}}{}_{\tilde a _{1}}).
$$
In this case we get
$$
{\delta V\over \delta \widehat{{\varPhi}}^{a_1}{}_{\tilde a _1}}=
{\delta V\over \delta {\varPhi}^a{}_{\tilde a _1}}
{\delta \widehat{{\varPhi}}^{a_1}{}_{\tilde a _1}\over
\delta \widehat{{\varPhi}}^{a_1}{}_{\tilde a _1}}
$$
and
$$
{\delta ^2V\over \delta \widehat{{\varPhi}}^{b_1}{}_{\tilde b _1}
\delta \widehat{{\varPhi}}^{a_1}{}_{\tilde a _1}}=
{\delta ^2V\over \delta \widehat{{\varPhi}}^b{}_{\tilde b }\delta
{\varPhi}^a{}_{\tilde a }}\cdot
{\delta {\varPhi}^b{}_{\tilde b }\over \delta \widehat{{\varPhi}}^{b_1}{}_{\tilde b
_1}}{\delta {\varPhi}^a{}_{\tilde a }\over \delta
\widehat{{\varPhi}}^{a_1}{}_{\tilde
a_1}}.
$$
In all the formulae containing the derivatives of $V$ we suppose to pass
from $\displaystyle{\delta\over \delta {\varPhi}} $ to
$\displaystyle{\delta\over \delta \widehat{{\varPhi}}} $ using above formulae (the same in the matrix below).

The matrix of the masses for Higgs' bosons can be obtained in a similar
way i.e.
$$
V({\varPhi} ) = V({\varPhi} ^{k}_{\crt}) + {1\over 2} 
{\delta ^{2}V\over \delta {\varPhi} ^{a}{}_{\tilde a }\delta 
{\varPhi} ^{b}_{\tilde b }}\bigg|_{{\varPhi}={\varPhi}^k_{\crt}}
\phi ^{a}_{\tilde a }\phi^{b}{}_{\tilde b }+ \ldots \eqno(8.28)
$$
The matrix
$$
m^{2}({\varPhi} ^{k}_{\crt})^{\tilde a }{}_{a}{}^{\tilde b }{}_{b} = 
- {\delta ^{2}V\over \delta {\varPhi} ^{a}{}_{\tilde a }\delta 
{\varPhi} ^{b}_{\tilde b }}\bigg|_{{\varPhi}={\varPhi}^k_{\crt}}\eqno(8.29)
$$
is positively defined and symmetric (see Appendix A for further details). Thus we can diagonalize it and we
get
$$
{\delta ^{2}V\over \delta {\varPhi} ^{a}{}_{\tilde a }\delta 
{\varPhi} ^{b}_{\tilde b }}\bigg|_{{\varPhi}={\varPhi}^k_{\crt}}
\phi ^{\tilde a }{}_{a}\phi ^{\tilde b }{}_{b}= - 
\sum^{\ell _{k}}_{j=1}m^{2}_{j} ({\varPhi}
^{k}_{\crt})^{a}{}_{\tilde a }
{\varPsi} ^{a}{}_{\tilde a }{\varPsi} ^{a}{}_{\tilde a },\eqno(8.30)
$$
where
$$
{\varPsi} ^{a}{}_{\tilde a } = \sum^{}_{b,\tilde b } A ^{\tilde b
}{}_{b}{}^{\tilde a }{}_{a}\phi_{\tilde b }{}^{b}.\eqno(8.31)
$$
The matrix $A^{\tilde a }{}_{a}{}^{\tilde b }{}_{b}$ depends on the critical point of the Higgs' potential.
For the mass matrix one has:
$$
(A^{-1})^{\tilde c }{}_{c}{}^{a}{}_{\tilde a }
m^{2}({\varPhi} ^{k}_{\crt})^{\tilde a }{}_{a}{}^{\tilde b }{}_{b} 
A^{\tilde d }{}_{d}{}^{b}{}_{\tilde b }= 
(m^{2}({\varPhi} ^{k}_{\crt})^{\tilde c }{}_{c})\delta ^{\tilde d
}{}_{\tilde c }\delta ^c{}_d.\eqno(8.32)
$$
The eigenvalue problem for $m^{2}({\varPhi} ^{k}_{\crt})$ can be posed as follows
$$
m^{2} ({\varPhi} ^{k}_{\crt})^{\tilde a }{}_a^{\tilde b }{}_b
X^{a}{}_{\tilde a } =m^{2} ({\varPhi} ^{k}_{\crt}) 
X^{\tilde b }{}_{b}.\eqno(8.33)
$$
The mass spectrum of Higgs' particles one gets from the secular
equation
$$
\det ([m^{2}({\varPhi} ^{k}_{\crt})]-m^{2}({\varPhi}
^{k}_{\crt})J) = 0.\eqno(8.34)
$$
One easily writes the shape of mass matrix:
$$\displaylines{
m^{2}({\varPhi} ^{k}_{\crt})^{\tilde a }{}_a{}^{\tilde r}{}_r
=-{1\over r^{2}} \cdot {\delta ^{2}V\over \delta H^{c}{}_{\tilde n
\tilde m } \delta H^{d}{}_{\tilde p \tilde q }} 
\biggl[\alpha _{s}\sqrt{{\hbar\over c}}C^{c}_{av}(\delta ^{\tilde a
}{}_{\tilde n }({\varPhi} ^{k}_{\crt})^{v}{}_{\tilde m }-
({\varPhi} ^{k}_{\crt})^{v}{}_{\tilde n }\delta ^{\tilde a }{}_{m})-
f^{\tilde a }{}_{\tilde n \tilde m }\delta ^{c}_{a}\biggr] \cdot\hfill \cr
\cdot \biggl[\alpha _{s}\sqrt{{\hbar \over c}}C^{d}_{ru}(
\delta ^{\tilde r }{}_{\tilde p }({\varPhi}
^{k}_{\crt})^{u}{}_{\tilde q }-({\varPhi} ^{k}_{\crt})^{u}{}_{\tilde
p }\delta^{\tilde r}{}_{\tilde q}-f^{\tilde r }{}_{\tilde p\tilde q}\biggr] +\cr
\hfill + {\alpha _{s}\over r^{2}}\sqrt{{\hbar\over c}}
\biggl({\delta V\over \delta H^{c}{}_{\tilde n \tilde m }}\biggr)
C^{c}_{ar}(\delta ^{\tilde a }{}_{\tilde n }\delta ^{\tilde r
}{}_{\tilde m }- \delta ^{\tilde a }{}_{\tilde m }\delta ^{\tilde r
}{}_{\tilde n })\quad\ (8.35)\cr}
$$
(see Appendix A for further details).
The diagonalization procedure of the matrix $m^{2}({\varPhi}
^{k}_{\crt})^{\tilde a }{}_a{}^{\tilde b }{}_b$ can be
achieved in the two following ways. The first way more theoretical is
as follows.

The matrix defines a quadratic form on the representation space---$N$
for Higgs' field. Moreover the space $N$ can be decomposed into Higgs'
multiplets $\un{m}_{j}$ and according this decomposition the matrix can be
written in a block diagonal form
$$
[m^{2}({\varPhi} ^{k}_{\crt})] = \sum^{}_{j} \oplus m^{2}_{j}
({\varPhi} ^{k}_{\crt}).\eqno(8.36)
$$
We can diagonalize every matrix $m^{2}_{j}({\varPhi} ^{k}_{\crt})$. Corresponding to the
multiplet $\un{m}_{j}$.

The second way is based on the following observation.

The matrix $m^{2}({\varPhi} ^{k}_{\crt})^{\tilde a }{}_a{}^{\tilde b
}{}_b$ can be transformed into a different matrix
$(n_{1}n)\times (n_{1}n)$ forming an index from two indices
$\widetilde{a}$ and $a$
$$
\ov{a} = \alpha a + \beta \widetilde{a} + \gamma, \eqno(8.37)
$$
where $\alpha $, $\beta $, $\gamma $ are integers. The new index $\ov
a$ should be unambigous. Thus
we must choose $\alpha $, $\beta $, $\gamma $ in such a way that for every
$\ov{a}\in N^{n_{1}n}_{1}$ the
equation (8.37) has only one solution for $a\in N^{n}_{1}$, 
$\widetilde{a}\in N^{n_{1}}_{1}$.

After this we can diagonalize $m^{2}({\varPhi} ^{k}_{\crt})$ as an ordinary matrix. In
sec.~17 we consider some expansion procedures for $m^{2}({\varPhi} ^{k}_{\crt})$ and the
second way is more convenient in this case.

What is the scale of these masses? It is easy to see that:
$$
m_{\tilde A} \cong {\alpha _{S}\over r} \biggl({\hbar \over c}\biggr),\eqno(8.38)
$$
where $m_{\tilde A}$ is the mass of a typical vector boson which obtained mass due
to the Higgs' mechanism. The parameter, radius $r$, defines the scale of
these masses as in Refs.~[75], [80]. The spectrum of masses of all
massive vector bosons depends, of course, on the details of the theory.
One can say the same about masses of scalar particles (Higgs'
particles). Due to decomposition (8.18), we get mass terms from the
expansion (8.20) and the scale of mass for massive scalars will be the
same as for massive vector bosons. 
We can also define a so called deconfinement parameter $\varLambda(G)$ of order
of a mass of typical Higgs' particle. $\varLambda(G)$ in our theory is of the
same order as $m_{\tilde A}$.
An interesting point will arise
here if some of these particles remain without a rest mass or not. This
has some importance in the hierarchy of mass scales and the hierarchy
of physical interactions (see Ref.~[89]). Moreover we expect massless
scalars, the so-called pseudo-Goldstone bosons. It is interesting to
notice that the scale of masses is the same in both cases $k = 0, 1$,
however there is a different spectrum of masses for the vector and
scalar particles and with different unbroken groups. It is easy to
notice that due to a factor $e^{-2{\varPsi} }$ in Eq.~(8.22) 
we will have to do with
an effective scale of masses depending on~${\varPsi} $.

Let us consider the following decomposition of the connection $\omega _{E}$
defined on the principal fibre bundle $Q(E,G)$ 
$$\displaylines{
\hfill \omega _{\rE} = \omega ^{0}_{\rE} + \sigma
_{\rE},\hfill\llap{(8.39)}\cr
\hfill \omega _{\rE}^0\in \fg, \quad\ \sigma _{\rE}\in
\fm,\hfill\llap{(8.40)}\cr}
$$
corresponding to the decomposition of the Lie algebra $\fg$
$$
\fg= \fg_{0}\dot{+} \fm.
$$
In this way we consider a reduction of $Q(E,G)$ to $Q_{0}(E,G)$ induced by an
embedding of $G_{0}$ into $G$. The form $\omega ^{0}_{E}$ is a connection defined on $Q_{0}(E,G)$
and $\sigma _{E}$ is a tensorial form defined on $Q(E,G)$. We suppose that the
reduction of the bundle $Q$ to $Q_{0}$ is possible. The condition for this is
quoted in sec.~12. In this way the transport of the connection is
possible, because $G_{0}$ is a closed subgroup of $G$. The form $\omega ^{0}_{E}$ corresponds
to the Yang--Mills' field (massless vector bosons) which remains after
symmetry breaking. The tensorial form $\sigma _{E}$ corresponds to massive vector
bosons.

One gets for the curvature form
$$
{\varOmega} _{\rE} = {\varOmega} ^{0}_{\rE} + D^{0}\sigma _{\rE} + 
[\sigma _{\rE},\sigma _{\rE}],\eqno(8.41)
$$
where ${\varOmega} ^{0}_{\rE}$ is a curvature form for $\omega ^{0}_{E}$ and $D^{0}$ means a covariant derivative
with respect to $\omega ^{0}_{\rE}$.
Thus
$$\displaylines{
\hfill {\varOmega} ^{0}_{\rE} = {1\over 2} \widetilde{H}^{\dah{i}}{}_{\mu\nu } \theta ^{\mu}\wedge \theta ^{\nu } \beta
^{i}{}_{\dah{i}} Y_{i},\hfill\llap{\hbox{(8.42a)}}\cr
\hfill\sigma _{\rE} = \sigma ^{\tilde a } Y_{\tilde a
},\hfill\llap{\hbox{(8.42b)}}\cr
\hfill e^{*}D^{0}\sigma _{\rE} = \mathop{\nabla_{[\mu}^0}\limits^{\gauge} 
\overline{\sigma }^{\tilde a }_{\nu ]} \overline{\theta }^{\mu}\wedge
\overline{\theta }^{\nu } Y_{\tilde a
},\hfill\llap{\hbox{(8.42c)}}\cr
\hfill \mathop{\nabla_{[\mu}^0}\limits^{\gauge} 
\overline{\sigma }^{\tilde a }_{\nu ]} = \partial _{[\mu}
\overline{\sigma }^{\tilde a }_{\nu ]} + f^{\tilde a }{}_{\tilde b i}
\beta ^i{}_{\dah{i}}\overline{\sigma }^{\dah{b}}_{[\nu }\ov{A}^{\dah{i}}_{\mu
]},\hfill\llap{\hbox{(8.42d)}}\cr
%\obraz{\noalign{\eject}}
\hfill e^{*}\sigma _{\rE} = \overline{\sigma }^{\tilde a }_{\nu }
\overline{\theta }^{\nu } Y_{\tilde a }= \overline{\sigma }^{i}_{\nu }
\overline{\theta }^{\nu } \beta ^{\tilde a }_{i} Y_{\tilde a
},\hfill\llap{\hbox{(8.42e)}}\cr
\hfill e^{*}\omega ^{0}_{\rE} = \widetilde{A}^{i}_{\mu} \beta ^{i}_{\dah{i}}
\overline{\theta }^{\mu} Y_{i},\hfill\llap{\hbox{(8.42f)}}\cr}
$$
The matrices $\beta ^{\tilde a }_{i}$, $\beta ^{i}_{\dah{i}}$ define an embedding of 
$\fg_{0}$ into $\fg$ ($G_{0}$ into $G$),
$\mathop{\nabla^0_{\mu}}\limits^{\gauge} $ means a gauge derivative with respect to the connection $\omega
^{0}_{\rE}$ and $e$ is a local section.

The above decomposition of $\omega _{E}$ corresponds to a case, where the symmetry
group is broken from $G$ to $G_{0}$, i.e. to the first type of the critical
points of the Higgs' potential. Moreover in the case of the second
type of the critical point we can repeat the decomposition of $\omega
_{\rE}$ according to the decomposition
$$
\fg = \fg'_0\dot{+}\fm',\eqno(8.43)
$$
where $\fg'_0$ corresponds to $G]_0$---a group which remains unbroken.
One gets
$$
\omega _{\rE} = \omega_{\rE}^{,o} + \sigma '_{\rE}
$$
and $Q(E,G)$ is reduced to $Q(E,G'_0)$.

One easily connects $\sigma _{\rE}$ and $\sigma'_{\rE}$ with 
$\widetilde{B}$ (i.e.\ fields with defined non-zero
rest mass, because $\widetilde{A}$ have not defined masses) fields
$$
e^{*}\sigma _{\rE} = {g\over \hbar c} \widetilde{B}^{i,}_{\mu}
\overline{\theta }^{\mu} Y_{i'},\eqno(8.44)
$$
where
$$
Y_{i'} = (A^{-1})^{i}_{i'} Y_{i} = (A^{-1})^{i}_{i'} \beta ^{\tilde a }_{i} 
Y_{\tilde a }.\eqno(8.45)
$$
The matrix A is defined by the Eq.~$(8.25*)$. The same holds for
$\sigma'_{\rE} $.

Let us consider the following gauge transformation i.e.\ a change of a
local section of $Q(E,G)$ from $e$ to $f$
$$
e(x) = \nad{U}{(k)}{}^{-1}(x)f(x),\eqno({*}{*}{*})
$$
where
$$\displaylines{
\nad{U}{(k)}(x) = \exp \Bigl(\sum_{\nad{a}{\nu}} \eta_{\nad{a}{\nu}}(x)
Y_{\nad{a}{\nu}}\Bigr),\cr
\hbox{for}\ k=0,\ Y_{\nad{a}{\nu }} \in \fm\ \hbox{i.e.}\ \nad{a}{\nu }=
\widetilde{a},\quad\ \hbox{for}\ k=1,\ Y_{\nad{a}{\nu }} \in \fm' .\cr}
$$
and $\nad{\eta}{(k)}_{\nad{a}{\nu }}(x)$ is a multiplet of scalar fields on $E$ transforming according
to the Ad $G_{0}$ (Ad $G'_0$). $Y_{\nad{a}{\nu }}$ span $\fm$ or 
$\fm'$ ($k=0$ or $k=1)$. Such a gauge
condition is called a ``unitary gauge".

Let us consider the following parametrization of the Higgs' field
$$
{\varPhi} ^{c}{}_{\tilde a } =\Ad_{\rG}(\nad{U}{(k)}(x))^{\tilde a '}_{a}
\cdot \Ad_{\rH}(\nad{U}{(k)}(x))^{c}_{c'} \cdot 
(({\varPhi})_{\crt}^k)^{c'}_{\tilde a '} + 
\nad{\phi}{(k)}{}^{c'}_{\tilde a '}(x)).\eqno({**}{**})
$$
For $k=0$ we suppose that $\nad{\phi}{(0)}{}^{c_{0}}_{\tilde a } =0$, 
where $c_{0}$ is an established index. For
$k=1$ it is more complex and strongly depends on the details of the
theory. Moreover this is always possible to give similar condition. The
nonzero $\nad{\phi}{(k)}{}^{c}_{\tilde a }$ fields are usually called surviving Higgs' fields.

Thus we can transform Higgs' and gauge fields i.e.
$$\eqalignno{
({\varPhi} ^{\nad{U}{(k)}})^c_{\tilde a }
&=\Ad_{H}(\nad{U}{(k)}{}^{-1})^{c}_{c'}\cdot
\Ad_{G}(\nad{U}{(k)}{}^{-1})^{\tilde a '}_{\tilde a }{\varPhi}
^{c'}_{\tilde a '} = ({\varPhi}^{k}_{\crt})^a_{\tilde a } +
\nad{\varphi}{(k)}{}^c_{\tilde a }(x),\cr
\widetilde{B}^{'i}_{\mu}(x) Y_{i}&=\Ad_{H}(\nad{U}{(k)}(x))^{i}_{j} 
\widetilde{B}^j_\mu(x)Y_{i} - {\hbar c\over g}\partial _{\mu} 
\nad{U}{(k)}(x) \nad{U}{(k)}{}^{-1}(x).\cr}
$$
One easily notice that
$$
[\nad{\phi}{(k)}{}^k_{\crt}]^{c}_{\tilde a } = 
[\nad{\eta}{(k)}{}^k_{\crt}]_{\tilde a } =0.\eqno(8.46)
$$
Moreover simple counting of the degrees of freedom on both sides of
the Eq.~$({**}{**})$ reveals that some of $\nad{\phi}{(k)}{}^{c}_{\tilde a }$ fields are not independent
(neglecting the fact that ${\varPhi} ^{c}_{\tilde a }$ satisfies a constraint).

Thus we should constrain $\nad{\phi}{(k)}{}^{c}_{\tilde a }$ in such a way that $(\dim G - \dim G_{0}) \dim H
= \dim \nad{G}{(k)}+ \rank (\nad{\phi}{(k)})$ where $\nad{G}{(0)} = G_{0}$
and $\nad{G}{(1)} = G'_0$.
One easily gets
$$
\mathop{\nabla_{\mu}^0}\limits^{\gauge} {\varPhi} ^{c}{}_{\tilde a } = \Ad_{\rH}
(\nad{U}{(k)} (x))^{c}_{c'}{\cdot} \Ad_{\rG}
(\nad{U}{(k)}{}^{-1} (x))^{\tilde a '}_{\tilde a } {\cdot }
\mathop{\nabla_{\mu}^0}\limits^{\gauge} (\nad{{\varPhi}}{(k)}{}^u)^{c'}_{\tilde a'}.
$$
On the level of the tensorial form $\sigma _{\rE}$ one establishes that
$$ 
f^{*}\nad{\sigma }{(k)}_{\rE} = \Ad_{\rG}
(\nad{U}{(k)}{}^{-1} (x))
(e^{*} \nad{\sigma }{(k)}_{\rE}) - d\nad{U}{(k)}(x){\cdot}
\nad{U}{(k)}{}^{-1}(x)
$$
and 
$$ 
{\cal L}_{\rYM}(\widetilde{A}) = {\cal L}_{\rYM}(\widetilde{B}) = 
{\cal L}_{\rYM}(\widetilde{B}').
$$
It is important to notice that we should consider a local section for
$\ell _{ij}$ i.e $a_{ij} = e^{*}\ell _{ij}$ in the lagrangian for the
Yang--Mills' field.

The fields $\widetilde{B}' $ are massive with the same masses as
$\widetilde{B}$. The important
point to notice is that the full lagrangian is still $G$---gauge
invariant. Moreover a choice of a particular value of ${\varPhi}
^{k}_{\crt}$ (which is $G_{0}$ invariant) reduces symmetry from $G$ to
$G_{0}$ (spontaneously). The fields
$\eta _{\nad{a}{\nu }}(x)$ disappear. They are ``eaten" by the gauge transformation and due
to this the massive vector fields have 3 polarization degrees of
freedom. Sometimes $\eta _{\nad{a}{\nu }}(x)$ are called ``would-be Goldstone bosons". In the
matrix of masses they corresponds to zero modes. It means the mass
matrix $m^{2}({\varPhi} ^{k}_{\crt})^{\tilde a }{}_a{}^{\tilde b
}{}_b$ should be transformed into $m^{,2}({\varPhi} ^{0}_{\crt})^{\tilde a }{}_a{}^{\tilde b}{}_b$ 
where $a,b\neq c_{0}$ for $k=0$ and for $k=1$ into 
$m^{2}({\varPhi} ^{1}_{\crt})^{\tilde a' }{}_{a'}{}^{\tilde b'}{}_{b'}
{\varLambda}^{\tilde a '}{}_a{}^{\tilde a}{}_{\tilde a '}
{\varLambda}^{b '}{}_b{}^{\tilde b}{}_{\tilde b '}$, where
${\varLambda}^{b'}{}_b{}^{\tilde b}{}_{\tilde a '} = 0$ for those
$\widetilde{a}$, $a$ for which $\nad{\phi}{(1)}{}^a_{\tilde a }$ is constrained to zero.
Moreover we can check it by a direct calculations of the eigenvalues
of $m^{2}({\varPhi} ^{k}_{\crt})^{\tilde a }{}_{a}{}^{\tilde b}{}_{b}$ 
that it has $(\dim G - \dim G_{0})$ $((\dim G - \dim G'_0))$ zero
eigenvalues corresponding to the ``eaten" Goldstone bosons. This result
can be obtained directly using invariancy of the classical vacuum with
respect to the action of the group $G_{0}$ or $G'_0$.

One of simplest condition to choose independent parametrization of ${\varPhi}
^{a}_{\tilde a }$ is to take such $\nad{\phi}{(k)}{}^a_{\tilde a }$ 
that 
$$ 
\sum_{\tilde a }\nad{\phi}{(k)}{}^{c_0}_{\tilde a }Y_{\tilde a }=
\nad{\phi}{(k)}{}^{c_0}
$$
and 
$$ 
\sum_{\check{a}}\nad{\eta}{(k)}_{\nad{a}{\nu }}Y_{\nad{a}{\nu } }=
\nad{\eta}{(k)}
$$
are orthogonal in $\fg$ i.e. $\widetilde{h}
(\nad{\phi}{(k)}{}^{c_0},\nad{\eta}{(k)})= 0$, where $c_{0}$ is an established
index.

For both cases if the vector boson $\widetilde{B}^{i}$ is massive after spontaneous
symmetry breaking and the Higgs' mechanism, the strength of the
Yang--Mills' field which corresponds to $\widetilde{B}^{i}$ has Yukawa-type behaviour
with short range
$$
\widetilde{H}^{i}_{\mu\nu }\sim {1\over R} e^{-(R/r)}.
$$
Due to equation (6.1) we have similarly for
$$
\widetilde{L}^{i}_{\mu\nu }\sim {1\over R} e^{-(R/r)}
$$
and for $\widetilde{M}^{i}{}_{\mu\nu}$ (see (6.3))
$$
\widetilde{M}^{i}_{\mu\nu }\sim {1\over R} e^{-(R/r)}
$$
Thus we get that for the Yang--Mills polarization induced by the
skew-symmetric part of the metric for ``broken", massive vector bosons
is of short range. Let us pass to the cosmological terms in both of
these cases.
$$
\lambda _{ck} = e^{(n+2){\varPsi} } {\alpha ^{2}_{S}\over \ell ^{2}_{pl}} 
\widetilde{R}(\widetilde{{\varGamma} }) + {e^{n{\varPsi} } \widetilde{\un{P}}
\over r^{2}}
+ e^{(n-2){\varPsi} } {4\over \hbar cr^{2}}\biggl({\ell ^{2}_{pl}\over 
r^{2}}\biggr)V({\varPhi} ^{k}_{\crt}),\quad\ k = 0, 1,\eqno(8.47)
$$
where $\widetilde{\un{P}}$ is defined by (6.21). These two terms are different and both
depend on the scalar field ${\varPsi} $. Using (8.20) one gets:
$$\displaylines{\quad
\lambda _{ck} = e^{(n+2){\varPsi} } {\alpha ^{2}_{S}\over \ell ^{2}_{pl}} 
\widetilde{R}(\widetilde{{\varGamma} }) + e^{n{\varPsi} }
{m^{2}_{\tilde A }\over \alpha ^{2}_{S}} \biggl({c\over \hbar}\biggr)^{2}
\widetilde{\un{P}}
+ 4e^{(n-2){\varPsi} } {m^{4}_{\tilde A }\over \alpha ^{4}_{S}} 
\biggl({\hbar^3\over c^{5}}\biggr) \ell ^{2}_{pl} V({\varPhi}
^{k}_{\crt}).\hfill(8.48)\cr}
$$
We get that:
$$
\widetilde{R}(\widetilde{{\varGamma} }) = {Q_{s}(\xi )\over P_{s+1}(\xi )},
\quad \hbox{or}\quad {Q_{s}(\xi )\over P_{s}(\xi )},\eqno(8.49)
$$
where $Q_{s}$ and $P_{s}$ are polynomials of the $s^{\text{\rm th}}$ order and $P_{s+1}$ is the
polynomial of $(s+1)^{\text{\rm th}}$ order with respect to $\xi $. 
$Q_{s}$ and $P_{s+1}(P_{s})$ have not common divisors.
$$
\ell _{ab} = h_{ab} + \xi k_{ab}.\eqno(8.50)
$$
In a similar way we can prove that:
$$
\widehat{\ov{R}}(\widehat{\overline{\varGamma}}) = 
{W_{k}(x,\zeta )\over V_{k+1}(x,\zeta )} ,\quad \hbox{or}\quad 
{W_{k}(x,\zeta )\over V_{k}(x,\zeta )},\quad\ x \in G/G_{0},\eqno(8.51)
$$
where $W_{k}(x,\zeta )$, $V_{k}(x,\zeta )$ are polynomials of the
$k^{\text{\rm th}}$ order with respect to
$\zeta $ with coefficients depending on $x\in G/G_{0}$ and $V_{k+1}(x,\zeta )$ is the polynomial
of the $(k+1)$ order with respect to $\zeta $ with coefficients depending on
$x\in G/G_{0}$. $W_{k}$ and $V_{k+1}(V_{k})$ have not common divisors.
$$
\widetilde{g}_{\tilde a \tilde b } = h^{0}_{\tilde a \tilde b } + \zeta 
k^{0}_{\tilde a \tilde b }.\eqno(8.52)
$$
However:
$$\eqalignno{
\widetilde{\un P} ={}& {\mmint_{M}
\sqrt{|\widetilde{g}|}d^{n_1}x\widehat{\ov{R}}(\widehat{\overline{\varGamma}})
\over \mmint_{M}\sqrt{|\widetilde{g}|}d^{n_1}x}
= {1\over V_{1}} \mmint_{M}\sqrt{|\widetilde{g}|}d^{n_1}x {W_{k}(x,\zeta )
\over V_{k+1}(x,\zeta )} =\cr
={}& {R_{r}(\zeta )\over S_{r+1}(\zeta )} \phi (\zeta ) ,\quad
\hbox{or}\quad {R_{r}(\zeta )\over S_{r}(\zeta )} \phi (\zeta
)&(8.53)\cr}
$$
and $\phi (\zeta )$ is a function of $\zeta $ where $R_{r}$, $S_{r}$ are polynomials of the
$r^{\text{\rm th}}$
order with respect to~$\zeta $ and $S_{r+1}$ is the polynomial of
$(r+1)^{\text{\rm st}}$ order
with respect to~$\zeta $. In both cases we have similar asymptotic behaviour
with respect to $\xi $ and $\zeta $ if the function $\phi $ is bounded.
$R_{r}$ and $S_{r+1}(S_{r})$ have not common divisors (see Appendix A for further details).
$$\eqalignno{
\widetilde{R}(\widetilde{{\varGamma} })&\sim {C_{1}\over \xi }\quad
\hbox{or}\quad \sim C_{1},&(8.54)\cr
\widetilde{\un{P}} &\sim {C_{2}\over \zeta }\quad \hbox{or}\quad \sim C_{2},&(8.55)\cr}
$$
where $C_{1}$ and $C_{2}$ are constants. If the polynomials $Q_{s}$ and $R_{r}$ have real
roots $\xi _{0}$ and $\zeta _{0}$ such that:
$$
Q_{s}(\xi _{0}) = 0\eqno(8.56)
$$
and
$$
R_{r}(\zeta _{0}) = 0\eqno(8.57)
$$
we get:
$$
\lambda _{ck}(\xi _{0},\zeta _{0}) = 4e^{(n-2){\varPsi} } 
{m^{4}_{\tilde A }\over \alpha ^{4}_{S}} \biggl({\hbar^{3}\over
c^{5}}\biggr) \ell ^{2}_{pl} V({\varPhi} ^{k}_{\crt}),\quad\ k = 0,
1.\eqno(8.58)
$$
In the case for sufficiently large $\xi $, $\zeta $ one gets:
$$
\lambda _{ck}(\xi ,\zeta ) = \biggl(e^{(n+2){\varPsi} }
{C'_1\over \xi }+e^{n{\varPsi} }{C'_2\over \zeta }\biggr) + 
4{m^{4}_{\tilde A }\over \alpha ^{4}_{S}} \biggl({\hbar^{3}\over
c^{5}}\biggr) \cdot e^{(n-2){\varPsi} } \ell ^{2}_{pl} 
V({\varPhi} ^{k}_{\crt})
$$
or
$$
\lambda _{ck}(\xi ,\zeta ) = (e^{(n+2){\varPsi} }
C'_1+e^{n{\varPsi} }C'_2) + 
4{m^{4}_{\tilde A }\over \alpha ^{4}_{S}} \biggl({\hbar^{3}\over
c^{5}}\biggr) \cdot e^{(n-2){\varPsi} } \ell ^{2}_{pl} 
V({\varPhi} ^{k}_{\crt}),\eqno(8.59)
$$
$k = 0, 1$; $C'_1, C'_2$ are constants.

Thus in some cases we are able to make the first part of $\lambda _{ck}$ as small
as we want. From the observational data point of view we know that the
cosmological constant is almost zero. Thus it occurs in the first or
second case (real roots of polynomials $Q_{s}$ and $R_{r}$ or in the limit of
large $\xi $ and $\zeta )$.
One gets:
$$\eqalignno{
\lambda _{0} &= 0,&(8.60)\cr
\lambda _{1} &= 4{m^{4}_{\tilde A }\over \alpha ^{4}_{S}} \biggl({\hbar^{3}\over
c^{5}}\biggr)\ell ^{2}_{pl} V({\varPhi} ^{1}_{\crt}).&(8.61)\cr}
$$
For the ``false vacuum" the cosmological constant is not zero. It is
zero only for the ``true vacuum"---absolute minimum of the potential $V$.
Thus we get that worlds corresponding to two critical points of the
potential $V$ are completely different with different unbroken groups,
different mass spectrum for intermediate bosons and different
cosmological constants. Thus the ``tunnel effect" from point $A$ to
$B$ (see Fig.~3) results in a transition between completely different
worlds.

In some cosmological models (see [81--85]) one introduces in a
phenomenological way the Higgs potential of this type with two critical
points: first (true vacuum) with zero cosmological constant and second
(false vacuum) with nonzero cosmological constant. If the model of the
universe is divided into some number of clusters with the ``false" and
the ``true" vacuum then evolution of clusters with different vacua will
be completely different. There are different laws of expansion for
cosmological models in General Relativity (in Moffat's theory too) for
two cases: with zero and nonzero cosmological constants (radiation and
de Sitter universes). Transition from point $A$ to $B$ results in the first
order phase transition in models of the universe. Clusters with ``true
vacuum" devour clusters with ``false vacuum" (see [81]). Thus we can
reproduce the inflationary scenario of the Universe (old scenario). If
the point $A$ on Fig.~3 is identical with the maximum of $V$ we can
reproduce the new inflationary scenario of the Universe (without the
``tunnel effect") as well (the so-called ``rolling penny''
scenario). We can also work with an extended inflationary
scenario, because our theory is equipped with a scalar field
${\varPsi}$, which plays the role of dilatation scalar field.

In Moffat's theory of gravitation there are some cosmological
models with the phase transition of the second order in the early
stages of the universe (see [90--91]). This occurs due to the
skewsymmetric part of the metric $g_{\mu\nu }$ on the space-time. It is easy to
see that if the metric on the homogeneous space $M = G/G_{0} $,
$g_{\tilde a \tilde b }$ is symmetric
$$
g_{\tilde a \tilde b } = h^{0}{}_{\tilde a \tilde b },\eqno(8.62)
$$
we never obtain a second local minimum for the potential $V$.

Let us calculate $\widetilde{\un{P}}$ for $M = S^{2} = \SO(3)/\SO(2)$ (a cosmological constant).
One gets
$$
\widetilde{\un P} = {1\over V_2}\mmint_{S^2}
\widehat{\ov{R}}(\widehat{\overline{\varGamma}})\sqrt{|\widetilde{g}|}\,dm = 
{4\pi\over V_2}\mint_{0}^\pi R(\theta ,\zeta )\sqrt{1+\zeta ^2\cos
^2\theta }\sin \theta \,d\theta.\eqno(8.63)
$$
One gets after some algebra (for $\zeta \neq 0)$
$$\displaylines{
\widetilde{\un P} (\zeta ) =\biggl\{{16|\zeta |^3(\zeta ^2+1)\over 3(2\zeta
^2+1)(1+\zeta ^2)^{5/2}}\biggl(\zeta ^2E\biggl({|\zeta |\over
\sqrt{\zeta ^2+1}}\biggr)-(2\zeta ^2+1)K\biggl({|\zeta |\over
\sqrt{\zeta ^2+1}}\biggr)\biggr)+\hfill\cr
\noalign{\vskip8pt}
\hfill+ 8\ln (|\zeta |\sqrt{\zeta ^2+1}) + {4(1+9\zeta ^2-8\zeta ^4)|\zeta
|^3\over 3(1+\zeta ^2)^{3/2}}\biggr\}
\bigg{/} (\ln(|\zeta |+\sqrt{\zeta ^2+1})+2\zeta ^2+1).\quad\
(8.64)\cr}
$$
For $V_{2} = \int_{S^2}\sqrt{|\widetilde{g}|}\,dm$ we get
$$
V_{2} = {2\pi\over |\zeta |}(\ln(|\zeta |+ \sqrt{\zeta ^2+1})
+ 2\zeta ^{2} + 1)\neq 0,\eqno(8.65)
$$
where 
$$\eqalignno{
K(k) &= \mint_{0}^{\pi/2}{d\theta \over \sqrt{1-k^2\sin \theta
}},&(8.65\text{\rm a})\cr
E(k) &= \mint_0^{\pi/2}\sqrt{1-k^2\sin \theta }\,d\theta,\quad\ 0\le k^{2}\le 1
&(8.65\text{\rm b})\cr}
$$
are first and second order elliptic integrals.

It is easy to see that for $\zeta =0$, $\widetilde{\un P} (0)=0$. The function
$\widetilde{\un P}$ is plotted on
Fig.~6. Let us look for zeros of $\un P$.

\obraz{
\midinsert \centerline{\epsfxsize\hsize \epsfbox{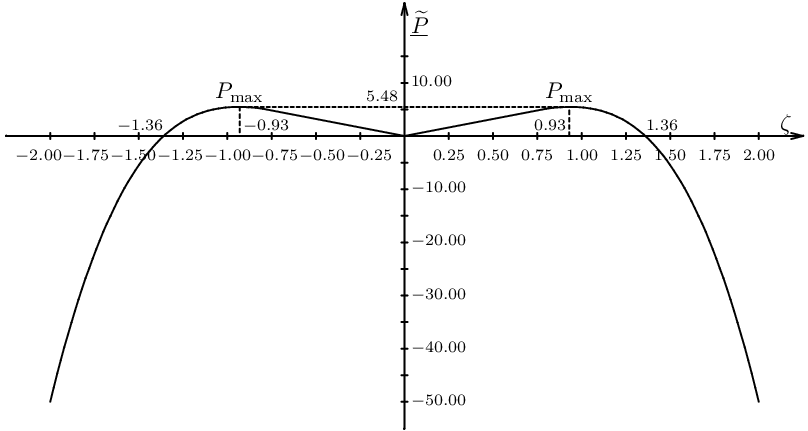}}
\centerline{\nine Fig. 6. The function $\underline{\widetilde P}$ plotted versus $\zeta$}
\endinsert

}$\widetilde{\un P}(\zeta _{0})=0$, $\zeta _{0}\neq 0$ is a transcendental equation and we should find a
solution using numerical methods. One gets $\zeta _{0} = \pm 1.36\ldots$
and $\widetilde{\un P}(\zeta )\to -\infty $
if $|\zeta |\to + \infty $. Let us notice that the integral (8.69) has been
calculated under the assumption $\zeta \neq 0$. For $\zeta = 0$ we should do it in a
different way. It is easier of course and corresponds to the symmetric
connection on $S^{2}$.

Let us consider a more general case given by Eq.~(6.49).

One gets in the case of the K\"ahlerian structure.
$$
\widetilde{\un P} = {1\over V_2}\mmint_{M}
\widehat{\ov{R}}(\widehat{\overline{\varGamma}})\sqrt{|\widetilde{g}|}\,dm(g)=
{1\over V_2}\mmint_{M} \sqrt{|\widetilde g|}\, g^{\tilde a \tilde b
}\widetilde{\ov{R}}_{\tilde a \tilde b }(\widetilde{\overline{\omega
}})\,dm(g).\eqno(8.66)
$$
Moreover $\widehat{\ov{R}}(\widehat{\overline{\varGamma}})$ is left-invariant with respect to the action of the
group $G$ on~$M$. Thus we have
$$
\widehat{\ov{R}}(\widehat{\overline{\varGamma}}(y)) =
\widehat{\ov{R}}(\widehat{\overline{\varGamma}}(g y)),\quad\ 
g\in G ,\ y \in M.\eqno(8.67)
$$
In particular
$$
\widehat{\ov{R}}(\widehat{\overline{\varGamma}}(\{\varepsilon G_0\})) =
\widehat{\ov{R}}(\widehat{\overline{\varGamma}}(y)).\eqno(8.68)
$$
Thus one gets
$$
\widetilde{\un P} = g^{\tilde a \tilde b }\widetilde{\ov{R}}_{\tilde a \tilde b
}(\widetilde{\overline{\omega }}).\eqno(8.69)
$$
Let us notice that in the case of $S^{2}$ equipped with the nonsymmetric
tensor coming from K\"ahlerian structure on $S^{2}$ one simply gets (see
Eq.~(6.60)).
$$
P = {4\over (1+\zeta ^2)}.\eqno(8.70)
$$
Thus for a large $\zeta $ we have
$$
P \sim {4\over \zeta ^2}.\eqno(8.71)
$$

The interesting point in our theory is a place of electromagnetic
field in the theory. Let us suppose that ${\bold U}(1)_{el}\subset G$ is a
subgroup of the gauge group and let us consider the case when this
group is broken to the ${\bold U}(1)_{el}$ only. The Higgs field
${\varPhi}_{\tilde a}^a(x)$ in this case takes the special value
$({\varPhi}_{\crt}^0)^a_{\tilde a}$ or $({\varPhi}'_{\crt})^a_{\tilde a}$
(depending on the appropriate minimum of the Higgs potential
$V({\varPhi})$. If the Higgs field takes value from the vacuum
manifold $G/G_0$ $(G/G'_0)$ we can suppose that
$$
\mathop{\nabla_{\mu}}\limits^{\gauge}{\varPhi}
^{a}_{\tilde a }=0.\eqno(8.72)
$$
We can consider a transformation
$$
{\varSigma}({\varPhi})\in G/G_0,\quad\
{\varSigma}:\widetilde{S}\to G/G_0\eqno(8.73)
$$
such that
$$
\mathop{\nabla_{\mu}}\limits^{\gauge}\widehat{{\varPhi}}^{c}_{\tilde a
}(x)=0.\eqno(8.74)
$$
Now we consider $G_0={\bold U}(1)_{\text{\rm el}}$. This is beyond our considerations,
because ${\bold U}(1)_{\text{el}}$ is not semisimple. However in that case we can
reduce the problem to much more simpler one. The Higgs field just
belongs to the adjoint representation of the group $G$ and we really
have to do with $\phi (x)=\phi ^{\dah{i}}(x)Y_{\dah{i}}$. Thus
we can consider a gauge field
$$
\widetilde{h}_{\dah{i}\dah{j}}\phi
^{\dah{i}}(x)\widetilde{\omega}^{\dah{i}}_E=\omega_{el},\quad\
\mathop{\nabla_{\mu}}\limits^{\gauge}\phi 
=0,\eqno(8.75)
$$
which is a one-from with values in $R$ (a ${\bold U}(1)$-group Lie algebra).
This form will be considered as an electromagnetic connection.
($\phi $ belongs to vacuum manifold $G/{\bold U}(1)$).

Let us notice that
$$
\pi_2(G/{\bold U}(1))=\pi_1({\bold U}(1))=Z.\eqno(8.77)
$$
This means that the topological type of the field is characterized by
an integer, the topological charge.

Thus we can define a strength $F_{\mu\nu}(x)$ of an electromagnetic
field.
$$
F_{\mu\nu}(x)=\widetilde{h}_{\dah{i}\dah{j}}\widetilde{H}^{\dah{i}}_{\mu\nu}
P^{\dah{j}}({\varSigma}(\phi )),\eqno(8.78)
$$
where
$$
{\varSigma}:\fg\to G/{\bold U}(1)\eqno(8.79)
$$
and takes the field $\phi $ to a vacuum manifolds. $P$ is a
generator for the stabilized ${\bold U}(1)_{\phi} $, $\phi \in G/{\bold U}(1)$
satisfying the normalization condition:
$$
\widetilde{h}_{\dah{i}\dah{j}}P^{\dah{i}}(\phi
)P^{\dah{j}}(\phi )=1.\eqno(8.80)
$$
Moreover, we can compute the electromagnetic field strength in a more
general way finding a gauge-invariant expression that coincide with
the previous one.

One consider the following expression
$$
F_{\mu\nu}(x)=\widetilde{h}_{\dah{i}\dah{j}}\widetilde{H}^{\dah{i}}_{\mu\nu}
P^{\dah{j}}({\varSigma}(\phi
))+\varkappa_{\dah{i}\dah{j}}(\phi (x))\mathop{\nabla_{\mu}}\limits^{\gauge}
\phi ^{\dah{i}}(x)\mathop{\nabla_{\nu}}\limits^{\gauge}\phi
^{\dah{j}}(x), \eqno(8.81)
$$
where
$$
\varkappa_{\dah{i}\dah{j}}=-\varkappa_{\dah{j}\dah{i}}.\eqno(8.82)
$$
This could be rewritten in a different form.
$$
F_{\mu\nu}(x)=\widetilde{h}_{\dah{i}\dah{j}}\widetilde{H}^{\dah{i}}_{\mu\nu}
P^{\dah{j}}({\varSigma}(\phi
))+\varkappa_{\phi }(\mathop{\nabla_{\mu}}\limits^{\gauge}
\phi ,\mathop{\nabla_{\nu}}\limits^{\gauge}\phi), \eqno(8.83)
$$
where
$$
\displaylines{\hfill \varkappa_{\phi
}(X,Z)=\varkappa_{\dah{i}\dah{j}}(\phi
)X^{\dah{i}}Z^{\dah{j}},\hfill\llap{(8.84)}\cr
\hfill X=X^{\dah{i}}Y_{\dah{i}},\quad\
Z=Z^{\dah{j}}Y_{\dah{j}}\hfill\llap{(8.85)}\cr}
$$
This form satisfies the following conditions:
$$\eqalignno{
\varkappa_{\phi }(X,Z)=&\varkappa_{{\varSigma}(\phi
)}({\varSigma}'X,{\varSigma}'Z),&(8.86)\cr
\varkappa_{\phi_0}(\Ad_G(A)\phi _0,\Ad_G(B)\phi _0)&
=-\widetilde{h}_{\dah{i}\dah{j}}[A,B]^{\dah{i}}P^{\dah{j}}(\phi_0),&
(8.87)\cr}
$$
$A, B\in \fg$ and $\phi _0$ is the critical value of the Higgs
field. ${\varSigma}'$ means a tangent transformation of
${\varSigma}$ at $\phi $ i.e.
$$
{\varSigma}'=\biggl({\partial {\varSigma}^{\dah{i}}(\phi )\over
\partial \phi ^{\dah{j}}}\biggr).\eqno(8.88)
$$
Let us notice that
$$
d\omega_{el}={1\over 2}F_{\mu\nu}\theta^{\mu}_{\wedge}\theta^{\nu}
$$
In the case of $G=\SU(2)$ one simply gets $\phi (x)\in A_1$,
$\phi =\phi ^{\dah{i}}X_{\dah{i}}$, and $\phi $ is a
three-dimensional vector in $R^3$ i.e.\ $\phi
=\overrightarrow{\phi }$ (see the second position of Ref.~[17]).

One writes
$${\varSigma}(\overrightarrow{\phi })=a{\overrightarrow{\phi }\over \|\phi
\|},\quad\ \|\phi \|=\sqrt{\phi _1^2+\phi _2^2+\phi
_3^2}.\eqno(8.89)
$$
We also get 
$$\eqalignno{
\overrightarrow{P}(\phi )&={\overrightarrow{\phi }\over \|\phi \|},&(8.90)\cr
{\varSigma}'(\phi )^i_j&={a\over \|\phi
\|}\biggl(\delta_j^i-{1\over \|\phi \|^2}\phi ^i\phi _j\biggr)
&(8.91)\cr}
$$
and
$$
\varkappa_{\phi }(X,Z)={1\over \|\phi
\|^3}\varepsilon_{\dah{i}\dah{j}\dah{k}}\phi
^{\dah{i}}X^{\dah{j}}Z^\dah{k}.\eqno(8.92)
$$
Recently have appeared some papers on Higgs'-bosonless Standard Model
(see the last two positions of the Ref.~[75]). In this approach
the remaining Higgs' boson is eliminated
from the theory (after spontaneous symmetry breaking) due to
conformal invariance of the theory. This interesting approach wants
to cure the problem with difficulty in Higgs' boson hunting (maybe
because of its very big mass). In our theory we can make the Higgs
boson masses very big and we can eliminate one of the remaining
Higgs' boson connecting it to the scalar field ${\varPsi}$ (but only one).
This makes our approach closer to the mentioned one.

This can be achieved in the following way. Let us suppose that one of
the surviving Higgs' field is $\varphi$ and let us connect it to the
${\varPsi}$ field by a transformation $\varphi=g({\varPsi})$, where
$g$ is a smooth function of one variable. In this way the Higgs'
field $\varphi$ disappears from the theory becoming a ${\varPsi}$
field connected to the variable gravitational ,,constant'' (an
effective one). If the field $\varphi$ is the field $H^0$ from
Weinberg--Salam Model, we can remove $H^0$ from the theory.
Let us remind to the reader that the Weinberg--Salam Model (a bosonic
part of the model) is covered by the theory (see Ref.~[75]). The form
of the function $g$ could be established by a requirement that the
full lagrangian of ${\varPsi}$ in a linear approximation looks
classical.

Let us consider a more general case of symmetry breaking i.e. $G$ to
$G_0$ or to $G'_0$ corresponding to two possible critical point of
the Higgs' potential. In this case we use a smooth transformation
$$
{\varSigma}:\widetilde S\to G/G_0\quad (\hbox{or}\ G/G'_0)\eqno(8.93)
$$
is such a way that
$$
\mathop{\nabla_{\nu}}\limits^{\gauge}{\varPhi}^c_{\tilde
a}=0.\eqno(8.94)
$$
Moreover due to decomposition (a reducive one) of the Lie algebra of
the group $G$
$$
\fg =\fg_0\dot{+}\fm =\fg'_0 \dot{+}\fm \eqno(8.95)
$$
we have a natural parametrization of the homogeneous space $G/G_0$
(or $G/G'_0$) in forms of the complement $\fm$\ $(\fm')$ in such a
way that:
$$
(x_5,x_6,\ldots,x_{n_1+4})\to \varphi \Bigl(g\exp\Bigl(\sum_{\tilde
a}x_{\tilde a}Y_{\tilde a}\Bigr)\Bigr),\eqno(8.96)
$$
which maps a neighbourhood of zero in $\fm(\fm')$ on a neighbourhood
of $\varphi(g)$ in $G/G_0$. An inverse mapping is a local coordinate
system in a neighbourhood of $\varphi(g)\in G/G_0$ on $M=G/G_0$.
$\varphi$ is a natural (canonical) map
$G\mathop{\to}\limits_{\varphi} M=G/G_0$. Let us call this map
$\psi$. One gets
$$
{\varSigma}^a({\varPhi}) = \psi({\varPhi}^a{}_{\tilde
a}),\eqno(8.97)
$$
where
$$
\psi^{-1}(x_5^a,x_6^a,\ldots, x_{n_1+4}^a)=
\varphi \Bigl(g\exp\Bigl(\sum_{\tilde a}{\varPhi}^a_{\tilde a}(x)Y_{\tilde
a}\Bigr)\Bigr),\quad\ x\in E,\ g\in G.\eqno(8.98)
$$
It is easy to see that for every $a$, ${\varSigma}^a({\varPhi})\in
G/G_0\ (G/G'_0)$. In this way we do not suppose
$$
\mathop{\nabla_{\mu}}\limits^{\gauge}{\varPhi}^a_{\tilde
a}=0.\eqno(8.99)
$$
Now we can construct a gauge covariant form of the gauge strength
$$
\widetilde{H}^{\dah{i}}_{\mu\nu}=h_{ij}\widetilde{H}^{i}_{\mu\nu}P^j
({\varSigma}({\varPhi})\mu'{}^{\dah{i}}_a)
+\varkappa^{\tilde a\tilde b}({\varPhi}(x))[(
\mathop{\nabla_{\mu}}\limits^{\gauge}{\varPhi}^a_{\tilde a})
(\mathop{\nabla_{\mu}}\limits^{\gauge}{\varPhi}^b_{\tilde b})
\mu'{}^{\dah{k}}_a\mu'{}^{\dah{j}}_b]
f^{\dah{i}}_{\dah{k}\dah{j}},\eqno(8.100)
$$
where $P^j({\varPhi}^a)$ are generators for the stabilizator of
$(G_0)_{{\varPhi}^a}$, ${\varPhi}^a\in G/G_0$ (the same for $G'_0$
case). Satisfying the normalization conditions
$$
\|P^j({\varPhi}^a)\|=1.\eqno(8.101)
$$
This is the gauge field strength which remains massless after the
symmetry breaking.

If we take under consideration 
$$ 
\pi_2(G/G_0)=\pi_1(G_0),\eqno(8.102)
$$
\looseness-10(the same for the $G'_0$ case) we can consider a nontrivial
topological configuration of the gauge field i.e. the so called
,,coloured'' magnetic monopoles. The gauge field strength defined
above is simply the curvature of the connection $\omega^0{}_E$ 
(or~$\omega^0{}'_E$). 
$$\displaylines{
\mathop{D}\limits^{\gauge}\widetilde{\omega}^0{}_E={1\over
2}\widetilde{H}^{\dah{i}}{}_{\mu\nu}\theta^\mu_{\wedge}\theta^{\nu}Y_{
\dah{i}}\quad \hbox{or}\quad
\mathop{D}\limits^{\gauge}\widetilde{\omega}^0{}'{}_E={1\over
2}\widetilde{H}'{}^{\dah{i}}{}_{\mu\nu}\theta^\mu_{\wedge}\theta^{\nu}Y_{
\dah{i}'}\cr
\noalign{\vskip4pt}
\biggl(\mathop{D}\limits^{\gauge}\widetilde{\omega}^0{}_E=d\widetilde{\omega}^0
{}_E+{1\over 2}[\omega^0{}_E,\omega^0{}_E]\biggr).\cr}
$$
and analogically for $\widetilde{\omega}^0{}'_{E}$.

The form $\varkappa^{\tilde a\tilde b}({\varPhi}(x))$ satisfies the
following conditions: 
$$ 
\varkappa^{\tilde a\tilde b}=\varkappa^{\tilde b\tilde
a}.\eqno(8.103)
$$
If we define a form 
$$ 
\varkappa({\varPhi}(x)(X,Z)=\sum_{\tilde a,\tilde b}
\varkappa^{\tilde a\tilde b}({\varPhi}(x))X^{\tilde a}Z^{\tilde
a}\eqno(8.104)
$$
the formula for the strength of the gauge field can be rewritten as
follows:
$$
\eqalignno{\widetilde{H}^{\dah{i}}_{\mu\nu}(x)={}&
\langle\widetilde{H}_{\mu\nu}(x),P({\varSigma}^a({\varPhi})\mu'{}^{\dah{i}}_a)
\rangle+\cr
&+\varkappa({\varPhi}(x))
(\mu'{}^{\dah{k}}_a\mathop{\nabla_{\mu}}\limits^{\gauge}{\varPhi}^a_{\tilde a},
\mu'{}^{\dah{j}}_b \mathop{\nabla_{\nu}}\limits^{\gauge}{\varPhi}^b_{\tilde b})
f^{\dah{i}}_{\dah{k}\dah{j}}.&(8.105)\cr}
$$
The form $\varkappa({\varPhi})$ can be found from the requirements:
$$
\varkappa({\varPhi})(X,Z)=\varkappa({\varSigma}{\varPhi})
({\varSigma}'X,{\varSigma}'Y),\eqno(8.106)
$$
where
${\varSigma}'=({\varSigma}^a)'=\biggl({\displaystyle{\partial
{\varSigma}^a\over \partial{\varPhi}^b_{\tilde a}}}\biggr)$ is
tangent map to ${\varSigma}$ 
$$\displaylines{\quad\
\varkappa
({\varPhi}^k_{\crt})(\Ad_G(A)^{\dah{k}'}_{\dah{k}}\mu'{}^{\dah{k}}_a
\Ad_G(A)^{\tilde{a} '}_{\tilde{a}
}({\varPhi}^k_{\crt})^a_{\tilde{a}'},
\Ad_G(B)^{\dah{j}'}_{\dah{j}}\mu'{}^{\dah{j}}_b
\Ad_G(B)^{\tilde{b}
'}_{\tilde{b}}({\varPhi}^k_{\crt})^b_{\tilde{b}'})=\hfill\cr
\hfill -\langle [A,B],p^i({\varPhi}^k_{\crt})\rangle,\quad\
k=0,1,\quad\ (8.108)\cr}
$$
where $\langle X,Y\rangle=h_{ij}X^iX^j$ and $A,B,X,Y\in \fg$.

We can consider in our theory topologically nontrivial field
configurations labeled by $\pi_2(G/G_0)$ or $\pi_2(G/G'_0)$. They
are ,,coloured'' magnetic charges. In the case of $G_0={\bold U}(1)$ we
have to do with ordinary magnetic charges.

Let us define a magnetic and a ,,coloured'' magnetic field
$$
\eqalignno{
\overrightarrow{H}(\overrightarrow{x}) &=
(F_{23}(\overrightarrow{x}),F_{31}(\overrightarrow{x}),
F_{12}(\overrightarrow{x})),&(8.109)\cr
\overrightarrow{H}^{\dah{i}}(\overrightarrow{x}) &=
(F_{23}^{\dah{i}}(\overrightarrow{x}),F_{31}^{\dah{i}}(\overrightarrow{x}),
F_{12}^{\dah{i}}(\overrightarrow{x}))&(8.110)\cr}
$$
and magnetic and coloured magnetic charge (monopole):
$$\eqalignno{
m &= {1\over 4\pi}\oint \overrightarrow{H}(\overrightarrow{x})
\,d\overrightarrow{S},&(8.111)\cr
m^{\dah{i}} &= {1\over 4\pi}\oint \overrightarrow{H}^{\dah{i}}(\overrightarrow{x})
\,d\overrightarrow{S},&(8.112)\cr}
$$
where the integrals are taken over a sphere infinitely far away i.e.
for a sphere at infinity. In this case we suppose that
$$\eqalignno{
&{\varPhi}^a_{\tilde{a} }(\overrightarrow{x})\in G/G_0\quad
\hbox{if}\quad \|\overrightarrow{x}\|\to \infty\quad \hbox{and}\cr
&\mathop{\nabla_{\mu}}\limits^{\gauge}{\varPhi}^a_{\tilde{a}
}(\overrightarrow{x})\to 0,\quad \hbox{if}\quad
\|\overrightarrow{x}\|\to \infty,\cr}
$$
where $\|\overrightarrow{x}\|=\sqrt{|x_1|^2+|x_2|^2+|x_3|^2}$.

For example we can get a nontrivial topological gauge configurations
for the following classical Lie groups 
$$\eeqalign{
&\SO(n),\ \pi_1(\SO(n))={\sym Z}_2,\quad\ &\hbox{for}\ &n>1,\cr
&{\bold U}(n),\ \pi_1({\bold U}(n))={\sym Z},\quad\ &\hbox{for}\ &n\ge
1.\cr}
$$
In the case of $\SU(n)$, $\Sp (n)$, and $\Spin (n)$, $n\ge 1$ we
get trivial gauge configurations, became
$$
\pi_1(\SU (n))=\pi_1(\Spin (n))=\pi_1(\Sp(n))=0.
$$
Thus we can expect ,,coloured'' magnetic charges (monopoles) for
$\SO(n+1)$ and ${\bold U}(n)$, $n\ge 1$ (equal to $G_0$ or $G'_0$).

Finally let us remind to the reader that in the case of $G_0$ group
(a ,,true'' vacuum case) we have to do with a trivial gauge
configuration of the Higgs' field ${\varPhi}^a_{\tilde{a} }(x)$
(i.e.\ ${\varPhi}^0_{\crt}$ case). In this case a part of a strength
of the multi-dimensional gauge field connected to Higgs' field {\bf
does vanish}. In the case of $G_0$-group (a ,,false'' vacuum care)
this part of the strength of the multidimensional gauge field {\bf
does not vanish}. Thus the ,,magnetic'' coloured charges (monopoles)
in both cases have different physical meanings. In some sense in the
second care they corresponds to something which could be called
,,exited'' magnetic monopoles (coloured ones) i.e. for
${\varPhi}^1_{\crt}$ case.

It is worth to say we suppose now that $G_0$ or $G'_0$ is simply
connected. The topological (magnetic) charges depend only on the
configurations of Higgs' fields ${\varPhi}^a_{\tilde{a}
}(\overrightarrow{x})$ and does not depend on the gauge field
$\widetilde{\omega }^0{}_E$ or $\widetilde{\omega }^0{}'_E$
($\widetilde{\omega }_{el}$ in the case of electromagnetic field).

We defined the topological type of the field by looking of the
asymptotic behaviour of the scalar field ${\varPhi}$. Thus the
topological type of the field is an element of
$\{S^2,G/G_0\}$ that is a homotopy class of mapping from $S^2$ into
the vacuum manifold. When $G_0$ is simply connected, the set
$\{S^2,G/G_0\}$ can be identified with $\pi_2(G/G_0)=\pi_1(G_0)$,
which is a finite group.

Thus there are $r$ topological charges, as many as there are magnetic
charges. Clearly, the magnetic charges completely determine the
integral values of the topological charges of a field. In addition to
integral topological charges, there may be charges taking values in a
finite group ${\sym Z}_m$ (i.e.\ ${\sym Z}_2)$, these do not
correspond to magnetic charges. They are coloured magnetic charges
(monopoles). In general we can consider topological charges for
$\widetilde{H}^i_{\mu\nu}$ field
$$
\chi (\widetilde{\omega }_E)={1\over 16\pi^2}\mmint_E\langle
\widetilde{H}\wedge \widetilde{H}\rangle,\eqno(8.113)
$$
where
$$
\langle \widetilde{H}\wedge \widetilde{H}\rangle = -{1\over
2}\widetilde{h}_{ij}\widetilde{H}^i_{\alpha \beta}\widetilde{H}^j_{\gamma
\delta }\overline{\theta }^d_{\wedge}\overline{\theta }^p_{\wedge}\overline{\theta
}^{\gamma }_{\wedge}\overline{\theta }^{\delta }
$$
and similarly for $\widetilde{H}^{\dah{i}}_{\mu\nu}$
$$
\widehat{\chi}(\widetilde{\omega }^0{}_E) = {1\over
16\pi^2}\mmint_E\langle \widehat{\widetilde{H}}\wedge
\widehat{\widetilde{H}}\rangle,\eqno(8.114)
$$
where
$$
\langle \widehat{\widetilde{H}}\wedge \widehat{\widetilde{H}}\rangle = 
-{1\over 2}\widetilde{h}_{\dah{i}\dah{j}}\widetilde{H}^{\dah{i}}_{\alpha \beta}
\widetilde{H}^{\dah{j}}_{\gamma
\delta }\overline{\theta }^\alpha _{\wedge}\overline{\theta }^\beta _{\wedge}\overline{\theta
}^{\gamma }_{\wedge}\overline{\theta }^{\delta }.
$$
The topological number of a field $\widetilde{\omega }_E$ (or
$\widetilde{\omega }^0{}_E$) on $E$ can be defined when the strength
of the fields $\widetilde{H}^{\dah{i}}_{\alpha \beta }$ or
$\widetilde{H}^i_{\alpha \beta }$ tends to zero sufficiently fast as
$\|x\|\to \infty$, $\|x\|^2=x_1^2+x_2^2+x_3^2+x_4^2$ (In this case we
take $E=R^4$). If the field $\widetilde{H}$ decays quickly
$\widetilde{\omega }_E$ coincides asymptotically with a pure gauge
field, that is it has asymptotics of $g^{-1}\,dg$, where $g(x)$ is a
$G$-valued function. In that case we have
$$\eqalignno{
\chi(\widetilde{\omega }_E)&=-{1\over 96\pi^2}\mmint_{S^3}\langle
\widetilde{\omega }_E\wedge [\widetilde{\omega }_E,\widetilde{\omega
}_E]\rangle\cr
&=-{1\over 96\pi^2}\mmint_{S^3}\langle
g^{-1}\,dg\wedge[g^{-1}\,dg,g^{-1}\,dg]\rangle,&(8.115)\cr}
$$
where $S^3$ is taken as a sphere at infinity. In this case the
topological charges are classified by $\pi_3(G)$, $\pi_3(G_0)$ or
$\pi_3(G'_0)$.

\vfill\eject

\null
\vskip36pt plus4pt minus4pt
\centerline{{\four\bf 9. Variational Principle. Equations of Fields.}}

\vskip3pt
\centerline{{\four\bf Interpretation of the Scalar Field ${\varPsi} $}}
\vskip24pt plus4pt minus4pt
\write0{9. Variational Principle. Equations of Fields.
Interpretation of the Scalar Field ${\varPsi} $ \the\pageno}

Let us consider Palatini variational principle for the action $S$.
$$
\delta S = 0.\eqno(9.1)
$$
It is easy to see that (9.1) is equivalent to
$$
\delta \mmint_{{\bold U}} L(g,\ov{W}, \widetilde{A}, {\varPsi} , 
{\varPhi})\sqrt{-g}d^4 x= 0 ,\quad\ {\bold U} \subset E.\eqno(9.2)
$$
We have the following independent quantities $g_{\mu\nu }$, 
$\ov{W}^{\lambda }{}_{\mu\nu }$, $\widetilde{\omega }_{\rE}$, ${\varPsi} $ and ${\varPhi} $.
We vary with respect to the independent quantities. After some
calculations one gets:
$$\displaylines{
\hfill \ov{R}_{\mu\nu }(\ov{W}) - {1\over 2} g_{\mu\nu
}\ov{R}(\ov{W})
= {8\pi K\over c^{4}} (\mathop{T_{\mu\nu }}\limits^{\gauge} +T_{\mu\nu }({\varPhi} )
+ \mathop{T_{\mu \nu }}\limits^{\scal} ({\varPsi} )+
\mathop{T_{\mu \nu }}\limits^{\Int} +g_{\mu \nu }{\varLambda }),\hfill\llap{(9.3)}\cr
\hfill\falkag^{[\mu \nu ]}{}_{,\nu } = 0,\hfill\llap{(9.4)}\cr
\hfill \overline{\nabla}_{\nu } g^{[\mu \nu ]} =
0,\hfill\llap{\hbox{(9.4a)}}\cr
\hfill g_{\mu \nu ,\sigma } - g_{\xi\nu }\overline{\varGamma}^{\xi}{}_{\mu
\sigma } - g _{\mu \xi}\overline{\varGamma}^{\xi}{}_{\sigma \nu }=
0,\hfill\llap{(9.5)}\cr
\noalign{\vskip2pt}
((n^{2}+2M)\widetilde{g}^{(\alpha \mu )}-n^{2}g^{\nu \mu }
g_{\delta \nu }\widetilde{g}^{(\alpha \delta )}) 
{\partial ^{2}{\varPsi} \over \partial x^{\alpha } \partial x^{\mu }}
+\hfill\cr
\noalign{\vskip2pt}
\hfill\eqalign{&+ {1\over \sqrt{-g}} \partial _{\mu }\biggl\{\sqrt{-g}\biggl[n^{2}
\widetilde{g}^{(\alpha \mu )} - {n^{2}\over 2} g_{\delta \nu }(
g^{\nu \alpha } \widetilde{g}^{(\mu \delta )} + g^{\nu \mu } 
\widetilde{g}^{(\mu \alpha )}) - 2M \widetilde{g}^{(\mu \alpha
)}\biggr]\biggr\}\times \cr
\noalign{\vskip2pt}
&\times {\partial {\varPsi} \over \partial x^{\alpha }} - 8\pi (n+2) 
e^{-(n+2){\varPsi} } {\cal L}_{\rYM}(\widetilde{A}) - 
{4e^{-2{\varPsi} }\over r^{2}} {\cal L}_{\kin}({\varPhi} ,\widetilde{A})
+\cr
\noalign{\vskip2pt}
&+ {(n-2)\over r^{4}} e^{(n-2){\varPsi} } V({\varPhi} ) + 
{4(n-2)\over r^{2}} e^{(n-2){\varPsi} } {\cal L}_{\Int}({\varPhi} ,
\widetilde{A}) +\cr
\noalign{\vskip2pt}
&- {n\over r^{2}} e^{n{\varPsi} } \widetilde{\un{P}} - 
{(n+2)\alpha ^{2}_{S}\over \ell ^{2}_{pl}} e^{(n+2){\varPsi} } 
\widetilde{R}(\widetilde{{\varGamma} }) = 0,\cr}\quad\ (9.6)\cr
\noalign{\vskip2pt}}
$$
$$\eqalignno{\noalign{\vskip-8pt}
\mathop{\nabla_{\mu}}\limits^{\gauge} (\widetilde{\ell
}_{ij}\widetilde{\falkaL}^{i\alpha \mu }) ={}& 2\falkag^{[\alpha
\beta ]} \mathop{\nabla_{\beta}}\limits^{\gauge}(\widetilde{h}_{ij}g^{[\mu \nu ]}
\widetilde{H}^{i}{}_{\mu \nu }) +\cr
\noalign{\vskip4pt}
&+ 2\sqrt{-g}\alpha _{S} {1\over \sqrt{\hbar c}} 
{e^{n{\varPsi} }\over r^{2}} \biggl[ \ell _{ab}g^{\tilde b
\tilde n } g^{\mu \alpha } L^{a}{}_{\mu \tilde b } 
({\varPhi} ^{d}{}_{\tilde c }C^{b}_{dc}\alpha ^{c}_{j}+{\varPhi}
^{b}{}_{\tilde a }f^{\tilde a }_{\tilde n j}) +\cr
\noalign{\vskip4pt}
&+ \biggl( {\delta L^{a}{}_{\beta \tilde b }\over \delta 
\mathop{\nabla_{\alpha}}\limits^{\gauge}{\varPhi} ^{w}_{\tilde v }}\biggr)
\ell _{ab}g^{\tilde b \tilde n } g^{\beta \mu } 
(\mathop{\nabla_{\mu}}\limits^{\gauge}{\varPhi} ^{b}{}_{\tilde n }) 
({\varPhi} ^{d}{}_{\tilde w }C^{w}_{dc}\alpha ^{c}_{j}+{\varPhi}
^{w}_{\tilde a }f^{\tilde a }{}_{\tilde n j}) \biggr]_{av}+\cr 
\noalign{\vskip4pt}
&+ 4\sqrt{-g} {e^{2n{\varPsi} }\over r^{2}} h_{ab} \mu ^{a}_{k} 
\widetilde{\ell }_{ij} \ell ^{ki} \widetilde{g}^{[\tilde a \tilde b ]} 
\mathop{\nabla_{\mu}}\limits^{\gauge}\biggl\{g^{[\mu \alpha ]} \times \cr
\noalign{\vskip4pt}
&\times \biggl[{1\over \alpha _{S}}\sqrt{\hbar c}C^{b}_{cd}
{\varPhi} ^{c}{}_{\tilde a }{\varPhi} ^{d}{}_{\tilde b }-
\alpha _{S}\biggl({\hbar \over c}\mu ^{b}_{\dah{i}}f^{\dah{i}}{}_{\tilde
a \tilde b }-\alpha _{S}{1\over \sqrt{\hbar c}}{\varPhi}
^{b}{}_{\tilde d }f^{\tilde d }{}_{\tilde a \tilde b
}\biggr)\biggr]\biggr\} +\cr
\noalign{\vskip4pt}
&+ (n+2) \partial _{\beta }{\varPsi} 
[\widetilde{\ell }_{ij}\widetilde{\falkaL}^{i\beta \alpha }-
2\falkag^{[\mu \nu ]}(\widetilde{h}_{ij}g^{[\mu \nu
]}\widetilde{H}^{i}{}_{\mu \nu })],&(9.7)\cr
\noalign{\vskip4pt}
\mathop{\nabla_{\mu}}\limits^{\gauge} (\ell _{ab}\falkaL^{a\mu }{}_{\tilde b })_{av}
={}& - \sqrt{-g} {e^{n{\varPsi} }\over 2r^{2}}\biggl\{\biggl({\delta V' \over \delta 
{\varPhi} ^{b}{}_{\tilde n}}\biggr) g_{\tilde b \tilde n}
+\phantom{\}}\cr
\noalign{\vskip4pt}
&- 2 \sqrt{-g} e^{n{\varPsi} } \mu ^{e}_{i} (\widetilde{H}^{i}{}_{\mu
\nu }g^{[\mu \nu ]}) h_{ed} \biggl( {2\over \alpha _{S}}\sqrt{\hbar
c}\un{g}^{[\tilde a \tilde n]} C^{d}_{cb} {\varPhi}
^{c}{}_{\tilde a } g{}_{\tilde b \tilde n} +\cr
\noalign{\vskip4pt}
&\phantom{\{} - \alpha _{S} {1\over \sqrt{\hbar
c}}g^{[\tilde c \tilde d ]} f^{\tilde n}{}_{\tilde c
\tilde d } g_{\tilde b \tilde n}\biggr) + 2 
\partial _{\mu }{\varPsi} \ell _{ab}\falkaL^{a\mu }{}_{\tilde b
}\biggr\}_{av},&(9.8)\cr}
$$
where
$$\eqalignno{
\mathop{T_{\alpha\beta}}\limits^{\gauge} ={}& - {\widetilde{\ell }_{ij}\over 4\pi }
\biggl\{g_{\gamma \beta } g^{\tau\rho} g^{\varepsilon \gamma } 
\widetilde{L}^{i}{}_{\rho\alpha }
\widetilde{L}^{j}{}_{\tau\varepsilon} - 2 g^{[\mu \nu ]} 
\widetilde{H}^{(i}_{\mu \nu } \widetilde{H}^{j)}_{\alpha \beta }
+\phantom{\}}\cr
\noalign{\vskip2pt}
&\phantom{\{}- {1\over 4} g_{\alpha \beta } [\widetilde{L}^{i\mu
\nu }\widetilde{H}^{j}{}_{\mu \nu }-2(g^{[\mu \nu
]}\widetilde{H}^{i}{}_{\mu \nu })(g^{[\gamma \sigma ]}
\widetilde{H}^{j}{}_{\gamma \sigma })]\biggr\}&(9.9)\cr}
$$
is the energy-momentum tensor for the gauge (Yang--Mills') field with
the zero trace.
$$\eqalignno{
\mathop{T_{\alpha \beta }}\limits^{\gauge} g^{\alpha \beta } ={}&
0,&(9.10)\cr
\noalign{\vskip2pt}
\mathop{T_{\alpha \beta }}\limits^{\scal} ({\varPsi} ) ={}& - {e^{(n+2){\varPsi} }\over 16\pi}
\biggl\{(g_{\varkappa \alpha }g_{\omega \beta }+g_{\omega \alpha }
g_{\varkappa \beta }) \times \phantom{\}}\cr
\noalign{\vskip2pt}
&\times \widetilde{g}^{(\gamma \varkappa )} \widetilde{g}^{(\nu
\omega )} \cdot \biggl[{n^{2}\over 2}(g^{\xi\mu }g_{\nu \xi}-
\delta ^{\mu }_{\nu }){\varPsi} _{,\mu }+\ov{M}{\varPsi} _{,\nu
}\biggr] {\varPsi} _{,\gamma } +\cr
\noalign{\vskip2pt}
&\phantom{\{}- g_{\alpha \beta } [\ov{M}\widetilde{g}^{(
\nu\mu )}+n^{2}g^{[\mu \nu ]}g_{\delta \mu }\widetilde{g}^{(\gamma
\delta )}{\varPsi} _{,\nu }{\varPsi} _{,\gamma }\biggr]\biggr\}&(9.11)\cr}
$$
is the energy-momentum tensor for the scalar field ${\varPsi} $ with nonzero 
trace
$$\displaylines{
\hfill\mathop{T_{\alpha \beta }}\limits^{\scal} ({\varPsi} ) 
g^{\alpha \beta } \neq 
0\hfill\llap{(9.12)}\cr
\hfill K = G_{\rN} e^{-(n+2){\varPsi} } = G_{\eff}({\varPsi}
).\hfill\llap{(9.13)}\cr}
$$
It plays the role of an effective gravitational constant. 
$$\eqalignno{
T_{\mu \nu }({\varPhi} ) ={}& {e^{(n-4){\varPsi} }\over 4\pi r^{2}} 
\ell _{ab}\un{g}^{\tilde b \tilde n} L^{a}{}_{\mu \tilde
b } \mathop{\nabla_{\nu}}\limits^{\gauge} {\varPhi} ^{b}{}_{\tilde n} +\cr
&- {1\over 2} g_{\mu \nu } \biggl(-{e^{2(n-2){\varPsi} }\over 8\pi r^{4}}
V({\varPhi} )
+ {e^{(n-4){\varPsi}}\over 4\pi r^{2}}\ell _{ab}
(g^{\tilde b \tilde n}g^{\alpha \beta }L^{a}{}_{\alpha
\tilde b }\mathop{\nabla_{\beta}}\limits^{\gauge} {\varPhi} ^{b}{}_{\tilde
n})_{av}\biggr).\qquad\ &(9.14)\cr}
$$
It is an energy-momentum tensor for the Higgs' field.
$$\displaylines{\eqalign{
\mathop{T_{\mu\nu }}\limits^{\Int} ={}& -{e^{2(n-2){\varPsi} }\over 2\pi r^{2}} h_{ab} 
\mu ^{a}_{i} \widetilde{H}^{i}{}_{\mu \nu
}\widetilde{\un{g}}^{[\tilde a \tilde b ]}\biggl({1\over \alpha _{S}}
\sqrt{\hbar c} C^{b}_{cd} {\varPhi} ^{c}{}_{\tilde a } {\varPhi}
^{d}{}_{\tilde b } +\cr
&- \alpha _{S} {1\over \sqrt{\hbar c}} \mu ^{b}_{\dah{i}} f^{\dah{i}}{}_{\tilde a \tilde b } - \alpha _{S} {1\over \sqrt{\hbar c}} 
{\varPhi} ^{b}{}_{\tilde d } f^{\tilde d }{}_{\tilde a \tilde b
}\biggr)+\cr
&+ {e^{2(n-2){\varPsi} }\over 4\pi r^{2}} g_{\mu \nu }\biggl[h_{ab} 
\mu ^{a}_{i} (\widetilde{H}^{i}_{\alpha \beta }g^{[\alpha \beta
]})\widetilde{\un{g}}^{[\tilde a \tilde b ]} \times \cr}\hfill\cr
\hfill\times \biggl({1\over \alpha _{S}}\sqrt{\hbar c}C^{b}_{cd}
{\varPhi} ^{c}_{\tilde a }{\varPhi} ^{d}_{\tilde b }-\alpha _{s}
{\hbar\over c}\mu ^{b}_{\dah{i}}f^{\dah{i}}{}_{\tilde a \tilde b }-
\alpha _{S}{1\over \sqrt{\hbar c}}{\varPhi} ^{b}{}_{\tilde d
}f^{\tilde d }{}_{\tilde a \tilde b }\biggr)\biggr].\quad\ (9.15)\cr}
$$
It is an energy-momentum tensor corresponding to the nonminimal
interaction term ${\cal L}_{\Int}(\widetilde{A},{\varPhi} )$.
$$\eqalignno{
{\varLambda} &= {1\over 16\pi G_{\rN}} \biggl({e^{(2n+4){\varPsi} } 
\alpha ^{2}_{S}\over \ell ^{2}_{pl}} \widetilde{R}(\widetilde{{\varGamma} })
+{e^{(n+2){\varPsi} }\over r^{2}}\widetilde{\un{P}}\biggr)
= 16\pi G_{\rN} e^{-(n+2){\varPsi}} \overline{\lambda }_{c}.&(9.16)\cr}
$$
It plays the role of the ``cosmological constant", which now depends on
the scalar field ${\varPsi} $. The quantity
${\displaystyle{\delta L^{a}_{\beta \tilde b }\over \delta
\mathop{\nabla_{\alpha}}\limits^{\gauge} {\varPhi} ^{w}_{\tilde v }}}$ satisfies the following 
equation:
$$\displaylines{
\hfill \ell _{dc} g_{\mu \beta } g^{\gamma \mu } {\delta L^{d}_{\gamma
\tilde a }\over \delta \mathop{\nabla_{\alpha}}\limits^{\gauge}{\varPhi}
^{w}_{\tilde v }} + \ell _{cd} g_{\tilde a \tilde m} g^{\tilde
m\tilde c } {\delta L^{d}_{\beta \tilde c }\over \delta
\mathop{\nabla_{\alpha}}\limits^{\gauge} {\varPhi} ^{w}_{\tilde v}}
= 2\ell _{cd} g_{\tilde a \tilde m} g^{\tilde m\tilde c } 
\delta ^{\alpha }{}_{\beta } \delta ^{\tilde v}{}_{\tilde c } \delta
^{d}_{w},\hfill\llap{(9.17)}\cr
\hfill \widetilde{\falkaL}^{i\mu \nu } = \sqrt{-g} g^{\beta \mu }
g^{\gamma \nu } \widetilde{L}^{i}_{\beta \gamma },\hfill\llap{(9.18)}\cr
\hfill \falkag^{[\mu \nu ]} = \sqrt{-g} g^{[\mu \nu ]},\hfill\llap{(9.18\rm
a)}\cr}
$$
$\widetilde{L}^{i}{}_{\beta \gamma }$ satisfies equation (9.1) and 
$L^{a}{}_{\beta \tilde b }$ the equation (9.4) ${\displaystyle{\delta V' 
\over \delta {\varPhi} ^{b}_{\tilde n}}}$ is
defined by the equation (8.10). The scalar field ${\varPhi}
^{c}_{\tilde a }$ satisfies
constraints (8.8). Let us pass to the interpretation of the field
equations (9.3--8). Equations (9.3), (9.4) and (9.5) are gravitational
equations from N.G.T. with the following matter sources: Yang--Mills'
field (in the nonsymmetric version), Higgs' field, scalar field ${\varPsi} $ with
a presence of the cosmological term depending on scalar field ${\varPsi} $.
Equation (9.6) is the equation for the scalar field ${\varPsi} $. This field is
of course chargeless, but it interacts with Yang--Mills' field and
Higgs' field due to some terms in (9.6). It interacts also with
cosmological terms, which effectively depend on ${\varPsi} $. This field due to
equation (9.13) has an interpretation as an effective gravitational
constant. Equation (9.7) is the equation for Yang--Mills' field. Now as
in the Nonsymmetric Kaluza--Klein Theory we have for this field two
tensors of strength $\widetilde{H}_{\mu \nu }$ and
$\widetilde{L}^{i}{}_{\mu \nu }$ (ordinary and an induction one) and
the nonsymmetric parts of metrics on $E$ and on $G$ induce the polarization
$\widetilde{M}^{i}{}_{\mu \nu }$. In the equation (9.7) we have sources connected to a
skewsymmetric part of metric $g_{\mu \nu }$ and to Higgs' field. Due to the
existence of a skewsymmetric part of metric $\ell _{ab}$ and 
$g_{\tilde a \tilde b }$ the current
connected to Higgs' field is more complicated. Equation (9.8) is an
equation for Higgs' field. We write this equation in terms of tensor 
$$
L^{a\mu }{}_{\tilde b } = g^{\alpha \mu } L^{a}{}_{\alpha \tilde b }.\eqno(9.19)
$$
This tensor plays a similar role for $\mathop{\nabla_{\alpha}}\limits^{\gauge}
{\varPhi} ^{a}_{\tilde b }$ as $\widetilde{L}^{i\mu }{}_{\alpha }$ for 
$\widetilde{H}^{i}{}_{\mu \alpha }$.Thus we
have in the theory an analogue of the polarization tensor
$M^{a}{}_{\alpha \tilde b }$ for Higgs' field 
$$
L^{a}{}_{\alpha \tilde b } = \mathop{\nabla_{\alpha}}\limits^{\gauge}{\varPhi}
^{a}_{\beta } - {4\pi \over c} M^{a}{}_{\alpha \tilde b }.\eqno(9.20)
$$
Let us consider the following two $\Ad_{H^-}$ type two-forms
$\widehat{L} = L^{a}{}_{\alpha \tilde b }\theta ^{\alpha }\wedge
\theta ^{\tilde b }X_{a}$
and $\widehat{M} = M^{a}{}_{\alpha \tilde b }\theta ^{\alpha }\wedge 
\theta ^{\tilde b }X_{a}$. One gets $\widehat{L}=
\widehat{{\varOmega}} - {\displaystyle{4\pi\over c}} \widehat{M} =
\widehat{{\varOmega}} - {\displaystyle{1\over 2}} \widehat{Q}$, where
$$
\widehat{Q} = \mathop{\nabla_{\alpha}}\limits^{\gauge} {\varPhi} ^{a}{}_{\tilde b
}\theta ^{\alpha }\wedge \theta ^{\alpha }X_{a},\quad\ \widehat{Q} = 
Q^{a}{}_{\alpha \tilde b }\theta ^{\alpha }\wedge \theta ^{b}X_{a}. 
$$
In this way we get a geometrical interpretation of the polarization 
2-form of the Higgs' field as a part of a torsion.

The field ${\varPsi} $ due to (9.13) is connected to the effective 
gravitational
constant. However, it also enters the definition of energy-momentum
tensors $T_{\mu \nu }({\varPhi} )$ and $\mathop{T_{\mu \nu }}\limits^{\Int}$. 
Thus it plays the role of the universal
factor. In the next section we deal with this field in details.

Let us notice that the left-hand side of Eq.~(9.7) can be rewritten in
the following way 
$$
\mathop{\nabla_{\mu}}\limits^{\gauge} (\widetilde{l}_{ij} \widetilde{\falkaL}^{i\alpha
\mu } ) = \sqrt{-g} \mathop{\widetilde{\overline{\nabla}}_{\mu}}\limits^{\gauge} 
(\widetilde{l}_{ij}\widetilde{L}^{i\alpha \mu }), 
$$
where $\mathop{\widetilde{\overline{\nabla}}_{\mu}}\limits^{\gauge}$ means a covariant derivative with respect to the connection
$\widetilde{\overline{\omega }}^{\alpha }{}_{\beta }$ on $E$ and $\omega
_{\rE}$ on $Q(E,G)$ at once. The Eqs.~(9.4--4a) are equivalent to
the second condition of Eq.~(5.5).

Let us consider a case with a trivial group $G_{0} = \{\varepsilon \}$. 
This means a
complete broken $G$ symmetry group and $M = G/G_{0} \simeq G$, 
$V \simeq E \times G$. Thus we
get that the Higgs' field ${\varPhi} ^{c}_{i}$ transforms according to the adjoint
representation of the group $H$ and $G$ and it does not obey any
constraints. They are satisfied trivially. Thus it transforms
according to $\Ad_{H} \otimes \Ad_{G}$. If we choose $H = G$ one gets
$\Ad_{G} \otimes \Ad_{G}$.

The Higgs' potential looks like:
$$\eqalignno{\noalign{\vskip-2pt}
V({\varPhi} ) ={}& {\ell _{ab}\over V_{G}}\mmint_G\sqrt{|\widetilde{\ell}|}
d^{n_{2}}x [ 2\widetilde{\ell }^{[ij]} (C^{a}_{cd}{\varPhi} ^{c}_{i}
{\varPhi} ^{d}_{j}-{\varPhi} ^{a}_{k}f^{k}_{ij})
\widetilde{\ell }^{[mn]} (C^{b}_{ef}{\varPhi} ^{e}_{m}
{\varPhi} ^{f}_{n}-{\varPhi} ^{b}_{p}f^{p}_{mn}) +\cr
&- \widetilde{\ell }^{im} \widetilde{\ell }^{jn} L^{a}{}_{ij} 
(C^{b}_{cd}{\varPhi} ^{c}_{m}{\varPhi} ^{d}_{n}-{\varPhi} ^{b}_{e}
f^{e}_{mn})].&(9.21)\cr\noalign{\vskip-2pt}}
$$
The kinetic part of the Higgs' field lagrangian is as follows: 
$$
{\cal L}_{\kin}(\mathop{\nabla}\limits^{\gauge} {\varPhi} ) = {1\over V_{G}}
\mmint_G\sqrt{|\widetilde{\ell}|} d^{n_{2}}x ( \ell _{ab} \widetilde{\ell }^{bn}
L^{a\mu }{}_{b} ^{g}\mathop{\nabla_{\mu}}\limits^{\gauge}{\varPhi} ^{b}_{n} ),
\eqno(9.22)
$$
where
$$
\ell _{dc} g_{\mu \beta } g^{\gamma \mu } L^{d}{}_{\gamma a} + \ell _{cd} 
\widetilde{\ell }_{am} \widetilde{\ell }^{mp} L^{d}{}_{\beta p} = 
2\ell _{cd} \widetilde{\ell }_{am} \widetilde{\ell }^{mp}
\mathop{\nabla_{\mu}}\limits^{\gauge}{\varPhi} ^{d}_{p}\eqno(9.23)
$$
and
$$
\ell _{dc} \widetilde{\ell }_{mb} \widetilde{\ell }^{pm} L^{d}{}_{pa} + 
\ell _{cd} \widetilde{\ell }_{am} \widetilde{\ell }^{mp} L^{d}{}_{bp} = 
2\ell _{cd} \widetilde{\ell }_{am} \widetilde{\ell }^{mp}
H^{d}{}_{bp}.\eqno(9.24)
$$
For $H^{d}{}_{bp}$ one gets:
$$
H^{d}{}_{bp} = C^{d}_{ac} {\varPhi} ^{a}_{b} {\varPhi} ^{c}_{p} - 
{\varPhi} ^{d}_{q} f^{q}_{bp}.\eqno(9.24\text{\rm a})
$$
In the above formulae $\widetilde{\ell }_{ij}$ is the same as in
sec.~6.

The nonsymmetric tensor on $G$ is left-invariant and on $H$ is
right-invariant.

The interaction term in the lagrangian is as follows:
$$
{\cal L}_{\Int}({\varPhi} ,\widetilde{A}) = h_{ab} \mu ^{a}_{i} 
\widetilde{H}^{i}\mmint_G{\sqrt{|\widetilde{\ell}|}\over V_G}
d^{n_2}x \widetilde{\ell }^{[pq]} \cdot (C^{b}_{cd}{\varPhi} ^{c}_{p}
{\varPhi} ^{d}_{q}-{\varPhi} ^{b}_{e}f^{e}_{pq}).\eqno(9.25)
$$
The cosmological term takes the shape:
$$
\lambda _{c} = {e^{(n+2){\varPsi} } \alpha ^{2}_{S}\over \ell ^{2}_{pl}} 
\widetilde{R}(\widetilde{{\varGamma} }) + {e^{n{\varPsi} }\over r^{2}} 
R_{G},\eqno(9.26)
$$
where $R_{G}$ is the scalar curvature of the connection defined on the group
$G$. In order to vanish it we take $\xi =\xi _{0}$ and $\zeta =\zeta _{0}$ 
such that $\widetilde{R}(\widetilde{{\varGamma} }(\xi _{0}))=0$ and
$R$ $(\zeta _{0}) = 0$.
$$
V_{G} = \mmint_G\sqrt{|\widetilde{\ell}|}\, d\mu _{G}(g) = \mmint_G\sqrt{|\widetilde{\ell}|}
\,d^{n_{2}}x.\eqno(9.27)
$$
The dimensional reduction procedure can be described in terms of ${\varOmega}
$, $Q$ (see~5.56--58).

In this way ${\varOmega} $ decomposes into $\widetilde{{\varOmega}}_{\rE}$, 
$\Check{{\varOmega}}$, $\widehat{{\varOmega}}$, and $\ov{L}$ into 
$\widetilde{L}$, $\widehat{L}$, $\Check{L}$ (see
paragraphs below formula (6.3), below (6.7) and below (9.20)).
Similarly we get a decomposition for $\ov{M}$ and $\ov{Q}$ into $\widetilde{M}$,
$\widehat{M}$, $\Check{M}$ and into $\widetilde{Q}$,
$\widehat{Q}$, $\Check{Q}$. 

According to Eqs.~(4.53--55) one gets for $\ov{L}$, $\ov{M}$ and $\ov{Q}$
$$\eqalignno{
\ov{L} &= \widetilde{L} \oplus \widehat{L} \oplus \Check{L},\cr
\ov{M} &= \widetilde{M} \oplus \widehat{M} \oplus \Check{M},&(9.28)\cr
\ov{Q} &= \widetilde{Q} \oplus \widehat{Q} \oplus \Check{Q},\cr}
$$
similarly as for ${\varOmega} $
$$
{\varOmega} = \widetilde{{\varOmega} } \oplus
\widehat{{\varOmega}} \oplus \Check{{\varOmega}}\eqno(9.29)
$$
(see Eq.~(4.52)), where 
$$
{\varOmega} _{\rE} = d\omega _{\rE} +{1\over 2}[\omega _{\rE},\omega
_{\rE}].\eqno(9.30)
$$

{\def\mPsi{\varPsi}
\def\mPhi{\varPhi}
\def\eq#1 {\eqno(\text{\rm9.#1})}
\let\al\aligned
\let\eal\endaligned
\let\ealn\eqalignno
\def\cL{\Cal L}
\def\gH{\frak h}
\def\gG{\frak g}
\def\gM{\frak m}
\def\hi{{\hat\imath}}
\def\hj{{\hat\jmath}}
\let\o\overline
\let\ul\underline
\let\d\delta
\let\f\varphi
\def\G{\varGamma}
\let\t\widetilde
\let\wh\widehat
\let\u\tilde
\def\Id{\text{Id}}
\def\rank{\operatorname{rank}}
\def\gag{\overset\text{\rm gauge}\to\nabla}
\def\({\left(}\def\){\right)}
\def\dr#1{_{\text{\rm#1}}}
\def\dpl{\mathbin{\overset.\to+}}
\def\ct{constant}
\def\co{cosmological}
\def\dt{dependent}
\def\m{manifold}
\def\s{symmetr}
\def\sb{symmetry breaking}
\def\q{quintessence}
\def\wrt{with respect to }

Let us incorporate in our scheme a hierarchy of a symmetry breaking. In order
to do this let us consider a case of the \m
$$
M=M_0\times M_1\times \ldots \times M_{k-1} \eq31
$$
where
$$
\ealn{
\dim M_i& = \o n_i, \quad i=0,1,2,\ldots,k-1\cr
\dim M&= \sum_{i=0}^{k-1} \o n_i, &(9.32)\cr
M_i&=G_{i+1}/G_i\,. &(9.33)}
$$
Every \m\ $M_i$ is a \m\ of vacuum states if the \s y is breaking from
$G_{i+1}$ to~$G_i$, $G_k=G$.

Thus
$$
G_0\subset G_1\subset G_2\subset \ldots \subset G_k=G. \eq34
$$
We will consider the situation when
$$
M\simeq G/G_0. \eq35
$$
This is a constraint in the theory. From the chain (9.34) one gets
$$
\gG_0\subset \gG_1\subset \ldots \subset \gG_k =\gG \eq36
$$
and
$$
\gG_{i+1}=\gG_i \dpl \gM_i, \quad i=0,1,\ldots,k. \eq37
$$
The relation (9.35) means that there is a diffeomorphism $g$ onto $G/G_0$
such that
$$
g: \prod_{i=0}^{k-1}\(G_{i+1}/G_i\)\to G/G_0\,. \eq38
$$
This diffeomorphism is a deformation of a product (9.31) in $G/G_0$. The
theory has been constructed for the case considered before with $G_0$
and~$G$. The multiplet of Higgs' fields $\mPhi$ breaks the \s y from $G$
to~$G_0$ (equivalently from $G$ to~$G_0'$ in the false vacuum case).
$\gG_i$~mean Lie algebras for groups $G_i$ and $\gM_i$ a complement in a
decomposition (9.37). On every \m~$M_i$ we introduce a radius $r_i$
(a~``size'' of a manifold) in such a
way that $r_i>r_{i+1}$. On the \m\ $G/G_0$ we define the radius~$r$ as
before. The diffeomorphism $g$ induces a contragradient transformation for a
Higgs field $\mPhi$ in such a way that
$$
g^\ast \mPhi=\(\mPhi_0,\mPhi_1,\ldots,\mPhi_{k-1}\) \eq39
$$
decomposes into fields $\mPhi_i$, $i=0,1,\ldots,k-1$.

In this way we get the following decomposition for a kinetic part of the
field~$\mPhi$ and for a potential of this field:
$$
\ealn{
\cL\dr{kin}(\gag\mPhi)&=\sum_{i=0}^{k-1} \cL\dr{kin}^i(\gag\mPhi_i)&(9.40)\cr
V(\mPhi)&=\sum_{i=0}^{k-1}V^i(\mPhi_i) &(9.41)\cr
\cL^i\dr{kin}(\gag\mPhi_i)&=\frac1{V_i}\intop_{M_i}
\sqrt{|\t g_i|}\,d^{\bar n_i}x\(l_{ab}g^{\u b_i\u n_i}_i
L^{a\mu}_{\u b_i}\gag_{\kern-6pt\mu} \mPhi^b_{i\u n_i}\)&(9.42)
}
$$
where
$$
\ealn{
V_i&=\intop_{M_i}\sqrt{|\t g_i|}\,d^{\bar n_i}x &(9.43)\cr
\t g_i&=\det\(g_{i\u b_i\u a_i}\). &(9.44)
}
$$
$g_{i\u b_i\u a_i}$ is a non\s ic tensor on a \m~$M_i$.
$$
\al
V^i(\mPhi_i)&=\frac{l_{ab}}{V_i} \intop_{M_i} \sqrt{|\t g_i|}\,d^{\bar n_i}x
\biggl[2g^{[\u m_i\u n_i]} \(C^a_{cd}\mPhi^c_{i\u m_i}\mPhi^d_{i\u n_i}
-\mu^a_{\hi_i}f^{\hi_i}_{\u m_i\u n_i}-\mPhi^a_{i\u e_i}
f^{\u e_i}_{\u m_i\u n_i}\)\\
&\times g^{[\u a_i\u b_i]}
\(C^b_{ef}\mPhi^e_{i\u a_i}\mPhi^f_{i\u b_i}
-\mu^b_{\hj_i} f^{\hj_i}_{\u a_i\u b_i}-\mPhi^b_{i\u a_i}
f^{\u d_i}_{\u a_i\u b_i}\)\\
&-g^{\u a_i\u m_i}_i g^{\u b_i\u n_i}_i L^a_{\u a_i\u b_i}
\(C^b_{cd}\mPhi^c_{i\u m_i}\mPhi^d_{\u n_i}
-\mu^b_{\hi_i}f^{\hi_i}_{\u m_i\u n_i}-\mPhi^b_{\u e_i}f^{\u e_i}
_{\u m_i\u n_i}\)\biggr],
\eal \eq45
$$
$f^{\hj_i}_{\u a_i\u b_i}$ are structure \ct s of the Lie algebra
$\gG_i$. 

The scheme of the \s y breaking acts as follows from the group $G_{i+1}$
to~$G_i$ ($G_i'$) (if the \s y has been broken up to $G_{i+1}$). The potential
$V^i(\mPhi_i)$ has a minimum (global or local) for $\mPhi^k\dr{$i$crt}$,
$k=0,1$. The value of the remaining part of the sum (9.41) for fields
$\mPhi_j$, $j<i$, is small for the scale of energy is much lower ($r_j>r_i$,
$j<i$). Thus the minimum of $V^i(\mPhi_i)$ is an approximate minimum of the
remaining part of the sum (9.41). In this way we have a descending chain of
truncations of the Higgs potential. This gives in principle a pattern of a 
\s y breaking. However, this is only an approximate \s y breaking. The real
\s y breaking is from $G$ to $G_0$ (or to $G_0'$ in a false vacuum case). The
important point here is the diffeomorphism~$g$.
$$
\ealn{
g^\ast \mPhi^b&=\(\mPhi^b_0,\mPhi^b_1,\ldots,\mPhi^b_{k-1}\)&(9.46)\cr
\mPhi^b_i&=\mPhi^b_{i\u a_i}\t \theta^{\u a_i}, 
\quad i=0,\ldots,k-1. &(9.47)
}
$$

The shape of $g$ is a true indicator of a reality of the \s y breaking
pattern. If
$$
g=\Id+\d g \eq48
$$
where $\d g$ is in some sense small and $\Id$ is an identity, the sums
(9.40--41) are close to the formulae (6.17) and (6.18). The smallness of $\d
g$ is a criterion of a practical application of the \s y breaking pattern
(9.34). It seems that there are a lot of possibilities for the condition
(9.38). Moreover, a smallness of $\d g$ plus some natural conditions for
groups~$G_i$ can narrow looking for grand unified models. Let us notice that
the decomposition of~$M$ results in decomposition of \co\ terms
$$
\ul{\t P}=\sum_{i=0}^{k-1} \ul{\t P}_i \eq49
$$
where
$$
\ul{\t P}_i=\frac1{r^2_iV_i}\intop_{M_i}\sqrt{|\t g_i|}\,
\wh{\o R}_i(\wh{\o\G}_i)\,d^{n_i}x \eq50
$$
where $\wh{\o \G}_i$ is a non\s ic connection on $M_i$ compatible with the
non\s ic tensor $g_{i\u a_i\u b_i}$ and $\wh{\o R}_i(\wh{\o \G}_i)$ its
curvature scalar. The scalar field $\mPsi$ is now a function on a product
of~$M_i$, 
$$
\mPsi=\mPsi(x,y_0,\ldots,y_{k-1}),\quad
y_i\in M_i,\ i=0,1,\ldots,k-1.
$$
The truncation procedure can be proceeded in several ways. Finally let us
notice that the energy scale of broken gauge bosons is fixed by a
radius~$r_i$ at any stage of the \s y breaking in our scheme. The possible
groups in our \s y breaking pattern can be taken from section~4 (see
4.62--4.66). 

Let us consider Eq.\ (9.39) in more details. One gets
$$
A^{\u a}_{i\u a_i}(y)\mPhi^b_{\u a}(y)=
\mPhi^b_{\u a_i}(y_i), \quad y\in M,\ y_i\in M_i \eq51
$$
where
$$
g^\ast(y)=\(A_0\Biggm|A_1\Biggm|A_2\Biggm|\ldots\Biggm| A_{k-1}\), \eq52
$$
$$
A_i=\(A^{\u a}_{i\u a_i}\)_{\u a=1,2,\ldots,n_1,\,\u a_i=1,2,\ldots,\bar
n_i}\,,\ i=0,1,2,\ldots,k \eq53
$$
is a matrix of Higgs' fields transformation.

According to our assumptions one gets also:
$$
\(\frac{r_i}r\)^2 g_{i\u a_i\u b_i}(y_i)=
A^{\u a}_{\u a_i}(y)A^{\u b}_{\u b_i}(y)g_{\u a\u b}(y). \eq54
$$
For $g$ is an invertible map we have $\det g^\ast(y)\ne0$.

We have also
$$
n_1=\sum_{i=0}^{k-1} \o n_i \eq55
$$
and
$$
\mPhi^b_{\u a}(y)=\sum_{i=0}^{k-1} \t A_{i\u a}^{\u a_i}(y)
\mPhi^b_{\u a_i}(y_i) \eq56
$$
or
$$
{g^\ast}^{-1}(y)=\(\vcenter{\offinterlineskip
\hbox to40pt{\hfil$\t A_0$\hfil \vrule height10pt depth4pt width0pt}
\hrule width40pt 
\hbox to40pt{\hfil$\t A_1$\hfil \vrule height10pt depth4pt width0pt}
\hrule width40pt 
\hbox to40pt{\hfil$\vdots$\hfil \vrule height10pt depth4pt width0pt}
\hrule width40pt 
\hbox to40pt{\hfil$\t A_{k-1}$\hfil \vrule height10pt depth4pt width0pt}
}\)\eq57
$$
$$
\t A_i=\(\t A^{\u a_i}_{i\u a}\)_{\u a_i=1,2,\ldots,\bar n_i,\,\u a=1,2,\ldots,
n_1} \eq58
$$
such that
$$
\ealn{
&g(y_0,\ldots,y_{k-1})=y &(9.59)\cr
&(y_0,y_1,\ldots,y_{k-1})=g^{-1}(y) &\text{(9.59a)}
}
$$

For an inverse tensor $g^{\u a\u b}$ one easily gets
$$
\(\frac{r_i^2}{r^2}\)g^{\u a\u b}=
\sum_{i=0}^{k-1} \t A^{\u a}_{i\u a_i} g^{\u a_i\u b_i}_i
A^{\u b}_{i\u b_i}\,. \eq60
$$
We have
$$
r^{2n_1}\det(g_{\u a\u b})=\prod_{i=0}^{k-1}
r^{2\bar n_i}_i\det(g_{i\u a_i\u b_i}). \eq61
$$
In this way we have for the measure
$$
d\mu(y)=\prod_{i=0}^{k-1}d\mu_i(y_i) \eq62
$$
where
$$
\ealn{
d\mu(y)&=\sqrt{\det g}\,r^{n_1}\,d^{n_1}y &(9.63)\cr
d\mu_i(y_i)&=\sqrt{\det g_i}\,r^{\bar n_i}_i\,d^{\bar n_i}y_i\,. &(9.64)
}
$$
In the case of $\cL\dr{int}(\mPhi,\t A)$ one gets
$$
\cL\dr{int}(\mPhi,\t A)=\sum_{i=0}^{k-1} \cL\dr{int}(\mPhi_i,\t A) \eq65
$$
where
$$
\cL\dr{int}(\mPhi_i,\t A)=h_{ab}\mu^a_{\ul i}\t H^{\ul i}\ul g^{[\u a_i\u b_i]}
_i \(C^b_{cd}\mPhi^c_{i\u a_i}\mPhi^d_{i\u b_i}
-\mu^b_{\hi}f^{\hi}_{\u a_i\u b_i}-\mPhi^b_{\u d_i}f^{\u d_i}_{\u a_i\u b_i}\)
\eq66
$$
where
$$
\ul g^{[\u a_i\u b_i]}_i = \frac 1{V_i}\intop_{M_i} \sqrt{|\t g_i|}\,d^{\bar n_i}x\,
g^{[\u a_i\u b_i]}_i, \quad i=0,1,2,\ldots,k-1. \eq67
$$
Moreover, to be in line in the full theory we should consider a chain of
groups $H_i$, $i=0,1,2,\ldots,k-1$, in such a way that
$$
H_0\subset H_1\subset H_2\subset \ldots \subset H_{k-1}=H. \eq68
$$
For every group $H_i$ we have the following assumptions
$$
G_i\subset H_i \eq69
$$
and $G_{i+1}$ is a centralizer of $G_i$ in $H_i$. Thus we should have
$$
G_i \otimes G_{i+1}\subset H_i, \quad i=0,1,2,\ldots,k-1. \eq70
$$
We know from elementary particles physics theory that
$$
\al
G_0&=U\dr{el}\otimes SU(3)\dr{color},\\
G_1&=SU(2)\dr{L} \otimes U(1)\dr{Y} \otimes SU(3)\dr{color}
\eal
$$
and that $G2$ is a group which plays the role of $H$ in the case of a \s y
breaking from $SU(2)\dr{L}\otimes U\dr{Y}(1)$ to $U\dr{el}(1)$. However, in this
case because of a factor $U(1)$, $M=S^2$. Thus $M_0=S^2$ and $G2\subset H_0$.

It seems that in a reality we have to do with two more stages of a \s y
breaking. Thus $k=3$. We have
$$
\ealn{
&M\simeq S^2 \times M_1 \times M_2 &(9.71)\cr
&M=G/(U(1)\otimes SU(3)) &(9.72)\cr
&M_1={G_1}\big/(SU(2)\times U(1)\times SU(3))\cr
&U(1)\otimes SU(3) \subset SU(2)\otimes U(1)\otimes SU(3)\subset
G_2 \otimes SU(3)\subset G_3=G &(9.73)
}
$$
and
$$
\displaylines{
\hfill G_1\subset H_1\subset H \hfill(9.74)\cr
\hfill U(1)\otimes SU(3)\otimes G\subset H \hfill\text{(9.74a)}\cr
\hfill \bigl(U(1)\otimes SU(2)\otimes SU(3)\bigr)\otimes G_2 \subset H_1
\hfill\text{(9.74b)}}
$$
and
$$
\displaylines{
\hfill G_2 \otimes G \subset H_2=H \hfill \text{(9.74c)}\cr
\hfill M_2=G/G_1. \hfill(9.75)
}
$$

We can take for $G$ $SU(5)$, $SO(10)$, $E6$ or $SU(6)$. Thus there are a lot
of choices for $G_2$, $H_1$ and~$H$.

We can suppose for a trial that
$$
G2\otimes SU(3) \subset H_0. \eq75a
$$
We have also some additional constraints
$$
\rank(G)\ge 4. \eq75b
$$
Thus
$$
\rank(H_0)\ge 4. \eq75c
$$
We can try with $F4=H_0$.

In the case of $H$
$$
\rank(H)\ge \rank(G)+3 \ge7. \eq75d
$$
Thus we can try with $E7$, $E8$
$$
\displaylines{
\rank(H_1)\ge \rank(G_2)+4\cr
\hfill \rank(H)\ge \rank(G_2)+\rank(G)\ge \rank(G_2)+4\ge
\rank(G)+4\ge8. \hfill \text{(9.75e)}
}
$$
In this way we have
$$
\rank(H)\ge 8. \eq75f
$$
Thus we can try with
$$
H=E8. \eq75g
$$
But in this case
$$
\rank(G_2)=\rank(G)=4.
$$
This seems to be nonrealistic. For instance, if $G=SO(10)$, $E6$,
$$
\al
&\rank(SO(10))=5\\
&\rank E6=6.
\eal
$$
In this case we get
$$
\al
&\rank(H)=9\\
&\rank(H)=10
\eal
$$
and $H$ could be $SU(10)$, $SO(18)$, $SO(20)$.

In this approach we try to consider additional dimensions connecting to the
\m~$M$ more seriously, i.e.\ as physical dimensions, additional space-like
dimensions. We remind to the reader that gauge-dimensions connecting to the
group~$H$ have different meaning. They are dimensions connected to local
gauge \s ies (or global) and they cannot be directly observed. Simply saying
we cannot travel along them. In the case of a \m~$M$ this possibility still
exists. However, the \m~$M$ is diffeomorphically equivalent to the product of
some \m s~$M_i$, $i=0,1,2,\ldots,k-1$, with some characteristic sizes~$r_i$. 

The radii $r_i$ represent energy scales of \s y breaking. The lowest energy
scale is a scale of weak interactions (Weinberg-Glashow-Salam model)
$r_0\simeq 10^{-16}$~cm. In this case this is a radius of a sphere~$S^2$. The
possibility of this ``travel'' will be considered in section 16. In this case
a metric on a \m~$M$ can be dependent on a point $x\in E$ (parametrically). 

It is interesting to ask on a stability of a \s y breaking pattern \wrt
quantum fluctuations. This difficult problem strongly depends on the details
of the model. Especially on the Higgs sector of the practical model. In order
to preserve this stability on every stage of the \s y breaking we should
consider remaining Higgs' fields (after \sb) with zero mass. According to
S.~Weinberg, they can stabilize the \sb\ in the range of energy
$$
\frac1{r_i}\(\frac{\hbar}c\)<E <\frac1{r_{i+1}}\(\frac{\hbar}c\),
\quad i=0,1,2,\ldots,k-1, \eq76
$$
i.e.\ for a \sb\ from $G_{i+1}$ to $G_i$.

In this place we can introduce deconfinement parameters $\varLambda(G_i)=
\frac{\alpha_s}{r_i}\bigl(\frac{\hbar}{c}\bigr)$ which introduce a scale of energy for a \sb.

It seems that in order to create a realistic grand unified model based on
non\s ic Kaluza--Klein (Jordan--Thiry) theory it is necessary to nivel \co\
terms. This could be achieved in some models due to choosing \ct s~$\xi$ and
$\zeta$ and~$\mu$. After this we can control the value of those terms, which
are considered as a selfinteraction potential of a scalar field~$\mPsi$. The
scalar field~$\mPsi$ can play in this context a role of a \q.

Let us notice that using (4.51) and (9.56) one gets
$$
\sum_{i=0}^{k-1} \t A^{\u a_i}_{i\u b} \mPhi^c_{\u a_i} f^{\u b}_{\hi\u a}
=C^c_{ab}\mu^a_{\hi}\sum_{i=0}^{k-1} \t A^{\u a_i}_{\t{ia}}\mPhi^b_{\u a_i}\,.
\eq77
$$

In this way we get constraints for Higgs' fields $\mPhi_0$, $\mPhi_1$, \dots,
$\mPhi_{k-1}$,
$$
\mPhi_i=\(\mPhi^b_{\u a_i}\), \quad i=0,1,\ldots,k-1.
$$
Solving these constraints we obtain some of Higgs' fields as functions of 
in\dt\ components. This could result in some cross terms in the potential
(9.41) between $\mPhi$'s with different~$i$. For example a term
$$
V(\mPhi'_i,\mPhi'_j),
$$
where $\mPhi'$ means in\dt\ fields. This can cause some problems in a
stability of \sb\ pattern against radiative corrections. This can be easily
seen from Eq.~(4.51) solved by in\dt~$\mPhi'$,
$$
\ealn{
\mPhi&=B\mPhi' &(9.78)\cr
\mPhi^c_{\u b}&=B^{c\u{\bar b}}_{\u b\bar c}{\mPhi'}^{\bar c}_{\u{\bar b}}
&(9.79)
}
$$
where $B$ is a linear operator transforming in\dt~$\mPhi'$ into $\mPhi$.

We can suppose for a trial a condition similar to~(4.51) for every
$i=0,\ldots, k-1$,
$$
\mPhi^{c_i}_{\u b_i}f^{\u b_i}_{\hi_i\u a_i}=
\mu^{a_i}_{\hi_i} \mPhi^{b_i}_{\u a_i} C^{c_i}_{a_ib_i} \eq80
$$
where $C^{c_i}_{a_ib_i}$ are structure \ct s for the Lie algebra $\gH_i$ of
the group~$H_i$. $f^{\u b_i}_{\hi_i\u a_i}$ are structure \ct s of the Lie
algebra $\gG_{i+1}$, $\hi_i$~are indices belonging to Lie algebra $\gG_i$ and
$\t a_i$ to the complement~$\gM_i$.

In this way
$$
\mPhi^c_{\u b_i} = \mPhi^{c_i}_{\u b_i} \d^c_{c_i}\,. \eq81
$$
In this case we should have a consistency between (9.80) and (9.77) which
impose constraints on $C,f,\mu$ and $C^i,f^i,\mu^i$ where $C^i,f^i,\mu^i$
refer to $H_i$,~$G_{i+1}$. Solving (9.80) via introducing in\dt\ fields
$\mPhi'_i$ one gets
$$
\mPhi^{c_i}_{i\u b_i} = B^{c_i\u{\bar b}_i}_{i\bar c_i\u b_i}
{\mPhi'_i}^{\bar c_i}_{\u{\bar b}_i}. \eq82
$$

Combining (9.79), (9.81--82) one gets
$$
B^{c\u{\bar b}}_{\u b\bar c}{\mPhi'}^{\bar c}_{\u{\bar b}}
=\sum_{i=0}^{k-1} \t A^{\u a_i}_{i\u b} \d^c_{c_i} B^{c_i\u{\bar b}_i}_{i\bar
c_i\u b_i}{\mPhi'_i}^{\bar c_i}_{\u{\bar b}_i}\,.\eq83
$$ 
Eq.\ (9.83) gives a relation between in\dt\ Higgs' fields $\mPhi'$ and
$\mPhi'_i$. Simultaneously it is a consistency condition between Eq.~(4.51)
and Eq.~(9.80). However, the condition (9.80) seems to be too strong and
probably it is necessary to solve a weaker condition (9.77) which goes
to the mentioned terms $V(\mPhi'_i,\mPhi'_j)$. The conditions (9.80) plus a
consistency (9.83) avoid those terms in the Higgs potential. This problem
demands more investigations.

It seems that the condition (9.38) could be too strong. In order to find a
more general condition we consider a simple example of~(9.34). Let
$G_0=\{e\}$ and $K=2$. In this case we have
$$
\ealn{
&\{e\}\subset G_1\subset G_2=G &(9.84)\cr
&M_0=G_1, \quad M_1=G/G_1 &(9.85)\cr
&g:G_1\times G/G_1 \to G. &(9.86)
}
$$
In this way $G_1\times G/G_1$ is diffeomorphically equivalent to~$G$.

Moreover, we can consider a fibre bundle with base space $G/G_1$ and a
structural group~$G_1$ with a bundle \m~$G$. This construction is known in
the theory of induced group representation (see~[51] and section~4). The
projection $\f:G\to G/G_1$ is defined by $\f(g)=\{gG_1\}$. The natural
extension of (9.86) is to consider a fibre bundle $(G,G/G_1,G_1,\f)$. In this
way we have in a place of~(9.86) a local condition
$$
g_U:G_1\times U\underset{\text{in}}\to\longrightarrow G \eq87
$$
where $U\subset G/G_1$ is an open set. Thus in a place of~(9.38) we consider
a local diffeomorphism
$$
g_U : M_0\times M_1\times \ldots \times
M_{k-1}\underset{\text{in}}\to\longrightarrow G/G_0\,. \eq88
$$
where
$$
U=U_0\times U_1\times \ldots \times U_{k-1},
$$
$U_i\subset M_i$, $i=0,1,2,\ldots,k-1$, are open sets. Moreover we should
define projectors $\f_i$, $i=0,1,2,\ldots,k-1$,
$$
\f_i: G/G_0 \to G_{i+1}/G_i, \eq89
$$
i.e.
$$
\f_i\(\{gG_0\}\)=\{g_{i+1}G_i\}, \eq90
$$
$$
g\in G,\quad g_{i+1}\in G_{i+1}, \quad 
G_0 \subset G_i \subset G_{i+1} \subset G
$$
in a unique way. This could give us a fibration of $G/G_0$ in $\prod\limits
_{i=0}^{k-1} \(G_{i+1}/G_i\)$.

For $g\in G_{i+1}$ we simply define
$$
\f_i\(\{gG_0\}\)=\{gG_i\}. \eq91
$$
If $g\in G$, $g\notin G_{i+1}$, we define
$$
\f_i\(\{gG_0\}\)=\{G_i\}. \eq92
$$
Thus in general
$$
\f_i\(\{gG_0\}\)=\{p(g)G_i\} \eq93
$$
where
$$
p(g)=\cases g, & g\in G_{i+1}\\
e,& g\notin G_{i+1}\,.\endcases \eq94
$$
Thus in a place of (9.38) we have to do with a structure
$$
\Bigl\{G/G_0, \prod_{i=0}^{k-1} \(G_{i+1}/G_i\), \f_0,\f_1,\ldots,\f_{k-1}
\Bigr\} \eq95
$$
such that
$$
g_U \circ \f_{|U} =\text{id} \eq96
$$
where
$$
\f_{|U}=\prod_{i=0}^{k-1} {\f_i}_{|U_i}\,. \eq97
$$

This generalizes (9.38) to the local conditions (9.88). Now we can repeat all
the considerations concerning a decomposition of Higgs' fields using local
diffeomorphisms $g_U$ ($g^\ast_U$) in the place of~$g$ ($g^\ast$). Let us
also notice that in the chain of groups it would be interesting to consider
as~$G_2$ 
$$
G_2=SU(2)\dr{L} \otimes SU(2)\dr{R} \otimes SU(4)
$$
suggested by Salam and Pati, where $SU(4)$ unifies $SU(3)\dr{color} \otimes
U(1)\dr{Y}$. This will be helpful in our future consideration concerning
extension to super\s ic groups, i.e.\ $U(2\,|\,2)$ which unifies $SU(2)\dr{L}
\otimes SU(2)\dr{R}$ to the super Lie group $U(2\,|\,2)$ considered by Mohapatra.
Such models on the phenomenological level incorporate fermions with a
possible extension to the super\s ic $SO(10)$ model. They give a natural
framework for lepton flavour mixing going to the neutrino oscillations
incorporating see-saw mechanism for mass generations of neutrinos. In such
approaches the see-saw mechanism
is coming from the grand unified models. Our
approach after incorporating \m s with anticommuting parameters, super Lie
groups, super Lie algebras and in general super\m s (superfibrebundles) can
be able to obtain this. However, it is necessary to develop a formalism (in
the language of super\m s, superfibrebundles, super Lie groups, super Lie
algebras) for non\s ic connections, non\s ic Kaluza--Klein (Jordan--Thiry)
theory. In particular we should construct an analogue of Einstein--Kaufmann
connection for super\m, a non\s ic Kaluza--Klein (Jordan--Thiry) theory for
superfibrebundle with super Lie group. In this way we should define first of
all a non\s ic tensor on a super Lie group and afterwards a non\s ic
metrization of a superfibrebundle.

Let us notice that on every stage of \s y breaking, i.e.\ from $G_{i+1}$
to~$G_i$, we have to do with group $G_i'$ (similar to the group~$G_0'$). Thus
we can have to do with a true and a false vacuum cases which may complicate a
pattern of a \s y breaking.

}
\vfill\eject

\null
\vskip36pt plus4pt minus4pt
\centerline{{\four\bf 10. Properties of the Scalar Field ${\varPsi} $.}}

\vskip4pt
\centerline{{\four\bf Cosmological Drifting of the Mass Scale}}
\vskip24pt plus2pt minus2pt
\write0{10. Properties of the Scalar Field ${\varPsi} $.
Cosmological Drifting of the Mass Scale \the\pageno}

In section 9 we get the equation for scalar field ${\varPsi}$ (see (9.6)).
It has the following shape:
$$
\widehat{\square } {\varPsi} + I({\varPsi} ) = 0,\eqno(10.1)
$$
where $\widehat{\square }$ is a wave operator (differential
linear operator of the second order) and $I({\varPsi} )$
describes interaction of ${\varPsi} $ with Higgs' field,
Yang--Mills' field and cosmological terms. In the terms
of the effective gravitational constant the field ${\varPsi} $
is defined:
$$
{\varPsi} = - {1\over (n+2)} \cdot \ln \bigg({G_{\eff}\over
G_{N}}\bigg).\eqno(10.2)
$$
Simultaneously the field ${\varPsi} $ enters the equation (8.21) and (8.22). It
means that there exists an effective scale of masses for broken gauge
bosons. It is
$$
m_{\eff} = e^{-{\varPsi} } m_{\tilde A } \simeq e^{-{\varPsi} } 
{\alpha _{S}\over r} {\hbar \over c},\eqno(10.3)
$$
The field ${\varPsi} $ enters also the effective cosmological constant for 
``false vacuum" case. We get:
$$
\lambda _{c1} = e^{(n-2){\varPsi} } {4\over \hbar cr^{2}}
\bigg({\ell ^{2}_{pl}\over r^{2}}\bigg)V({\varPhi} ^{1}_{\crt}).
\eqno(10.4)
$$
Using effective scale of masses one gets:
$$
\lambda _{c1} = 4e^{(n+2){\varPsi} } {m^{4}_{\eff}\over \alpha ^{4}_{S}} 
\bigg({\hbar^{3}\over c^{5}}\bigg) \ell ^{2}_{pl} V({\varPhi}
^{1}_{\crt}).\eqno(10.5)
$$
Let us suppose that we have a situation corresponding to one of the
minima of the potential $V({\varPhi} )$. It means that:
$$\eqalign{
{\cal L}_{\Int}(\widetilde{A},{\varPhi} ) &= 0,\cr
V({\varPhi} ) &= V({\varPhi} ^{k}_{\crt}),\cr
{\cal L}_{\kin}(\mathop{\nabla}\limits^{\gauge} {\varPhi}) &=
N^{\mu \nu } M^{2}_{ij}({\varPhi} ^{k}_{\crt})
\widetilde{A}^{i}_{\mu } \widetilde{A}^{j}_{\nu } ,\cr
{\cal L}_{\rYM} &= 0.\cr}\eqno(10.6)
$$
Let us suppose that the cosmological terms are negligible. In this
case we have for $I({\varPsi} )$:
$$
I({\varPsi} ) = {4e^{-2{\varPsi} }\over r^{2}} N^{\mu \nu }
M^{2}_{ij}({\varPhi} ^{k}_{\crt})
\widetilde{A}^{i}_{\mu } \widetilde{A}^{j}_{\nu } +
{(n-2)\over r^{4}} e^{(n-2){\varPsi} } V({\varPhi}^{k}_{\crt}).
\eqno(10.7)
$$
In the vacuum state the fields $\widetilde{A}^{i}_{\mu }$ and ${\varPsi} $ 
fluctuate only around this state. Thus
$$\eqalignno{
\widetilde{A}^{i}_{\mu } &= \widetilde{A}^{i}_{0\mu } + \delta
\widetilde{A}^{i}_{\mu },&(10.8)\cr
{\varPsi} &= {\varPsi} _{0} + \delta {\varPsi} .\cr}
$$
It means that:
$$\eqalignno{
I({\varPsi} ) \simeq{}& \bigg({4e^{-2{\varPsi}_0}\over r^2}
\eta ^{\mu \nu }M^{2}_{ij}({\varPhi}
^{k}_{\crt})\widetilde{A}^{i}_{0\mu }\widetilde{A}^{j}_{0\nu }+
{(n-2)\over r^{4}} e^{(n-2){\varPsi}_0}V({\varPhi}
^{k}_{\crt})\bigg) +\cr
\noalign{\vskip4pt}
&+ \bigg({-8e^{-2{\varPsi}_0 }\over r^{2}} \eta ^{\mu \nu }M^{2}_{ij}
({\varPhi} ^{k}_{\crt})\widetilde{A}^{i}_{0\mu
}\widetilde{A}^{j}_{0\nu }+ {(n-2)^{2}\over r^{4}}
e^{(n-2){\varPsi}_0 }V({\varPhi} ^{k}_{\crt})\bigg) \delta {\varPsi}=\cr
={}& d_{1} + d_{2} \delta {\varPsi} ,&(10.9)\cr}
$$
$d_{1} $, $d_{2}$ are constants and we put $N^{\mu \nu } = \ov{M}
\eta ^{\mu \nu }$ (in zero order of
approximation). And from (10.1) we get:
$$
\square (\delta {\varPsi} ) + \bigg({d_{1}\over \ov{M}}\bigg) + 
\bigg({d_{2}\over \ov{M}}\bigg) \delta {\varPsi} = 0.\eqno(10.10)
$$
It means that (10.1) becomes the Klein--Gordon equation for the
fluctuation of the scalar field ${\varPsi} $ and $\delta {\varPsi} $ is a massive field. Thus it
has Yukawa-type behaviour.
$$
\delta {\varPsi} \sim {1\over r} e^{-\alpha r} + \const,\quad\ 
\alpha > 0.\eqno(10.11)
$$
For the field ${\varPsi} $ we get:
$$
{\varPsi} \sim \const' + {1\over r} e^{-\alpha r} ,\quad\ \alpha >
0.\eqno(10.12)
$$
Due to this we preserve the weak equivalence principle for the
gravitational field. The scalar forces connected to the field ${\varPsi} $ are of
the short range. This is possible if $d_{2} > 0$. The first term in $d_{2}$ is
greater than zero and we have no troubles in the true vacuum case
$(V({\varPhi} ^{0}_{\crt}) = 0)$. In the case of the false vacuum the situation may be
different $(d_{2} \le 0)$. Thus the world built over the false vacuum state
can violate the weak equivalence principle. Summing up we get that in
the case of true vacuum the field ${\varPsi} $ is massive with short range. It is
in accordance with the weak equivalence principle, the universal fall
of all bodies in gravitational field. This property is connected, from
cosmological point of view of course, to space dependence of ${\varPsi}$ (or
$G_{\eff})$. It is interesting to ask what will happen if $G_{\eff} = 
G_{\eff}(t)$. It
means that ${\varPsi} = {\varPsi} (t)$ depends on the time $t$ only. Now it is a
cosmological time. In this case ${\varPsi} (t)$ plays the role of the
cosmological factor. Let us estimate the dependence on time for the
effective scale of masses, $m_{\eff}$, using some estimations for $G_{\eff}$. We
have (see [56], [57]):
$$\eqalignno{
\eta &= {{\varDelta} G_{\eff}\over G_{\eff} {\varDelta} t} = 
{10^{-11} - 10^{-10}\over yr},&(10.13)\cr
\noalign{\vskip4pt}
\eta &= {d\over dt} \ln G_{\eff}(t) = -(n+2) \bigg({d{\varPsi} \over
dt}\bigg).&(10.14)\cr}
$$
Thus we get
$$
{\varDelta} {\varPsi} = - {\eta \over (n+2)} {\varDelta} t.
\eqno(10.15)
$$
For the effective scale of mass one gets:
$$
\omega = {{\varDelta} m_{\eff}\over m_{\eff} {\varDelta} t} = 
{d\over dt} \ln m_{\eff}(t) = - {d\over dt} {\varPsi} (t) = - 
{{\varDelta} {\varPsi} \over {\varDelta} t}.\eqno(10.16)
$$
Thus we get
$$
\omega = {\eta \over (n+2)}\eqno(10.17)
$$
and
$$
\omega = {10^{-12} - 10^{-11}\over yr},\eqno(10.18)
$$
if we suppose that $n \ge 24$ ($n = \dim G$ and for $G = \SU(5)$ we can have
$n=24$). Thus we are able to connect changing of the effective
gravitational constant with very small drifting of the scale of masses
for the broken gauge bosons. This drifting is really very small and it
does not contradict experimental data. It is worth to notice that the
sign of changing of $G_{\eff}$ and $m_{\eff}$ is the same. If $G_{\eff}$ increases then
$m_{\eff}$ increases too and vice-versa. This behaviour is in agreement with
Mach's principle.

Let us remind to the reader the Dirac large number hypothesis. He gets that
$$
\frac{e^2}{G_Nm_em_p}\simeq 10^{40}, \eqno(10.19)
$$
where $m_e$, $m_p$ mean electron and proton mass, respectively, and
$$
\frac t{e^2/m_ec^3} \simeq 10^{40}, \eqno(10.20)
$$
where $t$ is an age of the Universe and $\frac{e^2}{m_ec^3}$ is a time needed
for light to pass a distance of a classical radius of an electron.

If those large numbers are equal forever one gets
$$
\frac{e^2}{G_Nm_em_p}\simeq \frac t{e^2/m_ec^3} \eqno(10.21)
$$
and we get an interesting cosmological dependence
$$
\frac 1t = \left(\frac{m_pc^3}{e^4}\right) Gm_e^2. \eqno(10.22)
$$

The mass of a proton is established by $\varLambda_{\text{\rm QCD}}$ and we do not
expect the cosmological drifting of $\varLambda_{\text{\rm QCD}}$. Thus
$m_p=\const$. However, in the case of~$m_e$ is different
$$
m_e \sim m_{\text{\rm EW\,eff}} \sim \frac1r e^{-\varPsi} \sim
m_{\text{\rm W}} e^{-\varPsi} \eqno(10.23)
$$
and
$$
G=G_{\text{\rm eff}} = G_Ne^{-(n+2)\varPsi}. \eqno(10.24)
$$
Thus we get
$$
\frac 1t=Ce^{-(n+4)\varPsi} \eqno(10.25)
$$
where $C$ is a constant.

And finally
$$
{\varPsi}(t)=\frac{1}{n+4} \ln(Ct), \eqno(10.26)
$$
which gives an evolution law for a scalar field~$\varPsi$. This could be
incorporated as an ansatz for some cosmological models, where
$$
C\simeq \frac{10^{20}}{t_{\text{\rm pl}}} \eqno(10.27)
$$
and $t_{\text{\rm pl}}$ is a Planck time.

\vfill\eject

\null
\vskip36pt plus4pt minus4pt
\centerline{{\four\bf 11. Fermion Number, ${\sym R}_{+}$ and ${\bold U}(1)_{\rF}$ Invariance}}
\vskip24pt plus2pt minus2pt
\write0{11. Fermion Number, ${\sym R}_{+}$ and ${\bold U}(1)_{\rF}$
Invariance \the\pageno}

In N.G.T. and in Einstein Unified Field Theory there exists a
gauge invariance for the 2-form of curvature
$$
\overline{\varOmega}^{\alpha }{}_{\beta }(\ov{W}) = d\ov{W}^{\alpha
}{}_{\beta } + \ov{W}^{\alpha }{}_{\beta } \wedge \ov{W}^{\gamma
}{}_{\beta }\eqno(11.1)
$$
such that
$$
\overline{\varOmega}^{\alpha }{}_{\beta }(\ov{W}) =
\overline{\varOmega}^{\alpha }{}_{\beta }(\ov{W}'),\eqno(11.2)
$$
where
$$
\ov{W}^{\alpha }{}_{\beta } \rightarrow \ov{W}'{}^{\alpha }{}_{\beta } =
\ov{W}^{\alpha }{}_{\beta } - {2\over 3} \delta ^{\alpha }{}_{\beta
}\, d\phi \eqno(11.3)
$$
and where $\phi $ is a function on $E$. This results as a gauge transformation
for the 1-form $\ov{W}$
$$
\ov{W}\rightarrow \ov{W}' = \ov{W} + d\phi .\eqno(11.4)
$$
It is easy to see that the 2-form of curvature for the connection
$W^{\tilde A }{}_{\tilde B }$ is also invariant under (11.4)
$$
{\varOmega} ^{\tilde A }{}_{\tilde B }(W) = {\varOmega} ^{\tilde A
}{}_{\tilde B }(W' ),\eqno(11.5)
$$
where
$$
W'{}^{\tilde A }{}_{\tilde B } = W^{\tilde A }{}_{\tilde B } - {4\over 3(m+2)} 
\delta ^{\tilde A }{}_{\tilde B } d\phi .\eqno(11.6)
$$
In N.G.T. (see [34], [47], [65], [66]) the vector field $\ov{W}_{\mu }$ is coupled
to the fermion current (fermion number plays in this theory the role of
the second gravitational charge). Thus (11.4) results in the
conservation law for the fermion current. Fermion current is
simultaneously the source of the skew-symmetric part of the metric (see
for details [34], [47], [65]). The fermion current for a Dirac
particle has the same shape as an electric current
$$
S^{\mu } \sim \overline{\varPsi} \gamma ^{\mu } {\varPsi}.\eqno(11.7)
$$
We have only a different coupling constant between $S^{\mu }$ and
$\ov{W}_{\mu }$.

One can ask what is the origin of the transformation (11.3). It
is very well known that the linear connection (coefficients of the
connection) $\ov{W}^{\alpha }{}_{\beta }$ obeys the following law of transformation: (the linear
connection is not a geometrical quantity of $\sigma $-type in a Shouten sense
as for example tensor, vector, spinor, 2-form of curvature, etc., see
Ref.~[59] for more details).
$$
A^{\mu }{}_{\nu }\ov{W}' {}^{\nu }{}_{\beta } = \ov{W}^{\mu }{}_{\nu } 
A^{\nu }{}_{\beta } + dA^{\mu }{}_{\beta },\eqno(11.8)
$$
if we change basic forms $\overline{\theta }^{\alpha }$ into $\overline{\theta
}'{}^{\alpha }$ such that
$$
\overline{\theta }^{\alpha } = A^{\alpha }{}_{\beta }\overline{\theta }'{}^{\beta
},\eqno(11.9)
$$
where $A^{\alpha }{}_{\beta }$ is a matrix valued function on the manifold
$E$ (space-time)
with values in $\GL (4,{\sym R})$, $\det (A^{\alpha }{}_{\beta }) \neq 
0$. If we restrict $\GL (4,{\sym R})$ to the
subgroup $\GL_{\downarrow} (4,{\sym R}) = \{(A^{\alpha }{}_{\beta }), \det
(A^{\alpha }{}_{\beta }) > 0 \}$ we may consider the
following transformation:
$$
A^{\alpha }{}_{\beta } = \delta ^{\alpha }{}_{\beta } e^{-{2\over 3}\phi }
\eqno(11.10)
$$
and we get from (11.8) Eq.~(11.3). It is easy to see that (11.10) is a
dilatation. Let us write every matrix $A$ from $\GL_{\downarrow} (4,{\sym R})$ in the form:
$$
A = e^{B}.\eqno(11.11)
$$
This is possible, because $\det A > 0$ and let us decompose $B$ into two
parts
$$
B^{\alpha }{}_{\beta } = (B^{\alpha }{}_{\beta }-b\delta ^{\alpha
}{}_{\beta }) + b\delta ^{\alpha }{}_{\beta } = B' {}^{\alpha
}{}_{\beta } - b\delta ^{\alpha }{}_{\beta },\eqno(11.12)
$$
where $b = 1/4B^{\alpha }{}_{\alpha }$ and $B' {}^{\alpha }{}_{\beta }$ is a 
traceless matrix. According to (11.12) one gets
$$
A = e^{b} e^{B' }.\eqno(11.13)
$$
But
$$
\det A = e^{\Tr B} = e^{4b}.\eqno(11.14)
$$
Comparing (11.13), (11.14) and (11.10) one gets:
$$
\phi = - {3\over 8} \ln (\det A),\eqno(11.15)
$$
where $A \in \GL_{\downarrow} (4,{\sym R})$. In this way we have the decomposition
$$
\GL_{\downarrow} (4,{\sym R}) = {\sym R}_{+} \otimes \SL (4,{\sym R}),
\eqno(11.16)
$$
where ${\sym R}_{+} = \{e^{\rho },\ \rho \in {\sym R}\}$ and $\SL 
(4,{\sym R}) = \{A \in \GL (4,{\sym R}),\ \det A = 1\}$. The
gauge transformation (11.3), (11.4) and (11.6) are the gauge
transformations corresponding to ${\sym R}_{+}$ in the decomposition (11.16).
Transformation (11.4) induces the following dilatation transformations
for spinor fields ${\varPsi} $ and $\overline{\varPsi}$ (see Ref.~[92])
$$\eqalignno{
{\varPsi} &\rightarrow {\varPsi} ' = {\varPsi} e^{-i\beta \phi
},&(11.17)\cr
\overline{\varPsi} &\rightarrow \overline{\varPsi} ' = \overline{\varPsi} 
e^{i\beta \phi },&(11.18)\cr}
$$
where $\beta $ is a constant connected to the universal coupling constant 
$a$ from Moffat's theory. In this way we have compactification of the
group ${\sym R}_{+}$ from the decomposition (11.16) (see Ref.~[92] for details) and
spinors transform according to the following group:
$$
{\bold U}(1)_{\rF} \otimes \SL (2,{\sym C}).\eqno(11.19)
$$
In the case of electromagnetic fields we have
$$
{\bold U}(1)_{el} \otimes \SL (2,{\sym C})\eqno(11.20)
$$
and the transformation law for ${\varPsi} $ and $\overline{\varPsi}$
$$\eqalignno{
{\varPsi} \rightarrow {\varPsi} ' &= {\varPsi} e^{-i{e\over
\hbar c}\chi },&(11.21)\cr
\overline{\varPsi} \rightarrow \overline{\varPsi}' &= \overline{\varPsi} 
e^{i{e\over \hbar c}\chi },&(11.22)\cr
A \rightarrow A' &= A + d\chi.&(11.23)\cr}
$$
The difference between (11.19) and (11.20) is significant, because
${\bold U}(1)_{\rF}$ is a compactification of ${\sym R}_{+}$ and ${\bold U}(1)_{el}$ is compact from the very
beginning. But there is a more fundamental difference, ${\sym
R}_{+}$ (dilatations) belong to space-time symmetries and ${\bold U}(1)_{el}$ to the
internal symmetry of the electromagnetic field. It is easy to see that
we can get (11.19) from the following scheme
$$\displaylines{\quad
\GL (4,{\sym R}) {\rightarrow} \GL_{\downarrow} (4,{\sym R}) 
{\rightarrow} {\sym R}_{+} {\otimes} \SL(4,{\sym R})
{\rightarrow} {\sym R}_{+} {\otimes} \SO (1,3) {\rightarrow} 
{\bold U}(1)_{\rF} {\otimes} \SL (2,{\sym C})\hfill (11.24)\cr}
$$
and ${\bold U}(1)_{\rF}$ is a compactification of ${\sym R}_{+}$ in the following sense:
$$
{\bold U}(1)_{\rF} = \bigg\{e^{-i\psi} , \psi = {3\over 8} \ln 
(\det (A^{\mu }{}_{\nu })) , \det (A^{\mu }{}_{\nu }) > 0\bigg\}.\eqno(11.25)
$$
Fermion number is independently connected to the internal symmetries of
elementary particles. We know that there are no forces with long range
connected to fermion number (weak equivalence principle). In N.G.T.
there are also no forces with long range connected to this charge. The
theory satisfies the weak equivalence principle --- universal fall of all
bodies. In the Newtonian approximation of N.G.T. there is no Coulomb
force connecting to the $\ov{W}_{\mu }$ potential or to the skew-symmetric part of
the metric $g_{[\mu \nu ]}$. On the other hand fermion number is conserved in a
similar manner to that of electric charge and it is usually connected
to the ${\bold U}(1)_{\rF}$ internal invariance.

In the Grand Unified Theory based on the SO(10) group (see [93]),
fermion number is one of the generators of the Lie algebra of SO(10),
$Y_{\rF}$, and $F=B-L$, where $B$ is the baryon number and $L$ the lepton number.
Thus we have the local symmetry ${\bold U}(1)_{\rF}$. After spontaneous symmetry
breaking the intermediate gauge boson corresponding to $Y_{F}$ in the
adjoint representation obtains a rest mass and we have no forces of
long range connecting to $F$.

This symmetry breaking is the spontaneous symmetry breaking and
the lagrangian of the theory is invariant under local transformations
of SO(10) (thus under ${\bold U}(1)_{F}$ too). Symmetry is breaking only due to
choosing a concrete representative of the vacuum, which is here
degenerated.

In the Grand Unified Theory based on $\SU(5)$ we have no ${\bold U}(1)_{\rF}$ local
invariance, we have only global ${\bold U}(1)_{\rF}$ invariance.

In the SO(10) G.U.T. we get after symmetry breaking ${\bold U}(1)_{\rF}$ global
invariance and the fermion number $F=B-L$ is conserved. Space-time
symmetries and internal symmetries are usually disconnected. Only some
geometrical approaches similar to dimensional reduction or Kaluza--Klein
theory or maybe supersymmetry are able to connect these symmetries
treating them as ``space-time" symmetries on many-dimensional manifolds.

In our theory we have this situation. We have a many-dimensional
manifold and after dimensional reduction we get gravity coupled to the
Yang--Mills' field and to the Higgs' field with spontaneous symmetry
breaking and the Higgs' mechanism. Simultaneously gravitation in our
theory is described by N.G.T. Thus we should try to connect ${\sym R}_{+}$
symmetry from N.G.T. with ${\bold U}(1)_{\rF}$ internal symmetry. There are two
possibilities for such relations.

1) Let us suppose that ${\bold U}(1)_{\rF}$ is a subgroup of the group $G$
$$
{\bold U}(1)_{\rF} \subset G\eqno(11.26)
$$
in such a way that $Y_{\rF} \in \fm$. This means that after symmetry breaking
the intermediate boson corresponding to $Y_{\rF}$ (for ``true" vacuum case)
obtains the rest mass and we will have no long range forces connecting
to $F$. This is similar to the SO(10) G.U.T.

Now let us suppose that there exists a homomorphism between
$\GL_{\downarrow} (4,{\sym R})$ and $G$
$$
\varepsilon : \GL_{\downarrow} (4,{\sym R}) \rightarrow G,\eqno(11.27)
$$
such that
$$
\varepsilon (A) = e^{-(i{3\over 8}\ln (\det A)Y_{\rF})},\eqno(11.28)
$$
where $A = (A^{\mu }{}_{\nu })$, $\det A > 0$.

(11.28) is really the homomorphism between $\GL_{\downarrow} (4,{\sym R})$ 
and ${\bold U}(1)_{\rF}$. It
projects $\GL_{\downarrow} (4,{\sym R})$ to a one-parameter subgroup of 
$G$. This has the
result that dilatations on the space-time from ${\sym R}_{+}$ induce internal
rotations of ${\bold U}(1)_{\rF}$.

2) Let us suppose that the group $H$ has the following structure
$$
{\bold U}(1)_{\rF}\not\subset_| G ,\quad\ G \otimes G_{0} \subset H
,\quad\ {\bold U}(1)_{\rF} \subset H.\eqno(11.29)
$$
Let there exists a homomorphism between $\GL_{\downarrow} (4,{\sym R})$ and $H$
$$
\eta : \GL_{\downarrow} (4,{\sym R}) \rightarrow H,\eqno(11.30)
$$
such that
$$
\eta (A) = e^{-(i{3\over 8}\ln (\det A)Y_{\rF})},\eqno(11.31)
$$
where $A = (A^{\mu }{}_{\nu })$, $\det A > 0$. Now dilatations from the space-time also
induce internal ${\bold U}(1)_{\rF}$ rotations. However the situation is different.
Now $Y_{\rF}$ does not belong to the Lie algebra of the group $G$ and due to the
Wang condition there is no gauge boson connecting to ${\bold
U}(1)_{\rF}$ (to fermion
number). There is no gauge field defined on space-time connecting to
$Y_{F}$. The ${\bold U}(1)_{\rF}$ symmetry acts only on the Higgs' multiplet and it is
really global. Thus we have a situation similar to SU(5) G.U.T.

Moreover the correct interpretation of $Y_{\rF}$ can be obtained if we
consider a fundamental representation of $G$ (if ${\bold U}(1)_{\rF} \subset G)$. In this
case we put fermion fields into the representation and eigenvalues of
$Y_{\rF}$ tell us what is the meaning of $F$. In the case of ${\bold
U}(1)_{\rF} \subset H$ and
${\bold U}(1)_{\rF} \not\subset G$ the interpretation of $F$ is more complex. We should consider
a fundamental representation of $H$ and try to span its space by fermion
fields looking for eigenvalues of $Y_{\rF}$. However the problem of fermion
fields in the Nonsymmetric Jordan--Thiry Theory seems to be interesting
and it will be done elsewhere.

We do not identify our group $G$ with SU(5) or SO(10) from G.U.T.
The relationship between these classical G.U.T.'s and dimensional
reduction procedures seems to be more complex (see Ref.~[80]). The
feature concerning the second possibility of the place of the ${\bold U}(1)_{\rF}$
group seems to be more general than for ${\bold U}(1)_{\rF}$ only. The dimensional
reduction scheme offers us a treatment of global symmetries in
elementary particle physics. Let $S$ be a group such that the group $H$
has a structure:
$$
S \subset H ,\quad\ S\not\subset G ,\quad\ G \otimes G_{0} \subset H.
\eqno(11.32)
$$
We have here $S$ in a place of ${\bold U}(1)_{\rF}$ in (11.29). Due to the Wang
condition we do not get any gauge field corresponding to the gauge
group $S$. This group is really global. Thus the dimensional reduction
scheme allows us to consider both global and local symmetries. Both
symmetries can be unified by the one group $H$. The group $S$ acts, of
course, on the Higgs' field ${\varPhi} $ and would be broken spontaneously,
without the Higgs' mechanism. The group $S$ is the symmetry group of the
potential $V$. Thus we should get after symmetry breaking from $G$ to $G_{0}$
massless scalars, the so-called pseudo-Goldstone bosons. The number of
those zero modes is equal or greater $\dim S$. They can get masses through
radiative corrections. The correct interpretation of generators of the
Lie algebra of $S$ can be obtained in a similar way as $Y_{\rF}$. In the next
section we consider in details the first possibility (1)).

\vfill\eject

\null
\vskip36pt plus24pt minus4pt
\centerline{{\four\bf 12. Fermion Number as the Second Gravitational Charge}}

\vskip4pt
\centerline{{\four\bf $\ov{W}_{\mu }$ and $\widetilde{A}^{\rF}{}_{\mu }$ Potentials}}
\vskip24pt plus2pt minus2pt
\write0{12. Fermion Number as the Second Gravitational Charge
$\overline{W}_{\mu }$ and $\widetilde{A}^{\rF}{}_{\mu }$ Potentials \the\pageno}

Let us consider the principal fibre bundle $\widehat{P}$ of frames over the
manifold $E$ (space-time) with the structural group
$$
\GL_{\downarrow} (4,{\sym R}) =\{(A^{\alpha }{}_{\beta }) ,\ \det
(A^{\alpha }{}_{\beta }) > 0\}.\eqno(12.1)
$$
Let $\widehat{\pi }$ be a projection $\widehat{\pi } : \widehat{P} \rightarrow
E$. For every $x \in E$ we have that
$$
F_{x} = \widehat{\pi }^{-1}(\{x\}) \cong \GL_{\downarrow} (4,{\sym R}),
\eqno(12.2)
$$
i.e., the fibre $F_{x}$ is equivalent to the structural group
$\GL_{\downarrow} (4,{\sym R})$. Let
$\cW$ be a connection on this bundle and let us decompose the Lie group
$\GL_{\downarrow} (4,{\sym R})$ into
$$
\GL _{\downarrow} (4,{\sym R}) = {\sym R}_{+} \otimes \SL (4,{\sym R}),
\eqno(12.3)
$$
where
$$
\SL (4,{\sym R}) = \{A \in \GL (4,{\sym R}) ,\ \det A = 1 \}
$$
and
$$
{\sym R}_{+} = \biggl\{e^{\rho } ,\ \rho = {1\over 4}\ln (\det A) ,\ A \in 
\GL _{\downarrow} (4,{\sym R})\biggr\}.
$$
The decomposition (12.3) can be used to construct two subbundles of
$\widehat{P}: \widehat{P}_{1}$ and $\widehat{P}_{2}$ see Fig.~7. They are defined over $E$ with structural groups
${\sym R}_{+}$ and $\SL (4,{\sym R})$. For $\SL (4,{\sym R})$ is a closed subgroup of $\GL _{\downarrow} (4,{\sym R})$, the
transport of the connection from $\widehat{P}$ to $\widehat{P}_{1}$ and $\widehat{P}_{2}$ is possible (see
Ref.~[59]).

\obraz{
\midinsert \centerline{\epsfxsize\hsize \epsfbox{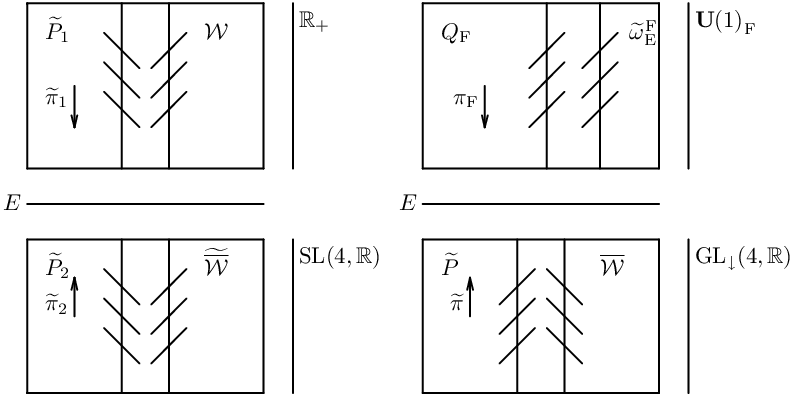}}
\centerline{\nine Fig. 7. Principal fibre bundles $\widehat{P}_1$,
$\widehat{P}_2$, $Q_{\text{\rm F}}$ and $\widehat{P}$}
\endinsert

}In general the condition for a reduction of the principal bundle $P(M,G)$
to a subbundle $P' (M,G' )$ where $G' $ is a Lie subgroup of $G$ is as follows.
There is an open covering $\{U_{\alpha }\}$ of $M$ with a set of transition functions
$\{{\varPsi} _{\beta \alpha }\}$ with values in $G' $. We suppose that this condition is satisfied
in our cases (see Ref.~[59]).

In this way we have, for the connection $\cW$, a decomposition
$$
\overline{\cW} = \widetilde{\cW} \oplus \widehat{\cW},\eqno(12.4)
$$
where $\widetilde{\cW}$ is a 1-form with values in the Lie algebra $\SL(4,{\sym R})$ and $\widehat{{\cal W}}$ is a
1-form with values in the Lie algebra of ${\sym R}_{+}$.
$\widetilde{\cW}$ is a connection on $\widehat{P}_{2}$ and
$\widehat{\cW}$ on $\widehat{P}_{1}$. Thus we have
$$
\widehat{\cW} = \cW Z,\eqno(12.5)
$$
where $\cW $ is a 1-form with real values and $Z$ is a generator of the Lie
algebra~${\sym R}_{+}$. 
Let us take a local section $f$ of the bundle $\widehat{P}$
$$
f : E \rightarrow \widehat{P}.\eqno(12.6)
$$
One gets:
$$
f^{*} \overline{\cW} = (\ov{W}^{\alpha }{}_{\beta \gamma } \overline{\theta
}^{\gamma }) X^{\beta }{}_{\alpha },\eqno(12.7)
$$
where $X^{\beta }{}_{\alpha }$, are generators of the Lie algebra of the group $\GL _{\downarrow} (4,{\sym R})$. We
can define the second connection $\widetilde{\cW}$ (coefficients of a connection 
$\widetilde{\cW}^{\alpha }{}_{\beta })$
according to Eq.~(12.7) in terms of $\overline{\cW}^{\alpha }{}_{\beta }$.

Let us consider the general transformation law for the
coefficients of the connection $\ov{W}^{\alpha }{}_{\beta }$
$$
A^{\lambda }{}_{\mu }\ov{W}' {}^{\mu }{}_{\nu } =\ov{W}^{\lambda
}{}_{\mu } A^{\mu }{}_{\nu } + dA^{\lambda }{}_{\mu },\eqno(12.8)
$$
where $A^{\lambda }{}_{\mu }$ is a matrix valued function from $\GL _{\downarrow} 
(4,{\sym R})$ and a frame transforms as
$$
\overline{\theta }^{\alpha } = A^{\alpha }{}_{\beta }\overline{\theta }'
{}^{\beta }.\eqno(12.9)
$$
If we choose
$$
A^{\alpha }{}_{\beta } = e^{-2/3\phi } \delta ^{\alpha }{}_{\beta }
\eqno(12.10)
$$
we get
$$
\ov{W}' {}^{\lambda }{}_{\mu } = \ov{W}^{\lambda }{}_{\mu } - 
{2\over 3} \delta ^{\lambda }{}_{\mu }\, d\phi\eqno(12.11)
$$
or
$$
\ov{W}' {}^{\lambda }{}_{\mu } = \overline{\omega }^{\lambda }{}_{\mu } - 
{2\over 3} \delta ^{\lambda }{}_{\mu } \ov{W}' ,\eqno(12.12)
$$
where
$$
\ov{W}' = \ov{W}+ d\phi \eqno(12.13)
$$
and
$$
\overline{\omega }^{\lambda }{}_{\mu } = f^{*}\widetilde{\cW}^{\lambda
}{}_{\mu },\quad\ \widetilde{\cW}= \widetilde{\cW}^{\lambda }{}_{\mu } 
X^{\mu }{}_{\lambda }.
$$
Thus it is possible to consider (12.10) and (12.13) as a gauge
transformation for the 1-form $\ov{W}$. This was done in Refs.~[34], [26],
[27], [47], [65], [66]. It is easy to see that the 2-form of the
curvature for the connection $\overline{\cW}^{\alpha }{}_{\beta }$, 
${\varOmega} ^{\alpha }{}_{\beta }(\overline{\cW})$ is invariant under this
transformation (see Refs.~[26], [27]).

We can call this transformation the Einstein-$\lambda $-transformation (see
Ref.~[3]). This is also true for the curvature of the 
$\cW ^{\tilde A}{}_{\tilde B}$ connection,
i.e. ${\varOmega} ^{\tilde A }{}_{\tilde B }$ 
(see Refs.~[26], [27]). 
Let us consider the decomposition
(12.4) for the connection $\overline{\cW}$ corresponding to the decomposition (12.3)
for the group $\GL _{\downarrow} (4,{\sym R})$.Thus for every local section $f$ of the bundle $\widehat{P}$
we get:
$$
f^{*}\overline{\cW}= \overline{\omega }^{\lambda }{}_{\mu } \widetilde{X}^{\mu
}{}_{\lambda } + \ov{W} Z,\eqno(12.14)
$$
$\widetilde{X}^{\mu }{}_{\lambda }$ are generators of the Lie algebra of
$\SL (4,{\sym R})$. This means that we
have chosen the decomposition of the Lie algebra of $\GL
_{\downarrow} (4,{\sym R})$ into
$\SL (4,{\sym R})$ and ${\sym R}$ (${\sym R}$ is a real axis). Thus for every section of $\widehat{P}$ or $\widehat{P}_{2}$
and $\widehat{P}_{1}$ one gets
$$\eqalignno{
f^{*}\widetilde{\cW} &= \overline{\omega }^{\lambda }{}_{\mu } 
\widetilde{X}^{\mu }{}_{\lambda },&(12.15)\cr
f^{*}\widehat{\cW } &= \ov{W}_{v}\overline{\theta }^{v} Z.&(12.16)\cr}
$$
Let us consider the decomposition (11.3) and compare it with formula
(11.15). We have
$$
\phi = - {3\over 8} \ln (\det A).\eqno(12.17)
$$
The group ${\sym R}_{+}$ is called the
 dilatation group. Thus the gauge
transformation (12.13) for the 1-form $\ov{W}$ is connected to the dilatation
subgroup of $\GL _{\downarrow} (4,{\sym R})$. Let us consider the 
principal bundle
$Q$ (see Fig.~2A) with the structural group~$G$ and with a projection 
$\pi _{E}$ over 
$E$ (space-time). On this bundle we have defined a 1-form of a connection
$\widetilde{\omega }_{\rE}$. (see Fig.~7).

Let ${\bold U}(1)_{\rF}$ be a subgroup of the group $G$ such that
$$
{\bold U}(1)_{\rF}\not\subset G_{0}\eqno(12.18)
$$
and let us call this group the fermion charge group. Eq.~(12.18) means
that this subgroup is broken spontaneously and the Higgs' mechanism
makes massive the gauge boson corresponding to this group. Let $Y_{\rF} \in \fg$
be a generator of ${\bold U}(1)_{\rF}$. For ${\bold U}(1)_{\rF}$ is broken spontaneously
$Y_{\rF} \in \fm$.
Let us suppose that there exists the following homomorphism between
$\GL _{\downarrow} (4,{\sym R})$ and $G$
$$
\sigma : \GL _{\downarrow} (4,{\sym R}) \rightarrow G,\eqno(12.19)
$$
such that
$$
\sigma (A) = e^{-(i{3\over 8}\ln (\det A)Y_{\rF})}.\eqno(12.20)
$$
In this way we project the group $\GL _{\downarrow} (4,{\sym R})$ on the 
${\bold U}(1)_{\rF}$ subgroup of $G$.
The Eq.~(12.20) is really the smooth homomorphism between ${\sym R}_{+}$ and 
${\bold U}(1)_{\rF}$.
Simultaneously (12.20) can be made a smooth morphism between typical
fibres of the bundles $\widehat{P}$ and $Q$. This suggests the following. Let us
consider the morphism of the bundles $\widehat{P}$ and $Q$
$$
{\varSigma} : \widehat{P} \rightarrow Q\quad \hbox{or}\quad ({\varSigma} _{1}
:\widehat{P}_{1}\rightarrow Q_{\rF}),\eqno(12.21)
$$
such that we have (12.20) and on the space-time $E$ (base manifold)
$$
\id _{\rE} : E \rightarrow E,\eqno(12.22)
$$
where $\id_{E}$ means the identity on $E$. The morphism ${\varSigma} $ 
has the following
meaning. The fibre bundle $\widehat{P}$ is reduced to the principal fibre bundle
$\widehat{P}_{1}$ over $E$ with a structural group ${\sym R}_{+}$. Simultaneously the principal
fibre bundle $Q$ is reduced to the ${\bold U}(1)_{\rF}$ fibre bundle over $E$ i.e.\ $Q_{F}$.
This is possible because of the property of the ${\bold U}(1)_{\rF}$ subgroup of $G$ and
the definition of ${\sym R}_{+}$ subgroup of $\GL _{\downarrow} (4,{\sym R})$. The transport of the
connections of both bundles to their subbundles is possible. We can
write our construction as a chain of morphisms:
$$
\widehat{P} \tor \widehat{P}_{1} \toS Q_{\rF} \leftor Q,
$$
where $r_{1}$ means a reduction of $\widehat{P}$ to $\widehat{P}_{1}$,
$r_{2} a$ reduction of $Q$ to $Q_{\rF}$ and ${\varSigma} _{1}$
a morphism induced by (12.20). Let us consider the contragradient
transformation for ${\varSigma}$ (a pull-back), ${\varSigma} ^{*}$ and apply it for $\omega _{E}$. One gets:
$$
{\varSigma} ^{*}(\widetilde{\omega }_{\rE}) = \widehat{\cW }\eqno(12.23)
$$
or
$$
{\varSigma} ^{*}(\widetilde{\omega }_{\rE}) = {\varSigma} ^{*}
(\widetilde{\omega }^{\rF}_{\rE} Y_{\rF}) = \cW Z.\eqno(12.24)
$$
Let us take a local section $f$ of $\widehat{P}$. One finds
$$
- {3\over 2} \widetilde{A}^{\rF}{}_{\mu } \overline{\theta }^{\mu } = 
\ov{W}_{\mu } \overline{\theta }^{\mu }.\eqno(12.25)
$$
Thus we get a relationship between the gauge field
$\widetilde{A}^{\rF}{}_{\mu }$, corresponding
to the fermion number $F$, and the field $\ov{W}_{\mu }$ which couples the fermion
current in the nonsymmetric theory of gravitation. For the curvature
of the connection $\widetilde{\omega }_{\rE}$, $\widetilde{{\varOmega}
}_{\rE}$ one gets:
$$
{\varSigma} ^{*}(\widetilde{{\varOmega} }_{\rE}) = 
{\varSigma} ^{*}(\widetilde{H}^{\rF}Y_{\rF}) = 
{\varSigma} ^{*}(d\widetilde{\omega }^{\rF}_{\rE}Y_{\rF}) = 
\widehat{{\varOmega} } = d\widehat{\cW } = d\cW Z.\eqno(12.26)
$$
One easily finds that:
$$
- {4\over 3}\ov{W}_{[\mu ,\nu ]} = \widetilde{F}^{\rF}{}_{\mu \nu } = 
\partial _{\mu }\widetilde{A}^{\rF}{}_{\nu } - \partial _{\nu }
\widetilde{A}^{\rF}{}_{\mu }.\eqno(12.26\text{\rm a})
$$
In this way, we can make the following statements:

1) local gauge invariance on the space-time $E$ connected to the ${\sym R}_{+}$
group is the same as the local gauge invariance ${\bold U}(1)_{\rF}$ for the fermion
number.

2) the gauge field $\widetilde{A}^{\rF}{}_{\mu }$ and $\ov{W}_{\mu }$ 
are the same. We have that
$$
\ov{W}^{\lambda }{}_{\mu } = \overline{\omega }^{\lambda }{}_{\mu } + 
\delta ^{\lambda }_{\mu } (\widetilde{A}^{\rF}{}_{\nu } \overline{\theta
}^{\nu })\eqno(12.27)
$$
and in the gravitational lagrangian
$$
-{4\over 3} \falkag^{[\mu \nu ]} \ov{W}_{[\mu ,\nu ]} = 
\falkag^{[\mu \nu ]} \widetilde{H}^{\rF}_{\mu \nu }\eqno(12.28)
$$
Due to the spontaneous symmetry breaking the field
$\widetilde{A}^{\rF}{}_{\mu }$ is massive and
has short distance. Thus the Lorentz force term which appears for
$W_{[\mu ,\nu ]}$ is of short distance
$$ 
{2\over 3} f \widetilde{g}^{(\mu \nu )} W_{[\mu ,\alpha ]} u^{\alpha } 
= - 2 f \widetilde{g}^{(\mu \nu )} \widetilde{H}^{\rF}{}_{\mu \alpha } 
u^{\alpha },\eqno(12.29)
$$
(see Ref.~[95]). In this way the weak equivalence principle is
satisfied. This is really impossible to do in pure N.G.T., where the
Lorentz force term (or Coriolis force term) appears as a term of long
distance forces and it is necessary to remove it using the so called
splitting of the conservation laws (see Refs.~[34], [47], [65]). We do
not need these additional assumptions, which are not of geometric
nature. Moreover, the splitting of the conservation laws is an
integral part of Moffat's theory. It seems that it provides a
contradiction with the field equations with electromagnetic or scalar
sources in N.G.T. (see Ref.~[34]). It seems that it is also in
contradiction with the Lagrange formulation of hydrodynamics. However
the splitting of the conservation laws is necessary in N.G.T. in order
to get the equation of motion for the test particle from
conservation laws (Bianchi identities) as in G.R.T. In this manner
Moffat fits the anomalous perihelion precession of Mercury and Icarus
in the presence of a non-zero quadrupole moment of the sun

Otherwise, he gets an equation with a Lorentz force term
(Coriolis force term) for $W_{[\mu,\nu]}$.

In some new formulations of N.G.T. (Ref.~[95]) J.W.~Moffat
abandons the splitting of the conservation laws getting equation of
motion for a test particle with Lorentz-like (Coriolis force term).

The additional force term in the equation of motion has been
connected to the ``fifth force" (Refs.~[54], [53]). However, it seems
that the fifth force term (if this force really exists) should be of
exponential decaying, not ${1\over r^{5}}$ as in Moffat's approach ([95]). In our
meaning the ``fifth force" is rather scalar than vector-like.

The conditions (12.23), (12.24), (12.25) are really constraints in
the theory. Thus we should add to the full lagrangian the term:
$$
{8\pi G_{N}\over c^{4}}\falkaL^{\mu} \biggl(\widetilde{A}^{F}{}_{\mu} + {2\over 3}
\ov{W}\biggr),\eqno(12.30)
$$
where
$$ 
\falkalam^\mu= \sqrt{-g}\lambda ^{\mu}\eqno(12.31)
$$
is a Lagrange multiplier and because of the gauge invariance $({\sym R}_{+}$ and
${\bold U}(1)_{F})$ we suppose
$$
\partial _{\mu}\falkalam^{\mu} = 0.\eqno(12.32)
$$
The Lagrange multiplier becomes a source for $g_{[\mu\nu]}$
$$
\falkag^{[\mu\nu]}{}_{,\nu} = - \kappa \falkalam^{\mu} = 4\pi a^{2}
\falkaS^{\mu},\quad\ \kappa = {8\pi G_{N}\over c^{4}}.\eqno(12.33)
$$
\looseness4Thus it has a natural physical interpretation as a 
fermion current. It
is coupled to the $\widetilde{A}^{F}_{\mu}$ 
and simultaneously to 
$\ov W_{\mu}$. In this way the
fermion charge plays the role of the second gravitational charge.
The field $\ov W_{\mu}$ disappears from the
theory. In the place of $\ov W_{\mu}$ 
we have $\widetilde{A}^{F}_{\mu}$ which is massive and due to
this the weak equivalence principle is satisfied. The Lorentz force
term
(Coriolis force term) for $\widetilde{A}^{F}_{\mu}$ is of short range. From the
Nonsymmetric Gravitational Theory we know that there are not any
Lorentz forces and Coulomb potentials connected to $g_{[\mu\nu]}$.
Thus the weak equivalence principle is satisfied
and we have two gravitational charges: a mass and a fermion charge.

In this way the argument raised by C. Will (see $4^{\text{\rm th}}$ of Ref.~[47]) does
not concern our case. The arguments is of course valid in the case of
N.G.T., because there $\ov{W}$ is not of a short range.

Let us notice that we choose the second possibility of the role of
${\bold U}(1)_{F}$ in our scheme from Ref.~[26] and sec.~11. This corresponds to
the situation similar for SO(10) G.U.T., i.e. the first possibility
from Ref.~[26] and sec.~11. It seems that this possibility is more
natural. Some possibilities concerning the intermediate stages of the
symmetry breaking will be considered elsewhere.

Finally we can say that the $SU(5)$ scheme of G.U.T. seems to be in
contradiction with experiment (proton decay). For this kind of scheme
we cannot proceed with the construction presented here. ${\bold U}(1)_{F}$ is
global and $Y_{F} \in $ (Lie algebra of the group $H)$, and $Y_{F} \not\in $.

Let us conclude that in this section we have obtained the physical
interpretation of the Einstein-$\lambda $-transformation (Ref.~[3]). It seems
that this transformation is a local ${\bold U}(1)_{\rF}$ gauge transformation
corresponding to the fermion-number generator in G.U.T. In this way
the problem seems to be solved. The Einstein Unified Field Theory in a
real version was used as a theory of the pure gravitational field. It
was necessary to combine it with the Yang--Mills' and Higgs' field via
the Nonsymmetric Jordan--Thiry Theory.

Let us consider the lagrangian of the theory and find his part
which contains the gauge field $\widetilde{A}^{\rF}{}_{\mu } = - 
{2\over 3}\ov{W}_{\mu }$. In order to do this we
consider Eq.~(6.1) with the $\widetilde{H}^{\rF}{}_{\mu \nu }$ contribution. One gets:
$$
\widetilde{Q}^{j}_{i} g_{\delta \beta } g^{\sigma \delta } 
\widetilde{L}^{i}{}_{\sigma \alpha } + g_{\alpha \delta } 
g^{\sigma \delta } \widetilde{L}^{j}{}_{\beta \sigma } = 
2g_{\alpha \delta } g^{\delta \sigma } \widetilde{H}^{j}{}_{\beta
\sigma },\eqno(12.34)
$$
where
$$
\widetilde{Q}^{j}_{i} = \widetilde{\ell }^{kj} \widetilde{\ell
}_{ik}.
$$
If $j=F$ we have
$$
\widetilde{Q}^{\rF}{}_{i} g_{\delta \beta } g^{\sigma \delta } 
\widetilde{L}^{i}{}_{\sigma \alpha } + g_{\alpha \delta } g^{\delta
\sigma } \widetilde{L}^{\rF}{}_{\beta \sigma } = 2g_{\alpha \delta } 
g^{\delta \sigma } \widetilde{H}^{\rF}{}_{\beta \sigma }.\eqno(12.35)
$$
For
$$ 
\widetilde{H}^{\rF}{}_{\mu \nu } = - {4\over 3} W_{[\mu ,\nu ]}
\eqno(12.36)
$$
one gets
$$
Q^{F}{}_{i} g_{\delta \beta } g^{\sigma \delta }
\widetilde{L}^{i}{}_{\sigma \alpha } + g_{\alpha \delta } g^{\delta
\sigma } \widetilde{L}^{\rF}{}_{\beta \delta } = - {8\over 3} 
g_{\alpha \delta } g^{\delta \sigma } W_{[\beta ,\sigma ]}.\eqno(12.37)
$$
This means that the field $W_{\mu }$ has a contribution to the full induction
tensor of the theory.

The Yang--Mills' field lagrangian with a direct contribution of
$W_{\mu }$ looks as follows:
$$
{\cal L}_{\rYM}(W) = - {1\over 6\pi } \widetilde{\ell }_{\rF\rF}
\biggl[ {8\over 3} ( g^{[\mu \nu ]}\ov{W}_{[\mu ,\nu ]})^{2} +
\widetilde{L}^{\rF\mu \nu }\ov{W}_{[\mu ,\nu ]}\biggr].\eqno(12.38)
$$
There is also a usual part of N.G.T. lagrangian containing 
$\ov{W}_{\mu }$ field i.e.
$$
{2\over 3} g^{[\mu \nu ]}\ov{W}_{[\mu ,\nu ]}.\eqno(12.39)
$$
We have also a contribution of the field $\ov{W}_{\mu }$ in the kinetic 
part of the
lagrangian for a Higgs' field if the gauge derivative of the Higgs'
field contains $\widetilde{A}^{\rF}{}_{\mu } = - {2\over
3}\ov{W}_{\mu }$ i.e.
$$\eqalignno{
e^{*}(\mathop{\nabla_{\mu}}\limits^{\gauge}{\varPhi} ^{a}_{\tilde a }) ={}& 
\partial _{\mu }{\varPhi} ^{a}_{\tilde a } + {g\over \hbar c} ( C^{a}_{de} 
\alpha ^{e}_{j} \widetilde{A}^{j}_{\mu } {\varPhi} ^{d}_{\tilde a } + 
f^{\tilde b }_{\tilde a j} \widetilde{A}^{j}_{\mu } {\varPhi}
^{a}_{\tilde b })=\cr
\noalign{\vskip4pt}
={}& \partial _{\mu }{\varPhi} ^{a}_{\tilde a} - {3g\over 2\hbar c} 
(C^{a}_{de}\alpha ^{e}_{\rF}\ov{W}_{\mu }{\varPhi} ^{a}_{\tilde a }+
f^{\tilde c }_{\tilde a \rF}\ov{W}_{\mu }{\varPhi} ^{a}_{\tilde c }) +\cr
\noalign{\vskip4pt}
&+ {g\over \hbar c} \sum^{}_{j\neq \rF}(C^{a}_{de}\alpha ^{e}_{j}
\widetilde{A}^{j}_{\mu }{\varPhi} ^{a}_{\tilde a }+f^{\tilde c
}_{\tilde a j}\widetilde{A}^{j}_{\mu }{\varPhi} ^{a}_{\tilde c })=\cr
={}& \mathop{\nabla_{\mu }^w}\limits^{\gauge}{\varPhi} ^{a}_{\tilde
a } + {g\over \hbar c} \sum^{}_{j\neq \rF}(C^{a}_{de}\alpha ^{e}_{j}
\widetilde{A}^{j}_{\mu }{\varPhi} ^{a}_{\tilde a }+f^{\tilde c
}_{\tilde a j}\widetilde{A}^{j}_{\mu }{\varPhi} ^{a}_{\tilde c
}).&(12.40)\cr}
$$
Thus one gets:
$$
{\cal L}^{W}_{\kin} (\mathop{\nabla_{\mu }}\limits^{\gauge}
{\varPhi}) = (\mathop{\nabla_{\mu }^w}\limits^{\gauge}{\varPhi}
^{a}_{\tilde a } L^{\mu \tilde a }{}_{a})_{av}.\eqno(12.41)
$$
Taking everything together we have:
$$\eqalignno{
{\cal L}(\ov{W}) ={}& {2\over 3} g^{[\mu \nu ]}\ov{W}_{[\mu ,\nu ]} - 
e^{-(n+2){\varPsi} } {\lambda ^{2}\over 3} \widetilde{\ell }_{\rF\rF} 
\biggl[ {8\over 3} ( g^{[\mu \nu ]}\ov{W}_{[\mu ,\nu ]})^{2} 
+ \widetilde{L}^{\rF\mu \nu }\ov{W}_{[\mu ,\nu]}\biggr] +\cr
& -{\lambda ^{2}\over 2r^{2}} e^{-2{\varPsi} }(\mathop{\nabla_{\mu }^w}\limits^{\gauge}
{\varPhi} ^{a}_{\tilde a } L^{\mu \tilde a }{}_{a})_{av}.&(12.42)\cr}
$$
However because of the Eq.~(6.1) and Eq.~(6.4) we have the 
$\ov{W}_{\mu }$ field
contribution via an induction tensor $\widetilde{L}^{i}{}_{\mu \nu }$ and 
$L^{\mu \tilde a }{}_{a}$ in the remaining
parts of ${\cal L}_{\rYM}$ and ${\cal L}^{\kin} (\mathop{\nabla_{\mu }}\limits^{\gauge}
{\varPhi})$. Finally let us rewrite the field
equations {\it modulo}\/ constraint $\widetilde{A}^{\rF}{}_{\mu } = 
- {2\over 3} \ov{W}_{\mu }$. 
$$
\eqalignno{
\ov{R}_{(\alpha \beta )}(\overline{\varGamma}) &= {8\pi K\over c^{4}} 
\biggl(\mathop{T}\limits^{\tot}{\!\!}_{(\alpha \beta )} - {1\over 2} g_{(\alpha \beta )}
T\biggr),&(12.43)\cr
\ov{R}_{[[\alpha \beta ],\lambda ]}(\overline{\varGamma}) &= {8\pi \over c^{4}}
\biggl( K \mathop{T}\limits^{\tot}{\!\!}_{[\alpha \beta ]} - {1\over 2} g_{[[\alpha
\beta ]} T \biggr){\!}_{,\lambda ]},&(12.44)\cr}
$$
where
$$
\Te^{\tot}{\!\!}_{\alpha \beta } = \Te^{\gauge}{\!\!}_{\alpha \beta } + 
\Te^{\scal}{\!\!}_{\alpha \beta } + T_{\alpha \beta }({\varPhi} ) + 
\Te^{\Int}{\!\!}_{\alpha \beta } + {\varLambda} g_{\alpha \beta
}\vadjust{\vskip-4pt}
\eqno(12.45)
$$
and
$$\displaylines{
\noalign{\vskip-4pt}
T = g^{\alpha \beta } \Te^{\tot}{\!\!}_{\alpha \beta },\cr
\hfill \falkag^{[\mu \nu ]}{}_{,\nu } = 4\pi \falkaS^{\mu
},\hfill\llap{(12.46)}\cr
\hfill g_{\mu \nu ,\sigma } - g_{\rho\nu
}\widetilde{{\varLambda}}^{\rho}{}_{\mu \sigma }- g_{\mu \rho}
\widetilde{{\varLambda}}^{\rho}{}_{\sigma \nu }=
0,\hfill\llap{(12.47)}\cr}
$$
where
$$\displaylines{\quad\
\falkaS^{\mu } = {3\lambda ^{2}\alpha ^{2}_{S}\over c\hbar
}\sqrt{-g} {\delta \over \delta \ov{W}_{\mu }} \biggl(
e^{-(n+2){\varPsi}} {\cal L}_{\rYM} + {e^{-2{\varPsi} }\over 4\pi r^{2}} 
{\cal L}^{\kin} (\mathop{\nabla}\limits^{\gauge} {\varPhi}) +\hfill\cr
\hfill- {e^{(n-2){\varPsi} }\over 2\pi r^{4}} {\cal L}_{\Int}({\varPhi} ,
\widetilde{A})\biggr),\quad\ (12.48)\cr
\hfill \widetilde{\overline{\varLambda}}_{\mu \sigma } =
\widetilde{{\varGamma}}^{\rho}{}_{\mu \sigma } + D^{\rho}{}_{\mu
\sigma }(S)\hfill\llap{(12.49)}\cr}
$$
and
$$
g_{\rho\nu } D^{\rho}{}_{\mu \sigma } + g_{\mu \rho}
D^{\rho}{}_{\sigma \nu } = {4\pi \over 3} S^{\rho} (g_{\mu \sigma }
g_{\rho\nu }-g_{\mu \rho}g_{\sigma \nu }+g_{\mu \nu }g_{[\sigma
\rho]}),\eqno(12.50)
$$
$\widetilde{\ov{R}}_{\alpha \beta }(\widetilde{\overline{\varGamma}})$ is a
Moffat--Ricci tensor for a connection $\widetilde{\overline{\omega }}^{\alpha
}{}_{\beta } = \widetilde{\overline{\varGamma}}^{\alpha }{}_{\beta \gamma } 
\theta ^{\gamma }$. It is
easy to see that we have sources for equation (9.4) and (9.5). This is
similar to the situation with the second pair of Maxwell equations in
the Nonsymmetric Kaluza--Klein Theory in comparison to the classical
Kaluza--Klein Theory.

The above equations have been obtained as results of variations
with respect to $g_{\mu \nu }$ and
$$
\ov{W}^{\lambda }{}_{\mu } = \widetilde{\overline{\omega }}^{\lambda }{}_{\mu } - 
{2\over 3} \delta ^{\lambda }_{\mu } \ov{W}= \widetilde{\overline{\omega} }^{\lambda
}{}_{\mu } + \delta ^{\lambda }_{\mu } \widetilde{A}^{\rF}.
$$
Moreover if we vary with respect to
$$
\widetilde{\omega }_{\rE} = \widetilde{\omega }^{\rF}_{\rE} X_{\rF} + 
\sum^{}_{i\neq \rF}\widetilde{\omega }^{i}_{\rE} X_{i}
$$
the variation with respect to the first part of $\omega $ is included in the
variation of $\ov{W}^{\lambda }{}_{\mu }$ because of
$$
e^{*}\widetilde{\omega }^{\rF}_{\rE} = \widetilde{A}^{\rF} = - {2\over 3}
\ov{W},\eqno(12.51)
$$
Thus we get Eq.~(9.7) for $j\neq F$ only. 

We can easily calculate the form of $S^{\mu }$. One gets
$$\eqalignno{
\falkaS^{\alpha } ={}& {-8G_{N}\hbar \over
c^{4}}\biggl\{\mathop{\nabla_{\mu }}\limits^{\gauge} (\widetilde{\ell
}_{i\rF}\widetilde{\falkaL}^{i\alpha \mu } ) + 2\falkag^{[\alpha
\beta ]} \mathop{\nabla_{\beta }}\limits^{\gauge} (\widetilde{h}_{i\rF}
g^{[\mu \nu ]}\widetilde{H}^{i}{}_{\mu \nu }) +\phantom{\}}\cr
&+ {2\sqrt{-g}\over r^{2}} \alpha _{S}\sqrt{\hbar c} e^{n{\varPsi} } 
\biggl[ \ell _{ab} g^{\tilde b \tilde n } g^{\mu \alpha } L^{a}{}_{\mu
\tilde b }({\varPhi} ^{d}_{\tilde n }C^{b}{}_{dc}\alpha ^{c}{}_{\rF}+
f^{\tilde e }_{n\tilde F}{\varPhi} ^{b}{}_{\tilde e } +\cr
\bez{\noalign{\eject}}
&+ \biggl( {\delta L^{a}_{\beta \tilde b }\over \delta
\mathop{\nabla_{\alpha }}\limits^{\gauge} {\varPhi} ^{w}_{\tilde
v}}\biggr) \ell _{ab} g^{\tilde b \tilde n }g^{\beta \mu } (\mathop{\nabla_{\mu }}\limits^{\gauge}
{\varPhi} ^{b}_{\tilde n }) ({\varPhi} ^{d}_{\tilde w}C^{w}_{dc}
\alpha ^{c}_{\rF}+f^{\tilde e }_{\tilde w \rF}{\varPhi} ^{w}_{\tilde
e })\biggr]_{av}+\cr
&+ 4\sqrt{-g}{e^{2n{\varPsi} }\over r^{2}} h_{ab} \mu ^{a}_{k} 
\widetilde{\ell }_{i\rF} \widetilde{\ell }^{ki}
\widetilde{g}^{[\tilde a \tilde b ]} \mathop{\nabla_{\mu }}\limits^{\gauge}
\biggl\{ g^{[\mu \alpha ]} \times \phantom{\}}\cr
&\times \phantom{\{} \biggl[{\alpha _{s}\over \sqrt{\hbar c}}C^{b}_{cd}
{\varPhi} ^{c}_{\tilde a } {\varPhi} ^{d}_{\tilde b }-{\hbar c\over \alpha _{s}}
\mu ^{b}_{\dah{i}}f^{\dah{i}}_{\tilde a \tilde b }-\phi ^{b}_{\tilde d }
f^{\tilde d }_{\tilde a \tilde b }\biggr]\biggr\} +\cr
&+\phantom{\{} (n+2) \partial _{\beta }{\varPsi} [\widetilde{\ell }_{i\rF}
\widetilde{\falkaL}^{i\beta \alpha }-2\falkag^{[\beta \alpha ]}
(\widetilde{h}_{i\rF}g^{[\mu \nu ]}\widetilde{H}^{i}{}_{\mu \nu
})]\biggr\}.&(12.52)\cr}
$$
The remaining equations i.e.\ Eqs.~(9.6), (9.8) are the same
substituting for $\widetilde{H}^{\rF}{}_{\mu \nu }$, $- {4\over
3}\ov{W}_{[\mu ,\nu ]}$ and for $\widetilde{A}^{\rF}{}_{\mu }$, $- {2\over 3}
\ov{W}_{\mu }$.

\vfill\eject

\null
\vskip36pt plus4pt minus4pt
\centerline{{\four\bf 13. Equation of Motion for a Test Particle }}

\vskip4pt
\centerline{{\four\bf in the Nonsymmetric Jordan--Thiry Theory}}
\vskip24pt plus2pt minus2pt
\write0{13. Equation of Motion for a Test Particle 
 in the Nonsymmetric Jordan--Thiry Theory \the\pageno}

Let us derive the equation of motion for a test particle in our
theory from the Bianchi identities in the hydrodynamic limit.

In N.G.T. one derives the following Bianchi identity (see Refs.\
[5], [97], [98])
$$
{1\over 2} ( G_{\rho\nu }\falkag^{\alpha \nu })_{,\alpha } + {1\over 2} 
( G_{\nu \rho}\falkag^{\nu \alpha })_{,\alpha } + {1\over 2}
\falkaG_{\varepsilon\nu } g^{\varepsilon\nu }{}_{,\rho} = 0,\eqno(13.1)
$$
where
$$\eqalignno{
G_{\rho\nu } &= \ov{R}_{\rho\nu }(\overline{\varGamma}) - {1\over 2} 
g_{\rho\nu } \ov{R}(\overline{\varGamma}),&(13.2)\cr
\falkaG_{\varepsilon\nu } &= \sqrt{-g} G_{\varepsilon\nu
},&(13.3)\cr}
$$
$\ov{R}_{\rho\nu }(\overline{\varGamma})$ is the Moffat--Ricci tensor for 
the connection $\overline{\omega }^{\alpha }{}_{\beta }$ and
$R(\overline{\varGamma})$ is
the scalar of curvature.

From Eq.~(9.3) one gets:
$$
R_{\alpha \rho}(\ov{W}) - {1\over 2} g_{\alpha \rho } R(\ov{W}) = 
{8\pi G\over c^{4}} e^{-(n+2){\varPsi} } \Te^{\tot}{\!\!}_{\alpha\beta},\eqno(13.4)
$$
where
$$
\Te^{\tot}{\!\!}_{\alpha\rho } = \Te^{\gauge}{\!\!}_{\alpha\rho } + T_{\alpha
\rho }({\varPhi} ) + \Te^{\scal}{\!\!}_{\alpha\rho } + g_{\alpha \rho } 
{\varLambda} + \Te^{\Int}{\!\!}_{\alpha\rho }.\eqno(13.5)
$$
If we suppose the constraints (12.30) we have to use an effective
energy-momentum tensor
$$
\Te^{\eff}{\!\!}_{\alpha\rho } = \Te^{\tot}{\!\!}_{\alpha\rho } + 
\Te^{\text{\rm constraints}}{\hskip-15pt}_{\alpha\rho },\eqno(13.6)
$$
where
$$
\Te^{\text{\rm constraints}}{\hskip-15pt}_{\alpha\rho } = e^{(n+2){\varPsi} } \biggl[
\lambda _{\alpha } \biggl(\ov{W}_{\rho } + {2\over 3}
\widetilde{A}^{\rF}{}_{\rho }\biggr) - {1\over 2} g_{\alpha \rho } 
\lambda ^{\mu } \biggl(\ov{W}_{\mu } + {2\over 3}
\widetilde{A}^{\rF}{}_{\mu }\biggr)\biggr].\eqno(13.7)
$$
Now Eq.~(13.4) takes the form:
$$
R_{\alpha \rho }(\ov{W}) - {1\over 2} g_{\alpha \rho } R(\ov{W}) = 
{8\pi G\over c^{4}} e^{-(n+2){\varPsi} } \Te^{\eff}{\!\!}_{\alpha\rho }.
\eqno(13.8)
$$
Simultaneously for Eq. (9.4) we have
$$
\falkag{}^{[\mu \nu ]}_{,\nu } = {3\over 2} \falkalam^{\mu } = 
4\pi a^{2} \falkaS^{\mu }.\eqno(13.9)
$$
From Eq.~(13.1) one derives:
$$
[e^{-(n+2){\varPsi}} (g^{\alpha \nu } \mathop{\falkaT}\limits^{\eff}{\!\!}_{\nu \rho }+
g^{\nu \alpha }\mathop{\falkaT}\limits^{\eff}{\!\!}_{\nu \rho })]_{,\alpha } + g^{\mu
\nu }{}_{,\rho } e^{-(n+2){\varPsi} } \Te^{\eff}{\!\!}_{\mu \nu }
+ {2\over 3} W_{[\rho ,\nu ]}\falkaS^{\nu } = 0.\quad\
\eqno(13.10)
$$
Let us suppose that:

1) The field configuration described by Eqs.~(9.3)--(9.8) and
constraints (12.30) correspond to a situation close to the ``true"
vacuum case.

2) The hydrodynamic limit for this field configuration.

From 1) we have (see Eq.~(10.12)):
$$\displaylines{
\hfill e^{-(n+2){\varPsi} } \simeq \const,\hfill\llap{(13.11)}\cr
\hfill {\varLambda} \simeq 0,\hfill\llap{(13.12)}\cr
\hfill - {2\over 3}\ov{W}_{[\mu ,\nu ]} = \widetilde{A}^{\rF}{}_{[\mu
,\nu ]} \simeq {1\over r^{2}} e^{-\beta r} \simeq 
0,\hfill\llap{(13.13)}\cr}
$$
where ${\displaystyle{1\over \beta }}$ is the range of the gauge 
field $\widetilde{A}^{\rF}{}_{\mu }$.

Thus from Eq.~(13.10) one gets:
$$
(g^{\alpha \nu } \mathop{\falkaT}\limits^{\eff}{\!\!}_{\rho \nu }+g^{\nu \alpha } 
\mathop{\falkaT}\limits^{\eff}{\!\!}_{\nu\rho })+g^{\mu \nu }{}_{,\rho } \mathop{\falkaT}\limits^{\eff}{\!\!}_{\mu
\nu } = 0.\eqno(13.14)
$$
2) means that our field configuration is described by the macroscopical
(averaged) hydrodynamic quantities: $u^{\alpha }$ (four-velocity of a fluid),
$p$ (pressure), $\rho$ (density), $T$ (temperature) and $F_{\mu
\nu }$ (macroscopic
electromagnetic field) which is the only gauge field with long range.
$$
\Te^{\eff}{\!\!}_{\rho \nu } = \Te^{\eff}{\!\!}_{\rho \nu }(u^{\alpha },p,\rho ,T,
F_{\mu \beta }).\eqno(13.15)
$$
Thus Eq.~(13.14) would be treated as an energy-momentum conservation
law in N.G.T. for phenomenological sources.

Let us derive the equation for a test particle from Eq.~(13.14).
In order to do this let us go to the limit of one uncharged particle.
This means:
$$\displaylines{
\hfill p = 0 ,\quad\ F_{\mu \nu } = 0,\hfill\llap{\hbox{(13.16a)}}\cr
\hfill T_{\rho \nu } = m_{0} g_{\rho \alpha } g_{\beta \nu }
u^{\alpha } u^{\beta },\hfill\llap{\hbox{(13.16b)}}\cr
\hfill du = d(\sqrt{-g}u_{\nu }\overline{\theta }^{\nu }) =
0.\hfill\llap{\hbox{(13.16c)}}\cr}
$$
The last equation gives:
$$
\falkau^{\alpha }{}_{,\alpha } = 0,\eqno(13.17)
$$
where $u^{\alpha }$ is a four-velocity of a test particle and $m_{0}$ its mass. Putting
Eqs.\ (13.16b) and (13.17) into (13.14) one gets after some algebra:
$$
{d^{2}x^{\gamma }\over dt^{2}} + {\gamma \brace \alpha \beta }
\biggl({dx^{\alpha }\over dt}\biggr) \biggl({dx^{\beta }\over dt}\biggr) =
0,\eqno(13.18)
$$
where $u^{\alpha } = {\displaystyle{dx^{\alpha }\over dt}}$ and ${\gamma
\brace \alpha \beta }$ are Christoffel symbols formed for the metric
$g_{(\alpha \beta )}$.

Thus we get that a test particle (without spin and electric
charge) moves along a geodetic line for the metric $g_{(\alpha
\beta )}$. The
Eq.~(13.6) fits the anomalous perihelion shift of Mercury and Icarus in
the presence of a nonzero mass quadrupole moment for the sun if we use
the symmetric part of the spherically symmetric, static solution in
N.G.T. (see Ref.~[45]). In this limit we suppose that Mercury and
Icarus can be treated as test particles in the gravitational field of
the sun.

It works also very well in the case of the anomalous periastron
movement of the closed binary system DI Hercules (DI Her i.e.\ HD175227),
which contradicts G.R.T. predictions up to $21\sigma$ (standard deviations)
(see Refs.~[67], [99]). The equation fits the observational data. In
the case of different closed binary systems (AG Per, AS Cam.$\alpha $Vir, V541
Cyg, V1143 Cyg, V889 Aql) with strange movement of the periastron the
equation can also fit the observational data. Using the equation under
above conditions we can also fit the data for $X$-ray bursters
(Ref.~[49]).

Let us notice that the Eq.~(13.15) could be treated as a
fundamental equation in any post-Newtonian approximation for N.G.T.
using the hydrodynamic formalism as in Ref.~[100].

Finally let us consider the Newtonian limit for ${\varPsi} $. Let us expand
$e^{-(n+2){\varPsi} }$ into a power series:
$$
e^{-(n+2){\varPsi} } \cong 1 - (n+2){\varPsi} + \ldots \eqno(13.19)
$$
The field ${\varPsi} $ has a Yukawa-type behaviour i.e.
$$
{\varPsi} \cong C' + {C\over r} e^{-\gamma r},\eqno(13.20)
$$
where $\gamma , C' , C = \const$ and $\gamma $, $C > 0$. One gets:
$$
e^{-(n+2){\varPsi} } \simeq (1+C' ) - {(n+2)C\over r} e^{-\gamma
r}.\eqno(13.21)
$$
We can consider a correction to the Newtonian potential coming from the
scalar field ${\varPsi} $. One has:
$$
V_{\eff }(r) = -G_{\eff } {M\over r} = -(\widetilde{G}e^{-(n+2){\varPsi}
}) {M\over r} = -G_{N} {M\over r} + {\alpha \over r^{2}} e^{-\gamma
r},\eqno(13.22)
$$
where $G_{N} = \widetilde{G}(1+C' )$ plays the role of the Newton constant and
$$
\alpha = \widetilde{G} C(n+2) > 0.\eqno(13.23)
$$
Thus the correction to the Newtonian potential has a repulsive
character. Putting Eq.~(13.18) into linearized equation of motion for a
test particle we can get a composi\-tion-dependent repulsive correction
with a short range. This could be considered as something similar to
the fifth force (Refs.~[53], [54]). We can also try to get an
equation of motion from the conservation laws under conditions (13.17)
and (13.19), i.e.
$$
( g^{\alpha \nu } \mathop{\falkaT}\limits^{\eff}{\!\!}_{\rho \nu } + g^{\nu \alpha }
\mathop{\falkaT}\limits^{\eff}{\!\!}_{\nu
\rho }) + g^{\mu \nu }{}_{,\rho } - (n+2){\varPsi} g^{\mu \nu
}{}_{,\rho } \mathop{\falkaT}\limits^{\eff}{\!\!}_{\mu \nu } = 0.\eqno(13.24)
$$

Recently J.~W.~Moffat and his coworkers have renamed ([10]) N.G.T. into
N.G.E.T. (Nonsymmetric Gravitational and Electromagnetic Theory), taking
the Lagrangian for the electromagnetism from Nonsymmetric
Kaluza--Klein Theory~([18]). They found nonsingular solutions in the
case for pure gravitational field with interesting physical
interpretations ([102], [103]).

\vfill\eject

\null
\vskip36pt plus4pt minus4pt
\centerline{{\four\bf 14. On a Cosmological Origin of the Mass}}

\vskip4pt
\centerline{{\four\bf of the Scalar Field ${\varPsi}$ (or $\rho$).}}

\vskip4pt
\centerline{{\four\bf Cosmological Models with Quintessence and Inflation}}
\vskip24pt plus4pt minus4pt
\write0{14. On a Cosmological Origin of the Mass
 of the Scalar Field ${\varPsi}$ (or $\rho$).
 Cosmological Models with Quintessence and Inflation \the\pageno}

Let us consider the lagrangian in our theory supposing that
Yang--Mills' and Higgs' fields are zero. One gets in terms of the field
${\varPsi}$ (see Eq.~(7.5)):
$$
{\cal L} = \bigg[\ov{R}(\ov{W})+ {\lambda ^{2}\over 4} \bigg(
-e^{(n-2){\varPsi} } {V(0)\over r^{4}} - {\cal L}_{\scal}({\varPsi} )
+{8\over \lambda ^{4}} e^{(n+2){\varPsi} } \widetilde{R}(
\widetilde{{\varGamma} })+{4e^{n{\varPsi} }\over \lambda ^{2}r^{2}}
\widetilde{\un{P}}\bigg)\bigg].\eqno(14.1)
$$
If we suppose that we have to do with a cosmological background i.e.
$\ov{R}(\ov{W}) = R_{0} = \const$ and ${\varPsi} $ is constant we get in terms of 
$\rho = e^{-{\varPsi} }\ (\rho >0)$
$$
{\cal L} = \bigg[R_{0}+{\lambda ^{2}\over 4\rho ^{n+2}}\bigg(-\rho ^{4}
{V(0)\over r^{4}}+{8\over \lambda ^{4}}\widetilde{R}(\widetilde{{\varGamma} })
+4\rho ^{2}{\widetilde{\un P}\over \lambda ^{2}r^{2}}\bigg)\bigg] = W(\rho ).\eqno(14.2)
$$
Eq.~(14.2) is a self-interacting potential for a scalar field $\rho $.  The
potential has a critical point if
$$
{dW\over d\rho } = 0,\eqno(14.3)
$$
i.e.
$$
- (n-1) \bigg({\lambda ^{2}V(0)\over 4r^{4}}\bigg) \rho ^{4} + n 
\bigg({\widetilde{\un P}\over 4r^{2}}\bigg) \rho ^{2} + 4(n+2) 
{\widetilde{R}(\widetilde{{\varGamma} })\over \lambda ^{2}} = 0.\eqno(14.4)
$$
Eq.~(14.4) has real positive solutions if
$$
\widetilde{\un P}{}^{2} \ge - \bigg({16(n-1)(n+2)\over n^{2}}\bigg) \bigg({\widetilde{R}
(\widetilde{{\varGamma} })V(0)\over \lambda ^{2}}\bigg)\eqno(14.5\text{\rm a})
$$
and
$$
{n\lambda ^{2}\widetilde{\un P}\over V(0)} \le {\sqrt{n^2\lambda
^2\widetilde{\un P}{}^2+16(n-1)(n+2)V(0)\widetilde{R}(\widetilde{{\varGamma}})}
\over V(0)}.\eqno(14.5\text{\rm b})
$$
In this case we have:
$$
\rho ^{\pm }_{0,1,2} = \pm {r\over \lambda } 
\sqrt{{\sqrt{n^2\lambda
^2{\widetilde{\un P}}{}^2+16(n-1)(n+2)V(0)\widetilde{R}(\widetilde{{\varGamma}})}\pm
n\lambda \widetilde{\un P}\over V(0)}}\eqno(14.6)
$$
i.e.\ in general two positive real roots. The potential $W$ has a minimum
at $\rho = \rho _{0}$ if
$$
{d^{2}W\over d\rho ^{2}}\bigg|_{\rho =\rho _{0}} > 0\eqno(14.7)
$$
i.e.
$$
\left.\left({-(n-1)(n-2) 
{\displaystyle{\lambda ^{2}V(0)\over 4r^{4}}} \rho ^{4} 
+ {\displaystyle{n(n+1)\un P\over 4r^{2}}} \rho ^{2} + (n+1)(n+2) 
{\displaystyle{4\widetilde{R}(\widetilde{{\varGamma} })\over \lambda^{2}}}
\over \rho ^{n+2}}\right)\right|_{\rho =\rho _{0}} > 0.\eqno(14.8)
$$
All of these conditions could be easily satisfied if 
$\widetilde{R}(\widetilde{{\varGamma} })=0$ and 
${\displaystyle{\widetilde{\un P}\over
V(0)}}>0$. %\break
The condition (14.8) can be satisfied for only 2 roots of Eq.~(14.4).
For $\widetilde{R}(\widetilde{{\varGamma} }) = 0$ Eq.~(14.6) has the shape:
$$
\rho ^{\pm }_{0} = \pm {r\over \lambda } \sqrt{{n\widetilde{\un P}\over
(n-1)V(0)}}.\eqno(14.6')
$$
In terms of a scalar field ${\varPsi} $ one gets:
$$
W(\rho ({\varPsi} )) = W(\rho _{0}) + {dW\over d{\varPsi} }
\bigg|_{{\varPsi} ={\varPsi} _{0}} {\varPsi} ' + {1\over 2} 
{d^{2}W\over d{\varPsi} ^{2}} \bigg|_{{\varPsi} ={\varPsi} _{0}} 
({\varPsi} ' )^{2} +\ldots ,\eqno(14.9)
$$
where
$$
{\varPsi} = -\ln \rho _{0} + {\varPsi} ' = {\varPsi} _{0} + {\varPsi} '.
$$
One finds
$$
W({\varPsi} ) = W(\rho _{0}) + {\rho ^{2}_{0}\over 2} {d^{2}W\over d\rho ^{2}}
\bigg|_{\rho =\rho _{0}} ({\varPsi} ' )^{2} +\ \hbox{higher order terms 
in}\ {\varPsi} '. \eqno(14.10)
$$
For (14.7) is satisfied we can write:
$$
\rho ^{2}_{0} {d^{2}W\over d\rho ^{2}}\bigg|_{\rho =\rho _{0}} = 
{\lambda ^{2}\over 4} m^{2}_{0}\ov{M}\eqno(14.11)
$$
and rewrite Eq.~(14.2) in terms of ${\varPsi} ' $. One easily gets:
$$
{\cal L} = R_{0} + {\lambda ^{2}\over 4} \bigg({\varLambda} _{0} + 
{m^{2}_{0}\over 2}\ov{M} ({\varPsi} ' )^{2} +\ \hbox{higher order
terms}\bigg),\eqno(14.12)
$$
where
$$
{\lambda ^{2}\over 4} {\varLambda} _{0} = W(\rho _{0}) - R_{0},
$$
is a new cosmological constant.

Thus one can write Eq.~(14.1) in terms of ${\varPsi} ' $ using Eq.~(14.12).
$$
{\cal L} = \ov{R}(\ov{W}) + {\lambda ^{2}\over 4} \bigg(-{\cal L}_{\scal}
({\varPsi} ' ) + \ov{M} {m^{2}_{0}\over 2} ({\varPsi} '
)^{2}
+\ \hbox{higher order terms in}\ {\varPsi} ' +
{\varLambda}\bigg).\eqno(14.13)
$$
Thus the cosmological background can induce a mass for the scalar field
${\varPsi} (\rho )$ via usual mechanism known in solid state physics if (14.5a),
(14.5b), (14.7) are satisfied. In this way a mass $m_{0}$ for ${\varPsi} ' $ has a
cosmological origin and the field ${\varPsi} $ or $\rho $ has a  short  range
(Yukawa-type) behaviour. Moreover we should redefine a field ${\varPsi} ' $.

The scalar field ${\varPsi} (\rho )$ is connected to the effective gravitational
constant and the mechanism presented in this section gives us a
cosmological reason for a slow change of the gravitational constant.
This mechanism gives also some explanation for the origin of the
possible ``fifth force" similar in spirit to the Fujii theory (see
Ref.~[96]). It is easy to see that if we choose $\rho _{0}>0$ the reflection
symmetry of $W(\rho )$ is spontaneously broken.

Let us notice that the value $R_{0}$ does not enter to the formulae for
$\rho _{0}$ and $m_{0}$. Thus we can abandon in principle a constancy of $R_{0}$ and
consider a slow changing in time of $\ov{R}$ i.e.
$$
\ov{R}(t) = R_{0} - 24\pi G_{N} d_{0} H_{0}(t-t_{0}),\eqno(14.14)
$$
where $R_{0}$ is the scalar curvature at the present epoch time
$t_{0}$, $d_{0}$ is a
present rest mass galaxies density and $H_{0}$ is the Hubble constant.

Finally let us calculate the mass of the scalar field ${\varPsi} ' $ and the
cosmological constant ${\varLambda} _{0}$.
$$
m_{0} = {4\over \lambda } \sqrt{
{-(n-2)(n-1){\displaystyle{\lambda ^2V(0)\over 4r^4}} \rho _0^4+
{\displaystyle{n(n+1)\widetilde{\un P}\over 4r^4}}\rho _0^2 +
(n+2)(n+3){\displaystyle{4\widetilde{R}(\widetilde{{\varGamma}})\over
\lambda ^2}}\over \ov{M}\rho _0^n}}\eqno(14.15)
$$
and
$$
{\varLambda} _{0} = {\rho ^{4}_{0} {\displaystyle{V(0)\over r^{4}}} + 
\rho ^{2}_{0} {\displaystyle{\widetilde{\un P}\over 4r^{2}\lambda ^{2}}} + 
{\displaystyle{4\widetilde{R}(\widetilde{{\varGamma} })\over \lambda^{2}}}
\over \rho ^{n+2}_{0}},\eqno(14.16)
$$
where $\rho _{0}$ is given by Eq.~(14.6). In the case 
$\widetilde{R}(\widetilde{{\varGamma} }) = 0$ one easily gets
$$
m_{0} = \bigg({\lambda \over r}^{{n\over 2}}\bigg)
{\sqrt{\widetilde{\un P}(n-2n-2)}\over 4\sqrt{|\ov{M}|}} 
\bigg({n\widetilde{\un P}\over (1-n)V(0)}\bigg)^{-{n\over 4}}\eqno(14.17)
$$
and a range
$$\displaylines{
\hfill r_{0} = {\hbar\over m_{0}c},\hfill\llap{\hbox{(14.17a)}}\cr
\hfill {\varLambda} _{0} = (\pm 1)^{n} {\lambda ^{n+2}\widetilde{\un P}\over (1-n)r^{n-2}} 
\bigg({n\widetilde{\un P}\over (n-1)V(0)}\bigg)^{-{n\over
2}}.\hfill\llap{\hbox{(14.18)}}\cr}
$$
We can calculate $V(0)$ using Eq.~(6.18). One gets:
$$\displaylines{
V(0) = {\ell _{ab}\over V_{2}}\mmint_{M}\sqrt{|\widetilde{g}|} d^{n_{1}}x 
( 2g^{[\tilde m \tilde n ]} g^{[\tilde b \tilde n ]} \mu
^{a}{}_{\dah{i}} \mu ^{b}{}_{\dah{j}} f^{\dah{i}}{}_{\tilde m \tilde n }
f^{\dah{j}}{}_{\tilde a \tilde b } 
- g^{\tilde a \tilde m} g^{\tilde b \tilde n } \mu ^{b}_{\dah{i}} 
f^{\dah{i}}_{\tilde m \tilde n } L^{a}{}_{\tilde a \tilde b
}(0)),\hfill(14.19)\cr}
$$
where
$$
\ell _{dc} g_{\tilde m \tilde b } g^{\tilde c \tilde m }
L^{d}{}_{\tilde c \tilde a }(0) + \ell _{cd} g_{\tilde a \tilde m } 
g^{\tilde m \tilde c } L^{d}{}_{\tilde b \tilde c }(0) = -2\ell _{cd} 
g_{\tilde a \tilde m } g^{\tilde m \tilde c } \mu ^{d}_{\dah{i}}
f^{\dah{i}}{}_{\tilde b \tilde c }.\eqno(14.20)
$$
In the symmetric case one easily finds:
$$
V(0) = -{1\over V_{2}}\mmint_{M}\sqrt{|\widetilde{g}|} d^{n_{1}}x
g^{\tilde a \tilde m } g^{\tilde b \tilde n } f^{\dah{i}}{}_{\tilde m
\tilde n } f^{\dah{j}}{}_{\tilde a \tilde b } h_{ab} \mu
^{a}_{\dah{j}} \mu
^{b}{}_{\dah{j}}.\eqno(14.21)
$$
Let us consider the effective gravitational constant $G_{\eff}$ close to the
minimum of $W(\rho )$. One gets:
$$
G_{\eff} = G_{0} e^{-(n+2){\varPsi} } = G_{N} e^{-(n+2)/\sqrt{|M|}
{\varPsi}'} \cong G_{N} \bigg(1-{\alpha \over
r}e^{-(r/r_0)}\bigg),\eqno(14.22)
$$
where
$$
G_{N} = G_{0} \rho ^{n+2}_{0}\eqno(14.23)
$$
and
$$
0 < \alpha = {n+2\over\sqrt{|M|}} \gamma,\eqno(14.24)
$$
$(\gamma $ is a positive constant).

{\def\eq#1 {\eqno(\text{\rm14.#1})}
\def\rf#1 {\text{(14.#1)}}
\let\al\aligned
\let\eal\endaligned
\let\ealn\eqalignno
\def\cL{\Cal L}
\def\gH{\frak h}
\def\gG{\frak g}
\def\gM{\frak m}
\def\hi{{\hat\imath}}
\def\hj{{\hat\jmath}}
\let\o\overline
\let\ul\underline
\let\a\alpha
\let\b\beta
\let\d\delta
\let\D\varDelta
\let\e\varepsilon
\let\f\varphi
\let\g\gamma
\let\G\varGamma
\let\k\kappa
\let\la\lambda
\let\La\varLambda
\let\om\omega
\let\O\varOmega
\let\mPi\varPi
\let\P\varPsi
\let\pa\partial
\let\t\widetilde
\let\u\tilde
\def\Sh{\operatorname{sinh}}
\def\Ch{\operatorname{cosh}}
\def\ctgh{\operatorname{ctgh}}
\def\tgh{\operatorname{tgh}}
\def\arctg{\operatorname{arctg}}
\def\ar{\operatorname{arccos}}
\def\ars{\operatorname{arsinh}}
\def\ns{\operatorname{ns}}
\def\sc{\operatorname{sc}}
\def\sh{\operatorname{sinh}}
\def\sn{\operatorname{sn}}
\def\cn{\operatorname{cn}}
\def\dn{\operatorname{dn}}
\def\sd{\operatorname{sd}}
\def\cd{\operatorname{cd}}
\def\nd{\operatorname{nd}}
\def\tov#1{\widetilde{\overline#1}}
\def\gag{\overset\text{\rm gauge}\to\nabla}
\def\br#1#2{\overset\text{\rm #1}\to{#2}}
\def\ddt{\frac d{dt}}
\def\({\left(}\def\){\right)}
\def\[{\left[}\def\]{\right]}
\def\kl{\left\{}\def\kr{\right\}}
\def\h#1#2{\,{\vphantom F}_#1F_#2}
\def\dr#1{_{\text{\rm#1}}}
\def\ur#1{^{\text{\rm#1}}}
\def\GR{General Relativity}
\def\cf{configuration}
\def\ct{constant}
\def\co{cosmological}
\def\dt{dependent}
\def\ef{effective}
\def\ev{evolution}
\def\ex{exponential}
\def\il{inflation}
\def\lg{lagrangian}
\def\gr{gravitational}
\def\pt{perturbation}
\def\q{quintessence}
\def\U{Universe}
\def\uP{\ul{\t P}}
\def\Mp{M\dr{pl}}

According to new observational data [104] concerning distances of type Ia
supernovae it seems that we need some kind of ``dark energy'' which drives
the evolution of the Universe. This ``dark energy'' can be considered as a
\co\ constant or more general as \co\ terms in field equations (in the
lagrangian). In the case of \co\ model this type of ``dark
energy''---``vacuum energy'' is a cause to accelerate the evolution of the
Universe (a~scale factor $R(t)$) (see Ref.~[105]). The \co\ constant is
negligible on the level of the Solar System and on the level of the Galaxy.
Moreover it can be important if we consider even nonrelativistic movement of
galaxies in a cluster of galaxies (see Ref.~[106]). In some papers considered
\co\ terms result in changing with time of a \co\ constant (see Ref.~[107]).
Some of them introduce additional scalar field (or fields) in order to give a
field-theoretical description of such an evolution of a \co\ ``constant''.
Those scalar fields are independent in general of the additional scalar
fields in inflationary models.

Thus the inflation field (or fields in multicomponent inflation, which can be\break
the same as some of Higgs' fields from G.U.T.-models) can be different from
those\break fields. In particular considering scalar-tensor theories of gravitation
results in so called quintessence models (see Ref.~[108]). Moreover in such
theories there is a natural field-theoretical background for an inconstant
``gravitational constant''. In such a way this quintessence field can be used
in twofold ways. First as a source of change in space and time of a
gravitational constant. Secondly as a source of \co\ terms leading to the
model of quintessence and a change in time of a \co\ ``constant''. In our
theory we have a natural occurrence of these phenomena due to the scalar
field~$\varPsi$ (or~$\rho$). Let us consider the \lg\ of our theory paying a
special attention to the part involving the scalar field~$\varPsi$, i.e.:
$$
\al
L&=\o R(\o W)+8\pi G_N\biggl(e^{-(n+2)\varPsi}\cL\dr{YM}(\t A)+
\frac{e^{-2\varPsi}}{4\pi r^2}\cL\dr{kin}(\gag\varPsi)\\
&-\frac{e^{(n-2)\varPsi}}{8\pi r^2}V(\varPhi)-\frac{e^{(n-2)\varPsi}}{2\pi r^4}\cL
\dr{int}(\varPhi,\t A)\biggr)\\
&-8\pi G_N\cL\dr{scal}(\varPsi)+e^{n\varPsi}\(\frac{e^{2\varPsi}\a^2_s}{l^2\dr{pl}}
\t R(\t \G)+\frac{\ul{\t P}}{r^2}\),
\eal \eq25
$$
where all the terms are defined by Eqs (6.16--22).
$$
\cL\dr{scal}(\varPsi)=\(\o M\t g^{(\g \nu )}+n^2g^{[\mu \nu ]}g_{\d\mu}
\t g^{(\d\g)}\)\varPsi_{,\nu }\varPsi_{,\g} 
$$
and $\o M$ is defined by Eq.~(6.27a). We put $c=\hbar=1$.

Now we rewrite the \lg~(14.25) in the following form.
$$
\al
L&=\o R(\o W)-8\pi G_N e^{-(n+2)\varPsi}L\dr{matter}\\
&-8\pi G_N\cL\dr{scal}(\varPsi)+e^{n\varPsi}\(\frac{e^{2\varPsi}}{l^2\dr{pl}}
\a^2_s\t R(\t \G)+\frac{\ul{\t P}}{r^2}\),
\eal \eq25a
$$
where in $L\dr{matter}$ we include all the terms from Eq.~(14.25) with
Yang-Mills' fields, Higgs' fields, their interactions and coupling to the
scalar field~$\varPsi$.

The effective \gr\ constant is defined by
$$
G\dr{eff}=G_N e^{-(n+2)\varPsi}
$$
in such a way that the \lg\ of the Yang-Mills field in $L\dr{matter}$ is
without any factor involving scalar field~$\varPsi$. If the scalar field~$\varPsi$
is constant (e.g. $\varPsi=0$) we can redefine all the fields in such a way that
we get ordinary (standard) \lg s for these fields.

Let us consider the situation after a spontaneous symmetry breaking and
simplify to the case of $g_{[\mu \nu ]}=0$. We get
$$
L=\t{\o R} - 8\pi G\dr{eff}L\dr{matter} -8\pi G_NL\dr{scal}(\varPsi)
+8\pi G_N U(\varPsi) \eq26
$$
where 
$$
L\dr{scal}=\o M g^{\g \nu }\varPsi_{,\nu }\cdot \varPsi_{,\g},\quad \o M>0 \eq27
$$
$$
8\pi G_N U(\varPsi)=-\frac{V(\varPhi^K\dr{crt})}{r^4}l^2\dr{pl}e^{(n-2)\varPsi}
+e^{(n+2)\varPsi}\(\frac{\a^2_s \t R(\t\G)}{l^2\dr{pl}}\)+
\frac{\ul{\t P}}{r^2}e^{n\varPsi}. \eq28
$$

We get the following equations
$$
\t{\o R}_{\mu \nu }-\frac12 \t{\o R}g_{\mu \nu }=8\pi G\dr{eff}
\br{matter}T_{\kern-9pt \mu \nu }+8\pi G_N\br{scal}T_{\kern-5pt\mu \nu }, \eq29
$$
where $\t{\o R}_{\mu \nu }$ and $\t{\o R}$ are a Ricci tensor and a scalar
curvature for a Riemannian geometry generated by $g_{\mu \nu }=g_{(\mu \nu )}$,
$$
\br{matter}T^{\mu\nu}=(p+\rho)u^\mu u^\nu - pg^{\mu \nu } \eq30
$$
is an energy-momentum tensor for a matter considered as a radiation plus a
dust. 
$$
8\pi G_N\br{scal}T_{\kern-5pt\mu \nu }=8\pi G_N \(\frac{\o M}2 g_{\mu\nu} \cdot(g^{\a\b}
\varPsi_{,\a}\cdot\varPsi_{,\b})-\o M\varPsi_{,\mu}\cdot
\varPsi_{,\nu }\)+g_{\mu \nu }\o\la_{cK}
\eq31
$$
where
$$
\o\la_{cK}=2e^{(n-2)\varPsi}\frac{m^4_{\t A}}{\a^4_s} l^2\dr{pl}V(\varPhi^K\dr{crt})
-e^{(n+2)\varPsi}\frac{\a^2_s}{2l^2\dr{pl}}\t R(\t\G) - e^{n\varPsi}\frac{m^2_{\t A}}
{2\a^2_s}\ul{\t P}, \eq32
$$
where $m_{\t A}$ is a scale of a mass for massive Yang-Mills' fields (after a
spontaneous symmetry breaking for a true vacuum case). It is convenient to
write 
$$
\o\la_{cK}=\frac{e^{(n-2)\varPsi}}2 \a_K - \frac{e^{n\varPsi}}2 \g -
\frac{e^{(n+2)\varPsi}}2 \b \eq33
$$
$$
\al
\a_K&=\a_K(\xi,\zeta,m_{\t A},\a_s)\\
\b&=\b(\xi,\a_s,m_{\t A})\\
\g&=\g(\zeta,m_{\t A},\a_s)
\eal \eq33a
$$
$K=0,1$ corresponds to true and false vacuum case, i.e.\ 
$$
V(\varPhi^0\dr{crt})=0,\ V(\varPhi^1\dr{crt})\ne 0, \eq34
$$
$$
\a_0=0,\ \a_1\ne0. \eq34a
$$
For a scalar field $\varPsi$ we have the following equation:
$$
\al
16\pi G_N \o M g^{\a\b}\(\tov\nabla_\a(\pa_\b\varPsi)\)&+(n-2)e^{(n-2)\varPsi}
\a_K-(n+2)e^{(n+2)\varPsi}\b\\
&- n e^{n\varPsi}\g - (n+2)G\dr{eff}T=0
\eal
\eq35
$$
where $T=\rho-3p$ is a trace of an energy-momentum tensor for a matter field.
For we are interested in \co\ models we take for a metric tensor a
Robertson-Walker metric:
$$
ds^2=dt^2 - R^2(t)\left[ \frac{dr^2}{1-kr^2} + r^2d\theta^2 + r^2\sin^2\theta
d\f^2\right], \quad k=-1,0,1 \eq36
$$
and we suppose that $\varPsi,\rho,p$ are functions of $t$ only.

One gets
$$
\al
\frac1{M^2\dr{pl}}\ddot \varPsi&=\frac3{M^2\dr{pl}} H\dot \varPsi
-\(\frac{n-2}{\o M}\)\a_Ke^{(n-2)\varPsi}+
\frac{(n+2)}{\o M}\b e^{(n+2)\varPsi}\\
&+\frac n{\o M}e^{n\varPsi}\g+\frac{(n+2)}{M^2\dr{pl}}e^{-(n+2)\varPsi}\cdot
(\rho-3p)
\eal
\eq37
$$
$$
M\dr{pl}=\(\sqrt{8\pi G_N}\)^{-1}=\frac{m\dr{pl}}{\sqrt{8\pi}}\,.
$$

In this case we get standard equations for a \co\ model adapted to our theory
$$
\frac{3\ddot R}R = -8\pi G\dr{eff}\(\frac12(\rho+3p)\)
-8\pi G_N\(\frac12(\rho_\varPsi+3p_\varPsi)\), \eq38
$$
$$
R\ddot R+2\dot R+2k= 8\pi G\dr{eff} \(\frac12(\rho-p)\)R^2+
8\pi G_N\(\frac12(\rho_\varPsi-p_\varPsi)\)R^2. \eq39
$$

Using Eqs (14.38--39) one easily gets
$$
H^2+k = \frac{8\pi G\dr{eff}}3 \rho + \frac{8\pi G_N}3 \rho_\varPsi \eq40
$$
where $H=\dfrac{\dot R}R$ is a Hubble constant
$$
\al
8\pi G_N\rho_\varPsi& =8\pi G_N \frac{\o M}2 {\dot \varPsi}^2+
\frac12\o\la_{cK}\\
8\pi G_N p_\varPsi& =8\pi G_N \frac{\o M}2 {\dot \varPsi}^2-
\frac12\o\la_{cK}.
\eal
\eq41
$$

Let us consider a \co\ model for a ``false vacuum'' case, i.e.\ without matter
and only with a scalar $\varPsi$ and a vacuum energy $V(\varPhi^1\dr{crt})\ne0$. In
this case $\rho=p=0$ and we get
$$
\eqalignno{
&2\o M\ddot \varPsi - \frac{6\o M\dot R}R \dot\varPsi - M^2\dr{pl}\frac{d\la_{c1}}{d\varPsi}=0
&(14.42)\cr
&{\dot R}^2+k = \frac13 \(\frac12\,\frac{\o M}{M^2\dr{pl}}{\dot\varPsi}^2
+\frac12\la_{c1}\)R^2 &(14.43)\cr
\noalign{\eject}
&\frac{3\ddot R}R = -\frac{\o M}{M^2\dr{pl}} {\dot \varPsi}^2+\frac12\la_{c1}.
&(14.44)
}
$$
Let us take
$$
\varPsi=\varPsi_1=\text{const.} \eq45
$$
Thus we get
$$
\frac{d\la_{c1}}{d\varPsi}(\varPsi_1)=0 \eq46
$$
and
$$
\eqalignno{
&(n+2)\b x^4_1 + n\g x^2_1 - (n-2)\a_1=0 &(14.47)\cr
&{\dot R}^2+k = \frac16 \la_{c1}(x_1)R^2 &(14.48)\cr
&\frac{3\ddot R}R = \frac12 \la_{c1}(x_1) &(14.49)
}
$$
$$
x_1=e^{\varPsi_1}. \eqno(*)
$$

One gets from Eq.\ (14.49)
$$
\o R(t)=R_0 e^{H_0 t} \eq50
$$
where
$$
H_0=\sqrt{\frac{\la_{c1}(x_1)}6} \eq51
$$
is Hubble constant (really constant). From Eq.~(14.47) we obtain
$$
x_1=\sqrt{\frac{-n\g+\sqrt{n^2\g^2+4(n^2-4)\a_1\b}}{2(n+2)\b}}. \eq52
$$
For $\a_1>0$, $\b>0$ we get
$$
\sqrt{n^2\g^2+4(n^2-4)\a_1\b} > n|\g| \eq53
$$
$$
\la_{c1}(x_1)=\frac{x_1^{n-2}}{2(n+2)^2\b}
\left[n\g^2 - \g\sqrt{n^2\g^2
+4(n^2-4)\a_1\b}+4\a_1\b(n+2)\right]
\eq54
$$
$$
H_0=\frac{x_1^{(n-2)/2}}{2(n+2)\sqrt{3\b}}
\left[n\g^2 - \g\sqrt{n^2\g^2
+4(n^2-4)\a_1\b}+4\a_1\b(n+2)\right]^{1/2}.
\eq55
$$
Thus we get an exponential expansion of the Universe. Using Eq.~(14.48) we
get also $k=0$ (i.e.\ a flatness of a space).

In this way we get de Sitter model of the Universe. This is of course a very
special solution to the Eqs~(14.42--44) with very special initial conditions
$$
\left.
\al
\o R(0)&=R_0\\
\frac{d\o R}{dt}(0)&=H_0R_0\\
\varPsi(0)&=\varPsi_1\\
\frac{d\varPsi}{dt}(0)&=0
\eal
\right\} \eq56
$$
Let us disturb the solution by a small perturbation and examine its
stability. Let
$$
\al
&\varPsi=\varPsi_1+\f,\quad |\f|,|\dot\f|\ll \varPsi_1\\
&R=\o R+\d R,\quad |\d R|,|\d \dot R|\ll \o R.
\eal \eq57
$$
One gets in a linear approximation
$$
\eqalignno{
&2\o M\ddot\f + 6\o MH_0\dot \f+M^2\dr{pl} \frac{d^2\la_{c1}}{d\varPsi^2}
(\varPsi_1)\f=0
&(14.58)\cr
&\frac{3\d\ddot R}{\o R}=-\o M\dot\f^2 &(14.59)\cr
&2\dot{\o R}\d\dot R+k= \frac16\o M\dot\f^2({\o R}^2+\o R\d R).&(14.60)
}
$$
One gets
$$
\f=\f_0e^{-\frac32H_0t}\sin\(\frac{\sqrt p}2t + \d\) \eq61
$$
where (we take for simplicity $M=\dfrac{\o M}{M^2\dr{pl}}$)
$$
p=-\D=\frac1{2M}\,\frac{d^2\la_{c1}}{d\varPsi^2}(\varPsi_1)-9H_0^2>0
\eq62
$$
if 
$$
\al
&M<M_0=\frac43\,\\
&\times\frac
{n(n+2)\g\sqrt{n^2\g^2+4(n^2-4)\a_1\b}-4(n+2)^2(n-1)\a_1\b-n^2(n+2)\g^2}
{n\g^2+4(n+2)\a_1\b-\g\sqrt{n^2\g^2+4(n^2-4)\a_1\b}}\,.
\eal
\eq63
$$

Thus we get an exponential decay of the solution $\f$, i.e.\ a damped 
oscillation around $\varPsi=\varPsi_1$. It means the solution $\varPsi=\varPsi_1$ is
stable against small \pt s of initial conditions for~$\varPsi$, i.e.\ 
$$
\eqalignno{
\varPsi(0)&=\varPsi_1+\f_0\sin\d &(14.64)\cr
\frac{d\varPsi}{dt}(0)&=\f_0\(-\tfrac32 H_0\sin\d+\cos\d\).& (14.65)
}
$$
Making $\f_0$ and $\d$ sufficiently small we can achieve smallness of \pt s
of initial conditions. The exponential decay of the solution to Eq.~(14.58)
can be satisfied also for some different conditions than Eq.~(14.63) if we
consider aperiodic case. However, we do not discuss it here.

Thus we get
$$
0\le \frac12 \o M\dot\f^2 \le a^2e^{-3H_0t} \eq66
$$
where $a^2$ is a constant.

From Eq.~(14.59) one gets
$$
0\le \d \ddot R\le a^2e^{-4H_ot}, \eq67
$$
which means that $\d\ddot R\sim0$ and $\d \dot R=O(\d R(0))$. It means 
$\d R=\d R(0)+\d \dot R(0)t$.
Moreover from Eq.~(14.60) we get that $k=0$ due to the fact that $\d R$ is a
linear function of~$t$. In this way we get
$$
R(t)\cong \o R(t)+\d R(0)+\d \dot R(0)t=
\(R_0+\frac{\d R(0)+\d \dot R(0)t}{e^{H_0t}}\)e^{H_0t}. \eq68
$$

This means that the small \pt s for initial conditions for~$R$ result in a
\pt\ of~$R_0$ in such a way that $R_0$ is perturbed by a quickly decaying
function. Thus our de Sitter solution is stable under small \pt s and is an
attractor for any small perturbed initial data. One can think that this
evolution will continue forever. However, we should remember that we are in a
``false'' vacuum regime and this configuration of Higgs' fields is unstable.
There is a stable configuration---a ``true'' vacuum case for which $\a_K=0$
($K=0$). For $\a=0$ the considered de Sitter evolution cannot be continued.
Thus we should take under consideration a second order phase transition 
in the configuration
of Higgs' fields from metastable state to stable state, from ``false'' vacuum
to ``true'' vacuum, from $\varPhi^1\dr{crt}$ to $\varPhi^0\dr{crt}$. In this case
the Higgs fields play a r\^ole of an order parameter. Let us suppose that the
scale time of this phase transition is small in comparison to $\dfrac1{H_0}$
and let it take place locally. We suppose locally a conservation of a density
of an energy. Thus
$$
\br{scal}T_{\kern-4pt 44}=\br{scal}T_{\kern-4pt 44}+\br{matter}T_{\kern-9pt 44}. \eq69
$$
We suppose also that the scalar field $\varPsi$ will be close to the new
equilibrium (new minimum for \co\ terms) and that the matter will consist of
a radiation only:
$$
\ealn{
\br{scal}T_{\kern-4pt 44}&=\frac12 \la_{c1}(x_1) &\text{(14.70a)}\cr
\br{scal}T_{\kern-4pt 44}+\br{matter}T_{\kern-9pt 44}&=\frac12 \la_{c0}(x_0)+8\pi G_N
\frac{\rho_r}{x_0^{n+2}}. &\text{(14.70b)}
}
$$

For $x_0=e^{\varPsi_0}$ is a new equilibrium point we have:
$$
\ealn{
&\frac{d\la_{c0}}{d\varPsi}(\varPsi_0)=0 &(14.71)\cr
&\la_{c0}=-e^{(n+2)\varPsi}\b - e^{n\varPsi}\g &(14.72)\cr
&\frac{d\la_0}{d\varPsi}=-e^{n\varPsi}\((n+2)\b e^{2\varPsi}+n\g\). &(14.73)
}
$$

One gets
$$
e^{\varPsi_0}=x_0=\sqrt{\frac{n|\g|}{(n+2)\b}} \eq74
$$
and supposing 
$$
\g<0 \eq75
$$
thus
$$
\la_{c0}(\varPsi_0)=\frac{2|\g|^{\frac n2+1}n^{\frac n2}}{\b^{\frac n2}
(n+2)^{\frac n2+1}}\,. \eq76
$$
From Eqs (14.70ab) one gets
$$
\al
\rho_r&=\frac1{16\pi G_N} \biggl\{
\frac{x_1^{n-2}}{(n+2)^2\b} \Bigl[n\g^2 - \g\sqrt{n^2\g^2+4(n^2-4)\a_1\b}
+2\a_1\b(n+4)\Bigr]\\
&-\frac{2|\g|^{\frac n2+1}}{\b^{\frac n2}(n+2)^{\frac n2+1}}\biggr\}
\frac{n^{\frac{n+2}2}\g^{\frac{n+2}2}}{(n+2)^{\frac{n+2}2}\b^{\frac{n+2}2}}\,.
\eal
\eq77
$$
This gives us matching condition for a second order phase transition and
simultaneously it is an initial condition for a new epoch of an evolution of
the Universe plus a condition $\dot\varPsi(t_r)=0$, $R(t_r)=\o R(t_r)$, where
$t_r$ is a time for a phase transition to occur. Notice we have simply
$\varPsi_0\ne \varPsi_1$. Thus we have to do with a discontinuity for a
field~$\varPsi$. Let us consider Hubble's constants for both phases of the
Universe 
$$
\al
H_0^2&=\frac{\la_{c1}(x_1)}6 \\
H_1^2&=\frac{\la_{c0}(x_0)}6 =\frac{|\g|^{\frac n2+1}n^{\frac n2}}{3\b^{\frac n2}
(n+2)^{\frac n2+1}}\\
H_0^2&\ne H_1^2
\eal
\eq79
$$
Summing up we get
$$
\al
R(t_r)&=\o R(t_r)=R_0 e^{+H_0t_r}\\
\dot \varPsi_1(t_r)&=\dot \varPsi_0(t_r)=0 \\
\varPsi_1&\ne \varPsi_0\\
H_0^2&\ne H_1^2
\eal
\eq80
$$

Thus we see that the second order phase transition in the \cf\ of Higgs'
fields results in the first order phase for an evolution of the Universe. We
get discontinuity for Hubble constants and values of scalar field before and
after phase transition.

Let us calculate deceleration parameters before and
after phase transition
$$
\ealn{
&q=-\frac{R\ddot R}{{\dot R}^2} &(14.81)\cr
&q=-1 &(14.82)
}
$$
before phase transition and
$$
q=-1 \eq83
$$
after phase transition. 

Let us come back to the Eqs (14.38), (14.40), (14.37). Let us rewrite
Eq.~(14.37) in the following form
$$
\o M\ddot \varPsi+\frac{6\o M\dot R}R \dot\varPsi + M^2\dr{pl}\frac{d\la_{c0}}
{d\varPsi}=0. \eq84
$$
(Remember we now have to do with a radiation for which the trace of an
energy-momentum tensor is zero.)

Let us suppose that $\dot\varPsi\approx0$. Thus Eq.~(14.84) simplifies 
$$
\frac1{M^2\dr{pl}} M\ddot \varPsi + \frac{d\la_{c0}}{d\varPsi}=0 \eq85
$$
and we get the first integral of motion
$$
\ealn{
& \frac{\o M}{2M^2\dr{pl}}\dot\varPsi^2+\la_{c0}=
\frac12\o \d =\text{const.} &(14.86)\cr
&\o M\dot\varPsi^2=M^2\dr{pl}(\o\d-2\la_{c0}).&\text{(14.86a)}
}
$$
One gets from Eqs (14.38) and (14.40)
$$
\ealn{
&\frac{3\ddot R}R = -\frac1{M^2\dr{pl}}\cdot \frac{\rho_r}{e^{(n+2)
\varPsi}}-\frac12\(\frac{2\o M}{M^2\dr{pl}} \dot\varPsi^2-\la_{c0}\) &(14.87)\cr
&\dot R^2+k=\frac1{3M^2\dr{pl}}\cdot\frac{\rho_r}{e^{(n+2)\varPsi}}R^2
+\frac13\(\frac{\o M}{2M^2\dr{pl}}\dot\varPsi^2+\frac12 \la_{c0}\)R^2.&(14.88)
}
$$
Using Eq (14.86a) we get:
$$
\ealn{
&\frac{3\ddot R}R = -\frac1{M^2\dr{pl}}\cdot \frac{\rho_r}{e^{(n+2)
\varPsi}}-\o\d+\frac52\la_{c0} &(14.89)\cr
&\dot R^2+k=\frac1{3M^2\dr{pl}}\cdot\frac{\rho_r}{e^{(n+2)\varPsi}}R^2
+\frac16(\o\d- \la_{c0})R^2.&(14.90)
}
$$
One can derive from Eqs (14.89--90)
$$
\ealn{
\frac d{dt}(R\dot R)&=-\frac16\o\d R^2+\frac23 \la_{c0}R^2-k &(14.91)\cr
\la_{c0}&=-e^{n\varPsi}(\b e^{2\varPsi}+\g). &(14.92)
}
$$

Let us take $k=0$ and $\d=0$ and let us change independent and dependent
variables in (14.91) using (14.92) and (14.86a). One gets
$$
2y^2(y^2-1)\frac{d^2f}{dy^2}+y\((n+4)y^2-(n+1)y-2\)\frac{df}{dy}
=\frac{4\o M}3 (y^2-1)f(y) \eq93
$$
where
$$
y=\sqrt{\frac\b{|\g|}}e^\varPsi \eq94
$$
and
$$
f=R^2. \eq95
$$
It is easy to see that $y^2>1$ for
$$
\frac{d\varPsi}{dt}=\pm \frac{M\dr{pl}}{\sqrt{\o M}}\cdot\frac{|\g|^{\frac{n+2}4}}
{\b^\frac n4}\sqrt{y^n(y^2-1)}. \eq96
$$
Moreover according to our assumptions $\dot\varPsi\approx0$ and this can be
achieved only if $0<y-1<\e$, where $\e$ is sufficiently small.

Thus for further investigations we take $z=y-1$, $y=z+1$. One gets
$$
\al
2z(z+2)(z+1)^2\frac{d^2f}{dz^2}&+(z+1)\((n+4)z^2+(n+7)z
-(n+3)\)\frac{df}{dz}\\
&-\frac{4\o M}3 z(z+2)f(z)=0. 
\eal
\eq97
$$
Let us consider Eqs (14.89--90) supposing $k=0$ and $\o\d=0$.

After eliminating $\ddot R$ (via differentiation of Eq.~(14.90) with respect
to time $t$) and changing independent and dependent variables one gets:
$$
\frac d{dy}\biggl[\(8\pi G_N\frac{\t\rho_r}{y^{n+2}} + \frac12 y^n(y^2-1)\)f^2
\biggr]=-\frac d{dy}(f^2)y^n(y^2-1) \eq98
$$
where
$$
\o\rho_r=\rho_r \frac{|\g|^n}{\b^{n+1}}\,. \eq99
$$
It is convenient for further investigations to consider a parameter
$$
\o r=\frac{4\o M}3\,. \eq100
$$
In such a way we get
$$
2y^2(y^2-1)\frac{d^2f}{dy^2}+y\((n+4)y^2-(n+1)y-2\)\frac{df}{dy}
- \o r(y^2-1)f(y) \eq101
$$
and
$$
\al
2z(z+2)(z+1)^2 \frac{d^2f}{dz^2}&+(z+1)\((n+4)z^2+(n+7)z-(n+3)\)
\frac{df}{dz}\\
&-\o rz(z+2)f(z)=0. 
\eal
\eq102
$$

One can transform Eq.~(14.98) into
$$
\frac d{dy}\left[\(
8\pi G_N \frac{\t\rho_r}{y^{n+2}}+\frac32 y^n(y^2-1)\)f^2\right]
=y^{n-1}\left[(n+2)y^2-n\right]f^2. \eq103
$$
Changing the independent variable from $y$ to $z=y-1$ one gets:
$$
\al
\frac d{dz}&\left[\(8\pi G_N\frac{\t\rho_r}{(z+1)^{n+2}}+
\frac32 (z+1)^nz(z+2)\)f^2\right]\\
&\qquad{}=(z+1)^{n-1}\left[(n+2)(z+1)^2-n\right]f^2. 
\eal
\eq104
$$
Let us come back to Eq.~(14.96) in order to find time-dependence of~$\varPsi$.
One gets
$$
\int\frac{d\varPsi}{\sqrt{\b e^{(n+2)\varPsi}-|\g|e^{n\varPsi}}}
=\pm \frac{t-t_1}{\sqrt{\o M}}\,M\dr{pl} \eq105
$$
or
$$
\int\frac{dx}{x\sqrt{\b x^{n+2}-|\g|x^n}}
=\pm \frac{t-t_1}{\sqrt{\o M}}\,M\dr{pl}. \eq105a
$$
If $n=2l$, where $l$ is a natural number, one gets for $\b>0$
$$
\al
\(\frac{|\g|}\b\)^{-\frac12}
&\Biggl\{ \sum_{k=1}^l\, \sum_{p=0}^k \frac{\binom kp(-2)^{k-p}}{2p}
\(\sqrt{\frac{\sqrt\b x-\sqrt{|\g|}}{\sqrt\b x+\sqrt{|\g|}}}\)^{2p+1}\\
&+\frac1{2\sqrt3} 
\ln\(\frac{\sqrt{\sqrt\b x-|\g|}-\sqrt{3(\sqrt\b x+|\g|)}}
{\sqrt{\sqrt\b x-|\g|}+\sqrt{3(\sqrt\b x+|\g|)}}\)\Biggr\}=
\pm \frac{(t-t_1)}{\sqrt{\o M}}\,M\dr{pl}. 
\eal
\eq106
$$

In this case it can be expressed in terms of elementary functions. For a
general case of $n$, $\b$ and~$\g$ one gets
$$
\frac{2x^{-\frac n2}\h12(\frac12,-\frac n4,(1-\frac n4);\frac{x^2\b}\g)}
{\sqrt\g n}=\pm\frac{(t-t_1)}{\sqrt{\o M}}\,M\dr{pl} \eq107
$$
where $\h12(a,b,c;z)$ is a hypergeometric function. For a small~$z$ (around
zero) one finds the following solution to Eq.~(14.102)
$$
f(z)=C_1z^\frac{(n+7)}4 \(1-\frac{(n+7)(5n+23)}{8(n+1)}z\)+
C_2\(1+\frac{\o r}{(n-1)}z^2\), \eq108
$$
$C_1,C_2=\text{const.}$ Thus
$$
\al
R(y)&=\Biggl[C_1(y-1)^\frac{(n+7)}4 \(1-\frac{(n+7)(5n+23)}{8(n+1)}(y-1)\)\\
&\qquad{}+C_2\(1+\frac{\o r}{(n-1)}(y-1)^2\)\Biggr]^{\frac12} 
\eal
\eq109
$$
where $y>1$ (but only a little).

Let us make some simplifications of the formulae taking under consideration
that $y-1$ is very small.

One gets
$$
\ealn{
f(y)&=C\(1+\frac{\o r}{(n-1)}(y-1)^2\), \quad C>0& (14.110)\cr
R(y)&=R_1\(1+\frac{\o r}{(n-1)}(y-1)^2\)^\frac12. &(14.111)
}
$$
After some calculations one gets
$$
\al
8\pi G_N\t\rho_r&=y^{2(n+1)}\biggl(-\frac{2(n+5)}{(n+4)(n-1)}y^4
+\frac{10\o r(n+2)}{(n-1)(n+4)}y^3\\
&+\frac{\(8\o r-(n-1)(n+4)\)}{2(n-1)(n+4)}y^2
-\frac{2\o r(n+3)}{(n^2-1)}y+ \frac{(\o r-n+1)}{(n-1)}\biggr)
\eal
\eq112
$$
(taking into account that $(y-1)$ is very small). Let us make some
simplifications in the formula (14.96) for $y$ close to~1. We find:
$$
\int \frac{dy}{\sqrt{y-1}}=\pm \frac{|\g|^\frac{n+4}4 \sqrt2}{\b^\frac n4
\sqrt{\o M}} (t-t_1)M\dr{pl} \eq113
$$
and finally
$$
y=\frac{|\g|^\frac{n+4}2}{2\o M \b^\frac n2}(t-t_1)^2+1 \eq114
$$
such that $y=1$ for $t=t_1$.

Let us pass to the minimum value for $\t\rho_r$ with respect to $y$ close
to~1. Thus we are looking for a minimum of a function
$$
\al
&\t\rho=\frac1{8\pi G_N} y^{2(n+1)}\biggl[
-\frac{2(n+5)}{(n+4)(n-1)}y^4
+\frac{10\o r(n+2)}{(n-1)(n+4)}y^3\\
&+\frac{\(8\o r-(n-1)(n+4)\)}{2(n-1)(n+4)}y^2
-\frac{2\o r(n+3)}{(n^2-1)}y+ \frac{(\o r-n+1)}{(n-1)}\biggr]
=\frac1{8\pi G_N}y^{2(n+1)}W_4(y)
\eal
\eq115
$$
for such a $y$ that is close to~1.

Let us write $y=1+\eta$, where $\eta$ is very small and develop $W_4(1+\eta)$
up to the second order in~$\eta$. One finds
$$
W_4(1+\eta)\simeq V_2(\eta) \eq116
$$
and
$$
V_2(\eta)=a\eta^2+b\eta+c \eq117
$$
where
$$
\ealn{
a&=\frac{\(-n^2+3n(20\o r-9)+4(32\o r-31)\)}{2(n+4)(n-1)} &(14.118)\cr
b&=\frac{\(-n^3-4n^2(7\o r-3)+5n(18\o r-11)+22(3\o r-2)\)}{(n+4)(n^2-1)} &(14.119)\cr
c&=\frac{\(-3n^3+2n^2(9\o r-8)+n(50\o r-29)+4(7\o r-4)\)}{2(n+4)(n^2-1)} &(14.120)
}
$$

Let us find minimum of $V_2(\eta)$ for $\eta>0$ such that $V_2(\eta)\ge0$.
For $a<0$, $b<0$, $c>0$ we get
$$
V_2(\eta\dr{min})=0 \eq121
$$
and
$$
\eta\dr{min}=\frac{b+\sqrt{b^2+4|a|c}}{2|a|}\,. \eq122
$$
In this case
$$
(1+\eta\dr{min})^{2(n+1)} W_4(1+\eta\dr{min})>0 \eq123
$$
and will be close to the real minimum of the function (14.115).
Simultaneously 
$$
\eta\dr{min}<1.
$$
For
$$
\t\rho_r=\frac{|\g|^n}{\b^{n+1}}\sigma T^4 \eq124
$$
(where $\sigma$ is a Stefan-Boltzmann constant) we get $T\dr{min}$ and we
call it a~$T_d$ (a decoupling temperature-decoupling of matter and
radiation). Thus $y_d=1+\eta\dr{min}$. $y_d$ is reached at a time $t_d$
(a~decoupling time)
$$
t_d=t_1 + \frac{\eta\dr{min}\o M\b^\frac n2}{2|\g|^{(\frac{n+4}2)}}\,M\dr{pl}. \eq125
$$

From that time ($t_d$) the evolution of the Universe will be driven by a
matter (with good approximation a dust matter) and the scalar field~$\varPsi$.
The radius of the Universe at $t=t_d$ is equal to
$$
R(t_d)=R_1\(1+\frac{\o r}{(n-1)}\eta^2\dr{min}\)^\frac12. \eq126
$$
The full field equations (generalized Friedmann equations) are as follows:
$$
\ealn{
\(\frac{\dot R}R\)^2 &= \frac{8\pi G\dr{eff}}3\rho_m+ \frac13\(\frac12
\frac{\o M}{M^2\dr{pl}}{\dot\varPsi}^2+\frac12 \la_{c0}\) &(14.127)\cr
\frac{3\ddot R}R &= \frac{8\pi G\dr{eff}}3\rho_m-\frac12\(2\frac{\o M}
{M^2\dr{pl}}{\dot\varPsi}^2-\la_{c0}\) &(14.128)\cr
-\frac{2\o M}{M^2\dr{pl}}\ddot \varPsi &-\frac{6\o M}{M^2\dr{pl}}
\cdot \frac{\dot R}R \dot\varPsi - \frac{d\la_{c0}}{d\varPsi}+(n+2)8\pi \rho_m
G\dr{eff}=0. &(14.129)
}
$$

Before we pass to those equations we answer the question what is a mass of
the scalar field~$\varPsi$ during the de Sitter phases and the radiation era. One
gets:
$$
\ealn{
-m_1^2&=\frac{M^2\dr{pl}}{2\o M}\frac{d^2\la_{c1}}{d\varPsi^2}(\varPsi_1)
&(14.130)\cr
-m_0^2&=\frac{M^2\dr{pl}}{2\o M}\frac{d^2\la_{c0}}{d\varPsi^2}(\varPsi_0).
&(14.131)
}
$$
The second question we answer is an evolution of the Universe during a period
from $t_r$ to $t_1$, i.e.\  when $y$ is changing from  $y=\sqrt{\frac n{n+2}}$
to $y=1$.

This will be simply the de Sitter evolution
$$
R(t)=R e^{H_1t} \eq132
$$
where
$$
H_1=\sqrt{\frac{\la_{c0}(x_0)}6}\,. \eq133
$$
This will be unstable evolution and will end at $t=t_1$, i.e.\ for
$\dot\varPsi\simeq0$ and value of the field~$\varPsi$ will be changed to the value
corresponding to $y=1$, i.e.\ 
$$
\varPsi(t_1)=\frac12 \ln\frac\b{|\g|}\,. \eq134
$$
Thus we have to do with two de Sitter phases of the evolution with two
different Hubble constants $H_1$ and~$H_0$. In the third phase (a radiation
era) the radius of the Universe is given by the formula
$$
R(t)=R_0e^{H_0t_r+H_1(t_1-t_r)} \cdot \(1+
\frac{|\g|^{n+4}}{3\o M(n-1)\b^n}(t-t_1)^4\)^\frac12
\eq135
$$
and such an evolution ends for $t=t_d$,
$$
R(t_d)=R_0\exp(H_0t_r+H_1(t_1-t_r)) \cdot \(1+\frac{\o r}{(n-1)}\eta\dr{min}^2\)
^\frac12. \eq136
$$
In both de Sitter phases the equation of state for the matter described by
the scalar field~$\varPsi$ is the same:
$$
\rho_\varPsi=-p_\varPsi. \eq137
$$
In the radiation era we get
$$
p_\varPsi=\tfrac43 \rho_\varPsi. \eq138
$$

Now we go to the matter dominated Universe and we change the notation
introducing a scalar field~$Q$ such that
$$
Q=\sqrt{\o M} \frac\varPsi{M\dr{pl}}\,. \eq139
$$
In this case we have
$$
\ealn{
\rho_Q&=\frac12{\dot Q}^2+U(Q) &\text{(14.140a)}\cr
p_Q&=\frac12{\dot Q}^2-U(Q) &\text{(14.140b)}
}
$$
where
$$
\ealn{
U(Q)&=\frac12\la_{c0}\(\frac {QM\dr{pl}}{\sqrt{\o M}}\)=-\frac\b2 e^{aQ}+
\frac12|\g|e^{bQ} &(14.141)\cr
&a=\frac{(n+2)}{\sqrt{\o M}}\,M\dr{pl}, \ b=\frac n{\sqrt{\o M}}\,M\dr{pl}. &(14.142)
}
$$

Let us write Eqs (14.127--129) in terms of $Q$:
$$
\ddot Q+3H\dot Q+U'(Q)-8\pi(n+2)G\dr{eff}\rho_m=0 \eq143
$$
where
$$
\ealn{
G\dr{eff}&=G_N e^{-\frac{(n+2)}{\sqrt{\o M}}M\dr{pl}Q} &(14.144)\cr
H^2&=\frac13\(8\pi G\dr{eff}\rho_m+\frac12{\dot Q}^2+U\)&(14.145)\cr
\frac{3\ddot R}R&= \frac12\(8\pi G\dr{eff}\rho_m+2({\dot Q}^2-U)\).&(14.146)
}
$$

Let us define an equation of state
$$
W_Q=\frac{p_Q}{\rho_Q}=\frac{\frac12{\dot Q}^2-U}{\frac12{\dot Q}^2+U}
\eq147
$$
and a variable
$$
x_Q=\frac{1+W_Q}{1-W_Q}=\frac{\frac12{\dot Q}^2}U, \eq148
$$
which is a ratio of a kinetic energy to a potential energy density for~$Q$.
It is interesting to combine (14.143) and (14.145) using (14.147)
and~(14.148). 

Let us define
$$
\ealn{
\O_Q&=\frac{\frac12 {\dot Q}^2+U}{3H^2} &(14.149)\cr
\O_m&=\frac{8\pi G\dr{eff}\rho_m}{3H^2}. &(14.150)\cr
}
$$
One gets
$$
\O_m+\O_Q=1 \eq151
$$
or
$$
8\pi G\dr{eff}\rho_m=3(1-\O_Q)H^2. \eq152
$$
We call $\rho_m$ an energy density of a background and for such a matter the
equation of state is
$$
W_m=\frac{p_m}{\rho_m}=0. \eq153
$$
It is natural to define
$$
\t\rho_m=8\pi G\dr{eff}\rho_m \eq154
$$
as an effective energy density of a background with the same equation of a
state as~(14.153).

Under an assumption of slow roll for $Q$, i.e.\ 
$$
\frac12{\dot Q}^2 \ll U(Q) \eq155
$$
one gets
$$
W_Q\simeq -1, \eq156
$$
i.e.\ an equation of state of a \co\ constant evolving in time. In this case
$x_Q \simeq 0$. This can give in principle an account for an acceleration of
an evolution of the Universe if we suppose those conditions for our
contemporary epoch.

Let us come back to the inflationary era in our model. For our Higgs' field
is multicomponent, i.e.\ it is a multiplet of Higgs' fields, we have to do
with so called multicomponent inflation (see Ref.~[109]). In this case we can
define slow-roll parameter for our model in equations for Higgs' fields.

One gets from Eq.~(9.8)
$$
\frac d{dt}\(L^{d0}_{\u b}\)_{av} - 3H\(L^{d0}_{\u b}\)_{av}=
-\frac{e^{n\varPsi_1}}{2r^2} l^{db}\left\{\frac{\d V'}{\d \varPhi^b_{\u n}}
g_{\u b\u n}\right\}_{av} \eq157
$$
where $L^d_{0\u a}=L^d_{t\u a}$ (a time component of $L^d_{\mu\u a}$) is
defined by
$$
l_{dc}L^d_{0\u a}+l_{cd}g_{\u a\u m}g^{\u m\u c}L^d_{0\u c}=
2l_{cd} g_{\u a\u m}g^{\u m\u c}\frac d{dt}\(\varPhi^d_{\u c}\),
\eq158
$$
$V'$, $\dfrac{\d V'}{\d \varPhi^b_{\u n}}$ and 
$\left\{\dfrac{\d V'}{\d \varPhi^b_{\u n}}\right\}_{av}$ are defined by Eqs
(8.6--12) and Eq.~(6.24).
$$
\ealn{
3H^2&=8\pi G_N \(\frac{e^{(n-2)\varPsi_1}}{r^4}V'(\varPhi)
-\frac{2e^{-2\varPsi_1}}{r^2}\cL\dr{kin}(\dot \varPhi)
+\frac12 \la_{c1}(\varPsi_1)\) & (14.159)\cr
\dot H&=8\pi G_N \(-\frac{2e^{-2\varPsi_1}}{r^2}\cL\dr{kin}(\dot \varPhi)\)
&(14.160)
}
$$
where $H$ is a Hubble constant and
$$
\cL\dr{kin}(\dot\varPhi)=l_{ab}\left\{g^{\u b\u n}L^a_{0\u b}
\frac d{dt}\varPhi^b_{\u n}\right\}_{av}. \eq161
$$
$\varPsi_1$ is a constant value for a field $\varPsi$. The slow-roll parameters are
defined by
$$
\ealn{
\e&=\frac{\l_{ab}\kl g^{\u b\u n}L^a_{0\u b}\ddt \varPhi^b_{\u n}\kr_{av}}
{H^2} &(14.162)\cr
\d&=\frac{\l_{ab}\kl g^{\u b\u n}\ddt\(L^a_{0\u b}\)\ddt \(\varPhi^b_{\u n}\)\kr_{av}}
{H^2\(\l_{ab}\kl g^{\u b\u n}L^a_{0\u b}\ddt \(\varPhi^b_{\u n}\)\kr_{av}\)}
&(14.163)
}
$$
with the assumptions of the slow roll
$$
\e=O(\eta),\ \d=O(\eta) \eq164
$$
for some small parameter $\eta$.

Under slow-roll conditions we have
$$
\ealn{
&3H\(L^d_{0\u b}\)_{av}+\frac{e^{n\varPsi_1}}{2r^2} l^{db}\(\frac{\d V'}{\d
\varPhi^b_{\u n}}g_{\u b\u n}\)_{av}\cong 0 &(14.165)\cr
&H^2\cong \frac{8\pi}{3M^2\dr{pl}}\kl\frac{e^{(n-2)\varPsi_1}}{r^4}V'(\varPhi)\kr
+\frac12\la_{c1}(\varPsi_1) &(14.166)
}
$$
and with a standard extra assumption
$$
\frac{\dot \d}{H}=O(\eta^2). \eq167
$$
We remind that the scalar field $\varPsi$ is the more important agent to drive
the inflation. However, the full amount of time the inflation takes place
depends on an evolution of Higgs' fields (under slow-roll approximation). The
evolution must be slow and starting for $\varPhi=\varPhi^1\dr{crt}$. It ends at
$\varPhi=\varPhi^0\dr{crt}$. 

Thus we have according to Eq.~(14.50)
$$
\o N=\ln\frac{R(t\dr{end})}{R(t\dr{initial})}=
H_0(t\dr{end}-t\dr{initial})
$$
where
$$
\ealn{
\varPhi(t\dr{end})&=\varPhi^0\dr{crt} &(14.169)\cr
\varPhi(t\dr{initial})&=\varPhi^1\dr{crt} &(14.170)\cr
t\dr{end}&=t_r
}
$$
(we omit indices for $\varPhi$).

Thus we should solve Eq.~(14.165) for initial condition (14.169) and with
an assumption $H=H_0$. We have of course a short period of an inflation with
$H=H_1$ from $t\dr{end}=t_r$ to $t=t_1$, i.e.\ 
$$
\o N\dr{tot}=\o N+\o N_1=H_0(t_r-t\dr{initial})+H_1(t_1-t_r). \eq171
$$

Let us remind to the reader that $\o N$ is called an amount of inflation. Let us
consider inflationary fluctuations of Higgs' fields in our theory. We are
ignoring \gr\ backreaction for the \co\ \ev\ is driven by the scalar field~$\varPsi$.
We write the Higgs field $\varPhi^a_{\u m}$ as a sum 
$$
\varPhi^a_{\u m}(\vec r,t)=\varPhi^a_{\u m}(t)+\d \varPhi^a_{\u m}(\vec r,t) \eq172
$$
where $\varPhi^a_{\u m}(t)$ is a solution of Eq.~(14.165) with boundary condition
(14.169--170) and $\d\varPhi^a_{\u m}(\vec r,t)$ is a small fluctuation which
will be written in terms of Fourier modes
$$
\d\varPhi^a_{\u m}(\vec r,t)=\sum_{\vec K} \d\varPhi^a_{\vec K\u m}(t)
e^{i\vec K\vec r}. \eq173
$$
The full field equation for Higgs' field in the de Sitter background
linearized for $\d\varPhi^a_{\vec K\u m}(t)$ can be written
$$
\al
\frac{d^2}{dt^2}\(M^d_{\vec K\u b}\)_{av}
&+3H_0\ddt \(M^d_{\vec K\u b}\)_{av}+\frac1{R^2}{\vec K}^2\(M^d_{\vec K}\)
_{av}\\
&=-\frac{e^{n\varPsi_1}}{2r^2} l^{db}\kl\frac{\d^2V'\(\varPhi^a_{\u b}(t)\)}
{\d\varPhi^b_{\u n}\d \varPhi^c_{\u m}}g_{\u b\u n}\kr
\d\varPhi^c_{\vec K\vec m}(t)
\eal
\eq174
$$
where
$$
l_{dc}M^d_{\vec K\u b}+l_{cd}g_{\u a\u m}g^{\u m\u c}M^d_{\vec K\u c}
=2l_{cd}g_{\u a\u m}g^{\u m\u c}\d\varPhi^d_{\vec K\u c}. \eq175
$$
Let us change \dt\ and in\dt\ variables.
$$
\ealn{
&d\tau=\frac{dt}{\o R(t)},\quad \o R(t)=R_0e^{H_0t} &(14.176)\cr
&R(\tau)=-\frac1{H_0\tau} &(14.177)\cr
&-\infty<\tau<0 \quad\hbox{(a conformal time)}&(14.178)
}
$$
and
$$
\d\varPhi^a_{\vec K\u m}=\frac{\chi^a_{\vec K\u m}}R\,. \eq179
$$
Simultaneously we neglect the term with the second derivative of the
potential. Eventually we get
$$
\frac{d^2}{d\tau^2}\(\t M^d_{\vec K\u b}\)_{av} - \frac2{\tau^2}
\frac d{d\tau}\(\t M^d_{\vec K\u b}\)_{av}+{\vec K}^2\(\t M^d_{\vec K\u
b}\)_{av}=0 
\eq180
$$
where
$$
l_{dc}\t M^d_{\vec K\u b}+l_{cd}g_{\u a\u m}g^{\u m\u c}\t M^d_{\vec K\u c}
=2l_{cd}g_{\u a\u m}g^{\u m\u c}\chi^d_{\vec K\u c}. \eq181
$$

Let us notice that $|\vec K\tau|$ is the ratio of the proper wavenumber
$|K|R^{-1}$ to the Hubble radius $\dfrac1{H_0}$. At early times $|K\tau|\gg1$,
the wavelength is small in comparison to Hubble radius and the mode
oscillates as in Minkowski space. However, if $\tau$ goes to zero,
$|K\tau|$~goes to zero too. The wavelength of the mode is stretched far
beyond the Hubble radius. It means the mode freezes. Thus the analyzes
of the modes are similar to those in classical symmetric inflationary \pt\ 
theory neglecting the fact that we have to do with multicomponent inflation
and that the structure of representation space of Higgs' fields should be
taken into account.

However, the form of Eq.~(14.180) is exactly the same as in one component
symmetric theory if we take $\t M^d_{\vec K\u a}$ a field to be considered.
The relation (14.181) between $\t M^d_{\vec K\u a}$ and $\chi^d_{\vec K\u a}$
is linear and under some simple conditions unambigous.

It is possible to find an exact solution to Eq.~(14.180) and Eq.~(14.181).
One gets
$$
\al
\chi^a_{\vec K\u m}&=C^a_{1\vec K\u m}\(\frac
{\tau|\vec K|\cos(\tau|\vec K|)-\sin(\tau|\vec K|)}\tau\)\\
&+C^a_{2\vec K\u m}\(\frac
{\tau|\vec K|\sin(\tau|\vec K|)+\cos(\tau|\vec K|)}\tau\) 
\eal
\eqno(*)
$$
and
$$
\al
M^d_{\vec K\u b}&=N^d_{1\vec K\u b}\(\frac
{\tau|\vec K|\cos(\tau|\vec K|)-\sin(\tau|\vec K|)}\tau\)\\
&+N^d_{2\vec K\u b}\(\frac
{\tau|\vec K|\sin(\tau|\vec K|)+\cos(\tau|\vec K|)}\tau\) 
\eal
\eqno(**)
$$
where
$$
l_{dc}N^d_{i\vec K\u b}+l_{cd}g_{\u a\u m}g^{\u m\u c}N^d_{i\vec K\u c}
=2l_{cd}g_{\u a\u m}g^{\u m\u c}C^d_{i\vec K\u c}, \quad i=1,2.
\eqno(*{*}*)
$$

In this way one obtains
$$
\d\varPhi^a_{\vec K\u m}=\frac1{R_0} e^{-H_0t}\chi^a_{\vec K\u m}
\(-\frac{e^{-H_0t}}{R_0H_0}\) \eqno(*{*}{*}{*})
$$
and using $(*)$ one easily gets
$$
\al
\d\varPhi^a_{\vec K\u m}&=\t C^a_{1\vec K\u m}
\(-\frac{|\vec K|}{R_0H_0}e^{-H_0t}\cos\(\frac{|\vec K|}{R_0H_0}e^{-H_0t}\)
+\sin\(\frac{|\vec K|}{R_0H_0}e^{-H_0t}\)\)\\
&+\t C^a_{2\vec K\u m}
\(\frac{|\vec K|}{R_0H_0}\sin\(\frac{|\vec K|}{R_0H_0}e^{-H_0t}\)
+\cos\(\frac{|\vec K|}{R_0H_0}e^{-H_0t}\)\) 
\eal
\eqno\text{(V$*$)}
$$
where
$$
\t C^a_{i\vec K\u m}=-H_0C^a_{i\vec K\u m}\,. \eqno\text{(VI$*$)}
$$
However, it is not necessary to use a full exact solution (V$*$) to
proceed an analysis which we gave before.

Thus the primordial value of $R_{\vec K}(t)$ is equal to
$$
R_{\vec K}=-\[\frac{H_0}{\ddt(\varPhi^d_{\u m})}\d\varPhi^d_{\vec K\u m}\]
_{\big| t=t^\ast} \eq182
$$
where $R_{\vec K}(t)$ is defined in \co\ \pt\ theory as
$$
R^{(3)}(t)=4\frac{\vec K^2}{R^2} R_{\vec K}(t) \eq183
$$
(see Ref.~[7]) and $t^\ast$ is such that $|\vec K|=R(t^\ast)H_0$, i.e.
$$
t^\ast=\frac1{H_0} \ln\biggl[\frac{|\vec K|}{R_0H_0}\biggr], 
\quad |\vec K|=K. \eq184
$$
The primordial value is of course time-in\dt. We suppose that fluctuations of
Higgs' fields (the initial values of them) are in\dt\ and Gaussian.

Thus we can repeat some classical results concerning a power spectrum of
primordial \pt s. One gets
$$
\[\bigl| \d\varPhi^d_{\vec K\u m}\bigr|^2\] = \(\frac{H_0}{2\pi}\)^2. \eq185
$$
Thus
$$
P_R(K) \cong \(\frac{H_0}{2\pi}\)^2 \sum_{\u m,d}
\(\frac{H_0}{\ddt(\varPhi^d_{\u m})}\)_{\big| t=t^\ast}^2
\eq186
$$
for $t^\ast$ given by Eq.~(14.184).

Taking a specific situation with concrete groups $H,G,G_0$ and constants 
$\xi,\zeta,r$ we can in principle calculate exactly $P_R(K)$ and a power
index~$n_s$. Thus using some specific software packages (i.e.\ CMBFAST code)
we can obtain theoretical curves of CMB (Cosmic Microwave Background)
anisotropy including polarization effects.

Let us give the following remark. The condition for slow-roll evolution
for~$\varPhi$ can be expressed in terms of the potential~$V'$. If those are
impossible to be satisfied for some models (depending on $G,G_0,H,G_0'$ and
parameters $\xi,\zeta$), we can employ a scenario with a tunnel effect from
``false'' to ``true'' vacuum and bubbles coalescence. In the last case the
time of an inflation (in the first de Sitter phase) depends on the
characteristic time of the coalescence.

Let us consider a more general model of the Universe filled with ordinary
(dust) matter, a radiation and a quintessence.

From Bianchi identity we have:
$$
\ddt \rho\dr{tot}+(\rho\dr{tot}+p\dr{tot})\frac{3\dot R}R=0 \eq187
$$
where
$$
\ealn{
\rho\dr{tot}&=\rho_Q + \t\rho_m+\t \rho_r &(14.188)\cr
p\dr{tot}&=p_Q+\t p_r. &(14.189)
}
$$
Let us suppose that the evolution of radiation, ordinary matter (barionic +
cold dark matter) and a \q\ are in\dt. In this way we get in\dt\ conservation
laws 
$$
\ealn{
&\ddt(R^3\t \rho_m)=0&(14.190)\cr
&\ddt(R^4\t \rho_r)=0&(14.191)
}
$$
and
$$
\dot\rho_Q+(\rho_Q+p_Q)\frac{3\dot R}R=0. \eq192
$$

One gets
$$
\ealn{
&\t \rho_m=\frac A{R^3},\quad A=\text{const.} &(14.193)\cr
&\t \rho_r=\frac B{R^4},\quad B=\text{const.} &(14.194)
}
$$
From Eq.~(14.192) we get
$$
\dot\rho_Q+(1+W_Q)\rho_Q\cdot\frac{3\dot R}R=0. \eq195
$$
Substituting Eqs (14.193--194) into Eqs (14.38) and (14.40), remembering Eqs
(14.139) and (14.141--142) one gets
$$
\ealn{
\(\frac{\dot R}R\)^2+k&= \frac1{3M^2\dr{pl}}\(\frac AR+\frac B{R^2}\)
+\frac1{3M^2\dr{pl}}\rho_Q R^2 &(14.196)\cr
\frac{3\ddot R}R&= -\frac1{M^2\dr{pl}}\(\frac A{2R^3}+\frac {2B}{3R^4}\)
-\frac1{2M^2\dr{pl}}(1+3W_Q)\rho_Q &(14.197)
}
$$
and from Eq.\ (14.143)
$$
\ddot Q+3H\dot Q-\frac{(n+2)A}{M^2\dr{pl}R^3}=0. \eq198
$$
However, now $Q$ and $U(Q)$ are redefined in such a way that
$\dfrac1{M^2\dr{pl}}$ factor appears before~$\rho_Q$. This is a simple
rescaling. We have of course
$$
\ealn{
\t\rho_m=e^{-(n+2)\varPsi}\rho_m&=e^{-aQ}\rho_m &(14.199)\cr
\t\rho_r=e^{-(n+2)\varPsi}\rho_r&=e^{-aQ}\rho_r. &(14.200)
}
$$
Thus we get
$$
\ealn{
&\rho_m=\frac A{R^3}e^{aQ} &(14.201)\cr
&\rho_r=\frac B{R^4}e^{aQ}. &(14.202)
}
$$

Let us consider Eqs (14.196--198) and Eq.~(14.192). Combining these equations
we get the following formula
$$
\frac{(n+2)A}{M^2\dr{pl}R^3}\sqrt{(1+W_Q)\rho_Q}=0 . \eq203
$$
The one way to satisfy this equation is to put $W_Q=-1$. It is easy to
see that if $A=0$ then Eq.~(14.203) is satisfied trivially. The condition
$A=0$ does not imply $B=0$ and we can still have $B\ne0$.
For such a solution $Q=Q_0=\text{const.}$ and $Q_0=\dfrac{\sqrt{\o M}}{M\dr{pl}}\varPsi_0$ found
earlier 
$$
e^{\varPsi_0}=x_0=\sqrt{\frac{n|\g|}{(n+2)\b}}=
\frac1{\a_s}\(\frac{m_{\u A}}{m\dr{pl}}\)\sqrt{\frac{n|\ul{\t P}|}
{(n+2)\t R(\t\G)}}. \eq204
$$
Thus
$$
e^{-aQ_0}=e^{-(n+2)\varPsi_0}=
\a_s^{n+2}\(\frac{m\dr{pl}}{m_{\u A}}\)^{n+2}\(\frac{n+2}n\)^\frac{n+2}2
\(\frac{\t R(\t\G)}{|\ul{\t P}|}\)^\frac{n+2}2 \eq205
$$
or
$$
e^{-aQ_0}=
\a_s^{n+2}\(\frac r{l\dr{pl}}\)\(\frac{n+2}n\)^\frac{n+2}2
\(\frac{\t R(\t\G)}{|\ul{\t P}|}\)^\frac{n+2}2. \eq206
$$

Let us put $W_Q=-1$ into Eqs (14.196--197). One gets
$$
\ealn{
{\dot R}^2+k&= \frac1{3M^2\dr{pl}}\(\frac AR+\frac B{R^2}
+\rho_Q R^2\) &(14.207)\cr
\frac{3\ddot R}R&= -\frac1{M^2\dr{pl}}\(\frac A{2R^3}+\frac {2B}{3R^4}
+\rho_Q\). &(14.208)
}
$$

From Eq.\ (14.207) one gets
$$
\frac{\pm R\,dR}{\sqrt{\frac{\rho_Q}{3M^2\dr{pl}}R^4
-kR^2+\frac{A}{3M^2\dr{pl}}R+\frac{B}{3M^2\dr{pl}}}}=dt \eq209
$$
if $A=0$,
$$
\frac{\pm R\,dR}{\sqrt{\frac{\rho_Q}{3M^2\dr{pl}}R^4
-kR^2+\frac{B}{3M^2\dr{pl}}}}=dt. \eq210
$$

Taking $x=R^2$ one gets
$$
\int\frac{dx}{\sqrt{\frac{\rho_Q}{3M^2\dr{pl}}x^2
-kx+\frac{B}{3M^2\dr{pl}}}}=\pm2(t-t_0) \eq211
$$
and finally
$$
\pm\int\frac{dy}{\sqrt{y^2-\frac{3kM^2\dr{pl}}{\sqrt{B\rho_Q}}y+1}}=
\frac{2\sqrt3\sqrt{\rho_Q}M\dr{pl}}B(t-t_0) \eq212
$$
where
$$
x=\sqrt{\frac B{\rho_Q}}y. \eq213
$$

For a flat case $k=0$ we find:
$$
\pm\int\frac{dy}{\sqrt{y^2+1}}=
\frac{2\sqrt3\sqrt{\rho_Q}M\dr{pl}}B(t-t_0). \eq214
$$
The last formula can be easily integrated and one gets
$$
R=\root 4 \of{\frac B{\rho_Q}} \sqrt{\Sh\(\frac{
2\sqrt3\sqrt{\rho_Q}M\dr{pl}}B(t-t_0)\)} \eq215
$$
where we take a sign $+$ in the integral on the left hand side of Eq.~(14.214).

Let us calculate a Hubble parameter (constant) for our model. One gets
$$
H=\frac{\dot R}R=\sqrt3 M\dr{pl}\(\frac{\sqrt{\rho_Q}}B\)
\ctgh\(\frac{2\sqrt3\sqrt{\rho_Q}M\dr{pl}}B(t-t_0)\). \eq216
$$
For a deceleration parameter one obtains
$$
q=-\frac{\ddot RR}{R^2}=-\frac{\(2\Sh^2(u)-\Ch(u)\)}
{\Ch(u)\Sh^\frac12(u)} \eq217
$$
where
$$
u=\frac{2\sqrt3\sqrt{\rho_Q}M\dr{pl}}B(t-t_0). \eq218
$$
Thus we get a spatially flat model of a Universe dominated by a radiation
which is expanding and an expansion is accelerating.

Let us take $k=0$ and $B=0$ in Eq.~(14.209). One gets
$$
\ealn{
&\frac{\pm R\,dR}{\sqrt{\frac{\rho_Q}{3M^2\dr{pl}}R^4
+\frac{A}{3M^2\dr{pl}}R}}=dt &(14.219)\cr
&R=\root3\of{\frac A{\rho_Q}}x,\quad x=R\root3\of{\frac{\rho_Q}A}\,.&(14.220)
}
$$
One finally gets
$$
\int\frac{x\,dx}{\sqrt{x(x^3+1)}}=
\pm\frac1{3M\dr{pl}}\sqrt{\rho_Q}(t-t_0). \eq221
$$
Let us notice that
$$
\ealn{
&H^2=\(\frac{\dot R}R\)^2=\frac1{3M^2\dr{pl}}\(\frac A{R^3}+\rho_Q\)>
\frac{\rho_Q}{M\dr{pl}} &(14.222)\cr
&\frac{\ddot R R}{{\dot R}^2}= \frac{(2\rho_QR^4-A)R}{2(A+\rho_QR^3)}\,.
&(14.223)
}
$$
Thus the model of the Universe is expanding and accelerating.

Let us consider the integral on the left hand side of Eq.~(14.221). One gets
after some tedious algebra:
$$
\al
&\int\frac{x\,dx}{\sqrt{x(x^3+1)}}=
\frac{\sqrt{\sqrt3 +2}}{12}\(9-2\sqrt3\)\ln W+
\frac{9\sqrt2}{52}\sqrt{4+\sqrt3}\(6\sqrt3+11\)\mPi\\
&-\frac{\sqrt6}{598}\(92\sqrt{4\sqrt3+3}\(141\sqrt3+272\)
+\(9\sqrt3-15\)\sqrt{598}\sqrt{16+9\sqrt3}\)F
\eal
\eq224
$$
where
$$
\al
&W=\left(-20-\dfrac{9\sqrt3}2+\dfrac{(16-7\sqrt3)}2
\(\dfrac{2x+\sqrt3-1}{2x-\sqrt3-1}\)^2 \vphantom{\sqrt{\sqrt{\Bigg[}}}\right.\\
&\left.\vphantom{\sqrt{\sqrt{\Bigg[}}}
{}+2\sqrt{6-\sqrt3}
\sqrt{\[\(\dfrac{(2x+\sqrt3-1)}{(2x-\sqrt3-1)}\)^2+7-4\sqrt3\]
\[\(\dfrac{(2x+\sqrt3-1)}{(2x-\sqrt3-1)}\)^2-1+\dfrac{\sqrt3}2\]}
\right)\\
&\qquad{}\times\(\(\dfrac{(2x+\sqrt3-1)}{(2x-\sqrt3-1)}\)^2-1\)^{-1},
\eal
\eq225
$$
$$
\mPi=\mPi\(\ar\(\(\frac{(2x+(\sqrt3-1))}{(2x-(\sqrt3+1))}\)\sqrt{2(2+\sqrt3)}\),
\(\sqrt3-1\),\sqrt{\frac{(5-2\sqrt3)}{13}}\) \eq226
$$
$$
F=F\(\ar\(\(\frac{(2x+(\sqrt3-1))}{(2x-(\sqrt3+1))}\)\sqrt{2(2+\sqrt3)}\),
\sqrt{\frac{(5-2\sqrt3)}{13}}\) \eq227
$$
where $\mPi$ is an elliptic integral of the third kind and $F$ is an elliptic
integral of the first kind
$$
\ealn{
F(u,k)&=\intop_0^u \frac{d\f}{\sqrt{1-k^2\sin^2\f}} &(14.228)\cr
k&=\sqrt{\frac{(5-2\sqrt3)}{13}} &(14.229)\cr
\mPi(u,n,k)&=\intop_0^u \frac{d\f}{(1-n\sin^2\f)\sqrt{1-k^2\sin^2\f}}
&(14.230)\cr
\noalign{\eject}
n&=\sqrt3-1&(14.231)
}
$$
and with the same modulus as $F(u,k)$
$$
k^2=\frac{5-2\sqrt3}{13}\,.
$$
However, the most important condition for an existence of a physical solution
is coming from the first part of a sum in the right hand side of
Eq.~(14.224). 

We should have
$$
W>0. \eq232
$$
In order to analyze condition (14.232) let us write $W$ in the following way
$$
W=\frac{-20-\frac{9\sqrt3}2+\frac{(16-7\sqrt3)}2y
+2\sqrt{6-\sqrt3}\sqrt{(y+7-4\sqrt3)\(y-1+\frac{\sqrt3}2\)}}{y-1}
\eq233
$$
where
$$
y=\(\frac{2x+\sqrt3-1}{2x-\sqrt3-1}\)^2. \eq234
$$
Obviously $y\ge 0$.

We have two possibilities:
$$
\displaylines{
\text{I} \hfill y>1 \hfill (14.235)
}
$$
and
$$
-20-\tfrac{9\sqrt3}2+\tfrac{(16-7\sqrt3)}2y
+2\sqrt{6-\sqrt3}\sqrt{(y+7-4\sqrt3)\(y-1+\tfrac{\sqrt3}2\)}>0
\eq236
$$
$$
\displaylines{
\text{II} \hfill y<1 \hfill (14.237)
}
$$
and
$$
-20-\tfrac{9\sqrt3}2+\tfrac{(16-7\sqrt3)}2y
+2\sqrt{6-\sqrt3}\sqrt{(y+7-4\sqrt3)\(y-1+\tfrac{\sqrt3}2\)}<0.
\eq238
$$

The first possibility can easily be solved:

$$\displaylines{
\text{I} \hfill y>4.0612 \hfill (14.239)
}
$$
or for the second possibility
$$\displaylines{
\text{II} \hfill \frac{2-\sqrt3}2 \le y < 1. \hfill (14.240)
}
$$
We have finally
$$
\ealn{
0.7916 &< x<\frac{\sqrt3+1}2\,,&(14.241)\cr
\frac{3\sqrt3-5}2 &< x < \frac12. &(14.242)}
$$
For 
$$
x>\frac{\sqrt3+1}2 \eq243
$$
we get
$$
x<3.072. \eq244
$$

Let us notice the following: the function $W$ has a finite limit at
$x=\frac{\sqrt3+1}2$. Thus we can remove a singularity at
$x=\frac{\sqrt3+1}2$. In this way we get
$$
0.7916 < x< 3.072 \eq245
$$
or
$$
\frac{3\sqrt3-5}2 < x < \frac12. \eq246
$$
This simply means that the solution exists only for
$$
0.7916 \root3\of{\frac A{\rho_Q}}< R < 3.072 \root3\of{\frac A{\rho_Q}}
\eq247
$$
or
$$
\frac{3\sqrt3-5}2 \root3\of{\frac A{\rho_Q}} <R< \frac12 \root3\of{\frac A{\rho_Q}}\,. \eq248
$$

Let us calculate a density of a \q\ energy:
$$
\al
\rho_Q&=M^2\dr{pl}U(Q_0)=-\frac12e^{n\varPsi_0}\(\b e^{2\varPsi_0}-|\g|\)\\
&=\(\frac n{n+2}\)^\frac n2 \(\frac{m_{\u A}}{\a_s^{2(n+1)}}\)
\(\frac{m_{\u A}}{m\dr{pl}}\)^n \(\frac {|\ul{\t P}|}{\t R(\t \G)}\)^\frac n2 |\ul{\t P}|.
\eal
\eq249
$$
An energy density of a matter is equal to
$$
\rho_m=\frac A{R^3} e^{aQ_0} \simeq \rho_Q e^{aQ_0} \eq250
$$
and
$$
\frac{\rho_m}{\rho_Q}\cong e^{aQ_0}=
\(\frac{1}{\a_s}\)^{n+2} \(\frac{m_{\u A}}{m\dr{pl}}\)^{n +2}
\(\frac n{n+2}\)^\frac {n+2}2 \(\frac {\t R(\t \G)}{|\ul{\t P}|}\)^\frac {n+2}2 .
\eq251
$$

Let us consider the behaviour of the ``effective'' \gr\ constant in our model
$$
\al
G\dr{eff}&=G_N e^{-(n+2)\varPsi_0}=G_N e^{-aQ_0}= G_N\(\frac1{x_0}\)^{n+2}\\
&=G_N \a_s^{2(n+2)}
\(\frac{m\dr{pl}}{m_{\u A}}\)^\frac{n+2}2
\(\frac{\t R(\t \G)}{|\ul{\t P}|}\)^\frac{n+2}2 \(\frac{n+2}n\)^\frac{n+2}2
\eal
\eq252
$$
Thus $G\dr{eff}$ is a constant. For we are living in that model we should
rescale the constant and consider $G\dr{eff}(Q_0)$ (Eq.~(14.252)) a Newton
constant. Thus
$$
G_N=G_0\a_s^{2(n+2)}\(\frac{m\dr{pl}}{m_{\u A}}\)^\frac{n+2}2
\(\frac{n+2}n\)^\frac{n+2}2 \(\frac{\t R(\t \G)}{|\ul{\t P}|}\)^\frac{n+2}2
\eq253
$$
where $G_0$ is a different constant responsible for a strength of \gr\
interactions in earlier epochs of an evolution of the Universe. It means we
should write
$$
G\dr{eff}=
\(G_N \a_s^{-2(n+2)}
\(\frac{m_{\u A}}{m\dr{pl}}\)^\frac{n+2}2
\(\frac n{n+2}\)^\frac{n+2}2 \(\frac {|\ul{\t P}|}{\t R(\t \G)}\)^\frac{n+2}2 \)\cdot
e^{-(n+2)\varPsi}. \eq254
$$
The obtained solution (i.e.\ (14.224)) is for $W_Q=-1$. However, this is not an
attractor of the dynamical equations. This is similar to the tracker solution
of Steinhardt (see Ref.~[111]). Moreover we cannot apply his method because
our equation for a scalar field~$\varPsi$ (or~$Q$) is different. It contains an
additional term with a trace of the energy-momentum tensor (matter +
radiation). Only with a radiation filled Universe our equation is the same (if
we identify $\t\rho_r$ with Steinhardt density of a matter). In this case we
could apply Steinhardt tracker solution method and apply his criterion for a
potential $U(Q)$. In our case $U(Q)$ is given by Eq.~(14.141)
$$
\ealn{
U'(Q)&=-\frac{M\dr{pl}}{2\sqrt{\o M}}\((n+2)\b e^{aQ} - n|\g|e^{bQ}\) &(14.255)\cr
U''(Q)&=-\frac{M\dr{pl}^2}{4\sqrt{\o M}}\((n+2)^2\b e^{aQ} - n^2|\g|e^{bQ}\). &(14.256)
}
$$

The Steinhardt criterion consists in finding
$$
\G=\frac{U''U}{(U')^2}=\frac
{\((n+2)^2\b e^{aQ} - n^2|\g|e^{bQ}\)\(\b e^{aQ}-|\g|e^{bQ}\)}
{\((n+2)\b e^{aQ} - n|\g|e^{bQ}\)^2}\,.
\eq257
$$
To get a tracker solution we need
$$
\G\ge 1. \eq258
$$
In our case
$$
\ealn{
\G&=1-\frac{4\b|\g|e^{(a+b)Q}}{\((n+2)\b e^{aQ}-n|\g|e^{bQ}\)^2} &(14.259)\cr
\G&\simeq 1- \frac{4|\g|}{(n+2)\b}e^{(b-a)Q}= 1-\frac{4|\g|}{\b(n+2)}
e^{-\frac{2M\dr{pl}}{\sqrt{\o M}}Q}=1-\frac{4|\g|}{\b(n+2)}
e^{-2\varPsi}.\qquad &(14.260)
}
$$

Thus asymptotically we are close to $\G=1$. Moreover the Steinhardt
criterion is only a sufficient condition. Our radiation filled Universe seems
to be unstable due to appearing of a matter (a~dust matter). Moreover a
quintessential period of a Universe model has a severe restrictions of a
value of a radius. Thus the solution with $W_Q=-1$ and a matter in\dt ly
evolving cannot evolve forever.

In order to answer a question what is a further evolution of the Universe we
come back to Einstein equations with a scalar field~$\varPsi$ and matter
sources. Supposing as usual a Robertson-Walker metric and a spatial flatness of
the metric we write once again a Bianchi identity
$$
\(\frac{\br{scal}T^{\mu \nu }}{M^2\dr{pl}}+\frac1{M^2\dr{pl}}
e^{-(n+2)\varPsi}\br{matter}T^{\mu \nu }\)_{;\nu }=0 \eq261
$$
where
$$
\br{matter}T^{\mu \nu }=\rho_m u^\mu u^\nu \eq262
$$
is an energy-momentum tensor for a dust matter. One gets
$$
\al
\frac1{M^2\dr{pl}} e^{-(n+2)\varPsi}&\(\rho_m+\dot \rho_m+\rho_m\frac{3\dot R}R
-(n+2)\rho_m\dot\varPsi\)\\
&+\frac{d\la_{c0}}{d\varPsi}\dot\varPsi
-\frac{\o M}{M^2\dr{pl}}\dot\varPsi \ddot\varPsi-\frac{\o M}{M^2\dr{pl}}{\dot\varPsi}^2
\frac{3\dot R}R=0.
\eal
\eq263
$$
We make the following ansatz concerning an evolution of a matter density
$$
\ealn{
\rho_m&=\rho_0e^{(n+2)\varPsi}&(14.264)\cr
\rho_0&=\text{const.} &(14.265)
}
$$
In that moment the scalar field $\varPsi$ and the matter $\rho_m$ interact
nontrivially. 

Using Eqs (14.264--265), (14.263) and an equation for a scalar field $\varPsi$
one obtains
$$
\frac{\rho_0}{M^2\dr{pl}}-\frac{(n+2)}{M^2\dr{pl}}\rho_0+
\frac{\rho_0}{M^2\dr{pl}}\frac{3\dot R}R=0. \eq266
$$
Using Eq.~(14.266) and Eq.~(14.127) one gets
$$
\ealn{
&\frac{d\varPsi}{dt}=\pm\frac{\sqrt2 M\dr{pl}}{\sqrt{3\o M}}
\sqrt{\(\frac{n+1}{M\dr{pl}}\)^2-\frac{3\rho_0}{M^2\dr{pl}}
-\frac32\la_{c0}} &(14.267)\cr
&\int \frac{dx}{x\sqrt{\d+3\b x^{n+2}+3\g x^n}}=\pm
\frac{M\dr{pl}}{\sqrt{\o M}}(t-t_0) &(14.268)
}
$$
where
$$
\ealn{
\d&=2\(\frac{n+1}{M\dr{pl}}\)^2 - \frac{3\rho_0}{M^2\dr{pl}} &(14.269)\cr
x&=e^\varPsi. &(14.270)
}
$$

Using Eq.~(14.268) and Eq.~(14.127) one gets
$$
\ddt \ln R=\frac{(n+1)}{M\dr{pl}} \eq271
$$
or
$$
R=R_0 e^{(\frac{n+1}{M\dr{pl}})(t-t_0)}.\eq271a
$$
Let us consider Eq.~(14.268) and let us suppose that $x<1$ (and practically
small). In this case we can simplify
$$
\int \frac{dx}{x\sqrt{\d+3\b x^{n+2}+3\g x^n}}\simeq
\int \frac{dx}{x\sqrt{\d+3\g x^n}} \eq272
$$
and finally we get
$$
\varPsi=\frac1{2n} \ln\(\frac\d{3|\g|}\)-\frac1n\sqrt{\frac{2\d}{3\o M}}M\dr{pl}
(t-t_0). \eq273
$$
Thus our prediction that $x<1$ and is small has been justified. Using
Eq.~(14.273) one easily gets
$$
\ealn{
p_\varPsi &= \frac{\d M^2\dr{pl}}{3n^2}+\frac{\sqrt{|\g|\d}}{2\sqrt3}
\exp\(-\sqrt{\frac{2\d}{3\o M}}M\dr{pl}(t-t_0)\) &(14.274)\cr
\rho_\varPsi &= \frac{\d M^2\dr{pl}}{3n^2}-\frac{\sqrt{|\g|\d}}{2\sqrt3}
\exp\(-\sqrt{\frac{2\d}{3\o M}}M\dr{pl}(t-t_0)\). &(14.275)
}
$$
It is easy to see that
$$
\frac{p_\varPsi}{\rho_\varPsi}\underset{t\to\infty}\to\longrightarrow 1. \eq276
$$
Thus in this model we have to do asymptotically (practically very quickly)
with a stiff matter equation of state:
$$
p=\rho. \eq277
$$

Simultaneously an energy density for a scalar field~$\varPsi$ is dominated by a
kinetic energy which is practically constant and
$$
\frac12{\dot Q}^2 \cong \frac{\d M^2\dr{pl}}{3n^2}\,. \eq278
$$
In this sense a dark matter generated by $\varPsi$ in the model is in some sense
$K$-essence (not a \q).

Let us calculate an energy density for a matter
$$
\rho_m=\rho_0\(\frac\d{3|\g|}\)^\frac{(n+2)}{2n}
\exp\(-\frac{(n+2)}n \sqrt{\frac{2\d}{3\o M}} M\dr{pl}(t-t_0)\). \eq279
$$
The effective \gr\ constant reads
$$
G\dr{eff}=G_N \(\frac{3|\g|}\d\)^\frac{(n+2)}n \exp\(
\(\frac{n+2}n\)\sqrt{\frac{2\d}{3\o M}}M\dr{pl}(t-t_0)\). \eq280
$$
It is easy to write $R$ as a function of $\varPsi$:
$$
R=R_0 \exp\(\frac{n+1}{M^2\dr{pl}}\sqrt {\o M} \intop_{x_0}^{e^\varPsi}
\frac{dx}{x\sqrt{\d+3\b x^{n+2}+3\g x^n}}\) \eq281
$$
where
$$
x_0=\(\frac{3|\g|}\d\)^{2n}. \eq282
$$
Eq.\ (14.281) can be easily reduced to the Eq.\ (14.271) using Eq.~(14.268).

Thus an energy density of matter goes to zero in an exponential way. The
effective \gr\ constant is growing exponentially. What does it mean for the
future of the Universe? First of all the \U\ will be very diluted after some
time of such an evolution. Secondly, a relative strength of \gr\ interactions
will be stronger. This means that if we take substrat particles as cluster of
galaxies then in a quite short time any cluster will be a lonely island in
the \U\ without any communication with other clusters. All the clusters will
be beyond a horizon. On the level of a single cluster the strength of \gr\
interactions will be stronger and eventually they collapse to form a black
hole. In an intermediate time a \gr\ dynamics on the level of galaxies'
clusters will be governed by a Newtonian dynamics (nonrelativistic) with a
\gr\ potential corrected by a \co\ constant. The \co\ constant induces a
positive pressure (a~stiff matter) and will be important up to a moment of
sufficiently large \gr\ constant. After a sufficiently long time only a
classical Newtonian gravity will drive a dynamics of galaxies' clusters. Let
us roughly estimate this effect. In order to do this let us suppose that
$$
\rho_\varPsi=p_\varPsi=\frac{\d M^2\dr{pl}}{3n^2}=\text{const.} \eq283
$$
In an energy momentum we get
$$
\br{scal}T^{\mu \nu }=2\cdot \frac{\d M^2\dr{pl}}{3n^2} u^\mu u^\nu - 
\frac{\d M^2\dr{pl}}{3n^2} g^{\mu \nu }. \eq284
$$

However, we have in the place of Newtonian constant $G_N$ a $G\dr{eff}$ which
is a function of \co\ time and in the equation for \gr\ field for~$g_{\mu \nu }$
we should put
$$
\frac1{(M\dr{pl}\ur{eff})^2}\br{scal}T^{\mu \nu }=
\frac\d{3n^2}\(\frac{M\dr{pl}}{M\dr{pl}\ur{eff}}\)^2
\(2u^\mu u^\nu -g^{\mu \nu }\) \eq285
$$
where
$$
\al
M\dr{pl}\ur{eff}=\frac1{\sqrt{8\pi G\dr{eff}}}&=
M\dr{pl}\(\frac{\a_s}{m\dr{pl}m_{\t A}}\)^{n+2}
\(\frac{2(n+1)^2-3\rho_0}{3|\uP|}\)^\frac{n+2}2\\
&\times\exp\[\(\frac12\(\frac{n+2}2\)\sqrt{\frac{2\d}{3\o M}}\)M\dr{pl}(t-t_0)\].
\eal
\eq286
$$
Thus
$$
\frac1{(M\dr{pl}\ur{eff})^2}\br{scal}T^{\mu \nu }=
\t Ae^{-\k (t-t_0)}\(2u^\mu u^\nu -g^{\mu \nu }\) \eq287
$$
where
$$
\ealn{
&\k=\frac12\(\(\frac{n+2}2\)\sqrt{\frac{2\d}{3\o M}}\) &(14.288)\cr
&\t A=\frac{3n^2}\d \(\frac{\a_s}{m\dr{pl}m_{\t A}}\)^{-(n+2)}
\(\frac{2(n+1)^2-3\rho_0}{3|\uP|}\)^{-(\frac{n+2}2)}. &(14.289)
}
$$

Consider the following case (the justification will be given below):
$$
\frac1{(M\dr{pl}\ur{eff})^2}\br{scal}T^{\mu \nu }=
-\t A e^{-\k(t-t_0)}g^{\mu \nu }. \eq290
$$
Now let us solve the following problem. Let $\o \La$ be a value 
$\t Ae^{-\k(\bar t-t_0)}$ in an established \co\ time $t=\o t$ and the same
for $\o G=G\dr{eff}(\o t)$. Let us consider a \gr\ field of a point
mass~$M_0$ static and spherically symmetric. Consider a metric
$$
ds^2=B(r)dt^2-\frac{dr^2}{B(r)}-r^2\(d\theta^2+\sin^2\theta \,d\varphi^2\)
\eq291
$$
in spherical coordinates and Einstein equations
$$
R_{\mu \nu }-\frac12g_{\mu \nu }R=-\o\La g_{\mu \nu } \eq292
$$
for the metric (14.291) where
$$
\o\La=\t Ae^{-\k(\bar t-t_0)}. \eq293
$$
Via a standard procedure one obtains
$$
B(r)=1-\frac{2\o GM_0}r -\frac{\o \La r^2}3. \eq294
$$
Thus an effective nonrelativistic Newton-like \gr\ potential has a shape
$$
V(r)=-\frac{\o GM_0}r - \frac{\o\La r^2}6\,.\eq295
$$
Remembering that $\o G$ is growing exponentially in time and $\o\La$ is
decaying in time exponentially we see that the second term is important only
for a short time. For a contemporary time
$$
\o\La\simeq 10^{-52} \text{m}^{-2} \eq296
$$
according to the Perlmutter data and it could be important on the level of a
size of galaxies' cluster. However, in a short time it will be negligible
even on that level. Let us calculate a Schwarzschild radius for a mass $M_0$
with an effective \gr\ constant $G\dr{eff}$. One gets
$$
\al
r\ur{eff}_g&=\sqrt{2M_0\o G}=\sqrt{2M_0 G\dr{eff}(\o t)}
=r_g\(\frac{m\dr{pl}m_{\u A}}{\a_s}\)^{n+2}
\\
&\times \(\frac{3|\uP|}{2(n+1)^2-3\rho_0}\)^\frac{n+2}2\exp
\[\(\frac12\(\frac{n+2}2\)\sqrt{\frac{2\d}{3\o M}}\)M\dr{pl}(t-t_0)\]
\eal
\eq297
$$

Thus $r\ur{eff}_g$ (calculated for an effective \gr\ constant $G\dr{eff}(\o
t)$) is growing exponentially. It means that after some time it will be of an
order of size of galaxies' cluster. Then galaxies' clusters collapse to form
black holes. The time of a collapse is easy to calculate.
$$
\ealn{
r\ur{eff}_g &= r\dr{cluster} &(14.298)\cr
t\dr{collapse} &=t_0+(n+2)\sqrt{6\o M}
\Biggl[\ln\(\dfrac{r\dr{cluster}}{r_g}\)
+(n+2)\ln\(\dfrac{m\dr{pl}m_{\t A}}{\a_s}\)\cr
&\quad{}+
\(\dfrac{n+2}2\)\ln\(\dfrac{3|\uP|}{2(n+1)^2-3\rho_0}\)\Biggr]
\cdot\[\sqrt\d (n+2)M\dr{pl}\]^{-1} &(14.299)
}
$$
where $r_g$ is a \gr\ radius of a mass of a galaxies' cluster with $G=G_N$. 

Let us calculate a Hubble constant and a deceleration parameter for our
model. One gets
$$
\ealn{
H&=\frac{\dot R}R =(n+1)M\dr{pl} &(14.300)\cr
q&=-\frac{\ddot RR}{{\dot R}^2}=-1.&(14.301)
}
$$
Summing up we get the following scenario of an evolution. The \U\ starts in a
``false'' vacuum state and its stable evolution is driven by a ``\co\
constant'' formed from $\t R(\t \G)$, $\o P$, $V(\varPhi^1\dr{crt})$ and $\varPsi_1$
being a minimal solution to an effective selfinteracting potential
for~$\varPsi$. This evolution is stable against small \pt\ for~$\varPsi$ and~$R$.
The evolution is \ex\ and a Hubble constant is calculable in terms of our
theory (the nonsymmetric Jordan--Thiry theory). This evolution ends at a
moment the ``false'' vacuum state changes into a ``true'' vacuum state. In
that moment an energy of a vacuum is released into radiation (and a matter).
The \U\ is reheating and a big-bang scenario begins at a very hot stage.
Moreover the evolution is still governed by a scalar field~$\varPsi$ which
attains a new minimum of an effective potential. The effective potential is
different and a new value of~$\varPsi$, a~$\varPsi_0$, is different too. The
evolution is \ex\ and a new Hubble constant~$H_1$ is also calculable in terms
of the underlying theory. In this background we have an evolution of a
multiplet of scalar fields~$\varPhi^a_{\u m}$ from $\varPhi^1\dr{crt}$ to
$\varPhi^0\dr{crt}$. This goes to a spectrum of fluctuations calculable in the
theory. After that time the field~$\varPsi$ is slowly changing $\dot\varPsi\approx
0$ and the temperature of the \U\ is going down. The evolution of a radiation
is in the second inflationary stage adiabatic and afterwards during next phase
($\dot\varPsi\simeq0$) nonadiabatic. The last means there is an energy exchange
between radiation and a scalar field~$\varPsi$. The total amount of an inflation
$\o N\dr{tot}$ is calculable in the theory. The evolution of a factor $R(t)$ is
governed by a simple elementary function of a small rise. After this the
radiation condenses into a matter (a~dust with zero pressure) and a matter
(a~dust), a radiation and a scalar field evolve adiabatically without
interaction among them. The evolution of a scalar field is governed by a \q,
i.e.\ $p_Q=W_Q\rho_Q$, where $W_Q\cong-1$. 

The \q\ field $Q$ is a normalized scalar field~$\varPsi$. During a radiation era
($\dot\varPsi\simeq0$) an equation of state for a scalar field~$\varPsi$ is
different: $p_\varPsi=\frac43\rho_\varPsi$ (similar to stiff matter equation of
state). The scale factor $R(t)$ evolves according to some elementary function
built from hyperbolic function in radiation + \q\ scenario and according to
complicated function formed from the first and the second elliptic integrals
(after reversing of those functions). It expands and accelerates. The
solution with radiation and \q\ has a restricted behaviour (Eq.~(14.247--248)).
The energy density of radiation is going to zero. Only a \q\ energy density
is constant. The energy of a \q\ is a potential energy dominated. However, the
solution cannot be continued up to the end for an interaction between a
matter and a scalar field starts to be important. The effective \gr\ constant
remains constant during the period. Let us remind to the reader that during
\il ary epochs the effective \gr\ constant was constant. Only during a
radiation era it was slightly changing. In order to keep the \U\ to evolve
forever it is necessary to find a solution with an interaction between a
matter and a scalar field~$\varPsi$. Using an appropriate ansatz for an \ev\ of
an energy density of a matter we get such a solution. The scale factor $R(t)$
is \ex ly expanding. 

An energy density of a matter is going to zero and an energy density of a
scalar field is approaching (\ex ly) a constant. However, now a scalar
field~$\varPsi$ forms a different form of a matter---a stiff matter with an
equation of a state $\rho_\varPsi\simeq p_\varPsi$ ($W_\varPsi\simeq1$). The effective
\gr\ constant is growing \ex ly in time. In contradistinction with the
previous period of an \ev\ the scalar field~$\varPsi$ forms a matter with a
kinetic energy dominance (i.e.\ $K$-essence). In the previous period we have
to do with a potential energy dominance (i.e.\ \q). 

Let us give some remarks
on a spatial curvature of a model of the \U. In the first period (\il ary
scenario) the spatial curvature has to be zero and any \pt s of initial
conditions cannot change it. In next periods it is natural to suppose that it
is not changed. Thus our model of a \U\ is spatially flat. In the moment of
change of a phase of an \ev\ we should apply a matching conditions for an
energy density and for a scale factor (a~radius of a model of a \U). However,
an \ev\ in such moments undergoes phase transitions for a law of an \ev\ of a
scalar field and a scale factor, which can be considered as a second order or
even a first order phase transition. It is worth to notice that in formulae
concerning Hubble constants, deceleration parameters, \gr\ constants
(effective) meet some parameters from unification theory of fundamental
interactions and from cosmology. It seems that we are on a right track to
give an account for large number hypothesis by P.~Dirac. This demands of
course some investigations especially in developing nonsymmetric Kaluza--Klein
(Jordan--Thiry) theory. It is also interesting to mention that the scalar
field~$\varPsi$ plays many r\^oles in our theory. It works as an \il\ field during
an \il\ epoch. (However Higgs' fields $\varPhi^a_{\u b}$ play an important
r\^ole in creating fluctuations spectrum.) It plays also a r\^ole of a \q\
and a $K$-essence, except its influence on an effective \gr\ constant and on
an effective scale of masses in unification of fundamental interactions. This
field is massive getting masses from several mechanisms. During \il\ phases
it acquires mass due to spontaneous symmetry breaking, different in both
epochs. It gets a mass from a \co\ background mentioned earlier in this
section. It can get a mass from interactions with Higgs' fields (see
sec.~10). Thus except its \co\ importance the fluctuations of a scalar field
around its \co\ value could be detected (in principle) as massive scalar
particles of a large mass. In this way an effective \gr\ potential in
nonrelativistic limit takes a form
$$
V(r)=-\frac{G_NM_0}r \(1-\frac\a r e^{-(\frac r{r_0})}\)e^{\k(t-t_0)}
-\frac{\t A r^2}6e^{-\k(t-t_0)}
\eq302
$$
where $\k,\t A$ are given by Eqs (14.288--289), $r_0$ is a range of massive
scalar $\d\varPsi$ (a~fluctuation of a scalar field~$\varPsi$ in a \co\
background). This could be detected as a tiny change of a Newton constant in
\gr\ law (e.g.\ the fifth force) on short distances. Let us notice that on
short distances and on short time scales an \ef\ \gr\ potential reads:
$$
V(r)=-\frac{G_NM_0}r\(1-\frac \a r \,e^{-(\frac r{r_0})}\). \eq303
$$
On large distances and short time scales 
$$
V(r)=-\frac{G_NM_0}r-\frac{\o \La r^2}6. \eq304
$$
On short distances and large time scales 
$$
V(r)=-\frac{G_NM_0}r\(1-\frac\a r\,e^{-(\frac r{r_0})}\)e^{\k(t-t_0)}. \eq305
$$
On intermediate distances and large time scales 
$$
V(r)=-\frac{G_NM_0}r e^{\k(t-t_0)}. \eq306
$$
The latest being a cause to collapse a cluster of galaxies into a black hole.
It seems also possible to consider an \ef\ coupling \ct~$\a_s\ur{eff}$
for~$\a_s$ enters~$m_{\u A}$ and~$m_{\u A}\ur{eff}$. However
$$
m\ur{eff}_{\u A} = e^{-\varPsi}\(\frac{\a_s}r\), \eq307
$$
$r$ is a radius of a manifold $G/G_0$ (e.g.\ a radius of a sphere~$S^2$). If
we rewrite Eq.~(14.307) as
$$
\frac{m\ur{eff}_{\u A}}{\a_s\ur{eff}} = \frac1r e^{-\varPsi}, \eq308
$$
it would be quite difficult to distinguish a drift of $m\ur{eff}_{\u A}$ from
$\a_s\ur{eff}$ drift. Only a quotient has a definite dependence on the scalar
field~$\varPsi$. Especially it is interesting to consider a drift of $\a\dr{em}$
(a~fine structure \ct) reported from several sources. However, without
extending our theory to include fermion (spinor) fields this is impossible.
Thus it is necessary to construct a nonsymmetric-geometric version with
supersymmetry and supergravity including noncommutative (anticommutative)
coordinates to settle the problem. The scalar field~$\varPsi$ can form an
infinite tower of massive scalar fields which could have important
astrophysical and high-energy consequences (see sec.~16).

Let us consider Eqs (4.51) and (4.56) and changing a base in the Lie algebra
$\gH$ we get
$$
\mu^a_{\hi}=\d^a_\hi \eq309
$$
and
$$
\a^c_i=\d^c_i. \eq310
$$
Simultaneously for $G\subset H$ ($\gG\subset \gH$) we get
$$
\ealn{
f^\hi_{\u a\u b}&= C^\hi_{\u a\u b}&\text{(14.311a)}\cr
f^{\u d}_{\u a\u b}&= C^{\u d}_{\u a\u b}&\text{(14.311b)}
}
$$
where $\u a,\u b,\u c$ refer to the complement $\gM$ in $\gG$ ($\gG=\gG_0
\dot+ \gM$), $\hi,\hj,\hat k$ to the subalgebra $\gG_0$ of~$\gG$. Let us
suppose a symmetry requirement
$$
[\gM,\gM]\subset\gG_0. \eq312
$$
From Eq.\ (14.312) it follows that
$$
C^{\u a}_{\u b\u c}=f^{\u a}_{\u b\u c}=0. \eq313
$$
Eq.\ (4.51) looks like
$$
\varPhi^c_{\u b}C^{\u b}_{\hi\u a}=\varPhi^b_{\u a}C^c_{\hi b}. \eq314
$$
Eq.\ (4.55)
$$
H^c_{\u a\u b}=C^c_{ab}\varPhi^a_{\u a}\varPhi^b_{\u b}- C^c_{\u a\u b}
-\varPhi^c_{\u d}C^{\u d}_{\u a\u b}. \eq315
$$
Let us come back to Eq.\ (14.157).

For our \il\ driving agent is a scalar field~$\varPsi$ we do not carry any
backreaction of Higgs' fields and we can consider some nonstandard scenarios
of an \ev\ of those fields. Thus we can consider an \ev\ which is not a
slow-roll \ev. Let us suppose that
$$
\ddt\(\varPhi^a_{\u m}\)\simeq0. \eq316
$$
In this way we get from (14.316)
$$
\ddt\({L^d}{}^0_{\u b}\)_{av}+\frac{e^{n\varPsi_1}}{2r^2}l^{db}\kl
\frac{\d V'}{\d\varPhi^b_{\u n}}g_{\u b\u n}\kr. \eq317
$$
The next step in simplification is such that we collapse all the degrees of
freedom of Higgs' fields into one
$$
\varPhi^a_{\u m}=\d^a_{\u m}\varPhi. \eq318
$$
In this way constraints (14.314) are satisfied trivially and Eq.~(14.315) is
as follows:
$$
H^c_{\u a\u b}=C^c_{\u a\u b}\(\varPhi^2-1-\varPhi\). \eq319
$$
In this way $V=V'$ (no constraints) and
$$
V=\frac A{\a^2_s}\(\a^2_s\varPhi^2-1-\a_s\varPhi\)^2 \eq320
$$
where
$$
A=\[l_{ab}\(2g^{[\u m\u n]}g^{[\u a\u b]}C^c_{\u m\u n}C^b_{\u a\u b}
-C^b_{\u m\u n}g^{\u a\u n}g^{\u b\u m}\t L^a_{\u a\u b}\)\]_{av} \eq321
$$
and
$$
\ealn{
&L^a_{\u a\u b}=\t L^a_{\u a\u b}\(\a_s\varPhi^2-\frac1{\a_s}-\varPhi\) &(14.322)\cr
&l_{dc}g_{\u m\u b}g^{\u c\u m}\t L^d_{\u c\u a}+
l_{cd}g_{\u a\u m}g^{\u m\u c}\t L^d_{\u b\u c}=
2l_{cd}g_{\u a\u m}g^{\u m\u c}C^d_{\u b\u c}. &(14.323)
}
$$

We can get the following identity
$$
l_{dc}g^{\u c\u q}g^{\u a\u p}\t L^d_{\u c\u a}C^c_{\u p\u q}+
l_{cd}g^{\u p\u c}g^{\u q\u b}\t L^d_{\u b\u c}C^c_{\u p\u q}=
2l_{cd}g^{\u p\u c}g^{\u q\u b}C^d_{\u b\u c}C^c_{\u p\u q}. \eq324
$$
Using Eq.\ (14.317) we get the following equation for $\varPhi$:
$$
\frac{d^2\varPhi}{dt^2} + \frac{e^{n\varPsi_1}}{2r^2}\frac{\d V}{d\varPhi}=0 \eq325
$$
if the following constraints are satisfied
$$
\int \sqrt{|\t g|}d^nx \(g_{\u q\u c}\t l^{\u p\u e}+l^{\u e\u n}
g_{\u c\u n}\d^{\u p}_{\u q}\)=2V_2\d^e_{\u c}\d^{\u p}_{\u q}. \eq326
$$
Thus from Eq.\ (14.325) one gets a first integral of motion
$$
\frac{{\dot\varPhi}^2}2+\frac{e^{n\varPsi_1}}{2r^2}V=\frac B2=\text{const.} 
\eq327
$$

From Eq.\ (14.327) one gets
$$
\intop_{\varPhi_0}^\varPhi \frac{d\varPhi}
{\sqrt{-\dfrac{e^{n\varPsi_1}}{r^2\a_s^2} A\(\a_s^2\varPhi^2-\a_s\varPhi-1\)^2+B}}
=\pm(t-t_0) \eq328
$$
and finally
$$
\intop_{\f_0}^\f \frac{d\f}{\sqrt{(a-\f^2+\f+1)(a+\f^2-\f-1)}}
=\pm\sqrt{\frac{Ae^{n\varPsi_1}}{r^2}}(t-t_0) \eq329
$$
where
$$
\ealn{
a&=\a_s\sqrt{\frac{Br^2}{Ae^{n\varPsi_1}}} &(14.330)\cr
\varPhi&=\frac1{\a_s}\f. &(14.331)
}
$$

In order to find an integral on the left hand side of Eq.~(14.329) and its
inverse function we use a uniformization theory of algebraic curves. Let
$$
f(\f)=(a-\f^2+\f+1)(a+\f^2-\f-1)=-\f^4+2\f^3+\f^2-2\f+(a^2-1). \eq332
$$
One easily notices that
$$
\f_0=\frac{1+\sqrt{5+4a}}2 \eq333
$$
is a root of the polynomial $f(\f)$ (a real root).

In this way one gets
$$
\f=\frac{f'(\f_0)}{4\(P(z)-\frac1{24}f''(\f_0)\)}+\f_0 \eq334
$$
where
$$
P(z)=P(z;g_2,g_3) \eq335
$$
is a $P$ Weierstrass function with invariants:
$$
\ealn{
&g_2=\frac{5-3a^2}{3} &(14.336)\cr
&g_3=\frac{a^2}{12}+\frac{10}{27}\,. &(14.337)
}
$$

One easily finds
$$
\ealn{
f'(\f_0)&=\tfrac12 (5+3a)\sqrt{5+4a} &(14.338)\cr
f''(\f_0)&=-2(5+6a) &(14.339)
}
$$
and
$$
z=\int_{\f_0}^\f [f(x)]^{-\frac12}\,dx. \eq340
$$
In this way one gets
$$
\varPhi(t)=\frac{(5+3a)\sqrt{5+4a}}{8\a_s\(P(z)+\frac1{12}(5+6a)\)}
+\frac{1+\sqrt{5+4a}}{2\a_s} \eq341
$$
where
$$
z=\pm\sqrt{\frac{Ae^{n\varPsi_1}}{r^2}}(t-t_0). \eq342
$$

Let us come back to the potential $V$ Eq.\ (14.320) and let us look for its
critical points. One gets
$$
\frac{dV}{d\varPhi}=\frac{2A}{\a_s}\(\a_s^2\varPhi^2-\a_s\varPhi-1\)(2\a_s\varPhi-1) \eq343
$$
and from
$$
\frac{dV}{d\varPhi}=0 \eq344
$$
easily finds
$$
\ealn{
\varPhi_1&=\frac{1}{2\a_s} &\text{(14.345a)}\cr
\varPhi_2&=\frac{1-\sqrt5}{2\a_s} &\text{(14.345b)}\cr
\varPhi_3&=\frac{1+\sqrt5}{2\a_s} &\text{(14.345c)}
}
$$
One obtains
$$
\ealn{
V(\varPhi_1)&=\frac{25A}{16\a^2_s} &\text{(14.346a)}\cr
V(\varPhi_2)&=0 &\text{(14.346b)}\cr
V(\varPhi_3)&=0. &\text{(14.346c)}
}
$$
It is easy to see that $\varPhi_1$ is a maximum and $\varPhi_2$ and~$\varPhi_3$ are
minima of the potential~$V$.

However, in our simplified model of an \ev\ of Higgs' fields we have not 
a second (local)
minimum. Thus in order to mimic a real situation we start an \ev\ of a Higgs
field from the value
$$
\varPhi=0. \eq347
$$
In this case
$$
V(0)=\frac A{\a_s^2}\,.\eq348
$$
Thus
$$
\ealn{
\varPhi(t\dr{initial})&=0 &(14.349)\cr
\varPhi(t\dr{end})&=\frac{1+\sqrt5}{2\a_s}\,. &(14.350)
}
$$
Coming back to Eq.\ (14.329) we get
$$
\intop_{0}^\frac{1+\sqrt5}{2} \frac{d\f}{\sqrt{(a-\f^2+\f+1)(a+\f^2-\f-1)}}
=\sqrt{\frac{Ae^{n\varPsi_1}}{r^2}}(t\dr{end}-t\dr{initial}). \eq351
$$

Thus the amount of \il\ obtained in our \ev\ is equal to
$$
\o N_0=H_0\sqrt{\frac {r^2}{Ae^{n\varPsi_1}}}
\intop_{0}^\frac{1+\sqrt5}{2} \frac{d\f}{\sqrt{(a-\f^2+\f+1)(a+\f^2-\f-1)}}\,.
\eq352
$$
Supposing simply
$$
a<\frac54 \eq353
$$
one gets
$$
\o N_0= H_0\sqrt{\frac {2r^2}{5Aae^{n\varPsi_1}}}
\Biggl(\intop_{0}^{\f_1}\frac{d\theta}{\sqrt{1-\(\frac{4a+5}{10}\)\sin^2\theta}}
-\intop_{0}^{\f_2}\frac{d\theta}{\sqrt{1-\(\frac{4a+5}{10}\)\sin^2\theta}}\Biggr)
\eq354
$$
where
$$
\ealn{
\cos^2\f_1&=\frac1{5+4a} &(14.355)\cr
\cos^2\f_2&=\frac5{5+4a} &(14.356)
}
$$
or
$$
\o N_0=H_0\sqrt{\frac {2r^2}{5Aae^{n\varPsi_1}}}
\intop_{\f_2}^{\f_1}\frac{d\theta}{\sqrt{1-\(\frac{4a+5}{10}\)\sin^2\theta}}
\eq357
$$
or
$$
\o N_0=\frac{H_0}{\sqrt{5\a_s}}\root4\of{\frac {4r^2}{ABe^{n\varPsi_1}}}
\intop_{\f_2}^{\f_1}\frac{d\theta}{\sqrt{1-\(\frac{4a+5}{10}\)\sin^2\theta}}\,.
\eq358
$$
In order to get 60-fold \il
$$
\o N_0\simeq 60 \eq359
$$
we should play with parameters in our theory. Let us notice that in our
simple model the constant $\a_1$ equals
$$
\a_1=m^2_{\u A}\(\frac{m_{\u A}}{m\dr{pl}}\)^2\(\frac A{\a^6_s}\)=
\frac1{r^2}\(\frac{l\dr{pl}}r\)^2\(\frac A{\a^6_s}\) \eq360
$$
(see Eqs (14.32--33a).

Let us consider a slow-roll dynamic of Higgs' field in this simplified model.
From Eq.~(14.165) one gets
$$
\ddt\(\varPhi^f_{\u p}\)=-\frac{e^{n\varPsi_1}}{12V_2r^2H_0}
\(l^{fb}g_{\u p\u n}+l^{bf}g_{\u n\u p}\)\frac{\d V'}{\d \varPhi^b_{\u n}}\,.
\eq361
$$
Using our simplifications from Eqs (14.309--319) and Eqs (14.300--306) one
gets 
$$
\frac{d\varPhi}{dt}=-\frac{e^{n\varPsi_1}}{12r^2H_0}\frac{dV}{d\varPhi} \eq362
$$
supposing a condition
$$
\d^f_{\u p}=l^{f\u n}g_{\u p\u n}+l^{\u nf}g_{\u n\u p} \eq363
$$
or
$$
\frac{d\varPhi}{dt}+\frac{e^{n\varPsi_1}A}{6\a_sr^2H_0}
\(\a_s^2\varPhi^2-\a_s\varPhi-1\)\(2\a_s\varPhi-1\)=0.
\eq364
$$
Changing the \dt\ variable into
$$
\f=\a_s\varPhi \eq365
$$
one gets
$$
\frac{d\f}{dt}+b\(\f^2-\f-1\)(2\f-1)=0 \eq366
$$
where
$$
b=\frac{e^{n\varPsi_1}A}{6r^2H_0}>0. \eq367
$$
From Eq.\ (14.366) one finds
$$
\frac{\f^2-\f-1}{(\f-\frac12)^2}=e^{-5b(t-t_0)} \eq368
$$
and finally
$$
\varPhi(t)=\frac1{2\a_s}\[1\pm\sqrt{\frac5
{1-e^{-5b(t-t_0)}}}\].
\eq369
$$

One easily notices that for $(t-t_0)\to\infty$
$$
\varPhi(t)\to \frac1{\a_s}\(\frac{1\pm\sqrt5}2\). \eq370
$$
It simply means that a slow-roll \il\ never ends.

If we suppose that an \il\ starts at $\f=3$, i.e.\ for
$$
t\dr{initial}=t_0+\frac1{5b}\ln\frac54\,, \eq371
$$
it is evident that to complete it we need an eternity.

Summing up we get in our simplified model a finite amount of an \il\ for a
nonstandard dynamic of a Higgs field and an infinite amount of \il\ for a
slow-roll dynamic. The truth probably is in a middle.

Probably it would be necessary to consider a full equation for a Higgs field
in a simplified model
$$
\frac{d^2\varPhi}{dt^2}-3H_0\frac{d\varPhi}{dt}-3bH_0
\(\a^2_s\varPhi^2-\a_s\varPhi-1\)\(2\a_s\varPhi-1\)=0 \eq372
$$
or
$$
\frac{d^2\f}{dt^2}-3H_0\frac{d\f}{dt}-3b\a_s H_0
\(\f^2-\f-1\)\(2\f-1\)=0. \eq372a
$$
Changing an in\dt\ variable from $t$ to $\tau=H_0t$ one gets
$$
\frac{d^2\f}{d\tau^2}-3\frac{d\f}{d\tau}-3\frac {b\a_s}{H_0}
\(\f^2-\f-1\)\(2\f-1\)=0. \eq372b
$$
Let
$$
\frac{3b\a_s}{H_0}=\o a \eq373
$$
and let us change a \dt\ variable $\f$ into
$$
\ealn{
z&=2\f-1 &(14.374)\cr
\f&=\frac{z+1}2\,. &(14.375)
}
$$
One gets
$$
\frac{d^2z}{d\tau^2}-3\frac{dz}{d\tau}-\frac{\o a}4 z^3-\frac{(15\o a)}2z=0. 
\eq376
$$

Now let us take a special value for $\o a=-\frac4{15}$ and let us change $z$ into~$r$:
$$
z=i\sqrt5r. \eq377
$$
One gets
$$
\frac{d^2r}{d\tau^2}+3\frac{dr}{d\tau}-2r^3+2r=0. \eq378
$$
This equation can be solved in general ([112])
$$
r(\tau)=iC_1e^\tau \sn(C_1e^\tau+C_2,-1). \eq379
$$
Thus
$$
z(\tau)=-\sqrt5C_1e^\tau \sn(C_1e^\tau+C_2,-1). \eq380
$$

$\sn(u,-1)$ is a Jacobi elliptic function with a modulus $K^2=-1$, $C_1$
and~$C_2$ are constants. In this way
$$
\varPhi(t)=\frac1{2\a_s}\(1-\sqrt5C_1e^{H_0t}\sn(C_1e^{H_0t}+C_2,-1)\). \eq381
$$
However, we are forced to put $\o a=-\frac4{15}$. Let us remind that
$$
\o a=\frac{e^{n\varPsi_1}}{2r^2H_0^2}A \eq382
$$
and we are supposing $A>0$. However (see Eq.~(14.321) for a definition
of~$A$), it is possible to consider $A<0$ only for a pleasure to play. Thus we
use formula (14.381) to consider a dynamics of a Higgs field. 
Let us start an \il\ for $t=0$ with $\varPhi=0$.
$$
\varPhi(0)=0. \eq383
$$
One finds
$$
\frac1{C_1\sqrt5}=\sn(C_1+C_2,-1) \eq384
$$
(i.e.\ $t\dr{initial}=0$).

Let
$$
\varPhi(t\dr{end})=\frac1{\a_s}\(\frac{1+\sqrt5}2\) \eq385
$$
(a true minimum---a ``true'' vacuum). One gets
$$
\frac{1}{2}=-C_1e^{H_0t\dr{end}}\sn\(C_1e^{H_0t\dr{end}}
+C_2,-1\). \eq386
$$

Let us calculate the derivative of $\varPhi(t)$. One gets
$$
\al
\frac{d\varPhi}{dt}&=-\frac{\sqrt5}{2\a_s}C_1H_0e^{H_0t}
\Bigl(\sn\(C_1e^{H_0t}+C_2,-1\)\\
&+\cn\(C_1e^{H_0t}+C_2,-1\)\cdot\dn\(C_1e^{H_0t}+C_2,-1\)\Bigr).
\eal
\eq387
$$

Let us suppose a slow movement of $\varPhi$ such that
$$
\frac{d\varPhi}{dt}(0)=0. \eq388
$$
From (14.388) one gets
$$
\sc(C_1+C_2,-1)=-\dn(C_1+C_2,-1). \eq389
$$

This gives us an equation for a sum of integration \ct s. Thus we can get
from Eq.~(14.389) and Eq.~(14.384) integration \ct s $C_1$ and~$C_2$ and from
Eq.~(14.386) $t\dr{end}$, which can give us an amount of \il
$$
\o N_0=H_0t\dr{end}. \eq390
$$
In some sense Eq.~(14.386) is an equation for an amount of \il
$$
\frac{1}{2}=-C_1e^{\bar N_0}\sn\(C_1e^{\bar N_0}+C_2,-1\). \eq391
$$

Using Eq.~(14.389) and Eq.~(14.384) one easily gets
$$
C_1=-\sqrt{\frac{5-\sqrt5}{10}} \simeq -0.5256. \eq392
$$
One gets
$$
C_1+C_2=\frac1{\sqrt2}\(F(90^\circ\backslash 45^\circ)
-F(32^\circ\backslash 45^\circ)\)=0.9431 \eq393
$$
where $F$ is an elliptic integral of the first kind and
$$
\arcsin\(\sqrt{\frac{5-\sqrt5}{10}}\)\simeq\arcsin(0.5256)\simeq32^\circ. 
\eq394
$$

Let us come back to Eq.~(14.391) and let us denote
$$
e^{\bar N_0}=x_0. \eq395
$$
One gets using (14.393) and (14.392)
$$
C_2\simeq 1.4687 \eq396
$$
and finally
$$
1.4866x_0-5.0354=F\(\f/45^\circ\),
\eq397
$$
where 
$$
\cos\f=\frac{1.9026}{x_0}\,. \eq398
$$

From Eq.~(14.397) one finds
$$
\frac{2.8284}{\cos\f}-5.0354=\intop_0^\f \frac{d\theta}
{\sqrt{1-\frac12\sin^2\theta}}\,. \eq399
$$
We come back to this equation later.

Thus we get the following results. The Higgs field evolves from a ``false''
vacuum value ($\varPhi=0$) to the ``true'' vacuum value ($\varPhi=\varPhi_0$)
completing a second order phase transition. The potential of selfinteracting
$\varPsi$~field changes and a new equilibrium $\varPsi_0$ is attained. However, the
field~$\varPsi$ evolves from~$\varPsi_0$ to new value a little different
from~$\varPsi_0$ and afterwards a radiation era starts. The change of the value
of a scalar field~$\varPsi$ is small in terms of a variable~$y$ (see Eq.~(14.94)
for a definition) only from $\sqrt{\frac n{n+2}}$ to~1 ($n$~is a natural number
$n=\dim H$, and $H$ is at least~$G2$, $\dim G2=14$). Let us consider an
\ev\ of a field~$\varPsi$ in this epoch. From Eq.~(14.84) we get
$$
\ddot\f+\om^2_0 \f=0 \eq400
$$
where
$$
\ealn{
\om^2_0&=\frac{M^2\dr{pl}}{2\o M}(n+2)\(\frac n{n+2}\)^\frac n2
|\g|\(\frac{|\g|}\b\)^\frac n2 &(14.401)\cr
\varPsi&=\varPsi_0+\f &(14.402)
}
$$
and we linearize Eq.~(14.84) around~$\varPsi_0$ neglecting a term with a first
derivative of~$\f$. Thus a scalar field~$\varPsi$ undergoes small oscillations
around the equilibrium~$\varPsi_0$. These oscillations cannot be too long for a
field should change from
$$
\varPsi_0=\ln\(\sqrt{\frac{n|\g|}{(n+2)\b}}\) \eq403
$$
to
$$
\o\varPsi=\ln\(\sqrt{\frac{|\g|}{\b}}\). \eq404
$$
In terms of a field $\f$ from $\f=0$ to $\f=\frac12\ln(1+\frac2n)\simeq
\frac1n$. Thus we have
$$
\f=\f_0\sin(\om_0t). \eq405
$$

For $t=0$, $\f=0$ the time $\D t$ to go from $\f=0$ to $\f=\frac1n$ is simply
equal to
$$
\D t=\frac{\left|\arcsin(\frac1{\f_0n})\right|}{\om_0}
\le \frac\pi{2\om_0}\,. \eq406
$$
Thus the full amount of an \il\ in this epoch is equal to
$$
N_1=\D tH_1=\frac{H_1}{\om_0}\left|\arcsin\(\frac1{\f_0n}\)\right|
\le \frac{\pi H_1}{2\om_0}=\frac{\pi M\dr{pl}}{2\sqrt{6\o M}(n+2)}\,.
\eq407
$$
In our notation
$$
t_1=t_r+\D t \eq408
$$
(see Eq.~(14.114)).

After a time $t_1$ the \U\ goes to the phase of radiation domination with a
strong interaction between a radiation and a scalar field~$\varPsi$ up to a
minimal temperature where an ordinary (a~dust) matter appears. This matter
evolves afterwards in\dt ly of a radiation and of a scalar field~$\varPsi$. The
scalar field now is evolving as a \q.

In order to get some comparison let us consider an \ev\ of a scalar
field~$\varPsi$ during the second de Sitter phase in a slow-roll approximation.
Using Eq.~(14.84) one gets
$$
\frac{dy}{dt}=\frac{\o By^\frac{(n-2)}2(y^2-(\frac n{n+2}))}
{\sqrt{1-y^2}} \eq409
$$
or
$$
\int \frac{\sqrt{1-y^2}\,dy}{y^{(\frac{n-2}2)}(y^2-(\frac n{n+2}))}
=\o B(t-t_0) \eq410
$$
where
$$
\o B=2(n+2)|\g| \(\frac n{n+2}\)^\frac n2\(\frac{|\g|}\b\)^\frac n2 \eq411
$$
and
$$
y=\sqrt{\frac\b{|\g|}}\cdot e^\varPsi. \eq412
$$

Let
$$
I=\int \frac{\sqrt{1-y^2}\,dy}{y^{(\frac{n-2}2)}(y^2-(\frac n{n+2}))}
\simeq \int \frac{\sqrt{1-y^2}\,dy}{(y^2-(\frac n{n+2}))}
\eq413
$$
for
$$
\ealn{
&\frac n{n+2} \le y^2 \le 1 &(14.414)\cr
&n>\dim G2=14. &(14.415)
}
$$
For the integral $I$ one finds:
$$
\al
I&=-\arcsin(y)+\frac1{\sqrt{n(n+2)}}\sqrt{1-y^2}\\
&+\(\sqrt{\frac 2n}+\sqrt{\frac n2}\)\frac1{n+2} \cdot
\ln\[\frac{2\[\(1+y\sqrt{\frac n{n+2}}\)\(y+\sqrt{\frac n{n+2}}\)
+\sqrt{\frac 2{n+2}}\sqrt{1-y^2}\]}{\(y+\sqrt{\frac n{n+2}}\)}\]\\
&-\frac1{n+2}\(\sqrt{\frac 2n}+\sqrt{\frac n2}\) \cdot
\ln\[\frac{2\[\(1-y\sqrt{\frac n{n+2}}\)\(y-\sqrt{\frac n{n+2}}\)
+\sqrt{\frac 2{n+2}}\sqrt{1-y^2}\]}{\(y-\sqrt{\frac n{n+2}}\)}\].
\eal
\eq416
$$
In order to find an amount of an \il\ let us find limits of~$I$ for $y=1$ and
$y=\sqrt{\frac n{n+2}}$. One finds
$$
\ealn{
\lim_{y\to1}I &= -\frac\pi2+\(\sqrt{\frac 2n}+\sqrt{\frac n2}\)
\frac{\ln\(2\(1+\sqrt{\frac n{n+2}}\)\)}{(n+2)}\cr
&\hphantom{{}=-\frac\pi2}-\(\sqrt{\frac 2n}+\sqrt{\frac n2}\)
\frac{\ln\(2\(1-\sqrt{\frac n{n+2}}\)\)}{(n+2)} &(14.417)\cr
\lim_{y\to\sqrt{\frac n{n+2}}} I &=-\infty. &(14.418)
}
$$
Thus we see that a slow-roll approximation offers an infinite in time
\ev\ of a field~$y$ from $\sqrt{\frac n{n+2}}$ to~1. It is similar to an
\ev\ of a Higgs field in some slow-roll approximation schemes. The \il\ never
ends. Let us come back to the Eq.~(14.135) and let us calculate a Hubble \ct\
and a deceleration parameter for this model
$$
\ealn{
H&=\frac{\dot R}R=\(\frac{2|\g|^{n+4}}{3\o M\b^n(n-1)}\)
\frac{(t-t_1)^3}{\(1+\frac{|\g|^{n+4}}{3\o M\b^n(n-1)}
(t-t_1)^4\)} &(14.419)\cr
q&=-\frac{\ddot R R}{{\dot R}^2}= -\frac23\(\frac1{(t-t_1)^{4}}
+\frac{|\g|^{n+4}}{9\o M\b^n}\). &(14.420)
}
$$

Let us compare Eq.~(14.419) with the similar equation for radiation dominated
\U\ in General Relativity
$$
H\dr{GR}=\frac2{3(t-t_1)} \eq421
$$
and let us calculate a speed up factor
$$
\D_1=\frac H{H\dr{GR}}=\frac{|\g|^{n+4}}{\o M\b^n(n-1)}
\frac{(t-t_1)^4}{\(1+\frac{|\g|^{n+4}}{3\o M\b^n(n-1)}
(t-t_1)^4\)} \eq422
$$
and let us do the same for a model Eq.~(14.215) (i.e.\ radiation dominated
with a \q). Using Eq.~(14.216) one gets
$$
\D_2=\frac H{H\dr{GR}} = \frac{3\sqrt3 m\dr{pl}\rho_Q(t-t_1)}
{2B\tgh\(\frac{2\sqrt3\rho_Q}B m\dr{pl}(t-t_1)\)}\,. \eq423
$$
Let us notice that for a $(t-t_1)\sim0$
$$
\D_1\simeq 0 \eq424
$$
and
$$
\lim_{t\to\infty}\D_1=+\infty. \eq425
$$

Thus at early stages model Eq.~(14.135) is slower than that in General
Relativity. For large time the behaviour depends on details of the theory. In
the case of model Eq.~(14.215) we have
$$
\lim_{t\to t_1}\D_2=\tfrac34 \eq426
$$
and
$$
\lim_{t\to\infty}\D_2=+\infty. \eq427
$$
Thus in the first case $\D$ is monotonically going from~0 to
$+\infty$ and in the second case from $\frac34$ to infinity.
According to modern ideas an expansion rate in early \U\ has an important
influence on a production of light elements, the so called primordial
abundance of light elements (see Ref.~[113]). If at a beginning of primordial
nucleosynthesis the \U\ expansion rate is slower than in~GR, then we have
${}^4$He underproduction. This can be balanced by considering a larger ratio
of a number density of barions to number density of photons (remember the
barion number is a conserved quantity)
$$
\eta=\frac{n_B}{n_\g} \eq428
$$
where $n_B,n_\g$---barion and photon density numbers for a high redshift
$z=10^{10}$. In contrast to the latest if an expansion rate became faster
than in~GR during nucleosynthesis process, those bigger~$\eta$ (traditionally
one uses the so called) $\eta_{10}$
$$
\eta_{10}=10^{10}\eta \eq429
$$
do not result in excessive burning of deuterium because this happens in a
shorter time. The standard model of big-bang nucleosynthesis (SBBN) demands
$$
3\le \eta_{10}\le 5.6. \eq430
$$
It seems in the light of observational data from cosmic microwave background
(CMB) (BOOMERANG and MAXIMA) and from the Lyman $\a$-forest that
$\eta_{10}=8.8\pm1.4$, significantly higher than the SBBM value (for CMB) and
$8.2\pm2$ or $12.3\pm2$ for Lyman $\a$-forest. Thus the second model could
help in principle to explain the data without modification of nuclear
reaction rates due to neutrino degeneracy or introducing new decaying
particles. As the \U\ expands, cools and becomes more dilute, the nuclear
reactions cease to create and destroy nuclei. The abundances of the light
nuclei formed during this epoch are determined by the competition between a
time available (an expansion rate) and a density of reactans: neutrons and
protons. 

The abundances of D, ${}^3$He, ${}^7$Li are limited by an expansion rate and
are determined by the competition between the nuclear production/destruction
rates and an (universal) expansion of the \U. The SBBN theory is based on a
flat radiation dominated \U\ model in General Relativity and found in a
laboratory nuclear reaction rates. Thus it is strongly constrained. Any
significant discrepancy between observed and calculated value of~$\eta_{10}$
(known as baryometry) could be very dangerous. Thus changing the ratio of an
universal expansion relative to the model of \GR\ can give some margin in a
primordial alchemy.

In our theory after two \il ary phases we have in principle two radiation
dominated phases described by Eq.~(14.135) and Eq.~(14.215) with speed up
factors (14.422) and~(14.423). However, the important point is to match those
models. We get
$$
\al
R(t_d)&=R_0\exp\(H_0t_r+H_1(t_1-t_r)\)\(1+\frac{\o r}{(n-1)}\eta^2\dr{min}\)^\frac12\\
&=\root4\of{\frac B{\rho_Q}}\(\sh\(\frac{2\sqrt3\sqrt{\rho_Q}}B
m\dr{pl}\(t_1-\o t_0+\frac{\sqrt{2\o M}\b^\frac n4}{|\g|^{\frac{n+4}4}}\eta\dr{min}^\frac12
\)\)\)^\frac12 
\eal
\eq431
$$
and
$$
\al
B&=\rho_r(t_d)R^3(t_d)=\rho_r\(t_1+
\frac{\sqrt{2\o M}\b^\frac n4}{|\g|^{\frac{n+4}4}}\eta\dr{min}^\frac12\)R^3\(t_1+
\frac{\sqrt{2\o M}\b^\frac n4}{|\g|^{\frac{n+4}4}}\eta\dr{min}^\frac12\)\\
&=
\rho\dr{min}R_0^3\exp\(3H_0t_r+3H_1(t_1-t_r)\)\(1+\frac{\o r}{(n+1)}
\eta\dr{min}^2\)^\frac32 
\eal
\eq432
$$
where $\rho\dr{min}$ is given by
$$
\rho\dr{min}=M^2\dr{pl}\frac{\b^{n+1}}{|\g|^n}\(1+\eta\dr{min}\)^{2(n+1)}
W_4(1+\eta\dr{min}), \eq433
$$
$W_4(y)$ is given by formula (14.115) and $\eta\dr{min}$ by formula (14.122).

The time $t_r=t\dr{end}$ in any \il\ model for an \ev\ of the Higgs field in
the first de Sitter phase. The time $t_1=t_r+\D t$ can be calculated from
Eq.~(14.406) 
$$
\D t=\frac{\arcsin(\frac1{\f_0n})}{\om_0} \eq434
$$
and is bounded by
$$
\D t\le \frac1{\om_0}=\frac1{M\dr{pl}|\g|^\frac12}
\sqrt{\frac{2\o M}{(n+2)}}\(\frac{n+2}n\)^\frac n4\(\frac\b{|\g|}\)^\frac n4.
\eq435
$$
From Eq.\ (14.431) we can get the constant $\o t_0$
$$
\o t_0=t_1+\frac{\sqrt{2\o M}\b^\frac n4}{|\g|^{\frac{n+4}4}}\eta\dr{min}
^\frac12 -T \eq436
$$
where
$$
T=\frac{B\ars\[R_0^2\sqrt{\frac{\rho_Q}B}\exp\(2H_0t_r+2H_1(t_1-t_r)\)
\(1+\frac{\o r}{(n-1)}\eta^2\dr{min}\)\]}
{2\sqrt3\sqrt{\rho_Q}M\dr{pl}} \eq437
$$
and $B$ is given by Eq.~(14.432).

Let us notice that for $t=t_d$ the \U\ undergoes a phase transition due to a
change in an equation of state for a scalar field~$\varPsi$ from
$p_\varPsi=\frac43\rho_\varPsi$ in a first radiation dominated phase to
$p_\varPsi=-\rho_\varPsi$ in a second radiation dominated phase (a~\q\ phase). 

A matter which appears in the second phase does not play any r\^ole in an
\ev\ of the \U. For $t=t_d$ we have a discontinuity in Hubble constants
(parameters) for both phases
$$
H(t_d)=\frac{2|\g|^{n+4}}{3\o M(n-1)\b^n} \cdot 
\frac{(t_d-t_1)^\frac32}{\(1+\frac{\o r}{(n-1)}\eta\dr{min}
^2\)} \eq438
$$
in the first phase and
$$
\o H(t_d)=\sqrt3 m\dr{pl} \(\frac{\sqrt{\rho_Q}}B\)
\ctgh\(\frac{2\sqrt3}B \sqrt{\rho_Q} m\dr{pl} T\) 
\eq439
$$
for the second one,
$$
H(t_d)\ne \o H(t_d) \eq440
$$
$$
\rho_Q=\frac12\la_{c0}(\varPsi_0)=
\frac{|\g|^{\frac n2+1}n^\frac n2}{\b^\frac n2(n+2)^{\frac n2+1}} . \eq441
$$

Let us notice that a speed up factor $\D_1$ changes from
$$
\D_1(t_1)=0 \eq442
$$
to
$$
\D_1(t_d)=\frac{|\g|^{n+4}}{\o M(n-1)\b^n}\cdot\frac
{(t_d-t_1)^4}{\(1+\frac{\o r}{(n-1)}
\eta^2\dr{min}\)} \eq443
$$
and an expansion rate can be slower than in \GR.

The speed up factor $\D_2$ has a more complicated behaviour than~$\D_1$. It
is natural to expect a rapid speed up of a rate of expansion resulting in
non-equilibrium nuclear synthesis of light elements. Thus the presented
scenario offers interesting possibilities for an \ev\ of early \U. Recently
some papers have appeared on scalar-tensor theories of gravitation exploiting
the idea of a speed up factor in primordial nucleosynthesis~[114].

Finally let us come back to the model (14.221), (14.224).
We cannot invert analytically the formula~(14.224). Moreover taking under
consideration Eqs (14.242--245) and making some natural simplifications we
come to the following approximative formulae for $R(t)$.

For $x<1.1969$ (Eq.\ (14.243))
$$
R(t)=\root3\of{\frac A{\rho_Q}}\(0.44 + 0.085 N\) \eq444
$$
where
$$
N=\exp\(0.374 \frac{\sqrt{\rho_Q}}{M\dr{pl}}(t-t_0)+157.93\). \eq445
$$
One can calculate a Hubble parameter and a deceleration parameter
and gets:
$$
\ealn{
H&=\frac{\dot R}R = \frac{\sqrt{\rho_Q}}{M\dr{pl}}
\frac{3.17N}{44+8.5N} & (14.446)\cr
-q&=\frac{\ddot R R}{{\dot R}^2}=\frac{5.21}{N}+1. &(14.447)
}
$$
We have $H>0$, and $q<0$. Thus the model expands and
accelerates. First of all we consider
$$
\rho_m=\frac A{R^3} e^{aQ} 
$$
and we take boundary for $R$. In this way one gets
$$
\rho_m=a_i \cdot \rho_Q e^{aQ}, \quad i=1,2,3,4, \eq448
$$
$$
\al
a_1&=2.015\\
a_2&=0.034\\
a_3&=1060\\
a_4&=8
\eal
\eq449
$$
If we reconsider a contemporary \gr\ \ct~$G_N$ as $G_Ne^{-(n+2)\varPsi_0}$ we
would get interesting ratios. Thus we get
$$
\frac{\rho_m}{\rho_Q}=0.034 \div 2.015 \hbox{\quad or\quad }8\div 1060 \eq450
$$
which for the first is an excellent agreement with recent data concerning an acceleration
of the \U. Solving Eq.~(14.445) for $N$ if $R=1.33\sqrt{\frac A{\rho_Q}}$
gives us
$$
\ealn{
N&=10.47 &(14.451)\cr
H&=\frac{\sqrt{\rho_Q}}{M\dr{pl}}\cdot 0.25 &(14.452)\cr
-q&=1.5. &(14.453)
}
$$

Using Eq.\ (14.445) and Eq.\ (14.452) we can estimate a time of our
contemporary epoch
$$
t=-\frac 1h \cdot10^{15}\text{yr} - t_0 \eq454
$$
where $h$ is a dimensionless Hubble parameter, $H=h\cdot H_0$ (we take for a
contemporary Hubble parameter $H_0=100\frac{\text{km/s}}{\text{Mps}}$), 
$0.7<h<1$, which seems to be too much.

We can also estimate a density of a \q
$$
\rho_Q=\frac{H^2}{G_N}\cdot 0.571=h^2\cdot 0.08\cdot 10^{-23}
\frac{\text{kg}}{\text{m}^3}=8h^2\cdot 10^{-29}
\frac{\text{g}}{\text{cm}^3}\,. \eq455
$$

Let us come back to the Eqs (14.334) and (14.341). In Eq.~(14.341) a Higgs
field is given by a $P$~Weierstrass function. This function can be expressed
by Jacobi elliptic function
$$
P(z)=e_3+(e_1-e_2)\ns^2\(z(e_1-e_2)^\frac12\) \eq456
$$
with
$$
K^2=\frac{e_2-e_3}{e_1-e_3} \eq457
$$
where $e_1,e_2,e_3$ are roots of the polynomial
$$
4x^3-g_2x-g_3. \eq458
$$
Thus we should solve a cubic equation
$$
x^3-\frac1{12} \(5-3a^2\)-\frac1{12} \(\frac{a^2}{4}+\frac{10}9\)=0.
\eq459
$$
We want Eq.\ (14.459) to have all real roots. Thus we need
$$
p=-\frac1{12} \(5-3a^2\)<0 \eq460
$$
and
$$
D=\frac{p^3}{27}+\frac{q^2}4<0 \eq461
$$
where
$$
q=-\frac1{12}\(\frac{a^2}4+\frac{10}9\).\eq462
$$
Both conditions (14.460) and (14.461) are satisfied if
$$
0<a<0.3115\,. \eq463
$$
In this case we get
$$
\ealn{
e_1&=\frac13 \sqrt{5-3a^2} \cos\frac\f3 &(14.464)\cr
e_2&=\frac13 \sqrt{5-3a^2} \cos\frac{\f+2\pi}3 &(14.465)\cr
e_3&=\frac13 \sqrt{5-3a^2} \cos\frac{\f+4\pi}3 &(14.466)
}
$$
where
$$
\cos\f=-\frac{(9a^2+40)}{12(5-3a^2)^{3/2}}\,. \eq467
$$
From (14.464--466) one gets
$$
\ealn{
&K^2=\frac{\sin\frac\f3}{\sin\frac{\f+2\pi}3} &(14.468)\cr
&e_1>e_2>e_3 &\text{(14.469)}
}
$$
Thus
$$
\varPhi(t)=\frac{(5+3a)\sqrt{5+4a}}{8\a_s(A_1\ns^2(u)+A_2)}
+\frac{\(1+\sqrt{5+4a}\)}{2\a_s} \eq470
$$
where
$$
\ealn{
u&=\root4\of{\frac{Ae^{n\varPsi_1}\(5Ae^{n\varPsi_1}-3\a_s^2r^2\)}{r^4}}
\sqrt{\sin\frac{\f+\pi}3}(t-t_0) &(14.471)\cr
A_1&=e_1-e_2=\sqrt{\frac53-a^2}\,\sin\frac{\f+\pi}3 &(14.472)\cr
A_2&=e_3+\frac1{12}(5+6a)=\sqrt{\frac53-a^2}\,
\cos\frac{\f+4\pi}3+\frac1{12}(5+6a)&(14.473)
}
$$
and
$$
\ealn{
\cos\f&=-\frac{\sqrt A e^{\frac n2\varPsi_1}
\(9\a_s^2Br^2+40Ae^{n\varPsi_1}\)}{12\(5Ae^{n\varPsi_1}-3\a_s^2Br^2\)^{3/2}}
&(14.474)\cr
&0<\sqrt{\frac{\a_s^2Br^2}{Ae^{n\varPsi_1}}}<0.3115 &(14.475)
}
$$
and $a$ is given by Eq.\ (14.330).

The function $\ns(u,K)$ is an elliptic Jacobi function with a modulus~$K$
given by Eq.~(14.468)
$$
\ealn{
\ns(u,K)&=\frac1{\sn(u,K)} &(14.476)\cr
\sn(u,K)&=\operatorname{sinam}(u,K). &(14.477)
}
$$
In order to justify our treatment of Eqs (14.287) and (14.290) we consider
the full field equations
$$
R_{\mu \nu }-\frac12g_{\mu \nu }R=2\o\La u_\mu u_\nu - \o\La g_{\mu \nu } \eq478
$$
(where $\o\La$ is given by Eq.\ (14.293)).

However, in this case we consider $\o\La$ arbitrary. Let us consider static and
spherically symmetric metric
$$
ds^2=e^v\, dt^2-e^\la\, dr^2-r^2\(d\theta^2+\sin^2\theta\,d\f^2\) \eq479
$$
where $v=v(r)$ and $\la=\la(r)$.

From Eqs (14.478--479) one gets
$$
\ealn{
r^2\o\La &= e^{-\la}(rv'+1)-1 &\text{(14.480)} \cr
r^2\o\La &= e^{-\la}(r\la'-1)+1 &\text{(14.481)} \cr
\o\La' &= -\o\La v' &\text{(14.482)}
}
$$
where $'$ means derivation with respect to~$r$. Summing up (14.480) and
(14.481) one gets
$$
e^{-\la}(\la'+v')=2\o\La r. \eq483
$$
Moreover in our case $\o\La$ is very small. Thus we get approximately
$$
\frac{d\o\La}{dr}\simeq 0 \eq484
$$
and
$$
\la' \cong -v'. \eq485
$$
In this way we go to the solution (14.294) for a small $\o\La$ which is our
case. In this way $\la=-v$ and
$$
B(r)=e^v=e^{-\la}. \eq486
$$

Let us check a consistency of our solution. In order to do this we consider
Eqs (14.480--482) in full. One gets from Eq.~(14.482)
$$
\o\La=e^{-\mu_0}e^{-v}. \eq487
$$
For $\o\La$ is very small we should suppose that $\mu_0$ is positive and very
large ($\mu_0\to\infty$). From Eqs~(14.480--481) one gets
$$
\frac d{dr}(\la+v)=2e^{-\mu_0}\cdot r \cdot e^{(\la-v)}. \eq488
$$
Supposing that
$$
\la(r_0)+v(r_0)=0 \eq489
$$
for an established $r_0>0$ ($r_0$ is greater than a Schwarzschild radius of
the mass~$M_0$ from the solution (14.294)). Taking sufficiently big~$\mu_0$
we get approximately
$$
\frac d{dr}(\la+v)\simeq0 \eq490
$$
and of course
$$
\la\cong -v. \eq491
$$

Let us come back to the Eq.\ (14.399) and consider it for
$$
\f=k\pi+\tfrac\pi2 -\e, \quad k=0,\pm1,\pm2,\ldots, \eq492
$$
where $\e$ is considered to be small. One gets
$$
\frac{2.8284(-1)^k}{\sin\e}-5.0354=
\intop_0^{\frac\pi2-\e} \frac{d\theta}{\sqrt{1-\frac12\sin^2\theta}}
+2kK\(\tfrac12\), \eq493
$$
where
$$
K\(\tfrac12\)=\intop_0^{\frac\pi2}\frac{d\theta}{\sqrt{1-\frac12\sin^2\theta}}
=1.8541 \eq494
$$
is a full elliptic integral of the first kind for a modulus equal $\frac12$.
For a small~$\e$ we can write 
$$
\ealn{
\sin\e &\simeq \e &(14.495)\cr
\intop_0^{\frac\pi2-\e} \frac{d\theta}{\sqrt{1-\frac12\sin^2\theta}}
&\cong K\(\tfrac12\) - \sqrt2\, \e.&(14.496)}
$$

Thus one gets for $k=2l$ 
$$
\frac{2.8284}{\e}-5.0354=-\sqrt2\,\e+(4l+1)1.8541. \eq497
$$
It is easy to notice that we can realize a 60-fold \il\ in our model for
$$
x_0=\frac{1.9026}{|\e|} \eq497a
$$
can be arbitrary big for sufficiently big $l$.

Let us take $l=25\cdot 10^{24}$. In this case we have for $\e$ an equation
$$
\e^2-\bigl((4l+1)1.3110+5.0354\bigr)\e+2=0 \eq498
$$
or
$$
\e^2-\(10^{26}+6.0354\)\e+2=0, \eq498a
$$
i.e.
$$
\e^2-10^{26}\e+2=0, \eq498b
$$
and finally
$$
\ealn{
|\e|&\simeq 10^{-26} &(14.499)\cr
x_0&\simeq 1.9026\cdot 10^{26} &(14.500)\cr
\ln x_0&\simeq 60.51 &(14.501)
}
$$
which gives us a 60-fold \il.

Moreover the Eq.\ (14.399) has an infinite number of solutions for $0\leq \psi
\leq\frac\pi2$, $\f=k\pi+\psi$,
$$
\frac{2.8284}{\cos\psi}-5.0354=\intop_0^\psi \frac{d\theta}
{\sqrt{1-\frac12\sin^2\theta}}+7.5164\cdot l \eq502
$$
with
$$
x_0=\frac{1.9026}{\cos\psi},\quad
l=0,1,2,3,\ldots,\quad k=2l. \eq503
$$
One can find roots of Eq.\ (14.502) for 

\noindent
$l=5$
$$
\al
x_0&=16.83\\
\ln x_0&=2.82,
\eal \eq504a
$$
$l=50$
$$
\al
x_0&=0.24\cdot 10^3\\
\ln x_0&=5.48
\eal \eq504b
$$
and $\ln x_0$ for $l=5\cdot10^{24}$. In the last case one finds using
70-digit arithmetics
$$
\ln x_0\simeq 58.479\ldots \eq505
$$
which gives us an almost 60-fold \il. In general an amount of \il
$$
\o N_0=\ln x_0 \eq506
$$
is a function of $l=0,1,2,\ldots$ and probably can be connected to the Dirac
large number hypothesis. 

Finally let us take $l=10^n$ where $n>10$. In this case one can solve Eq.\
(14.498a) and get
$$
|\e|\simeq \frac2{[(4\cdot10^n+1)1.3110+5.0354]}\,. \eq507
$$
Thus for $x_0$ we find
$$
x_0\simeq 5\cdot 10^n \eq508
$$
and
$$
\o N_0(n) \simeq 1.60+2.30n. \eq509
$$
In this way, for large $l$, $\o N_0$ is a linear function of a logarithm
of~$l$, 
$$
\al
&\o N_0(10)\simeq 24.6, \ \o N_0(20)\simeq 47.6,\ \o N_0(24)\simeq 56.8,\\
&\o N_0(25)\simeq 59.1, \ \o N_0(26)\simeq 61.4
\eal \eq510
$$
or
$$
\o N_0(\log_{10} l)=1.6+\ln l. \eq511
$$
It is easy to see that Eq.\ (14.511) is an excellent approximation even for
$l=50$. 

Let us come back to the Eq.\ (14.186) in order to find a power spectrum for
our simplified model of \il. Using Eqs (14.318), (14.387), (14.392) and
(14.396) one gets after some algebra
$$
P_R(K) \cong 2.8944 \(\frac{H_0}{2\pi}\)^2 n_1\a_s^2
f\(\frac K{R_0H_0}\) \eq512
$$
where
$$\al
f(x)&=x^{-2}\Bigl(\sn(1.4687-0.5256x,-1)\\
&\qquad{}+\cn(1.4687-0.5256x,-1)
\dn(1.4687-0.5256x,-1)\Bigr)^{-2}. 
\eal \eq513
$$
Using some relations among elliptic functions one gets
$$
f(x)=\frac1{x^2} \cdot \frac{2\dn^4(u,\frac12)}
{\(\sn(u,\frac12)\dn(u,\frac12)+\sqrt2\,\cn(u,\frac12)\)^2}
\eq514
$$
where
$$
u=2.0770-0.7433x. \eq515
$$

Let us consider a more general situation for the Eq.\ (14.388), i.e.
$$
\frac{d\varPhi}{dt}(0)=h \eq516
$$
where $h\ne0$. In this way one gets
$$
h=-\frac{\sqrt5}{\a_s} C_1H_0\bigl(\sn(C_1+C_2,-1)
+\cn(C_1+C_2,-1)\dn(C_1+C_2,-1)\bigr). \eq517
$$
Using Eq.\ (14.384) one gets
$$
\al
C_2&=\frac{K(\frac12)}{\sqrt2}-\frac{\sqrt{10}}
{\sqrt{\sqrt{\bigl(1+\frac{\a_sh}{H_0}\bigr)^4+4} - 
\bigl(1+\frac{\a_sh}{H_0}\bigr)^2}}\\
&-\frac1{\sqrt2}\int_0^
{\arccos\sqrt{\frac{\sqrt{\(1+\frac{\a_sh}{H_0}\)^4+4} -
\(1+\frac{\a_sh}{H_0}\)^2}2}}
\frac{d\f}{\sqrt{1-\frac12\sin^2\f}}
\eal \eq518
$$
and
$$
|C_1|=\frac{\sqrt{10}}{\sqrt{\sqrt{\bigl(1+\frac{\a_sh}{H_0}\bigr)^4+4} -
\bigl(1+\frac{\a_sh}{H_0}\bigr)^2}}\,. \eq519
$$

Let us come back to the Eq.\ (14.386) to find an amount of \il\ for $C_2$
and~$C_1$ given by Eqs (14.518--519). One gets
$$
\sqrt2 C_2 - K\(\tfrac12\) = \intop_0^{\arccos\bigl(\frac1{2C_1e^{N_0}}\bigr)}
\frac{d\theta}{\sqrt{1-\frac12\sin^2\theta}} - \sqrt2 C_1 e^{N_0}, \eq520
$$
or using a natural substitution
$$
\cos\f=\frac1{2C_1e^{N_0}}\,, \eq521
$$
$$
\sqrt2 C_2 - K\(\tfrac12\) = \intop_0^\f
\frac{d\theta}{\sqrt{1-\frac12\sin^2\theta}} - \frac{\sqrt2}{2\cos\f}\,. \eq522
$$
Taking as usual
$$
\f=2l\pi+\tfrac\pi2-\e, \quad l=0,1,2,\ldots, \eq523
$$
one gets for small $\e$
$$
\e\simeq \frac1{\sqrt2 (4l+2)K(\frac12)-2C_2} \simeq 
\frac1{(4l+2)\sqrt2 K(\frac12)}\eq524
$$
(for large $l$) and eventually
$$
\al
&e^{N_0}\simeq \frac{4\sqrt2 K(\frac12)}{C_1}l\\
&N_0=\ln\(\frac{4\sqrt2 K(\frac12)}{C_1}\)+\ln l.
\eal \eq525
$$

Let us calculate $\frac{d\varPhi}{dt}$ for $t=t^\ast=\frac1{H_0}
\ln\bigl(\frac K{R_0H_0}\bigr)$. One gets
$$
\al
\frac{d\varPhi}{dt}(t^\ast)&=
-\frac{\sqrt5}{2\a_s}C_1H_0 \(\frac K{R_0H_0}\)
\biggl(\sn\Bigl(C_1\Bigl(\frac K{R_0H_0}\Bigr)+C_2,-1\Bigr)\\
&+\cn\Bigl(C_1\Bigl(\frac K{R_0H_0}\Bigr)+C_2,-1\Bigr)
\dn\Bigl(C_1\Bigl(\frac K{R_0H_0}\Bigr)+C_2,-1\Bigr)\biggr).
\eal \eq526
$$
Thus we can write down a $P_R(K)$ function.
$$
P_R(K)=\(\frac{H_0}{2\pi}\)^2 \cdot \frac{4\a_s^2n_1}{5C_1^2}\,
f\(\frac K{R_0H_0}\) \eq527
$$
where
$$
f(x)=\frac1{x^2}\bigl(\sn(C_1x+C_2,-1)+\cn(C_1x+C_2,-1)\dn(C_1x+C_2,-1)\bigr)
^{-2}.
\eq528
$$
Using some relation among elliptic functions one finds
$$
P_R(K)=\(\frac{H_0}{2\pi}\)^2 \cdot \frac{4\a_s^2n_1}{5C_1^2}\,
g\(\frac K{R_0H_0}\) \eq529
$$
where
$$
g(x)=\frac1{x^2} \cdot \frac2
{\(\sd(u,\frac12)+\sqrt2 \cd(u,\frac12)\nd(u,\frac12)\)^2} 
\eq530
$$
and
$$
u=\sqrt2 C_1x+\sqrt2 C_2\,. \eq531
$$

Moreover we can reparametrize (14.530--531) in the following way
$$
u=\sqrt2 C_1x+K(\tfrac12)-C_1\sqrt2-
\intop_0^{\arccos\(\frac{\sqrt5}{C_1}\)}
\frac{d\f}{\sqrt{1-\frac12\sin^2\f}} \eq532
$$
where
$$
\ealn{
\frac{\a_sh}{H_0}&=-1\pm \frac1{|C_1|}\sqrt{\frac{C_1^4-25}5} &(14.533)\cr
|C_1|&>\sqrt5. &(14.534)
}
$$
For large $C_1$ one gets
$$
\ealn{
&u\cong \sqrt2 C_1(x-1) &(14.535)\cr
&\frac{\a_sh}{H_0}\simeq \frac{C_1}{\sqrt5}\,. &(14.536)
}
$$

Using Eq.\ (14.525),
$$
u=\frac{4\sqrt2 K(\frac12)l}{e^{N_0}}\,(x-1). \eq537
$$
If we take large $C_1$ (it means, a large $h$)
$$
h=\frac{C_1H_0}{\sqrt5 \a_s} \eq538
$$
and simultaneously sufficiently large $l$ we can achieve a 60-fold \il\ with
a function $P_R(K)$ given by the formula (14.530). Large $C_1$ means here
$C_1\simeq 100$, large $l$ means $l\simeq 10^{25}$.

Let us calculate the spectral index for our $P_R(K)$ function, i.e.
$$
n_s(K)-1 = \frac{d\ln P_R(K)}{d\ln K}\,.\eq539
$$
One gets
$$
n_s(K)-1 = -2-2C_1x\,\frac{\sqrt2 \cd(u,\frac12)-\sd^3(u,\frac12)}
{\sd(u,\frac12)+\sqrt2 \cd(u,\frac12)\nd(u,\frac12)} \eq540
$$
where
$$
\al
u=\sqrt2 (C_1x+C_2)&=\frac{\sqrt2}{R_0H_0}\,(C_1K+C_2R_0H_0)\\
x&=\frac K{R_0H_0}
\eal
\eq541
$$
A very interesting characteristic of $P_R(K)$ is also $\dfrac{dn_s(K)}{d\ln
K}$. One gets
$$
\al
\frac{dn_s(K)}{d\ln K}&=
-2C_1x\,\frac{\sqrt2 \cd(u,\frac12)-\sd^3(u,\frac12)}
{\sd(u,\frac12)+\sqrt2 \cd(u,\frac12)\nd(u,\frac12)}\\
&{}-2C_1^2x^2\sd^2(u,\tfrac12)-2C_1^2x^2
\frac{\(\sqrt2 \cd(u,\frac12)-\sd^3(u,\frac12)\)^2}
{\(\sd(u,\frac12)+\sqrt2 \cd(u,\frac12)\nd(u,\frac12)\)^2}
\eal
\eq542
$$
where $u$ and $x$ are given by Eq.\ (14.541).

Let us take for a trial $C_1=-0.5256$ and $C_2=1.4687$. In this case one
finds 
$$
n_s(K)-1 = -2+1.0512x\,\frac{\sqrt2 \cd(u,\frac12)-\sd^3(u,\frac12)}
{\sd(u,\frac12)+\sqrt2 \cd(u,\frac12)\nd(u,\frac12)} \eq543
$$
$$
\al
\frac{dn_s(K)}{d\ln K}&=
1.0512x\,\frac{\sqrt2 \cd(u,\frac12)-\sd^3(u,\frac12)}
{\sd(u,\frac12)+\sqrt2 \cd(u,\frac12)\nd(u,\frac12)}\\
&{}-0.5525x^2\sd^2(u,\tfrac12)-0.5525x^2
\frac{\(\sqrt2 \cd(u,\frac12)-\sd^3(u,\frac12)\)^2}
{\(\sd(u,\frac12)+ \sqrt2\cd(u,\frac12)\nd(u,\frac12)\)^2}
\eal
\eq544
$$
$$
\al
u&=2.0771-0.7433x\\
x&=\frac K{R_0H_0}
\eal
\eq544a
$$

The interesting point is to find $n_s(K)\simeq1$ (a~flat power spectrum). One
gets
$$
\sd(u,\tfrac12)+ \sqrt2\cd(u,\tfrac12)\nd(u,\tfrac12)=
C_1x\(\sd^3(u,\tfrac12)-\sqrt2 \cd(u,\tfrac12)\) \eq545
$$
and
$$
\sd(u,\tfrac12)+ \sqrt2\cd(u,\tfrac12)\nd(u,\tfrac12)\ne0. \eq546
$$
Using (14.545--546) one gets
$$
\frac{dn_s}{d\ln K} = -2C_1^2x^2\sd^2(u,\tfrac12) \eq547
$$
if Eqs (14.545--546) are satisfied.

In the case of special $C_1$ and $C_2$ one gets
$$
\frac{dn_s}{d\ln K} = -0.5525x^2\sd^2(u,\tfrac12) \eq548
$$
where $x,u$ are given by Eq.\ (14.544a).

Let us reparametrize the Eq.\ (14.545). One gets
$$
\sd(u,\tfrac12)+ \sqrt2\cd(u,\tfrac12)\nd(u,\tfrac12)=
\frac{u-\sqrt2 C_2}{\sqrt2}\(\sd^3(u,\tfrac12)-\sqrt2 \cd(u,\tfrac12)\). \eq549
$$
For sufficiently big $u$ one gets (the equation has infinite number of roots)
$$
u=u_1+2lK(\tfrac12),\qquad l=0,\pm1,\pm2,\ldots \eq550
$$
where $u_1$ satisfies the equation
$$
\sn(u_1,\tfrac12)=\frac13\(1+\root3\of5
\(\root3\of{65+\sqrt{20357}}-\root3\of{\sqrt{20357}-65}\)\)^\frac12
\simeq 0.65\ldots \eq551
$$
Moreover for $x=\frac K{R_0H_0}>0$ we have the condition
$$
\frac{u_1+2lK(\frac12)-\sqrt2 C_2}{\sqrt2C_1}>0. \eq552
$$

One finds
$$
\sd^2(u_1,\tfrac12)=0.53\ldots \eq553
$$
Thus
$$
\al
\frac{dn_s}{d\ln K}\(u_1+2lK(\tfrac12)\) &\simeq
-0.53\(\frac{u_1+2lK(\frac12)-\sqrt2 C_2}{\sqrt2C_1}\)^2,\\
 l&=0,\pm1,\pm2,\ldots 
\eal
\eq554
$$
One gets
$$
\ealn{
&\arcsin0.65 \simeq 40^\circ.54 &(14.555)\cr
&u_1\cong \intop_0^{40^\circ.54}\frac{d\theta}{\sqrt{1-\frac12 \sin^2\theta}}
=F(40^\circ.54/45^\circ)\cong 0.73\ldots &(14.556)
}
$$
Taking special value of $C_1$ and $C_2$ one gets
$$
x\cong 1.81 - 4.68l \eq557
$$
where $l=0,\pm1,\pm2,\ldots$. For $x>0$ it is easy to see that $l$~should be
nonpositive. Practically $l$~should be a negative integer
$$
l<-3 \eq558
$$
and
$$
\frac{dn_s}{d\ln K} \cong -0.29 (1.81-4.68l)^2. \eq559
$$

Thus for 
$$
K_l=R_0H_0(1.81-4.68l)
$$
we get
$$
n_s(K_l)\cong1. \eq560
$$
The important range of $\ln K$, $\D\ln K$ is about 10.

Thus
$$
n_s(K)\simeq 1\pm2.9(1.81-4.68l)^2. \eq561
$$
Taking $l=-3$ one gets
$$
-300 < n_s(K) < 300 \eq562
$$
and
$$
P_R(K) \sim K^{1-n_s(K)}
$$
is not flat in the range considered.

Moreover we can improve the results considering Eq.~(14.554) for large~$C_1$.
For large $C_1$ one gets
$$
\frac{C_2}{C_1}\simeq -1. \eq563
$$
Thus we find
$$
\frac{dn_s}{d\ln K} \simeq -1.06\(u_1+2lK(\tfrac12)+C_1\)^2. \eq564
$$
Taking large value of $C_1$ in such a way that
$$
C_1=-2lK(\tfrac12)-u_1+\e \eq565
$$
where $\e$ is a small number,
$$
\e\simeq 10^{-n}, 
$$
one gets
$$
\frac{dn_s}{d\ln K}\simeq -1.06\cdot 10^{-2n}. \eq566
$$
The last condition means that we should take
$$
h\cong \frac{H_0|C_1|}{\sqrt5 \a_s}= 
\frac{H_0}{\sqrt5 \a_s}\,\(0.73-2lK(\tfrac12)-10^{-n}\) \eq567
$$
where $l$ is an integer, $l\simeq-100$.

Thus for some special values of $h$ we can get arbitrarily small
$\dfrac{dn_s}{d\ln K}$ which means we can achieve a flat power spectrum in
the range considered ($\D\ln K\sim 10$),
$$
n_s(K)=1\pm10^{-(2n-1)}, \eq568
$$
$n$ arbitrarily big.

Let us consider the value of $K$ coresponding to our value of $n_s(K)=1$. One
gets 
$$
x=\frac{0.73+2lK(\frac12)-\sqrt2 C_2}{\sqrt2 C_1}\,. \eq569
$$
Using our assumption on a large $C_1$ and Eq.\ (14.565) one finds
$$
x\cong \(1-\frac1{\sqrt2}\)-\frac \e {2lK(\frac12)} \eq570
$$
or (for $\e$ is small and $l$ quite big)
$$
x\simeq 0.293 \eq571
$$
and
$$
K\simeq 0.3R_0H_0 \eq572
$$
with the range $\D\ln K\sim10$. It means that
$$
0.3e^{-10} \le \frac K{R_0H_0} \le 0.3\cdot e^{10} \eq573
$$
which gives us a full \co ly interesting region
$$
10^{-4} \le \frac K{R_0H_0} \le 10^4. \eq574
$$

We consider some \co\ consequences of the Nonsymmetric
Kaluza--Klein (Jordan--Thiry) Theory. Especially we use the scalar
field~$\P$ appearing in order to get \co\ models with a \q\ and phase
transitions. We consider a dynamics of Higgs' fields with various
approximations and models of \il.

Eventually we develop a toy model of this dynamics to obtain an amount of
\il\ and $P_R(K)$ function (a~spectral function for fluctuations). We
calculate a spectral index $n_s(K)$ and $\dfrac{dn_s}{d\ln K}$ for this
model (see Ref.~[115]). 

\smallskip
Finally let us consider in this section two problems. First we match our
solutions of an \ev\ of the \U, i.e.\ both de Sitter phases, radiation
dominated \U, matter dominated \U\ and K-essence \U\ (kinetic energy
dominated \q). In order to simplify calculations we match the first de Sitter
phase to radiation dominated \U\ using some issues from the second de Sitter
phase. Secondly we write down solution to the \ev\ of \U\ with a \co\ \ct\
(from the second de Sitter phase), with a radiation and with a matter
(a~dust). All of these material ingredients are evolving in\dt ly in this
case, similarly as a matter (dust) and a \q\ for the model \rf224 . 

Let us consider Eq.\ \rf50 \ and let us match it to Eq.~\rf215 . One gets
$$
\ealn{
R_1&=\root4\of{\frac B{\rho_Q}} \sqrt{\sh\(\frac{2\sqrt3\sqrt{\rho_Q}M\dr{pl}}
{B} t\dr{end}\)} &\rf575 \cr
R_1&=R_0e^{H_0t\dr{end}} &\rf576 
}
$$
where
$$
t\dr{end}=\frac{N_0}{H_0} \eq577
$$
is the time of the end of the first de Sitter phase. ($N_0$ is the amount of
an \il.) 

Simultaneously we need a local conservation of an energy, i.e.
$$
\t \rho_r(R_1)=\frac B{R_1^4} \eq578
$$
where
$$
\t \rho_r(R_1)=6\(H_0^2-H_1^2\) \eq579
$$
(see the formula (19.57)).
After some calculations one gets
$$
B=\frac{2\sqrt3 M\dr{pl}\sqrt{\rho_Q}t\dr{end}}{\ars\(\sqrt{\frac{\rho_Q}
{\u\rho_r}}\)}\,. \eq580
$$

Now we need match a solution with a radiation dominated period to the
solution with matter dominated period (with the same \co\ \ct---\q\ energy
density). One gets
$$
\frac A{R_2^3}+\rho_Q= \frac B{R_2^4}+\rho_Q \eq581
$$
and finally
$$
A=\frac B{R_2} \eq582
$$
where
$$
R_2=0.7916\root3\of{\frac A{\rho_Q}}=\root4\of{\frac B{\rho_0}}
\sqrt{\sh\(\frac{2\sqrt3\sqrt{\rho_Q}M\dr{pl}}{B}\,t_1\)}\,, \eq583
$$
$t_1$ is a time to match (see \rf247 ). Let us consider
$$
N_1=H_1\(t_1-t\dr{end}\) \eqno(\ast)
$$
as an amount of an \il\ during the second de Sitter phase. In this way one
gets 
$$
t_1=\frac{N_1}{H_1}+\frac{N_0}{H_0} \eq584
$$
and finally
$$
\ealn{
A&=2.015\cdot \frac{M\dr{pl}^{3/4}\rho_Q^{5/8}\(\frac{N_0}{H_0}\)^{3/4}}
{\ars^{4/3}\sqrt{\frac{\rho_Q}{\u \rho_r}}}\,
\sh^{3/2}\(\sqrt{\frac{\rho_Q}{\t\rho_r}}\(\frac{N_1}{N_0}\,
\frac{H_0}{H_1}+1\)\) &\rf585 \cr
B&=\frac{2\sqrt3\,M\dr{pl}\sqrt{\rho_Q}}{\ars\(\sqrt{\frac{\rho_Q}{\u
\rho_r}}\)} \,\frac{N_0}{H_0}\,. &\rf586
}
$$
However, we should match also a time. In this way we get from Eqs \rf444--445
$$
t_0=\frac{418.48}{\sqrt{\rho_Q}}\,M\dr{pl}+\frac{N_1}{H_1} \eq587
$$
or using the value $\rho_Q=\la_{c0}=10^{-52}\frac1{{\text m}^2}$ we get
$$
t_0=0.447\cdot 10^{14}\text{yr}+\frac{N_1}{H_1}\,. \eq588
$$

If $R_0$ is an initial value of ``a radius'' of the \U\ we get
$$
\al
R_0=R_1e^{-N_0}
&=\root4\of{\frac
{2\sqrt3\,M\dr{pl}\sqrt{\rho_Q}}{\t\rho_r\ars\(\sqrt{\frac{\rho_Q}{\u
\rho_r}}\)}}
\cdot \exp\(-\(N_0+\frac{n+2}4 \P_0\)\)\cr
&=\root4\of{\frac
{2\sqrt3\,M\dr{pl}\sqrt{\rho_Q}}{\t\rho_r\ars\(\sqrt{\frac{\rho_Q}{\u
\rho_r}}\)}}\cdot\(\frac{(n+2)\b}{n|\g|}\)^{(n+2)/8}e^{-N_0}. 
\eal
\eq589
$$
The only one ingredient from the second de Sitter phase is using the formula
$(\ast)$. Now we should match the solution \rf444--445 \ with conditions
\rf247 \ to the solution \rf281 , \rf273 , \rf279 . We should match a
field~$\P$, a density of an energy, a radius and a time. One gets
$$
\ealn{
&R_3=\o R_0 \exp\(\frac{n+1}{M\dr{pl}}(t_2-\o t_0)\)=
3.07\root3\of {\frac A{\rho_Q}} &\rf590 \cr
&\P(t_2)=\frac1{2n}\ln\(\frac\d{3|\g|}\)-\frac1n \sqrt{\frac{2\d}{3|\o M|}}
\,M\dr{pl}(t_2-\o t_0)=\P_0 &\rf591 \cr
&\rho_Q+\frac A{R_3^3}=\rho_0e^{(n+2)\P_0}+\rho_\P &\rf592 
}
$$
where
$$
\ealn{
&\rho_\P=\frac{\d M^2\dr{pl}}{3n^2}-\frac{\sqrt{|\g|\d}}{2\sqrt3}
\exp\(-\sqrt{\frac{2\d}{3\o M}}\, M\dr{pl}(t_2-\o t_0)\) &\rf593 \cr
&\d=2\,\frac{n+1}{M^2\dr{pl}}-\frac{3\rho_0}{M^2\dr{pl}} &\rf594 \cr
&\o t_0\ne0. &\rf594a
}
$$

One finds:
$$
t_2=\sqrt{\frac{3\o M}{2\d}}\,\frac1{M\dr{pl}}\ln\(\sqrt{
\frac{\d(n+2)^n\b^n}{3|\g|^{n+1}n^n}}\)+\o t_0 \eq595
$$
and $t_2$ is large,
$$
\ealn{
&\rho_\P\cong \frac{\d M^2\dr{pl}}{3n^2}=\frac{M^2\dr{pl}}{3n^2}
\(2\,\frac{n+1}{M^2\dr{pl}}-\frac{3\rho_0}{M^2\dr{pl}}\) &\rf596 \cr
&\rho_0=\frac{n^2\la_{c0}-3(n+1)}{n^2x_0^{n+2}-1}\,. &\rf597
}
$$
From
$$
\al
\frac A{R^3_3}&=-\rho_Q+\rho_0\(x_0^{n+2}-\frac1{n^2}\)+\frac{3(n+1)}{n^2}\\
R_3&=3.072\root3\of{\frac A{\rho_Q}}
\eal
\eq598
$$
one obtains
$$
\d=\frac1{M^2\dr{pl}}\(2(n+1)-3\,\frac{\la_{c0}n^2-3(n+1)}{n^2x_0^{n+2}-1}\).
\eq599
$$
From Eqs \rf247 , \rf444 \ and \rf445 \ we get
$$
\ealn{
&t_2=\frac{N_1}{H_1}+\frac{5.38}{\sqrt{\rho_Q}}\,M\dr{pl} &\rf600 \cr
&\o t_0=\frac{N_1}{H_1}+\frac{5.38}{\sqrt{\rho_Q}}\,M\dr{pl}
-\sqrt{\frac{3\o M}{2\d}} \, \frac1{M\dr{pl}} \ln\(\sqrt{
\frac{\d(n+2)^n\b^n}{3|\g|^{n+1}n^n}}\)
&\rf601 \cr
&M\dr{pl}\frac{5.38}{\sqrt{\rho_Q}}=5.64\cdot 10^{10}\,\text{yr}. &\rf602
}
$$
Finally we find $\o R_0$:
$$
\o R_0=R_3\exp\(-\frac{n+1}{M\dr{pl}}(t_2-\o t_0)\)=
3.072\root3\of{\frac A{\rho_Q}} \exp\(-\frac{n+1}{M\dr{pl}}(t_2-\o t_0)\).
\eq603
$$

In this way we match an \ev\ of the \U\ from the very beginning up to
\hbox{K-essence} dominance. We find all the \ct s. Let us notice that the solution
\rf247 , \rf444--445 \ has a quite known behaviour in the theory of ordinary
nonlinear differential equations. The solution cannot be extended after some
value of the \dt\ variable (and also in\dt). One says the solution dies. Due
to this interesting behaviour we obtain physical consequences. Matter and a \q\
cannot evolve in\dt ly. We get also a ratio between a density of matter and a
\q\ which agrees with observational data and helps (in principle) to
calculate a time of our contemporary epoch.

If we use Eqs\ \rf444 \ and \rf445 \ we can calculate a time of our
contemporary epoch, i.e.\ a time when the ratio of a density of a matter to a
\q\ density
is~$\frac37$ (this is this ratio measured now). One gets
$$
\o t\dr{contemporary}=8.97\,\frac{M\dr{pl}}{\sqrt{\rho_Q}}+\frac{N_1}{H_1}
=9.42\cdot 10^{10}\,\text{yr}+\frac{N_1}{H_1}\,. \eqno\text{(I)}
$$
In the same way using Eqs \rf444 , \rf445 \ and \rf446 \ we calculate a
Hubble parameter for our contemporary epoch.
$$
H\dr{contemporary}=h\dr{contemporary}\frac{100\frac{\text km}{\text s}}
{\text{Mpc}} \eqno\text{(II)}
$$
where
$$
h\dr{contemporary}=1.21. \eqno\text{(IIa)}
$$
Both values are a little too big. 

Moreover one can improve the results. However, the age of the \U\ is a little
longer,
$$
t\dr{age of the \U}=t\dr{contemporary}+t_1=
9.42\cdot 10^{10}\,\text{yr}+2\,\frac{N_1}{H_1}+\frac{N_0}{H_0}\,. \eqno
\text{(Ia)}
$$
In general one gets
$$
H=\frac{\sqrt{\rho_Q}}{M\dr{pl}}\,\frac{32.27\root3\of{\frac{\rho_Q}{\rho_m}}
-16.38}{14.96\root3\of{\frac{\rho_Q}{\rho_m}}-0.03}
$$
and
$$
-q=\frac{5.21}{11.76\root3\of{\frac{\rho_Q}{\rho_m}}-5.17}+1
$$
where $\frac{\rho_Q}{\rho_m}$ is a ratio of a \q\ energy density to a matter
(dust) energy density, or
$$
h=0.9115\,\frac{32.27\root3\of{\frac{\rho_Q}{\rho_m}}
-16.38}{14.96\root3\of{\frac{\rho_Q}{\rho_m}}-0.03}\,.
$$

Finally, let us express an age of the \U\ in terms of the ratio
$\frac{\rho_Q}{\rho_m}$. One gets
$$
t\(\frac{\rho_Q}{\rho_m}\)=2.667 \ln\(31.87\root3\of{\frac{\rho_Q}{\rho_m}}
-14.03\)\frac{M\dr{pl}}{\rho_Q} + 2\(\frac{N_1}{H_1}\)+\(\frac{N_0}{H_0}\)\,
$$
or
$$
t\(\frac{\rho_Q}{\rho_m}\)=2.82\cdot 10^{10}\,\text{yr}\cdot
 \ln\(31.87\root3\of{\frac{\rho_Q}{\rho_m}}
-14.03\)\frac{M\dr{pl}}{\rho_Q} + 2\(\frac{N_1}{H_1}\)+\(\frac{N_0}{H_0}\)\,.
$$
For $\frac{\rho_m}{\rho_Q}=0.0034$ one gets
$$
t(294.12)=14.91\cdot 10^{10}\,\text{yr}+
2\(\frac{N_1}{H_1}\)+\(\frac{N_0}{H_0}\)\,. 
$$
For $\frac{\rho_m}{\rho_Q}=2.015$ 
$$
t(0.496)=6.76\cdot 10^{10}\,\text{yr}+
2\(\frac{N_1}{H_1}\)+\(\frac{N_0}{H_0}\)\,. 
$$
$$
\D t=t(294.12)-t(0.496)=8.15\cdot 10^{10}\,\text{yr}.
$$
The last number is a duration time of the epoch for $\rho_Q$ is equal to the
contemporary value of a \co\ \ct.

Let us come back to the Eq.\ \rf209 \ and consider it in the flat case $K=0$.
One gets
$$
\frac{R\,dR}{\sqrt{\frac{\rho_Q}{3\Mp^2}\cdot R^4+\frac A{3\Mp^2}\cdot R
+\frac B{3\Mp^2}}}=\pm dt. \eq604
$$
We change $R$ into $x$:
$$
R=\root3\of{\frac A{\rho_Q}}\,x \eq605
$$
and finally get
$$
\pm\int\frac{x\,dx}{\sqrt{x^4+x+a}}=\sqrt{\frac{\rho_Q}{3\Mp^2}}\,(t-t_0)
\eq606
$$
where
$$
0<a=\frac BA \root3\of{\frac{\rho_Q}A}\,. \eq607
$$

This is an \ev\ of a flat \U\ with radiation, matter (dust) and a density of
a \q\ (this is a \co\ \ct\ $\la_{c0}$). The integral on LHS of Eq.~\rf606 \
is an elliptic integral and can be calculated. The properties of the result
of calculations strongly depend on the value of~$a$. Let us notice that the
integral we have calculated here has $a=0$ (no radiation).

The \ev\ of a \q, a matter and a radiation is here in\dt. One gets the
following results for
$$
I=\int\frac{x\,dx}{\sqrt{x^4+x+a}}\,. \eq608
$$

In general there are two cases:
$$
\displaylines{
\text{I\hphantom{I}} \hfill B_2>0 \hfill \rf609 \cr
\text{II} \hfill B_2<0 \hfill \rf610
}
$$
where
$$
B_2=\frac{\sqrt{12b_1^4-a}-2b_1^2-\sqrt{4b_1^4-a}}
{2\sqrt{12b_1^4-a}} \eq611
$$
and $b_1>0$ is a solution of the cubic equation
$$
\displaylines{
\hfill y^3-ay-\tfrac18=0 \hfill \rf612 \cr
\hfill y=2b_1^4 \hfill \rf613
}
$$
Eq.\ \rf612 \ can be solved by using the Cardano formulae in two cases

\item{1)} $D>0$,
\item{2)} $D<0$,

$$
D=\frac{27-256a^3}{27\cdot256}\,. \eq614
$$
In case 1) one gets
$$
\al
b_1&=\frac12 \root4\of{\root3\of{4+\tfrac12\sqrt{27-256a^3}}
+ \root3\of{4-\tfrac12\sqrt{27-256a^3}}}>0\\
0&<a<\frac3{4\root3\of4}=0.472470393\ldots
\eal \eq615
$$
In case 2) one gets
$$
b_1=\root4\of{\frac a3}\sqrt{\cos\frac\f3} \eq616
$$
where
$$
\displaylines{
\hfill \cos\f=\frac3{16a}\sqrt{\frac3a} \hfill \rf617 \cr
\hfill a>\frac3{4\root3\of4}\,, \quad \f\in\langle0,\tfrac\pi2\rangle. \hfill 
\rf618
}
$$
Condition I can be transformed into
$$
4b_1^2<a+1 \eq619
$$
for both cases 1) and 2). In case 1) this condition has no solution. It means
we have always $B_2<0$. In case~2) we have always \rf619 . It means
$$
B_2>0 \eq620
$$
for an equation
$$
\frac1{\root4\of4}\,\frac{\cos^2\(\frac{\f}3\)}{\cos^{4/3}\f}=
\frac3{4\root3\of4 \cos^{2/3}\f}+1. \eq621
$$
has no solution in $(0,\frac\pi2)$ and in
$(\frac{3\pi}2, 2\pi)$.

The fact that in case 1) we have always
$$
4b_1^2>a+1 \eq622
$$
is caused by a nonexistence of real solutions of an equation
$$
4(a+1)^{1/2}=\root3\of{4+\tfrac12\sqrt{27-256a^3}}
+\root3\of{4-\tfrac12\sqrt{27-256a^3}} \eq623
$$
where
$$
0<a<\frac3{4\root3\of4}\,. \eq624
$$
Thus for sufficiently small $a$ we always have case~I (in a limit $a=0$, no
radiation also).

One can express $b_1$ and $a$ in terms of $\f$:
$$
\ealn{
b_1^4&=\frac3{4\root3\of4}\,\frac{\cos^2(\frac\f3)}{\cos^{2/3}\f} &\rf625 \cr
a&=\frac3{4\root3\of4}\cos^{-2/3}\f. &\rf626
}
$$
Thus one obtains in case I
$$
\ealn{
&I=\frac12 \ln \frac {e(\frac{x-\a}{x-\b})^2 +f
+2\sqrt{\bigl((\frac{x-\a}{x-\b})^2+c\bigr)\bigl((\frac{x-\a}{x-\b})^2+d\bigr)(1+c)(1+d)}}
{(\frac{x-\a}{x-\b})^2-1}\cr
&\quad{}+K_1(b_1,a)\mPi\(\arctg\[P(b_1,a)\,\frac{x-\a}{x-\b}\],n_1,q_1\)\cr
&\quad{}+L_1(b_1,a)F\(\arctg\[P(b_1,a)\,\frac{x-\a}{x-\b}\],q_1\)
&\rf627 \cr
\noalign{\vskip3pt plus2pt}
&P_1(b_1,a)=\(\frac{\sqrt{12b_1^4-a}-2b_1^2-\sqrt{4b_1^4-a}}
{\sqrt{12b_1^4-a}+2b_1^2+\sqrt{4b_1^4-a}}\)^{1/2} &\rf628 \cr
\noalign{\vskip3pt plus2pt}
&q_1=\frac{\sqrt6 |b_1|}{\(3b_1^2+\sqrt{12b_1^4-a}\)^{1/2}} &\rf629 \cr
\noalign{\vskip3pt plus2pt}
&K_1(b_1,a)=\(\frac{2b_1^2+\sqrt{4b_1^4-a}+\sqrt{12b_1^4-a}}
{\sqrt{12b_1^4-a}-2b_1^2-\sqrt{4b_1^4-a}}\)^{3/2} &\rf630 \cr
\noalign{\vskip3pt plus2pt}
&n_1=\frac{2\sqrt{12b_1^4-a}}{2b_1^2+\sqrt{4b_1^4-a}+\sqrt{12b_1^4-a}}
&\rf631 \cr
%\noalign{\vskip3pt plus2pt}
\noalign{\eject}
&L_1(b_1,a)=\(M(b_1,a)+\frac{2b_1^2+\sqrt{4b_1^4-a}+\sqrt{12b_1^4-a}}
{2\sqrt{12b_1^4-a}}\)\frac1{\(3b_1^2+\sqrt{12b_1^4-a}\)^{1/2}}\cr
&\quad{}\times\frac
{2(12b_1^4-a)^{1/2}\(\sqrt{4b_1^4-a}-b_1^2\)^{1/2}}
{\(2b_1^2+\sqrt{4b_1^4-a}-\sqrt{12b_1^4-a}\)
\(2b_1^2+\sqrt{4b_1^4-a}+\sqrt{12b_1^4-a}\)^{1/2}} &\rf632 \cr
\noalign{\vskip3pt plus2pt}
&M(b_1,a)=\frac\b{\a-\b}\cr
&\quad{}=\biggl((6b_1^4+a)\sqrt{4b_1^4-a}+2b_1^2a+
\sqrt{192b_1^{12}-112b_1^8a+20b_1^4a+a^3}\cr
&\qquad{}+\sqrt{576b_1^{12}-240b_1^8a+28b_1^4a-a^3}
+6b_1^2\sqrt{48b_1^8-16b_1^4a+a^2}\cr
&\qquad{}+(a-4b_1^4)\sqrt{12b_1^4-a}\biggr)^{1/2}\cr
&\quad{}\times\Biggl(\biggl((12b_1^4-a)\sqrt{4b_1^4-a}
+4b_1^4\sqrt{12b_1^4-a}-4b_1^2a+24b_1^6+4b_1^4\sqrt{12b_1^4-a}\cr
&\qquad\quad{}-\sqrt{192b_1^{12}-112b_1^8a+20b_1^4a+a^3}
-\sqrt{576b_1^{12}-240b_1^8a+28b_1^4a-a^3}\biggr)^{1/2}\cr
&\qquad{}-\biggl(8b_1^2\sqrt{48b_1^8-16b_1^4a+a^2}+2(6b_1^4-a)\sqrt{4b_1^4-a}
-4b_1^4\sqrt{12b_1^4-a}\cr
&\qquad\quad{}+4b_1^6+6b_1^2a+\sqrt{192b_1^{12}-112b_1^8a+20b_1^4a+a^3}\cr
&\qquad\quad{}+\sqrt{576b_1^{12}-240b_1^8a+28b_1^4a-a^3}\biggr)^{1/2}\Biggr)^{-1}.
&\rf633 \cr
\noalign{\vskip3pt plus2pt}
\a&=\sqrt{\frac{\sqrt{48b_1^8-16b_1^4a+a^2}+6b_1^2\sqrt{4b_1^4-a}
+2b_1^2\sqrt{12b_1^4-a}+a}
{2b_1^2+\sqrt{4b_1^4-a}+\sqrt{12b_1^4-a}}}>0 &\rf634 \cr
\noalign{\vskip3pt plus2pt}
\b&=\sqrt{\frac{\sqrt{48b_1^8-16b_1^4a+a^2}+6b_1^2\sqrt{4b_1^4-a}
-2b_1^2\sqrt{12b_1^4-a}+a}
{2b_1^2-\sqrt{4b_1^4-a}-\sqrt{12b_1^4-a}}}>0 &\rf635 \cr
\noalign{\vskip3pt plus2pt}
c&=\frac{\(3b_1^2+\sqrt{12b_1^4-a}\)\(2b_1^2-\sqrt{4b_1^4-a}-\sqrt{12b_1^4-a}\)}
{\(\sqrt{4b_1^4-a}+2b_1^2\)\(\sqrt{12b_1^4-a}-3b_1^2\)} \cr
\noalign{\vskip3pt plus2pt}
d&=\frac{\sqrt{12b_1^4-a}+2b_1^2+\sqrt{4b_1^4-a}}
{\sqrt{12b_1^4-a}-2b_1^2-\sqrt{4b_1^4-a}}\,,\quad
|d|>1 &\rf636 \cr
\noalign{\eject}
%\noalign{\vskip3pt plus2pt}
e&=\frac{2\sqrt{12b_1^4-a}\(\sqrt{12b_1^4-a}+\sqrt{4b_1^4-a}-4b_1^2\)}
{\(\sqrt{4b_1^4-a}+2b_1^2+\sqrt{12b_1^4-a}\)\(\sqrt{12b_1^4-a}-3b_1^2\)}
&\rf637 \cr
\noalign{\vskip3pt plus2pt}
f&=\frac{2\sqrt{12b_1^4-a}}
{\(\sqrt{12b_1^4-a}-3b_1^2\)\(\sqrt{12b_1^4-a}-2b_1^2-\sqrt{4b_1^4-a}\)}\cr
&\quad{}\times\(2b_1^2+\sqrt{4b_1^4-a}+\sqrt{12b_1^4-a}\)^{-2}\cr
&\quad{}\times\Bigl(2(14b_1^4-a)\sqrt{4b_1^4-a}-4b_1^2a
+2\sqrt{192b_1^{12}-112b_1^8a+20b_1^4a+a^3}\cr
&\qquad{}+2(b_1^4-a)\sqrt{12b_1^4-a}-2\sqrt{576b_1^{12}-240b_1^8a+28b_1^4a-a^3}
\Bigr). &\rf638
}
$$

Case II
$$
\al
I&=\frac12 \ln \frac {e(\frac{x-\a}{x-\b})^2 +f
+2\sqrt{\bigl((\frac{x-\a}{x-\b})^2+c\bigr)\bigl((\frac{x-\a}{x-\b})^2+d\bigr)(1+c)(1+d)}}
{(\frac{x-\a}{x-\b})^2-1}\\
&+K_2(b_1,a)\mPi\(\arccos\[P_2(b_1,a)\,\frac{x-\a}{x-\b}\],n_2,q_2\)\\
&+L_2(b_1,a)F\(\arccos\[P_2(b_1,a)\,\frac{x-\a}{x-\b}\],q_2\)
\eal \eq639
$$
$$
\ealn{
&q_2=\frac1{q_1}=\frac{\(3b_1^2+\sqrt{12b_1^4-a}\)^{1/2}}{\sqrt6 |b_1|}
&\rf640 \cr 
\noalign{\vskip3pt}
&n_2=\frac{2b_1^2+\sqrt{4b_1^4-a}+\sqrt{12b_1^4-a}}{2\sqrt{12b_1^4-a}}
&\rf641 \cr
\noalign{\vskip3pt}
&K_2(b_1,a)=\frac
{\(2b_1^2+\sqrt{4b_1^4-a}-\sqrt{12b_1^4-a}\)\sqrt{12b_1^4-a}
\(\sqrt{4b_1^4-a}-b_1\)^{1/2}}
{\(\sqrt{12b_1^4-a}+2b_1^2+\sqrt{4b_1^4-a}\)\(\sqrt{12b_1^4-a}
-3b_1^2\)^{3/2}}\hskip30pt &\rf642 \cr
\noalign{\vskip3pt}
&P_2(b_1,a)=\(\frac{2b_1^2+\sqrt{4b_1^4-a}-\sqrt{12b_1^4-a}}
{\sqrt{12b_1^4-a}+2b_1^2+\sqrt{4b_1^4-a}}\)^{1/2} &\rf643 \cr
\noalign{\vskip3pt}
&L_2(b_1,a)=M(b_1,a)\cr
&\quad{}-\frac
{4(12b_1^4-a)\(\sqrt{4b_1^4-a}-b_1^2\)^{1/2}
\(\sqrt{12b_1^4-a}-2b_1^2-\sqrt{4b_1^4-a}\)}
{\(3b_1^2-\sqrt{12b_1^4-a}\)^{3/2}\(2b_1^2+\sqrt{4b_1^4-a}+\sqrt{12b_1^4-a}\)}\,.
&\rf644
}
$$

The function \rf606 \ cannot be inverted globally. It can be inverted only locally in
some intervals where the solution lives. The solution dies on the end of an
interval being reborning on beginning of a next interval. All the intervals
can be obtained from the condition
$$
\o W>0 \eq645
$$
where
$$
\o W=\frac {e(\frac{x-\a}{x-\b})^2 +f+2\sqrt{\bigl((\frac{x-\a}{x-\b})^2
+c\bigr)\bigl((\frac{x-\a}{x-\b})^2+d\bigr)(1+c)(1+d)}}
{(\frac{x-\a}{x-\b})^2-1} \eq646
$$
and
$$
\left|\frac{x-\a}{x-\b}\right|\ne1 \eq647
$$

 In the case I $c>0$, $d>0$.

In the case II $c>0$, $d<0$.

Thus one can get very interesting behaviour from the physical point of view,
because every end of an interval of living (existence) of the solution of an
\ev\ equation means a start of nontrivial
interaction between a matter, a radiation and a \q. In all of these intervals
one can find an approximate inverse function, i.e.\ one can write
$$
R=F(t) \eq648
$$
where
$$
F(t_1)=R_1<R<R_2=F(t_2) \eq649
$$
and
$$
t_1<t<t_2. \eq650
$$

Moreover we do not develop this project here in details. $F$ and $\mPi$ are
usual elliptic integrals of the first and of the third kind given by the
formulae \rf228 \ and \rf230 .

For future convenience of this project we find roots of the polynomial
$$
P(x)=x^4+x+a, \quad a>0. \eq651
$$
First of all according to the Ferrari method we should solve a resolvent
equation which is a cubic equation. One gets
$$
z^3-4az-1=0. \eq652
$$
The discriminant of Eq.\ \rf652 \ reads
$$
\o D=\frac{27-256a^3}{108}
$$
and we have two cases
$$
\ealn{
&\o D>0,\quad 0<a<\frac3{4\root3\of4}, &\rf653 \cr
&\o D<0, \quad a>\frac3{4\root3\of4}, &\rf654
}
$$

In the case \rf653 \ one gets
$$
\ealn{
z_1&=\root3\of{\frac{\sqrt{27-256a^3}+3\sqrt3}{6\sqrt3}}
-\root3\of{\frac{\sqrt{27-256a^3}-3\sqrt3}{6\sqrt3}} &\rf655 \cr
z_{2,3}&=\frac12 \(\root3\of{\frac{\sqrt{27-256a^3}-3\sqrt3}{6\sqrt3}}
-\root3\of{\frac{\sqrt{27-256a^3}+3\sqrt3}{6\sqrt3}}\)\cr
&\pm\frac{i\sqrt3}2 \(\root3\of{\frac{\sqrt{27-256a^3}-3\sqrt3}{6\sqrt3}}
+\root3\of{\frac{\sqrt{27-256a^3}+3\sqrt3}{6\sqrt3}}\) &\rf656
}
$$

In the second case \rf654 
$$
\ealn{
z_1&=4\sqrt{\frac a3}\cos\(\frac{\o\f}3\) &\rf657 \cr
z_2&=4\sqrt{\frac a3}\cos\(\frac{\o\f}3+\frac{2\pi}3\) &\rf658 \cr
z_3&=4\sqrt{\frac a3}\cos\(\frac{\o\f}3+\frac{4\pi}3\) &\rf659 
}
$$
where $\cos\o\f=\sqrt{\frac3{16a}}$.

We can also distinguish the special case $\o D=0$ ($a=\frac3{4\root3\of4}$):
$$
\ealn{
z_1&=\root3\of4 & \rf660 \cr
z_{2,3}&=-\tfrac12\root3\of4 & \rf661
}
$$

In the first case we get one real and two complex conjugated roots, in the
second three real roots, in the third case two different real roots (one of
them is double).

According to the Ferrari method one gets
$$
\ealn{
2x_1&=\sqrt{z_1}+\sqrt{z_2}+\sqrt{z_3} &\rf662a \cr
2x_2&=\sqrt{z_1}-\sqrt{z_2}-\sqrt{z_3} &\rf662b \cr
2x_3&=-\sqrt{z_1}+\sqrt{z_2}-\sqrt{z_3} &\rf662c \cr
2x_4&=-\sqrt{z_1}+\sqrt{z_2}+\sqrt{z_3} &\rf662d 
}
$$
$$
z_1z_2z_3=1. \eq663
$$

In the future analysis of an \ev\ of our model of the \U\ there are three
important cases for $z_1$, $z_2$ and~$z_3$:

\item{A.} All $z_1,z_2,z_3$ are real and positive. In this case all $x_1,x_2,
x_3$ and~$x_4$ are real.

\item{B.} One root (e.g.\ $z_1$) is positive, two remaining ($z_2,z_3$) are
negative. In this case we have two pairs of conjugate roots.

\item{C.} One root (e.g.\ $z_1$) is positive, two remaining ($z_2,z_3$) are
complex conjugate. In this case we have two real roots and one pair of
complex conjugate roots.

This analysis is quite important because only for positive real roots the
integral~$I$ can have singular points. It means in A and C case. This gives
us some restrictions on the parameter~$a$.

Let us notice that in the case of $a=0$ (considered before) we have to do
with the case~C. Fortunately one real root is zero and the second real root
is negative ($x_2=-1$).

Let us notice that our case with $a=0$ is in some sense exceptional from the
point of view of a general theory. In that case some of the formulae are
singular and because of this it was considered separately.

Let us notice that for sufficiently big $a$ the polynomial \rf651 \ has no real
roots. This is true for $a>\frac3{4\root3\of4}$. In this case the integral
\rf608 \ has no singular points. This is case~II.

Let us notice a simple relation between $D$ and $\o D$, $\o D=64D$. This
connects Eq.~\rf652 \ to Eq.~\rf612 \ and gives the same restriction for~$a$.
}

\vfill\eject

\null
\vskip36pt plus4pt minus4pt
\centerline{{\four\bf 15. Geodetic Equations on the Manifold $P$}}
\vskip24pt plus2pt minus2pt
\write0{15. Geodetic Equations on the Manifold $P$ \the\pageno}

Let us consider geodetic equation for a curve ${\varGamma} \subset P$ 
on the manifold
$P$ i.e.\ a nonsymmetrically metrized fibre bundle over $E \times G/G_{0}$ 
with
respect to the connection $\omega ^{\tilde{\rA}}{}_{\tilde{\rB}}$. 
Such equations have a usual
interpretation as test particle equations of motion. One gets ($\nabla
_{u}u=0$, where $u$ is tangent to ${\varGamma} $):
$$
u^{\tilde{\rA} } \nabla _{\tilde{\rA} }u^{\tilde{\rB} }=0.\eqno(15.1)
$$
Using formulae (15.1), (5.8), (5.9), (5.11) one gets:
$$\displaylines{
{\ov{D}u^{\alpha }\over d\tau} + \bigg({q^{c}\over m_{0}}\bigg) 
u^{\beta }\bigg[\ell _{cd} g^{\alpha \delta } H^{d}{}_{\beta \delta } - 
{1\over 2} (\ell _{cd}g^{\alpha \delta }-\ell _{dc}g^{\delta \alpha
}) L^{d}{}_{\beta \delta } +\hfill\cr
\noalign{\vskip4pt}
- {\|q\|^2\over 8m^{2}_{0}} \widetilde{g}^{(\alpha \beta )} 
\bigg({1\over \rho ^{2}}\bigg)_{,\beta } + \bigg({q^{c}\over
m_{0}}\bigg)u^{\tilde{b}}\bigg[\ell _{cd} g^{\alpha \delta }
\mathop{\nabla_{\delta }}\limits^{\gauge}{\varPhi}
^{d}{}_{\tilde{b} } +\cr
\noalign{\vskip4pt}
\hfill - {1\over 2} (\ell _{cd}g^{\alpha \delta }-\ell _{dc}g^{\delta
\alpha }) L^{d}{}_{\tilde{b}\delta}\bigg]= 0,\quad\ (15.2)\cr
\noalign{\vskip4pt}
{\widetilde{D}u^{\tilde{a} }\over d\tau} + {1\over r^{2}} \bigg({q^{c}\over 
m_{0}}\bigg) u^{\beta } \bigg[\ell _{cd}g^{\tilde{a} \tilde d
}\mathop{\nabla_\beta}\limits^{\gauge} {\varPhi} ^{d}_{\tilde d }-
{1\over 2}(\ell _{cd}g^{\tilde{a} \tilde d }-\ell _{dc}g^{\tilde d
\tilde{a} }) L^{d}{}_{\beta \tilde d }\bigg] +\hfill\cr
\noalign{\vskip4pt}
+ {1\over r^{2}} \bigg({q^{c}\over m_{0}}\bigg) u^{\tilde{b}}\bigg[\ell _{cd} 
g^{\tilde{a} \tilde d } (C^{d}_{ab} {\varPhi} ^{a}_{\tilde d } 
{\varPhi} ^{b}_{\tilde{b} } - \mu ^{d}_{\dah{i}} f^{\dah{i}}{}_{\tilde
d \tilde{b} } - {\varPhi} ^{d}_{\tilde e } f^{\tilde e }_{\tilde d
\tilde{b} }) +\cr
\noalign{\vskip4pt}
\hfill - {1\over 2} (\ell _{cd} g^{\tilde{a} \tilde d } - \ell _{dc} 
g^{\tilde d \tilde{a} }) L^{d}{}_{\tilde{b} \tilde d }\bigg]=
0,\quad\ (15.3)\cr
\noalign{\vskip4pt}
\hfill {d\over d\tau} \bigg({q^{b}\over m_{0}}\bigg) =
0,\hfill\llap{(15.4)}\cr}
$$
where $\ov{D}$ means a covariant derivative along a line with respect to the
connection $\overline{\omega }^{\alpha }{}_{\beta }$ on $E$, $\widetilde{D}$ means a covariant derivative along a line with
respect to the connection $\widetilde{\omega }^{\tilde{a}
}{}_{\tilde{b} }$ on $G/G_{0}$ ($r=\const$).
$$
u^{\tilde{\rA} } = ( u^{\alpha }, u^{\tilde{a} }, u^{a} ) = (\hor (u),
\ver (u))\eqno(15.5)
$$
and
$$
2\rho ^{2}u^{a} = {q^{a}\over m_{o}}\quad\ (q =
m_{0}\widehat{\varkappa }(\ver u(\tau))),
$$
$u^{\alpha }$ --- is a four-velocity of a test particle, $q^{a}$ its colour (isotopic)
charge, $u^{\tilde{a} } a$ charge associated with a Higgs' field. This charge
transforms according to the properties of a complement $\fm$ with respect
to $G_{0}$ and $G$. One has $\hor (u) = (P_{\rE}(\hor (u), P_{\rM=G/G_{0}}
(\hor(u))))$ where $P_{\rE}$ is
a projection on $E$ and $P_{M=G/G_{0}}$ on $M$ in the Cartesian product 
$V = E \times M = E \times G/G_{0}$.

Eq.~(15.2) describes a movement of a test particle in a gravitational,
gauge and Higgs' field. We should consider (of course) physical
world-lines on $E$ i.e. time-like or null-like.

Eq.~(15.3) is an equation for a charge associated with Higgs' field.
This charge describes a coupling between a test particle and a Higgs'
field.

Eq.~(15.4) has a usual meaning (a constancy of a colour (isotopic)
charge).

In order to understand the significance of a new term in equation
of motion for a test particle with nonzero colour and Higgs' charges
let us suppose that:
$$
g_{\alpha \beta } = g_{\beta \alpha } ,\quad \ell _{cd} = \ell _{dc} =
h_{cd},\quad g_{\tilde{a} \tilde{b} } = h^{0}{}_{\tilde{a}
\tilde{b} } = g_{\tilde{b} \tilde{a} },
$$
i.e.\ that we have to do with a symmetric case. We suppose also that
$\rho =\const$. One easily gets:
$$\displaylines{
\hfill {\ov{D} u^{\alpha }\over d\tau} + \bigg({q^{c}\over
m_{0}}\bigg) h_{cd} g^{\alpha \delta } H^{d}{}_{\beta \delta } 
u^{\beta } + \bigg({q^{c}\over m_{0}}\bigg) h_{cd} g^{\alpha \delta } 
\mathop{\nabla_\delta }\limits^{\gauge} {\varPhi} ^{d}_{\tilde{b} } 
u^{\tilde{b} } = 0,\hfill \llap{(15.6)}\cr
\quad\ {\widetilde{D}u^{\tilde{a} }\over d\tau} + \bigg({q^{c}\over
m_{0}}\bigg) h_{cd} h^{0\tilde{a} \tilde d } {\De^{\gauge}{\varPhi}
^{d}_{\tilde d }\over d\tau} +\hfill\cr
\hfill + \bigg({q^{c}\over m_{0}}\bigg) h_{cd} h^{0\tilde{a} \tilde d } 
(C^{d}_{ab} {\varPhi} ^{a}_{\tilde d } {\varPhi} ^{b}_{\tilde{b} } - 
\mu ^{d}_{\dah{i}} f^{\dah{i}}_{\tilde d \tilde{b} } - {\varPhi}
^{d}_{\tilde e } f^{\tilde e }_{\tilde d \tilde{b}} ) u^{\tilde{b} } =
0.\quad\ (15.7)\cr}
$$
In Eq.~(15.6) we have an additional term:
$$
\bigg({q^{c}\over m_{0}}\bigg) h_{cd} g^{\alpha \delta }
\mathop{\nabla_\delta }\limits^{\gauge} {\varPhi} ^{d}_{\tilde{b} } 
u^{\tilde{b} },\eqno(15.8)
$$
which plays the role of Lorentz force term for a Higgs' field. The
test particle couples Higgs' (its gauge derivative) via a Higgs' charge
$u^{\tilde{a} }$. Propagation properties of the Higgs' charge are described by the
Eq.~(15.7). They are much more complex than that of a colour charge. In
general this charge is not constant during a motion. We put:
$$
{\De^{\gauge}\over d\tau} {\varPhi} ^{d}_{\tilde d } = 
\mathop{\nabla_\beta }\limits^{\gauge} {\varPhi} ^{d}_{\tilde d } 
u^{\beta }.\eqno(15.9)
$$
It is interesting to consider Eqs.\ (15.6) and (15.7) in the case of the
completely broken group $G$ i.e.
$$\eqalignno{
{\ov{D}u^{\alpha }\over d\tau} + \bigg({q^{c}\over m_{0}}\bigg) h_{cd} 
g^{\alpha \delta } H^{d}_{\beta \delta } u^{\beta } +\bigg({q^{c}\over
m_{0}}\bigg) h_{cd} g^{\alpha \delta } \mathop{\nabla_\delta }\limits^{\gauge}
{\varPhi} ^{d}_{b} u^{b} = 0,\quad &(15.10)\cr
{du^{b}\over d\tau} + {1\over r^{2}} \bigg({q^{c}\over
m_{0}}\bigg) h_{cd} \widetilde{h}^{pq} {\mathop{D}\limits^{\gauge}
{\varPhi} ^{d}_{q}\over d\tau} +
{1\over r^{2}} \bigg({q^{c}\over m_{0}}\bigg) h_{cd} 
\widetilde{h}^{bq} (C^{d}_{ap}{\varPhi} ^{a}_{q}{\varPhi} ^{p}_{k}-
{\varPhi} _{e}f^{e}_{qk}) u^{k} = 0.\quad &(15.11)\cr}
$$
In this case $\widetilde{h}_{ab}$ is a Killing--Cartan tensor on $G$,
$u^{b}$ is an Ad-type
quantity with respect to $G$. Moreover $u^{b}$ is not a Yang--Mills' gauge
independent charge. This indicates that a charge which couples a test
particle to the Higgs' field has completely different properties than
that known for Yang--Mills' field.

Moreover for a vacuum state of the Higgs' potential one gets:
$$\displaylines{
\hfill {\ov{D} u^{\alpha }\over d\tau} + \bigg({q^{c}\over
m_{0}}\bigg) h_{cd} g^{\alpha \delta } H^{d}_{\beta \delta } u^{\beta } =
0,\hfill\llap{(15.12)}\cr
\hfill {du^{b}\over d\tau} = 0\hfill\llap{(15.13)}\cr}
$$
and $u^{b}$ decouples from equations of motion.

If the Higgs' field is closed to the vacuum configuration the
Eq.~(15.13) is not satisfied, but instead of it
$$
{d\over d\tau} \bigg(u^{b}+\bigg({q^{c}\over m_{0}}\bigg) h_{cd} 
\widetilde{h}^{bq} {\varPhi} ^{d}_{q}\bigg)\cong 0\eqno(15.14)
$$
and we can define a ``constant" Higgs' field charge:
$$
g^{b} = u^{b} + ({q^{c}\over m_{0}}) h_{cd} \widetilde{h}^{bq} {\varPhi} ^{d}_{q}
\eqno(15.15)
$$
and substitute it to Eq.~(15.10). Thus we get:
$$\displaylines{
{\ov{D}u^{\alpha }\over d\tau} + \bigg({q^{c}\over m_{0}}\bigg) h_{cd} 
g^{\alpha \delta } H^{d}_{\beta \delta } u^{\beta } + \bigg({q^{c}g^{b}\over 
m_{0}}\bigg) h_{cd} g^{\alpha \delta }
\mathop{\nabla_{\delta}}\limits^{\gauge} {\varPhi} ^{d}_{b}
+\hfill\cr
\hfill - \bigg({q^{c}q^{e}\over m_{0}}\bigg) h_{ef} \widetilde{h}^{bq} 
{\varPhi} ^{f}_{q} h_{cd} g^{\alpha \delta } \mathop{\nabla_{\delta}}\limits^{\gauge}
{\varPhi} ^{d}_{b} = 0.\quad\ (15.16)\cr}
$$
The Eq.~(15.2) has the following first integral of motion:
$$
\varkappa (u(\tau),u(\tau)) = g_{(\alpha \beta )} u^{\alpha }
u^{\beta } + r^{2} g_{\tilde{a} \tilde{b} } u^{\tilde{a} }
u^{\tilde{b} } + \rho ^{2} \ell _{ab} u^{a} u^{b} = \const.\eqno(15.17)
$$
Introducing the following notation:
$$
\widetilde{u}^{2} = - g_{\tilde{a} \tilde{b} } u^{\tilde{a} }
u^{\tilde{b} }\vadjust{\vskip-4pt}\eqno(15.18)
$$
one gets: %\vadjust{\vskip-4pt}
$$
g_{(\alpha \beta )} u^{\alpha } u^{\beta } - r^{2}\widetilde{u}^{2} 
- {\|q\|^{2}\over 4\rho ^{2}} =\const.\eqno(15.18')
$$
However in general the length of the charge $u^{\tilde{b} }$ is not 
constant even in
the symmetric case. For a constant field $\rho = \const$, one gets:
$$
g_{(\alpha \beta )} u^{\alpha } u^{\beta } = r^{2}\widetilde{u}^{2} +
\const .\eqno(15.19)
$$
Thus the interpretation of the Eq.~(15.3) demands more investigations.

Finally let us consider Eqs.~(15.2--3) in a more general case for the
scalar field $\rho = \rho (x,y)$. In this case we get one more term in
Eq.~(15.3) involving the field $\rho$ $(\rho \in {\sym C}^{1}$ 
$(E\times M))$ i.e.:
$$
- {\|q\|^{2}\over 8m^{2}_{0}r^{2}} \widetilde{g}^{(\tilde{a}
\tilde{b})} \bigg({1\over \rho ^{2}}\bigg)_{,\tilde{b} }\eqno(15.20)
$$
and the term in Eq.~(15.2):
$$
- {\|q\|\over 8m^{2}_{0}} \widetilde{g}^{(\alpha \beta )} \bigg({1\over 
\rho ^{2}}\bigg)_{,\beta }\eqno(15.21)
$$
depends on the point $y \in M$.

If we suppose that:
$$
\rho (x,y) = \rho _{0}(x) \rho _{1}(y),
$$
then we get
$$
- {\|q\|^{2}\over 8m^{2}_{0}r^{2}\rho ^{2}_{0}}
\widetilde{g}^{(\tilde{a} \tilde{b} )}\bigg({1\over \rho
^{2}_{1}}\bigg)_{,\tilde{b} }\eqno(15.20\text{\rm a})
$$
and
$$
- {\|q\|^{2}\over 8m^{2}_{0}\rho^{2}_{1}} \widetilde{g}^{(\alpha
\beta )} \bigg({1\over \rho ^{2}_{0}}\bigg)_{,\beta }.\eqno(15.21\text{\rm a})
$$
In the case of the expansion of $\rho $ into generalized harmonics one gets:
$$
{\|q\|^{2}\over 4m^{2}_{0}r^{2}\rho ^{3}}
\widetilde{g}^{(\tilde{a} \tilde{b} )}\Big(\sum^{\infty }_{k=1}\rho _{k}(x) 
\chi _{k,\tilde{b} }(y)\Big)\eqno(15.20\text{\rm b})
$$
and
$$
{\|q\|^{2}\over 4m^{2}_{0}\rho ^{3}} \widetilde{g}^{(\alpha \beta )}
\Big(\rho _{0,\beta } + \sum^{\infty }_{k=1}\rho _{k,\beta } 
\chi _{k}(y)\Big).\eqno(15.21\text{\rm b})
$$
The last two terms describe an interaction of a test particle with a
tower of scalar fields $\rho _{k}(x)$, $k=0,1,2,\ldots $ The truncation procedure
for the field $\rho $ substitutes $\rho _{0}$ in place of $\rho $. We can give an
interpretation of normal coordinates as velocities, gauge-independent
charges and Higgs' charges for test particles. We can consider in our
theory geodetic deviation equations. In this way we can derive
``tidal-forces" for the Higgs' field additionally to gravitational and
Yang--Mills' ones. 

Eventually let us remark that in the Kaluza--Klein Theory we can treat
the Higgs' field as a part of a multidimensional Yang--Mills' field.
However the Lorentz-like force term for the Higgs' field seems to be
more complex.

{\def\eq#1 {\eqno(\text{\rm15.#1})}
\def\rf#1 {\text{(15.#1)}}
\let\al\aligned
\let\eal\endaligned
\let\ealn\eqalignno
\def\cL{\Cal L}
\let\o\overline
\let\ul\underline
\let\a\alpha
\let\b\beta
\let\d\delta
\let\e\varepsilon
\let\D\Delta
\let\g\gamma
\let\G\varGamma
\let\la\lambda
\let\La\varLambda
\let\n\nabla
\let\om\omega
\let\F\varPhi
\let\P\varPsi
\let\s\sigma
\let\z\zeta
\let\pa\partial
\let\t\widetilde
\let\h\widehat
\let\u\tilde
\def\uP{\t{\ul P}}
\def\RG{\t R(\t\G)}
\def\arctg{\operatorname{arctg}}
\def\ctg{\operatorname{ctg}}
\def\tg{\operatorname{tg}}
\def\Li{\operatorname{Li}}
\def\Re{\operatorname{Re}}
\def\br#1#2{\overset\text{\rm #1}\to{#2}}
\def\({\left(}\def\){\right)}
\def\[{\left[}\def\]{\right]}
\def\dr#1{_{\text{\rm#1}}}
\def\gi#1{^{\text{\rm#1}}}
\def\ct{constant}
\def\co{cosmological}
\def\dt{dependent}
\def\ef{effective}
\def\ex{exponential}
\def\lg{lagrangian}
\def\gr{gravitational}
\def\q{quintessence}
\def\wrt{with respect to }
\def\nos{\noalign{\vskip2pt plus2pt}}

\def\tga#1 {\t\g^{(#1)}}
\def\tge#1{\t g^{(#1)}}
\def\id#1{_{\lower1pt\hbox{$\scriptstyle#1$}}}
\def\rp#1{\rho\id{,#1}}
\def\dw{{2\rho^2m_0}}
\def\hl#1#2{H^{#1}_{#2}-L^{#1}_{#2}}
\def\ga#1{\br{gauge}{\n_{#1}}}
\def\gl#1#2#3{\ga{#2}\F^{#1}_{#3}-L^{#1}_{#2#3}}
\def\lu#1#2#3#4{L^{#1}_{#2#3}u^{#3}+L^{#1}_{#2#4}u^{#4}}
\def\lv#1#2#3#4{L^{#1}_{#2#3}u^{#2}+L^{#1}_{#4#3}u^{#4}}
\def\fw#1{\(\frac{#1}\dw\)}
\def\qm#1{\(\frac{q^{#1}}{m_0}\)}

\vfuzz6.6pt
\vbadness10000

Let us consider a geodetic deviation equation in our theory
$$
u^{\u B}\n_{\u B}v^{\u A}+ R^{\u A}{}_{\u C\u M\u B}u^{\u C}\z^{\u M}u^{\u B}
-Q^{\u N}{}\id{\u M\u B}\n_{\u N}u^{\u A}\z^{\u M}u^{\u B}=0
\eq22
$$
and for $u^{\u B}$ Eq.~\rf1 \ is satisfied, where 
$$
u^{\u A}=\frac{dx^{\u A}}{d\tau}\,,\quad v^{\u A}=\frac{d\z^{\u A}}{d\tau}\,.
$$
In this way we consider a generalization of the geodetic deviation equation
to the $(n+n_1+4)$-dimensional case in non-Riemannian geometry.

$R^{\u A}{}_{\u C\u M\u B}$ is a curvature tensor for a connection $\om^{\u
A}{}_{\u B}$ on~$P$ and $Q^{\u A}{}\id{\u C\u B}$ is a tensor of a torsion for
$\om^{\u A}{}_{\u B}$, $\n_{\u N}$ is a covariant derivative \wrt $\om^{\u
A}{}_{\u B}$. 

One gets using formulae from sec.~5 and Eqs (14.56) and (14.57) from the
fourth point of Ref.~[18]
$$
\al
&u^B \o\n_Bv^A + \o R^A{}_{BMN}u^B\z^Mu^N - \o Q{}^N{}\id{MB}\o\n_Nu^A\z^Mu^B\\
\nos
&\quad{}
+\(\frac{q^b}{2m_0\rho^2}\)l_{bd}\g^{AB}\(2H^d_{NB}- L^d_{NB}\)v^N
-\rho^2 v^n l_{dn}\g^{DA}L^d_{DB}u^B \\
\nos
&\quad{}
- \(\frac{q^n}{2m_0\rho^2}\)\t \g^{(AB)}\rp{B}l_{bn}v^b\\
\nos
&\quad{}
+2\rho^2\(l_{cd}\g^{AP}\(2H^d_{P[M}-L^d_{P[M}\)L^c_{N]B}
+l_{db}\g^{DA}L^d_{DB}H^b_{NM}\)u^B\z^Mu^N\\
\nos
&\quad{}
+\frac{q^b}{2m_0\rho^2} \Bigl[2\o\n_{[M}\(\rho^2l_{db}\g^{AB}\(2H^d_{N]B}
-L^d_{N]B}\)\)\\
\nos
&\quad{}
+\rho^2 l_{db}\g^{AB}\(2H^d_{CB}-L^d_{CB}\)\o Q{}^C{}\id{MN}
-\rho \t\g^{(AB)}\rp{B}l_{bc} H^c{}_{MN}\\
\nos
&\quad{}
+2\rho l_{bd}\g^{AB}\(2H^d_{[M|B|}-L^d_{[M|B|}\)\g\id{|D|N]}
\t\g^{(DC)}\rp{C}\Bigr]\z^Mu^N\\
\nos
&\quad{}
+\Bigl\{\rho l_{bd}\g^{AP}\g\id{BD}\t\g^{(DZ)}\rp{Z}\(H^d_{MP}-L^d_{MP}\)
-\o\n_M\(\rho^2l_{db}\g^{DA}L^d_{DB}\)\\
\nos
&\quad{}
-\rho \t\g^{(AP)}\rp{P}l_{cd}L^c{}_{BM}\Bigr\}
u^B\(\z^M\,\frac{q^b}{2\rho^2m_0}-\z^bu^M\)\\
\nos
&\quad{}
+\biggl(2\rho^4 l_{d[b}l_{|e|f]}\g^{DA}\g^{ZC}L^d_{DC}L^e_{ZB}
+\t\g^{(AP)}\rp{P}\g\id{BD}\t\g^{(DZ)}\rp{Z}l_{[bf]}\\
\eal
\eq23
$$

$$
\al
&\quad{}
+\frac{\rho^2}2\,l_{dp}\g^{DA}L^d_{DB}C^P_{bf}\biggr)u^B
\(\frac{\z^bq^f}{2m_0\rho^2}\)\\
\nos
&\quad{}
+\biggl\{\o\n_M\(\rho\tga AB \rp B\)l_{ba}
+\t\n_a\(\rho^2l_{bd}\g^{AB}\(2H^d_{MB}-L^d_{MB}\)\)\\
\nos
&\quad{}
-\rho^4 l_{da}l_{bf}\g^{DA}\g^{CB}L^d_{DC}\(2H^d_{MB}-L^d_{MB}\)\\
\nos
&\quad{}
-\tga AZ \rp Z \g\id{DM}\tga DN \rp N l_{ba}\biggr\}
\(\frac{q^b}{2m_0\rho^2}\) \(\z^au^M-\z^M\,\frac{q^a}{2m_0\rho^2}\)\\
\nos
&\quad{}
+\(\rho\tga AB \rp B l_{bk}C^k{}_{ac} + 2\rho^3\tga CB \rp B \g^{DA}
L^d_{DC}l_{d[a}l_{|b|c]}\)\(\frac{q^b\z^aq^c}{4\rho^4m_0^2}\)\\
\nos
&\quad{}
-\(2\rho^2 l_{bd}\g^{NB}H^d_{CB}+\rho^2\(l_{bd}\g^{NB}+l_{db}\g^{BN}\)
L^d_{BC}\)\\
\nos
&\qquad{}
\times\(\o\n_N u^A +\frac12\(l_{fd}\g^{AB}\(2H^d_{NB}-L^d_{NB}\)
\(\frac{q^f}{m_0}\)\)\)\(v^C\,\frac{q^b}{2\rho^2m_0} - v^b u^C\)\\
\nos
&\quad{}
-2\mu\rho\tga NB \rp B k_{bc}\(\o\n_N u^A + \frac12\,l_{df}\g^{AC}
\(2H^d_{NC}-L^d_{NC}\)\(\frac{q^f}{m_0}\)\)\(\frac{q^cv^b}{2m_0\rho}\)\\
\nos
&\quad{}
+2\(H^n_{MN}-L^n_{NM}\)\(\rho^2l_{dn}\g^{DA}L^d_{DB}u^B
+\frac1{2\rho^2}\,\tga AB \rp B l_{bn}\(\frac{q^b}{m_0}\)\)v^Mu^N\\
\nos
&\quad{}
-\(\g\id{ZM}\tga AC \rp C \rho l_{bm}\g^{DZ}L^b_{DB}u^B
+\frac1{2\rho^2}\,\tga AB \rp B l_{bm}\(\frac{q^b}{m_0}\)\)\\
\nos
&\qquad{}
\times\(v^M\(\frac{q^m}{2\rho^2m_0}\)-v^mu^M\)\\
\nos
&\quad{}
-\frac1{2\rho^2}\,\t Q{}^n{}\id{mb}\(\rho^2l_{dn}\g^{DA}L^d_{DB}u^B
-\frac1{2\rho}\,\tga AB \rp B l_{pn}\(\frac{q^p}{m_0}\)\)v^m\(\frac{q^b}{m_0}\)\\
&=0
\eal
\eq23\sevenrm cont.
$$
and
$$
\al
&\frac{dv^a}{d\tau}+ \t R^a{}_{bcd}\(\frac{q^b}\dw\)\z^c\(\frac{q^d}\dw\)
-\t Q{}^a{}\id{mb}\t\n_n\(\frac{q^a}\dw\)v^m\(\frac{q^b}\dw\)\\
\nos
&\quad{}
+\biggl(2\rho^2l_{bd}\g^{CB}\(2\hl d{[N|B|}\)L^a_{|C|M]}
+\frac1\rho \,\d^a{}_b\g\id{DA}\tga DC \rp C \o Q{}^A{}\id{MN}\\
\nos
&\quad{}
-2\d^a{}_b\n_{[N}\(\frac1\rho\,\g\id{|D|M]}\tga DC \rp C\)\biggr)
\(\frac{q^b\z^Mu^N}\dw\)\\
\nos
&\quad{}
+\(\rho\tga CB \rp B l_{bc}L^a_{CM}
+\rho l_{bd}\tga BN \(2\hl d{MB}\)\d^a{}_c\)
\eal
\eq24
$$
$$
\al
\nos
&\qquad{}
\times\(\frac{q^b}\dw\)\(\z^cu^M-\z^M\(\frac{q^c}\dw\)\)\\
&\quad{}
+\biggl(\t\n_bL^a_{BC}-\o\n_C \(\frac1\rho\,\g\id{BD}\tga DN \rp N \)\d^a{}_b
+\rho^2 l_{db}\g^{DM}L^a_{MC}L^d_{DB}\\
\nos
&\quad{}
-\frac1{\rho^2}\,\g\id{DC}\tga DE \rp E \g\id{BN}\tga NM \rp M \d^a{}_b\biggr)
u^B\(\z^bu^C-\z^C\(\frac{q^b}\dw\)\)\\
\nos
&\quad{}
+\(-\frac1\rho\, \g\id{BD}\tga DC \rp C C^a{}_{bc}+2\rho\g\id{CN}\tga NM
\rp M g^{DC}L^d_{DB} l_{d[b}\d^a{}_{c]}\)u^B\(\frac{\z^bq^c}{2\rho^4m_0}\)\\
\nos
&\quad{}
-2\g\id{CD}\tga DM \rp M \tga CB \rp B l_{b[c}\d^a{}_{d]}\(\frac{q^b\z^dq^c}{4\rho^2m_0^2}\)\\
\nos
&\quad{}
+\biggl(2\o\n_{[M}L^a_{|B|N]}+L^a_{BC}\o Q{}^C{}\id{MN}
+\frac2\rho\,\g\id{BD}\tga DC \rp C H^a_{MN}\\
\nos
&\qquad{}
+\frac2\rho\,\g\id{C[M}L^a_{|B|N]}\tga DC \rp C\biggr) u^B\z^Mu^N\\
\nos
&\quad{}
-\frac1{2\rho}\,\t Q{}^a{}\id{mb}\g\id{BD}\tga DC \rp C u^Bv^m\(\frac{q^b}{m_0}\)\\
\nos
&\quad{}
-\t Q{}^N{}\id{MZ}u^Zv^M\(-\frac1{\rho^3}\,\rp N\(\frac{q^a}{m_0}\)
+L^a_{BN}u^B+\frac1{2\rho^3}\,\g\id{DN}\tga DC \rp C \(\frac{q^c}{m_0}\)\)\\
\nos
&\quad{}
-\frac1\rho\,\g\id{ZM}\tga ZP \rp P\(\t\n_b\(\frac{q^a}\dw\)
+\frac1\rho\,\g\id{BD}\tga DC \rp C u^B\d^a{}_b\)\\
\nos
&\qquad{}
\times\(v^M\(\frac{q^b}\dw\)-v^bu^M\)\\
\nos
&\quad{}
-2\(\hl n{MN}\)\(\t\n_n\(\frac{q^a}\dw\)
+\frac1\rho\,\g\id{BD}\tga DC \rp C u^B\d^a{}_n\)v^Mu^N\\
\nos
&\quad{}
-2\mu \rho \tga NM \rp M k_{bc}\biggl(-\frac1{\rho^2}\(\frac{q^a}{m_0}\)\rp N
+L^a_{CN}u^C \\
\nos
&\quad{}
+\frac1{2\rho^3}\,\g\id{DN}\tga DC \rp C \(\frac{q^a}{m_0}\)\biggr)
\(\frac{v^bq^c}\dw\)\\
\nos
&\quad{}
-\(2\rho^3 l_{bd}\g^{NB}H^d_{CB}+\rho^2\(l_{bd}\g^{NB}+l_{db}\g^{BN}\)
L^d_{BC}\)\\
\nos
&\qquad{}
\times\(-\frac1{\rho^3}\(\frac{q^a}{m_0}\)\rp N + L^a_{CN}u^C
+\frac1{2\rho^3}\,\g\id{DN}\tga DC \rp C \(\frac{q^a}{m_0}\)\)\\
\nos
&\qquad{}
\times \(v^C\(\frac{q^b}\dw\)-v^bu^C\)\\
\eal
\eq24\sevenrm cont.
$$
$$
\al
&\quad{}
+\frac1{2\rho^3}\,\g\id{BD} \tga DC \rp C v^B\(\frac{q^a}{m_0}\)
+L^a_{BN}v^Bu^N\\
\nos
&\quad{}
+\frac1\rho\,\g\id{DN}\tga DC \rp C v^au^N\\
\nos
&=0
\eal
\eq24\sevenrm cont.
$$
where in both Eqs \rf23--24 \ $\o Q{}^A{}\id{MN}$ is a torsion on $E\times M$
for a connection $\o\om^A{}_B$ defined here, $\o R^A{}_{BCD}$ is a tensor of
a curvature for this connection, $\o\n_N$ is a covariant derivative for this
connection. A field $\rho$ is defined on $E\times M$, $\rho=\rho(x,y)$.

$\t\n_n$ is a covariant derivative \wrt a connection on~$H$ (a nonsymmetric
connection on a group manifold~$H$), $\t Q{}^a{}\id{bc}$ is a torsion of this
connection, $\t R^a{}_{bcd}$ a tensor of a curvature. $H^a_{MN}$ is a
curvature of a connection $\om$ on fibre bundle defined over base manifold
$E\times M$, $L^a_{MN}$ is defined by Eq.~(5.13). We have
$$
\ealn{
u^{\u A}&=(u^A,u^a) &\rf 25 \cr
v^{\u A}&=(v^A,v^a). &\rf 26 \cr
\z^{\u A}&=(\z^A,\z^a) &\rf 27
}
$$
and of course
$$
\ealn{
v^A&=\frac{d\z^A}{d\tau} &\rf 28 \cr
v^a&=\frac{d\z^a}{d\tau} &\rf 29
}
$$

$\tga AB $ is defined by Eq.~(5.12). In this way we have geodetic deviation 
equation on~$P$. Moreover, we need a dimensional reduction procedure to
obtain them in terms of Yang--Mills' and Higgs' fields. It is easy to see
that 
$$
\frac d{d\tau}q^b=0 \eq30
$$
or
$$
q^b=\text{const.} \eq30a
$$

Let us consider Eq.\ \rf 23 \ according to sections 5 and 6. In this way one
gets for $A=\a$ (i.e.\ for space-time~$E$):
$$
\al
&u^\b\o\n_\b v^\a +\o R^\a{}_\b u^\b \z^\mu u^\nu
-\o Q{}^\nu{}_{\mu\b}\(\o\n_\nu u^\a\)\z^\mu u^\b\\
\nos
&\quad{}
+\(\frac{q^b}\dw\)l_{bd}g^{\a\b}\[\(2\hl d{\nu\b}\)v^\nu
+\(2\ga\b \F^d_{\u n}-L^d_{\b\u n}\)v^{\u n}\]\\
&\quad{}
-\rho^2v^n l_{dn}g^{\d\a}\(L^d_{\d\b}u^\b+L^d_{d\u b}u^{\u b}\)
-\(\frac{q^n}\dw\)\t g^{(\a\b)}\rp\b l_{bn}v^b
\eal
\eq31
$$
$$
\al
&\quad{}
+2\rho^2\Biggl[\(l_{cd}g^{\a\om}\(2\hl d{\om[\mu}\)L^c_{\nu]\b}
+l_{db}g^{\d\a}L^d_{\d\b}H^b_{\mu\nu}\)\z^\mu u^\nu u^\b\\
\nos
&\qquad{}
+\(\frac12\,l_{cd}g^{\a\om}\(2\ga\om \F^d_{\u m}-L^d_{\om\u m}\)L^c_{\nu\b}
-l_{db}g^{\d\a}L^d_{\d\b}\ga \nu \F^b_{\u m}\)\z^{\u m}u^\nu u^\b\\
\nos
&\qquad{}
+\(\frac12\,l_{cd}g^{\a\om}\(2\hl d{\om\mu}\)L^c_{\u n\b}
+l_{db}g^{\d\a}L^d_{\d\b}\ga\mu \F^b_{\u n}\)\z^\mu u^{\u n}u^\b\\
\nos
&\qquad{}
+\(\frac12\,l_{cd}g^{\a\om}\(2\hl d{\om\mu}\)L^c_{\nu\u b}
+l_{db}g^{\d\a}L^d_{\d\u b}H^b_{\mu\nu}\)\z^\mu u^\nu u^{\u b}\\
\nos
&\qquad{}
+\(\frac12\,l_{cd}g^{\a\om}\(2\ga\om \F^d_{\u m}-L^d_{\om\u m}\)L^c_{\u n\u b}
+l_{db}g^{\d\a}L^d_{\d\u b}H^b_{\u m\u n}\)\z^{\u m}u^{\u n}u^{\u b}\\
\nos
&\qquad{}
+\(\frac12\,l_{cd}g^{\a\om}\(2\ga\om \F^d_{\u m}-L^d_{\om \u m}\)L^c_{\nu\u b}
-l_{db}g^{\d\a}L^d_{\d\u b}\ga\nu \F^b_{\u m}\)\z^{\u m}u^\nu u^{\u b}\\
\nos
&\qquad{}
+\(\frac12\,l_{cd}g^{\a\om}\(2\hl d{\om\mu}\)L^c_{\u n\u b}
+l_{db}g^{\d\a}L^d_{\d\u b}\ga\mu \F^b_{\u n}\)\z^\mu u^{\u n}u^{\u b}\\
\nos
&\qquad{}
+\(\frac12\,l_{cd}g^{\a\om}\(2\ga\om \F^d_{\u m}-L^d_{\om\u m}\)L^c_{\u n\b}
+l_{db}g^{\d\a}L^d_{\d\b}H^b_{\u m\u n}\)\z^{\u m}u^{\u n}u^\b\Biggr]\\
\nos
&\quad{}
+\frac{q^b}\dw \Biggl\{\biggl[
2\o\n_{[\mu}\(\rho^2 l_{db}g^{\a\b}\(2\hl d{\nu]\b}\)\)\\
&\qquad\quad{}
+\rho^2 l_{db}g^{\a\b}\(2\hl d{\g\b}\)\o Q{}^\g{}\id{\mu\nu}
-\rho\t g^{(\a\b)}\rp\b l_{bc}H^c_{\mu\nu}\\
&\qquad\quad{}
+2\rho l_{bd}g^{\a\b}\(2\hl d{[\mu|\b|}\)g\id{|\d|\nu]}\t g^{(\d\g)}\rp\g
\biggr]\z^\mu u^\nu\\
\nos
&\qquad{}
+\biggl[2\t\n_{[\u m}\(\rho^2 l_{db}g^{\a\b}\(-2\ga\b \F^d_{\u n}+L^d_{\b\u n}\)\)\\
&\qquad\quad{}
+\rho^2 l_{db}g^{\a\b}\(2\hl d{\g\b}\)Q^\g\id{\u m\u n}\\
&\qquad\quad{}
-\rho l_{bd}g^{\a\b}\(2\ga\b \F^d_{\u m}-L^d_{\b\u m}\)
g\id{\u d\u n}\t g^{(\u d\u c)}\rp{\u c}\biggr]\z^{\u m}u^{\u n}\\
\nos
&\qquad{}
+\biggl[\t\n_{\u m}\(\rho^2 l_{db}g^{\a\b}\(2\hl d{\nu\b}\)\)\\
&\qquad\quad{}
+\rho^2 l_{db}g^{\a\b}\[\(2\hl d{\g\b}\)Q^\g{}\id{\u m\nu}
+\(-2\ga\b \F^d_{\u c}+L^d_{\b\u c}\)Q^{\u c}{}_{\u m\nu}\]\\
&\qquad\quad{}
+\rho\t g^{(\a\b)}\rp\b l_{bc}\ga\nu \F^c_{\u m}\\
&\qquad\quad{}
+\rho l_{bd}g^{\a\b}\(-\ga\b \F^d_{\u m}+L^d_{\b\u m}\)
g\id{\d\nu}\t g^{(\d\g)}\rp\g\biggr]\z^{\u m}u^\nu
\eal
\eq31\sevenrm cont.
$$

$$
\al
&\qquad{}
+\Biggl[\o\n_\mu\(\rho^2 l_{db}g^{\a\b}\(-2\ga\b \F^d_{\u n}+L^d_{\b\u n}\)\)\\
&\qquad\quad{}
+\rho^2 l_{db}g^{\a\b}\biggl[\(2\hl d{\g\b}\)Q^\g{}\id{\mu\u n}\\
&\qquad\quad{}
+\(-2\ga\b \F^d_{\u c}+L^d_{\b\u c}\)Q^{\u c}{}\id{\mu\u n}\biggr]
-\rho\t g^{(\a\b)}\rp \b l_{bc}\ga\mu \F^c_{\u n}\\
&\qquad\quad{}
+2\rho l_{bd}g^{\a\b}\(2\hl d{\b\mu}\)g\id{\u d\u n}\t g^{(\u n\u c)}\rp c\Biggr]
\z^\mu u^{\u n}\Biggr\}\\
\nos
&\quad{}
+\biggl\{\rho l_{bd}g^{\a\om}g\id{\b\d}\tge{\d\z} \rp\z\(2\hl d{\mu\om}\)
-\o\n_\mu\(\rho^2 l_{db}g^{\d\a}L^d_{\d\b}\)\\
&\qquad{}
-\rho \tge{\a\om} \rp \om l_{cd}L^c_{\b\mu}\biggr\}
u^\b\(\z^\mu\,\frac{q^b}\dw -\z^b u^\mu\)\\
\nos
&\quad{}
+\biggl(2\rho^4 l_{d[b}l_{|e|f]}g^{\d\a}
\(g^{\z\g}L^d_{\d\g}L^e_{\z\b}
+\frac1{r^2}\,g^{\u z\u c}L^d_{\d\u c}L^e_{\u z\b}\)\\
&\qquad{}
+\tge{\a\om}\rp\om g\id{\b\d}\tge{\d\z}\rp\z l_{[bf]}
+\frac{\rho^2}2\,l_{dp}g^{\d\a}L^d_{\d\b}C^p_{bf}\biggr)u^\b
\(\frac{\z^bq^f}\dw\)\\
\nos
&\quad{}
+\biggl(2\rho^4l_{d[b}l_{|e|f]}g^{\d\a}\(g^{\z\g}L^d_{\d\g}L^e_{\z\u b}
+\frac1{r^2}g^{\u z\u c}L^d_{\d\u v}L^e_{\u z\u b}\)\\
&\qquad{}
+\mu\tge{\a\om}\rp \om g^{\u b\u d}\tge{\u d\u z}\rp{\u z}k_{bf}
+\frac{\rho^2}2 l_{dp}g^{\d\a}L^d_{\d\u b}C^p{}_{bf}\biggr)
u^{\u b}\(\frac{\z^bq^f}\dw\)\\
\nos
&\quad{}
+\biggl\{\o\n_\mu\(\rho \tge{\a\b}\rp\b\)l_{ba}
+\t\n_a\(\rho^2l_{bd}g^{\a\b}\(2\hl d{\mu\b}\)\)\\
&\qquad{}
-\rho^4l_{dc}l_{bf}g\id{\d\a}g^{\g\b} L^d_{\d\g}\(2\hl f{\mu\b}\)\\
&\qquad{}
-\frac{\rho^4}{r^2}l_{dc}l_{bf}g^{\d\a}g^{\u c\u b} L^d_{\d\u c}
\(2\ga\mu \F^f_{\u b}-L^f_{\mu\u b}\)\\
&\qquad{}
-\tge{\a\z}\rp\z g\id{\d\mu}\tge{\d\nu}\rp\nu l_{bc}\biggr\}
\(\frac{q^b}\dw\)\(\z^au^\mu -\z^\mu\,\frac{q^a}\dw\)\\
&\quad{}
+\biggl\{\h{\o\n}_{\u m}\(\rho \tge{\a\b}\rp\b\)l_{ba}
-\t\n_a\(\rho^2l_{bd}g^{\a\b}\(2\ga\b \F^d_{\u m}-L^d_{\b\u m}\)\)\\
&\qquad{}
+\rho^4l_{dc}l_{bf}g^{\d\a}g\id{\g\b}L^d_{\d\g}\(2\ga\b \F^f_{\u m}-L^f_{\b\u
m}\) \\
&\qquad{}
-\frac{\rho^4}{r^2}\,l_{dc}l_{bf}g^{\d\a}\tge{\u c\u b}L^d_{\d\u c}
\(2\hl f{\u m\u b}\)\\
\eal
\eq31\sevenrm cont.
$$

$$
\al
&\qquad{}
-r^2\tge{\a\z}\rp\z g\id{\u d\u m}\tge{\u d\u n}\rp{\u n}l_{bc}\biggr\}
\(\frac{q^b}\dw\)\(\z^au^{\u m}-\z^{\u m}\,\frac{q^a}\dw\)\\
\nos
&\quad{}
+\biggl(\rho\tge{\a\b}\rp\b l_{bk}C^k_{ac}
+2\rho^3\tge{\g\b}\rp\b g^{\d\a} L^d_{\d\g}l_{d[a}l_{|b|c]}\\
&\qquad{}
+\frac{2\rho^3}{r^2}\,\tge{\a\b}\rp\b g^{\u d\u a} 
L^d_{\u d\u c}l_{d[a}l_{|b|c]}\biggr)\(\frac{q^b\z^aq^c}{4\rho^2m_0^2}\)\\
\nos
&\quad{}
-\biggl(2\rho^2l_{bd}g^{\a\b}H^d_{\g\b}
+\rho^2\(l_{bd}g^{\nu\b}+l_{db}g^{\b\nu}\) L^d_{\b\g}\\
&\qquad{}
\times\(\o\n_\nu u^\a +\frac12\,l_{bf}g^{\a\b}\(2\hl d{\nu\b}\)\(\frac{q^f}{m_0}\)\)
\biggr)\(u^\g\,\frac{q^b}\dw - v^bu^\g\)\\
\nos
&\quad{}
-\biggl(\frac2{r^2}\,\rho^2 l_{bd}\tge{\u n\u b}\ga\g \F^d_{\u b}
+\frac{\rho^2}{r^2}\(l_{bd}g^{\u n\u b}+l_{db}g^{\u b\u n}\) L^d_{\g\u b}\\
&\qquad{}
\times\(\h{\o \n}_{\u n}u^\a+\frac12\,l_{bf}g^{\a\b}
\(2\ga\b \F^d_{\u n}-L^d_{\b\u n}\)\)\(\frac{q^f}{m_0}\)\biggr)
\(u^\g \,\frac{q^b}\dw-v^bu^\g\)\\
\nos
&\quad{}
-2\mu\rho \tge{\nu\b}\rp\b k_{bc}
\(\o\n_\nu u^\a+\frac12\(l_{df}g^{\a\g}\(2\hl d{\nu\g}\)\(\frac{q^f}{m_0}\)\)\)
\(\frac{q^cv^b}\dw\)\\
&\quad{}
-\frac{2\mu\rho}{r^2}\,\tge{\u n\u b}\rp{\u b}k_{bc}
\(\h{\o\n}_{\u n} u^\a - \frac12\(l_{df}g^{\a\g}
\(2\ga\g \F^d_{\u n}-L^d_{\g\u n}\)\(\frac{q^f}{m_0}\)\)\)
\(\frac{q^cv^b}\dw\)\\
\nos
&\quad{}
+\(2\hl n{\mu\nu}\)\biggl(\rho^2l_{dn}g^{\d\a}\(L^d_{\d\b}u^\b+L^d_{\d\u b}u^{\u
b}\)
+\frac1{2\rho^2}\,\tge{\a\b}\rp\b l_{bn}\(\frac{q^b}{m_0}\)\biggr)v^\mu u^\nu\\
\nos
&\quad{}
+\(2\hl n{\u m\u n}\)\(\rho^2 l_{dn}g^{\d\a}\(L^d_{\d\b}u^\b
+L^d_{\d\u b}u^{\u b}\)
+\frac1{2\rho^2}\tge{\a\b}\rp\b l_{bn}\(\frac{q^b}{m_0}\)\)v^{\u m}u^{\u n}\\
\nos
&\quad{}
+2\(\ga\mu \F^n_{\u n}-L^n_{\mu\u n}\)\(\rho^2l_{dn}g^{\d\a}
\(L^d_{\d\b}u^\b+L^d_{\d\u b}u^{\u b}\)
+\frac1{2\rho^2}\tge{\a\b}\rp\b l_{bn}\(\frac{q^b}{m_0}\)\)v^\mu u^{\u n}\\
\nos
&\quad{}
-2\(\ga\nu \F^n_{\u m}-L^n_{\nu\u m}\)\(\rho^2l_{dn}g^{\d\a}
\(L^d_{\d\b}u^\b+L^d_{\d\u b}u^{\u b}\)
+\frac1{2\rho^2}\tge{\a\b}\rp\b l_{bn}\(\frac{q^b}{m_0}\)\)v^{\u m}u^\mu\\
\nos
&\quad{}
-g\id{\z\mu}\tge{\a\g}\rp\g \biggl(\rho l_{bm}\g^{\d\z}\(L^b_{\d\b}u^\b
+L^b_{\d\u b}u^{\u b}\)\\
&\qquad{}
+\frac1{2\rho^2}\,\tge{\z\b}\rp\b l_{bm}\(\frac{q^b}{m_0}\)\biggr)
\(v^\mu\(\frac{q^m}\dw\)-v^m u^\mu\)\\
\nos
&\quad{}
-g\id{\u z\u m}\tge{\a\g}\rp\g \(\frac{\rho}{r^2}\,l_{bm}g^{\u d\u z}
\(L^b_{\u d\b}u^\b+L^d_{\u d\u b}u^{\u b}\)
+\frac1{2r^2\rho^2}\tge{\u z\u b}\rp{\u b} l_{bm}\(\frac{q^b}{m_0}\)\)
\eal
\eq31\sevenrm cont.
$$

$$
\al
&\qquad{}
\times\(v^{\u z}\(\frac{q^m}\dw\)-v^mu^{\u z}\)
-\frac1{2\rho^2}\,\t Q{}^n{}\id{mb}\biggl(\rho^2l_{dn}g^{\d\a}
\(L^d_{\d\b}u^\b+L^d_{\d\u b}u^{\u b}\)\\
&\qquad{}
-\frac1{2\rho}\,\tge{\a\b}l_{pn}\(\frac{q^p}{m_0}\)\biggr)v^m\(\frac{q^b}{m_0}\)\\
&=0
\eal
\eq31\sevenrm cont.
$$
where $H^a_{\u m\u n}$ are given by Eq.~(4.55) and $\br{gauge}\n$ means a
derivative \wrt the connection $\t\om_E$ on a bundle $P$ (over a
space-time~$E$). $\o Q{}^\nu{}\id{\mu\b}$ means a torsion on the space-time~$E$
\wrt a connection ${\o\om^\a}_\b$ and $\o\n_\mu$ a covariant derivative \wrt
this connection. ${\o R^\a}_{\b\mu\nu}$ means a curvature tensor calculated
for this connection. $\h{\o\n}_{\u m}$ means a covariant derivative \wrt a 
connection $\h{\o\om}^{\u m}{}_{\u n}$ on~$M$, $\t\n_a$ means a covariant
derivative \wrt a connection ${\t\om^a}_b$ on a group manifold~$H$.

$H^a_{\mu\nu}$ is a curvature of a connection $\t\om_E$ given by Eq.~(4.53), 
$L^a_{\mu\nu}$ is given by Eq.~(5.22), $L^a_{\b\u n}, L^a_{\u a\u b}$ are
given by Eq.~(5.23) and Eq.~(5.24), respectively.

Let us consider Eq.~\rf23 \ for $A=\u a$ (i.e.\ for a manifold~$M$). One gets
$$
\al
&u^\b \o\n_\b v^{\u a}+u^{\u b}\t \n_{\u b}v^{\u a}
+\h{\o R}{}^{\u a}{}_{\u b\u m\u n}u^{\u b}\z^{\u m}u^{\u n}
-\h{\o Q}{}^{\u n}{}_{\u m\u b}\h{\o\n}_{\u n}u^{\u a}\z^{\u m}u^{\u b}\\
\nos
&\quad{}
+\frac1{r^2}\fw{q^b}l_{bd}g^{\u a\u b}\(2\gl d\mu{\u b}\)v^\mu\\
\nos
&\quad{}
+\frac1{r^2}\fw{q^b}l_{bd}g^{\u a\u b}\(2\hl b{\u n\u b}\)v^{\u b}\\
\nos
&\quad{}
-\frac{\rho^2}{r^2}\,v^nl_{dn}g^{\u d\u a}\(\lu d{\u d}{\u b}\b\)
-\fw{q^n}\tge{\u a\u b}\rp{\u b}l_{bn}v^b\\
\nos
&\quad{}
-\frac{2\rho^2}{r^2}\(l_{cd}g^{\u a\u p}\(2\gl d{[\mu}{|\u p|}\)L^d_{v]\b}
+l_{db}g^{\u d\u a}L^d_{\b\u d}H^b_{\mu\nu}\)u^\b \z^\mu u^\nu\\
\nos
&\quad{}
+\frac{2\rho^2}{r^2}\(l_{cd}g^{\u a\u p}\(2\hl d{\u p[\u m}\)L^c_{\u N]\b}
+l_{db}g^{\u d\u a}L^d_{\u d\b}H^b_{\u m\u n}\)u^\b\z^{\u m}u^{\u n}\\
\nos
&\quad{}
+\frac{2\rho^2}{r^2}\(l_{cd}g^{\u a\u p}\(2\hl d{\u p[\u m}\)L^c_{\u n]\u b}
+l_{db}g^{\u d\u a}L^d_{\u d\u b}H^b_{\u m\u n}\)u^{\u b}\z^{\u m}u^{\u n}\\
\nos
&\quad{}
-\frac{2\rho^2}{r^2}\(\frac12\,l_{cd}g^{\u a\u p}\(2\gl d\mu{\u p}\)L^c_{\u n\u b}
-l_{db}g^{\u d\u a}L^d_{\u d\u b}\ga\mu \F^d_{\u n}\)u^{\u b}\z^\mu u^{\u n}\\
\nos
&\quad{}
+\frac{2\rho^2}{r^2}\(\frac12\,l_{cd}g^{\u a\u p}\(2H^d_{\u p\u m}-L^d_{\u m\u p}\)
L^c_{\nu\u b}
-l_{db}g^{\u d\u a}L^d_{\u d\u b}\ga\nu \F^d_{\u m}\)u^{\u b}\z^{\u m}u^\nu\\
\nos
&\quad{}
-\frac{2\rho^2}{r^2}\(l_{cd}g^{\u a\u p}\(2\gl d{[\mu}{|\u p|}\)L^c_{\nu]\u b}
-l_{db}g^{\u d\u a}L^d_{\u d\u b}H^d_{\mu\nu}\)u^{\u b}\z^\mu u^\nu
\eal
\eq32
$$

$$
\al
&\quad{}
-\frac{2\rho^2}{r^2}\(\frac12\,l_{cd}g^{\u a\u p}
\(2\ga\mu \F^d_{\u p}-L^d_{\u p\mu}\)L^c_{\u n\b}
-l_{db}g^{\u d\u a}L^d_{\u d\b}\ga\mu \F^b_{\u n}\)u^\b \z^\mu u^{\u n}\\
\nos
&\quad{}
+\frac{2\rho^2}{r^2}\(\frac12\,l_{cd}g^{\u a\u p}\(2\hl d{\u p\u
m}\)L^c_{\nu\b} 
-l_{db}g^{\u d\u a}L^d_{\u d\b}\ga\nu \F^b_{\u m}\)u^\b \z^{\u m}u^\nu\\
\nos
&\quad{}
+\frac{q^b}{2m_0r^2\rho^2}\biggl[2\o\n_{[\mu}\(\rho^2l_{db}g^{\u a\u b}
\(2\gl d{\nu]}{\u b}\)\)\\
&\qquad{}
+\rho^2 l_{db}g^{\u a\u b}\(2\gl d\g{\u b}\)\o Q{}^\g{}\id{\mu\nu}
-\rho \tge{\u a\u b}\rp{\u b}l_{bc}H^c_{\mu\nu}\biggr]\z^\mu u^\nu\\
\nos
&\quad{}
+\frac{q^b}{2m_0r^2\rho^2}\biggl[\o\n_\mu\(\rho^2l_{db}g^{\u a\u b}
\(2\hl d{\u n\u b}\)\)\\
&\qquad{}
-\rho \tge{\u a\u b}\rp{\u b}l_{bc}\ga\mu \F^c_{\u n}
+\rho l_{bd}g^{\u a\u b}\(2\gl d\mu{\u b}\)g\id{\u d\u n}\tge{\u d\u c}\rho_{\u c}\biggr]
\z^\mu u^{\u n}\\
\nos
&\quad{}
+\frac{q^b}{2r^2m_0\rho^2}\biggl[2\h{\o\n}_{[\u m}\(\rho^2l_{db}g^{\u a\u b}
\(2\hl d{\u n]\u b}\)\)\\
&\qquad{}
+\rho^2l_{db}g^{\u a\u b}\(2\hl d{\u c\u b}\)\h{\o Q}{}^{\u c}{}\id{\u m\u n}
-\rho \tge{\u a\u b}\rp{\u b}l_{bc}H^c_{\u m\u n}\\
&\qquad{}
+2\rho l_{bd}g^{\u a\u b}\(2\hl d{[\u m|\u b|}\)g\id{|\u d|\u n]}\tge{\u d\u c}
\rp{\u c}\biggr]\z^{\u m}u^{\u n}\\
\nos
&\quad{}
+\frac1{r^2}\biggl\{\rho l_{bd}g^{\u a\u p}g\id{\b\d}\tge{\d\z}\rp\z
\(2\gl d\mu{\u p}\)
-\o\n_\mu\(\rho^2 l_{db}g^{\u d\u a}L^d_{\u d\b}\)\\
&\qquad{}
-\rho \tge{\u a\u p}\rp{\u p}l_{cd}L^c_{\b\mu}\biggr\}
u^\b\(\z^\mu\fw{q^b}-\z^b u^\mu\)\\
&\quad{}
+\frac1{r^2}\biggl\{\rho l_{bd}g^{\u a\u p}g\id{\u b\u d}\tge{\u d\u z}\rp{\u z}
\(2\gl d\mu{\u p}\)
-\o\n_\mu\(\rho^2l_{db}g^{\u d\u a}L^d_{\u d\u b}\)\\
&\qquad{}
+\rho \tge{\u a\u p}\rp{\u p}l_{cd}L^c_{\mu\u b}\biggr\}
u^{\u b}\(\z^\mu\fw{q^b}-\z^b u^\mu\)\\
\nos
&\quad{}
+\frac1{r^2}\biggl\{\rho l_{bd}g^{\u a\u p}g\id{\u b\u d}\tge{\u d\u z}\rp{\u z}
\(2\hl d{\u m\u p}\)
-\h{\o\n}_{\u m}\(\rho^2l_{db}g^{\u d\u a}L^d_{\u d\u b}\)\\
&\qquad{}
-\rho \tge{\u a\u p}\rp {\u p}l_{cd}L^c_{\u b\u m}\biggr\}
u^{\u b}\(\z^{\u m}\fw{q^b}-\z^b u^{\u m}\)\\
\nos
&\quad{}
+\frac1{r^2}\biggl(2\rho^4l_{d[b}l_{|e|f]}g^{\u d\u a}
\(\frac1{r^2}\,g^{\u z\u c}L^d_{\u d\u c}L^e_{\u z\b}
+g^{\z\g}L^d_{\u d\g}L^e_{\z\b}\)\\
&\qquad{}
+\tge{\u a\u p}\rp{\u p}g\id{\u b\u d}\tge{\u d\u z}\rp{\u z} l_{[bf]}
+\frac{\rho^2}2\,l_{dp}g^{\u d\u a}L^d_{\u d\b}C^p{}_{bf}\biggr)
u^\b \fw{\z^b q^f}
\eal
\eq32\sevenrm cont.
$$

$$
\al
&\quad{}
+\frac1{r^2}\biggl\{\o\n_\mu\(\rho \tge{\u a\u b}\rp{\u b}\)l_{ba}
+\t\n_a\(\rho^2 l_{bd}g^{\u a\u b}\(2\gl d\mu{\u b}\)\)\\
&\qquad{}
-\rho^4l_{dc}l_{bf}g^{\u d\u a}\biggl(\frac1{r^2}g^{\u c\u b}L^d_{\u d\u c}
\(2\gl f\mu{\u b}\)\\
&\qquad{}
+g^{\g\b}L^d_{\u d\g}\(2\hl f{\mu\b}\)\biggr)\\
&\qquad{}
-\tge{\u a\u z}\rp{\u z}g\id{\d\mu}\tge{\d\nu}\rp\nu l_{bc}\biggr\}
\fw{q^b}\(\z^a u^\mu-\z^\mu\fw{q^a}\)\\
%\nos
&\quad{}
+\frac1{r^2}\biggl\{\h{\o\n}_{\u m}\(\rho^2\tge{\u a\u b}\rp{\u b}\)l_{ba}
+\t\n_a\(\rho^2l_{bd}g^{\u a\u b}\(2\hl d{\u m\u b}\)\)\\
&\qquad{}
-\rho^2l_{dc}l_{bf}g^{\u d\u a}
\(\frac1{r^2}\,g^{\u c\u b}L^d_{\u d\u c}\(2\hl f{\u m\u b}\)
-g^{\g\b}L^d_{\u d\g}\(2\gl f\b{\u m}\)\)\\
&\qquad\quad{}
-\tge{\u a\u z}\rp{\u z}g\id{\u d\u m}\tge{\u d\u n}\rp{\u n}l_{bc}\biggr\}
\fw{q^b}\(\z^au^{\u m}-\z^{\u m}\fw{q^a}\)\\
%\nos
&\quad{}
+\frac1{r^2}\biggl(\rho \tge{\u a\u b}\rp{\u b}l_{bk}C^k{}_{ac}\\
&\qquad{}
+2\rho^3 g^{\u d\u a}\(\frac1{r^2}\,L^d_{\u d\u c}\tge{\u c\u b}\rp{\u b}
+L^d_{\u d\g}\tge{\g\b}\rp\b\)l_{d[a}l_{|b|c]}\biggr)\(\frac{q^b\z^aq^c}{4q^4m_0^2}\)\\
%\nos
&\quad{}
-\biggl\{2\rho^2l_{bd}\biggl[\frac1{r^2}
\(g^{\u n\u b}H^d_{\u c\u b}+\rho^2\(l_{bd}g^{\u n\u b}+l_{db}g^{\u b\u n}\)
L^d_{\u b\u c}\)\\
&\qquad{}
\times\(\h{\o\n}_{\u n}u^{\u a}
+\frac1{r^2}\,l_{fd}g^{\u a\u b}\(2\hl d{\u n\u b}\)\(\frac{q^f}{m_0}\)\)\\
&\qquad{}
+\(-g^{\nu\b}\ga\b \F^d_{\u
c}+\rho^2\(l_{bd}g^{\nu\b}+l_{db}g^{\b\nu}\)L^d_{\b\u c}\)\\
&\qquad{}
\times\(\o\n_\nu u^{\u a}+\frac1{r^2}\,l_{fd}g^{\u a\u b}\(2\gl d\nu{\u b}\)
\(\frac{q^f}{m_0}\)\)\biggr]\biggr\}
\(u^{\u c}\fw{q^b}-v^bu^{\u c}\)\\
%\nos
&\quad{}
-\biggl\{2\rho^2 l_{bd}\biggl[\frac1{r^2}\(g^{\u n\u b}\ga\g \F^d_{\u b}
+\rho^2\(l_{bd}g^{\u n\u b}+l_{db}g^{\u b\u n}\)L^d_{\u b\b}\)\\
&\qquad\quad{}
\times\(\h{\o\n}_{\u n}u^{\u a}+\frac1{r^2}\,l_{fd}g^{\u a\u b}
\(2\hl d{\u n\u b}\)\(\frac{q^f}{m_0}\)\)\\
&\qquad{}
+\(g^{\nu\b}H^d_{\b\g}+\rho^2\(l_{bd}g^{\nu\b}+l_{db}g^{\b\nu}\)L^d_{\b\g}\)\\
&\qquad\quad{}
\times\(\o\n_\nu u^{\u a }+\frac1{r^2}\,l_{fd}g^{\u a\u b}
\(2\gl d\nu{\u b}\)\(\frac{q^f}{m_0}\)\)\biggr]\biggr\}
\(u^\g\fw{q^b}-v^bu^\g\)\\
%\nos
&\quad{}
-\mu\biggl[\tge{\u n\u b}\rp{\u b}k_{bc}\(\h{\o\n}_{\u n}u^{\u a}
+\frac12\,l_{df}g^{\u a\u c}\(2\hl d{\u n\u c}\)\(\frac{q^f}{m_0}\)\)
\eal
\eq32\sevenrm cont.
$$

$$
\al
&\qquad{}
+\tge{\nu\b}\rp\b k_{bc}\(\o\n_\nu u^{\u a}
+\frac12\,l_{df}g^{\u a\u c}\(2\gl d\nu{\u c}\)\(\frac{q^f}{m_0}\)\)\biggr]
\(\frac{q^cv^b}{m_0}\)\\
\nos
&\quad{}
+\frac2{r^2}\(\rho^2 l_{dn}g^{\u d\u a}\(\lu d{\u d}{\u b}\b\)
+\frac1{2\rho^2}\,\tge{\u a\u b}\rp{\u b}l_{bn}\(\frac{q^b}{m_0}\)\)\\
&\qquad{}
\times\biggl(\(\hl n{\mu\nu}\)v^\mu u^\nu+\(\hl n{\u m\u n}\)v^{\u m}u^{\u n}\\
&\qquad{}
+\(\gl n\mu{\u n}\)v^\mu u^{\u n}
-\(\gl n\nu{\u m}\)v^{\u m}u^\nu\biggr)\\
\nos
&\quad{}
-\frac1{r^2}\(g\id{\z\mu}\tge{\u a\u c}\rp{\u c} \rho l_{bn}g^{\d\z}
\(\lu b\d\b{\u b}\)
+\frac1{2\rho^2}\,\tge{\u a\u b}\rp{\u b}l_{bm}\(\frac{q^b}{m_0}\)\)\\
&\qquad{}
\times\(v^\mu\fw{q^m}-v^m u^\mu\)\\
\nos
&\quad{}
-\frac1{r^2}\(g\id{\u z\u m}\tge{\u a\u c}\rp{\u c}\rho l_{bn}g^{\u d\u z}
\(\lu d{\u d}\b{\u b}\)
+\frac1{2\rho^2}\,\tge{\u a\u b}\rp{\u b} l_{bm}\(\frac{q^b}{m_0}\)\)\\
&\qquad{}
\times\(v^{\u m}\fw{q^m}-v^mu^{\u m}\)\\
\nos
&\quad{}
-\frac1{2r^2\rho^2}\,\t Q{}^n{}\id{mb}\(\rho^2l_{dn}g^{\u d\u a}
\(\lu d{\u d}\b{\u b}\)
-\frac1{2\rho}\,\tge{\u a\u b}\rp{\u b}l_{pn}\(\frac{q^p}{m_0}\)\)\\
&\qquad{}
\times v^m\(\frac{q^b}{m_0}\)\\
&=0
\eal
\eq32\sevenrm cont.
$$
where $\h{\o R}{}^{\u a}{}_{\u b\u m\u n}$ means a curvature tensor of a
connection $\h{\o\om}{}^{\u a}{}_{\u b}$ on~$M$ and $\h{\o Q}{}^{\u n}{}_{\u
m\u b}$ its tensor of torsion. $\tge{\u a\u b}$ and $\tge{\a\b}$ are defined
similarly to~$\tga AB $. We have
$$
\ealn{
\z^A&=(\z^\a,\z^{\u a}) & \rf33 \cr
v^A&=(v^\a,v^{\u a}) & \rf34 \cr
u^A&=(u^\a,u^{\u a}). & \rf35 
}
$$

Let us pass to the Eq.\ \rf24 \ and develop it according to our scheme. One
gets:
$$
\al
&\frac{dv^a}{d\tau}+\t R^a{}_{bcd}\fw{q^b}\z^c\fw{q^d}
-\t Q{}^c{}_{mb}\t\n_c\fw{q^a}v^m\fw{q^b}\\
&\quad{}
+\Biggl(2\rho^2l_{bd}g^{\g\d}\(2\hl d{[\nu|\d|}\)L^a_{|\g|\mu]}
+\frac1\rho\,\d^a{}_b g\id{\d\a}\tge{\d\g}\rp\g \o Q{}^\a{}_{\mu\nu}
\eal
\eq36
$$

$$
\al
&\qquad{}
-2\d^a{}_b \o\n_{[\nu}\(\frac1\rho\,g\id{|\d|\mu]}\tge{\d\g}\rp\g\)\\
&\qquad{}
+\(\frac1{r^2}g^{\u c\u d}\(2\gl d{[\nu}{|\u d|}\)L^a_{|\u d|\mu]}\)\Biggr)
\fw{q^b\z^\mu u^\nu}\\
\nos
&\quad{}
+\Biggl(2\rho^2l_{bd}g^{\g\d}\(-\ga\d \F^d_{\u n}+\frac12\,L^d_{\d\u
n}\)L^a_{\g\mu} 
-\d^a{}_b \h{\o\n}_{\u n}\(\frac1\rho\, g\id{\d\mu}\tge{\d\g}\rp\g\)\\
&\qquad{}
+\frac1{2r^2}\(g^{\u c\u d}\(2\hl d{\u n\u d}\)L^a_{\u d\mu}\)\Biggr)
\fw{q^b\z^\mu u^{\u n}}\\
\nos
&\quad{}
+2\rho^2l_{bd}\biggl(g^{\g\d}\(-\ga\d \F^d_{\u n}+\frac12\,L^d_{\d\u n}\)L^a_{\g\u
m}\\
&\qquad{}
+\frac1\rho\,\d^a{}_b g\id{\u d\u a}\tge{\u d\u c}\rp{\u c}
\h{\o Q}{}^{\u a}{}_{\u m\u n}
-2\d^a{}_b \h{\o\n}_{[\u n}\(\frac1\rho\,g\id{|\u d|\u m]}
\tge{\u d\u c}\rp{\u c}\)\\
&\qquad{}
+\frac1{r^2}\,g^{\u c\u d}\(2\hl d{[\u n|\u d|}\)L^a_{|\u c|\u m]}\biggr)
\fw{q^b\z^{\u m}u^{\u n}}\\
\nos
&\quad{}
+2\rho^2l_{bd}\biggl(\frac12\,g^{\g\d}\(2\hl d{\nu\d}\)L^a_{\g\u m}
-\d^a{}_b \o\n_\nu\(\frac1\rho\, g\id{\u d\u m}\tge{\u d\u c}\rp {\u c}\)\\
&\qquad{}
+\frac1{2r^2}\,g^{\u c\u d}\(2\gl d\nu{\u d}\)L^a_{\g\u m}\biggr)
\fw{q^b\z^{\u m}u^\nu}\\
\nos
&\quad{}
+\biggl(\rho \tge{\g\b}\rp\b l_{bc}L^a_{\g\mu}
+\frac1{r^2}\,\rho \tge{\u c\u b}\rp{\u b}l_{bc}L^a_{\u b\mu}
+\rho l_{bd}\tge{\b\nu}\rp\nu\(2\hl d{\mu\b}\)\d^a{}_c\\
&\qquad{}
+\frac1{r^2}\,\rho l_{bd}\tge{\u b\u n}\rp{\u n}
\(2\gl d\mu{\u b}\)\d^a{}_c\biggr)\fw{q^b}\(\z^c u^\mu -\z^\mu\fw{q^c}\)\\
\nos
&\quad{}
+\biggl(\rho\tge{\g\b}\rp\b l_{bc}L^a_{\g\u m}
+\frac1{r^2}\,\rho\tge{\u c\u b}\rp{\u b}l_{bc}L^a_{\u b\u m}
-\rho l_{bd}\tge{\b\nu}\rp\nu\(2\gl d\b{\u m}\)\d^a{}_c\\
&\qquad{}
+\frac1{r^2}\,\rho l_{bd}\tge{\u b\u n}\rp{\u n}\(2\hl d{\u m\u b}\)\d^a{}_c\biggr)
\fw{q^b}\(\z^cu^{\u m}-\z^{\u m}\fw{q^c}\)\\
\nos
&\quad{}
+\biggl(\t \n_b L^a_{\b\g}-\o\n_\g\(\frac1\rho\,g\id{\b\d}\tge{\d\nu}
\rp\nu\)\d^a{}_b
+\rho^2l_{db}g^{\d\mu}L^a_{\mu\g}L^d_{\d\b}\\
&\qquad{}
-\frac1{\rho^2}\,g\id{\d\g}\tge{\d\b}\rp\b g\id{\b\nu}\tge{\nu\mu}\rp\mu\d^a{}_b\biggr)
u^\b\(\z^b u^\mu-\z^\mu\fw{q^b}\)\\
&\quad{}
+\biggl(\t\n_b L^a_{\b\u c}-\h{\o\n}_{\u c}
\(\frac1\rho\,g\id{\b\d}\tge{\d\nu}\rp\nu\)\d^a{}_b
+\frac1{r^2}\,l_{db}g^{\u d\u m}L^a_{\u m\u c}L^d_{\u d\b}
\eal
\eq36\sevenrm cont.
$$

$$
\al
&\qquad{}
+\rho^2l_{db}g^{\d\mu}L^a_{\mu\u c}L^d_{\d\b}
-\frac1{\rho^2}\,g\id{\u d\u c}\tge{\u d\u b}\rp{\u b}
g\id{\b\nu}\tge{\nu\mu}\rp{\mu}\d^a{}_b\biggr)u^\b
\(\z^bu^{\u c}-\z^{\u c}\fw{q^b}\)\\
\nos
&\quad{}
+\biggl(\t\n_b L^a_{\u b\u c}-\h{\o\n}_{\u c}\(\frac1\rho\,g\id{\u b\u d}
\tge{\u d\u n}\rp{\u n}\)\d^a{}_b
+\rho^2l_{db}g^{\d\mu}L^a_{\mu\u c}L^d_{\d\u b}\\
&\qquad{}
+\rho^2l_{db}g^{\u d\u m}L^a_{\u m\u c}L^d_{\u d\u b}
-\frac1{\rho^2}\,g\id{\u d\u c}\tge{\u d\u e}\rp{\u e}
g\id{\u b\u n}\tge{\u n\u m}\rp{\u m}\d^a{}_b\biggr)
u^b\(\z^b u^{\u c}-\z^{\u c}\fw{q^b}\)\\
\nos
&\quad{}
+\biggl(\t\n_bL^a_{\u b\g}-\o \n_\g\(\frac1\rho\,g\id{\u b\u d}
\tge{\u d\u n}\rp{\u n}\)\d^a{}_b
+\rho^2l_{db}g^{\d\mu}L^a_{\mu\g}L^d_{\d\u b}\\
&\qquad{}
+\rho^2l_{db}g^{\u d\u m}L^a_{\u m\g}L^d_{\u d\u b}
-\frac1{\rho^2}\,g\id{\d\g}\tge{\d\e}\rp\e g\id{\u b\u n}\tge{\u n\u m}
\rp{\u m}\d^a{}_b\biggr)u^{\u b}\(\z^bu^\g-\z^\g\fw{q^b}\)\\
\nos
&\quad{}
+\biggl(-\frac1\rho\,g\id{\b\d}\tge{\d\g}\rp\g C^a{}_{bc}\\
&\qquad{}
+2\rho\(g\id{\g\nu}\tge{\nu\mu}\rp\mu g^{\d\g}L^d_{\d\b}
+\frac1{r^2}\,g\id{\u c\u n}\tge{\u n\u m}\rp{\u m}g^{\u d\u c}L^d_{\u d\b}\)
l_{d[b}\d^a{}_{c]}\biggr)\fw{\z^bq^c}u^\b\\
\nos
&\quad{}
+\biggl(-\frac1\rho\,g\id{\u b\u d}\tge{\u d\u c}\rp{\u c}C^a{}_{bc}\\
&\qquad{}
+2\rho\(g\id{\g\nu}\tge{\nu\mu}\rp\mu g^{\d\g}L^d_{\d\u b}
+\frac1{r^2}\,g\id{\u c\u n}\tge{\u n\u m}\rp{\u m}g^{\u d\u c}L^d_{\u d\u b}\)
l_{d[b}\d^a_{c]}\biggr)\fw{\z^bq^c} u^{\u b}\\
\nos
&\quad{}
-2\(g\id{\g\d}\tge{\d\mu}\rp\mu \tge{\g\b}\rp\b
+\frac1{r^2}\,g\id{\u c\u d}\tge{\u d\u m}\rp{\u m}
\tge{\u c\u b}\rp{\u b}\)l_{b[c}\d^a{}_{d]}
\(\frac{q^b\z^d q^c}{4\rho^4m_0^2}\)\\
\nos
&\quad{}
+\(2\o\n_{[\mu}L^a_{|\b|\nu]}+L^a_{\b\g}\o Q{}^\g{}\id{\mu\nu}
+\frac2\rho\,g\id{\b\d}\tge{\d\g}\rp\g H^a_{\mu\nu}
+\frac2\rho\,g\id{\d[\mu}L^a_{|\b|\nu]}\tge{\d\g}\rp\g\)u^\b\z^\mu u^\nu\\
\nos
&\quad{}
+\(2\h{\o\n}_{[\u m}L^a_{|\u b|\u n]}
+L^a_{\u b\u c}\h{\o Q}{}^{\u c}{}\id{\u m\u n}
+\frac2\rho\,g\id{\u b\u d}\tge{\u d\u c}\rp{\u c}H^a_{\u m\u n}
+\frac2\rho\,g\id{\u d[\u m}L^a_{|\u b|\u n]}\tge{\u d\u c}\rp{\u c}\)
u^{\u b}\z^{\u m}u^{\u n}\\
&\quad{}
+\(2\o\n_{[\mu}L^a_{|\u b|\nu]}+L^a_{\u b\g}\o Q{}^\g{}_{\mu\nu}
+\frac2\rho\,g\id{\u b\u d}\tge{\u d\u c}\rp{\u c}H^a_{\mu\nu}
+\frac2\rho\,g\id{\d[\mu}L^a_{|\u b|\nu]}\tge{\d\g}\rp\g\)
u^{\u b}\z^\mu u^\nu\\
&\quad{}
+\(\h{\o\n}_{\u m}L^a_{\b\nu}
-\frac2\rho\,g\id{\b\d}\tge{\d\g}\rp\g \ga\nu \F^a_{\u m}
+\frac1\rho\,g\id{\u d\u m}L^a_{\b\nu}\tge{\u d\u c}\rp{\u c}\)
u^\b \z^{\u m}u^\nu\\
\nos
&\quad{}
+\(\o\n_\nu L^a_{\b\u n}
+\frac2\rho\,g\id{\b\d}\tge{\d\g}\rp\g \ga\mu \F^a_{\u n}
+\frac1{\rho r^2}\,g\id{\d\mu}L^a_{\b\u n}\tge{\d\g}\rp\g\)u^\b\z^\mu u^{\u n}\\
\nos
&\quad{}
+\(2\h{\o\n}_{[\u m}L^a_{|\b|\u n]}
+L^a_{\b\u c}\h{\o Q}{}^{\u c}{}\id{\u m\u n}
+\frac2\rho\,g\id{\b\d}\tge{\d\g}\rp\g H^a_{\u m\u n}
+\frac2\rho\,g\id{\u d[\u m}L^a_{|\b|\u n]}\tge{\u d\u c}\rp{\u c}\)
u^\b \z^{\u m}u^{\u n}\\
\nos
&\quad{}
+\(\o\n_\mu L^a_{\u b\u n}
+\frac2\rho\,g\id{\u b\u d}\tge{\u d\u c}\rp{\u c}\ga\mu \F^a_{\u n}
+\frac1\rho\,g\id{\d\mu}L^a_{\u b\u n}\tge{\d\g}\rp\g\)u^{\u b}\z^\mu u^{\u n}
\eal
\eq36\sevenrm cont.
$$

$$
\al
&\quad{}
+\(\h{\o\n}_{\u m}L^a_{\u b\nu}
-\frac2\rho\,g\id{\u b\u d}\tge{\u d\u c}\rp{\u c}\ga\nu \F^a_{\u m}
+\frac1\rho\,g\id{\u d\u m}L^a_{\u b\nu}\tge{\d\u c}\rp{\u c}\)
u^{\u b}\z^{\u m}u^\nu\\
\nos
&\quad{}
-\frac1{2\rho}\,\t Q^a{}\id{mb}v^m\qm b\(g\id{\b\d}\tge{\d\g}\rp\g u^\b
+g\id{\u b\u d}\tge{\u d\u c}\rp{\u c}u^{\u b}\)\\
\nos
&\quad{}
+\h{\o Q}{}^{\u n}{}\id{\u m\u z}u^{\u z}v^{\u m}
\(\frac{\rp{\u n}}{\rho^3}\qm a
-\(L^a_{\b\u n}u^\b+L^a_{\u b\u n}u^{\u b}\)
-\frac1{2\rho^3}g\id{\u d\u n}\tge{\u d\u c}\rp{\u c}\qm a\)\\
\nos
&\quad{}
+\o Q{}^\nu{}\id{\mu\z}u^\z v^\mu 
\(\frac{\rp \nu}{\rho^3}\qm a
-\(L^a_{\b\nu}u^\b+L^a_{\u b\nu} u^{\u b}\)
-\frac1{2\rho^3}\,g\id{\d\nu}\tge{\d\g}\rp\g \qm a\)\\
\nos
&\quad{}
-\frac1\rho\,g\id{\z\mu}\tge{\z\om}\rp\om\biggl(\t\n_b\fw{q^a}\\
&\qquad{}
+\frac1\rho\(g\id{\b\d}\tge{\d\g}\rp\g u^\b
+g\id{\u b\u d}\tge{\u d\u c}\rp{\u c}u^{\u b}\)\d^a{}_b\biggr)
\(v^\mu\fw{q^b}-v^b u^\mu\)\\
\nos
&\quad{}
-\frac1\rho\,g_{\u z\u m}\tge{\u z\u p}\rp{\u p}\biggl(\t\n_b\fw{q^a}\\
&\qquad{}
+\frac1\rho\(g\id{\u b\u d}\tge{\u d\u c}\rp{\u c}u^{\u b}
+g_{\b\d}\tge{\d\g}\rp\g u^\b\)\d^a{}_b\biggr)
\(v^{\u m}\fw{q^b}-v^bu^{\u m}\)\\
\nos
&\quad{}
-2\(\hl n{\mu\nu}\)\(\t\n_n\fw{q^a}
+\frac1\rho\(g\id{\b\d}\tge{\d\g}\rp\g u^\b
+g\id{\u b\u d}\tge{\u d\u c}\rp{\u c}u^{\u b}\)\d^a{}_n\)v^\mu u^\nu\\
\nos
&\quad{}
-2\(\hl n{\u m\u n}\)\(\t\n_n\fw{q^a}
+\frac1\rho\(g\id{\b\d}\tge{\d\g}\rp\g u^\b
+g\id{\u b\u d}\tge{\u d\u c}\rp{\u c}u^{\u b}\)\d^a{}_n\)v^{\u m}u^{\u n}\\
\nos
&\quad{}
-2\(\gl n\mu{\u n}\)\(\t\n_n\fw{q^a}
+\frac1\rho\(g\id{\b\d}\tge{\d\g}\rp\g u^\b
+g\id{\u b\u d}\tge{\u d\u c}\rp {\u c}u^{\u b}\)\)v^\mu u^{\u n}\\
\nos
&\quad{}
+2\(\gl n\nu{\u m}\)\(\t\n_n\fw{q^a}
+\frac1\rho\(g\id{\b\d}\tge{\d\g}\rp\g u^\b
+g\id{\u b\u d}\tge{\u d\u c}\rp{\u c}u^{\u b}\)\)u^\nu v^{\u m}\\
\nos
&\quad{}
-\Biggl\{\biggl[2\mu\rho \tge{\nu\mu}\rp\mu k_{bc}\biggl(-\frac1{\rho^2}\qm a\rp\nu
+\(\lv a\g\nu{\u c}\)\\
&\qquad\quad{}
+\frac1{2\rho^3}\,g\id{\d\nu}\tge{\d\g}\rp\g\qm a\biggr)\biggr]\\
&\qquad{}
+\biggl[2\mu\rho \tge{\u n\u m}\rp{\u m}k_{bc}\biggl(-\frac1{\rho^2}\qm a\rp{\u n}
+\(\lv a\g{\u n}{\u c}\)\\
&\qquad\quad{}
+\frac1{2\rho^3}\,g\id{\u d\u n}\tge{\u d\u c}\rp{\u c}\qm a\biggr)\biggr]
\Biggr\}\fw{v^bq^c}
\eal
\eq36\sevenrm cont.
$$

$$
\al
&\quad{}
-\(2\rho^3l_{bd}g^{\nu\b}H^d_{\g\b}
+\rho^2\(l_{bd}g^{\nu\b}+l_{db}g^{\b\nu}\)L^d_{\b\g}\)\\
&\qquad{}
\times\(-\frac1{\rho^3}\qm a\rp\nu +\lv a\d\nu{\u d}
+\frac1{2\rho^3}\,g\id{\d\nu}\tge{\d\e}\rp\e \qm a\)\\
&\qquad{}
\times\(v^\g\fw{q^b}-v^bu^\g\)\\
\nos
&\quad{}
-\(2\rho^3l_{bd}g^{\u n\u b}\ga\g \F^d_{\u b}
+\rho^2\(l_{bd}g^{\u n\u b}+l_{db}g^{\u b\u n}\)L^d_{\u b\g}\)\\
&\qquad{}
\times\(-\frac1{\rho^3}\qm a\rp{\u n}+\lv a\a{\u n}{\u c}
+\frac1{2\rho^3}\,g\id{\u d\u n}\tge{\u d\u c}\rp{\u c}\qm a\)\\
&\qquad{}
\times\(v^\g\qm b-v^b u^\g\)\\
\nos
&\quad{}
-\(2\rho^3l_{bd}g^{\u n\u b}H^d_{\u c\u b}
+\rho^2\(l_{bd}g^{\u n\u b}+l_{db}g^{\u b\u n}\)L^d_{\u b\u c}\)\\
&\qquad{}
\times\(-\frac1{\rho^3}\qm a\rp{\u n}+\lv a{\u c}{\u n}\a
+\frac1{2\rho^3}\,g_{\u d\u n}\tge{\u d\u e}\rp{\u e}\qm a\)\\
&\qquad{}
\times\(v^{\u c}\qm b-v^b u^{\u c}\)\\
\nos
&\quad{}
-\(-2\rho^3l_{bd}g^{\nu\b}\ga\b\F^d_{\u c}
+\rho^2\(l_{bd}g^{\nu\b}+l_{db}g^{\b\nu}\)L^d_{\b\u c}\)\\
&\qquad{}
\times\(-\frac1{\rho^3}\qm a\rp \nu +\lv a{\u e}\nu\a
+\frac1{2\rho^3}\,g\id{\d\nu}\tge{\d\g}\rp\g\qm a\)\\
&\qquad{}
\times\(v^{\u c}\qm b-v^bu^{\u c}\)\\
\nos
&\quad{}
+\frac1{2\rho^3}\qm a\(g\id{\b\d}\tge{\d\g}\rp\g v^\b
+g\id{\u b\u d}\tge{\u d\u c}\rp{\u c}v^{\u b}\)\\
\nos
&\quad{}
+L^a_{\b\nu}v^\b u^\nu +L^a_{\u b\u n}v^{\u b}u^{\u n}
+L^a_{\u b\nu}v^{\u b}u^\nu +L^a_{\b\u n}v^{\b}u^{\u n}\\
&\quad{}
+\frac1\rho\,g_{\d\nu}\tge{\d\g}\rp\g v^au^\nu
+\frac1\rho\,g_{\u d\u n}\tge{\u d\u c}\rp{\u c} v^au^{\u n}\\
&=0
\eal
\eq36\sevenrm cont.
$$
where $\t R{}^a{}_{bcd}$ is a tensor of a curvature for a connection 
$\t \om{}^a{}_b$ on a group~$H$ and $\t Q{}^a{}_{bc}$ is a tensor of a
torsion for this connection. It is worth to notice that $Q^\g{}\id{\u m\nu}$,
$Q^\g{}\id{\mu\u n}$ are equal to zero. Moreover for a convenience of a
future consideration it is good to keep them there (i.e.\ in Eq.~\rf32 ).

Let us notice that $u^\a,u^\b$ satisfy equations \rf2 \ and \rf 3 . In this
way we get geodetic deviation equations in our theory. If we suppose that we
are close to the minimum of a Higgs' potential we should put in our equation
$H^n_{\u a\u b}=0$. What is the physical meaning of geodetic deviation
equations? In order to find it let us consider a flow of geodesic $\G(\s)$,
$\s\in U\subset {\Bbb R}$ given by
$$
\ealn{
x^{\u A}&=x^{\u A}(\tau,\s), &\rf 37 \cr
\z^{\u A}&=\(\frac{\pa x^{\a}}{\pa \s},\frac{\pa x^{\u a}}{\pa \s},
\frac{\pa x^{a}}{\pa \s}\)_{\big| \s=\s_0}, &\rf 38 \cr
&\s_0\in U.}
$$
$\s$ is a parameter such that for every $\s_1\ne\s_2$, $x^A(\tau,\s_1)$ and
$x^A(\tau,\s_2)$ are different geodesics. One can say that we have a family
of geodesic curves, $\G(\s)$. The geodesic considered here is $\G(\s_0)$,
i.e.\ for $\s=\s_0$.

Thus
$$
\frac{dx^a}{d\tau}=u^a=\frac1{2\rho^2}\qm a \eq39
$$
where $\rho=\rho(\tau,\s)=\rho(x(\tau,\s))$.

Thus one gets
$$
x^a=\frac12\qm a(\s)\intop_{\tau_0}^\tau \frac{d\tau}{\rho^2(\tau,\s)}
+x_0^a(\s) \eq40
$$
and
$$
\ealn{
\z^a&=\frac{\pa}{\pa\s}\biggl[\frac{q^a}{m_0}(\s)
\intop_{\tau_0}^\tau \frac{d\tau}{\rho^2(\tau,\s)}\biggr]
_{\big| \s=\s_0} + \frac{dx^a_0}{d\s} &\rf 41 \cr
v^a&=\frac{d\z^a}{d\tau}=\frac{\pa}{\pa\tau}
\(\frac{q^a}{m_0}(\s)\,\frac1{\rho^2(\tau,\s)}\)_{\big| \s=\s_0}. &\rf 42
}
$$
$v^a$ is (of course) an Ad-type quantity. In this way we get
$$
\frac{dv^a}{d\tau}=-\frac{\pa}{\pa\s}
\(\frac{q^a}{m_0}(\s)\,\frac1{\rho^3}\,\frac{\pa\rho}{\pa\tau}\)_{\big|
\s=\s_0}. \eq 43 
$$

In this way Eq.\ \rf 22 \ together with Eq.\ \rf 1 \ give us an
interpretation of geodetic deviation equations in our theory.
They are analogous to the deviation equations for charged particles moving in
nonabelian Yang-Mills' field and Higgs' field and nonsymmetric \gr \ field as
well.

Let us give some remarks on a physical interpretation of the vector 
$\z^{\u A}=(\z^\a,\z^{\u a},\z^a)$. The vector $\z^{\u A}$---``geodesic
separation''---is a displacement (tangent vector) from point on a fiducial
geodesic to a point on nearby geodesic characterized by the same value of an
affine parameter~$\tau$. Thus $v^{\u A}=(v^\a,v^{\u a},v^a)$ means a relative
``velocity'' and $u^{\u B}\n_{\u B}v^{\u A}$---a relative
``acceleration''---equals, according to Eq.~\rf22 \ terms with a curvature
and torsion. Thus we get ``tidal forces'' in our theory
($(n+n_1+4)$-dimensional case), for charged (in a nonabelian gauge
field sense and Higgs' field) test particles. In this equation we get \gr\
``tidal forces'' from NGT, Yang-Mills' ``tidal forces'', Higgs' field ``tidal
forces'' and additional effects which can be treated as
gravito-Yang-Mills-Higgs tidal forces. The scalar field~$\rho$ is also a
source of additional ``tidal forces''.

These new effects are ``interference effects'' between \gr, nonabelian
Yang-Mills' interactions and Higgs' field. We can project our equation on a
space-time~$E$ (they are defined on a bundle manifold~$P$), taking any local
section~$e$ of the bundle~$P$. In this way we get gauge dependent charges
$Q^a$ (Yang-Mills' charged-colour charges) and gauge dependent $\o v^a$. We
can substitute $\o v^a$ into Eq.~\rf 36 . However, we should substitute in
the place of $\frac{dv^a}{d\tau}$ the expression
$$
\frac{d\o v^a}{d\tau}-C^a{}_{bc}A^c_\mu u^\mu \o v^b
-C^a{}_{bc}\F^c_{\u b}\o v^{\u b} \eq44
$$
where
$$
e^\ast \om=A^a_\mu \o\theta{}^\mu X_a+\F^a_{\u b}\t\theta{}^{\u b}X_a\,. \eq45
$$

Finally, let us remark that Eq.\ \rf31 \ represents tidal
gravito-Yang-Mills-Higgs forces and Eq.~\rf36 \ is a relative change of 
$(\frac{q^a}{m_0})(\s)$ for different test particles via $v^a$ (or $(\frac{Q^a}{m_0})(r)$ via
$\o v^a$).

Eq.\ \rf32 \ is a relative change of a Higgs charge $u^{\u b}$ for different
test particles via~$v^{\u b}$. The interpretation of this equation is much
more complex for the ``Higgs charge'' $u^{\u b}$ is not ``conserved'' during
a motion of a test particle. The existence of a ``constant Higgs charge''
introduced here can help us in physical understanding of these interesting
phenomena. Our gravito-Yang-Mills-Higgs-scalar tidal forces are a strict and
natural generalization of gravito-electromagnetic-scalar tidal forces from
the Nonsymmetric Kaluza--Klein (Jordan--Thiry) Theory (in an electromagnetic
case) and gravito-Yang-Mills-scalar tidal forces from the Nonsymmetric
Kaluza--Klein (Jordan--Thiry) Theory in a general nonabelian case (see
Ref.~[18]).

}\vfill\eject

\null
\vskip36pt plus4pt minus4pt
\centerline{{\four\bf 16. A Tower of Scalar Fields}}

\vskip4pt
\centerline{{\four\bf in the Nonsymmetric-Nonabelian
Jordan--Thiry Theory.}}

\vskip4pt
\centerline{{\four\bf The Warp Factor}}
\vskip24pt plus2pt minus2pt
\write0{16. A Tower of Scalar Fields
in the Nonsymmetric-Nonabelian Jordan--Thiry Theory.
 The Warp Factor \the\pageno}

In section~5 we mention a more general case of the scalar field $\rho 
= \rho (x,y)$, $x\in E$, $y\in M$. Moreover it is more convenient to consider the
field ${\varPsi} $ in place of $\rho $.
$$
\rho = e^{-{\varPsi} }.\eqno(16.1)
$$
In this case one finds:
$$
{\varPsi} (x,y) = \sum^{\infty }_{k=0}{\varPsi} _{k}(x) \chi
_{k}(y).\eqno(16.2)
$$
Let us find a condition for the function $\rho (x,y)$ such that 
${\varPsi} (x,y)$ is an
$L^{2}( M,dm,\sqrt{|\widetilde{g}|} )$-valued function of $x\in E$. 
From (16.1) one finds that
$\ln (\rho )$ is an $L^{2}(M,dm,\sqrt{|\widetilde{g}|})$-valued function i.e.:
$$
\mmint_{M} (\ln (\rho ))^{2} \sqrt{|\widetilde{g}|} dm(y) <
\infty.\eqno(16.2\text{\rm a})
$$
Thus the sufficient condition for $\rho $ is that $\rho $ and $\rho ^{-1}$ are bounded on
$M-A$, where $m(A)=0$. It means that $\widetilde{\rho }(x), 
\widetilde{\rho }^{-1}(x) \in L^{\infty }(M,dm,\sqrt{|\widetilde{g}|})$
and $\|\widetilde{\rho }(x)\|_{\infty } =\break
\esssup_{y\in M}|\rho
(x,y)|< \infty $. The convergence $\widetilde{\rho }_{n}(x,\cdot ) 
\rightarrow \widetilde{\rho }(x,\cdot )$
with respect to the norm $\|\cdot \|_{\infty }$ means a uniform limit. We have the same
for $\widetilde{\rho }^{-1}(x)$. We can also consider a decomposition of $\rho (x,y)$ into a
series of generalized harmonics on $M$ i.e.
$$
\rho (x,y)\sim \rho _{0}(x) + \sum^{\infty }_{k=1}\rho _{k}(x) \chi _{k}(y)
\eqno(16.2\text{\rm b})
$$
such that
$$\eqalign{
\rho _{k}(x) &= {1\over V_{2}}\mmint_{M}\sqrt{|\widetilde{g}|}\rho (x,y) 
\chi _{k}(y) dm(y),\cr
\rho _{0}(x) &= {1\over V_{2}}\mmint_{M}\sqrt{|\widetilde{g}|}\rho (x,y) 
dm(y).\cr}\eqno(16.2\text{\rm c})
$$
If ${\displaystyle\sum^{\infty }_{k=0}}\rho ^{2}_{k}(x) < \infty $, $x\in E$ the series 
on the right-hand side of (16.2b)
converges to the function $\rho (x,y)$ in
$L^{2}(M,dm,\sqrt{|\widetilde{g}|})$ sense. The limit
can be understood in a uniform sense if $\rho (x,y)$ is continuous and
$V(x)=\Var (\widetilde{\rho }(x)) < \infty $ in $M$ for every $x\in E$.

Moreover it is interesting to know properties of $\widetilde{{\varPsi}
}({x})$ if $\widetilde{\rho}({x})$ is an %\break
$L^{2}({M}{,}{dm}{,}\sqrt{|\widetilde{g}|})$-valued function. One gets:
$$
\|\widetilde{\rho }(x)\|^{2} \le V_{2} + \sum^{\infty }_{n=1}{2^{n}\over n!} 
(\|\widetilde{{\varPsi} }(x)\|_{n})^{n} < \infty,\eqno(16.2\text{\rm d})
$$
where
$$
\|{\varPsi} (x)\|_{n} = \Big(\mmint_{M}\sqrt{|\widetilde{g}|}
|{\varPsi} (x,y)|^{n} dm(y)\Big)^{1/n}
$$
is an $L^{n}(M,dm,\sqrt{|\widetilde{g}|})$ norm. Let us suppose that for 
every $n\in {\sym N}^{\infty }_{1}$, $\widetilde{{\varPsi} }(x)$
is an $L^{n}(M,dm,\sqrt{|\widetilde{g}|})$-valued function. For
$$
V_{2} = \mmint_{M} \sqrt{|\widetilde{g}|} dm(y) < \infty 
$$
one gets
$$
\|\widetilde{{\varPsi} }(x)\|_{n} \le \|\widetilde{{\varPsi} }(x)\|_{n' } 
V^{1/n-1/n' }_{2},
$$
if $1 \le n \le n' \le \infty $ and
$$\eqalignno{
L^{\infty }(M,dm,\sqrt{|\widetilde{g}|}) &\subset 
L^{n'}(M,dm,\sqrt{|\widetilde{g}|}) \subset 
L^{n}(M,dm,\sqrt{|\widetilde{g}|})\subset\cr
&\subset L^{1}(M,dm,\sqrt{|\widetilde{g}|}) =
L(M,dm,\sqrt{|\widetilde{g}|}).\cr}
$$
Thus if $n' \rightarrow \infty $ one obtains
$$
\|\widetilde{{\varPsi} }(x)\|_{n} \le \|\widetilde{{\varPsi} }(x)\|_{\infty }
V^{1/n}_{2} = (\esssup_{y\in M} |{\varPsi} (x,y)|)
V^{1/n}_{2}.\eqno(16.2\text{\rm e})
$$
From (16.2d) and (16.2e) we have
$$
\|\widetilde{\rho }(x)\|^{2} \le V_{2} e^{2\|{\varPsi}
(\tilde{x})\|_{\infty}}.\eqno (16.2\text{\rm f})
$$
Thus if $\widetilde{{\varPsi} }(x) \in L^{\infty
}(M,dm,\sqrt{|\widetilde{g}|}) \widetilde{\rho }(x)$ is an
$L^{2}(M,dm,\sqrt{|\widetilde{g}|})$-valued
function . This means that ${\varPsi} (x,y)$ is bounded on $M-A$ where
$m(A)=0$. In
this way we get an interesting duality. The sufficient condition for
$\widetilde{{\varPsi} }(x)$ to be an
$L^{2}(M,dm,\sqrt{|\widetilde{g}|})$-valued function is $\widetilde{\rho }(x), 
\widetilde{\rho }^{-1}(x) \in 
L^{\infty}(M,dm,\sqrt{|\widetilde{g}|})$ and the sufficient condition for $\widetilde{\rho }(x)$ to be an
$L^{2}(M,dm,\sqrt{|\widetilde{g}|})$ is $\widetilde{{\varPsi} }(x) \in 
L^{\infty}(M,dm,\sqrt{|\widetilde{g}|})$. In the second case we can try
an expansion of ${\varPsi} (x,y)$ into a series of generalized harmonics similarly
as for $\rho$ (i.e.\ Eqs.~(16.2b--c).

The $(n_{1}+4)$-dimensional lagrangian for the scalar field ${\varPsi} $ looks like:
$$
{\cal L}_{\scal}({\varPsi} ) = (\ov{M}\widetilde{\gamma }^{(CM)} 
{\varPsi} _{,C} {\varPsi} _{,M} + n^{2} \gamma ^{[MN]} \gamma _{DM} 
\widetilde{\gamma }^{(DC)} {\varPsi} _{,N} {\varPsi} _{,C})\eqno(16.3)
$$
or %\vadjust{\eject}
$$\eqalignno{
{\cal L}_{\scal}({\varPsi} ) ={}& (\ov{M}\widetilde{g}^{(\gamma \mu )} 
{\varPsi} _{,\gamma } {\varPsi} _{,\mu } + n^{2} g^{[\mu \nu ]} 
g_{\delta \mu } \widetilde{g}^{(\delta \gamma )} {\varPsi} _{,\nu } 
{\varPsi} _{,\gamma }) +\cr
&+ {1\over r^{2}} (\ov{M}\widetilde{g}^{(\tilde c \tilde m )}
{\varPsi} _{,\tilde c }{\varPsi} _{,\tilde m }+n^{2}g^{[\tilde m
\tilde n ]}g_{\tilde d \tilde m }\widetilde{g}^{(\tilde d \tilde c )}
{\varPsi} _{,\tilde n }{\varPsi} _{,\tilde c }) 
=\widetilde{{\cal L}}_{\scal}({\varPsi} ) + Q({\varPsi}
).\qquad\ &(16.4)\cr}
$$
Moreover we should average over $M$ and $H$.

Thus one gets using (16.2), (16.4) and (5.50):
$$\displaylines{
{1\over V_{1}V_{2}}\mmint_{M} ({\cal L}_{\scal}({\varPsi} )+Q({\varPsi} )
)\sqrt{|\widetilde{g}|} \sqrt{|\ell|} dm(y) d\mu _{H}(h) =\hfill\cr
\hfill = \sum^{\infty }_{k=0} {\cal L}_{\scal}({\varPsi} _{k}) + {1\over 2}
\sum^{\infty }_{k,l=1} \widehat{M}_{kl} {\varPsi} _{k} {\varPsi}
_{l},\quad\ (16.5)\cr}
$$
where ${\cal L}_{\scal}({\varPsi} _{k})$ is given by the formula (7.6) i.e. is a usual
lagrangian for the scalar field ${\varPsi} $ in our theory and
$$
{1\over 2} \widehat{M}_{kl} = {(-1)\over V_{2}r^{2}}
\mmint_{M}(\ov{M}\widetilde{g}^{(\tilde c \tilde m )} +
n^{2}g^{[\tilde m \tilde n ]} g_{\tilde d \tilde m }
\widetilde{g}^{(\tilde d \tilde c )}) \chi _{k,\tilde c } 
\chi _{l,\tilde n }\sqrt{|\widetilde{g}|}\,dm(y) < \infty \eqno(16.6)
$$
and $\widehat{M}_{kl} = \widehat{M}_{lk} $, $k,l=1,2\ldots $

It is easy to see that we get a tower of fields ${\varPsi} _{k}$ such that according
to (16.6) the field ${\varPsi} _{0}$ is massless and remaining fields are
massive with a scale of a mass ${\displaystyle{1\over r^{2}}}$.

The infinite dimensional quadratic form
$$
{1\over 2} \sum^{\infty }_{k,l=1}\widehat{M}_{kl} {\varPsi} _{k}(x) 
{\varPsi} _{l}(x)
$$
is convergent in $\ell ^{2}$ for every $x$. This is easily satisfied if the
derivatives of $\chi _{k}$ belong to
$L^{2}(M,dm,\sqrt{|\widetilde{g}|})$ and $g_{\tilde{a} \tilde{b} }$ are continuous
functions on $M$, which is always satisfied for our case $(M$ is compact
without boundaries and $g_{\tilde{a} \tilde{b} }$ defined in section~4). 
For $L^{2}(M,dm,\sqrt{|\widetilde{g}|})$
is unitary equivalent to $L^{2}(M,dm)$ we can consider $\chi _{k}$ eigenfunctions
of Beltrami--Laplace operator on $M$.

In some cases we can diagonalize the infinite dimensional matrix $\widehat{M}_{kl}$
getting some new fields %\vadjust{\vskip-2pt}
$$ %\vadjust{\vskip-2pt}
{\varPsi}'_{k'}(x) = \sum^{\infty }_{k=1}A_{k'k} {\varPsi} _{k}(x),
\eqno(16.7)
$$
such that %\vadjust{\vskip-2pt}
$$
b_{k} \delta _{kl} = \sum^{\infty }_{k',l'=1}A_{kk'} A_{ll'} 
\widehat{M}_{k'l'},\eqno(16.8)
$$
where $A = (A_{k'k})$ is a unitary operator in $\ell ^{2}$ space i.e.
$$ %\vadjust{\vskip-2pt}
\|Aa\|= \|a\|\eqno(16.9)
$$
and $A(\ell ^{2}) = \ell ^{2}$.

The diagonalization procedure can be achieved if $\widehat{M}_{kl}$ is a symmetric
unbounded (Hermitian) operator in $\ell ^{2}$. This is equivalent to
$$
\widehat{M}_{kl} = \widehat{M}_{lk}\eqno(16.10)
$$
and $M(\ell ^{2}) = \ell ^{2}$.
Thus we get for the lagrangian:
$$
{\cal L}_{\scal}({\varPsi} ) = \sum^{\infty }_{k=0}{\cal L}_{\scal}
({\varPsi}'_k) - \sum^{\infty }_{k=1} {b_{k}\over 2}
({\varPsi}'_{k})^{2},\eqno(16.11)
$$
$\bigg({\displaystyle{b_{k}\over \ov{M}}}\bigg)$ have an interpretation as 
$m^{2}_{k}$ for ${\varPsi}'_k $. If $\widehat{M}_{kl}$ is a negatively
defined operator in $\ell ^{2}$ we have for every $k\ b_{k} \ge 0$ if $M_{kl}$ is invertible
$b_{k} > 0$ for every $k$. The latest condition means that every
${\varPsi}'_k $ is
massive with ${\displaystyle{1\over r^{2}}}$ as a scale of the mass. Let us write down a
self-interaction term for ${\varPsi}'_k $ coming from cosmological terms. One gets:
$$
V(\{{\varPsi}'_k\}) =
{1\over V_{2}} \mmint_{M}\sqrt{|\widetilde{g}|}
dm(y) \bigg({\alpha ^{2}_{S}\over \ell ^{2}_{pl}} \widetilde{R}
(\widetilde{{\varGamma} }) e^{(n+2){\varPsi} (x,y)} + 
{e^{n{\varPsi} (x,y)}\over r^{2}} \widehat{\ov{R}}
(\widehat{\overline{\varGamma}})\bigg),\eqno(16.12)
$$
where
$$
{\varPsi} (x,y) = {\varPsi} _{0}(x) + \sum^{\infty }_{k=1}
{\varPsi} _{k}(x) \chi _{k}(y)\eqno(16.13)
$$
and
$$
{\varPsi}'_k = \sum^{\infty }_{k=1}A_{kk'} {\varPsi} _{k'}.\eqno(16.14)
$$
The field ${\varPsi} _{0}(x)$ can get a mass from a cosmological 
background. If we
suppose ${\varPsi} = {\varPsi} (x)$ it means really that ${\varPsi} (x) = {\varPsi} _{0}(x)$ i.e. a truncation.
In this more general case for the field $\rho$ (or ${\varPsi} )$ one can easily get
similar formulae for the remaining part of the lagrangian involving
this field i.e. averaging over the manifold $M$.

All the factors in front of lagrangian terms are exponential functions
of ${\varPsi} $. The sufficient condition to make all the integration over $M$
convergent is to suppose that $\widetilde{{\varPsi} }(x) \in 
L^{2}(M,dm,\sqrt{|\widetilde{g}|}) \cap L^{\infty
}(M,dm,\sqrt{|\widetilde{g}|})$ i.e. $\widetilde{{\varPsi} }(x)$ is $L^{\infty }$-valued function.

In the case of a completely broken group $G$ we should put $G$ for
$M$.

Let us give a more rigorous justification for an intuitive procedure
concerning diagonalization of the infinite matrix $\widehat{M}$.

Let us consider a bilinear form in
$L^{2}(M,dm,\sqrt{|\widetilde{g}|}) = L^{2} \simeq L^{2}(M,dm)$
$$
\widehat{M}(f,g) = {1\over V_{2}r^{2}} \mmint_M (\ov{M}
\widetilde{g}^{(\tilde c \tilde m )}+n^{2}g^{[\tilde m (\tilde n ]}
g_{\tilde d \tilde n }\widetilde{g}^{(|\tilde d |\tilde c ))}) 
f_{,\tilde c } g_{,\tilde m }\sqrt{|\widetilde{g}|}
dm(y),\eqno(16.15)
$$
for $f,g \in L^{2}(M,dm,\sqrt{|\widetilde{g}|}) \cap {\sym
C}^{1}(M)$.

This form is always defined for $f$ and $g$, because $g_{\tilde{a}
\tilde{b} }$ are smooth
functions on a compact manifold and $f_{,\tilde c } , g_{,\tilde m } 
\in L^{2}(M,dm,\sqrt{|\widetilde{g}|})$ (if
they exist). ${\sym C}^{1}(M)$ is linearly dense in
$L^{2}(M,dm,\sqrt{|\widetilde{g}|})$. Let us
suppose that the tensor $P^{(\tilde c \tilde m )}$, where
$$
P^{\tilde c \tilde m } = (\ov{M}\widetilde{g}^{\tilde c \tilde m } + 
n^{2} g^{[\tilde m \tilde n ]} g_{\tilde d \tilde n }
\widetilde{g}^{(\tilde d \tilde c )})
$$
is invertible and negatively defined.

One notices that $P^{\tilde c \tilde m }$ is $G$-invariant on $M$. 
The same is of course true
for $P_{(\tilde c \tilde m )}$ being an inverse tensor of 
$\widetilde{P}^{(\tilde c \tilde m )}$. Moreover every
$G$-invariant metric is induced by a scalar product $<.,.>$ on 
$\fg /\fg _{0} = \fm$
which is invariant under the action of $\Ad_{G_0}$ on $\fm$. Thus $P$ is induced by
such a scalar product. The tensor $P_{\tilde c \tilde m }$ induces on $M$ an additional
Einstein--Kaufmann geometry compatible with it. We can repeat some
consideration concerning symmetric and skewsymmetric forms on $M$ from
section~4. If $P_{[\tilde d \tilde{b} ]}$ is not degenerate and 
$\widehat{\overline{\nabla}}_{\tilde{a} }P_{[\tilde d \tilde{b} ]} = 0$ we can get an
almost complex structure on $M$ induced by $P_{\tilde c \tilde m } = 
P_{(\tilde c \tilde m )}+ P_{[\tilde c \tilde m ]}$ (if $\dim M$ is
even).

Let us consider a linear functional on $L^{2} \cap {\sym C}^{1}(M) = D$
$$
F_{f}(g) = \widehat{M}(f,g) ,\quad\ g,f \in D \subset L^{2},\eqno(16.16)
$$
$F_{f} \in (L^{2})^{*} = L^{2}$ (Riesz theorem).
Thus for every $f$ we have a functional $F$.

Thus we can define a linear operator on $L^{2}$ such that
$$
L(f) = F_{f} ,\hbox{ or }(Lf,g) = \widehat{M}(f,g).\eqno(16.17)
$$
It is easy to see that if $f_{1} = f_{2}$ then $F_{f_{1}} = F_{f_{2}}$. Thus $L$ is well
defined. Simultaneously we find that $\Ker L = \text{\rm
 constant}$ functions on $M \cong 
{\sym R}^{1}$. This is a subspace of $L^{2}$. Thus we can consider $L$ a quotient space
of ${\cal H}$ and denote it $\widehat{L}$, where
$$
{\cal H} = L^2/(\Ker L).\eqno(16.18)
$$
Let $F_{f_{1}} = F_{f_{2}}$ then one gets
$$
\bigwedge _{g\in L^2-(\Ker L)} \widehat{M}(f_{1},g) =
\widehat{M}(f_{2},g).\eqno(16.19)
$$
Using (16.15) one gets that $f_{1,\tilde c } = f_{2,\tilde c }$ {\it
modulo}\/ a set of a zero
measure. Thus $[f_{1}] = [f_{2}] \in {\cal H}$. One can easily check that $\widehat{L}$ is
symmetric because $\widehat{M}(f,g) = \widehat{M}(g,f)$.

The operator $\widehat{L}$ is unbounded on $D$ and it is negatively strictly defined
i.e.
$$
(-\widehat{L}(f),f) \ge C\|f\|^{2},\eqno(16.20)
$$
$C$ is a positive constant.

For this we can apply Friedrichs' theory for $\widehat{L}$. We denote the
Friedrichs' space for $\widehat{L}$, $L^{2}_{0}(M)$ and the scalar product in $L^{2}_{0}(M)$ is
given by
$$
(f,g)_{\rF} = (-\widehat{L}f,g),\eqno(16.21)
$$
($D/(\Ker L)$ is dense in $L^{2}_{0}(M)$ in a sense of the norm $\|\cdot
\|_{F}$). Let $\widetilde{L}=\overline{\widehat{L}}$ (a
closure of $\widetilde{L})$ denotes a Hermitian extension of $\widehat{L}$ 
in $L^{2}_{0}(M)$, which
exists due to Friedrichs' theorem and it is invertible. Moreover $\widetilde{L}^{-1}$
is compact.

For this we can proceed a spectral decomposition of 
$\widetilde{L}$. Such an
operator has a pure point unbounded spectrum on a negative part of a
real axis and eigenvectors (eigenfunctions) span $L^{2}_{0}(M)$.
One gets
$$
\widetilde{L} = - \sum^{\infty }_{k=1}\mu ^{2}_{k} P_{k} ,\quad\ 
\mu _{k}\neq 0 ,\ \mu _{k}\rightarrow \infty ,\eqno(16.22)
$$
where $P^{2}_{k} = 1$, $P_{k} P_{l} = \delta _{kl}I$ are projective operators in $L^{2}_{0}(M)$. The sum
is understood in a weak topology at a point.

The above considerations justify an intuition of a diagonalization
procedure for an infinite matrix $\widehat{M}_{kl}$.

The existence of the tower of scalar fields in our theory can help
in renormalization problems. The fields can work as regulator fields
if we change some parameters in our theory.

The Friedrichs' theory gives us the following properties of
eigenvectors for~$\widetilde{L}$,
$$
\widetilde{L} \zeta _{k} = -\mu ^{2}_{k} \zeta _{k}\quad\ k=1,2\ldots 
\eqno(16.23)
$$
They are normalized and orthogonal:
$$\eqalignno{
\mmint_{M} \zeta ^{2}_{k}(y) \sqrt{|\widetilde{g}|} dm(y) &=1,&(16.24)\cr
\mmint_{M} \zeta _{k}(y) \zeta _{l}(y) \sqrt{|\widetilde{g}|} dm(y) &=
0\quad \hbox{if}\ k\neq l.&(16.25)\cr}
$$
If the spectrum is degenerated i.e. $\mu ^{2}_{k}$ is repeated $q$ times $(q$ is a
natural number)
$$
\mu ^{2}_{k} = \mu ^{2}_{k+1} = \ldots = \mu
^{2}_{k+q-1},\eqno(16.26)
$$
then $\zeta _{k}$, $\zeta _{k+1}$, \dots , $\zeta _{k+q-1}$ form an algebraic basis of a finite
dimensional linear space of functions $f \in L^{2}_{0}(M) \cap {\sym C}^{2}(M)$ satisfying
Eq.~(16.26).  After the  orthogonalization  procedure  of
$\{\zeta _{k},\zeta _{k+1},\ldots ,\zeta _{k+q-1}\}$ (for example Schmidt procedure) we can
define projectors:
$$
P_{k} = P_{\zeta _{k}}\hbox{ for every }k=1,2\ldots 
$$
The orthonormal set of eigenfunctions $\zeta _{k}$, $k=1,2\ldots $ is complete in
$L^{2}_{0}(M)$. Thus
$$
I = \sum^{\infty }_{k=1}P_{k} = \sum^{\infty }_{k=1}P_{\zeta _{k}},
\eqno(16.27)
$$
where $I$ is an identity operator in $L^{2}_{0}(M)$. Thus we can expand the field
${\varPsi} $ into a complete set of functions $\zeta _{k}$
$$\displaylines{\hfill
{\varPsi} (x,y) = {\varPsi} _{0}(x) + \sum^{\infty }_{k=1}{\varPsi} _{k}
(x) \zeta _{k}(y),\hfill\llap{(16.28)}\cr
\hbox{such that}\hfill
\sum^{\infty }_{k=1}{\varPsi} ^{2}_{k}(x) < \infty ,\hfill\llap{(16.29)}\cr}
$$
for $x\in E$.

Let us find conditions of a diagonalization of the infinite matrix $\widehat{M}_{kl}$.
It means we are looking for conditions of the existence of the unitary
transformation from the basis $\chi _{k}$, $k=1,2\ldots $ to the basis $\zeta
_{k}$, $k=1,2\ldots $ in
${\cal H}$ (or in $L^{2}_{0}(M))$. Later we give those conditions.

For $\{\zeta _{k}\}$ is complete one always gets
$$
\chi _{k} = \sum^{\infty }_{l=1}A_{kl} \zeta _{l}\quad\ \hbox{for every }k
\eqno(16.30)
$$
and
$$
\sum^{\infty }_{l=1}A^{2}_{kl} < \infty.\eqno(16.31)
$$
Moreover one has
$$
A_{kl} = (\chi _{k}, \zeta _{l}).\eqno(16.32)
$$
For the same reasons we have
$$\displaylines{
\hfill \zeta _{k} = \sum^{\infty }_{l=1}B_{kl} \chi
_{l},\hfill\llap{(16.33)}\cr
\hfill \sum^{\infty }_{l=1}B^{2}_{kl} < \infty\hfill \llap{(16.33$'$)}\cr}
$$
and
$$
B_{kl} = (\zeta _{k}, \chi _{l}).\eqno(16.34)
$$
One gets
$$
A_{kl} = B_{lk}\eqno(16.35)
$$
and
$$
\chi _{k} = \sum^{\infty }_{l=1}A_{kl} \sum^{\infty }_{p=1}B_{lp} \chi _{p} = 
\sum^{\infty }_{p=1}\Big(\sum^{\infty }_{l=1} A_{kl} B_{lp}\Big)\chi
_{p}.
\eqno(16.36)
$$
This is possible only if
$$
\sum^{\infty }_{l=1}A_{kl} B_{lp} = \sum^{\infty }_{l=1}A_{kl} A_{pl} = \delta
_{kp}.\eqno(16.37)
$$
Thus $A_{kl}$ is an isometry in the $\ell ^{2}$-space and $B_{kl}$ either. Moreover $A_{kl}$
and $B_{kl}$ are invertible and A is an inverse operation of $B$. This means
A is an unitary transformation. The transformation A diagonalizes the
infinite dimensional matrix $\widehat{M}_{kl}$. The only one condition for the
existence of $A$ is the following. The set $\{\chi _{k}\}$ or $\{\eta _{k}\}$ are complete
bases in $L^{2}_{0}(M)$.

Let us consider the operator $\widetilde{L}$ in more details and find its shape for
$f\in {\sym C}^{2}(M)/\break \Ker L)$. The operator $\widetilde{L}$ can be considered a Gateaux derivative of
the quadratic form $\widehat{M}$ in $L^{2}_{0}(M)$. One finds:
$$\displaylines{\hfill
\widetilde{L}f = {1\over \sqrt{|\widetilde{g}|}} (\sqrt{|\widetilde{g}|}
\ov{p}^{\tilde{a} \tilde{b} }f_{,\tilde{a}})_{,\tilde{a} }
= (\ln \sqrt{|\widetilde{g}|}_{,\tilde{b} }\ov{p}^{\tilde{a}
\tilde{b} } f_{,\tilde{a} } + \ov{p}^{\tilde{a} \tilde{b}
}{}_{,\tilde{b} } f_{,\tilde{a} } + \ov{p}^{\tilde{a} \tilde{b} } 
f_{,\tilde{a} ,\tilde{b} }),\hfill\llap{(16.38)}\cr\noalign{\vskip-6pt}}
$$
where\vadjust{\vskip-6pt}
$$
\ov{p}^{\tilde c \tilde m } =\ov{p}^{\tilde m \tilde c }
=\bigg[
\widetilde{g}^{(\tilde c \tilde m )} + {n^{2}\over 2} ( g^{[\tilde m
\tilde n ]} g_{\tilde d \tilde n } \widetilde{g}^{(\tilde d \tilde c )} 
+ g^{[\tilde c \tilde n ]} g_{\tilde d \tilde n }
\widetilde{g}^{(\tilde d \tilde m )})\bigg]\eqno(16.39)
$$
The operator $\widetilde{L}$ slightly differs from the Beltrami--Laplace operator
defined on $(M,\ov{p})$.
$$
\ov{p}= \ov{p}_{\tilde{a} \tilde{b} } \theta ^{\tilde{a} } \otimes 
\theta ^{\tilde{b} },\quad\ \ov{p}_{\tilde{a} \tilde{b} }
\ov{p}^{\tilde{a} \tilde c } = \delta ^{\tilde c }_{\tilde{b} }
$$
because $\det (\ov{p}_{\tilde{a} \tilde{b} } ) \neq \det
(g_{\tilde{a} \tilde{b} }) = \widetilde{g}$.
It differs also from the Beltrami--Laplace operator on $(M,h^{0})$.

Moreover $\widetilde{L}$ is a left-invariant operator on $C^{2}(M)$ of the second order.
Supposing the reductive decomposition of $\fg=\fg_{0} \dot{+}\fm$ we have a 
complete description of an algebra of $G$-invariant operators on $M$,
$D(G/G_{0})$ in terms of Lie algebra $\fg$ and $\fg_{0}$. The algebra is 
commutative if $M$ is a
symmetric space. Let $M$ be a symmetric space (of compact type of course)
and let rank $(M) = K$. The algebra $D(G/G_{0})$ has finite numbers of
generators i.e $D_{1}, D_{2}\ldots D_{K}$. In this case every 
$D \in D(G/G_{0})$ is a
symmetric polynomial of $D_{i}$, $i = 1,2\ldots$ $K, D = W(D_{1}, D_{2},\ldots 
D_{K})$. Notice that degrees of $D_{i}$, $i = 1,2\ldots K$ $d_{i} = 1,2\ldots 
K$ are canonically
established by $G$.

Thus $\Delta= W_{1}(D_{1},\ldots 
D_{K})$ and $\widetilde{L} = W_{2}(D_{1}\ldots D_{K},\zeta )$ and $W_{1} \neq W_{2} W_{1}(\ldots 
) = W_{2}(\ldots ,0)$.

The existence of the unitary operator $A$ in $\ell ^{2}$-space means that $\widehat{L}$ and
$\widehat{\Delta }=\Delta |_{{\cal H}}$ have the same Friedrichs' theory i.e. the same $L^{2}_{0}(M)$. For $M$ is
compact without boundaries and $h^{0}{}_{\tilde{a} \tilde{b} } 
\theta ^{\tilde{a} } \otimes \theta ^{\tilde{b} }$,
$\widetilde{g}_{\tilde{a} \tilde{b} } \theta ^{\tilde{a} } \otimes 
\theta ^{\tilde{b} }$, $p_{\tilde{a} \tilde{b} } \theta ^{\tilde{a} } 
\otimes \theta ^{\tilde{b} }$ are smooth functions on $M$ one gets:
$$
\ov{m}_{2}(-\widehat{\Delta }f,f) \le (-\widehat{L}f,f) \le \ov{m}_{1}
(-\widehat{\Delta }f,f),\eqno(16.40)
$$
$\ov{m}_{1}$, $\ov{m}_{2}$ are positive constants.

This means that $L^{2}_{0}(M)$ for $\widehat{\Delta }$ and $\widehat{L}$ are equivalent and the conditions
we need to imposed for $g_{\tilde{a} \tilde{b} }$ and $p_{\tilde{a}
\tilde{b} }$ are
$$
\det ( g_{\tilde{a} \tilde{b} } ) \neq 0\quad \hbox{and}\quad \det
(\ov{p}_{\tilde{a} \tilde{b} }) \neq 0.
$$
These are the only conditions for the existence of the unitary operator
A in $\ell ^{2}$-space.

Let us consider $M = S^{2}$ with the nonsymmetric tensor (6.58). One gets
$$
\widetilde{L} f = \bigg(\ov{M} + {n^2\zeta ^2\over (\zeta^2+1)}\bigg)
\Delta f,\eqno(16.41)
$$
where $\Delta$ is an ordinary Beltrami--Laplace operator on $S^{2}$ 
in spherical
coordinates. In this case we do not need any procedure described above,
because eigenfunction for $\widetilde{L}$ are simply spherical functions $Y_{\ell m}(\theta ,\varphi )$ and
$$
\widetilde{L} Y_{\ell m} = - \bigg(\ov{M} + {n^2\zeta ^2\over
(\zeta^2+1)}\bigg)\ell (\ell + 1)Y_{\ell m}.\eqno(16.42)
$$
The mass spectrum is degenerated for $m = -\ell $, $-(\ell - 1)\ldots 
 0\ldots \ell - 1,\ell $, $\ell = 1,2,3,\ldots $ and
$$
\ov{m}(\ell ,m) = {1\over r}\sqrt{\bigg(1+{\displaystyle{n^2\zeta ^2\over
\ov{M}(\zeta^2+1)}}\bigg)\ell(\ell+1)},\quad\ \ell = 1,2,3.\eqno(16.43)
$$
Let us notice the following fact: if the homogeneous space $M = G/G_{0}$ is a
two-point homogeneous space the operator $\widetilde{L}$ is proportional to the
Beltrami--Laplace operator on $M$. This fact is coming from
left-invariancy of $\widetilde{L}$ and from this fact that
$\widetilde{L}$ is of the second order
differential operator for $f \in C^{(2)}(M)$. In general this is not true.
Moreover for $\zeta = 0$
$$
\widetilde{L} = \ov{M}\Delta.\eqno(16.44)
$$
Let us notice that for $S^{2}$ (see Eq.~(8.71)) we need $\zeta \to\infty $ 
in order to kill
cosmological constant. Thus $\ov{m}(\ell ,m) \simeq {1\over
r}\sqrt{2(\ell +1)\ell}$. We have the following
two-point homogeneous spaces $\SU(p+1)/S(U(1)\times U(p))$, $\SO(n +
1)/\SO(n)$, $(f_{u(-52)}$, so(9)) of compact type, which can serve as the manifold $M =
G/G_{0}$. Only the first one is Hermitian (K\"ahlerian) and in this case one
gets
$$
\widetilde{L}= \bigg(\ov{M} + {n^2\zeta ^2\over
(\zeta^2+1)}\bigg)\Delta\eqno(16.45)
$$
the same formula as for $S^{2}$. Moreover for every two-point symmetric
space $\widetilde{L}$ is proportional to $\Delta$ (Beltrami--Laplace operator)
$$\eqalignno{
\widetilde{L} &= \fg (\zeta )\Delta,&(16.45\text{\rm a})\cr
\fg(0) &= \ov{M}.&(16.45\text{\rm b})\cr}
$$
It is interesting to find $L^{2}_{0}(M)$ in terms of irreducible spaces of
representations of the group $G$. In the case of $S^{2} = 
\SO(3)/\SO(2)$ one simply gets
$$
L^{2}{}_{0}(M) =\sum^{\infty }_{i=1}\oplus H^{\ell },\eqno(16.46)
$$
where $H^{\ell }$ is space of an irreducible representation of the group SO(3)
$$
\dim H^{\ell } = 2\ell + 1.
$$
In this case a dynamical group of $\widetilde{L}$ is SO(3,1) (the so called spectrum
generation group) SO(3)$\subset $SO(3,1).

${\displaystyle\sum^{\infty }_{i=0}}\oplus H^{\ell } = L^{2}(M)
=L^{2}{}_{0}(M) \oplus R$ is a representation space of a unitary
representation of SO(3,1).

In general the situation is more complex.

Let $G$ be a simple compact Lie group and let $G_{0}$ be its closed subgroup
such that it is a semisimple or $G_{0} = U(1)\otimes G_{0}$, where $G_{0}$ is semisimple.

Let $C = h_{ab}Y^{a}Y^{b}$ be a Casimir operator of $G$ and let $C_{0}$ be a Casimir
operator of $G_{0}$ (or $G_{0}'$). Let us suppose a reductive decomposition of
the Lie algebra $\fg$, $\fg = \fg_{0} \dot{+} \fm$ (or $\fg_{0}{}'$) and consider an operator
$$
- D = C - C_{0} = h_{\tilde{a} \tilde{b} }Y^{\tilde{a} }Y^{\tilde{b}
}.\eqno(16.47)
$$
This operator acts in the complement $\fm$, which is diffeomorphic to
$\Tan_{0}(G/G_{0}) $, $0 = \varphi (\varepsilon )$.

Let us define a left-invariant differential operator $\Delta$ on $M$
corresponding to $D$, via a pull-back of the left action of the group $G$
on $M$. Let us find eigenfunctions of this operator and its eigenvalues.
This can be done as follows. Let $H_{\bar{\lambda}_k}$ be invariant spaces of
irreducible representations of $G$ corresponding to the value of the
Casimir operator, $C$, $\overline{\lambda }_{k}$ and $H_{\bar{\mu }_{\ell }}$ be 
invariant spaces of irreducible
representations of $G_{0}$ corresponding to the value of the Casimir
operator $C_{0}$, $\overline{\mu}_{\ell }$. Both groups are compact and those representations
are finite-dimensional and unitary. All invariant spaces $H_{\bar{\lambda}_k}$
and $H_{\bar{\lambda}_{\ell }}$ are Hilbert spaces. One gets
$$
L^{2}(M) = \sum_{\bar{\lambda}_k} \oplus H_{\bar{\lambda}_k} = 
\sum_{\bar{\lambda}_k} \oplus \sum_{\bar{\mu}_{\ell } \in \bar{\lambda}_k}
\oplus b(\bar{\lambda}_k,\bar{\mu}_{\ell })H_{\bar{\mu}_{\ell }},\eqno(16.48)
$$
where $\overline{\mu}_{\ell } \in \overline{\lambda}_k$, means that $H_{\bar{\lambda}_k}$
is decomposed into some irreducible representation spaces $\overline{\mu}_{\ell }$
of the subgroup of $G$, $G_{0}$. $b(\overline{\lambda}_k,\overline{\mu}_{\ell })$ 
means a multiplicity of $H_{\bar{\mu}_{\ell }}$ in $H_{\bar{\lambda}_k}$.
Thus one gets
$$
\Delta f_{k,\ell ,\ell '} =\overline{\eta} (k,\ell )f_{k,\ell ,\ell '},
\eqno(16.49)
$$
where
$$
\overline{\eta}(k,\ell ) = \overline{\lambda}_k - \overline{\mu}_{\ell } < 0,\eqno(16.50)
$$
$f_{k,\ell ,\ell '} \in L^{2}(M)$ and $f_{k,\ell ,\ell '} \in H^{(\ell ')}
_{\bar{\mu}_{\ell }}$, $\overline{\mu}_{\ell } \in \overline{\lambda}_k$.
$$
H^{1}_{\bar{\mu}_{\ell }} \oplus \ldots \oplus H^{\ell '}\oplus \ldots 
\oplus H^{b(\bar{\lambda}_k,\bar{\mu}_{\ell })}_{\bar{\mu}_{\ell }} 
\subset H_{\bar{\lambda}_k}.
$$
The dimension of the space corresponding to $\eta _{(k,\ell )}$ can be easily
calculated.
$$
\dim H_{\bar{\eta} (k,\ell )} = b(\overline{\lambda}_k,\overline{\mu}_{\ell })\dim 
H_{\bar{\mu}_{\ell }}.\eqno(16.51)
$$
Thus $\overline{\eta}(k,\ell )$ is in general degenerated. The most important results is
this that
$$
L^{2}(M) =\sum^{}_{\bar{\lambda }_{k}}\oplus H_{\bar{\lambda
}_{k}},\quad\ \overline{\lambda }\ne 0\eqno(16.52)
$$
or
$$
L^{2}(M) = L^{2}{}_{0}(M)\oplus H_{0},\quad\ \dim H_{0} = 1,\ H_{0} \simeq 
R',\eqno(16.52\text{\rm a})
$$
or
$$
L^{2}(M) = L^{2}{}_{0}(M)/R'.\eqno(16.52\text{\rm b})
$$
In the case of a $U(1)$ factor we get similarly
$$
\Delta f_{(k,\ell ,\ell ',m)} =\overline{\eta}(k,\ell )f_{(k,\ell ,\ell',m)},
\eqno(16.53)
$$
where
$$
f_{(k,\ell ,\ell ',m)} = f_{k,\ell ,\ell '}\fg_{m},\quad\ m = 0,\pm 1,\pm 2,\ldots,
\eqno(16.54)
$$
$\fg_{m}$ is a function of a one-dimensional irreducible representation of
$U(1)$, $m = 0$, $\pm 1$, $\pm 2,\ldots $. Thus the construction of eigenfunctions 
of $\Delta$ is known
from representation theory of $G$ and $G_{0}$ ($G'_0$) and the spectrum is known
as well. The interesting problem which arises here is as follows. What
is the group (noncompact in general) for which $L^{2}(M)$ $(L^{2}_{0}(M))$ is the
invariant space of a unitary, irreducible representation. In other
words what is an analogue of SO(3,1) for $\Delta$ on $S^{2}$ in a general case.
This group (if exists) we call dynamical group of $\Delta$, $\widehat{G}$ or a spectrum
generating group. We suppose that such a group is minimal for the above
requirement. In this way $G$ must be maximally compact subgroup of this
group, $G \subset \widehat{G}$. Thus $L^{2}(M)$ is a space of a unitary representation of
$\widehat{G}$ (which is up to now unknown) $T: \widehat{G} \to L(L^{2}(M),L^{2}(M))$ such that
$$
T_{|G} = \sum^{}_{\lambda _{k}} \oplus H_{\bar{\lambda }_{k}},\eqno(16.55)
$$
where the sum is over all the irreducible representation of $G$ with
multiplicity equals one. Such an representation of $\widehat{G}$, $T$ is called
maximally degenerated (or most degenerated). Such a situation is
possible only if $G$ is a maximal compact subgroup of $\widehat{G}$. %\vadjust{\eject}

Thus the dynamical group $\widehat{G}$ for $\Delta$ is defined as follows:

\vskip5pt
\item{1.} $G \subset \widehat{G}$ and is a maximal compact subgroup
$\widehat{G}$ ($\widehat{G}$ is noncompact, of course)

\item{2.} $\widehat{G}/G$ is a symmetric irreducible space of noncompact type, such that
for a given $G$ its rank is minimal.

\item{3.} The maximally degenerated, unitary, irreducible, infinite
dimensional representation of $\widehat{G}$, $T$ restricted to $G$ is equivalent to
the simple sum of all irreducible representation of the group $G$
(which are finite dimensional).

\vskip5pt
\noindent In our case $G$ is simple and compact. Thus we have the following
possibilities:
$$\vbox{%\tabskip10pt plus5pt minus5pt
\halign{\strut\tabskip10pt#\hfil&\hfil#\hfil&#\hfil\tabskip0pt\cr
\omit\hfill {\bf Space}\hfill &&\omit\hfill {\bf Rank}\hfill\cr
$\SL(n,R)/\SO(n)$&,&$n-1$\cr
($\gote_{\text{\bf 6}(\text{\bf 6})},sp(4))$&,&6\cr
($\gote_{\text{\bf 6}(-\text{\bf 26})},\gf_{\text{\bf 4}})$&,&2\cr
($\gote_{\text{\bf 7}(\text{\bf 7})},su(8))$&,&7\cr
($\gote_{\text{\bf 8}(\text{\bf 8})},so(16))$&,&8\cr
($f_{\text{\bf 4}(-\text{\bf 20})},so(9))$&,&1\cr}}
$$
Except the listed above there are also some additional.
$$\vbox{\offinterlineskip\tabskip10pt plus5pt minus5pt
\halign {#\hfil&\hfil#\hfil&#\hfil\tabskip0pt\cr
SO$_{0}(p,1)$/SO$(p)$&,&1\cr}}
$$
SO$_{0}(3,1)$ is a Lorentz group and this is our example with $S^{2}$.
$$\vbox{\offinterlineskip\tabskip10pt plus5pt minus5pt
\halign {#\hfil&\hfil#\hfil&#\hfil\tabskip0pt\cr
SU$(p,1)$/U$(p)$&,&1\cr}}
$$
These are the only possibilities for which we have pairs $(\widehat{G},G)$
suspected to be a dynamical for $\Delta$ on $G/G_{0}$. Thus we need maximally
degenerated representations of SO$_{0}(p,1)$, SU(p,1), SL$(n,R)$,
$\gote_{\text{\bf 6}(\text{\bf 6})}$, $\gote_{\text{\bf 6}(-\text{\bf 26})}$
$\gote_{\text{\bf 7}(\text{\bf 7})}$, $\gote_{\text{\bf 8}(\text{\bf 8})}$, $\gote_{\text{\bf
4}(-\text{\bf 20})}$ and their decompositions, after a restrictions to
the maximally compact subgroup, to the irreducible representations of
those subgroups.

Let us consider SO$_{\text{\bf 0}}(p,1)$. The most degenerate discrete series
consists of
$$\widehat{T}(L),\ L = -\bigg\{{1\over 2}(p + 1) -4\bigg\}, -
\bigg\{{1\over 2}(p + 1) - 4\bigg\} + 1,
- \bigg\{{1\over 2}(p + 1) - 4\bigg\}+ 2, \ldots 
$$
in $L^{2}(H^{p,1},\mu )$.

$H^{p,1}$ --- means a hyperboloid in $R^{p+1}$,
${\displaystyle\sum^{p}_{i=1}}(x^{i})^{2} - (x^{p+1})^{2} = 1$ and $\mu $ is a
measure on $H^{p,1}$, quasiinvariant with respect to the action of the group
SO$_{\text{\bf 0}}(p)$ on $H^{p,1}$. In this case after a restriction of $\widehat{T}(L)$ to the
subgroup SO(p) every representation (finite dimensional, unitary,
irreducible) enters with a multiplicity equal one (see $2^{\text{\rm nd}}$ position of
Ref.~[62]).
$$
\widehat{T}_{\fg}(L)_{|G} = \sum_{\ell [p/2]}\oplus \widehat{T}^{\ell
[p/2]},\eqno(16.56)
$$
where $\widehat{T}^{\ell [p/2]}$ are symmetric finite-dimensional representations of SO(p)
determined by the highest weight
$$\displaylines{
m= [\ell _{[p/2]},0,0\ldots ],\cr
\ell _{[p/2]} = L + 2n, n = 0,1,2,\ldots \cr}
$$
Thus in the case of $G = \SO(p)$, $\widehat{G} = \SO(p,1)$.

In the case of $SU(p,1)$ the situation is even easier. Let us come back
to the operator $\widetilde{L}$. It is a $G$-invariant operator on $M$ of the second
order.

Thus it is a linear combination of $\ell = \rank M$, independent operators of
on $M$, left-invariant. If the rank of $M$ is 1 it is proportional to the
Beltrami--Laplace operator on $M$. Thus we have
$$
\widetilde{L} = \ov{M} \sum^{\ell -1}_{i=0} \fg_{i}(\zeta
)\Delta_{i},\eqno(16.57)
$$
such that $\Delta_{0} = \Delta$ (Beltrami--Laplace operator on $M)$ and
$$\displaylines{\hfill
\fg_{0}(0) = 1\hfill\llap{(16.58)}\cr
\fg_{i}(0) = 0,\quad\ i = 1,2,\ldots \ell -1,\cr}
$$
$\Delta_{i}$, $i =1,2,\ldots,\ell -1$ are remaining $G$-invariant differential 
operators on $M$.
Moreover if they commute we can find the spectrum of $\widetilde{L}$ using
eigenfunctions of $\Delta$ finding spectrum of masses for the tower of scalar
fields ${\varPsi} _{k}$. The matrix $M_{ke}$ can be diagonalized in $L^{2}(M,dm)$ which is
equivalent to $L^{2}(\sqrt{\widetilde{g}}, M, dm)$ and to Friedrichs' Hilbert space for $\Delta$.
Thus the operator has the same spectrum in $L_0{}^{2}(M)$, moreover the
eigenfunctions differ in their form from eigenfunctions of
Beltrami--Laplace operator due to different scalar product in $L_{0}{}^2(M)$.
$$
m_{k} = {1\over 2} \sqrt{-2{\displaystyle\sum^{\ell-1}_{i=0}}
\mu_{i}\fg_{i}(\zeta)}.\eqno(16.59)
$$
The spectrum generating group $\widehat{G}$ is the same for $\widetilde{L}$ as for the
Beltrami--Laplace operator on $M$.
$$
\widehat{G} \supset G \supset G_{0}.
$$
Let us remind to the reader that the maximally degenerated
representation means that all the Casimir operators are polynomials of
the Casimir operator of the lowest order. In our case of the Casimir of
the $2^{\text{\rm nd}}$ order. Our group $\widehat{G}$ is a spectrum generated group for the tower
of scalar field ${\varPsi} _{k}(x)$, $({\varPsi} (x,y)$ on $M\times E)$.

{\def\eqn#1 {\eqno(\text{\rm16.#1})}
\def\rf#1 {\text{(16.#1)}}
\let\al\aligned
\let\eal\endaligned
\let\ealn\eqalignno
\def\cL{\Cal L}
\let\o\overline
\let\ul\underline
\let\a\alpha
\let\b\beta
\let\d\delta
\let\e\varepsilon
\let\D\Delta
\let\g\gamma
\let\G\varGamma
\let\la\lambda
\let\La\varLambda
\let\n\nabla
\let\P\varPsi
\let\F\varPhi
\let\z\zeta
\let\pa\partial
\let\t\widetilde
\newbox\smbox
\def\smsh#1{\setbox\smbox=\hbox{#1}\ht\smbox=0pt \dp\smbox=0pt \box\smbox}
\mathchardef\simm="0218
\def\ut#1{\underset \raise2pt\hbox{$\scriptstyle\simm$}\to#1 }
\let\u\tilde
\let\h\widehat
\def\uP{\t{\ul P}}
\def\RG{\t R(\t\G)}
\def\arctg{\operatorname{arctg}}
\def\ctg{\operatorname{ctg}}
\def\tg{\operatorname{tg}}
\def\Li{\operatorname{Li}}
\def\Re{\operatorname{Re}}
\def\br#1#2{\overset\text{\rm #1}\to{#2}}
\def\({\left(}\def\){\right)}
\def\[{\left[}\def\]{\right]}
\def\dr#1{_{\text{\rm#1}}}
\def\gi#1{^{\text{\rm#1}}}
\def\ct{constant}
\def\co{cosmological}
\def\dt{dependent}
\def\ef{effective}
\def\ex{exponential}
\def\lg{lagrangian}
\def\gr{gravitational}
\def\q{quintessence}
\def\wrt{with respect to }
\def\dint{-\kern-11pt\intop}
\def\tge#1{\t g^{(#1)}}
\def\id#1{_{\lower1pt\hbox{$\scriptstyle#1$}}}
\def\ga#1{\br{gauge}#1}

Let us come back to the Eqs (5.17--19) and let us release the condition that
$\rho$ is in\dt\ of~$y$. Thus $\rho=\rho(x,y)$, $y\in G/G_0$. In
Eq.~(5.34) we get in a place of
$$
\frac{\la^2}4 \rho^{n-2} \(\o M \t g^{(\g\mu)}\rho_{,\g}\rho_{,\mu}
+n^2g^{[\mu\nu]}g_{\d\mu}\t g^{(\d\g)}\)\rho_{,\nu}\cdot \rho_{,\g}
\eqn60
$$
the formula
$$
\frac{\la^2}4 \rho^{n-2} \(\o M \t \g^{(CM)}\rho_{,C}\rho_{,M}
+n^2\g^{[MN]}\g_{DM}\t \g^{(DC)}\rho_{,N}\cdot \rho_{,C}\).
\eqn61
$$
This $\rho$ has nothing to do with a density of energy considered below.

Now let us repeat the procedure from Section~7, i.e. the redefinition of
$g_{\mu\nu}$ and~$\rho$. We get the formula (7.5), but in a place of
$\cL\dr{scal}(\varPsi)$ we get the formula (16.3). Simultaneously
$\varPsi=\varPsi(x,y)$, $x\in E$, $y\in G/G_0$ and the metric on a space-time $E$
depends on $y\in G/G_0$, i.e.
$$
g_{\mu\nu}=g_{\mu\nu}(x,y), \quad x\in E, \quad y\in G/G_0. \eqn62
$$
Thus $g_{\mu\nu}$ is parametrized by a point of $G/G_0$. Simultaneously we
can interpret a dependence on higher dimensions as an existence of a tower of
scalar fields $\varPsi_K$ (see Eqs (16.4--8), Eqs (16.11--14) and a discussion
below).

The interesting point will be to find physical consequences of this
dependence for~$g_{\mu\nu}$. This can be achieved by considering \co\
solutions of the theory. Thus let us come to Eq.\ (7.5) supposing the \lg\ of
matter fields is written as $L\dr{matter}$, $g_{\mu\nu}=g_{\nu\mu}$ and
depends on $y\in G/G_0$. We get
$$
\al
L\sqrt{-g}\sqrt{|\t g|} &= \sqrt{-g}\sqrt{|\t g|}
\biggl(\o R(\t{\o \G})+\frac{\la^2}4 e^{-(n+2)\varPsi} L\dr{matter}\\
&+\frac{\la^2}4 \cL\dr{scal}(\varPsi)+ \frac1{\la^2}e^{(n+2)\varPsi}\t R(\t \G)
+\frac1{r^2}e^{n\varPsi}\ul{\t P}\biggr).
\eal \eqn63
$$

According to the standard interpretation of the constant $\la$ we have
$$
\frac{\la^2}4 \sim \frac1{m^2\dr{pl}} \eqn64
$$
and
$$
\frac1{r^2}\sim m^2_{\u A} \eqn65
$$
where $m_{\u A}$ is a scale of a mass of broken gauge bosons in our theory
and $m\dr{pl}$ is a Planck mass, $c=1$, $\hbar=1$.

Thus one gets form variation principle with respect to $g_{\mu\nu}$
and~$\varPsi$ 
$$
\al
\t{\o R}_{\mu\nu} - \frac12 \t{\o R}g_{\mu\nu}&=
\frac{e^{-(n+2)\varPsi}}{m^2\dr{pl}}\,
\br{matter}T_{\kern-9pt \mu \nu }+\frac1{m^2\dr{pl}}
\br{scal}T_{\kern-5pt\mu \nu }\\
&+\(\frac{m^2\dr{pl}}8 e^{(n+2)\varPsi}\t R(\t \G)+m^2_Ae^{n\varPsi}\ul{\t P}\)g_{\mu\nu}
\eal
\eqn66
$$
$$
\al
\o Mg^{\mu\nu}\t{\o \nabla}_\mu \t{\o\nabla}_\nu\varPsi
&+\frac1{r^2}\t L\varPsi + \frac{m\dr{pl}^4}4 (n+2)e^{(n+2)\varPsi}\t R(\t\G)\\
&+(m_{\u A}m\dr{pl})^2n e^{n\varPsi}\ul{\t P}-(n+2)Te^{-(n+2)\varPsi}=0
\eal \eqn67
$$
where $\t{\o R}_{\mu\nu}$ and $\t{\o R}$ are a Ricci tensor and a scalar
curvature of a Riemannian geometry induced by a metric $g_{\mu\nu}$ (which
can depend on $y\in G/G_0$).

$\t L$ is an operator on $G/G_0$ defined by Eq.\ (16.38), $\t{\o\nabla}_\mu$
a covariant derivative with respect to a connection induced by $g_{\mu\nu}$
(a~Riemannian one).
$$
\br{matter}T^{\mu\nu}=(p+\varrho)u^\mu u^\nu- pg^{\mu\nu}. \eqn68
$$
Let us consider a simple model with a metric
$$
ds^2_4=e^{2v(y)}\,dt^2- d{\vec r}^{\,2} \eqn69
$$
and let us suppose that $G/G_0=S^2=SO(3)/SO(2)$. Thus
$$
ds^2_2=r^2\(d\chi^2+\sin^2\chi\,d\chi^2\), \eqn70
$$
$$
(\chi,\la)=y,\quad x\in \Bigl\langle0,\frac\pi2\Bigr),\quad \la\in\langle0,2\pi).
$$
In this case $v=v(\chi,\la)$ and
$$
g_{[\u a,\u b]}=g_{[56]}=-\zeta\sin\chi, \eqn71
$$
$\t L$ is defined by Eq.\ (16.41),
$$
\varPsi=\varPsi(\vec r,t,\chi,\la), \quad \varrho=\varrho(\vec r,t,\chi,\la),
\quad p=p(\vec r,t,\chi,\la).\eqn72
$$

Finally let us come to the toy model for which we suppose $\chi=\frac\pi2$
and we get \ef ly a 5-dimensional world $E\times S^1$, i.e.
$$
ds^2_5 = e^{2v(\la)}\,dt^2 - d{\vec r}^{\,2} - r^2\,d\la^2. \eqn73
$$
One easily gets
$$
\ealn{
R_{44}&=\(v''+(v')^2\)e^v &(16.74)\cr
R_{55}&=r^2\(v''+(v')^2\)e^{-v} &(16.75)
}
$$
where $'$ means a derivative with respect to $\la$. For simplicity we suppose
$\varrho=p=0$ and $\varPsi=\varPsi(\la)$ (does not depend on $\vec r$ and~$t$). In
this way
$$
\t L=\(\o M+\frac{n^2\zeta^2}{\zeta^2+1}\)\frac{\pa^2}{\pa\la^2}\,. \eqn76
$$
From Eq.\ (16.63) one obtains
$$
\(\frac{d^2}{d\la^2}e^v\)e^{-v} = \(\frac{m^2\dr{pl}}8 e^{(n+2)\varPsi}
\t R(\t \G)+ m^2_{\u A} e^{n\varPsi}\ul{\t P}\). \eqn77
$$
$$
\al
m^2_{\u A}\(\o M+\frac{n^2\zeta^2}{\zeta^2+1}\)\frac{d^2}{d\la^2}\varPsi
&+\frac{m\dr{pl}^4}4 (n+2)e^{(n+2)\varPsi}\t R(\t \G)\\
&+(m_{\t A}m\dr{pl})^2
ne^{n\varPsi}\ul{\t P}=0. 
\eal \eqn78
$$
The last equation can be transformed into
$$
\frac{d^2\varPsi}{d\la^2}=Ae^{(n+2)\varPsi}+Be^{n\varPsi} \eqn79
$$
where
$$
\ealn{
A&=-\frac{m^4\dr{pl}(n+2)}{4m^2_A(\o M+\frac{n^2\zeta^2}{\zeta^2+1})}
\t R(\t\G) &(16.80)\cr
B&=-\frac{m^2\dr{pl}n}{(\o M+\frac{n^2\zeta^2}{\zeta^2+1})}\ul{\t P}.
&(16.81)
}
$$
Supposing $\varPsi=\varPsi_0=\text{const.}$ one gets
$$
\ealn{
&Ae^{2\varPsi_0}+B=0 &(16.82)\cr
&e^{\varPsi_0}=\sqrt{-\frac BA}=\frac{4m_{\u A}}{m\dr{pl}}
\sqrt{\frac{n|\ul{\t P}|}{(n+2)\t R(\t\G)}}\,. &(16.83)
}
$$
Thus from Eq.\ (16.77) one gets ($z=e^v$)
$$
\frac{d^2z}{d\la^2}=\t Az \eqn84
$$
where
$$
\al
\t A&=\frac{m^2\dr{pl}}8 e^{(n+2)\varPsi_0}\t R(\t\G)+
m^2_A e^{n\varPsi_0}\ul{\t P}\\ &=
m^2_{\t A}\(\frac{4m_{\u A}}{\a_sm\dr{pl}}\)^n
\(\(\frac n{n+2}\)\frac{|\ul{\t P}|}{\t R(\t\G)}\)^\frac n2\(\frac{n-2}{n+2}\)
|\ul{\t P}|>0. \eal
\eqn85
$$
And eventually one gets
$$
z=z_0e^{\sqrt{\u A}\la}+z_0'e^{-\sqrt{\u A}\la}. \eqn86
$$
Taking $z_0'=0$ we get
$$
e^{2v}=z_0^2e^{2\sqrt{\u A}\la}. \eqn87
$$
Simply rescaling a time in the metric (16.73) we finally get
$$
ds^2_5=e^{2\sqrt{\u A}\la}\,dt^2 - d{\vec r}^{\,2} - r^2\,d\la^2. \eqn88
$$

In this way we get a funny toy model. We are confined on 3-dimensional
brane in a 4-dimensional euclidean space for a $\la=0$. Moreover $\la$ is
changing from~0 to~$2\pi$ resulting in some interesting possibilities of
communication and travel in extra dimensions. This is due to a fact that an 
\ef\ speed of light depends on~$\la$ in an \ex\ way:
$$
c\dr{eff}=e^{\sqrt{\u A}\la}. \eqn89
$$
If we move in $\la$ direction (even a little, remember $r=\frac1{m_{\u A}}$)
from~0 to $\la_0<2\pi$ and after this move in space direction from $\vec r_0$
to~$\vec r_1$ and again from~$\la_0$ to~0 (we \ef ly travel from $\vec r_0$
to~$\vec r_1$), then we can be in a point~$\vec r_1$ even very distant
from~$\vec r_0$ ($L=|\vec r_1-\vec r_0|$) in a much shorter time than~$\frac
Lc$ (where $c$ is the velocity of light taken to be equal~1). In some sense
we travel in hyperspace from {\it Star Wars} or {\it Wing Commander}.

Thus we consider a metric (see Ref. [116])
$$
ds^2_5=e^{2\sqrt{\u A}\la}\,dt^2 - d{\vec r}^{\,2} - r^2\,d\la^2. \eqn90
$$
We get the following geodetic equations for a signal travelling in~$\vec r$,
$\la$~direction with initial velocity at $\la=0$, $\frac{d\la}{dt}(0)=u$,
$\frac{d\vec r}{dt}(0)=\vec v$,
$$
\ealn{
\frac{d\vec r}{dt}&=\vec v\,e^{\sqrt{\u A}\la} &(16.91)\cr
\(\frac{d\la}{dt}\)^2&=e^{\sqrt{\u A}\la}-(1-u^2)e^{2\sqrt{\u A}\la}.&(16.92)
}
$$
One integrates
$$
\vec r(\la)=\vec r_0-\vec v_0\cdot \frac{2(1-u^2)^{3/2}}{\sqrt{\u A}}
\arctg\(\sqrt{1-\frac1{(1-u^2)^{1/2}}e^{-\sqrt{\u A}\la}}\), \eqn93
$$
$$
\la=\frac1{\sqrt{\u A}}\ln\(\frac{4(1-u^2)^{5/2}}{\u A(t-t_0)^2+4(1-u)^3}\)
\eqn94
$$
and
$$
\vec r(t)=\vec r_0 - \vec v_0\frac{(1-u^2)^{3/2}}{\sqrt{\u A}}\arctg
\(\sqrt{1+\frac{4(1-u^2)^2}{\u A(t-t_0)^2+4(1-u^2)^3}}\), \eqn95
$$
$$
0<\la<\min\[2\pi,\frac1{2\sqrt{\u A}}\ln\(\frac1{1-u^2}\)\]. \eqn96
$$

Let us consider a signal travelling along an axis $x$ from 0 to~$x_0$. The
time of this travel is simply~$x_0$. But now we can consider a travel from
zero to~$\la_0$ (in $\la$ direction) and after this from $(0,\la_0)$
to~$(x_0,\la_0)$ and after to~$(x_0,0)$. The time of this travel is as
follows: 
$$
t=2t_1+t_2 \eqn97
$$
where
$$
t_1=r \intop_0^{\la_0}\frac{d\la}{e^{\sqrt{\u A}\la}}
=\frac r{\sqrt{\u A}}\(1-e^{-\sqrt{\u A}\la_0}\) \eqn98
$$
and
$$
t_2=x_0e^{-\sqrt{\u A}\la_0}. \eqn99
$$
For $r$ is of order of a scale of G.U.T. the first term is negligible and
$$
t=x_0e^{-\sqrt{\u A}\la_0}. \eqn100
$$
Taking maximal value of $\la_0=2\pi$ we get
$$
t=x_0e^{-2\pi\sqrt{\u A}}. \eqn101
$$
Thus we achieve really shorter time of a travelling signal through the fifth 
dimension.

Let us consider the more general situation when we are travelling along a
time-like curve via fifth dimension from $\vec r=\vec r_0$ to $\vec r=\vec
r_1$. Let the parametric equations of the curve have the following shape:
$$
\displaylines{
\hphantom{(16.103)}\hfill\al
\vec r&=\vec r(\xi)\\
t&=t(\xi)\\
\la&=\la(\xi).
\eal
\hfill (16.102)\cr
\noalign{\vskip2pt}
0\le \xi\le 1,\cr
\noalign{\vskip2pt}
\la(0)=0, \ \la(1)=\la_0,\cr
\noalign{\vskip2pt}
\hphantom{(16.103)}\hfill\al
\vec r(0)&=\vec r_0\\
\vec r(1)&=\vec r_1.
\eal
\hfill (16.103)
}
$$
Let a tangent vector to the curve be $(\dot t(\xi),\dot {\vec r}(\xi),
\dot \la(\xi))$, where dot ``$\dot{\phantom t}$'' means a derivative \wrt
$\xi$. For a total time measured by a local clock of an observer one gets 
$$
\D t=\intop_0^1 d\xi\(e^{2\sqrt{\u A}\la(\xi)}
\(\frac{dt}{d\xi}\)^2 - \(\frac{d\vec r}{d\xi}\)^2 - r^2\(\frac{d\la}{d\xi}\)^2
\)^\frac12. \eqn104
$$
Let us consider a three segment curve such that
\item{1)} a straight line from $\vec r=0$, $\la=0$ to $\vec r=0$, $\la=\la_0$,
\item{2)} a straight line from $\vec r=0$, $\la=\la_0$ to $\vec r=\vec r_0$, $\la=\la_0$,
\item{3)} a straight line from $\vec r=\vec r_0$, $\la=\la_0$ to $\vec r=\vec
r_0$, $\la=0$.

The additional assumptions are such that the traveller travels with a
constant velocity $v_\la$ on the first segment and on the third segment and
with a velocity $\vec v$ on the second one.

Thus we have a parametrization:

1) On the first segment
$$
\al
\vec r&=0\\
t&=\frac{\la_0}{v_\la}\,\xi, \quad 0\le \xi\le 1\\
\la&=\xi\la_0.
\eal
\eqn105
$$

2) On the second segment
$$
\al
\vec r(\xi)&=\vec r_0\xi\\
t&=\frac{|\vec r_0|}{|\vec v|}\,\xi, \quad 0\le \xi\le 1\\
\la&=\la_0.
\eal
\eqn106
$$

3) On the third segment
$$
\al
\vec r&=\vec r_0\\
t&=\frac{\la_0}{|\vec v|}\,(1-\xi), \quad 0\le \xi\le 1\\
\la&=(1-\xi)\la_0.
\eal
\eqn107
$$

From Eq.\ (16.104) one easily gets
$$
\ealn{
\D t&=\D t_1 + \D t_2 &(16.108)\cr
\D t_1&=\frac{2r}{\sqrt{\u A}}\Biggl(
\frac1{\sqrt{v_\la \la_0}}\(\sqrt{e^{2\sqrt{\u A}\la_0}-v_\la \la_0}
-\sqrt{1-v_\la \la_0}\)\cr
&+\arctg\Biggl(\sqrt{\frac{e^{2\sqrt{\u A}\la_0}-v_\la \la_0}{v_\la \la_0}}\Biggr)
-\arctg\(\sqrt{\frac{1-v_\la \la_0}{v_\la \la_0}}\)\Biggr)&(16.109)\cr
\D t_2&=\frac{|\vec r_0|}{|\vec v|} \sqrt{e^{2\sqrt{\u A}\la_0}-|\vec v|^2}
&(16.110)
}
$$

For $r$ is of order of an inverse scale of G.U.T., the first term in Eq.\
(16.108) is negligible and
$$
\D t \simeq \frac{|\vec r_0|}{|\vec v|} \sqrt{e^{2\sqrt{\u A}\la_0} -|\vec
v|^2}. \eqn111a
$$
The traveller passing from $\vec r=0$ to $\vec r=\vec r_0$ (having $\la=0$
during his travel) reached $\vec r=\vec r_0$ at time
$$
\D t'=\frac{|\vec r_0|}{|\vec v|} \(1 -|\vec v|^2\)^{1/2}. \eqn111b
$$
The first traveller has the following condition for $\vec v$:
$$
|\vec v|\le e^{\sqrt{\u A}\la_0}, \quad 0\le \la_0\le 2\pi. \eqn112a
$$
The second
$$
|\vec v|\le 1. \eqn112b
$$
Thus we see that we can travel quicker taking a bigger $\vec v$.

For an unmoving observer the time needed for a travel in the first case is
$$
\al
&\D t=\frac{2r\la_0}{v_\la}+\frac{|\vec r_0|}{|\vec v|}\simeq \frac{|\vec
r_0|} {|\vec v|},\\
&|\vec v|\le e^{\sqrt{\u a}\la_0}, \quad 0\le \la_0\le 2\pi.
\eal
\eqn111a$*$
$$
In the second case
$$
\al
\D t'&=\frac{|\vec r_0|}{|\vec v|},\\
|\vec v|&\le 1
\eal
\eqn111b$*$
$$
Thus in some sense we get a dilatation of time:
$$
\D t'=\(e^{2\sqrt{\u A}\la_0}-{\vec v}^{\,2}\)^{1/2}\D t \eqn111a$**$
$$
in the first case,
$$
\D t'=(1-{\vec v}^{\,2})^{1/2}\D t \eqn111b$**$
$$
in the second case.

Let us come back to the Eqs (16.94--96). We see that a particle coming along
a geodesic is confined in a shell surrounding a brane in the fifth dimension
$$
0\le \la\le \frac r{\sqrt{\u A}}\ln\(\frac1{(1-u^2)^{1/2}}\). \eqn113
$$
The particle oscillates in the shell. Moreover $0\le \la \le 2\pi$, thus if
$\la$ excesses $2\pi$, we take for a new value $\la-2\pi$ (i.e.\
modulo~$2\pi$). Usually for~$u$ not close to~1 this is a really thin shell
of order $r$ (a~length of a unification scale).

Let us come back to our toy model for a cone with $\t A=0$. In this case an
equation of cone in 5-dimensional Minkowski space is
$$
x^2+y^2+z^2+r^2\la^2=t^2. \eqn114
$$
An interesting question is what is an analog for (16.114) if $\u A\ne 0$. One
easily gets that
$$
x^2+y^2+z^2+r^2\la^2=t^2r^2\la^2{\t A}^2e^{2\sqrt{\u A}\la}. \eqn115
$$

There is no simple way to get Eq.\ (16.114) from Eq.\ (16.115). Moreover if
we change a coordinate~$t$ into $rt\la\t A$ one gets
$$
x^2+y^2+z^2+r^2\la^2=t^2e^{2\sqrt{\u A}\la}. \eqn116
$$
and for $\t A=0$ we get Eq. (16.114).

Let us notice that Eq.\ (16.116) is necessary to be considered for $\la$
modulo~$2\pi$. It means even $\la$ could be any real number in the equation
we should put in a place of~$\la$, $[\la]$, where $[\la]$ is defined as
follows 
$$
\la=[\la]+2\pi n, \quad [\la]\in \langle0,2\pi),
$$
$n$ is an integer and $\la\in(-\infty,+\infty)$. In this way in a place of
Eq.~(16.116) we get
$$
x^2+y^2+z^2+r^2[\la]^2=t^2e^{2\sqrt{\u A}[\la]}. \eqn116$*$
$$
The last equation defines an interesting 4-dimensional hypersurface in
5-dimensional space (${\Bbb R}^5$).

For $\la=0$ we get a light cone on a brane (4-dimensional Minkowski space).
For branes with $0<\la=\la_0<2\pi$ we have a ``cone''
$$
x^2+y^2+z^2=t^2e^{2\sqrt{\u A}\la_0}-\la_0^2 \eqn117
$$
which is a 3-dimensional hyperboloid.

In our model we get in a quite natural way a warp factor known from
Ref.~[117]. Moreover we have a different reason to get it. The reason of this
warp factor is a many dimensional dependence of a scalar field
$\varPsi=\varPsi(x,\la)$. In particular higher-dimensional excitations of~$\varPsi$
forming a tower of massive scalar fields. In this model $\varPsi$ can be
developed in a Fourier series
$$
\varPsi(x,\la)=\sum_{m=0}^\infty \psi^1_m(x)\cos(m\la)+
\sum_{m=1}^\infty \psi^2_m(x)\sin(m\la) \eqn118
$$
where
$$
\ealn{
\psi^1_m(x)&=\frac1{\pi}\intop_0^{2\pi}\varPsi(x,\la)\cos(m\la)\,d\la
&(16.119)\cr 
\psi^2_m(x)&=\frac1{\pi}\intop_0^{2\pi}\varPsi(x,\la)\sin(m\la)\,d\la
&(16.120)\cr 
&m=1,2,\ldots\cr
\frac{\psi^1_0(x)}2&=\frac1{\pi}\intop_0^{2\pi}\varPsi(x,\la)\,d\la. &(16.121)
}
$$

Thus an interesting point will be to find a spectrum of mass for this tower. 
This is simply
$$
m_m^2=m_{\u A}^2 \(\o M+\frac{n^2\zeta^2}{\zeta^2+1}\)m^2, \qquad
m=0,1,2,\ldots \eqn122
$$
Now we come to the calculation of a content of a warp factor (a content in
terms of a tower of scalar fields). Thus we calculate
$$
\ealn{
\psi^1_m&=\frac1{\pi}\intop_0^{2\pi} e^{2\sqrt{\u A}\la}\sin(m\la)\,d\la,
\qquad m=1,2,3,\ldots &(16.123)\cr
\psi^2_m&=\frac1{\pi}\intop_0^{2\pi} e^{2\sqrt{\u A}\la}\cos(m\la)\,d\la,
\qquad m=1,2,3,\ldots &(16.124)\cr
\frac12 \psi^1_0&=\frac1{\pi}\intop_0^{2\pi} e^{2\sqrt{\u A}\la}\,d\la.
&(16.125)
}
$$
One gets
$$
\ealn{
\frac12\psi^1_0&=\frac1{2\pi\sqrt{\t A}}\(e^{4\pi\sqrt{\u A}}-1\) &(16.126)\cr
\psi^1_m&=-\frac m{\pi}\frac{(e^{4\pi\sqrt{\u A}}-1)}{(4\t A+m^2)}
&(16.127)\cr
\psi^2_m&=\frac{2\sqrt{\t A}(e^{4\pi\sqrt{\u A}}-1)}{\pi(4\t A+m^2)}\,.
&(16.128)
}
$$
The simple question is what is the energy to excite the warp factor in terms of
a tower of scalar fields. The answer is as follows:
$$
E\dr{wf}=\intop_0^{2\pi} \br{scal}T_{\kern-5pt 44}\bigl(e^{2\sqrt{\u A}\la}
\bigr)\,d\la \eqn129
$$
where $\br{scal}T_{\kern-5pt 44}$ is a time component of an energy-momentum
tensor for a scalar field calculated for $\varPsi=e^{2\sqrt{\u A}\la}$.

One gets
$$
\ealn{
&\br{scal}T_{\kern-5pt 44}=\frac12 \(\frac{d\varPsi}{d\la}\)^2 + 
\frac12 \la_{c_0}(\varPsi),
\quad \la_{c_0}(\varPsi)=\frac12|\g|e^{n\varPsi}-\frac\b2 e^{(n+2)\varPsi} &(16.130)\cr
&\intop_0^{2\pi}\frac12 \(\frac d{d\la}e^{2\sqrt{\u A}\la}\)^2\,d\la=
\frac{\sqrt{\t A}}2\(e^{8\pi\sqrt{\u A}}-1\) &(16.131)
}
$$
and
$$
\frac12 \la_{c_0}\bigl(e^{2\sqrt{\u A}\la}\bigr)=
-\frac12\, e^{n(e^{2\sqrt{\u A}\la})}
\(\b e^{2(e^{2\sqrt{\u A}\la})}-|\g|\), \eqn132
$$
where
$$
\al
\g&=\frac{m^2_{\u A}}{\a_s^2}\ul{\t P}=\frac1{r^2}\ul{\t P}, \\
\b&=\frac{\a_s^2}{l\dr{pl}^2}\t R(\t\G)={\a_s^2}m\dr{pl}^2\t R(\t\G) .
\eal
$$

One easily gets
$$
\al
E\dr{wf}&=\frac{\sqrt{\t A}}2 \bigl(e^{8\pi\sqrt{\u A}}-1\bigr)
-\frac1{4\sqrt{\t A}} \Biggl[|\g|\Biggl(\Li
\Biggl(\exp\biggl(\frac{e^{4\pi\sqrt{\u A}}}n\biggr)\Biggr)
-\Li\bigl(\root n\of e\bigr)\Biggr)\\
&-\b\Biggl(\Li\Biggl(\exp\biggl(\frac{e^{4\pi\sqrt{\u A}}}{(n+2)}\biggr)\Biggr)-\Li\bigl(
\root{n+2}\of e\bigr)\Biggr)\Biggr]
\eal \eqn133
$$
where
$$
\Li(x)=\dint_0^x\frac{dz}{\ln z},\quad x>1, \eqn134
$$
is an integral logarithm. The integral is considered in the sense of a
principal value.

Let us consider Eq.\ (16.133) using asymptotic formula for $\Li(x)$ for large~$x$:
$$
\Li(x) \sim x\biggl[\frac1{\ln x}+\sum_{m=1}^\infty
\frac{m!}{\ln^{m+1}x}\biggr].
$$
One gets
$$
\al
E\dr{wf}&=\frac{\sqrt{\t A}}2 \(e^{8\pi\sqrt{\u A}}-1\)-
\frac{e^{-4\pi\sqrt{\u A}}}{4\sqrt{\t A}}\Biggl(n|\g|\exp
\biggr(\frac{e^{4\pi\sqrt{\u A}}}n\biggl)\\
&-(n+2)\b\exp\biggl(\frac{e^{4\pi\sqrt{\u A}}}{n+2}\biggr)\Biggr).
\eal \eqn133a
$$

However, let us think quantum-mechanically in a following direction. Let us
create (maybe in an accelerator as some kind of resonances) scalar particles
of mass $m_n$ (Eq.~(16.122)) with an amount of $|\psi^1_m|^2+
|\psi^2_m|^2$. In this way we create the full warp factor (for they due to
interference will create it as a Fourier series). The question is what is an
energy needed to create those particles. The answer is simply
$$
\o E=\sum_{m=1}^\infty \(|\psi^1_m|^2+|\psi^2_m|^2\)m_m. \eqn135
$$
One easily gets
$$
\o E=\frac{m_{\u A}}{\pi^2}\(e^{4\pi\sqrt{\u A}}-1\)^2\(\o M+\frac{n^2\zeta^2}
{\zeta^2+1}\)^\frac12 \sum_{m=1}^\infty \frac m{4\t A+m^2}\,.\eqn136
$$
The series $\sum_{m=1}^\infty \frac m{4\u A+m^2}$ is divergent for
$$
\frac m{4\t A+m^2}\simeq \frac 1m\,. \eqn137
$$
Thus we need a regularization technique to sum it. We use $\zeta$-function
regularization technique similar as in Casimir-effect theory. Let us
introduce a $\zeta_{\u A}$-function on a complex plane
$$
\zeta_{\u A}=\sum_{m=1}^\infty \(\frac m{4\t A+m^2}\)^s \eqn138
$$
for $\Re(s)\ge1+\d$, $\d>0$.

This function can be extended analytically on a whole complex plane.
Obviously $\zeta_{\u A}$ has a pole for $s=1$ (as a Riemann $\zeta$
function). Thus we should regularize it at $s=1$. One gets
$$
\o E=\frac{m_{\u A}}{\pi^2}\(e^{4\pi\sqrt{\u A}}-1\)^2\(\o M+\frac{n^2\zeta^2}
{\zeta^2+1}\)^\frac12 \zeta^r_{\u A}(1)\eqn139
$$
where $\zeta^r_{\u A}$ means a regularized $\zeta_{\u A}$.

Let us introduce a $\zeta$-Riemann function
$$
\zeta\dr{R}(s)=\sum_{m=1}^\infty \frac1{m^s},
\qquad \Re(s)\ge1+\d,\ \d>0. \eqn140
$$

One can write
$$
\hbox{``}\zeta_{\u A}(1)\hbox{''}=
\hbox{``}\zeta_{\text{R}}(1)\hbox{''}+4\t A\sum_{m=1}^\infty \frac1{m(4\t A+m^2)}
\eqn141
$$
or
$$
\hbox{``}\zeta_{\u A}(1)\hbox{''}=
\hbox{``}\zeta_{\text{R}}(1)\hbox{''}+4\t A\zeta\dr{R}(3)
-16{\t A}^2\sum_{m=1}^\infty \frac1{m^3(4{\t A}^2+m^2)}\,. \eqn142
$$
The series in Eqs (16.141--142) are convergent. Moreover $\zeta\dr R(s)$ has
a pole at $s=1$ and
$$
\lim_{s\to1}\[\zeta\dr R(s)-\frac1{s-1}\]=\g_E \eqn143
$$
where $\g_E$ is an Euler constant.

Thus we can regularize $\zeta_{\u A}(s)$ in such a way that
$$
\zeta^r_{\u A}(s)=\zeta\dr{R}(s)-\frac1{s-1}+4\t A\sum_{m=1}^\infty
\frac1{m(4\t A+m^2)}\,. \eqn144
$$
Thus
$$
\zeta^r_{\u A}(1)=\g_E + 4\t A\sum_{m=1}^\infty \frac1{m(4\t A+m^2)}\,.
\eqn145
$$
The series
$$
\sum_{m=1}^\infty \frac1{m(4\t A+m^2)} \eqn146
$$
is convergent for every $z^2=-4\t A$ and defines an analytic function
$$
f(z)=\sum_{m=1}^\infty \frac1{m(m^2-z^2)}\,. \eqn147
$$
Using some properties of an expansion in simple fraction for
$$
\psi(x)=\frac d{dx}\log \G(x) \eqn148
$$
where $\G(x)$ is an Euler $\G$-function, one gets
$$
\ealn{
\psi(x)&=-\g_E - \sum_{m=0}^\infty \(\frac1{m+x}-\frac1{m+1}\)&(16.149)\cr
\zeta^r_{\u A}(1)&=-\frac12\(\psi(2\sqrt{\t A})+\psi(-2\sqrt{\t A})\)
&(16.150)
}
$$
or
$$
\zeta^r_{\u A}(1)=\frac1{4\sqrt{\t A}}+\frac\pi2 \ctg2\sqrt{\t A}\pi \eqn151
$$
(where we use $\G(z)\G(-z)=-\frac \pi{z\sin\pi z}$).

And finally
$$
\o E=\frac{m_{\u A}}{8\pi^2}\(\o M+\frac{n^2\zeta^2}{\zeta^2+1}\)
\(e^{4\pi\sqrt{\u A}}-1\)^2\(\frac1{2\sqrt{\t A}}+\pi\ctg(2\sqrt{\t A}\pi)\).
\eqn152
$$
It is easy to see that $\o E=0$ for $\t A=0$,
$$
\sqrt{\t A}=m_{\u A}\(\frac{4\sqrt n\, m_{\u A}}{\a_s\sqrt{n+2}\,m\dr{pl}}\)^\frac
n2 \(\frac{|\ul{\t P}|}{\t R(\t \G)}\)^\frac n4
\(\frac{n-2}{n+2}\,|\ul{\t P}|\)^\frac12. \eqn153
$$
For we use a dimensionless coordinate $\la$ (not $\t \la=r\cdot \la=\frac
\la{m_{\u A}}$), we can omit a factor $m_{\u A}$ in front of the right-hand
side in the formula (16.153).

It is interesting to notice that for $\t A=0$, $e^{2\sqrt{\u A}}=1$ and we
have to do with a Fourier expansion of a constant mode only ($m=0$).
This mode is a massless mode. However, a massless mode can obtain a mass from
some different mechanism (as in Section 14). Thus in some sense we have to do
with an energy
$$
\frac14 \Bigl|\psi^1_0(\t A=0)\Bigr|^2 m_\varPsi=m_\varPsi\,.\eqn154
$$

One gets the following formula
$$
m_\psi=m\dr{pl} \sqrt{\frac{|\ul{\t P}|(n+2)}{(\o M+\frac{n^2\zeta^2}{\zeta^2+1})}}
\(\frac{4\sqrt n\,\sqrt{|\ul{\t P}|}m_{\u A}}{\a_s\sqrt{n+2}\,\sqrt{\t R(\t\G)}\,
m\dr{pl}}\)^\frac n2.\eqn155
$$
Thus it seems that in order to create a warp factor $e^{2\sqrt{\u A}\la}$ in
a front of $dt^2$ we should deposite an energy of a tower of scalar particles
$$
\t E=E+m_\psi\,.\eqn156
$$
Even a factor equal to 1 is not free. This is of course supported by a
classical field formula (Eq.~(16.134)). For $\t A=0$ we get
$$
E\dr{wf}=-\pi\cdot e^n(\b e^2-|\g|) \eqn157
$$
or
$$
E\dr{wf}=-\frac{\pi e^n}{\a_s^2}\(e^2\a_s^4m\dr{pl}^2 \t R(\t \G)-m^2_{\u A}
|\ul{\t P}|\). \eqn158
$$

To be honest we should multiply $E\dr{wf}$ and $E$ by $V_3$, where $V_3$ is a
volume of a space and a result will be divergent. Moreover, if we want to
excite a warp factor only locally, $V_3$~can be finite and the result also
finite. 

Let us come back to the Eq.\ (16.63). The variational principle based on
(16.63) \lg\ is really in $(n_1+4)$-dimensional space $E\times G/G_0$, where
a size of a compact space $M=G/G_0$ is~$r$. In this way the gravity
lives on $(n_1+4)$-dimensional manifold, where $n_1$ dimensions are curled
into a compact space ([117]). The scalar field $\varPsi$ lives also on
$(n_1+4)$-dimensional manifold. The matter is 4-dimensional. Thus we can
repeat some conclusions from Randall-Sundrum ([118]) and Arkani-Hamed, Savas
Dimopoulos and Dvali ([119]) theory. Similarly as in their case we have
gravity for two regions ($V(R)$ is a Newtonian \gr\ potential).

1) $R\ll r$
$$
V(R)\cong \frac{m_1\cdot m_2}{m^{n_1+2}\dr{pl$(n_1+4)$}}\,
\frac1{R^{n_1+1}}\,, \eqn159
$$

2) $R\gg r$
$$
V(R)\cong \frac{m_1\cdot m_2}{m^{n_1+2}_{\text{pl}(n_1+4)}r^{n_1}}\,
\frac1{R}\,, \eqn160
$$
where $m\dr{pl$(n_1+4)$}$ is a $(n_1+4)$-dimensional Planck's mass. So
$$
m^2\dr{pl}=m^{n_1+2}_{\text{pl}(n_1+4)}\cdot r^{n_1}=
\frac{m^{n_1+2}_{\text{pl}(n_1+4)}}{m^{n_1}_{\u A}}=
\(\frac{m_{\text{pl}(n_1+4)}}{m_{\u A}}\)^{n_1}\cdot m^2_{\text{pl}
(n_1+4)}\,. \eqn161
$$
Thus
$$
m_{\text{pl}(n_1+4)}=(m\dr{pl})^\frac2{n_1+2}\cdot (m_{\u A})^\frac
{n_1}{n_1+2}. \eqn162
$$

Eqs (16.161--162) give us a constraint on parameters in our theory and
establish a real strength of \gr\ interactions given by
$m_{\text{pl}(n_1+4)}$. 

However, in our case we have to do with a scalar-tensor theory of gravity.
Thus our \gr\ \ct\ is \ef
$$
G\dr{eff}=G_Ne^{-(n+2)\varPsi}=G_N\rho^{n+2}. \eqn163
$$
In this case we have to do with \ef\ ``Planck's masses'' in $n_1+4$ and
4~dimensions 
$$
\ealn{
m^2_{\text{pl}(n_1+4)}&\longrightarrow m^2_{\text{pl}(n_1+4)}
\cdot e^{(n+2)\varPsi_{(n_1+4)}} &(16.164)\cr
m^2_{\text{pl}}&\longrightarrow m^2_{\text{pl}}
\cdot e^{(n+2)\varPsi} &(16.165)
}
$$
where $\varPsi_{(n_1+4)}$ is a scalar field $\varPsi$ for $R\ll r$ and $\varPsi$ is
for $R\gg r$.

In this way Eq.\ (16.161) reads
$$
m^2\dr{pl} e^{(n+2)\varPsi}=
\(\frac{m_{\text{pl}(n_1+4)}e^{\frac{(n+2)}2 \varPsi_{(n_1+4)}}}
{m_{\u A}}\)^{n_1}\cdot
m^2_{\text{pl}(n_1+4)}\cdot e^{(n+2)\varPsi_{(n_1+4)}}. \eqn166
$$
For our contemporary epoch we can take $e^{(n+2)\varPsi}\simeq1$.

If we take as in the case of Ref.\ [119]
$$
m_{\text{pl}(n_1+4)}\simeq m\dr{EW}, \eqn167
$$
where $m\dr{EW}$ is an electro-weak energy scale. We get
$$
\(\frac{m\dr{pl}}{m\dr{EW}}\)^2 = \(\frac{m\dr{EW}}{m_{\u A}}\)^{n_1}
\cdot e^{\frac12 (n+2)(n_1+2)\varPsi_{(n_1+4)}}. \eqn168
$$
For $R\ll r$ the field $\varPsi_{(n_1+4)}$ has the following behaviour
$$
\varPsi_{(n_1+4)}\cong \varPsi_0 +\frac \a{R^{n_1+1}}\,. \eqn169
$$

Thus we can approximate $\varPsi_{(n_1+4)}$ by $\varPsi_0$ where $\varPsi_0$ is a
critical point of a selfinteracting potential for~$\varPsi$. In this way one gets
$$
\(\frac{m\dr{pl}}{m\dr{EW}}\)\(\frac{m_{\u A}}{m\dr{EW}}\)^{n_1}
=\exp\(\frac{(n+2)(n_1+2)}2\,\varPsi_0\) \eqn170
$$
where $\varPsi_0$ is given by Eq.\ (14.204). Thus one gets:
$$
\al
&\(\frac{m\dr{pl}}{m\dr{EW}}\)\(\frac{m_{\u A}}{m\dr{EW}}\)^{n_1}
\(\frac{m\dr{pl}}{m_{\u A}}\)^\frac{(n+2)(n_1+2)}2\\
&\qquad{}=
\(\frac1{\a_s}\)^\frac{(n+2)(n_1+2)}2
\(\frac{n|\ul{\t P}|}{(n+2)|\t R(\t\G)|}\)^\frac{(n+2)(n_1+2)}4. 
\eal
\eqn171
$$
It is easy to see that we can achieve a solution for a hierarchy problem if
$\biggl|\dfrac{\ul{\t P}}{\t R(\t \G)}\biggr|$ is sufficiently big, e.g. taking $\t R(\t\G)$
sufficiently small and/or $\ul{\t P}$ sufficiently big playing with \ct s $\xi$
and~$\zeta$. In this way a real strength of gravity will be the same as
electro-weak interactions and the hierarchy problem has been reduced to the
problem of smallness of a \co\ \ct\ (i.e.\ $\t R(\t\G)\simeq0$).

Let us consider Eq.\ (16.152) in order to find such a value of~$\t A$ for
which $\o E$ is minimal. One finds that
$$
\o E=0 \eqn172
$$
for
$$
\t A=\frac{x^2}{4\pi^2} \eqn173
$$
where $x$ satisfies an equation
$$
x=-\tg x,\quad x>0. \eqn174
$$
Eq.\ (16.174) has an infinite number of roots. One can find some roots of
Eq.~(16.174) for $x>0$. We get
$$
\al
x_1&=4.913\dots\\
x_2&=7.978\dots\\
x_3&=11.086\dots\\
x_4&=14.207\dots\\
\eal \eqn175
$$
We are interested in large roots (if we want to have $c\dr{eff}$ considerably
big). 

In this case one finds
$$
x=\e+\tfrac\pi2(2l+1) \eqn176
$$
where $\e>0$ is small and $l=1,2,\ldots$. Moreover, to be in line in our
approximation, $l$~should be large. One gets
$$
\tfrac\pi2(2l+1)+\e=\ctg\e. \eqn177
$$
For $\e$ is small, one gets
$$
\ctg\e \simeq \frac1\e\,. \eqn178
$$

Finally one gets 
$$
\e^2+(2l+1)\tfrac\pi2-1=0 \eqn179
$$
and
$$
\ealn{
\e&\simeq \frac2{(2l+1)\pi} &(16.180)\cr
x&=\frac\pi2\,(2l+1)+\frac2{(2l+1)\pi} &\text{(16.180a)}\cr
\t A&=\frac1{4\pi^2}\(\frac\pi2 \,(2l+1)+\frac2{(2l+1)\pi}\)^2.&(16.181)
}
$$
It is easy to notice that our approximation (16.180a) is very good even for
small~$l$, e.g.\ for $l=4$ we get $x_4$ (Eq.~(16.175)).

In this way an \ef\ velocity of light can be arbitrarily large,
$$
c\dr{eff}\gi{max}=c\exp\(\frac12\(\frac\pi2(2l+1)+\frac2{(2l+1)\pi}\)\)\eqn182
$$
(i.e.\ for $\la=2\pi$). However, in order to get such large $c\dr{eff}$ we
should match Eq.~(16.153) with (16.181). One gets
$$
\frac\pi2\,(2l+1)+\frac2{(2l+1)\pi}
=2\pi\(\frac{4\sqrt n\, m_{\u A}}{\a_s\sqrt{n+2}\,m\dr{pl}}\)^\frac
n2 \(\frac{|\ul{\t P}|}{\t R(\t \G)}\)^\frac n4
\(\frac{n-2}{n+2}\,|\ul{\t P}|\)^\frac12. \eqn183
$$

In this way we get some interesting relations between parameters in our
theory. To have a large $c\dr{eff}$ we need to make $|\ul{\t P}|$ large and
$\t R(\t \G)$ small. In this case only the first term of the left-hand side
of (16.183) is important and we should play with the ratio $\(\dfrac
{m_{\u A}}{\a_sm\dr{pl}}\)$ in order to get the relation
$$
(2l+1)
=4\(\frac{4\sqrt n\, m_{\u A}}{\a_s\sqrt{n+2}\,m\dr{pl}}\)^\frac
n2 \(\frac{|\ul{\t P}|}{\t R(\t \G)}\)^\frac n4
\(\frac{n-2}{n+2}\,|\ul{\t P}|\)^\frac12 \eqn184
$$
where $l$ is a big natural number. In this way we can travel in hyperspace
(through extra dimension) almost for free. The energy to excite the warp factor
is zero (or almost zero). Thus it could be created even from a fluctuation in
our Universe.

Let us notice that Eq.\ (16.174) is in some sense universal for a warp factor.
It means that it is the same for any realization of the Nonsymmetric
Jordan-Thiry (Kaluza-Klein) Theory.

However, if we take Eq.\ (16.133) and if we demand
$$
E\dr{wf}=0, \eqn185
$$
we get some roots (they exist) depending on $n$, $\b$ and $|\g|$. Thus the
roots, some values of $\t A>0$ depend on details of the theory, they are not
universal as in the case of Eqs~(16.172--174). In this case the Eq.~(16.153) is
in some sense a selfconsistency condition for a theory and more restrictive.
The fluctuations of a tower of scalar fields in the case of $\sqrt{\t
A}=\frac14(2l+1)$ are as follows
$$
\ealn{
\frac12\psi^1_0 &=\frac{2\(e^{\pi(2l+1)}-1\)}{2l+1} &(16.186)\cr
\psi^1_m&=-\frac m\pi \cdot \frac{2\(e^{\pi(2l+1)}-1\)}{\((2l+1)^2+4m^2\)}
&(16.187)\cr
\psi^2_m&=\frac{2(2l+1)\(e^{\pi(2l+1)}-1\)}{\pi\((2l+1)^2+4m^2\)}\,. &(16.188)
}
$$
In the case of very large $l$ (which is really considered) one gets
$$
\ealn{
\frac12\psi^1_0&=\frac{e^{\pi(2l+1)}}{l} &\text{(16.186a)}\cr
\psi^1_m&=-\frac m{\pi}\cdot \frac{e^{\pi(2l+1)}}{4(l^2+m^2)}
&\text{(16.187a)}\cr 
\psi^2_m&=\frac{l\cdot e^{\pi(2l+1)}}{\pi(l^2+m^2)} &\text{(16.188a)}\cr
&m=1,2,\ldots
}
$$
According to our calculations the energy of the fluctuations (excitations)
(16.186a--188a) is zero.

Thus the only difficult problem to get a warp factor is to excite a tower of
scalar fields coherently in a shape of (16.186--188a) or to wait for such a
fluctuation. In the first case it is possible to use some techniques from
quantum optics applied to the scalar fields~$\psi_m$. In our case we get
$$
\ealn{
\psi_0&=\frac{e^{\pi(2l+1)}}{l} &(16.189)\cr
\psi_m&=\frac{e^{\pi(2l+1)}\sqrt{16l^2+m^2}}{4\pi(l^2+m^2)} \sin(m\la+\d_m)
e^{im_mt} &(16.190)
}
$$
where
$$
\tg \d_m=-\frac m{4l} \eqn191
$$
and $m_m$ is given by the formula (16.122).

For $\psi_0$ is \ct\ in time and does not contribute to the total energy of a
warp factor (only to the energy of a warp factor equal to~1), we consider
only 
$$
\psi(\la,t)=\frac{e^{\pi(2l+1)}}{4\pi}
\biggl[\sum_{m=1}^\infty \frac{\sqrt{16l^2+m^2}}{l^2+m^2}
\sin(m\la+\d_m)e^{im_mt}+\frac{4\pi}{l}\biggr].
\eqn192
$$
For $t=0$ we get
$$
\psi(\la,0)=e^{\pi(2l+1)\la} \eqn193
$$
and the wave packet (16.192) will disperse. Thus we should keep the wave
packet to not decohere in such a way that
$$
\psi(\la,t)=\frac{e^{\pi(2l+1)}}{8\pi}
\biggl[\sum_{m=1}^\infty \frac{\sqrt{16l^2+m^2}}{l^2+m^2}
\sin(m\la+\d_m)+\frac{4\pi}{l}\biggr].
\eqn194
$$
It means $\psi(\la,t)$ should be a pure zero mode for all the time $t\ge0$
(see Ref.~[120]).

Let us come back to the Eq.\ (16.184) (earlier to Eq.~(16.153)) taking under
consideration that in a real model $m_{\u A}=\frac{\a_s}r$). One can rewrite
Eq.~(16.184) in the following way:
$$
\(\frac{2l+1}{4\a_s}\)^{4/n}\(\frac{n+2}{(n-2)|\uP|}\)^{2/n}
\(\frac{\a_s\sqrt{n+2}\,m\dr{pl}}{4\sqrt n\, m_{\u A}}\)^2= \frac{|\uP|}
{\RG} \eqn195
$$
or
$$
\RG=\(\frac{4\a_s}{2l+1}\)^{4/n}\(\frac{(n-2)|\uP|}{n+2}\)^{2/n}
\(\frac{4\sqrt n\, m_{\u A}}{\a_s\sqrt{n+2}\,m\dr{pl}}\)^2 |\uP|=
2F(\z). \eqn196
$$

We want to find some conditions for~$\mu$ and~$\z$ (or~$\xi$ and~$\z$) in
order to satisfy Eq.~(16.153) for large~$l$ in a special model with $H=G2$,
$\dim G2=n=14$ for $M=S^2$ and $\RG$ calculated for $G=SO(3)$. In this case
one finds
$$
F(\z)=3^{2/7}\cdot 7 \cdot \(\frac{\a_s}{l}\)^{4/7} \(\frac{m_{\u A}}
{\a_s m\dr{pl}}\)^2 |\uP|^{15/7} \eqn197
$$
where $\uP$ is given by the formula (8.64). 

We calculate $\RG$ for a group $SO(3)$ in the fourth point of Ref.~[18]
(Eq.~(7.21)) and we get
$$
\RG=\frac{2(2\mu^3+7\mu^2+5\mu+20)}{(\mu^2+4)^2}\,. \eqn198
$$
The polynomial
$$
W(\mu)=2\mu^3+7\mu^2+5\mu+20 \eqn199
$$ 
possesses only one real root
$$
\mu_0=-\frac{\root3\of{1108+3\sqrt{135645}}}{6}-\frac76-\frac{19}
{6\cdot\root3\of{1108+3\sqrt{135645}}}=-3.581552661\ldots.\eqn200
$$ 
It is interesting to notice that $W(-3.581552661)=2.5\cdot10^{-9}$ and for
70-digit approximation of~$\mu_0$, $\t \mu$ equal to
$$
-3.581552661076733712599740215045436907383569800816123632201827285932446,
$$
we have
$$
W(\t\mu)=0.1\cdot 10^{-67}.
$$ 
The interesting point in $\t\mu$ is that any truncation of~$\t\mu$ (it means,
removing some digits from the end of this number) results in growing $W$,
i.e., 
$$
0<W(\t\mu_n)<W(\t\mu_{n-1}),
$$
where $\t\mu_n$ means a truncation of~$\t\mu$ with only $n$~digits, $n\le70$,
$\t\mu_{70}=\t\mu$. This can help us in some approximation for $\RG>0$,
close to zero.

Substituting (16.198) into Eq.\ (16.196) we arrive to the following equation:
$$
-F(\z)\mu^4+2\mu^3+\mu^2(7-4F(\z))+5\mu+4(-2F(\z)+5)=0. \eqn201
$$
Considering $F(\z)$ small we can write a root of this polynomial as
$$
\mu=\mu_0+\e \eqn202
$$
where $\mu_0$ is the root of the polynomial (16.199) and $\e$ is a small
correction. In this way one gets the solution
$$
\mu=-3.581552661\ldots +47\(\frac{\a_s}{l}\)^{4/7}\(\frac{m_{\u A}}
{m\dr{pl}}\)^2q(\z) \eqn203
$$
where $q(\z)$ is given by the formula
$$
\al
q(\z)&=\Biggl
| \frac{16|\z|^3(\z^2+1)}{3(2\z^2+1)(1+\z^2)^{5/2}}
\(\z^2E\(\frac{|\z|}{\sqrt{\z^2+1}}\)-(2\z^2+1)K\(\frac{|\z|}{\sqrt{\z^2+1}}\)\)\cr
&\qquad{}+8\ln\(|\z|\sqrt{\z^2+1}\)+\frac{4(1+9\z^2-8\z^4)|\z|^3}
{3(1+\z^2)^{3/2}}\Biggr|^\frac{15}7\cr
&\times\[\ln\(|\z|+\sqrt{\z^2+1}\)+2\z^2+1\]^{-\frac{15}7},
\eal
\eqn204
$$
$|\z|>|\z_0|=1.36$ (see Fig.\ 6), and $K$ and $E$ are given by the formulae
(8.65a) and (8.65b).

In this way taking sufficiently large $l$ we can choose $\mu$ such that
$\RG>0$ and quite arbitrary~$\z$ ($\uP<0$) to satisfy a zero energy condition
for an excitation of a warp factor.

Taking $m_{\u A}=m\dr{EW}\simeq80\,$GeV, $m\dr{pl}\cong2.4\cdot
10^{18}\,$GeV, $\a_s=\sqrt{\a\dr{em}}=\frac1{\sqrt{137}}$, one gets from
Eq.~(16.203) 
$$
\mu=-3.581552661\ldots+4.1\cdot10^{-35}\frac{q(\z)}{l^{4/7}} \eqn205
$$
which justifies our approximation for $\e$ and simultaneously gives an
account for a smallness of a \co\ \ct\ $\t R(\t\G(\mu))\simeq0$. Thus it
seems that in this simple toy model we can achieve a large $c\dr{eff}$
without considering large \co\ \ct. 

Let us give some numerical estimation for a time needed to travel a distance
of $200\,$Mps
$$
t\dr{travel}=\frac L{c\dr{eff}\gi{max}}=\frac Lc\cdot e^{-(2l+1)\pi/4}.
$$
One gets for $L=200\,$Mps,
$t\dr{travel}=2\cdot 10^{13}\,\text{s}\cdot e^{-l\pi/2}$
and for $l=100$, $t\dr{travel}\cong12\,$ns.

Let us notice that for $l=100$, $\mu$ from Eq.~(16.205) is equal to
$$
\mu=-3.581552661\ldots + 5.94\cdot 10^{-44}q(\z). \eqn206
$$
$q(\z)$ cannot be too large (it is a part of \co\ term $q(\z)=|\uP|^{15/7}$).
If $\uP$ is of order $0.1$, $q$ is of order $10^{-16}$; if $\uP$ is of order
$1.1$, $q$~is of order~$0.6$. Thus $\mu$ is very close to~$\mu_0$ for some
reasonable values of~$|\uP|$. It is interesting to ask how to develop this
model to more dimensional case. It means, to the manifold~$M$ which is not a
circle. It is easy to see that in this case we should consider a warp factor
which depends on more coordinates taking under consideration some ansatzes. 
For example we can consider a warp factor in a shape
$$
\exp\Bigl(\sum_{i=1}^m \sqrt{\t A_i}\,f_i(y)\Bigr) \eqn207
$$
where $f_i(y)=x_i$, $i=1,2,\ldots,m=\frac12n_1(n_1+1)$ are parametric
equations of the manifold~$M$ in $m$-dimensional euclidean space and
$$
ds^2_M=\biggl(\sum_{k=1}^m \(\frac{\pa f^k}{\pa y^i}\)
\(\frac{\pa f^k}{\pa y^j}\)\biggr) dy^i \otimes dy^j \eqn207a
$$
is a line element of the manifold $M$. In the case of the manifold $S^2$ it
could be
$$
\al
f_1&=\cos\theta\sin\la\cr
f_2&=\cos\theta\cos\la\cr
f_3&=\sin\theta\cr
m&=3
\eal
\eqn208
$$
and a warp factor takes the form
$$
\al
&\exp\(\sqrt{\t A_1}\cos\theta \sin\la+\sqrt{\t A_2}\cos\theta \cos\la
+\sqrt{\t A_3}\sin\theta \), \cr
&\theta\in\langle0,\pi),\ \la\in\langle0,2\pi).
\eal \eqn209
$$

Afterwards we should develop a warp factor in a series of spherical harmonics
on the manifold~$M$ and calculate an energy of a tower of scalar particles
connected with this development. We should of course regularize the series
(divergent) using $\z$-functions techniques. In this case we should use
$\z$-functions connected with the Beltrami-Laplace operator on the
manifold~$M$. Afterwards we should find a zero energy condition for special
types of \ct s $\t A_i$, $i=1,2,\ldots,m$. We can of course calculate the
energy of an excitation of the warp factor using some formulae from classical
field theory. In the case of $M=S^2$ we have to do with an ordinary Laplace
operator on~$S^2$ and with ordinary spherical harmonics with spectrum
given by Eq.~(16.14). The Fourier analysis of (16.209) can be proceeded
on~$S^2$. 

Moreover a toy model with $S^1$ can be extended to the \co\ background in
such a way that 
$$
ds^2_5=e^{\sqrt{\u A}\,\la}\,dt^2 - R^2(t)(dx^2+dy^2+dz^2)-r^2d\la^2. \eqn210
$$

It seems that from practical (calculational) point of view it is easier to
consider a metric of the form
$$
ds^2_5=dt^2-e^{-\sqrt{\u A}\,\la}R^2(t)(dx^2+dy^2+dz^2)- r^2\,d\la^2 \eqn211
$$
where $R(t)$ is a scale factor of the Universe. For further investigations we
should also consider more general metrics, i.e.
$$
\al
ds^2_6&=dt^2-\exp\(-\sqrt{\t A}\bigl(\cos\theta(\sin\la+\cos\la)+\sin\theta\bigr)\)\\
&\times R^2(t)(dx^2+dy^2+dz^2)- r^2\(d\theta^2+\sin^2\theta\, d\la^2\)
\eal \eqn212
$$
or even
$$
\al
ds^2_{4+n_1}&=dt^2-\exp\(-\sum_{i=1}^m \sqrt{\t A_1}f_i(y)\)
\cdot R^2(t)\,\frac{dx^2+dy^2+dz^2}{1-k(x^2+y^2+z^2)}\\
&- r^2\biggl(\biggl(\sum_{k=1}^m \(\frac{\pa f^k}{\pa y^i}\)\(\frac{\pa f^k}
{\pa y^j}\)\biggr)\cdot dy^i\otimes dy^j\biggr)
\eal \eqn213
$$
where
$$
m=\tfrac12\,n_1(n_1+1). \eqn214
$$

Let us notice that our toy model with a travel in a hyperspace (this is
really a travel along dimensions of a vacuum state manifold) can go to some
``acausal properties''. This is evident if we consider two signals: one sent in
an ordinary way with a speed of light and a second via the fifth dimension to
the same point. The second signal will appear earlier than the first one at
the point. Thus it will affect this point earlier. Effectively it means a
superluminal propagation of signals in a Minkowski space (or in
Friedmann-Robertson-Walker Universe). In this way we can construct a
time-machine using this solution (even if it does not introduce tachyons in a
4-dimensional space-time), i.e.\ a space-time where there are closed
time-like (or null) curves.

Let us notice that given a superluminal signal we can always find a reference
frame in which it is travelling backwards in time. Suppose we send a signal
from~$A$ to~$B$ (at time $t=0$) at an \ef\ speed $u\dr{eff}>1$ in frame~$S_1$
(it means we send a signal through a path described here earlier and an \ef\
speed $u\dr{eff}>1$) with coordinates $(t,x)$ (we consider only one space
dimension). \hbox{In a frame~$S_2$ moving \wrt $S_1$} with velocity
$v>\frac1{u\dr{eff}}$, the signal travels backwards in $t'$ time. This
follows from the Lorentz transformation, i.e.
$$
\ealn{
t'_B&=\frac{t_B(1-vu\dr{eff})}{\sqrt{1-v^2}} &\rf215 \cr
x'_B&=\frac{t_B(1-\frac v{u\dr{eff}})}{\sqrt{1-v^2}}\,. &\rf216
}
$$
We require both $x'_B>0$, $t'_B<0$, that is $\frac1{u\dr{eff}}<v<u\dr{eff}$
which is possible only if $u\dr{eff}>1$.

However this by itself does not mean a violation of a causality. For that we
require that the signal can be returned from~$B$ to a point in the past of
light cone of~$A$. Moreover, if we return the signal from~$B$ to~$C$ with the
same \ef\ speed (it means we send a signal through the fifth dimension with
the same \ef\ speed using the warp factor) in frame~$S_1$, then it arrives
at~$C$ in the future cone of~$B$. The situation is physically equivalent in
the Lorentz boosted frame~$S_2$---the return signal travels forward in
time~$t'$ and arrives at~$C$ in the future cone of~$A$. This is a
frame-independent statement. If a backwards-in-time signal $AC$ is possible
in frame~$S_1$, then a return signal sent with the same speed $u\dr{eff}$
will arrive at a point~$D$ in the past light cone of~$A$ creating a closed
time-like curve $ACDA$. In a Minkowski space (a~space for $\la=0$, or in
general for $\la=\la_0=\text{const.}$), local Lorentz invariance results in
that if a superluminal signal such as $AB$ is possible, then so is one
of~$AC$ for it is just given by an appropriate Lorentz boost. The existence
of a global inertial frames in a Minkowski space guarantees the existence of
the return signal~$CD$.

Thus we get unacceptable closed time loops. Moreover the theory for the
five-dimensional world is still causal, as in five-dimensional General
Relativity. Let us notice that in a \co\ background described by a metric \rf
211 \ the situation will be very similar. We have here also global inertial
frame for $\la=0$, and similar possibility to send a superluminal signal
through fifth dimension. However, in this case we should change a shape of
Lorentz transformation to include a scale factor $R(t)$ which slightly
complicates considerations. Moreover in order to create a warp factor we need
a zero energy condition. If this condition is not satisfied we need an
infinite amount of an energy to create it. The zero energy condition imposes
some restriction on parameters which are involved in our unified theory.
Maybe it is impossible to satisfy these conditions in our world. In this way
it would be an example of Hawking's chronology protection conjecture: ``{\sl
the laws of physics do not allow the appearance of closed timelike
curves\/}''. However, if a zero energy condition is satisfied in a realistic
Nonsymmetric Kaluza--Klein (Jordan--Thiry) Theory (it means in such a theory
which is consistent with a phenomenology of a contemporary elementary
particle physics) then we meet possibility of \ef\ closed time-like loops
(even the higher dimensions are involved). In this case we should look for
some weaker conjectures (especially in an expanding Universe, not just in a
Minkowski space). Such weaker condition protects us from some different
causal paradoxes as the following paradox: a highly energetic signal can
destroy a laboratory before it was created. In some sense it means some kind
of selfconsistency conditions.

Thus what is a prescription to construct a time-machine in our theory? It is
enough to have a 5-dimensional (or ($n_1+4$)-dimensional) space-time with a
warp factor described here earlier. Afterwards we need two frames $S_1$
and~$S_2$ in relative motion with relative velocity $v>\frac1{u\dr{eff}}$
where $u\dr{eff}$ is an \ef\ superluminal velocity $u\dr{eff}>1$. This could
be arranged e.g.\ as two spaceships with relative motion ($A$~and~$C$).
Sending a signal with \ef\ velocity $u\dr{eff}$ from $A$  to~$C$ we get the
signal in a past of~$C$. Afterwards we send a signal (through the fifth
dimension) from $C$ to~$A$. In this way we arrive in a past of~$A$.

If we want to travel into a future we should reverse a procedure with
$$
-1<v<0.
$$
In both cases we use an \ef\ time-like loop, constructed due to a warp factor
in a five-dimensional space-time. Using this loop several times we can go to
the future or to the past as far as we want. (As in a TVP spectacle
{\it Through the Fifth Dimension\/}.) An interesting point which can
be raised consists in an \ef ness of such a travel. It means how much time we
need to get e.g. 1\,s to past or 1\,s to future. It depends on a relative
velocity $v$ and~$u\dr{eff}$. The \ef ness can be easily calculated.
$$
\eta=\frac{u\dr{eff}v}{\sqrt{1-v^2}}\,.\eqn217
$$
If we use our estimation for $c\dr{eff}$ (see Eq.\ \rf182 ), we get
$$
\eta=\frac{v}{\sqrt{1-v^2}}\, e^{(2l+1)\pi/4} \eqn218
$$
for $l$ large. It is easy to see that $\eta\to\infty$ if $v\to1$ and if $l
\to\infty$. 

Let us take $v=\frac1{10}$ and $l=100$. In this case one gets
$$
\eta\approx 10^{67}. \eqn219
$$
What does this number mean? It simply means that in order to go 1\,s into the
past or into the future we loose $10^{-67}$\,s of our life.

However, to be honest, we should use as $u\dr{eff}$ a velocity different from
$c\dr{eff}$. In this case a time needed to travel through the fifth dimension
really matters. Let a speed along the fifth dimension be equal to the
velocity of light (or close to it). In this case one gets
$$
u\dr{eff}=\frac{c\dr{eff}}{1+2(\frac{r_0}L)c\dr{eff}} \eqn220
$$
where $r_0\simeq 10^{-16}$\,cm, $L$ is the distance between $A$ and~$B$,
e.g.\ 1000\,km. In this case $u\dr{eff}$ is smaller,
$$
u\dr{eff}=10^{26} \eqn221
$$
and
$$
\eta\simeq 10^{25}. \eqn222
$$
In this case we can estimate how much time we loose to go 2000 years back or
forward. It is $6.3\cdot10^{-15}$\,s. However, if we want to see dinos we
should go back 100\,mln\,yr and we loose $10^{-10}$\,s. Of course in the
formula for $u\dr{eff}$ we can use smaller~$L$ and a velocity in the place
$c\dr{eff}$ can be smaller too. We can travel more comfortably changing
smoothly a velocity. Any time a time to loose to travel back or forward in
time for our time machine is very small. In some sense it is very \ef\ (if it
is possible to construct).

\def\cor{\gi{corrected}}
Let us notice that the warp factor (or $\u A$) is fixed by the \ct s of the
theory (especially by a \co\ \ct\ $\la_{c0}$). In this way only in some
special cases we get a zero energy condition. However, in the case of the
warp factor existence we have to do with a tower of scalar fields (not only
with a \q). The \q\ field is a zero mode of this tower. Thus in the case of
an existing of a tower of scalar fields we will have to do with a corrected
\co\ \ct, which is really an averaged \co\ term over $S^1$ (in general
over~$M$) 
$$
\la\cor_{c0}=\frac1{2\pi}\,\int_0^{2\pi} \(|\g|-\b e^{2\P(\la)}\)
e^{n\P(\la)}\,d\la \eqn223
$$
where $\P(\la)$ is given by Eq.\ \rf118 . In this way according to our
considerations concerning an energy of excitation
$$
\la\cor_{c0}=E\dr{wf}\,. \eqn224
$$
However, in the formula for $E\dr{wf}$ we dropped an energy for a zero mode
for a tower of scalar fields (i.e., for a \q). Thus
$$
\la\gi{tot}_{c0}=\la_{c0}(\P_0)+\la\cor_{c0}. \eqn225
$$
If the zero energy condition for a warp factor is satisfied, we get
$$
\la\gi{tot}_{c0} =\la_{c0} \eqn226
$$
and we do not need any tunning of parameters in the theory. However, now we
should write down full 5-dimensional equations in our theory ($n_1+4$ in
general) to look for more general solutions mentioned here earlier. 

In Fig.\ 8 we draw a time-like loop in our theory together with a realistic
path in five-dimensional space-time with a warp factor corresponding to the
loop. 

\obraz{
\midinsert \centerline{\epsfbox{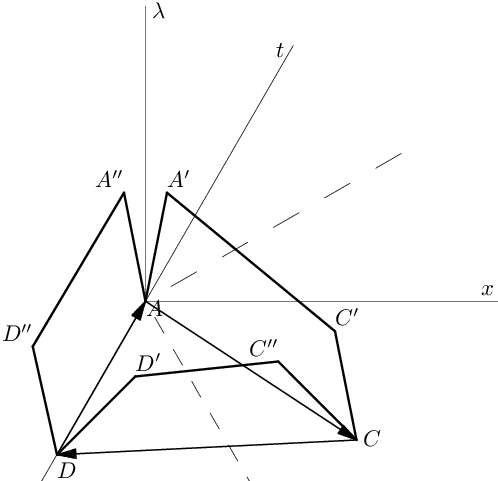}} 
\centerline{\nine Fig. 8. A closed time-like curve (a time-like loop)
$ACDA$ in a Minkowski space with a warp factor.}
\vskip-1pt
\centerline{\nine $AA'C'CC''D'D''A''A$ means a realistic path in a
5-dimensional space-time with a warp factor.} \endinsert
}

Let us consider a $(n_1+4)$ formalism of the \lg\ in our theory given by
Eq.~(7.5). In this case we do not suppose that a scalar field does not depend
on a point of a manifold~$M$. In general $\P=\P(x,y)$, $x\in E$, $y\in M$.
Simultaneously we should suppose that $g\id{\mu\nu}=g\id{\mu\nu}(x,y)$ for we use
the transformation (7.4). From the variation principle
$$
\d \intop_U d^4x\,d^{n_1}y\,L(g,\o W,\t A,\P,\F)\sqrt{-g}\sqrt{\t g}=0,
\quad U\subset E\times M \eqn227
$$
we get equations (9.3$\div$9.18a) with some differences connected with the
fact that now we do not average over a manifold~$M$. It means that in
definitions for $V(\F)$ there are no integrals and the same is true for~$\t
P$. Now in the place of~$\t P$ we have simply $\h{\o R}(\h{\o\G})$ (i.e.\ a
scalar curvature for a metric $g\id{\u a\u b}$). In the place of $\t g^{[\u a\u
b]}$ we have simply $g^{[\u a\u b]}$ and so on. Moreover the only one
``material source'' which lives in $(n_1+4)$-dimensional space-time is a
scalar field~$\P$. For this reason in a \lg\ of this field we have the
formula \rf4 \ (or \rf 3 ). In this way a full set of field equations should
be supplemented with
$$
\ealn{
R_{(\mu\u a)}&=\frac{8\pi K}{c^4}\, \br{scal}T_{\kern-5pt(\mu\u a)}(\P) &\rf228 \cr
R_{(\u a\u b)}&=\frac{8\pi K}{c^4}\, \br{scal}T_{\kern-5pt(\u a\u b)}(\P) &\rf229 
}
$$
and
$$
g\id{\mu\nu,\u s}-\G^\xi_{\mu \u s}g\id{\xi\nu}-\G^\xi_{\u s\nu}g\id{\mu\xi}=0
\eqn230 
$$
where $\br{scal}T_{\kern-5pt(\mu\u a)}$, $\br{scal}T_{\kern-5pt(\u a\u b)}$
are remaining parts of an energy-momentum tensor for a scalar field~$\P$ and
$\G^\xi_{\mu\u s}, \G^\xi_{\u s\nu}$ are remaining coefficients of a
connection. $R_{\mu\nu}$ are constructed from $\o W^\a{}_\b$ connection and
$R_{(\mu\u a)}$, $R_{(\u a\u b)}$ are constructed from the connection induced
by relations (9.5), (6.25), (9.4) and the condition
$$
\G^\xi_{\xi\u s}=\G^\xi_{\u s\xi}. \eqn231
$$

A metric tensor has a general form
$$
ds^2_{n_1+4}=g\id{(\mu\nu)}(x,y)\,dx^\mu \otimes dx^\nu
- r^2g\id{(\u a\u b)}\theta^{\u a}\otimes \theta^{\u b}. \eqn232
$$
We have the formula
$$
\al
\br{scal}T_{\kern-5pt AB}(\P)&=-\frac{e^{(n+2)\P}}{16\pi}
\biggl\{\(g\id{KA}g\id{ZB}+g\id{ZA}g\id{KB}\)\\
&\quad{}\times\tge{CK}\tge{NZ}
\[\frac{n^2}2\(g^{SM}g\id{NS}-\d^M{}_N\)+\o M\P_{,N}\]\P_{,C}\\
&\quad{}-g\id{AB}\[\o M \o g^{(NM)}+n^2g^{[MN]}g\id{[DM]}\tge{CD}\P_{,N}\P_{,C}\]\biggr\}
\eal
\eqn233
$$

$g\id{AB}$ tensor is here in the place of $\g\id{AB}$ in order to stress that
$g\id{AB}$ depends on $x$ and~$y$. We suppose however that $g\id{AB}$ has the
following structure
$$
g\id{AB}=\left(\matrix g\id{\mu\nu}&\vrule height10pt depth4pt&0\cr
\noalign{\hrule}
0&\vrule height10pt depth4pt& r^2g\id{\u a\u b}
\endmatrix \right). \eqn234
$$
This tensor is of course nonsymmetric, $g=\det(g\id{AB})$.

These field equations are fundamental equations to find our solutions. For we
are looking for some \co\ solutions (with warp factor), we suppose
additionally that
$$
\ealn{
g\id{\mu\nu}&=g\id{(\mu\nu)}(x,y) &\rf235 \cr
g\id{\u a\u b}&=g\id{\u a\u b}(y) &\rf235a
}
$$
The metric on a manifold $M$ is still nonsymmetric.

We will also suppose that the total energy momentum tensor (except for a
scalar field) is given by a sum of several hydrodynamics energy-momentum
tensors including a dust and a radiation. We get five-dimensional field
equations for $M=S^1$ and \hbox{6-dimensional} for $M=S^2$ (a 6-dimensional
Weinberg--Salam model). Moreover, there is a different approach to derive 
$(n_1+4)$-dimensional equations in the Nonsymmetric Kaluza--Klein
(Jordan--Thiry) Theory.

Let us consider an $(n_1+4)$ generalization of 4-dimensional field equations
in the Nonsymmetric Kaluza--Klein (Jordan--Thiry) Theory. According to the
second point of Ref.~[18] we get (see Eqs (4.8.3)$\div$(4.8.14)):
$$
\displaylines{
\hfill \o R_{MN}(\o W)-\frac12\,g\id{MN}\o R(\o W)=
8\pi K\(\br{gauge}T_{\kern-8pt MN}+\br{scal}T_{\kern-5pt MN}
+g\id{MN}F\)  \hfill \rf236 \cr
\hfill {\ut g}^{[MN]}{}_{,N}=0 \hfill \rf237 \cr
\hfill g\id{MN,S}-g\id{ZN}\o\G{}^Z{}_{MS}-g\id{MZ}\o\G{}^Z{}_{SN}=0 \hfill \rf238
\cr
\indent \ga\n_{\kern-5pt M}\(l_{ab}\ut L^{aAM}\)=2\ut g^{[AB]}
\ga\n_{\kern-5pt B}\(h_{ab}\ut g^{[MN]}
H^a_{MN}\)\hfill\cr
\hfill{}+(n+2)\pa_B \P\[l_{ab}L^{aBA}-2g^{[BA]}\(h_{ab}g^{[MN]}H^a_{MN}
\)\] \indent \rf239 \cr
\quad\[(n^2+2\o M)\tge{AM}-n^2g^{NM}\tge{AS}\]\frac{\pa^2\P}{\pa x^A\pa x^M}
+\frac1{\sqrt{-g}}\,\pa_M\biggl\{\sqrt{-g}\biggl(n^2\tge{MA}\hfill\cr
{}-\frac{n^2}2\,g\id{DN}\(g^{NA}\tge{MD}+g^{NM}\tge{AD}\)
-2\o M\tge{MA}\biggr)\biggr\}\frac{\pa\P}{\pa x^A}\cr
\hfill{}-8\pi(n+2)\pi e^{-(n+2)\P}\(\cL_{YM}-2F\)=0 \indent \rf240 
}
$$
where
$$
\al
\br{gauge}T_{\kern-5pt AB}&=-\frac{l_{ab}}{4\pi}
\biggl\{g\id{CB}g^{TZ}g^{EC}L^a_{ZA}L^b_{TE}-2g^{[MN]}H^{(a}_{MN}H^{b)}_{AB}\\
&-\frac14\,g\id{AB}\[L^{aMN}H^b_{MN}-2\(g^{[MN]}H^a_{MN}\)
\(g^{[CS]}H^b_{[CS]}\)\]\biggr\}
\eal
\eqn241
$$
is the energy momentum tensor for the $(n_1+4)$-dimensional gauge
(Yang-Mills) field, $\br{scal}T_{\kern-5pt AB}(\P)$ is given by the Eq.~\rf233
$$
\ealn{
K&=G_{(n_1+4)}e^{-(n+2)\P} &\rf242 \cr
F&=e^{2(n+2)\P}\,\frac{\t R(\t \G)}{16\pi} &\rf 243 \cr
\ut L^{aMN}&=\sqrt{-g}\,g^{BM}g^{CA}L^a_{BC} &\rf244 \cr
\ut g^{[MN]}&=\sqrt{-g}\,g^{[MN]} &\rf245
}
$$
where 
$$
G_{(n_1+4)}=\frac1{M\dr{pl$(n_1+4)$}}
$$
is an $(n_1+4)$-dimensional \gr\ \ct, $\ga\n_{\kern-8pt N}$ means a gauge
derivative for a connection defined on~$P$ (over an $(n_1+4)$-dimensional
base),
$$
\displaylines{
\hfill l_{dc}g_{MB}g^{CM}L^d_{CA}+l_{cd}g_{AM}g^{MC}L^d_{BC}
=2l_{cd}g_{AM}g^{MC}H^d_{BC} \hfill \rf246 \cr
\obraz{\noalign{\eject}}
\hfill \cL_{YM}=\frac1{8\pi}\,l_{cd}\(H^cH^d-L^{cMN}H^d_{MN}\) \hfill \rf247 \cr
\hfill H^c=g^{[AB]}H^c_{AB} \hfill \rf248
}
$$
$H^c_{AB}$ is a curvature of a connection $\omega$ on a fibre bundle $P$ (over
an $(n_1+4)$-dimensional base).

Now let us suppose that a base manifold is a Cartesian product $E\times M$,
$M=G/G_0$ (as in sec.~4) and let us suppose a symmetry conditions on a
connection $\omega$ over $E\times M$ ($\dim M=n_1$), i.e.\ an invariancy
of~$\omega$ \wrt the action of the group~$G$ on $E\times G/G_0$ (see sec.~4
for details). Simultaneously we suppose that $g\id{AB}$ has a form given by
Eq.~\rf 234 , where $g\id{\u a\u b}$ is defined and examined in sec.~4. In
this way we can reduce the field equation to simpler form described above.

The connection $\o W{}^A{}_B$ is given by the equation
$$
\o W{}^A{}_B=\o\omega^A{}_B - \frac2{(n_1+3)}\,\o W\d^A{}_B \eqn249
$$
where
$$
\displaylines{
\hfill \o\omega^A{}_B=\o\G{}^A{}_{BC}\theta^C \hfill \rf250 \cr
\hfill \o W=\o W_C \theta^C = \frac12\(\o W{}^S{}_{CS}-\o W^S{}_{SC}\)\theta^C
\hfill \rf251
}
$$

The dimension reduction procedure (described in sec.~4) reduces
$(n_1+4)$-dimen\-sional Yang-Mills' field to a 4-dimensional Higgs' field.
$R_{AB}$~is a Ricci (Moffat--Ricci) tensor for the connection $\o W{}^A{}_B$. 
In our approximation we can consider those fields as 4-dimensional matter
fields (i.e.~hydrodynamics energy-momentum tensors). Remembering that only
$g\id{\mu\nu}=g\id{\nu\mu}(x,y)$ and $\P=\P(x,y)$ are functions depending on
$y\in M$ we can try to solve our equations for metrics defined earlier.
Eventually let us remark that both approaches described here are not strictly
equivalent. Moreover for the class of problems considered here they are
equivalent. 

}
\vfill\eject

\null
\vskip36pt plus4pt minus4pt
\centerline{{\four\bf 17. The Approximation Procedure for the Lagrangian}}

\vskip4pt 
\centerline{{\four\bf of the Higgs Field}}

\vskip4pt
\centerline{{\four\bf in the Nonsymmetric Jordan--Thiry Theory}}
\vskip24pt plus2pt minus2pt
\write0{17. The Approximation Procedure for the Lagrangian
 of the Higgs Field in the Nonsymmetric Jordan--Thiry Theory \the\pageno}

In this section we develop the approximation procedure for the
lagrangian involving Higgs' field up to the second order of expansion
in $\zeta $ and $h_{\mu \nu } = g_{\mu \nu } - \eta _{\mu \nu }$. We also deal with geodetic equations on the
Jordan-Thiry manifold in the first order of approximation with respect
to $\zeta $ and $h_{\mu \nu }$ for terms coupled to Higgs' field. 
We find also the field equations in linear approximation and an influence of $\ov W_\mu$ 
and electromagnetic fields in this approximation. In this section we use
$p_{ab}$ to denote the Killing--Cartan tensor.

Let us consider the lagrangian for the Higgs' field i.e.:
$$
{\cal L}({\varPhi} ,{\varPsi} ) = {2\over r^{2}} e^{-2{\varPsi} } 
{\cal L}_{\kin} \big(\Nabla^{\gauge}{\varPhi}\big) - {1\over r^{4}} 
e^{(n-2){\varPsi} } V({\varPhi} ) 
- {4\over r^{4}} e^{(n-2){\varPsi} } {\cal L}_{\Int}({\varPhi}
,\widetilde{A})\eqno(17.1)
$$
where
$$\eqalignno{
{\cal L}_{\kin} \big(\Nabla^{\gauge}{\varPhi}\big) &= \ell _{ab} 
\big(g^{\tilde{b} \tilde n } g^{\alpha \mu } L^{a}{}_{\tilde{\alpha}} 
\mathop{\nabla_{\mu}}\limits^{\gauge}{\varPhi}^{b}_{\tilde n
}\big)_{av}&(17.2\text{\rm a})\cr
V({\varPhi} ) &= 2\ell _{ab} (\un{g}^{([\tilde m \tilde n
][\tilde{a} \tilde{b} ])} H^{a}_{\tilde m \tilde n } H^{b}_{\tilde{a}
\tilde{b} } - \un{g}^{\tilde{a} \tilde m \tilde{b} \tilde n }
L^{a}_{\tilde{a} \tilde{b} } H^{b}_{\tilde m \tilde n }
)_{av}&(17.2\text{\rm b})\cr
{\cal L}_{\Int} ({\varPhi} ,\widetilde{A}) &= p_{ab} \mu ^{a}{}_{i} 
\widetilde{H}^{i} \un{g}^{[\tilde{a} \tilde{b} ]} H^{b}_{\tilde{a}
\tilde{b} }&(17.2\text{\rm c})\cr}
$$
$\un{g}^{([\tilde m \tilde n ][\tilde{a} \tilde{b} ])}$,
$g^{\tilde{a} \tilde m \tilde{b} \tilde n }$ and $g^{[\tilde{a}
\tilde{b} ]}$ are given by the formula (6.24) and $H^{b}_{\tilde{a}
\tilde{b} }$ by the formula~(4.55).

Moreover for $L^{a}{}_{\alpha \tilde{b} }$ and $L^{a}{}_{\tilde{a}
\tilde{b} }$ we have
$$\eqalignno{
&\ell _{dc} g_{\beta \mu } g^{\gamma \beta } L^{d}{}_{\gamma
\tilde{a} } + \ell _{cd} g_{\tilde{a} \tilde m } g^{\tilde m \tilde c } 
L^{d}{}_{\mu \tilde c } = 2\ell _{cd} g_{\tilde{a} \tilde m }
g^{\tilde m \tilde c } \mathop{\nabla_\mu }\limits^{\gauge} {\varPhi}
^{d}_{\tilde c }&(17.3\text{\rm a})\cr
&\ell _{dc} g_{\tilde m \tilde{b} } g^{\tilde c \tilde m }
L^{d}{}_{\tilde c \tilde{a} } + \ell _{cd} g_{\tilde{a} \tilde m } 
g^{\tilde m \tilde c } L^{d}{}_{\tilde{b} \tilde c } = 2\ell _{cd}
g_{\tilde{a} \tilde m } g^{\tilde m \tilde c } H^{d}_{\tilde{b}
\tilde c }&(17.3\text{\rm b})\cr}
$$
Let us consider an expansion for $g_{\tilde{a} \tilde{b} }$ and $
g^{\tilde{a} \tilde{b} }$
$$\displaylines{
g_{\tilde{a} \tilde{b} } = h^{0}{}_{\tilde{a} \tilde{b} } + \zeta 
k^{0}{}_{\tilde{a} \tilde{b} }\cr
\hfill g^{\tilde m \tilde{a} } g_{\tilde n \tilde{a} } = ( h^{0\tilde
m \tilde{a} } + h^{1\tilde m \tilde{a} } + h^{2\tilde m \tilde{a} } +\ldots) 
\cdot ( h^{0}{}_{\tilde n \tilde{a} } + \zeta k^{0}{}_{\tilde n
\tilde{a} }) = \delta ^{\tilde m }_{\tilde n }\hfill\llap{(17.4)}\cr}
$$
From (17.4) we get:
$$\displaylines{
\hfill h^{1\tilde m \tilde n } = -\zeta h^{0\tilde m \tilde{a} } h^{0\tilde
n \tilde{b} } k^{0}{}_{\tilde{b} \tilde{a} }\hfill\llap{(17.5)}\cr
\hfill h^{2\tilde m \tilde n } = -\zeta h^{0\tilde n \tilde{b} } 
h^{1\tilde m \tilde{a} } k^{0}{}_{\tilde{b} \tilde{a} } = \zeta ^{2} 
h^{0\tilde n \tilde{b} } h^{0\tilde m \tilde e } h^{0\tilde{a} \tilde
s } k^{0}{}_{\tilde s \tilde e } k^{0}{}_{\tilde{b} \tilde{a}
}\hfill\llap{(17.6)}\cr}
$$
and
$$
g^{\tilde m \tilde n } = h^{0\tilde m \tilde n } - \zeta 
h^{0\tilde m \tilde{a} } h^{0\tilde n \tilde{b} } k^{0}{}_{\tilde{b}
\tilde{a} }
+ \zeta ^{2} h^{0\tilde n \tilde{b} } h^{0\tilde m \tilde e }
h^{0\tilde{a} \tilde{s} } k^{0}{}_{\tilde s \tilde e }
k^{0}{}_{\tilde{b} \tilde{a} } + \ldots \eqno(17.7)
$$
Let us write the Eq.~(17.3b) in a more convenient form:
$$\displaylines{\quad\
\ell _{dc} ( \delta ^{\tilde c }_{\tilde{b} } - 2\zeta 
k^{0}{}_{\tilde{b} \tilde d } \widetilde{g}^{\tilde c \tilde d }) 
L^{d}{}_{\tilde c \tilde{a} } + \ell _{cd} ( \delta ^{\tilde c
}_{\tilde{a} } - 2\zeta k^{0}{}_{\tilde d \tilde{a} }
\widetilde{g}^{\tilde d \tilde c }) L^{d}{}_{\tilde{b} \tilde c }
=\hfill\cr
\hfill = 2 (\delta ^{\tilde c }_{\tilde{a} } - 2\zeta 
k^{0}{}_{\tilde d \tilde{a} } \widetilde{g}^{\tilde d \tilde c } ) 
H^{d}{}_{\tilde{b} \tilde c } \ell _{cd}.\quad\ (17.8)\cr}
$$
Let us expand $L^{d}{}_{\tilde{b} \tilde c }$ into a power series with respect to $\zeta $:
$$
L^{d}{}_{\tilde{b} \tilde c } = L^{\nad{d}{(0)}}{}_{\tilde{b} \tilde c} + 
L^{\nad{d}{(1)}}{}_{\tilde{b} \tilde c} + 
L^{\nad{d}{(2)}}{}_{\tilde{b} \tilde c} +\ldots\eqno(17.9)
$$
Using (17.5) and (17.6) one gets up to the second order:
$$\eqalignno{
\ell _{dc} L^{d}{}_{\tilde{b} \tilde{a} } + \ell _{cd}
L^{d}{}_{\tilde{b} \tilde{a} } ={}& 2\ell _{cd} H^{d}{}_{\tilde{b}
\tilde{a} } - 4\zeta k^{0}{}_{\tilde d \tilde{a} } \ell _{cd}
H^{d}{}_{\tilde{b} \tilde c } \cdot \cr
\noalign{\vskip2pt}
&\cdot ( h^{0\tilde d \tilde c } - \zeta h^{0\tilde d \tilde t} 
h^{0\tilde d \tilde s } k^{0}{}_{\tilde s\tilde t} + \zeta ^{2}
h^{0\tilde d \tilde e } h^{0\tilde c \tilde r} h^{0\tilde s\tilde t} 
k^{0}{}_{\tilde s \tilde e } k^{0}{}_{\tilde r\tilde t}) +\cr
\noalign{\vskip2pt}
&+ 2\zeta ( h^{0\tilde c \tilde d } - \zeta h^{0\tilde c \tilde s } 
h^{0\tilde d \tilde t} k^{0}{}_{\tilde t\tilde s} + \zeta ^{2}
h^{0\tilde c \tilde r} h^{0\tilde d \tilde e } h^{0\tilde s\tilde t}
k^{0}{}_{\tilde s\tilde r} k^{0}{}_{\tilde e \tilde t}) \cdot\cr 
\noalign{\vskip2pt}
&\cdot ( k^{0}{}_{\tilde{b} \tilde d } \ell _{cd}
L^{d}{}_{\tilde c \tilde{a} } + k^{0}{}_{\tilde c \tilde{a} } \ell _{cd} 
L^{d}{}_{\tilde{b} \tilde d })&(17.10)\cr}
$$
One easily finds:
$$\eqalignno{
L^{\nad{a}{(0)}}{}_{\tilde{b} \tilde a} ={}& p^{ab} \ell _{bc}
H^{c}{}_{\tilde{b} \tilde{a} } = H^{a}{}_{\tilde{b} \tilde{a} } + \xi p^{ab} 
k_{bc} H^{c}{}_{\tilde{b} \tilde{a} }\qquad\quad\ &(17.11)\cr
L^{\nad{a}{(1)}}{}_{\tilde{b} \tilde a} ={}& \zeta h^{0\tilde d
\tilde c } p^{ac} p^{bd} \ell _{be} (k^{0}{}_{\tilde d \tilde{a} } \ell _{cd} 
H^{e}{}_{\tilde{b} \tilde{a} } + k^{0}{}_{\tilde{b} \tilde d } \ell _{dc} 
H^{e}{}_{\tilde c \tilde{a} })
- 2\zeta k^{0}{}_{\tilde d \tilde{a} } p^{ac} \ell _{cd}
H^{d}{}_{\tilde{b} \tilde c } h^{0\tilde d \tilde c }\qquad\quad\ &(17.12)\cr}
$$
or
$$\eqalignno{
L^{\nad{a}{(1)}}{}_{\tilde{b} \tilde a} ={}& \zeta (\delta ^{a}_{e}-\xi ^{2}
p^{ab}k_{bc}p^{cd}k_{de})\cdot h^{0\tilde c \tilde d
}(k^{0}{}_{\tilde{b} \tilde d } H^{e}{}_{\tilde c \tilde{a} } -
k^{0}{}_{\tilde{a} \tilde d } H^{e}{}_{\tilde c \tilde{b}})&(17.13)\cr
L^{\nad{a}{(2)}}{}_{\tilde{b} \tilde a} ={}& \zeta ^{2} \xi 
h^{0\tilde c \tilde d } h^{0\tilde t\tilde s} (k^{a}_{+b}k^{b}_{-c}) p^{cr} 
k_{rd} \times \cr
&\times ( k^{0}{}_{\tilde{b} \tilde d } H^{d}{}_{\tilde{s} \tilde{b} } 
k^{0}{}_{\tilde{a} \tilde{b} } - k^{0}{}_{\tilde c \tilde t}
k^{0}{}_{\tilde d \tilde{b} } H^{d}{}_{\tilde{s} \tilde{a} } +
2k^{0}{}_{\tilde{b} \tilde t} k^{0}{}_{\tilde d \tilde{a} }
H^{d}{}_{\tilde s \tilde c })&(17.14)\cr}
$$
where
$$
k^{a}_{\pm b} = \delta ^{a}_{b} \pm \xi p^{ar} k_{rb}\eqno(17.15)
$$
Let us consider the potential for the Higgs' field $V({\varPhi} )$ and expand it
into a power series with respect to $\zeta $.
$$
V({\varPhi} ) = \nad{V}{(0)}({\varPhi}) + \nad{V}{(1)}({\varPhi}) + 
\nad{V}{(2)}({\varPhi}) + \ldots \eqno(17.16)
$$
Using (17.12--14) and (17.2b) one gets after some calculations:
$$\eqalignno{
\nad{V}{(0)}({\varPhi} ) ={}&\mmint_M {\sqrt{h^0}\over V_{2}} dm(y) 
(p_{cb}+\xi ^{2}k_{cr}p^{rs}k_{sb}) H^{c}{}_{\tilde{b} \tilde c } 
H^{b}{}_{\tilde m \tilde{a} } h^{0\tilde{b} \tilde m } h^{0\tilde c
\tilde{a} }&(17.17)\cr 
\nad{V}{(1)}({\varPhi} ) ={}&\mmint_M {\sqrt{h^0}\over V_{2}}
2\zeta \xi dm(y) (p_{ab}+\xi k_{ab}) h^{0\tilde{b} \tilde{s} }
h^{0\tilde m \tilde t} h^{0\tilde c \tilde{a} } p^{ar}\cr
&\times (k_{rs}+\xi k_{rp}p^{pq}k_{qs}) k^{0}{}_{\tilde
t\tilde s} H^{s}{}_{\tilde{b} \tilde c } H^{b}{}_{\tilde m \tilde{a}
}&(17.18)\cr
\nad{V}{(2)}({\varPhi} ) ={}&{\zeta ^2\over V_2} \mmint_M \sqrt{h^0}
dm(y) (p_{ab}+\xi k_{ab}) H^{s}{}_{\tilde{b} \tilde c }
H^{b}{}_{\tilde m \tilde{a} } k^{0}{}_{\tilde t\tilde s}
k^{0}{}_{\tilde d \tilde e } \times \cr
&\times \bigg[ (2h^{o\tilde{b} \tilde e } h^{o\tilde m
\tilde t} h^{o\tilde s \tilde d } h^{o\tilde{a} \tilde c } +
h^{o\tilde{b} \tilde{s} } h^{o\tilde m \tilde t} h^{o\tilde c \tilde
e } h^{o\tilde{a} \tilde d } ) k^{a}_{+s} +\cr
&+ ( h^{0\tilde e \tilde s } h^{o\tilde m \tilde t} h^{o\tilde c
\tilde{a} } h^{o\tilde{b} \tilde d } - h^{o\tilde d \tilde s }
h^{o\tilde m \tilde t} h^{o\tilde c \tilde{a} } h^{o\tilde{b} \tilde
e } +\cr
&+ h^{o\tilde c \tilde s } h^{o\tilde m \tilde t} h^{o\tilde e
\tilde{a} } h^{o\tilde{b} \tilde d } - h^{o\tilde c \tilde s }
h^{o\tilde m \tilde t} h^{o\tilde d \tilde{a} } h^{o\tilde{b} \tilde
e }) k^{a}_{+r} k^{r}_{-s} +\cr
&+ {\xi \over 2} ( h^{o\tilde c \tilde m } h^{o\tilde t\tilde{a} } 
h^{o\tilde d \tilde{b}} h^{o\tilde e \tilde s } + h^{o\tilde c \tilde
m } h^{o\tilde t\tilde{a} } h^{o\tilde e \tilde{b} } h^{o\tilde d
\tilde s } +\cr
&- h^{o\tilde c \tilde m } h^{o\tilde{s} \tilde{a} } h^{o\tilde d
\tilde{b} } h^{o\tilde e \tilde t} - h^{o\tilde c \tilde m }
h^{o\tilde{s} \tilde{a} } h^{o\tilde e \tilde{b} } h^{o\tilde d
\tilde t} )k^{a}_{+r} k^{r}_{-s} +\cr
&+ {\xi \over 2} ( h^{o\tilde c \tilde m } h^{o\tilde t\tilde{a} } 
h^{o\tilde d \tilde{b}} h^{o\tilde e \tilde s } + h^{o\tilde c \tilde
m } h^{o\tilde t\tilde{a} } h^{o\tilde e \tilde{b} } h^{o\tilde d
\tilde s } +\cr
\noalign{\vskip1pt}
&- h^{o\tilde c \tilde m } h^{o\tilde{s} \tilde{a} } h^{o\tilde d
\tilde{b} } h^{o\tilde e \tilde t} - h^{o\tilde c \tilde m }
h^{o\tilde{s} \tilde{a} } h^{o\tilde e \tilde{b} } h^{o\tilde d
\tilde t} )(k^{a}_{+r} k^{r}_{-s})p^{pq}k_{qs} +\cr
\noalign{\vskip1pt}
&+ {\xi \over 2} ( h^{o\tilde c \tilde d } h^{o\tilde t\tilde{b} } 
h^{o\tilde s \tilde m } h^{o\tilde e \tilde{a} } - h^{o\tilde c
\tilde d } h^{o\tilde{s} \tilde{b} } h^{o\tilde t\tilde m }
h^{o\tilde e \tilde{a} } +\cr
\noalign{\vskip1pt}
&- h^{o\tilde c \tilde e } h^{o\tilde t\tilde{b} } h^{o\tilde s
\tilde m } h^{o\tilde d \tilde{a} } + h^{o\tilde c \tilde e }
h^{o\tilde{s} \tilde{b} } h^{o\tilde t\tilde m } h^{o\tilde d
\tilde{a} } ) (k^{a}_{+r}k^{r}_{-p}) p^{pq} k_{qs} +\cr
\noalign{\vskip1pt}
&- {\xi \over 2} ( h^{o\tilde s \tilde m } h^{o\tilde{a} \tilde t} 
h^{o\tilde{b} \tilde e } h^{o\tilde c \tilde d } - h^{o\tilde m
\tilde t} h^{o\tilde{a} \tilde{s} } h^{o\tilde{b} \tilde e }
h^{o\tilde c \tilde d } +\cr
\noalign{\vskip1pt}
&- h^{o\tilde s \tilde m } h^{o\tilde{a} \tilde t} 
h^{o\tilde{b} \tilde e } h^{o\tilde c \tilde e } + h^{o\tilde m
\tilde t} h^{o\tilde{a} \tilde{s} } h^{o\tilde{b} \tilde d }
h^{o\tilde c \tilde e })\delta ^{a}_{s}\bigg]&(17.19)\cr}
$$
Up to the first order of approximation in $\zeta $ one gets:
$$\eqalignno{
V({\varPhi} ) ={}& \mmint_M [(p_{cb}+\xi ^{2}k_{cr}p^{rs}k_{sb}) 
H^{c}{}_{\tilde{b} \tilde c } H^{b}{}_{\tilde m \tilde{a} }
h^{o\tilde{b} \tilde m } h^{o\tilde c \tilde{a} } +\cr
\noalign{\vskip1pt}
&+ 2 \xi \zeta p^{ar} k_{rs} (p_{ab}+\xi k_{ab}) (\delta ^{s}_{p}+\xi p^{sq}
k_{qp}) \times \cr
\noalign{\vskip1pt}
&\times h^{o\tilde{b} \tilde{s} } h^{o\tilde m \tilde t} h^{o\tilde
c \tilde{a} } k^{o}{}_{\tilde t\tilde s} H^{p}{}_{\tilde m \tilde c
} H^{b}{}_{\tilde m \tilde c }]{\sqrt{|h^0|}\over V_{2}}
dm(y).&(17.20)\cr}
$$
\looseness3It is 
easy to see that $V({\varPhi} )$ has only the first type of the critical
point corresponding to the trivial gauge configuration for the Higgs'
field up to the first order of expansion with respect to $\zeta $.
$\nad{V}{(2)}({\varPhi} )$
gives a contribution to the Higgs' potential resulting in the second
type of critical point for the potential corresponding to the
nontrivial gauge configuration of the Higgs' field. 

Let us consider Eq.~(17.3a) and rewrite it in a more convenient form.
$$\displaylines{
\ell _{dc} (\delta ^{\gamma }_{\beta }-2g_{[\beta \delta ]}
g^{\gamma \delta }) L^{d}{}_{\gamma \tilde{a} } + \ell _{cd} 
( \delta ^{\tilde c }_{\tilde{a} } - 2\zeta k^{o}{}_{\tilde d
\tilde{a} } \widetilde{g}^{\tilde d \tilde c }) L^{d}{}_{\beta \tilde
c } =\hfill\cr
\hfill = 2\ell _{cd} ( \delta ^{\tilde c }_{\tilde{a} } - 2\zeta 
k^{o}{}_{\tilde d \tilde{a} } g^{\tilde d \tilde c } )
\mathop{\nabla_\beta }\limits^{\gauge} {\varPhi} ^{d}_{\tilde c
}.\quad\ (17.21)\cr}
$$
Let us expand $L^{d}{}_{\mu \tilde c }$ into a power series with respect to $\zeta $ and with
respect to $h_{\mu \nu } = g_{\mu \nu } - \eta _{\mu \nu }$
$$\displaylines{
\hfill L^{d}{}_{\mu \tilde c } = L^{\nad{d}{(0)}}{}_{\mu \tilde c } + 
L^{\nad{d}{(1)}}{}_{\mu \tilde c }+ L^{\nad{d}{(0)}}{}_{\mu \tilde c }+\ldots,
\hfill\llap{(17.22)}\cr
g^{\mu \alpha }g_{\nu \alpha } = (\nad{h}{1}{}^{\mu \alpha
}+\nad{h}{1}{}^{\mu \alpha }+\nad{h}{2}{}^{\mu \alpha }+\ldots ).\cr}
$$
Using (17.21), (17.5), (17.6) one gets up to the second order in $\zeta $ and
$h_{\mu \nu }$. 
$$\displaylines{
\ell _{dc} L^{d}{}_{\beta \tilde{a} } + \ell _{cd} L^{d}{}_{\beta
\tilde{a} } - 2\zeta k^{o}{}_{\tilde d \tilde{a} } ( h^{o\tilde d
\tilde c } - \zeta h^{o\tilde d \tilde p} h^{o\tilde c \tilde{b} } 
k^{o}{}_{\tilde{b} \tilde p} 
+ \zeta ^{2} h^{o\tilde c \tilde{b} } h^{o\tilde d \tilde e } 
h^{o\tilde p\tilde s} k^{o}{}_{\tilde s \tilde e } k^{o}{}_{\tilde{b}
\tilde p} ) \ell _{cd} L^{d}{}_{\beta \tilde c } +\cr
- 2h_{[\beta \delta ]} (\eta ^{\gamma \delta }-\eta ^{\gamma
\alpha }\eta ^{\delta \mu }h_{\mu \alpha }+\eta ^{\gamma \mu }\eta
^{\delta \sigma }\eta ^{\alpha \nu }h_{\nu \mu }h_{\delta \alpha }) 
\ell _{dc} L^{d}{}_{\gamma \tilde{a} } =\cr
= 2\ell _{cd} \mathop{\nabla_\beta }\limits^{\gauge}
{\varPhi} ^{d}_{\tilde{a} } - 4\zeta \ell _{cd} k^{o}{}_{\tilde d
\tilde{a} } (h^{o\tilde d \tilde c } - \zeta h^{o\tilde d \tilde p} 
h^{o\tilde c \tilde{b} } k^{o}{}_{\tilde{b} \tilde p} +\cr
\hfill + \xi ^{2} h^{o\tilde c \tilde{b} } h^{o\tilde d \tilde e } 
h^{o\tilde p\tilde s} k^{o}{}_{\tilde s \tilde e } k^{o}{}_{\tilde{b}
\tilde p} )\mathop{\nabla_\beta }\limits^{\gauge} {\varPhi}
^{b}_{\tilde c }.\quad\ (17.23)\cr}
$$
From (17.23) one obtains:
$$\eqalignno{
\nad{L}{(0)}{}^a{}_{\mu \tilde c } ={}& p^{ab} \ell _{bc} 
\mathop{\nabla_\mu}\limits^{\gauge} {\varPhi} ^{c}_{\tilde c } = 
\mathop{\nabla_\mu}\limits^{\gauge} {\varPhi} ^{a}_{\tilde c } + \xi p^{ab} 
k_{bc} \mathop{\nabla_\mu}\limits^{\gauge} {\varPhi} ^{c}_{\tilde c
},&(17.24)\cr
\nad{L}{(1)}{}^1{}_{\beta \tilde{a} } ={}& \zeta k^{o}{}_{\tilde d
\tilde{a} } h^{o\tilde d \tilde c } p^{fc} (\ell _{cd}p^{db}\ell _{bp}-2
\ell _{cp}) \mathop{\nabla_\beta }\limits^{\gauge} {\varPhi}
^{p}_{\tilde c } +\cr
&+ \eta ^{\gamma \delta } h_{[\beta \delta ]} p^{fc} p^{db} \ell _{bp} 
\ell_{cd} \mathop{\nabla_\gamma }\limits^{\gauge} {\varPhi}
^{p}_{\tilde{a} },&(17.25)\cr
\nad{L}{(2)}{}^m{}_{\beta \tilde{a} } ={}& \zeta ^{2} p^{cm}
k^{o}{}_{\tilde d \tilde{a} } \ell _{cd} h^{o\tilde d \tilde p}
h^{o\tilde c \tilde q }[-k^{o}{}_{\tilde q\tilde p} p^{db} \ell _{bf} 
\mathop{\nabla_\beta }\limits^{\gauge} {\varPhi} ^{f}_{\tilde c }+\cr
&+ k^{o}{}_{\tilde c \tilde p} ( p^{dn} \ell _{nq} p^{qp} 
\ell _{bp} \mathop{\nabla_\beta }\limits^{\gauge}{\varPhi}
^{p}_{\tilde{q} } + p^{dq} \ell_{qp} \mathop{\nabla_\beta }\limits^{\gauge}
{\varPhi} ^{p}_{\tilde{q} } + 2 \mathop{\nabla_\beta }\limits^{\gauge}
{\varPhi} ^{d}_{\tilde c } )]+\cr
&+ \zeta p^{cm} \eta ^{\gamma \delta } h_{[\beta \delta ]}
k^{o}{}_{\tilde d \tilde{a} } h^{o\tilde d \tilde c } ( \ell _{cd} p^{nd} 
p^{qb} \ell _{bp} \ell _{nq} +\cr
&+ \ell _{dc} p^{dq} \ell _{qa} p^{ab} \ell _{bp} + p^{dq} \ell _{dc} 
\ell _{qp} ) \mathop{\nabla_\gamma }\limits^{\gauge} {\varPhi}
^{p}_{\tilde{q} } +\cr
&+ h_{[\beta \delta ]} \eta ^{\delta \mu } \eta ^{\gamma \alpha } 
p^{cm} \ell _{dc} \ell _{bp} (h_{\mu \alpha }p^{db}+h_{[\mu \alpha ]}
p^{dq} p^{ab} \ell _{qa}) \mathop{\nabla_\gamma }\limits^{\gauge}
{\varPhi} ^{p}_{\tilde{a} }.&(17.26)\cr}
$$
Let us expand ${\cal L}_{\kin} (\mathop{\nabla}\limits^{\gauge}
{\varPhi})$ into a power series with respect to $\zeta $ and
$h_{\mu \nu }$.
$$
{\cal L}_{\kin} (\mathop{\nabla}\limits^{\gauge} {\varPhi}) =
\nad{{\cal L}}{(0)}{}_{\kin} (\mathop{\nabla}\limits^{\gauge}
{\varPhi}) + \nad{{\cal L}}{(1)}{}_{\kin} (\mathop{\nabla}\limits^{\gauge}
{\varPhi} ) + \nad{{\cal L}}{(2)}{}_{\kin} (\mathop{\nabla}\limits^{\gauge}
{\varPhi} ) + \ldots \eqno(17.27)
$$
Using (17.2a) and (17.24--26) one gets: 
$$\eqalignno{
\nad{{\cal L}}{(0)}{}_{\kin}
(\mathop{\nabla}\limits^{\gauge}{\varPhi}) ={}& {1\over V_{2}}
\mmint_M \sqrt{|h^0|} dm(y) \eta ^{\alpha \mu } h^{0\tilde{b} \tilde
n } (\ell _{ab}p^{af}\ell _{fc})\mathop{\nabla_\alpha }\limits^{\gauge}{\varPhi}
^{c}_{\tilde{b} } \mathop{\nabla_\mu }\limits^{\gauge} {\varPhi}
^{b}_{\tilde n } =\cr
= \mmint_M& [\eta ^{\alpha \mu } h^{o\tilde{b} \tilde n } 
(p_{cb}+\xi ^{2}p^{af}k_{ab}k_{fc}) \mathop{\nabla_\alpha }\limits^{\gauge}
{\varPhi} ^{c}_{\tilde{b} } \mathop{\nabla_\mu }\limits^{\gauge}
{\varPhi} ^{b}_{\tilde n } \sqrt{|h^0|} dm(y),&(17.28)\cr
\nad{{\cal L}}{(1)}{}_{\kin}
(\mathop{\nabla}\limits^{\gauge}{\varPhi} ) ={}& {1\over V_{2}}
\mmint_M\{-\zeta \eta ^{\alpha \mu } (p_{cb}+\xi
^{2}p^{af}k_{ab}k_{fc}) k^{o}{}_{\tilde d \tilde{b} } h^{o\tilde d
\tilde n } h^{o\tilde{b} \tilde m } \times \cr
&\times (\mathop{\nabla_\alpha }\limits^{\gauge}
{\varPhi} ^{c}_{\tilde m } \mathop{\nabla_\mu }\limits^{\gauge}
{\varPhi} ^{b}_{\tilde n } + p^{ce} \ell _{ed} p^{db} \ell _{bp} 
\mathop{\nabla_\alpha }\limits^{\gauge} {\varPhi} ^{p}_{\tilde m } 
\mathop{\nabla_\mu }\limits^{\gauge} {\varPhi} ^{b}_{\tilde n } +\cr
&- 2 (p_{pb}+\xi ^{2}p^{qg}k_{qb}k_{gd}) p^{pi} \ell _{id} 
\mathop{\nabla_\mu }\limits^{\gauge} {\varPhi} ^{b}_{\tilde n } 
\mathop{\nabla_\alpha }\limits^{\gauge}{\varPhi} ^{d}_{\tilde m}]+\cr
&- (p_{cb}+\xi ^{2}p^{af}k_{ab}k_{fc}) \eta ^{\alpha \mu } \eta
^{\gamma \delta } h^{o\tilde{b} \tilde n } \times \cr
&\times [ h_{\delta \alpha } \mathop{\nabla_\mu }\limits^{\gauge}
{\varPhi} ^{c}_{\tilde{b} } \mathop{\nabla_\gamma }\limits^{\gauge}
{\varPhi} ^{b}_{\tilde n } + h_{[\delta \alpha ]} p^{ce} (p_{ep}+\xi ^{2}
p^{df}k_{dp}k_{fe}) \times \cr
&\times \mathop{\nabla_\mu }\limits^{\gauge}
{\varPhi} ^{b}_{\tilde n } \mathop{\nabla_\gamma }\limits^{\gauge}
{\varPhi} ^{p}_{\tilde{b} } ]\}\sqrt{|h^0|} dm(y),&(17.29)\cr
\nad{{\cal L}}{(2)}_{\kin}(\mathop{\nabla}\limits^{\gauge}{\varPhi}) ={}&
\mmint_M{\sqrt{|h^0|}\over V_{2}} dm(y) \times \cr
\noalign{\vskip0pt plus 1pt}
&\times [\zeta ^{2} ( \eta ^{\alpha \mu } \ell _{ab} 
p^{ad} \ell _{dc} h^{o\tilde{b} \tilde c } h^{o\tilde n \tilde s } 
h^{o\tilde{a} \tilde p} k^{o}{}_{\tilde p\tilde c }
k^{o}{}_{\tilde{s} \tilde{a}} \mathop{\nabla_\mu }\limits^{\gauge}
{\varPhi} ^{b}_{\tilde n } \mathop{\nabla_\alpha }\limits^{\gauge}
{\varPhi} ^{c}_{\tilde{b} } +\cr
\noalign{\vskip0pt plus 1pt}
&+ \eta ^{\alpha \mu } h^{o\tilde{b} \tilde r} h^{o\tilde n \tilde s } 
h^{o\tilde d \tilde c } k^{o}{}_{\tilde d \tilde{b} } k^{o}{}_{\tilde
r\tilde s} \ell _{ab} p^{ac} \ell _{cd} p^{dq} \ell _{qp}
\mathop{\nabla_\alpha }\limits^{\gauge}{\varPhi} ^{p}_{\tilde c } 
\mathop{\nabla_\mu }\limits^{\gauge}{\varPhi} ^{b}_{\tilde n } +\cr
\noalign{\vskip0pt plus 1pt}
&+ 2\eta ^{\alpha \mu } h^{o\tilde{b} \tilde r} h^{o\tilde m \tilde s } 
h^{o\tilde d \tilde c } k^{o}{}_{\tilde s\tilde r} k^{o}{}_{\tilde d
\tilde{b} } \ell _{ab} p^{ac} \ell _{cd} \mathop{\nabla_\alpha }\limits^{\gauge}
{\varPhi} ^{d}_{\tilde c } \mathop{\nabla_\mu }\limits^{\gauge}
{\varPhi} ^{b}_{\tilde n } ) +\cr
\noalign{\vskip0pt plus 1pt}
&+ \zeta ( 2\eta ^{\alpha \rho } \eta ^{\mu \beta } h_{\beta \rho } 
h^{o\tilde{b} \tilde n } h^{o\tilde d\tilde c } k^{o}{}_{\tilde d
\tilde{b} } \ell _{ab} p^{ac} \ell _{cd} \mathop{\nabla_\mu }\limits^{\gauge}
{\varPhi} ^{b}_{\tilde n } \mathop{\nabla_\alpha }\limits^{\gauge}
{\varPhi} ^{d}_{\tilde c } +\cr
\noalign{\vskip0pt plus 1pt}
&+ \eta ^{\alpha \rho } \eta ^{\mu \beta } h_{\beta \rho } 
h^{o\tilde{b} \tilde a } h^{o\tilde n\tilde r } k^{o}{}_{\tilde r
\tilde{a} } \ell _{ab} p^{ac} \ell _{cd} \mathop{\nabla_\mu }\limits^{\gauge}
{\varPhi} ^{b}_{\tilde n } \mathop{\nabla_\alpha }\limits^{\gauge}
{\varPhi} ^{c}_{\tilde c } +\cr
\noalign{\vskip0pt plus 1pt}
&- \eta ^{\alpha \rho } \eta ^{\mu \beta } h_{\beta \rho } 
h^{o\tilde{b} \tilde n } h^{o\tilde d\tilde c } k^{o}{}_{\tilde d
\tilde{b} } \ell _{ab} p^{ac} \ell _{cd} \ell_{dq}\mathop{\nabla_\mu }\limits^{\gauge}
{\varPhi} ^{b}_{\tilde n } \mathop{\nabla_\alpha }\limits^{\gauge}
{\varPhi} ^{p}_{\tilde c } +\cr
\noalign{\vskip0pt plus 1pt}
&- \eta ^{\alpha \mu } \eta ^{\gamma \delta } h_{[\alpha \delta ]} 
h^{o\tilde{b} \tilde r } h^{o\tilde m\tilde s } k^{o}{}_{\tilde s
\tilde{r} } \ell _{ab} p^{ac} p^{dq} \ell_{qp}\ell _{cd} 
\mathop{\nabla_\gamma }\limits^{\gauge}
{\varPhi} ^{p}_{\tilde b } \mathop{\nabla_\mu }\limits^{\gauge}
{\varPhi} ^{b}_{\tilde m } +\cr
\noalign{\vskip0pt plus 1pt}
&+ \eta ^{\alpha \mu } \eta ^{\gamma \delta } h_{[\alpha \delta ]} 
h^{o\tilde{b} \tilde n } h^{o\tilde d\tilde c } k^{o}{}_{\tilde d
\tilde{b} } \ell _{ab} p^{ca} \ell _{cd}
p^{nd}p^{qm}\ell_{mp}\ell_{nq}\times\cr
\noalign{\vskip0pt plus 1pt}
&\times \mathop{\nabla_\mu }\limits^{\gauge}
{\varPhi} ^{b}_{\tilde n } \mathop{\nabla_\gamma }\limits^{\gauge}
{\varPhi} ^{p}_{\tilde c } +\cr
\noalign{\vskip0pt plus 1pt}
&+ \eta ^{\alpha \mu } \eta ^{\gamma \delta } h_{[\alpha \delta ]} 
h^{o\tilde{b} \tilde n } h^{o\tilde d\tilde c } k^{o}{}_{\tilde a
\tilde{b} } \ell _{ab} p^{ca} \ell _{dc}
p^{dq}\ell_{qn}p^{nm}\ell_{mp}\times\cr
\noalign{\vskip0pt plus 1pt}
&\times \mathop{\nabla_\mu }\limits^{\gauge}
{\varPhi} ^{b}_{\tilde n } \mathop{\nabla_\gamma }\limits^{\gauge}
{\varPhi} ^{p}_{\tilde c } +\cr
\noalign{\vskip0pt plus 1pt}
&+ \eta ^{\alpha \mu } \eta ^{\gamma \delta } h_{[\alpha \delta ]} 
h^{o\tilde{b} \tilde n } h^{o\tilde d\tilde c } k^{o}{}_{\tilde d
\tilde{b} } \ell _{ab} p^{ca} \ell _{dc}
p^{dq}\ell_{qp} \mathop{\nabla_\mu }\limits^{\gauge}
{\varPhi} ^{b}_{\tilde n } \mathop{\nabla_\gamma }\limits^{\gauge}
{\varPhi} ^{p}_{\tilde c }) +\cr
\noalign{\vskip0pt plus 1pt}
&+ h^{o\tilde n \tilde{b} }(\eta ^{\alpha \mu } \eta^{\delta \nu }
\eta ^{\gamma \rho } h_{[\alpha \delta ]} h_{(\nu \rho )}
\ell _{ab} p^{ca} \ell _{dc}\ell_{np}
p^{dq}p^{nm}\ell_{qm}\times\cr
\noalign{\eject}
&\times \mathop{\nabla_\mu }\limits^{\gauge}
{\varPhi} ^{b}_{\tilde n } \mathop{\nabla_\gamma }\limits^{\gauge}
{\varPhi} ^{p}_{\tilde b } +\cr
\noalign{\vskip0pt plus 1pt}
&+ 2\eta ^{\alpha \mu } \eta ^{\delta \nu } \eta ^{\gamma \rho } 
h_{[\alpha \delta ]} h_{[\nu \rho ]} \ell _{ab} p^{ca} \ell _{dc} 
\ell _{np} p^{dq} p^{mn} \ell _{qm} \times \cr
&\times \mathop{\nabla_\mu }\limits^{\gauge} {\varPhi} ^{b}_{\tilde
n } \mathop{\nabla_\gamma }\limits^{\gauge}{\varPhi} ^{p}_{\tilde{b} }+\cr
\noalign{\vskip0pt plus 1pt}
&+ \eta ^{\alpha \gamma } \eta ^{\mu\sigma } \eta ^{\rho \beta } 
h_{\beta \gamma } h_{\sigma \rho } \ell _{ab} p^{\hbox{ad}} \ell _{dc}
\mathop{\nabla_\mu }\limits^{\gauge} {\varPhi} ^{b}_{\tilde n } 
\mathop{\nabla_\alpha }\limits^{\gauge}{\varPhi} ^{c}_{\tilde c}+\cr
\noalign{\vskip0pt plus 1pt}
&- \eta ^{\alpha \rho } \eta ^{\mu \beta } \eta ^{\gamma \delta } 
h_{\beta \rho } h_{[\gamma \delta ]} \ell _{ab} p^{ac} p^{dq} \ell _{qp} 
\ell _{cd} \mathop{\nabla_\mu }\limits^{\gauge}{\varPhi}
^{b}_{\tilde n } \mathop{\nabla_\alpha }\limits^{\gauge}
{\varPhi} ^{p}_{\tilde{b} })].&(17.30)\cr}
$$
It is easy to see that the skewon field $h_{[\mu \nu ]}$ enters the lagrangian
${\cal L}_{\kin}(\mathop{\nabla}\limits^{\gauge}){\varPhi})$ in the 
first order of approximation similarly as for the
Yang--Mills' field. Moreover it couples here the gauge derivative of
the Higgs' field.

Let us consider ${\cal L}_{\Int}({\varPhi} ,\widetilde{A})$. One gets:
$$
\nad{{\cal L}}{(0)}_{\Int}({\varPhi} ,\widetilde{A}) =
\nad{{\cal L}}{(1)}_{\Int}({\varPhi} ,\widetilde{A}) = 0\eqno(17.31)
$$
and
$$\displaylines{\quad\
\nad{{\cal L}}{(2)}_{\Int}({\varPhi} ,\widetilde{A}) = {\zeta \over V_{2}} 
\mmint_M \sqrt{|h^0|} dm(y) \times \hfill\cr
\hfill \times (p_{ab} \mu ^{a}_{i} \eta ^{\mu \alpha } 
\eta ^{\nu \beta } h_{\beta \alpha } h^{o\tilde m \tilde{a} } 
h^{o\tilde n \tilde{b} } k^{o}{}_{\tilde{b}\tilde{a} } 
\widetilde{H}^{i}_{\mu \nu } H^{b}{}_{\tilde m \tilde n }).\quad\
(17.32)\cr}
$$
The nonminimal interaction term enters from the second order of
approximation. (17.32) gives us an interpretation of a constant $\zeta $.
It plays the role of a coupling constant between the skewon field
$h_{[\mu \nu ]}$
and the Higgs' field. The coupling constant $\zeta $ is simultaneously
connected to a part of the cosmological constant in our theory. Thus
we get an interesting relation between cosmological constant and
coupling constants $\zeta $ and $\xi $ between skewon field
$h_{[\tilde m \nu ]}$ and Higgs' and Yang--Mills' fields.

Thus we get $V({\varPhi} )$, ${\cal L}_{\kin}
(\Nabla^{\gauge}{\varPhi})$, ${\cal L}_{\Int}({\varPhi}
,\widetilde{A})$ up to the second order in
an approximation with respect to $\zeta $ and $h_{\mu \nu } = 
g_{\mu \nu } - \eta _{\tilde m \nu }$. The higher
order terms can be obtained in a similar way.

Let us give a remark concerning a convergence of series appearing
here. They are double power series with respect to $\zeta $ and
$h_{\mu \nu }$ and
they converge for sufficiently small $\zeta $ and $h_{\mu \nu }$. However all the
functions of $\zeta $ and $h_{\mu \nu }$ considered here (i.e. 
$L^{a}{}_{\tilde{b} \tilde c }$, $L^{a}{}_{\alpha \tilde{a} }$, 
$g^{\tilde{a} \tilde{b} }$, $g^{\mu \nu }$,
${\cal L}_{\kin}(\Nabla^{\gauge}{\varPhi})$, $V({\varPsi} )$, 
${\cal L}_{\Int}({\varPhi} ,\widetilde{A}))$ are well defined for any $\zeta $ and
$h_{\mu \nu }$.
They are rational functions of these variables. Moreover the exact
form of all of these functions is hard to get.

Let us consider the terms involving Higgs' field in the geodetic
equation in the first order of approximation with respect to $\zeta $ and
$h_{\mu \nu }$
i.e.\ the fourth term in Eq.~(15.2) and Eq.~(15.3). One gets:
$$\displaylines{
\bigg({q^{c}\over m_{o}}\bigg) u^{\tilde{b} }[\eta ^{\alpha \delta } 
\ell _{cd} \mathop{\nabla_\delta }^{\gauge}{\varPhi} ^{d}_{\tilde{b} } - \xi 
\eta ^{\alpha \delta } k_{cd} p^{db} \ell _{bf}\mathop{\nabla_\delta }^{\gauge}
{\varPhi} ^{f}_{\tilde{b} } +\hfill\cr
- \zeta \xi \eta ^{\alpha \delta } k_{cd} k^{o}{}_{\tilde d
\tilde{b} } h^{o\tilde d \tilde c } p^{df} (\ell _{fe}p^{eq}\ell _{qp}-
2\ell _{fp}) \mathop{\nabla_\delta }^{\gauge} {\varPhi} ^{p}_{\tilde c}+\cr
 - \eta ^{\alpha \rho } \eta ^{\delta \beta } h_{\beta \rho } 
\ell _{cd} \mathop{\nabla_\delta }^{\gauge} {\varPhi}
^{d}_{\tilde{b} } - {1\over 2} (\ell _{cd}\eta ^{\alpha \delta }
\eta ^{\rho \beta }-\ell _{dc}\eta ^{\delta \rho }\eta ^{\alpha \beta
}) \times \cr
\hfill\times h_{\beta \rho } p^{db} \ell _{bf} \mathop{\nabla_\delta }^{\gauge}
{\varPhi} ^{f}_{\tilde{b} } - \xi \eta ^{\alpha \delta } 
\eta ^{\gamma \rho } h_{[\delta \rho ]} k_{cd} p^{df} p^{eb} \ell _{fe} 
\ell _{bp} \mathop{\nabla_\gamma }^{\gauge}{\varPhi}
^{p}_{\tilde{b}}].\quad\ (17.33)\cr}
$$
This is the fourth term in Eq.~(15.2) i.e.\ the term involving the
Higgs' field in the fist order of approximation. Let us write the
equation for the Higgs' charge in the first order of approximation.
One easily finds: 
$$\displaylines{
{\widetilde{D}u^{\tilde{a} }\over d\tau} + {1\over r^{2}} \bigg({q^{c}\over 
m_{0}}\bigg) u^{\beta }\{\ell _{cd} h^{o\tilde{a} \tilde d }
\mathop{\nabla_{\beta }}\limits^{\gauge} {\varPhi} ^{d}_{\tilde d } - 
\xi k_{cd} h^{o\tilde{a} \tilde d } p^{db} \ell _{bp} 
\mathop{\nabla_{\beta }}\limits^{\gauge}{\varPhi} ^{p}_{\tilde d }
+\hfill\phantom{\}}\hfill\cr
\noalign{\vskip1pt plus 1pt}
\hfill\eqalign{&+ {1\over 2} \zeta [( \ell _{cd} h^{o\tilde{a} \tilde n } h^{o\tilde
d \tilde m } - \ell _{dc} h^{o\tilde d \tilde n } h^{o\tilde{a}
\tilde m } ) k^{o}{}_{\tilde m \tilde n } p^{db} \ell _{bp} 
\mathop{\nabla_{\beta }}\limits^{\gauge}{\varPhi} ^{p}_{\tilde d }+\cr
\noalign{\vskip0pt plus 1pt}
&- 2 h^{o\tilde{a} \tilde n } h^{o\tilde d \tilde m } k^{o}{}_{\tilde
m \tilde n } \mathop{\nabla_{\beta }}\limits^{\gauge}{\varPhi}
^{d}_{\tilde d }]+\cr
\noalign{\vskip0pt plus 1pt}
&- \zeta \xi k_{cd} h^{o\tilde{a} \tilde d } k^{o}{}_{\tilde f\tilde
d } h^{o\tilde f\tilde c } p^{fe} (\ell _{ep}p^{pr}\ell _{rq}-2\ell
_{eq}) \mathop{\nabla_{\beta }}\limits^{\gauge}{\varPhi}
^{q}_{\tilde c } +\cr
\noalign{\vskip0pt plus 1pt}
&\phantom{\{}- \xi \eta ^{\gamma \delta } h_{[\beta \delta ]} k_{cd} p^{dr} p^{sp} 
\ell _{pf} \ell _{rs} \mathop{\nabla_{\beta }}\limits^{\gauge}{\varPhi}
^{f}_{\tilde d }\}+\cr
\noalign{\vskip1pt plus 1pt}
&+ {1\over r^{2}} \bigg({q^{c}\over m_{o}}\bigg) u^{\tilde{b} }\{\ell _{cd} 
h^{o\tilde{a} \tilde d} \mathop{\nabla_{\beta }}\limits^{\gauge}
H^{d}_{\tilde d \tilde{b} } - \xi k_{cd} h^{o\tilde{a} \tilde d } p^{df} 
\ell _{fe} H^{e}_{\tilde{b} \tilde d } +\cr
\noalign{\vskip1pt plus 1pt}
&+ {1\over 2} \zeta [(\ell _{cd} h^{o\tilde{a} \tilde n } h^{o\tilde d
\tilde m } - \ell _{dc} h^{o\tilde d \tilde n } h^{o\tilde{a} \tilde
m } ) k^{o}{}_{\tilde m \tilde n } p^{df} \ell _{fe} H^{e}_{\tilde{b}
\tilde d } +\cr
\noalign{\vskip1pt plus 1pt}
&- 2 h^{o\tilde{a} \tilde n } h^{o\tilde d \tilde m } k^{o}{}_{\tilde
m \tilde n } H^{d}_{\tilde d \tilde{b} }]+\cr
\noalign{\vskip1pt plus 1pt}
&- \zeta \xi k_{cd} h^{o\tilde{a} \tilde d }(h^{o\tilde f\tilde c } 
p^{dp} p^{bq} \ell _{be} ( k^{o}{}_{\tilde f\tilde d } \ell _{pq} 
H^{e}_{\tilde{b} \tilde c } + k^{o}{}_{\tilde{b} \tilde f} \ell _{qp} 
H^{e}_{\tilde c \tilde d }) +\cr
\noalign{\vskip1pt plus 1pt}
&- 2 k^{o}{}_{\tilde e \tilde d } h^{o\tilde e \tilde c } p^{dp} \ell _{pq} 
H^{q}_{\tilde{b} \tilde c }))\}.\cr}\quad\ (17.34)\cr}
$$
Eventually one easily writes down the explicit form of the mass matrix
for massive vector bosons (i.e. $M^{2}{}_{ij}({\varPhi} ^{k}_{\crt})$ from
(8.21--22) in zeroth
order of expansion with respect~$\zeta $.
$$\displaylines{
\nad{M}{(0)}{\!}^{2}{}_{ij}({\varPhi} ^{k}_{\crt }) = 
- \bigg( {\alpha _{s}\over r} \bigg)^{2}\bigg({\hbar\over c}\bigg)^{2} 
h^{o\tilde{b} \tilde n }(\ell _{ab}p^{af}\ell _{fc}) \cdot \hfill\cr
\noalign{\vskip2pt}
\hfill \cdot (({\varPhi} ^{k}_{\crt })^{p}_{\tilde{b} }C^{c}_{pq}
\alpha ^{q}_{j}+({\varPhi} ^{k}_{\crt })^{c}_{\tilde p}f^{\tilde p
}_{\tilde b j})\cdot (({\varPhi} ^{k}_{\crt })^{p}_{\tilde{b} } C^{b}_{ms}
\alpha ^{s}_{i} + ({\varPhi} ^{k}_{\crt})^b_{\tilde m }f^{\tilde m
}_{\tilde n i}).\quad\ (17.35)\cr}
$$
The next terms i.e. the first and the second can be obtained similarly
from Eq.~(17.28--30). Thus we get an expansion with respect to $\zeta $
$$
M^{2}{}_{ij}({\varPhi} ^{k}_{\crt }) = \nad{M}{(0)}{\!}^{2}{}_{ij}
({\varPhi} ^{k}_{\crt }) + \nad{M}{(1)}{\!}^{2}{}_{ij}
({\varPhi} ^{k}_{\crt }) + \nad{M}{(2)}{\!}^{2}{}_{ij}({\varPhi} ^{k}_{\crt })
+ \ldots ,\eqno(17.36)
$$
where 
$$
\eqalignno{\nad{M}{(1)}{\!}^{2}{}_{ij}({\varPhi}^k_{\crt})={}&
\zeta \mmint_M \bigg({\alpha _s\over r}\bigg)^2\bigg({\hbar \over
c}\bigg)^2[(p_{cb}+\xi ^2p^{af}k_{ab}k_{fc})\cdot k^o{}_{\tilde d
\tilde{b} }h^{o\tilde d \tilde n }h^{o\tilde{b} \tilde m }\cdot\cr
&\cdot (({\varPhi}^k_{\crt})^p_{\tilde m }C^c_{pq} \alpha ^q_j +
({\varPhi}^k_{\crt})^c_{\tilde p}f^{\tilde p}_{\tilde m j})\cdot 
(({\varPhi}^k_{\crt})^s_{\tilde n }C^c_{sr} \alpha ^r_i +
({\varPhi}^k_{\crt})^b_{\tilde q}f^{\tilde q}_{\tilde n i})+\cr
&+p^{ce}\ell_{ed}p^{db}\ell_{bp}\cdot\cr
&\cdot (({\varPhi}^k_{\crt})^u_{\tilde m }C^p_{uw} \alpha ^w_j +
({\varPhi}^k_{\crt})^p_{\tilde u}f^{\tilde u}_{\tilde m j})\cdot 
(({\varPhi}^k_{\crt})^v_{\tilde n }C^b_{vz} \alpha ^z_i +
({\varPhi}^k_{\crt})^b_{\tilde v}f^{\tilde v}_{\tilde n i})+\cr
&-2(p_{pb}+\zeta ^2 p^{qg}k_{qb}h_{gp})p^{pl}\ell_{ld}\cdot\cr
&\cdot (({\varPhi}^k_{\crt})^w_{\tilde n }C^b_{wq} \alpha ^q_j +
({\varPhi}^k_{\crt})^b_{\tilde v}f^{\tilde v}_{\tilde n j})\cdot 
(({\varPhi}^k_{\crt})^z_{\tilde n }C^d_{zv} \alpha ^v_i +
({\varPhi}^k_{\crt})^d_{\tilde p}f^{\tilde p}_{\tilde n
i})]\cdot\cr
&\cdot {\sqrt{|h^0|}\over V_2}\,dm (y),& (17.37)\cr
\nad{M}{(2)}{\!}^{2}{}_{ij}({\varPhi}^k_{\crt})={}&
\mmint_M\zeta ^2\bigg({\alpha _s\over r}\bigg)^2\bigg({\hbar \over
c}\bigg)^2[(\ell_{ab}+ p^{ad}\ell_{dc} h^{o\tilde b \tilde c }
h^{o\tilde{n} \tilde s }h^{o\tilde{a} \tilde p}k^o{}_{\tilde p\tilde
c }k^o{}_{\tilde{s} \tilde{a}} \cdot\cr
&\cdot (({\varPhi}^k_{\crt})^u_{\tilde n }C^b_{uq} \alpha ^q_j +
({\varPhi}^k_{\crt})^b_{\tilde m}f^{\tilde m}_{\tilde n j})\cdot 
(({\varPhi}^k_{\crt})^v_{\tilde b }C^c_{vw} \alpha ^w_i +
({\varPhi}^k_{\crt})^c_{\tilde q}f^{\tilde q}_{\tilde b i})+\cr
&+h^{o\tilde b \tilde r }
h^{o\tilde{n} \tilde s }h^{o\tilde{d} \tilde c}k^o{}_{\tilde d\tilde
b }k^o{}_{\tilde{r} \tilde{s}} \ell_{ab}p^{ce}\ell_{cd}h^{dq}\ell_{qp}\cdot\cr
&\cdot (({\varPhi}^k_{\crt})^r_{\tilde c }C^p_{rf} \alpha ^f_j +
({\varPhi}^k_{\crt})^p_{\tilde q}f^{\tilde q}_{\tilde c j})\cdot 
(({\varPhi}^k_{\crt})^s_{\tilde n }C^b_{sg} \alpha ^g_i +
({\varPhi}^k_{\crt})^b_{\tilde m}f^{\tilde m}_{\tilde n i})+\cr
&-2 h^{o\tilde b \tilde r }
h^{o\tilde{m} \tilde s }h^{o\tilde{d} \tilde c}k^o{}_{\tilde s\tilde
r }k^o{}_{\tilde{d} \tilde{b}}\ell_{ab}p^{ac}\ell_{cd}\cdot\cr
&\cdot (({\varPhi}^k_{\crt})^q_{\tilde c }C^d_{qp} \alpha ^p_j +
({\varPhi}^k_{\crt})^d_{\tilde m}f^{\tilde m}_{\tilde c j})\cdot 
(({\varPhi}^k_{\crt})^e_{\tilde n }C^b_{ef} \alpha ^f_i +
({\varPhi}^k_{\crt})^b_{\tilde p}f^{\tilde p}_{\tilde n i})]\cdot\cr
&\cdot {\sqrt{|h^0|}\over V_2}\,dm (y).&(17.38)\cr}
$$
After a diagonalization procedure for $M^{2}{}_{ij}({\varPhi} ^{k}_{\crt })$ we get a mass spectrum
for vector bosons. For $k=0$ in the ``true" vacuum case and for $k=1$ for
the ``false" vacuum case.
      
Let us consider the mass matrix for Higgs' particles (see
sec.~8)
$$
m^2({\varPhi}^k_{\crt})^{\tilde{a} }{}_a{}^{\tilde r}{}_r = {1\over
4}\bigg({\delta ^2V\over \delta {\varPhi}^a_{\tilde{a} }\delta
{\varPhi}^r_{\tilde
r}}\bigg|_{{\varPhi}={\varPhi}^k_{\crt}}\bigg)\eqno(17.39)
$$
and expand it in a power series with respect to $\zeta $. One gets
$$
m^2({\varPhi}^k_{\crt})^{\tilde{a} }{}_a{}^{\tilde r}{}_r =
\nad{m}{(0)}{\!}^2({\varPhi}^k_{\crt})^{\tilde{a} }{}_a{}^{\tilde
r}{}_r+\nad{m}{(1)}{\!}^2({\varPhi}^k_{\crt})^{\tilde{a} }{}_a{}^{\tilde
r}{}_r+\nad{m}{(2)}{\!}^2({\varPhi}^k_{\crt})^{\tilde{a} }{}_a{}^{\tilde
r}{}_r+\ldots \eqno(17.40)
$$
One finds
$$\eqalignno{
\nad{m}{(0)}{}^2({\varPhi}^k_{\crt})^{\tilde{a} }{}_a{}^{\tilde
r}{}_r={}&{1\over r^4}\bigg\{(p_{cd}+\xi ^2p^{rs}k_{cr}k_{sd})\cdot
h^{o\tilde n \tilde p}h^{\tilde m \tilde q}\cdot\cr
\noalign{\vskip0pt plus 1pt}
&\cdot \bigg({\alpha _s\over \sqrt{\hbar c}} C^c_{av}(\delta
^{\tilde{a} }_{\tilde n }[{\varPhi}^k_{\crt}]^v_{\tilde m
}-[{\varPhi}^k_{\crt}]^v_{\tilde n }\delta ^{\tilde{a} }_{\tilde m
})-f^{\tilde{a} }_{\tilde n \tilde m }\delta ^c_a\bigg)\cdot\cr
\noalign{\vskip0pt plus 1pt}
&\cdot\bigg({\alpha _s\over \sqrt{\hbar c}} C^d_{ru}(\delta
^{\tilde{r} }_{\tilde p}[{\varPhi}^k_{\crt}]^u_{\tilde q
}-[{\varPhi}^k_{\crt}]^u_{\tilde p}\delta ^{\tilde{r} }_{\tilde q
})-f^{\tilde{r} }_{\tilde p \tilde q }\delta ^u_d\bigg)+\cr
\noalign{\vskip0pt plus 1pt}
&+{2\alpha _s\over \sqrt{\hbar c}} C^c_{ar}(\delta ^{\tilde{a}}_{\tilde n }
\delta ^{\tilde{r}}_{\tilde m }-\delta ^{\tilde{a}}_{\tilde m }\delta
^{\tilde{r}}_{\tilde n })(p_{cd}+\xi ^2p^{rs}k_{cr}k_{sd})h^{o\tilde
p\tilde n }h^{o\tilde q\tilde m }\cdot\cr
\noalign{\vskip0pt plus 1pt}
&\cdot \bigg({\alpha _s\over \sqrt{\hbar c}} C^b_{zx}
[{\varPhi}^k_{\crt}]^z_{\tilde p}[{\varPhi}^k_{\crt}]^x_{\tilde q}
-[{\varPhi}^{\tilde k}_{\crt}]^b_{\tilde z } f^{\tilde{z} }_{\tilde p \tilde q }
-{\sqrt{\hbar c}\over \alpha _s}\mu _{\dah{i}}^b f^{\dah{i}}_{\tilde
p\tilde q}\bigg)\bigg\},&(17.41)\cr
\noalign{\vskip0pt plus 1pt}
\nad{m}{(1)}{}^2({\varPhi}^k_{\crt})^{\tilde{a} }{}_a{}^{\tilde r}{}_r={}&
{\zeta \xi \over r^4}\mmint_M\bigg\{\bigg({\alpha _s\over \sqrt{\hbar c}} 
C^c_{av}(\delta^{\tilde{a} }_{\tilde n }[{\varPhi}^k_{\crt}]^v_{\tilde m
}-[{\varPhi}^k_{\crt}]^v_{\tilde n }\delta ^{\tilde{a} }_{\tilde m
})-f^{\tilde{a} }_{\tilde n \tilde m }\delta ^c_a\bigg)\cdot\cr
\noalign{\vskip0pt plus 1pt}
&\cdot\bigg({\alpha _s\over \sqrt{\hbar c}} C^d_{ru}(\delta
^{\tilde{r} }_{\tilde p}[{\varPhi}^k_{\crt}]^u_{\tilde q
}-[{\varPhi}^k_{\crt}]^u_{\tilde p}\delta ^{\tilde{r} }_{\tilde q
})-f^{\tilde{r} }_{\tilde p \tilde q }\delta ^u_d\bigg)\cdot\cr
\noalign{\vskip0pt plus 1pt}
&\cdot h^{o\tilde n \tilde s}h^{o\tilde p\tilde t}h^{o\tilde n \tilde
q}p_{cd}p^{ar}(k_{rs}+\xi k_{rp}p^{pq}k_{qs})k^o{}_{\tilde t\tilde s}+\cr
&+{\alpha _s\over \sqrt{\hbar c}} C^c_{ar}(\delta ^{\tilde{a}}_{\tilde n }
\delta ^{\tilde{r}}_{\tilde m }-\delta ^{\tilde{a}}_{\tilde m }\delta
^{\tilde{r}}_{\tilde n })p^{ar}k^o{}_{\tilde t\tilde s}\cdot\cr
\noalign{\vskip0pt plus 1pt}
&\cdot [\ell_{bc} h^{o\tilde a \tilde m}h^{o\tilde b\tilde s}
h^{o\tilde n \tilde t}(k_{rs}+\xi k_{rp}p^{pq}k_{qs})+\cr
\noalign{\vskip0pt plus 1pt}
&+\ell_{ab}h^{o\tilde m \tilde s}h^{o\tilde b\tilde t}
h^{o\tilde n \tilde a}(k_{rc}+\xi k_{rp}p^{pq}k_{qc})]\cdot\cr
\noalign{\vskip0pt plus 1pt}
&\cdot \bigg({\alpha _s\over \sqrt{\hbar c}} C^b_{zx}
[{\varPhi}^k_{\crt}]^z_{\tilde b}[{\varPhi}^k_{\crt}]^x_{\tilde a}
-[{\varPhi}^{\tilde k}_{\crt}]^b_{\tilde z } f^{\tilde{z} }_{\tilde
b \tilde a } -{\sqrt{\hbar c}\over \alpha _s}\mu _{\dah{i}}^b 
f^{\dah{i}}_{\tilde p\tilde q}\bigg)\bigg\}\cdot\cr
\noalign{\vskip0pt plus 1pt}
&\cdot {\sqrt{|h^0|}\over V_2}dm(y), &(17.42)\cr
\noalign{\vskip0pt plus 1pt}
\nad{m}{(2)}{}^2({\varPhi}^k_{\crt})_a^{\tilde{a}}{}_r{}^{\tilde r}={}&
-\mmint_M{\zeta ^2 \over r^4}\bigg\{{\alpha _s\over \sqrt{\hbar c}} 
C^c_{av}(\delta^{\tilde{a} }_{\tilde n }[{\varPhi}^k_{\crt}]^v_{\tilde m
}-[{\varPhi}^k_{\crt}]^v_{\tilde n }\delta ^{\tilde{a} }_{\tilde m
})-f^{\tilde{a} }_{\tilde n \tilde m }\delta ^c_a\bigg)\cdot\cr
&\cdot\bigg(\alpha _s\sqrt{{\hbar\over c}} C^d_{ru}(\delta
^{\tilde{r} }_{\tilde p}[{\varPhi}^k_{\crt}]^u_{\tilde q
}-[{\varPhi}^k_{\crt}]^u_{\tilde p}\delta ^{\tilde{r} }_{\tilde q
})-f^{\tilde{r} }_{\tilde p \tilde q }\delta ^d_r\bigg)\cdot\cr
\noalign{\vskip0pt plus 1pt}
&\cdot \ell_{ac}k^o{}_{\tilde t\tilde s}k^o{}_{\tilde d \tilde e
}\cdot\cr
&\cdot \bigg[(2h^{o\tilde p\tilde q}h^{o\tilde m \tilde t}h^{o\tilde s \tilde d}
h^{o\tilde n \tilde c }+h^{o\tilde p\tilde s}h^{o\tilde m \tilde
t}h^{o\tilde q\tilde e }h^{o\tilde n \tilde d })k^a{}_{+d}+\cr
\noalign{\vskip0pt plus 1pt}
&+(h^{o\tilde c\tilde s}h^{o\tilde m \tilde t}h^{o\tilde q \tilde n}
h^{o\tilde p \tilde d}-h^{o\tilde d\tilde s}h^{o\tilde m \tilde
t}h^{o\tilde q\tilde n }h^{o\tilde p \tilde c }+\cr
\noalign{\vskip0pt plus 1pt}
&+h^{o\tilde q\tilde s}h^{o\tilde m \tilde
t}h^{o\tilde d\tilde n }h^{o\tilde p \tilde c })k^a{}_{+d}k^s{}_{-d}+\cr
\noalign{\vskip0pt plus 1pt}
&+{\xi \over 2}(h^{o\tilde q\tilde m}h^{o\tilde t \tilde n}
h^{o\tilde d \tilde p}h^{o\tilde e \tilde s}+h^{o\tilde q\tilde m}
h^{o\tilde t \tilde n}h^{o\tilde e\tilde p }h^{o\tilde d \tilde s }+\cr
\noalign{\eject}
&-h^{o\tilde q\tilde m}h^{o\tilde s \tilde n}h^{o\tilde d\tilde p}
h^{o\tilde e \tilde t}-h^{o\tilde q\tilde m}h^{o\tilde s \tilde n}
h^{o\tilde e\tilde p}h^{o\tilde d \tilde t})k^a{}_{+s}k^s{}_{-d}+\cr
\noalign{\vskip0pt plus 1pt}
&+{\xi \over 2}(h^{o\tilde q\tilde m}h^{o\tilde t \tilde n}
h^{o\tilde d \tilde p}h^{o\tilde e \tilde s}+h^{o\tilde q\tilde m}
h^{o\tilde t \tilde n}h^{o\tilde e\tilde p }h^{o\tilde d \tilde s }+\cr
\noalign{\vskip0pt plus 1pt}
&+h^{o\tilde q\tilde m}h^{o\tilde s \tilde n}h^{o\tilde d\tilde p}
h^{o\tilde e \tilde t}-h^{o\tilde q\tilde m}h^{o\tilde s \tilde n}
h^{o\tilde e\tilde p}h^{o\tilde d \tilde t})(k^a{}_{+s}k^s{}_{-d})p^{pq}
k_{qd}+\cr
\noalign{\vskip0pt plus 1pt}
&-{\xi \over 2}(h^{o\tilde m\tilde s}h^{o\tilde n \tilde t}
h^{o\tilde p \tilde p}h^{o\tilde q \tilde d}-h^{o\tilde m\tilde t}
h^{o\tilde n \tilde s}h^{o\tilde p\tilde e}h^{o\tilde q \tilde d}+\cr
\noalign{\vskip0pt plus 1pt}
&-h^{o\tilde m\tilde s}h^{o\tilde n \tilde t}h^{o\tilde p\tilde e}
h^{o\tilde q \tilde e}+h^{o\tilde m\tilde t}h^{o\tilde n \tilde s}
h^{o\tilde p\tilde d}h^{o\tilde q \tilde e})\delta ^a_d\bigg]+\cr
\noalign{\vskip0pt plus 1pt}
&+{\alpha _s\over \sqrt{\hbar c}} C^c_{ar}(\delta ^{\tilde{a}}_{\tilde n }
\delta ^{\tilde{r}}_{\tilde m }-\delta ^{\tilde{a}}_{\tilde m }\delta
^{\tilde{r}}_{\tilde n })k^o{}_{\tilde t\tilde s}k^o{}_{\tilde d
\tilde e }\cdot\cr
\noalign{\vskip0pt plus 1pt}
&\cdot \bigg\{\bigg[(2h^{o\tilde b\tilde c}h^{o\tilde m \tilde t}
h^{o\tilde s \tilde d}h^{o\tilde n \tilde c }+h^{o\tilde b\tilde s}
h^{o\tilde m \tilde t}h^{o\tilde c\tilde e }h^{o\tilde n \tilde d})k^a{}_{+b}+\cr
&+(h^{o\tilde e\tilde s}h^{o\tilde m \tilde t}
h^{o\tilde c \tilde n}h^{o\tilde b \tilde d}-h^{o\tilde d\tilde s}
h^{o\tilde m \tilde t}h^{o\tilde c\tilde n }h^{o\tilde b \tilde e }+\cr
\noalign{\vskip0pt plus 1pt}
&+h^{o\tilde c\tilde s}h^{o\tilde m \tilde t}h^{o\tilde e\tilde n}
h^{o\tilde b \tilde d}-h^{o\tilde c\tilde s}h^{o\tilde m \tilde t}
h^{o\tilde d\tilde n}h^{o\tilde b \tilde e})k^a{}_{+r}k^r{}_{-b}+\cr
\noalign{\vskip0pt plus 1pt}
&+{\xi \over 2}(h^{o\tilde c\tilde m}h^{o\tilde t \tilde n}
h^{o\tilde d \tilde b}h^{o\tilde e \tilde s}+h^{o\tilde c\tilde m}
h^{o\tilde t \tilde n}h^{o\tilde e\tilde b }h^{o\tilde d \tilde s }+\cr
\noalign{\vskip0pt plus 1pt}
&-h^{o\tilde c\tilde m}h^{o\tilde s \tilde n}h^{o\tilde d\tilde b}
h^{o\tilde e \tilde t}-h^{o\tilde c\tilde m}h^{o\tilde s \tilde n}
h^{o\tilde e\tilde b}h^{o\tilde d \tilde t})k^a{}_{+r}k^r{}_{-b}+\cr
\noalign{\vskip0pt plus 1pt}
&+{\xi \over 2}(h^{o\tilde c\tilde m}h^{o\tilde t \tilde n}
h^{o\tilde d \tilde b}h^{o\tilde e \tilde s}+h^{o\tilde c\tilde m}
h^{o\tilde t \tilde n}h^{o\tilde e\tilde b }h^{o\tilde d \tilde s }+\cr
\noalign{\vskip0pt plus 1pt}
&-h^{o\tilde c\tilde m}h^{o\tilde s \tilde n}h^{o\tilde d\tilde b}
h^{o\tilde e \tilde t}-h^{o\tilde c\tilde m}h^{o\tilde s \tilde n}
h^{o\tilde e\tilde b}h^{o\tilde d \tilde t})(k^a{}_{+s}k^r{}_{-p})p^{pq}
k_{qd}+\cr
\noalign{\vskip0pt plus 1pt}
&+{\xi \over 2}(h^{o\tilde c\tilde d}h^{o\tilde t \tilde b}
h^{o\tilde s \tilde m}h^{o\tilde e \tilde n}-h^{o\tilde c\tilde d}
h^{o\tilde s \tilde b}h^{o\tilde t\tilde m}h^{o\tilde e \tilde n }+\cr
\noalign{\vskip0pt plus 1pt}
&+h^{o\tilde c\tilde e}h^{o\tilde t \tilde b}h^{o\tilde s\tilde m}
h^{o\tilde d \tilde n}+h^{o\tilde c\tilde e}h^{o\tilde s \tilde b}
h^{o\tilde t\tilde m}h^{o\tilde d \tilde n})(k^a{}_{+r}k^s{}_{-d})p^{pq}
k_{qb}+\cr
\noalign{\vskip0pt plus 1pt}
&-{\xi \over 2}(h^{o\tilde m\tilde s}h^{o\tilde n \tilde t}
h^{o\tilde b \tilde s}h^{o\tilde c \tilde d}-h^{o\tilde m\tilde t}
h^{o\tilde n \tilde s}h^{o\tilde b\tilde e}h^{o\tilde c \tilde d}+\cr
\noalign{\vskip0pt plus 1pt}
&-h^{o\tilde m\tilde s}h^{o\tilde n \tilde t}h^{o\tilde b\tilde d}
h^{o\tilde c \tilde e}-h^{o\tilde m\tilde t}h^{o\tilde n \tilde s}
h^{o\tilde b\tilde d}h^{o\tilde c \tilde e})\delta
^a_d\bigg]\ell_{ac}+\cr
\noalign{\vskip0pt plus 1pt}
&-\ell_{ab}((2h^{o\tilde h\tilde e}h^{o\tilde c \tilde t}h^{o\tilde s \tilde d}
h^{o\tilde b \tilde m}+h^{o\tilde n\tilde s}h^{o\tilde c \tilde
t}h^{o\tilde m\tilde e }h^{o\tilde b \tilde d })k^a{}_{+c}+\cr
\noalign{\vskip0pt plus 1pt}
&+(h^{o\tilde e\tilde s}h^{o\tilde c \tilde t}h^{o\tilde m \tilde b}
h^{o\tilde n \tilde d}-h^{o\tilde d\tilde s}h^{o\tilde c \tilde
t}h^{o\tilde m\tilde b }h^{o\tilde n \tilde e }+\cr
\noalign{\vskip0pt plus 1pt}
&+(h^{o\tilde e\tilde s}h^{o\tilde c \tilde t}
h^{o\tilde m \tilde b}h^{o\tilde n \tilde d}-h^{o\tilde d\tilde s}
h^{o\tilde c \tilde t}h^{o\tilde m\tilde b}h^{o\tilde n \tilde e }+\cr
\noalign{\vskip0pt plus 1pt}
&+h^{o\tilde m\tilde s}h^{o\tilde c \tilde t}h^{o\tilde e\tilde b}
h^{o\tilde n \tilde d}-h^{o\tilde m\tilde s}h^{o\tilde c \tilde t}
h^{o\tilde d\tilde b}h^{o\tilde n \tilde e})k^a{}_{+r}k^r{}_{-c}+\cr
\noalign{\vskip0pt plus 1pt}
&+{\xi \over 2}(h^{o\tilde c\tilde m}h^{o\tilde t \tilde b}
h^{o\tilde d \tilde n}h^{o\tilde e \tilde b}+h^{o\tilde c\tilde m}
h^{o\tilde t \tilde b}h^{o\tilde d\tilde n }h^{o\tilde e \tilde s }+\cr
\noalign{\vskip0pt plus 1pt}
&-h^{o\tilde c\tilde m}h^{o\tilde s \tilde b}h^{o\tilde d\tilde n}
h^{o\tilde e \tilde t}-h^{o\tilde c\tilde m}h^{o\tilde s \tilde b}
h^{o\tilde e\tilde n}h^{o\tilde d \tilde t})k^a{}_{+r}k^r{}_{-c}+\cr
\noalign{\vskip0pt plus 1pt}
&+{\xi \over 2}(-h^{o\tilde m\tilde d}h^{o\tilde t \tilde n}
h^{o\tilde s \tilde c}h^{o\tilde e \tilde b}+h^{o\tilde m\tilde d}
h^{o\tilde s \tilde n}h^{o\tilde t\tilde c}h^{o\tilde e \tilde b }+\cr
\noalign{\eject}
&-h^{o\tilde m\tilde e}h^{o\tilde t \tilde n}h^{o\tilde s\tilde c}
h^{o\tilde d \tilde b}+h^{o\tilde m\tilde e}h^{o\tilde s \tilde n}
h^{o\tilde t\tilde c}h^{o\tilde d \tilde b})(k^a{}_{+s}k^r{}_{-p})p^{pq}
k_{qc}+\cr
\noalign{\vskip0pt plus 1pt}
&+{\xi \over 2}(-h^{o\tilde c\tilde m}h^{o\tilde t \tilde b}
h^{o\tilde d \tilde n}h^{o\tilde e \tilde s}+h^{o\tilde c\tilde m}
h^{o\tilde t \tilde b}h^{o\tilde e\tilde n}h^{o\tilde d \tilde s }+\cr
\noalign{\vskip0pt plus 1pt}
&+h^{o\tilde c\tilde m}h^{o\tilde s \tilde b}h^{o\tilde d\tilde n}
h^{o\tilde s \tilde t}-h^{o\tilde c\tilde m}h^{o\tilde s \tilde b}
h^{o\tilde e\tilde n}h^{o\tilde d \tilde t})(k^a{}_{+r}k^r{}_{-p})p^{pq}
k_{qc}+\cr
\noalign{\vskip0pt plus 1pt}
&-{\xi \over 2}(h^{o\tilde c\tilde s}h^{o\tilde b \tilde t}
h^{o\tilde n \tilde s}h^{o\tilde m \tilde d}-h^{o\tilde c\tilde t}
h^{o\tilde b \tilde s}h^{o\tilde n\tilde e}h^{o\tilde m \tilde d}+\cr
\noalign{\vskip0pt plus 1pt}
&-h^{o\tilde c\tilde s}h^{o\tilde b \tilde t}h^{o\tilde n\tilde d}
h^{o\tilde m \tilde e}+h^{o\tilde c\tilde t}h^{o\tilde b \tilde s}
h^{o\tilde n\tilde d}h^{o\tilde m \tilde e})\delta ^d_c\bigg)\cr
\noalign{\vskip0pt plus 1pt}
&\cdot \bigg({\alpha _s\over \sqrt{\hbar c}} C^b_{uv}
[{\varPhi}^k_{\crt}]^z_{\tilde c}[{\varPhi}^k_{\crt}]^v_{\tilde b}
-[{\varPhi}^{\tilde k}_{\crt}]^b_{\tilde z } f^{\tilde{z} }_{\tilde
c \tilde b} -{\sqrt{\hbar c}\over \alpha _s}\mu _{\dah{i}}^b 
f^{\dah{i}}_{\tilde p\tilde q}\bigg)\bigg\}\bigg\}\cdot\cr
\noalign{\vskip0pt plus 1pt}
&\cdot {\sqrt{|h^0|}\over V_2}dm(y). &(17.43)\cr}
$$
After a diagonalization of the matrix $m^{2}({\varPhi} ^{k}_{\crt
})^{\tilde{a}}{}_a{}^{\tilde{b} }{}_b$ we get a mass
spectrum for Higgs' particles. $k=0$ corresponds to the ``true" vacuum case
and $k=1$ to the ``false" vacuum case. In both cases i.e. for vector
bosons and Higgs' particles this is the mass spectrum up to the second
order of expansion with respect to~$\xi $. Next orders of expansion can be
obtained in a similar way. The lagrangian involving Higgs' field
depends on the scalar field ${\varPsi} $.

Let us consider the full lagrangian of the theory in the second order of
approximation with respect to $h_{[\mu \nu ]}$ and $\ov W_{\mu }$.
Thus let us take this part of Yang--Mill's and Higgs' lagrangian
which depends on $\ov W_{\mu }$ only. We remind to the reader that we
suppose in section 12 an identification $\widetilde A_{\mu
}^F=-{\displaystyle{2\over 3}} \ov W_{\mu }$ and $\widetilde H_{\mu
\nu }^F=-{\displaystyle{4\over 3}}\ov W_{[\mu ,\nu ]}$.

Let us take the Yang--Mill's lagrangian in the zeroth order
$$\eqalignno{
\nad{{\cal L}}{(0)}_{YM} ={}& {2\over 9\pi}[\widetilde p_{FF}+\widetilde
{\xi}^2\widetilde k_{Fr}\widetilde p^{rs}\widetilde k_{sF}]\ov W_{[\beta,
\gamma ]}\ov W_{[\mu, \alpha ]}\eta ^{\beta \mu }\eta ^{\gamma \alpha
}+\cr
&+ {1\over 12\pi}[\widetilde p_{Fb}+\widetilde
{\xi}^2\widetilde k_{Fr}\widetilde p^{rs}\widetilde k_{sb}]\ov W_{[\beta,
\gamma ]}H^b_{[\mu, \alpha ]}\eta ^{\beta \mu }\eta ^{\gamma
\alpha}&(17.44)\cr}
$$
and in the first order:
$$
\aligned
\nad{{\cal L}}{(1)}_{YM} &= -{4\over 9\pi}(\widetilde p_{FF}+\widetilde
{\xi}^2\widetilde k_{Fr}\widetilde p^{rs}\widetilde k_{sF})
\eta ^{\beta \sigma }\eta ^{\mu \tau}\eta ^{\gamma
\alpha}h_{(\tau\sigma)}\ov W_{[\beta,\gamma ]}\ov W_{[\mu, \alpha ]}+\cr
&+ {2\over 3\pi}(\widetilde p_{Fb}+\widetilde
{\xi}^2\widetilde k_{Fr}\widetilde p^{rs}\widetilde k_{sb})
\eta ^{\beta \sigma }\eta ^{\mu \tau}\eta
^{\gamma\alpha}h_{(\tau\sigma )}\ov W_{[\beta,\gamma ]}H^b_{[\mu, \alpha]}+\cr
&-{4\widetilde{\xi}\over 3\pi}(\widetilde p_{aF}+\widetilde\xi \widetilde
k_{aF})\widetilde p^{ar}\widetilde k_{rs}(\delta ^s{}_{F}+\widetilde
{\xi}\widetilde p^{sq}\widetilde k_{qF})\eta ^{\beta \sigma }\eta
^{\mu \tau}\eta ^{\gamma \alpha }h_{[\tau\sigma ]}\ov W_{[\beta,\gamma
]}\ov W_{[\mu ,\alpha ]}+\cr
&-{\widetilde{\xi}\over 3\pi}(\widetilde p_{aF}+\widetilde\xi \widetilde
k_{aF})\widetilde p^{ar}\widetilde k_{rs}(\delta ^s{}_{p}+\widetilde
{\xi}\widetilde p^{sq}\widetilde k_{qp})\eta ^{\beta \sigma }\eta
^{\mu \tau}\eta ^{\gamma \alpha }h_{[\tau\sigma ]}\ov W_{[\beta,\gamma]}
\ov W_{[\mu ,\alpha ]}+
\endaligned 
$$
$$\aligned
&-{4\widetilde{\xi}\over 3\pi}(\widetilde p_{ab}+\widetilde\xi \widetilde
k_{ab})\widetilde p^{ar}\widetilde k_{rs}(\delta ^s{}_{F}+\widetilde
{\xi}\widetilde p^{sq}\widetilde k_{SF})\eta ^{\beta \sigma }\eta
^{\mu \tau}\eta ^{\gamma \alpha }h_{[\tau\sigma ]}\ov W_{[\beta,\gamma]}
\ov W^b_{[\mu ,\alpha ]},
\endaligned \eqno(17.45)
$$
where $\widetilde {\ell}_{ab}=\widetilde p_{ab}+\widetilde
{\xi}\widetilde k_{\tilde a\tilde b}$ is the nonsymmetric right-invariant tensor
on the gauge group $G$ (being a subgroup of $H$). The field $\ov
W_{\mu }$ appears also in ${\cal L}_{\Int}({\varPhi}, \widetilde
A)$. One gets
$$\displaylines{
\nad{{\cal L}}{(2)}({\varPhi}, \widetilde A)=-{4\widetilde\xi \over
3V_2}\mmint_{M}\sqrt{|h^0|}\,dm(y)\cdot\hfill\cr
\hfill\cdot \{\widetilde p_{ab}\mu ^a{}_{F}\eta ^{\mu \alpha }\eta
^{\alpha \beta }h_{\beta \alpha }h^{o\tilde m\tilde a}h^{o\tilde n\tilde
b}k^o{}_{\tilde b\tilde a}\ov W_{[\mu ,\nu ]}H^b{}_{\tilde m\tilde
n}\}.\quad\ (17.46)\cr}
$$
Moreover, the field $\widetilde A^F{}_{\mu }=-{\displaystyle{2\over
3}}\ov W_{\mu }$ couples the Higgs field via the gauge derivative.
Thus we should consider ${\cal L}_{\kin}(\mathop{\nabla}\limits^{\gauge}{\varPhi})$ up to the
second order of~$\ov W_{\mu }$. Moreover, we do not consider in this
case the full lagrangian ${\cal L}_{\kin}(\mathop{\nabla}\limits^{\gauge}{\varPhi})$
because according to our assumption the $\widetilde A^F{}_{\mu }$
field becomes massive after a symmetry breaking from $G$ to $G_0$ and
it is enough to consider a mass term for this field in the
lagrangian
$$
{1\over 2}\eta ^{\mu \nu }m^2_F\widetilde A^F{}_{\mu }\widetilde
A^F_{\nu }\widetilde p_{FF}={2\over 9}m^2_F\widetilde p_{FF}\eta
^{\mu \nu }\ov W_{\mu }\ov W_{\nu },\eqno(17.47)
$$
where $m_F$ is the mass of boson corresponding to the broken
$\widetilde A^F{}_{\mu }$ field (the ${\bold U}(1)_F$ subgroup $G$,
${\bold U}(1)_F\not\subset G_0$). The remaining part of the full
lagrangian of the theory containing $\ov W_{\mu }$ is the term
$$
{2\over 3}g^{[\mu \nu ]}\ov W_{[\mu ,\nu ]}.\eqno(17.48)
$$
Taking this term up to the second order with respect to $h_{(\mu
\nu)}$, $h_{[\mu \nu ]}$ and $\ov W_{\mu }$ one gets:
$$
-{2\over 3}\eta ^{[\mu |\alpha |}\eta ^{\nu ]}h_{\beta \alpha }
\ov W_{[\mu ,\nu ]}.\eqno(17.49)
$$
We should add of course a quadratic approximation of $R_4(\widetilde
{\overline{\varGamma}})$ in N.G.T. to have full lagrangian containing $\ov
W_{\mu }$. This is known in N.G.T. The Nonsymmetric--Nonabelian
Jordan--Theory (Kaluza--Klein) Theory with Higgs' mechanism gives us
several additional couplings between $h_{[\mu \nu ]}$, $h_{(\mu \nu
)}$ and $\ov W_{\mu }$, which are absent in pure N.G.T. It seems that
they can help in solving some serious problems in N.G.T. raised
recently by Damour (see Ref.~[121]).

In order to have a more clear situation we take only this terms which
does not depend on the gauge field (broken or unbroken) and which are at
least quadratic in $\ov W_{\mu }$, $h_{[\mu \nu ]}$, $h_{(\mu \nu
)}$. One gets:
$$
\displaylines{\quad\ 
{\cal L}_{\text{\rm quadratic}}(\ov W)={2\over 9\pi}(\widetilde p_{FF}+\widetilde
{\xi}^2\widetilde k_{Fr}\widetilde p^{rs}\widetilde k_{sF})
\eta ^{\beta \mu }\eta ^{\gamma\alpha}\ov W_{[\beta,\gamma ]}\ov W_{[\mu, \alpha
]}+\hfill\cr
\hfill -{2\over 9}\widetilde p_{FF}m^2_F\eta ^{\mu \nu }\ov W_{\mu
}\ov W_{\nu }-{2\over 3}\eta ^{\nu \beta \eta \mu \alpha }h_{[\beta,
\alpha ]}\ov W_{[\mu ,\nu ]}.\quad\ (17.50)\cr}
$$
Thus we get the lagrangian which is a Proca-field lagrangian
interacting with $h_{[\beta \alpha ]}$.

Let us consider linearization of Eqs.~$(12.44\div 49)$ with respect to
$h_{(\mu \nu )}$, $h_{[\mu \nu ]}$ and $\ov W_{\mu }$ and let us
neglect all Yang--Mill's field and Higgs' field except $\ov W_{\mu }$
field $(\widetilde A^F{}_{\mu })$. One easily gets
$$\displaylines{\quad\
\widetilde R_{\mu \rho}+{1\over 4}F^\lambda {}_{\mu \rho ,\lambda
}+\bigg(\alpha +{5\over 6}\bigg)h_{[\mu \lambda ],\rho
,}{}^{\lambda}+\hfill\cr
\hfill-\bigg(\alpha +{1\over 2}\bigg)h_{[\rho \xi],}{}^{\lambda
}{}_{,\mu }+{1\over 2}h_{[\rho \mu ]},^\lambda ,_{\lambda }+
{2\over 3}(1+4\alpha )\ov W_{[\mu \rho ]}=ah_{\mu \rho }.\quad\
(17.51)\cr}
$$
Notice that we omit the energy---momentum tensor for $\ov W_{\mu }$,
because it is of the second order with respect to $\ov W_{\mu }$,
$ah_{\mu \rho }$ is a cosmological term with $a$ being a constant, $\widetilde R_{\mu \rho }$ is the
usual Ricci tensor for $h_{(\mu \nu )}$ in the linear approximation
$$
F_{\lambda \mu \rho }=3h_{[[\lambda \mu ],\rho ]}.\eqno(17.52)
$$
In the formula (17.51) we take for a full Ricci tensor the following
sum.
$$
R_{\mu \nu }(\ov W)=R^\lambda {}_{\mu \nu \lambda }+\alpha R^\lambda
{}_{\chi\mu \nu }.\eqno(17.53)
$$
In the work we are using $\alpha ={\displaystyle{1\over 2}}$.

The most interesting case is the interaction of the skewon field
$h_{[\mu \nu ]}$ with~$\ov W_{\mu }$.

The lagrangian for the $\ov W_{\mu }$ field looks as follows
$$
L(\ov W) = -{A^2\over 8\pi}\ov W_{[\mu ,\nu ]}\ov W^{[\mu ,\nu
]}+B^2\ov W_{\mu }\ov W^{\mu }
+{2\over 3}(1+4\alpha )h^{[\mu \nu ]}\ov W_{[\mu ,\nu ]},
\eqno(17.54)
$$
where $A$ and $B$ are constants and
$$
h^{[\mu \nu ]}=\eta ^{\mu \alpha }\eta ^{\nu \beta }h_{[\alpha \beta
]},\eqno(17.55)
$$
$\eta _{\mu \nu }$ is a Minkowski's tensor.

The constants $A$ and $B$ depends on the details of the theory
presented before (i.e. groups, constants $\xi$, $\mu $, $\zeta$).

Now we redefine the field $\ov W_{\mu }$ introducing a new field
$V_{\mu }$ 
$$ 
V_{\mu }={1\over 2}A\ov W_{\mu }.\eqno(17.56)
$$
One gets: 
$$ 
L(V)=-{1\over 8\pi}V_{\mu \nu }V^{\mu \nu }+{m^2\over 4\pi}V_{\mu
}V^\mu -{2\over 3}g_Vh^{[\mu \nu ]}V_{\mu \nu },\eqno(17.57)
$$
where\vadjust{\vskip-8pt}
$$
V_{\mu \nu }=\partial_{\mu }V_\nu -\partial_\nu V_{\mu }\eqno(17.58)
$$
and $m_V$ is a mass of the Proca field $V_\mu $, and $g_V$ is a
coupling constant, $g_V>0$. The equation for the field $V_\mu $ is as
follows:
$$ 
(-\eta ^{\mu \lambda }\square +\partial^\mu \partial^\lambda -\eta
^{\mu \chi}m^2_V)V_\lambda ={8\pi\over 3}g_V\partial_\nu h^{[\nu \mu
]}.\eqno(17.59)
$$
The equation can be rewritten in the following form:
$$
\partial_\nu V^{\nu \mu }+m^2_VV^\mu =-{8\pi\over 3}g_V\partial_\nu
h^{[\nu \mu ]}.\eqno(17.60)
$$
Now we come back to Eq.~(17.51) and write them for the symmetric and
antisymmetric part eliminating $\ov W_{\mu }$ field. One gets:
$$\displaylines{
\widetilde R_{\mu \rho }+2\bigg(\alpha +{5\over
6}\bigg)(h_{[\mu \lambda ],\rho ,}{}^{\lambda }+h_{[\rho \lambda
],\mu ,}{}^\lambda )+\hfill\cr
\hfill -{1\over 2}\bigg(\alpha +{1\over
2}\bigg)(h_{[\rho \lambda ],}{}^{\lambda }{}_{,\mu }+h_{[\mu
\lambda],}{}^\lambda{}_{,\rho } )=ah_{(\mu \rho )},\quad\ (17.61)\cr
F^{\lambda }_{[\mu \rho ,|\lambda |,\gamma ]}+
2\bigg(\alpha +{5\over 6}\bigg)(h_{[[\mu |\lambda|],\rho ,}{}^{\lambda
}{}{,\gamma ]}+h_{[[\rho |\lambda|],\mu ,}{}^\lambda{}_{,\gamma ]}
)+\hfill\cr
\hfill -2\bigg(\alpha +{1\over 2}\bigg)(h_{[[\rho
|\lambda|],}{}^{|\lambda|}{}_{,\mu ,\gamma ]}-h_{[[\mu
|\lambda|],}{}^{|\lambda|}{}_{,\rho ,\gamma ]})+{1\over 2}h_{[[\rho
\mu ],}{}^{|\lambda |}{}_{,|\lambda |,\gamma ]}={1\over 3}aF_{\mu
\rho \gamma }.\quad\ (17.62)\cr}
$$
The Eq.~(17.61) describes linear G.R.T. with skewon sources and
Eq.~(17.62) is the equation for skewon field which couples to the
$V_{\mu }$ field in Eq.~(17.60). For $\widetilde R_{\mu \rho }$ we
have a simple formula
$$
\widetilde R_{\mu \rho }=-\square \ov h_{\mu \rho },\eqno(17.63)
$$
where $\ov h_{\mu \rho }=h_{(\mu \rho )}-{1\over 2}\eta ^{\alpha
\beta }h_{(\alpha \beta )}\eta _{\mu \rho }$, and $\ov h^{\mu \nu
}{}_{,\nu }=0$.

Thus the Eq.~(17.61) takes the form
$$
\displaylines{
-\square \ov h_{\mu \rho }+{1\over 2}\bigg(\alpha +{5\over
6}\bigg)(h_{[\mu \lambda ],\rho ,}{}^{\lambda }+h_{[\rho \lambda
],\mu ,}{}^{\lambda})+\hfill\cr
\hfill -{1\over 2}\bigg(\alpha +{1\over
2}\bigg)(h_{[\rho \lambda ],}{}^{\lambda }{}_{,\mu }+h_{[\mu \lambda],}
{}^{\lambda}{}_{,\rho })=a\ov h_{\mu \rho }-{a\over 2}\eta ^{\alpha
\beta }\ov h_{(\alpha \beta )}\eta _{\mu \rho },\quad\ (17.64)\cr}
$$
where $\ov h^{\mu \rho }{}_{,\rho }=0$, $\ov h_{\mu \rho }=\ov h_{\rho
\mu }$.

Moreover, it seems that we should consider also some nonlinear terms
connected to $V_\mu $ as a source of gravitational field in (17.64).
It seems also that we should include an electromagnetic field, which
remains massless after symmetry breaking and the scalar field
${\varPsi}$. In this way we consider lagrangians and
energy---momentum tensors for $V_{\mu }$, $A_\mu $ and ${\varPsi}$.

In the lowest order the lagrangians for electromagnetic and
scalar fields take forms
$$\eqalignno{
{\cal L}_{\text{\rm em}} &= -{1\over 8\pi} F_{\mu \nu }F^{\mu \nu },&(17.65)\cr
{\cal L}_{\scal} &= {1\over 8\pi}\bigg(\ov M \eta ^{\mu \nu
}{\varPsi}_{,\mu }{\varPsi}_{,\nu }+{m^2\over
2}{\varPsi}^2\bigg),&(17.66)\cr}
$$
where $\ov M=(\ell^{[dc]}\ell_{[dc]}-3n(n-1))$.

According to our considerations the scalar field ${\varPsi}$ obtains
a non-zero mass. Moreover, we should redefine the field ${\varPsi}$
in such a way that
$$
{\cal L}_{\scal}={1\over 8\pi}\eta ^{\mu \nu }\varphi _{,\mu }\varphi
_{,\nu }+{m^2_{\varphi} \over 16\pi}\varphi ^2,\eqno(17.67)
$$
where $m_{\varphi}$ is a mass of the field $\varphi$.

It is easy to see that
$$
\varphi =\ov M{\varPsi}.\eqno(17.68)
$$
According to the Eq.~(12.44--50) we really need the quantity
$$
\ov T_{\mu \rho }=T_{\mu \rho }-{1\over 2}\eta _{\mu \rho }(\eta
^{\alpha \beta }T_{\alpha \beta} ).\eqno(17.69)
$$
One easily finds for $V_\mu $ and electromagnetic fields
$$\eqalignno{
\ov{T}_{\alpha \rho } ={}& {1\over 4\pi}\bigg[\eta ^{\tau\mu }V_{\mu
\alpha }V_{\tau\rho }-{1\over 4}\eta _{\mu \rho }(V_{\mu \beta
}V^{\mu \beta })+m^2_V V_\alpha V_\rho \bigg],&(17.70)\cr
\mathop{\ov{T}}\limits^{\text{\rm em}}{\!}_{\alpha \rho }={}&{1\over 4\pi}
\bigg(\eta ^{\tau\mu }F_{\mu
\alpha }F_{\tau\rho }-{1\over 4}\eta _{\alpha \rho }(F_{\mu \nu }
F^{\mu \nu })\bigg).&(17.71)\cr}
$$

For scalar field $\varphi $ we have
$$
\mathop{\ov{T}}\limits^{\scal}{\!}_{\alpha \rho }={1\over 4\pi}
\bigg(\varphi _{,\alpha }\varphi _{,\rho }-{1\over 4}\eta
_{\alpha \rho }(\varphi _{,\mu }\varphi ^{,\mu })+{m^2_{\varphi} \over
4}\eta _{\alpha \rho }\varphi ^2\bigg).\eqno(17.72)
$$
We should add of course to the set of equations of fields, equations
for electromagnetic and scalar fields. Thus one gets:
$$\displaylines{
-\square \ov h_{\mu \rho }+{1\over 2}\bigg(\alpha +{5\over 6}\bigg)
(h_{[\mu \lambda ],\rho ,}{}^{\lambda }+h_{[\rho \lambda],\mu ,}{}^{\lambda})
-{1\over 2}\bigg(\alpha +{1\over
2}\bigg)(h_{[\rho \lambda ],}{}^{\lambda }{}_{,\mu }+h_{[\mu \lambda],}
{}^{\lambda}{}_{,\rho })=\hfill\cr
=a\ov h_{\mu \rho }-{a\over 2}\eta ^{\alpha
\beta }\ov h_{(\alpha \beta )}\eta _{\mu \rho }
+{\varkappa\over 4\pi}\bigg(\eta ^{\tau \beta }(F_{\beta \mu }F_{\tau
\rho }+V_{\beta \mu }V_{\tau \rho })+\varphi _{,\mu }\varphi _{,\rho}
-{1\over 4}\eta _{\mu \rho }(V_{\alpha \beta }V^{\alpha \beta
}+\cr
\noalign{\eject}
\hfill +F_{\alpha \beta }F^{\alpha \beta }+\varphi _{,\alpha }\varphi
_{,}{}^{\alpha }
-m^2_\varphi \varphi ^2)+m^2_V V_\mu V_\rho \bigg),\quad\
(17.73)\cr
\hfill \partial_{\mu }F^{\mu \nu }=0,\quad\ F_{[\rho \lambda ,\nu
]}=0,\hfill\llap{(17.74)}\cr
\hfill (\square + m^2_\varphi )\varphi =0,\hfill\llap{(17.75)}\cr
\hfill \partial_{\nu }V^{\nu \mu }+m^2_V V^\mu =-{8\pi\over
3}g_V\partial_\nu h^{[\nu \mu ]},\quad\ V_{[\mu \nu ,\lambda ]}=0,
\hfill\llap{(17.76)}\cr
F^\lambda _{[\mu \rho ,|\lambda |,\gamma ]}+
2\bigg(\alpha +{5\over 6}\bigg)(h_{[[\mu |\lambda|],\rho ,}{}^{|\lambda
|}{}{,\gamma ]}+h_{[[\rho |\lambda|],\mu ,}{}^{|\lambda|}{}_{,\gamma ]}
)+\hfill\cr
\hfill -2\bigg(\alpha +{1\over 2}\bigg)(h_{[[\rho
|\lambda|],}{}^{|\lambda|}{}_{,\mu ,\gamma ]}-h_{[[\mu
|\lambda|],}{}^{|\lambda|}{}_{,\rho ,\gamma ]})+{1\over 2}h_{[[\rho
\mu ],}{}^{|\lambda |}{}_{,|\lambda |,\gamma ]}={1\over 3}aF_{\mu
\rho \gamma },\quad\ (17.77)\cr}
$$
where 
$$\eqalignno{
F_{\mu \rho \gamma } &= 3h_{[[\mu \rho ],\gamma ]},&(17.78)\cr
\ov h_{\mu\rho} &= h_{(\mu \rho )} -{1\over 2}\eta ^{\alpha \beta }h_{(\alpha
\beta )}\eta _{\mu \rho },&(17.79)\cr
\ov h &= \eta ^{\mu \rho }\ov h_{\mu \rho }&(17.80)\cr}
$$
and
$$
\ov h^{\mu \nu }{}_{,\nu }=0\eqno(17.81)
$$
and $\varkappa = {\displaystyle{8\pi G_N\over c^4}}$.

The Eqs.~$(17.73\div 17.81)$ are full set of equations for coupled
fields in our approximation.

Let us notice that in Eq.~(17.77) we have on the right-hand ride
a term ${1\over 3}a F_{\mu\rho\gamma}$. This term induces a mass
term for the field $h_{[\mu\nu]}$. In this way the skewon field
$h_{[\mu\nu]}$ is massive (due to cosmological term) and some
problems raised in Ref.~[121] are not applicable. 

Let us recall that the fields $V_{\mu}$, $\varphi$ are
also massive. $h_{[\mu\nu]}$, $V_{\mu}$,
$\varphi$ are additional fields which are connected to
gravitational interaction (except $h_{(\mu\nu)}$-field). They are
massive and in linear approximation they are of short range
(having a Yukawa-type behaviour). Due to this we have no
problems with gravitational radiation from closed binary
systems. Only the famous quadrupole radiation formula enters the
equation for a secular changing of an orbital period. This is
exactly the same as in G.R.T.~([122]).

Recently J. W. Moffat has proposed a reformulation of N.G.T. ([123]).
It seems that many of additional phenomenological terms in his 
gravitational lagrangian can be obtained from our theory. In particular
in the approximation presented here they are coming from an
effective interactions with massive multiplet of Higgs fields.
In our theory Higgs' field is a part of multidimensional gauge 
field in nonsymmetric Kaluza--Klein (Jordan--Thiry) theory.
In some sense this is a little similar to Connes Noncommutative Geometry
([124]). Moreover in order to be closer to this approach we should 
extend our formalism to include supersymmetry and supergravity in
a nonsymmetric version of our theory.

The interesting point in our approach is to find a quantum version of this
classical field theory. It simply means how to quantize the theory? It seems
that the best way to achieve it is to use Ashtekar approach of canonical
quantization ([125]). In his approach applied to gravity there is a
possibility to extend the formalism to include gauge fields and Higgs'
fields. In such a way we can work with a group $SU(2)\otimes G(SL(2,{\sym C})
\otimes G)$ in the place of $SU(2)(SL(2,{\sym C}))$.

{\def\co{cosmological}
\def\ct{constant}
\let\u\widetilde
\let\G\Gamma
\def\La{Lagrangian}
\def\br#1#2{\overset\text{\rm #1}\to{#2}}
Let us notice that there is a different possibility to quantize our theory
(even in a perturbative regime). This possibility is to consider the theory
as an effective one to some energy scale. In this way our Lagrangian is an
``effective \La''. It describes low energy interactions and we are using only
those degrees of freedom and interactions appropriate for those energies.
Whether or not is this theory fundamental, at low energy it can be described
by an effective theory. The lowest energy \La\ can be used to determine the
propagators and low energy vertices, and the rest can be treated as
perturbations. When we quantize the theory (introducing propagators for
graviton, skewon, photon, gluons, higsons, scalarons, intermediate bosons and
vertices for them) and calculate loops, the usual ultraviolet divergences can
be absorbed into definitions of renormalized couplings in the most general
\La. 

Those effects can be left-over in the amplitudes for long distance
propagation (low energy). Roughly speaking we can consider a full \La\ of the
theory as a sum of local \La s (the most general local effective \La)
absorbing infinities into all couplings' \ct s. Afterwards we use an energy
expansion and order the most general \La\ in powers of the low energy scale
divided by the high energy scale. If we want to quantize a full theory with a
tower of scalar fields, we can use those massive fields as regulator fields
in some kind of Pauli-Villars' renormalization scheme.  In this way we can
get some interesting results similar to J.~Donoghue's approach in
quantization of General Relativity (see Ref.~[126]). In some sense
J.~Donoghue's approach is an attempt to renormalize nonrenormalizable theory.
In both approaches to quantization of this unified theory we can use
generalized functions of J.~F.~Colombeau (see Ref.~[127]).

The interesting problem to consider in our theory is to find a reason to
existence of a \co\ \ct\ ($\lambda_{c0}=\overline\varLambda$, see Eq.\ (14.296)). A
natural assumption is to put both \co\ terms to zero. This is possible by
taking special values of $\mu$ and $\zeta$ such that
$$
\u R(\u\G(\mu_0))=\u P(\zeta_0)=0. \eqno(17.82)
$$
In this way we control a value of a \co\ \ct. In order to get a specific
value of $\lambda_{c0}$ we should take some values of $\mu$ and $\zeta$
different from $\mu_0$ and~$\zeta_0$. For $\mu$ and~$\zeta$ enter some
observables connected to elementary interactions and other \co\ predictions
(see sections 18 and 19), we should be very careful. In this way we have to
do with some kind of fine tunning of $\mu$ and~$\zeta$. It seems moreover
that the non-zero \co\ terms are connected to a breaking of some kind of
invariance in the theory. Probably to a conformal invariance.

In order to examine this conjecture let us linearize full field equations of
our theory (Eqs (9.3)$\div$(9.26)). It is easy to see that those (linearized)
equations are conformally invariant if both \co\ terms are zero and $r\to
\infty$. We linearize $g_{\mu\nu}$ around $\eta_{\mu\nu}$ and $\varPsi$
around~$\varPsi_0$. Equations for Yang-Mills field are not linearized. In a
general case in order to get  a scale invariance we should have a condition
$$
\br{tot}T_{\alpha\beta}g^{\alpha\beta}=0 \eqno(17.83)
$$
where $\br{tot}T_{\alpha\beta}$ is a total energy momentum tensor (including
\co\ terms). For this we should develop a theory to include fermions in order
to get more conclusive results. Thus we should extend this approach to
supermanifolds with anticommuting coordinates and to superfibrebundles with
super Lie groups (algebras).

}\vfill\eject

{\def\eq#1 {\eqno(\text{\rm18.#1})}
\let\al\aligned
\let\eal\endaligned
\let\ealn\eqalignno
\let\dsl\displaylines
\def\cL{\Cal L}
\let\o\overline
\let\ul\underline
\let\a\alpha
\let\b\beta
\let\d\delta
\let\e\varepsilon
\let\f\varphi
\let\D\Delta
\let\g\gamma
\let\G\varGamma
\let\P\varPsi
\let\F\varPhi
\let\la\lambda
\let\La\varLambda
\let\z\zeta
\def\gh{{\frak h}}
\def\gg{{\frak g}}
\def\gm{{\frak m}}
\let\t\widetilde
\let\u\tilde
\def\Sh{\operatorname{sh}}
\def\Ch{\operatorname{ch}}
\def\ctgh{\operatorname{ctgh}}
\def\tgh{\operatorname{tgh}}
\def\ar{\operatorname{arccos}}
\def\ars{\operatorname{arsh}}
\def\ns{\operatorname{ns}}
\def\sc{\operatorname{sc}}
\def\sh{\operatorname{sh}}
\def\sn{\operatorname{sn}}
\def\cn{\operatorname{cn}}
\def\dn{\operatorname{dn}}
\def\sd{\operatorname{sd}}
\def\cd{\operatorname{cd}}
\def\nd{\operatorname{nd}}
\def\arctg{\operatorname{arctg}}
\def\ctg{\operatorname{ctg}}
\def\tg{\operatorname{tg}}
\def\Li{\operatorname{Li}}
\def\Re{\operatorname{Re}}
\def\br#1#2{\overset\text{\rm #1}\to{#2}}
\def\2{^{\text{\rm II}}}
\def\ddt{\frac d{dt}}
\def\uP{\ul{\t P}{}}
\def\hG{\widehat{\overline\G}}
\def\RG{\t R(\t\G)}
\def\({\left(}\def\){\right)}
\def\[{\left[}\def\]{\right]}
\def\dr#1{_{\text{\rm#1}}}
\def\gi#1{^{\text{\rm#1}}}
\def\ct{constant}
\def\co{cosmological}
\def\dt{dependent}
\def\ef{effective}
\def\ex{exponential}
\def\gr{gravitational}
\def\il{inflation}
\def\lg{lagrangian}
\def\q{quintessence}
\def\pa{parameter}
\def\tr{transform}
\def\wrt{with respect to }
\def\KK{Kaluza--Klein}
\def\JT{Jordan--Thiry}
\def\nos{nonsymmetric}
\def\up#1{\uppercase{#1}}
\def\eu{\expandafter\up}
\def\dint{-\kern-11pt\intop}
\def\wiel{2\mu^3+7\mu^2+5\mu+20}
\def\lon{\ln\(|\z|+\sqrt{\z^2+1}\)+2\z^2+1}

\null
\vskip36pt plus4pt minus4pt
\centerline{{\four\bf 18. On some \eu\co\ Consequences}}

\vskip4pt 
\centerline{{\four\bf of the Nonsymmetric}}

\vskip4pt 
\centerline{{\four\bf Kaluza--Klein (Jordan--Thiry) Theory}}
\vskip24pt plus2pt minus2pt
\write0{18. On some \eu\co\ Consequences
of the Nonsymmetric
Kaluza--Klein (Jordan--Thiry) Theory \the\pageno}

We consider some consequences of the \eu\nos\ \KK
\ (\JT) Theory. We calculate primordial fluctuation spectrum functions, 
spectral indices, and first derivatives of spectral indices.

Let us consider Eq.\ (14.341).
$$
\varPhi(t)=\frac{(5+3a)\sqrt{5+4a}}{8\a_s(A_1\ns^2(u)+A_2)}
+\frac{\(1+\sqrt{5+4a}\)}{2\a_s} \eq1
$$
where
$$
\ealn{
u&=\root4\of{\frac{Ae^{n\varPsi_1}\(5Ae^{n\varPsi_1}-3\a_s^2r\)}{r^2}}
\cdot\sqrt{\sin\frac{\f+\pi}3}\,(t-t_0) &(18.2)\cr
A_1&=\sqrt{\frac53-a^2}\,\sin\frac{\f+\pi}3 &(18.3)\cr
A_2&=\sqrt{\frac53-a^2}\,\cos\frac{\f+4\pi}3+\frac1{12}(5+6a)&(18.4)
}
$$
and
$$
\ealn{
\cos\f&=-\frac{\sqrt A e^{\frac n2\varPsi_1}
\(9\a_s^2Br^2+40Ae^{n\varPsi_1}\)}{12\(5Ae^{n\varPsi_1}-3\a_s^2Br^2\)^{3/2}}
&(18.5)\cr
&0<a=\sqrt{\frac{\a_s^2Br^2}{Ae^{n\varPsi_1}}}<0.3115, &(18.6)
}
$$
$\ns$ is the Jacobi elliptic function for the modulus
$$
m^2=\frac{\sin\(\frac\f3\)}{\sin\(\frac{\f+2\pi}3\)}\,, \eq7
$$
$B$ is an integration \ct, $r$ is a radius of manifold of vacuum states (a
scale of energy), $\a_s$ a coupling \ct. $A$ is given by the equation (14.321)
and is fixed by the details of the theory. $\P_1$~is a
critical value of a scalar field~$\P$. $n$~is a dimension of
the group~$H$. Eq.~(18.1) gives a solution to the equation of an
evolution of the Higgs field under some special assumptions (in the \co\
background). Due to this solution we are able to calculate $P_R(K)$---a
spectrum of primordial fluctuations in the Universe.

According to the Eq.~(14.186)
$$
P_R(K) = \(\frac{H_0}{2\pi}\)^2 \sum_{\u m,d}
\(\frac{H_0}{\ddt(\varPhi^d_{\u m})}\)_{\big| t=t^\ast}^2.
\eq8
$$
Using Eq.~(18.1) and the assumption
$$
\F^d_{\u m}(t)=\d^d_{\u m}\F(t) \eq9
$$
one gets for $t=t^*=\frac1{H_0}\ln\frac K{R_0H_0}$
$$
P_R(K)=P_0\,\frac{\(A_1+A_2\sn^2(v)\)^2\sc^2(v)}{\dn^2(v)} \eq10
$$
where $H_0$ is a Hubble \ct
$$
\ealn{
\hskip-6pt P_0&=\frac{12H_0^4 \a_s^2 n_1 A^{3/2}e^{(3/2)n\P_1}\cdot
\sin^{-3}\(\frac{\f+\pi}3\)\cdot\pi^{-2}} 
{\(5\sqrt{Ae^{n\P_1}}+3\a_s\sqrt{Br^2}\)
\(5A\sqrt{Ae^{n\P_1}}+4\a_s\sqrt{Br^2}\)\(5Ae^{n\P_1}-3\a_s^2Br^2\)}\qquad &(18.11)\cr
\hskip-6pt v&=\(\ln K-t_0H_0-\ln(H_0R_0)\)\frac1{H_0}
\root4\of{\frac{Ae^{n\P_1}\(5Ae^{n\P_1}-3\a_s^2Br^2\)}{r^4}}
\sqrt{\sin\frac{\f+\pi}3}\,,\qquad &(18.12)
}
$$
$n_1$ means a dimension of the manifold $M=G/G_0$ (a vacuum states manifold),
$K^2={\vec K}^2$ means as usual the square of the length of a fluctuations
wave vector. We calculate a spectral index of a power spectrum
$$\ealn{
n_s(K)-1&=\frac{d\ln P_R(K)}{d\ln K}\,, &(18.13)\cr
n_s(K)-1&=2C\(\frac{\dn(v)}{\cn(v)\sn(v)}
-m^2\frac{\sn(v)\cn(v)}{\dn(v)}+ \frac{2A_2\sn(v)\cn(v)\dn(v)}
{A_1+A_2\sn^2(v)}\)\qquad &(18.14)
}
$$
and an important parameter
$$
\al
\frac{dn_s(K)}{d\ln K}&=2C^2\Biggl[
\(\sn^2(v)\dn(v)-\cn^2(v)\dn(v)-m^2\sn(v)\cn(v)\)\cr
&\qquad\qquad {}\times\(\frac1{\cn(v)\sn(v)}+\frac{m^2}{\dn(v)}\)\cr
&+2A_2\frac{\cn^2(v)\dn^2(v)-\sn^2(v)\dn^2(v)-m^2\sn^2(v)\cn^2(v)}
{A_1+A_2\sn^2(v)}\cr
&+4A_2^2\frac{\sn^2(v)\cn^2(v)\dn^2(v)}{\(A_1+A_2\sn^2(v)\)^2}\Biggr]
\eal
\eq15
$$
where $v$ is given by the formula (18.12) and all the Jacobi elliptic functions
in the formulae (18.10), (18.14--15) have the same modulus
$$
m^2=\frac{\sin\frac\f3}{\sin\frac{\f+2\pi}3},
$$
$$
C=\frac1{H_0} \root4\of{\frac{Ae^{n\P_1}\(5Ae^{n\P_1}-3\a_s^2Br^2\)}{r^4}}
\sqrt{\sin\frac{\f+\pi}3}\,. \eq16
$$

In this theory we have some arbitrary parameters which can be translated to
parameters $A_1$, $A_2$, $C$ and~$m$. Thus we can consider Eqs (18.10), (18.14--15) 
as parametrized by $P_0$, $A_1$, $A_2$, $C$ and~$m$. Simultaneously $v$
can be rewritten as
$$
v=C\(\ln K-K_0'\) \eq17
$$
where $K_0'$ is an arbitrary parameter. Let us remind that in our theory we
have arbitrary \pa s: $r$, $\zeta$, $\xi$, $\a_s$ (they could be fixed by
some data from elementary particle physics, however, they are free as some
\pa s of the theory) and integration \ct s $B$ and~$t_0$ ($R_0$~also in some
sense). The interesting idea for future research consists in applying CMBFAST
or CMBEASY
software [128] to calculate all the \co\ CMB characteristics with mentioned \pa
s considered as free. Afterwards we should fit them to existing \co\ data~[104] as
good as possible in order to find some constraints for \pa s from our theory.

Let us consider the function $P_R(K)$ from Eq.~(18.10). It is easy to see that
the function
$$
f(v)=\frac{\(A_1+A_2\sn^2(v)\)^2 \sc^2(v)}{\dn^2(v)} \eq18
$$
is a periodic function
$$
f(v)=f(v+2K(m^2)l), \quad l=0,\pm1,\pm2,\ldots, \eq19
$$
$$
K(m^2)=\intop_0^{\pi/2} \frac{d\theta}{\sqrt{1-m^2\sin^2\theta}}\,. \eq20
$$
Thus we get
$$
\ealn{
P_R(K)&=P_R(Ke^{l\eta}), &(18.21)\cr
\eta&=\frac{2K(m^2)H_0\root4\of{\frac {r^4}{Ae^n\P_1\(5Ae^{n\P_1}-3\a_s^2Br^2\)}}}
{\sqrt{\sin\frac{\f+\pi}3}} &(18.22)
}
$$
or
$$
\eta=\frac{2K(m^2)}C. \eq23
$$
Thus it is enough to consider $P_R(K)$ in the interval
$$
e^{-\eta/2}\le K\le e^{+\eta/2} \eq24
$$
(a little similar to the first Brillouin zone). In order to have an
interesting \co\ region we should have
$$
\eta \simeq 20 \eq25
$$
which gives us a constraint
$$
\frac{2K(m^2)}C=\frac{2K(m^2)H_0}{\sqrt{\sin\frac{\f+\pi}3}}
\root4\of{\frac {r^4}{Ae^{n\P_1}\(5Ae^{n\P_1}-3\a_s^2Br^2\)}}
\approx10.\eq26
$$
Moreover we should consider a region with $n_s(K)\cong1$.
This can be satisfied if
$$
\al
&\dn^2(v)\(A_1+A_2\sn^2(v)\)-m^2\sn^2(v)\cn^2(v)\(A_1+A_2\sn^2(v)\)\cr
&\qquad{}+2A_2\sn^2(v)\cn^2(v)\dn^2(v)=0 
\eal
\eq27
$$
with the condition
$$
\sn(v)\cn(v)\dn(v)\(A_1+A_2\sn^2(v)\)\ne0. \eq28
$$

Using some simple relations among Jacobi elliptic functions and introducing a
new variable
$$
x=\sn^2(v) \eq29
$$
one gets a cubic equation
$$
x^3+\frac{A_1m^2-4A_2m^2-2A_2}{3A_2m^2}\,x^2
+\frac{3A_2-2A_1m^2}{3A_2m^2}\,x+\frac{A_1}{3A_2m^2}=0 \eq30
$$
or
$$
x^3+\frac{\rho m^2-4m^2-2}{3m^2}\,x^2+\frac{3-2\rho m^2}{3m^2}\,x+
\frac\rho{3m^2}=0, \eq31
$$
where
$$
\rho=\frac{A_1}{A_2}=\frac{\sqrt{\frac53-a^2}\,\sin\frac{\f+\pi}3}
{\sqrt{\frac53-a^2}\,\cos\frac{\f+4\pi}3+\frac{5+6a}{12}}\,. \eq32
$$

This equation can be reduced to
$$
y^3+py+q=0 \eq33
$$
where
$$
\ealn{
p&=\frac{-m^4(\rho^2+10\rho+16)+m^2(4\rho+11)-4}{27m^4} &(18.34)\cr
q&=\frac{2m^6(\rho^3+15\rho^2+92\rho-64)+6m^4(49+25\rho-2\rho^2)+6m^2
(43\rho-64)-16}{27^2m^6}\qquad\quad &(18.35)\cr
x&=y-\(\frac{\rho m^2-4m^2-2}{9m^2}\). &(18.36)
}
$$
The discriminant of Eq.\ (18.33) can be written as
$$
D=\frac{q^2}{4}+\frac{p^3}{27} \eq37
$$
and one gets after some tedious algebra
$$
\al
D(\rho,m)=\frac1{27^4 m^{10}}\Bigl[
&-m^{10}\(387\rho^4+7376\rho^3+29456\rho^2+114944\rho+258048\)\cr
&{}+m^8\(243\rho^4+6660\rho^3+19062\rho^2+15000\rho+116352\)\cr
&{}+m^6\(246\rho^4+2292\rho^3-4716\rho^2+3299\rho+2473\)\cr
&{}+m^4\(144\rho^3+1512\rho^2+10476\rho+19251\)\cr
&{}+m^2\(-1548\rho^3+3584\rho^2+35190\rho-26460\)\cr
&{}+48\(75-39\rho\)\Bigr].
\eal
\eq38
$$

We use the Cardano formulae to find a root $x_0$. Afterwards we get
$$
v_\pm = \pm\intop_0^{\arcsin\sqrt{x_0}} \frac{d\theta}{\sqrt{1-m^2\sin^2
\theta}}\,. \eq39
$$
Thus we need $0< x_0< 1$. 

It is easy to see that every
$$
v_l=v_\pm+2K(m^2)l, \qquad l=0,\pm1,\pm2,\ldots, \eq40
$$
satisfies Eq.\ (18.27). 

If $D(m,\rho)>0$, we have only one real root
$$
x_0=\root3\of{-\frac q2+\sqrt D}+ \root3\of{-\frac q2-\sqrt D}
-\(\frac{\rho m^2-4m^2-2}{9m^2}\) \eq41
$$
where $q$ is given by Eq.\ (18.35) and $D$ by Eq.\ (18.38).
Moreover we need $0< x_0< 1$. In order to find a criterion let us transform
Eq.~(18.31) via
$$
x=\frac{1+w}{1-w}\,. \eq42
$$
This transformation maps a unit circle on a complex plane $|x|<1$ into $\Re
w<0$. One gets
$$
\al
&w^3+\frac{13m^2+\rho m^2-1+3\rho}{7m^2-3\rho m^2+5-\rho}\,w^2\cr
&\qquad{}+\frac{5m^2+3\rho m^2-5-3\rho}{7m^2-3\rho m^2+5-\rho}\,w
+\frac{\rho+1-m^2-\rho m^2}{7m^2-3\rho m^2+5-\rho}=0. 
\eal \eq43
$$
It is easy to see that the map (18.42) transforms roots of Eq.~(18.31) into roots
of Eq.~(18.43) and real roots into real roots. Especially roots inside the unit
circle into roots with negative real part.
Thus if Eq.~(18.31) possesses only one real root inside the unit circle then
Eq.~(18.43) possesses only one real root with the negative real part. Let $w_0$
be such a root of Eq.~(18.43) and $z$ and $\o z$ be the remaining roots. One gets
$$
\al
&(w-w_0)(w-z)(w-\o z)\cr
&\qquad{}=w^3-\(2\Re(z)+w_0\)w^2+
\(|z|^2+2\Re(z)w_0\)w-w_0|z|^2=0.
\eal \eq44
$$
Thus we only need 
$$
-w_0|z|^2>0, \eq45
$$
$$
x_0=\frac{1+w_0}{1-w_0}, \eq46
$$
which means
$$
\frac{\rho+1-m^2-\rho m^2}{7m^2-3\rho m^2+5-\rho}>0. \eq47
$$

The last condition means also that Eq.~(18.31) has a real root inside the unit
circle, i.e.~$x_0$. Now we only need to satisfy $x_0>0$, which simply means
$$
\frac \rho{3m^2}<0 \eq48
$$
or (if $m^2>0$)
$$
\rho<0. \eq49
$$
In this way we get conditions
$$
(\rho+1-m^2-\rho m^2)(7m^2-3\rho m^2+5-\rho)>0, \eq50
$$
$$
\rho<0. \eq51
$$

From (18.50--51) we get two solutions:
$$
\dsl{
\text{I} \hfill 0>\rho>-1,\quad 1>m^2>\frac{5-\rho}{7-3\rho} \hfill (18.52)\cr
\text{II} \hfill \rho<-1, \quad 0<m^2<\frac{5-\rho}{7-3\rho}\,. \hfill (18.53)}
$$
Simultaneously we have
$$
D(\rho,m)>0. \eq54
$$
Under these conditions our solution given by Eq.~(18.41) is non-negative with
modulus smaller than one.

In this way we have
$$
n_s(K)=1 \eq55
$$
for
$$
\ealn{
\ln K&=\frac{v_\pm+2K(m^2)l}{C}+K_0' &(18.56)\cr
K&=K_0 \exp\(\frac{v_\pm+2K(m^2)l}C\) &(18.57)
}
$$
where $K_0'=\ln K_0$.

Moreover we are interested in a deviation from a flat power spectrum. Thus we
consider
$$
n_s(K)=1\pm\frac{dn_s(K)}{d\ln K}\,\D\ln K \eq58
$$
where $\D\ln K\sim10$ and $\frac{dn_s(K)}{d\ln K}$ is calculated for $\ln K$
given by the formula (18.55). According to the formula (18.15) $\frac{dn_s(K)}{d\ln
K}$  is proportional to $C^2$.

Thus using (18.26) we get
$$
\frac{dn_s(K)}{d\ln K} \,\D\ln K \sim 10\(\frac{K(m^2)}{10}\)^2 \sim 0.1
\eq59
$$
which gives us rough estimate of a deviation from a flat power spectrum.

For $K$ given by Eq.\ (18.57) one gets
$$
\al
\frac{dn_s(K)}{d\ln K}&=2C^2 \biggl[\e\(\e_0x_0\sqrt{1-m^2x_0}
-\(1+x_0(m^2-1)\)\sqrt{1-x_0}\)\cr
&\qquad{}\times\frac{\sqrt{1-m^2x_0}+\e\e_0m^2\sqrt{x_0(1-x_0)}}{\sqrt{x_0(1-x_0)(1-m^2x_0)}}\cr
&+\frac{1+3m^2x_0^2-x_0(2m^2+1)}{\rho+x_0}+
\frac{4x_0(1-x_0)(1-m^2x_0)}{(\rho+x_0)^2}\biggr]
\eal
\eq60
$$
where
$$
\e_0^2=\e^2=1. \eq61
$$

It is interesting to estimate a range of $K$. One can consider
$$
K_0\exp\(\frac{v_+-K(m^2)}C\) \le K \le K_0 \exp\(\frac{v_++K(m^2)}C\).
\eq62
$$

Let us consider Eq.\ (18.58). If we find conditions for $\frac{dn_s(K)}{d\ln K}$
to be zero, we get a local flat spectrum (around $K$ given by Eq.~(18.57)). One
gets using (18.29) and $(18.60)$
$$
\al
&\(\e_0x_0\sqrt{1-m^2x_0}-\(1+x_0(m^2-1)\)\sqrt{1-x_0}\)\cr
&\qquad{}\times
\frac{\sqrt{1-m^2x_0}+\e\e_0m^2\sqrt{x_0(1-x_0)}}{\sqrt{x_0(1-x_0)(1-m^2x_0)}}\cr
&\quad{}=
\frac{x_0(2m^2+1)-1-3m^2x_0^2}{\rho+x_0}+
\frac{4x_0(x_0-1)(1-m^2x_0)}{(\rho+x_0)^2}\,. 
\eal
\eq63
$$

For $x_0=x_0(m,\rho)$ (see Eq.\ (18.41)) is a known function of $m$ and~$\rho$,
the condition $(18.63)$ can be considered as a constraint for $m$ and~$\rho$.
Together with (18.52), (18.53) and~(18.54) it can give an interesting range for
parameters $m$ and~$\rho$. In this range the spectrum function (18.10) looks flat
(around $K$ given by Eq.~(18.57)).

Let us consider a formula for an evolution of Higgs' field given by Eq.\ 
(14.369):
$$
\varPhi(t)=\frac1{2\a_s}\(1\pm\sqrt{\frac5{1-e^{-5b(t-t_0)}}}\)
\eq64
$$
where $b=\frac{e^{n\P_1}A}{6r^2H_0}>0$,
$$
\lim_{t\to\infty}\F(t)=\frac1{\a_s}\(\frac{\pm\sqrt5+1}2\).
$$
It is easy to see that for
$$
\ealn{
t_{\max}-t_0&=\frac2b, &(18.66)\cr
\F\(\frac2b+t_0\)&\simeq \frac1{\a_s}\(\frac{1\pm\sqrt5}2\). &(18.67)
}
$$
Thus we can consider from practical point of view that an \il\ has been
completed for $t=t_0+\frac2b$. The \il\ starts for
$$
t\dr{initial}=t_0+\frac1{5b}\ln\frac54\,.\eq68
$$

Let us calculate $P_R(K)$, a spectral function for this kind of evolution of
Higgs' field. One gets from Eqs (18.8) and~(18.9)
$$
\ealn{
P_R(K)&=P_0\frac{\(\frac K{K_0}\)^{5{\bar\mu}}}{\(\(\frac
K{K_0}\)^{\bar\mu}-1\)^3}\,,&(18.69)\cr 
\noalign{\eject}
P_0&=\frac{4H_0^4\a_s^2n_1}{125\pi^2b^2} &(18.70)\cr
K_0&=R_0H_0e^{t_0H_0}. &(18.71)
}
$$
For a spectral index we get
$$
\ealn{
\frac{d\ln P_R(K)}{d\ln K}&=n_s(K)-1={\o\mu}\,\frac
{2\(\frac K{K_0}\)^{\bar\mu}-5}{\(\frac K{K_0}\)^{\bar\mu}-1} &(18.72)\cr
{\o\mu}&=\frac{5b}{H_0}. &(18.73)
}
$$
For the first derivative of $n_s$
$$
\frac{dn_s}{d\ln K}=\frac{8{\o\mu}^2\(\frac K{K_0}\)^{\bar\mu}}{\(\(\frac K{K_0}\)^{\bar\mu}
-1\)^2}\,. \eq74
$$

An interesting question is to find the range of $K$. Taking
$$
t\dr{initial}^\ast=\frac1{H_0}\(\frac{K\dr{initial}}{R_0H_0}\)=
t_0+\frac1{5b}\ln\frac54=t\dr{initial} \eq75
$$
one gets
$$
\ealn{
K\dr{initial}&=K_0\(\frac54\)^{1/{\bar\mu}} &(18.76)\cr
t^\ast_{\max}&=\frac1{H_0}\(\frac{K_{\max}}{R_0H_0}\)=t_0+\frac2b=t\dr{max} &(18.77)\cr
K_{\max}&=K_0 e^{2H_0/b}=K_0e^{10/{\bar\mu}} &(18.78)
}
$$
and finally
$$
K_0\(\frac54\)^{1/{\bar\mu}}<K<K_0 e^{10/{\bar\mu}}. \eq79
$$

Let us find $\t K$ such that
$$
n_s(\t K)=1. \eq80
$$
One gets
$$
\t K=K_0\(\frac52\)^{1/{\bar\mu}} \eq81
$$
$$
K_0\(\frac54\)^{1/{\bar\mu}}<\t K<K_0 e^{10/{\bar\mu}}. \eq82
$$
For such a $\t K$ one gets
$$
\ealn{
\frac{dn_s(\t K)}{d\ln K}&=5{\o\mu}^2 &(18.83)\cr
n_s(\t K)&\cong 1\pm 5{\o\mu}^2\D\ln K. &(18.84)
}
$$
If we take as usual $\D\ln K\simeq 10$, we get
$$
n_s(K)\cong 1\pm\frac{1250}{H_0^2}\,b^2=
1\pm\frac{625}{18r^4H_0^4}\,e^{2n\P_1}A^2. \eq85
$$
We can try to make it sufficiently close to 1 taking a big $H_0$.

In this case we have parametrized a spectral function and a spectral index by
only three parameters ${\o\mu}$, $P_0$ and~$K_0$, from which only ${\o\mu}$ matters.
Taking $K$ from the range
$$
1.2^{1/{\bar\mu}} < \frac K{K_0} < 10^{4.3/{\bar\mu}} \eq86
$$
we can calculate in principle all CMB characteristics (see Ref.~[128]) and to
fit to observational data (as good as possible) parameters ${\o\mu}$
(and~$K_0$). This will be a subject of future investigations.

Finally (in order to give a complete set of spectral functions) let us come
back to some formulae (see Eqs (14.529--531)):
$$
\ealn{
P_R(K)&=P_0 g\(\frac K{R_0H_0}\) &(18.87)\cr
g(x)&=\frac1{x^2} \cdot \frac2
{\(\sd(u,\frac12)+\sqrt2 \cd(u,\frac12)\nd(u,\frac12)\)^2} 
\cr
u&=\sqrt2 C_1x+K\(\tfrac12\)-C_1\sqrt2 
-\intop_0^{\arccos(\sqrt5/C_1)}\frac{d\f}{\sqrt{1-\frac12\sin^2\f}}
&(18.88)\cr
P_0&=\(\frac{H_0}{2\pi}\)^2\cdot \frac{4\a_s^2n_1}{5C_1^2} &(18.89)\cr
n_s(K)-1&=-2-C_1x\cdot
\frac{\sqrt2 \cd(u,\frac12)-\sd^2(u,\frac12)}
{\sd(u,\frac12)+\sqrt2 \cd(u,\frac12)\nd(u,\frac12)} &(18.90)\cr
\frac{dn_s(K)}{d\ln K}&=-2C_1x \cdot
\frac{\sqrt2 \cd(u,\frac12)-\sd^2(u,\frac12)}
{\sd(u,\frac12)+\sqrt2 \cd(u,\frac12)\nd(u,\frac12)} 
-2C_1^2x^2\sd^2(u,\frac12)\cr
&-2C_1x^2\cdot
\frac{\(\sqrt2 \cd(u,\frac12)-\sd^3(u,\tfrac12)\)^2}
{\(\sd(u,\frac12)+\sqrt2 \cd(u,\frac12)\nd(u,\frac12)\)^2} &(18.91)
}
$$

In this case independent parameters are $C_1$ and $K_0=R_0H_0$. The amount of
\il\ in three cases can be calculated. In the first case (function (18.18)) it
is given by the corrected formulae (14.357--358), i.e.
$$
\o N_0=H_0\sqrt{\frac {2r^2}{5Aae^{n\varPsi_1}}}
\intop_{\f_2}^{\f_1}\frac{d\theta}{\sqrt{1-\(\frac{4a+5}{10}\)\sin^2\theta}}
\eq92
$$
or
$$
\o N_0=\frac {H_0} {\sqrt{5\a_s}}\root4\of{\frac {4r^2}{ABe^{n\varPsi_1}}}
\intop_{\f_2}^{\f_1}\frac{d\theta}{\sqrt{1-\(\frac{4a+5}{10}\)\sin^2\theta}}
\eq93
$$
$$
\cos^2\f_1=\frac1{5+4a},\quad \cos^2\f_2=\frac5{5+4a}\,. \eq94
$$

It is interesting to notice that we can calculate $\o N_0$ from Eq.~(18.92) for
simplified Weinberg-Salam model (see the first point of Ref.~[75] in the \nos\ version).
One gets
$$
\al
\o N_0&=J\cdot\sqrt{\frac2{15}}\(\frac{m_{\u A}}{m\dr{pl}\a_s}\)^{7/2}
2^{-27/4}\cr
&\times
\frac{\(7g(\z,\mu)+\sqrt{49g^2(\z,\mu)+384h(\z,\mu)}\)^{5/4}}
{\(2\mu^3+7\mu^2+5\mu+20\)^{7/4}|\z|^{1/4}(\mu^2+4)^{1/2}}\cr
&\times 
\frac{(1+\z^2)^{1/4}\(49g^2(\z,\mu)+g(\z,\mu)(\mu^2+4)
\sqrt{49g^2(\z,\mu)+384h(\z,\mu)}\)^{1/2}}
{\(\ln\(|\z|+\sqrt{\z^2+1}\)+2\z^2+1\)^{3/2}
\((5\z^2+4)\ln2-2\mu^2(1+\z^2)\)^{1/4}}
\eal
\eq92a
$$
where
$$
\ealn{
f(\z)&=\frac{16|\z|^3(\z^2+1)}{3(2\z^2+1)(1+\z^2)^{5/2}}
\(\z^2E\(\frac{|\z|}{\sqrt{\z^2+1}}\)-(2\z^2+1)K\(\frac{|\z|}{\sqrt{1+\z^2}}\)\)\cr
&+8\ln\(|\z|\sqrt{\z^2+1}\)
+\frac{4(1+9\z^2-8\z^4)|\z|^3}{3(1+\z^2)^{3/2}} &(18.95)\cr
g(\z,\mu)&=-f(\z)(\mu^2+4) &(18.96)\cr
h(\z,\mu)&=|\z|(2\mu^3+7\mu^2+5\mu+20)(1+\z^2)^{-1}\cr
&\times \((5\z^2+4)\ln2-2\mu^2(1+\z^2)\)
\(\ln\(|\z|+\sqrt{\z^2+1}\)+2\z^2+1\) &(18.97)\cr
J&=\intop_{\f_2}^{\f_1}\frac{d\theta}{\sqrt{1-\(\frac{4a+5}{10}\)\sin^2\theta}}
&(18.98)\cr
\noalign{\eject}
a&=\(\frac{m\dr{pl}}{m_{\u A}}\)^7 2^{35/2}\a_s^{16}B^{1/2}\frac1{m_{\u A}}
\cr
&\times\(\frac{(2\mu^3+7\mu^2+5\mu+20)}
{(\mu^2+4)\(7g(\z,\mu)+\sqrt{49g^2(\z,\mu)+384h(\z,\mu)}\)}\)^{7/2}\cr
&\times \frac{(1+\z^2)^{1/2}\(\ln\(|\z|+\sqrt{\z^2+1}\)+2\z^2+1\)^4}
{|\z|^{1/2}\((5\z^2+4)\ln2 - 2\mu^2(1+\z^2)\)^{1/2}} &(18.99)
}
$$
and $a$ satisfies the condition (18.6), which is a condition for an integration
\ct~$B$, ${B>0}$. All of these formulae are subject to conditions
$$
\ealn{
\mu>\mu_0=-3.581552661\ldots \qquad&(\t R(\t\G)>0) &\text{(18.100a)}\cr
|\z|>\z_0=1.36\dots \qquad&(\t P(\z)<0) &\text{(18.100b)}\cr
\mu^2<\frac{(5\z^2+4)\ln2}{2(1+\z^2)} \qquad &(A>0). &\text{(18.100c)}
}
$$

In the second case considered here the amount of an \il\ is infinite (in
principle). Moreover, it can be given by a finite formula
$$
\al
N&=H_0\(t\dr{end}-t\dr{initial}\)=H_0\(t\dr{max}-t\dr{initial}\)\cr
&=
H_0\(\frac2b-\frac1{5b}\ln\frac54\)=\frac{H_0}b\(2-\frac1{5}\ln\frac54\)
\cong 1.96\,\frac{H_0}b\,.
\eal
\eq101
$$

In the third case the amount of an \il\ is given by the formula (14.520)
$$
\sqrt2 C_2 - K\(\tfrac12\) = \intop_0^{\arccos\bigl(\frac1{2C_1e^{N_0}}\bigr)}
\frac{d\theta}{\sqrt{1-\frac12\sin^2\theta}} - \sqrt2 C_1 e^{N_0}, \eq102
$$
which is a transcendental equation for $N_0$.

All of those $P_R(K)$ functions will be treated by numerical methods from
Ref.~[128] and afterwards we compare the results with observational data.
This will be the first stage of an application of our theory in
cosmology. Afterwards we try to estimate intrinsic parameters of our theory
and to start a multicomponent \il.

\vskip4pt plus2pt
Now we calculate a total amount of an \il\ during two de Sitter phases in our
\co\ models and corresponding masses of \q\ particles.

We consider a total amount of an \il\ in \co\ models.
In our approach we get two de Sitter phases governed by two 
different Hubble \ct s $H_0$ and~$H_1$.
We underline that we have to do with phase transitions: of the second
order in the configuration of Higgs' field and of the first order in the
evolution of the Universe.

The Hubble \ct\ in the first phase is given by the formula ($\hbar=c=1$)
$$
\al
H_0&=\frac
{\(-n\g+\sqrt{n^2\g^2+4(n^2-4)\a_1\b}\)^{(n-2)/4}}
{2^{(n+2)/4}(n+2)^{(n+2)/4}\sqrt3\b^{n/4}}\cr
&\qquad{}\times\(n^2\g^2-\g\sqrt{n^2\g^2+4(n^2-4)\a_1\b}+4\a_1\b(n+2)\)^{1/2},
\eal
\eq103
$$
where
$$
\ealn{
\g&=\frac{m^2_{\u A}}{\a_s^2}\uP=\frac1{r^2}\uP &(18.104)\cr
\b&=\frac{\a_s^2}{l\dr{pl}^2}\t R(\t\G)={\a_s^2}m\dr{pl}^2\t R(\t\G) &(18.105)\cr
\a_1&=\frac{\a_s^2}{r^2}\(\frac{\a_s l\dr{pl}}r\)^2 \(\frac A{\a_s^6}\)=
m^2_{\u A}\(\frac{m_{\u A}}{m\dr{pl}}\)^2 \(\frac A{\a_s^6}\). &(18.106)
}
$$
$\a_s$ is a coupling \ct, $m\dr{pl}$ and $l\dr{pl}$ are Planck's mass and
Planck's length, respectively, $m_{\u A}$ and~$r$ are a scale of an energy (a
scale of a mass of broken gauge bosons). 

$\t R(\t\G)$ is a scalar of a curvature of a connection on a group
manifold~$H$. 
$$
\uP=\frac1{V_2}\intop_M R(\hG)\sqrt{|\t g|}\,d^{n_1}x, \eq107
$$
where $V_2$ is a measure of a manifold $M$ (a vacuum states manifold),
$R(\hG)$ is a scalar of a curvature for a connection defined on~$M$, $\t
g=\det(g_{\u a\u b})$.
$$
A=\frac1{V_2}\intop_M
\(l_{ab}\(g^{[\u m\u n]} g^{[\u a\u b]} C^b_{\u a\u b} C^a_{\u m\u n}
-C^b_{\u m\u n}g^{\u a\u n}g^{\u b\u n}\u L^a_{\u a\u b}\)\)
\sqrt{\u g}\,d^{n_1}x,  \eq108
$$
$$
l_{dc}g_{\u m\u b}g^{\u c\u b}\t L^d_{\u c\u a}+l_{cd}g_{\u a\u m}g^{\u m\u c}
\t L^d_{\u b\u c}=2l_{cd}g^{\u a\u m}g^{\u m\u c}C^d_{\u b\u c}. \eq109
$$
$C^d_{\u b\u c}$ are structure \ct s for a Lie algebra $\gh$ (a Lie algebra
of a group~$H$).

Using Eqs (18.104--106) one gets
$$
\al
H_0&=\(\frac{m_{\u A}}{\a_s}\)\(\frac{m_{\u A}}{\a_s^2m\dr{pl}}\)^{n/2} \frac
{\(-n\uP+\sqrt{n^2{\uP}^2+4(n^2-4)\t R(\t\G)A}\)^{(n-2)/4}}
{2^{(n+2)/4} (n+2)^{(n+2)/4}\sqrt3({\t R}(\t\G))^{n/4}}\cr
&\qquad{}\times\(n^2{\uP}^2-\uP\sqrt{n^2{\uP^2}+4(n^2-4)A\t R(\t\G)}
+4(n+2)A\t R(\t\G)\)^{1/2},
\eal
\eq103a
$$
In the case of more general models (two different types of critical points
for Higgs' field configuration) we take for $A$
$$
A=4\a_s^2V(\F\dr{crt}^1) \eq110
$$
where $V(\F\dr{crt}^1)$ is a value of Higgs' potential for $\F\dr{crt}^1$
(a~metastable state Higgs' field configuration).

In the case of a Hubble \ct\ $H_1$ we get
$$
H_1=\frac{|\g|^{(n+2)/4}n^{n/4}}{\sqrt3 \b^{n/4} (n+2)^{(n+2)/4}} \eq111
$$
or
$$
H_1=\(\frac{m_{\u A}}{m\dr{pl}}\)^{n/2}\frac{m_{\u A}}{\a_s^{n+1}}\,
\frac{|\uP|^{(n+2)/4}n^{n/4}}{\sqrt3  (n+2)^{(n+2)/4}(\t R(\t \G))^{n/4}} \eq111a
$$
In this case we suppose that $\g<0$ and $\a_1, \b>0$. This is possible
because $\g$ is a function of a \ct~$\zeta$ and can have any sign. In the
case of~$\b$ it is the same. $\b$~is a function of a \ct~$\xi$ ($\mu$~here).
This is because of properties of $\t R(\t\G)$ and $\uP$.

We consider here several consequences of Higgs' field dynamics on
primordial fluctuations (perturbations) spectrum. We find three different
functions and examine their properties. We write down also an equation for an
amount of an \il\ in the first de Sitter phase for these functions.
They are calculable for functions (18.10) and~(18.87). In
the case of a function~(18.69) we get under some practical
assumption 
$$
N_0=1.96\,\frac{H_0}b=1.96\,\frac{6r^2H_0^2}{e^{n\P_1}A}=
\frac{1.96}{\a_s^4}\(\frac{m_{\u A}}{m\dr{pl}}\)^2 F(\b,\g,\a_1)
\eq112
$$
where $F(\b,\g,\a_1)$ is given by the formula
$$
F(\b,\g,\a_1)=
\frac{\(n^2\g^2-\g\sqrt{n^2\g^2+4(n^2-4)\a_1\b}+4\a_1\b(n+2)\)}
{(n+2)\a_1\(-n\g+\sqrt{n^2\g^2+4(n^2-4)\a_1\b}\)}
\eq113
$$
or
$$
N_0=1.96\,\(\frac{m_{\u A}}{m\dr{pl}}\)^4
F(\t R(\t\G),\uP,A). \eq113a
$$

Now let us come to the second de Sitter phase. During this phase the \il\ is
driven by the field~$\P$ which changes (slowly $\dot\P\approx0$) from $\P_0$
to $\frac12\ln\(\frac\b{|\g|}\)$. We develop in Sec.~14 two approximation
schemes for an evolution of~$\P$. In the case of a harmonic oscillation we
calculate an amount of an \il. We find also a different scheme---the
so called slow-roll approximation which offers an infinite in time evolution
of~$\P$. 

Moreover from practical point of view we can suppose that an evolution
starts if $\P$ is very close to $\frac12\ln\(\frac{n|\g|}{(n+2)\b}\)$. In
term of a variable 
$$
y=\sqrt{\frac\b{|\g|}}\,e^\P \eq114
$$
$y$ is close to $\sqrt{\frac n{n+2}}$. Let $y$ be
$$
y=\sqrt{\frac n{n+2}}+\e \eq115
$$
where $\e>0$ is very small. Let us take 
$$
\e=\frac1{n(n+1)(n+2)}\,. \eq116
$$
We remind that $n\ge14$. Thus 
$$
\e<3\cdot10^{-4}. \eq117
$$

Let us consider the formula (14.416) and let us calculate the
limit 
$$
\lim_{y\to \sqrt{\frac n{n+2}}+\e}  I. \eq118
$$
One gets after some simplifications (taking under consideration the fact that
$n>14$)
$$
\lim_{y\to \sqrt{\frac n{n+2}}+\e}I=
-\frac\pi2-\frac1{2\sqrt{2n}}\ln2. \eq119
$$
From the other side we have
$$
\lim_{y\to1}I=-\frac\pi2+\frac1{\sqrt{2n}}\ln\(n+1+\sqrt{n(n+2)}\). \eq120
$$
We have
$$
I=\o B(t-t_0) \eq121
$$
where 
$$
\o B=4\sqrt{2\pi}(n+2)|\g|\(\frac n{n+2}\)^{n/2}\(\frac{|\g|}\b\)^{n/2}
\frac1{m\dr{pl}}\sqrt{\o M} \eq122
$$
or
$$
\o B=4\sqrt{2\pi}\(\frac{m_{\u A}}{m\dr{pl}}\)^{n+1} (n+2) 
\(\frac n{n+2}\)^{n/2}
\frac{m_{\u A}}{\a_s^{2(n+1)}} |\uP|
\(\frac{|\uP|}{\RG}\)^{n/2}\frac1{\sqrt{\o M}}\,.
\eq122a
$$
Writing
$$
\ealn{
I(1)&=\o B(t\2\dr{end}-t_0) &(18.123)\cr
I\(\sqrt{\frac n{n+2}}+\e\)&=\o B(t\2\dr{initial}-t_0) &(18.124)
}
$$
where $t\2\dr{initial}$ and $t\2\dr{end}$ mean an initial and end time of the
second de Sitter phase, one gets
$$
\D t\2=t\2\dr{end}-t\2\dr{initial}=
\frac1{\,\o B\,}\,\frac1{\sqrt{2n}}\ln\(\sqrt2\(n+1+\sqrt{n(n+2)}\)\) \eq125
$$
where $\D t\2$ means a period of time of the second de Sitter phase. Thus the
amount of \il\ for this phase is
$$
N_1=H_1\D
t\2=\frac{H_1}{\,\o B\,}\,\frac{\ln\(\sqrt2\(n+1+\sqrt{n(n+2)}\)\)}{\sqrt{2n}}\,. \eq126
$$
 
Using formulae (18.122) and (18.111) one gets
$$
N_1=\sqrt{2\pi}\(\frac\b{|\g|}\)^{n/4}\(\frac{n+2}n\)^{n/4}
\frac{\ln\(\sqrt2\(n+1+\sqrt{n(n+2)}\)\)}{\sqrt{6n|\g|}(n+2)^3}
\cdot \frac{m\dr{pl}}{\sqrt{\o M}} \eq127
$$
or
$$
\al
N_1&=\(\frac{2\sqrt{2\pi}}{\sqrt{\o M}}\)\(\frac{m\dr{pl}}{m_{\u A}}\)^{(n+2)/2}
\(\frac{\RG}{|\uP|}\)^{n/4}\(\frac{n+2}n\)^{n/4}\cr
&\times\frac{\ln\(\sqrt2\(n+1+\sqrt{n(n+2)}\)\)}{2\sqrt6 \a_s^{n-1} |\uP|^{1/2}(n+2)^3}
\,.
\eal \eq127a
$$

For large $n$ one finds
$$
N_1\simeq \sqrt{2\pi} \sqrt{\frac e6} \(\frac\b{|\g|}\)^{n/4}
\frac{\ln n}{|\g|^{1/2}n^{7/2}} \cdot \frac{m\dr{pl}}{\sqrt{\o M}} \eq128
$$
or 
$$
N_1\simeq \sqrt{\frac e6} \(\frac{\RG}{\uP}\)^{n/4}
\(\frac{m\dr{pl}}{m_{\u A}}\)^{(n+2)/2}\(\frac{\sqrt{2\pi}}{\sqrt{\o M}}\)
\frac{\ln n}{\a_s^{n-1}|\uP|^{1/2}n^{7/2}}\,. \eq128a
$$
The total amount of an \il\ considered in the paper is
$$
N\dr{tot}=N_0+N_1 \eq129
$$
and should be fixed to ${}\sim60$.

In order to give an example of these calculations let us consider a
six-dimensional Weinberg-Salam model (see Ref.~[75], the first point). It is of course a
bosonic part of this model. In this case $H=G2$, $\dim H=14$, $M=S^2$
(two-dimensional sphere). This will be of course the model in \eu\nos\ \KK\
(\JT) Theory. In order to simplify the calculations we take
$$
\t R(\t \G)=\frac{2(2\mu^3+7\mu^2+5\mu+20)}{(\mu^2+4)^2} \eq130
$$
(see formula (7.21) from the fourth point of Ref.~[18]).

The scalar curvature has been calculated here for $H=SO(3)\simeq SU(2)$.
Moreover it gives a taste of the full theory
$$
\al
&\uP=\uP(\z)\cr
&=\biggl\{\frac{16|\z|^3(\z^2+1)}{3(2\z^2+1)(1+\z^2)^{5/2}}
\(\z^2E\(\frac{|\z|}{\sqrt{\z^2+1}}\)-(2\z^2+1)K\(\frac{|\z|}{\sqrt{\z^2+1}}\)\)\cr
&+8\ln\(|\z|\sqrt{\z^2+1}\)
+\frac{4(1+9\z^2-8\z^4)|\z|^3}{3(1+\z^2)^{3/2}}\biggr\}
\biggm/ \(\ln\(|\z|+\sqrt{\z^2+1}\)+2\z^2+1\) 
\eal
\eq131
$$
(see formula (8.64)).

For $V_2$ one gets
$$
V_{2} = \frac{2\pi}{|\z|}\(\ln\(|\z|+\sqrt{\zeta ^2+1}\)
+ 2\zeta ^{2} + 1\)\eq132
$$
(see formula (8.65)), where
$$\eqalignno{
K(k) &= \intop_{0}^{\pi/2}\frac{d\theta}{\sqrt{1-k^2\sin^2 \theta}}\,,&(18.133)\cr
E(k) &= \intop_0^{\pi/2}\sqrt{1-k^2\sin^2 \theta }\,d\theta.&(18.134)
}
$$

$\uP$ has been calculated for an Einstein-Kaufmann connection defined
on~$S^2$. We need $\uP<0$, thus in our case
$$
|\z|>\z_0=1.36\dots \eq135
$$
(see Fig.~6).

In order to write the formula in this case we need to calculate $A$ from
Eq.~(18.108). We should calculate $\t L^d_{\u b\u c}$ from Eq.~(18.109). In our case
this is quite easy for $S^2$ is 2-dimensional and $\t a,\t b=1,2$ ($\t a,
\t b=5,6$ if we embed $S^2$ as a vacuum state manifold in the full theory). 

In this
case one easily finds
$$
\t L^c_{\u a\u b}=h^{ce}l_{ed}C^d_{\u a\u b} \eq136
$$
where $C^d_{\u a\u b}$ are structure \ct s of the group $H$ in such a way
that $\t a,\t b$ correspond to the complement $\gm$, $\gg=\gg_0 \mathrel{\dot+} \gm$,
$$
\gg \subset \gh \eq137
$$
where $\gg=A_1$, $l_{ab}=h_{ab}+\mu k_{ab}$, $h^{ef}h_{fd}=\d^e_d$.
Using the exact form of the \nos\ tensor on $SO(3)$ and the \nos\ tensor
on~$S^2$ (see formulae (2.24a), (6.31)) one gets
$$
\al
\o A&=l_{ab}\(g^{[\u m\u n]}g^{[\u a\u b]}C^a_{\u m\u n}C^b_{\u a\u b}
-C^b_{\u m\u n}g^{\u a\u n}g^{\u b\u m}\t L^a_{\u a\u b}\)\cr
&=-\(\frac{\z^2}{(1+\z^2)^2\sin^2\theta} 
+ \frac{4+\mu^2\sin^2\theta}{(1+\z^2)\sin^2\theta}\) 
\eal \eq138
$$
and finally
$$
\ealn{
A&=\frac1{V_2}\intop_{S^2} \o A\sqrt{\t g}\,d\theta\,d\f=
\frac1{V_2}\intop_0^{2\pi} \intop_0^\pi \,d\f\,d\theta\,\o A\sqrt{\t g}, &(18.139)\cr
A&=\frac{|\z|\[(5\z^2+4)\ln 2 - 2\mu^2(1+\z^2)\]}
{(1+\z^2)\(\ln\(|\z|+\sqrt{\z^2+1}\)+2\z^2+1\)}&(18.140)
}
$$
where the integral $\int_0^\pi \frac{d\theta}{\sin\theta}$ has been
calculated in the sense of a principal value:
$$
\intop_0^\pi \frac{d\theta}{\sin\theta}=\lim_{\e\to0^+}
\intop_\e^{\pi-\e}\frac{d\theta}{\sin\theta}=-\ln2. \eq141
$$
Using Eqs (18.130--131), (18.140) one writes the formula (18.103a) in the form ($n=14$)
$$
\al
\hskip-10pt H_0&=\frac1{2^{19}\sqrt3}\(\frac{m_{\u A}}{\a_s^2m\dr{pl}}\)^7
\cdot \frac{m_{\u A}}{\a_s}\cr
&\times\frac{\(7g(\z,\mu)+\sqrt{49g^2(\z,\mu)+384h(\z,\mu)}\)^3(\mu^2+4)^3}
{\(\ln\(|\z|+\sqrt{\z^2+1}\)+2\z^2+1\)\(2\mu^3+7\mu^2+5\mu+20\)^{7/2}}\cr
&\times\(98g^2(\z,\mu)+(\mu^2+4)g(\z,\mu)\sqrt{49g^2(\z,\mu)+384h(\z,\mu)}
+32H(\z,\mu)\)^{1/2}
\eal
\eq142
$$
where
$$
\ealn{
g(\z,\mu)&=-f(\z)(\mu^2+4) &(18.143)\cr
h(\z,\mu)&=|\z|(2\mu^3+7\mu^2+5\mu+20)
\((5\z^2+4)\ln2-2\mu^2(1+\z^2)\)\cr
&\times\frac{\(\ln\(|\z|+\sqrt{\z^2+1}\)+2\z^2+1\)}{(1+\z^2)}&\text{(18.144a)}\cr
H(\z,\mu)&=h(\z,\mu)\(\ln\(|\z|+\sqrt{\z^2+1}\)+2\z^2+1\)&\text{(18.144b)}\cr
f(\z)&=\frac{16|\z|^3(\z^2+1)}{3(2\z^2+1)(1+\z^2)^{5/2}}
\(\z^2E\(\frac{|\z|}{\sqrt{\z^2+1}}\)-(2\z^2+1)K\(\frac{|\z|}{\sqrt{\z^2+1}}\)\)\cr
&+8\ln\(|\z|\sqrt{\z^2+1}\)
+\frac{4(1+9\z^2-8\z^4)|\z|^3}{3(1+\z^2)^{3/2}}\,.&(18.145)
}
$$

From Eq.\ (18.111a) one gets
$$
\al
H_1&=\sqrt{\frac23}\(\frac{m_{\u A}}{m\dr{pl}}\)^7
\frac{m_{\u A}}{\a_s^{15}}\cdot\frac{7^7}{2^{13}}\cr
&\times\frac{g^4(\z,\mu)(\mu^2+4)^4}{\(2\mu^3+7\mu^2+5\mu+20\)^{7/2}}
\cdot\(\ln\(|\z|+\sqrt{\z^2+1}\)+2\z^2+1\)^{-4}. 
\eal \eq146
$$
From Eq.\ (18.113a) one finds
$$
\al
N_0&=0.12\(\frac{m_{\u A}}{m\dr{pl}}\)^{4}\cr
&\times\frac{\(7g(\z,\mu)+\sqrt{49g^2(\z,\mu)+384h(\z,\mu)}\)^{-1}(\mu^2+4)^2
(1+\z^2)}
{|\z|\((5\z^2+4)\ln2-2\mu^2(1+\z^2)\)}\cr
&\times\(98g^2(\z,\mu)+(\mu^2+4)g(\z,\mu)\sqrt{49g^2(\z,\mu)+384h(\z,\mu)}
+32H(\z,\mu)\).
\eal
\eq147
$$

Equation (18.122a) is transformed into
$$
\al
\o B&=\(\frac{m_{\u A}}{m\dr{pl}}\)^{15} \(\frac {7^7
m_{\u A}}{2^{16}\a_s^{30}}\) \(\frac{|f(\z)|}{2\mu^3+7\mu^2+5\mu+20}\)^7
\cdot\frac1{\sqrt{|\o M|}}\cr
&\times \frac{|f(\z)|(\mu^2+4)^6}{\(\ln\(|\z|+\sqrt{\z^2+1}\)+2\z^2+1\)^8}\,.
\eal
\eq148
$$
And finally Eq.\ (18.127a) gives
$$
\al
N_1&\simeq 5.47\cdot 10^{4}
\(\frac{m\dr{pl}}{m_{\u A}}\)^9
\(\frac{|f(\z)|}{2\mu^3+7\mu^2+5\mu+20}\)^{7/2}\cdot \frac{\a_s}{\sqrt{|\o M|}}\cr
&\times
\frac{\(\ln\(|\z|+\sqrt{\z^2+1}\)+2\z^2+1\)^{3/2}}{|f(\z)|^{1/2}(\mu^2+4)^3}
\,.
\eal
\eq149
$$

Let us notice that according to our assumption $f(\z)<0$ ($\uP<0$), 
$2\mu^3+7\mu^2+5\mu+20>0$ ($\t R(\t\G)>0$), and
$$
\frac{(5\z^2+4)\ln 2}{2(1+\z^2)}>\mu^2 \eq150
$$
(from $A>0$). 

The function on the left hand side of the inequality (18.150) is rising for
${\z>0}$ and falling for ${\z<0}$, having a minimum at ${\z=0}$ equal to
$2\ln2$. For $\z\to\pm\infty$ it is going to $\frac52 \ln2$. For a specific
value $\z=\z_0=\pm1.36$ (see Eq.~(18.135)), it is equal to $1.61$. Thus we get
$$
|\mu|\le \sqrt{2\ln2}\approx 1.17741. \eq151
$$
Moreover the polynomial $W(\mu)=2\mu^3+7\mu^2+5\mu+20$ possesses one real root
$$
\mu_0=-\frac{\root3\of{1108+3\sqrt{135645}}}{6}-\frac76-\frac{19}
{6\cdot\root3\of{1108+3\sqrt{135645}}}=-3.581552661\ldots.
$$ 

Thus we can
have $\t R(\t\G)>0$ for $\mu>\mu_0$ and in the region given by~(18.151).
Let us notice that taking sufficiently big $|\z|$ we can make $N_0$ in~(18.147)
arbiratrily big, i.e.\ about~60. From the other side if we take $|\z|$
sufficiently big, $N_1$ can also be arbitrarily large (i.e.\ $\sim60$). Thus
it seems that in this simple example it is enough to consider arbitrarily
big~$\z$ in order to get large amount of an \il. 

It is interesting to find in this simplified model a \ct~$a$ (see Eq.~(18.6)).
One gets after some algebra
$$
\al
a^2&=\frac{\a_s^4B}{Am_{\u A}^2e^{n\P_1}}=
\(\frac{m\dr{pl}}{m_{\u A}}\)^{14}
2^{35}\a_s^{32}\(\frac B{m_{\u A}^2}\)\cr
&\times \(\frac{2\mu^3+7\mu^2+5\mu+20}
{(\mu^2+4)\(7g(\z,\mu)+\sqrt{49g^2(\z,\mu)+384h(\z,\mu)}\)}\)^7\cr
&\times \frac{(1+\z^2)\(\ln\(|\z|+\sqrt{\z^2+1}\)+2\z^2+1\)^8}
{|\z|\((5\z^2+4)\ln2-2\mu^2(1+\z^2)\)}
\eal
\eq153
$$
(we put $n=14$).

The condition 
$$
0<a^2<0.09703 \eq154
$$
gives a constraint on an integration \ct~$B$. In this case $A>0$. However we
consider in Sec.~14 a special type of a dynamics of Higgs' field with $A<0$.
In this case we have the following constraint imposed on the \ct~$\o a$.
One gets after some algebra
$$
\al
\o a&=\frac{e^{n\P_1}m_{\u A}^2A}{2H_0^2\a_s^2}\cr
&=\frac{192\(7g(\z,\mu)
+\sqrt{49g^2(\z,\mu)+384h(\z,\mu)}\)\(\ln\(|\z|+\sqrt{\z^2+1}\)+2\z^2+1\)}
{98g^2(\z,\mu)+(\mu^2+4)g(\z,\mu)\sqrt{49g^2(\z,\mu)+384h(\z,\mu)}
+32H(\z,\mu)}\cr
&\times\frac{|\z|(\mu^2+4)\((5\z^2+4)\ln2-2\mu^2(1+\z^2)\)}{1+\z^2}\,.
\eal
\eq155
$$
Moreover in this case we should put
$$
\o a = -\frac4{15}, \eq156
$$
i.e.\ $A<0$ and we get
$$
\al
&720 \(7g(\z,\mu)
+\sqrt{49g^2(\z,\mu)+384h(\z,\mu)}\)\(\ln\(|\z|+\sqrt{\z^2+1}\)+2\z^2+1\)\cr
&\quad{}\times |\z|(\mu^2+4)\(2\mu^2(1+\z^2)-(5\z^2+4)\ln2\)\cr
&=\(98g^2(\z,\mu)+(\mu^2+4)g(\z,\mu)\sqrt{49g^2(\z,\mu)+384h(\z,\mu)}
+32H(\z,\mu)\)(1+\z^2).
\eal
\eq157
$$

Eq.\ (18.157) gives a constraint among parameters of the theory. If Eq.~(18.157) is
satisfied, we have the dynamics of Higgs' field described by Eq.~(14.381).

In this way we can consider the spectral function (14.107) and all
the consequences coming from it. We need of course some supplementary
conditions 
$$
\frac12\cdot\frac{5\z^2+4}{1+\z^2}\cdot \ln2<\mu^2 \eq158
$$
and as usual
$$
\dsl{
\hfill f(\z)<0 \hfill (18.159)\cr
\hfill 2\mu^3+7\mu^2+5\mu+20>0. \hfill(18.160)
}
$$
In the case when condition (18.154) is satisfied we can consider a different
dynamics of Higgs' field described by Eq.~(18.1) leading to the
spectral function (18.10) with all the consequences of this
function. In this case we have condition (18.150) and conditions (18.159) and~(18.160).
They are easily satisfied.

The programme of research given here will give an additional
constraint and could (in principle) lead to the realistic theory with more
complicated groups and patterns of symmetry breaking. 

Let us notice that in our theory we get \co\ terms. These terms are
described by \ct s $\b,\g$ and~$\a_1$ which are proportional to $\t R(\t\G)$,
$\uP$ and~$A$. The importance of a \co\ \ct\ is now obvious. Thus it is
necessary to control these terms. They depend on \ct s $\z$ and~$\mu$. The
first step in order to control them is to find conditions when they are equal
to zero. In the simplified model we have found these conditions:
$$
\ealn{
\b(\mu_0=-3.581552661\ldots)&=0 &(18.161)\cr
\g(\z=\pm1.36\ldots)&=0 &(18.162)
}
$$
and
$$
\a_1=0 \eq163
$$
for
$$
\mu=\pm\sqrt{\frac{\ln2(5\z^2+4)}{2(\z^2+1)}}\,. \eq164
$$

In this way we control the sign of $A$, $\uP$ and $\RG$ playing with \ct s
$\z$ and~$\mu$:
$$
\ealn{
\RG>0 \quad &\hbox{for }\mu>\mu_0=-3.581552661\ldots &(18.165)\cr
\uP<0 \quad &\hbox{for }|\z|>1.36. &(18.166)
}
$$
The sign of $A$ is also controllable. All of these results give us
interesting \co\ consequences.

It is interesting to find some conditions on stability of the first de Sitter
evolution of the Universe. In Sec.~14 we find a criterion
$$
M>M_0 \eq167
$$
where
$$
\al
&M_0=\frac43\cr
&\times\frac{(n+2)\g\sqrt{n^2\g^2+4(n^2-4)\a_1\b}-4(n+2)^2(n-1)\a_1\b
-n^2(n+2)\g^2}
{n\g^2+4(n+2)\a_1\b-\g\sqrt{n^2\g^2+4(n^2-4)\a_1\b}} 
\eal \eq168
$$
or
$$
\al
&M_0=\frac43\cr
&\times\frac{(n+2)\uP\sqrt{n^2\uP^2+4(n^2-4)A\RG}-4(n+2)^2(n-1)A\RG
-n^2(n+2)\uP^2}
{n\uP^2+4(n+2)A\RG-\uP\sqrt{n^2\uP^2+4(n^2-4)A\RG}}\,, 
\eal \eq168a
$$
If
$$
\uP>0,\ \RG>0,\ A>0, \eq169
$$
then
$$
M_0>0. \eq170
$$
Moreover if
$$
\uP<0,\ \RG>0,\ A>0, \eq171
$$
the stability condition is ($M_0<0$)
$$
M<0. \eq172
$$

Moreover for $A<0$ the condition $M_0<0$ is not trivial and we should have
$$
\kern-12pt 4(n+2)^2(n-1)|A|\RG> 
(n+2)|\uP|\(n^2|\uP|+\sqrt{n^2\uP^2-4(n^2-4)|A|\RG}\). \eq173
$$
In the case of $SO(3)$ group and $S^2$ vacuum states manifold (with $n=14$)
we have
$$
M(SO(3))=-\frac{2(36+7\mu^2)}{(4+\mu^2)M^2\dr{pl}}<0 \eq174
$$
(see the equation on p.~260 of the second point of Ref.~[18]).

Thus if $M_0<0$ the first de Sitter evolution of the Universe is stable. It
means it happens if condition (18.173) is satisfied. One finds:
$$
\al
&\frac{416|\z|}{1+\z^2}\bigl(2\mu^2(1+\z^2)-(5\z^2+4)\ln2\bigr)\cdot
(2\mu^3+7\mu^2+5\mu+20)\cr
&\qquad{}>\frac{g(\z,\mu)\(49g(\z,\mu)+\sqrt{49g^2(\z,\mu)-384|h(\z,\mu)|}\)}
{\ln\(\(|\z|+\sqrt{\z^2+1}\)+2\z^2+1\)}\,.
\eal \eq175
$$
We have also supplementary conditions
$$
\ealn{
&|\z|>|\z_0|=1.36\ldots&(18.176)\cr
&\mu>\mu_0=-3.581552661\ldots &(18.177)\cr
&\mu^2>\frac{\ln2}2\cdot\frac{5\z^2+4}{\z^2+1}\,. &(18.178)
}
$$

During the second de Sitter phase the evolution of the Universe is unstable
against small perturbation of initial data. Moreover it can be made stable if
$M<0$. However, the second de Sitter phase in our model should be unstable
because it ends with a radiation epoch. The instability of this phase in this
particular case is caused by some physical processes beyond \co\ models.

The interesting point in the theory is a mass of the scalar field (a scalar
particle) during both de Sitter phases. However, in that case we should
consider a \q\ field~$Q$
$$
Q=\sqrt{|\o M|}\frac\P{M\dr{pl}}\,. \eq179
$$
In both de Sitter phases we can consider small oscillations $q_k$ of the \q\
field around an equilibrium
$$
Q=Q_k+q_k, \qquad k=0,1. \eq180
$$
For these small oscillations one finds
$$
{\o m}_k^2=-\frac12 \, \frac{d^2\la_{ck}}{d\P^2}(\P_k) \eq181
$$
or
$$
\al
{\o m}^2_1&=\frac
{-\(-n\g+\sqrt{n^2\g^2+4(n^2-4)\a_1\b}\)^{(n-2)/2}}
{2^{(n+4)/2}(n+2)^{n/2}\b^{n/2}}\cr
&\times \(n\g\sqrt{n^2\g^2+4(n^2-4)\a_1\b}-2\g^2n^2
-4(n+2)(n-3)\a_1\b\)
\eal
$$
$$
\al
&=\(\frac{m_{\u A}}{\a_sm\dr{pl}}\)^n\(\frac{m_{\u A}}{\a_s}\)^2
(n+2)^{-n/2}2^{-(n+4)/2}\cr
&\times\(-\uP+\sqrt{n^2\uP^2+4(n^2-4)A\t R(\t\G)}\)^{(n-2)/2}\cdot
\frac1{\(\t R(\t\G)\)^{n/2}}\cr
&\times\(2n^2\uP^2+4(n+2)(n-3)A\t R(\t\G)
-n\uP\sqrt{n^2\uP^2+4(n^2-4)A\t R(\t\G)}\)
\eal
\eq182
$$
$$
\al
m_0^2&=\frac12 n|\g|\(\frac n{n+2}\)^{n/2}\(\frac{|\g|}{\b}\)^{n/2}\cr
&=\frac n2 \cdot \(\frac{m_{\u A}}{\a_s^2m\dr{pl}}\)^n
\(\frac n{n+2}\)^{n/2} \(\frac{m_{\u A}}{\a_s}\)^2\cdot |\uP|\cdot
\(\frac{|\uP|}{\t R(\t\G)}\)^{n/2}.
\eal
\eq183
$$

Using our simplified model for $\RG$, $\uP$ and~$A$ one gets
$$
\al
m_1^2&=\(\frac{m_{\u A}}{\a_s}\)^2 \(\frac{m_{\u A}}{\a_s^2m\dr{pl}}\)^{14}
\cdot 2^{-36}\cr
&\times\frac{\(7g(\z,\mu)+\sqrt{49g^2(\z,\mu)+384h(\z,\mu)}\)^6(\mu^2+4)^8}
{\(\ln\(|\z|+\sqrt{\z^2+1}\)+2\z^2+1\)^6\(2\mu^3+7\mu^2+5\mu+20\)^7}\cr
&\times\(7g(\z,\mu)\sqrt{49g^2(\z,\mu)+384h(\z,\mu)}-98g^2(\z,\mu)
-352h(\z,\mu)\).
\eal
\eq184
$$
Taking $m_{\u A}\simeq m\dr{EW}=80\,$GeV (an electro-weak energy scale) and
$m\dr{pl}\simeq 2.4\cdot 10^{18}\,$GeV, $\a_s^2=\a\dr{em}=\frac1{137}$, one
gets 
$$
m_1\simeq 3.34\cdot 10^{-105}m\dr{EW}G(\z,\mu) \eq185
$$
where
$$
\al
G(\z,\mu)&=\frac{\(7g(\z,\mu)+\sqrt{49g^2(\z,\mu)+384h(\z,\mu)}\)^3(\mu^2+4)^4}
{\(\ln\(|\z|+\sqrt{\z^2+1}\)+2\z^2+1\)^3\(2\mu^3+7\mu^2+5\mu+20\)^{7/2}}\cr
&\times\(7g(\z,\mu)\sqrt{49g^2(\z,\mu)+384h(\z,\mu)}-98g^2(\z,\mu)
-352h(\z,\mu)\)^{1/2}.
\eal
\eq186
$$
In the second de Sitter phase one gets
$$
\al
m_0^2&=\frac14 \(\frac74\)^8 \(\frac{m_{\u A}}{\a_s^2 m\dr{pl}}\)^{14}
\(\frac{m^2_{\u A}}{\a_s^2}\)
\(\frac{\mu^2+4}{2\mu^3+7\mu^2+5\mu+20}\)^7\cr
&\times\(\frac{g(\z,\mu)}{\ln\(|\z|+\sqrt{\z^2+1}\)+2\z^2+1}\)^8. 
\eal \eq187
$$

Taking as before for $m_{\u A}$ an electro-weak energy scale and for $\a_s^2=
\frac1{137}$, one gets
$$
m_0\cong 10^{-102}m\dr{EW}F(\z,\mu) \eq188
$$
where
$$
F(\z,\mu)=\frac{(\mu^2+4)^{7/2}(g(\z,\mu))^4}
{\(\wiel\)^{7/2} \(\lon\)^4}\,. \eq189
$$
According to modern ideas a mass of scalar-\q\ particle should be smaller
than $10^{-3}\,$eV. For the \co\ \ct\ from the second de Sitter phase is the
same for our contemporary epoch, $m_0$~is also a mass of a scalar \q\
particle for our epoch.

It is interesting to ask what is a number of scalar-\q\ particle per unit
volume during both de Sitter phases of the evolution of the Universe (if a
\q\ has been deposed as particles which is not so obvious). (It can be
deposed in a form of a classical scalar field.)
One gets
$$
\al
\kern-10pt n_1&=\frac{\rho^1_Q}{m_1}=\frac{6H_0^2}{m_1}\cr
&=\(\frac{m_{\u A}}{\a_s}\)\(\frac{m_{\u A}}{\a^s m\dr{pl}}\)^{n/2}
(2(n+2))^{-(n+4)/4}\cdot 2^5\cr
&\times \frac
{n^2\uP^2-\uP\sqrt{n^2\uP^2+4(n^2-4)A\RG}+4(n+2)A\RG}
{\(2n^2\uP^2+4(n+2)(n-3)A\RG
-n\uP\sqrt{n^2\uP^2+4(n^2-4)A\RG)}\)^{1/2}}\cr
&\times\(-n\uP + \sqrt{n^2\uP^2+4(n^2-4)A\RG}\)^{(n-2)/4}.
\eal \eq190
$$
In the second de Sitter phase one gets
$$
\al
n_0&=\frac{\rho_Q}{m_0}=\frac{6H_1^2}{m_0}\cr
&=\(\frac{m_{\u A}}{m\dr{pl}}\)^{n/2}\(\frac{m_{\u A}}{\a_s^n}\)
\(\frac n{n+2}\)^{(n+4)/4}|\uP|^{1/2}\(\frac{|\uP|}{\RG}\)^{n/4}.
\eal
\eq191
$$
Using our simplified model with $n=14$ one gets from (18.190)
$$
\al
\kern-7pt n_1&
=2^{-19}\(\frac{m_{\u A}}{\a_sm\dr{pl}}\)^7\(\frac{m_{\u A}}{\a^s }\)
\frac{\(7g(\z,\mu)+\sqrt{49g^2(\z,\mu)+384h(\z,\mu)}\)^3}
{\(\wiel\)^{7/2}}\cr
&\times \frac{(\mu^2+4)^3\(98g^2(\z,\mu)+(\mu^2+4)
\sqrt{49g^2(\z,\mu)+384h(\z,\mu)}+32H(\z,\mu)\)}
{\(7g(\z,\mu)\sqrt{49g^2(\z,\mu)+384h(\z,\mu)}-98g^2(\z,\mu)-352h(\z,\mu)\)
^{1/2}}\cr
&\times\(\lon\)^{-1}
\eal
\eq192
$$
and from the formula (18.191)
$$
\al
n_0&=\(\frac{m_{\u A}}{m\dr{pl}}\)^7\(\frac{m_{\u A}}{\a_s^{14}}\)
\(\frac {7^{9/2}}{2^{17}}\)\cr
&\times\frac{|f(\z)|^{1/2}}{\(\lon\)^{15/2}}
\cdot \(\frac{|f(\z)|(\mu^2+4)^2}{\wiel}\)^7.
\eal
\eq193
$$

For the \co\ \ct\ from the second de Sitter phase is the same for our
contemporary epoch, $n_0$~is also a number of scalar \q\ particles for our
epoch. Taking as usual $m_{\u A}=m\dr{EW}$ and $\a_s^2=\frac1{137}$ one gets
$$
\ealn{
n_1&\simeq 9\cdot 10^{-106}m\dr{EW}\cdot K(\z,\mu) &(18.194)\cr
n_0&\simeq 6.2 \cdot 10^{-96}m\dr{EW}\cdot L(\z,\mu), &(18.195)}
$$
where
$$
\ealn{
K(\z,\mu)&=\frac{\(7g(\z,\mu)+\sqrt{49g^2(\z,\mu)+384h(\z,\mu)}\)^3(\mu^2+4)^3}
{\(\wiel\)^{7/2}}\cr
&\times \frac{\(98g^2(\z,\mu)+(\mu^2+4)
\sqrt{49g^2(\z,\mu)+384h(\z,\mu)}+32H(\z,\mu)\)}
{\(7g(\z,\mu)\sqrt{49g^2(\z,\mu)+384h(\z,\mu)}-98g^2(\z,\mu)-352h(\z,\mu)\)
^{1/2}},\qquad\quad &(18.196)\cr
L(\z,\mu)&=\frac{|f(\z)|^{1/2}}{\(\lon\)^{15/2}}
\cdot \(\frac{|f(\z)|(\mu^2+4)^2}{\wiel}\)^7,&(18.197)
}
$$
$f(\z)$, $g(\z,\mu)$, $h(\z,\mu)$, $H(\z,\mu)$ are given by the formulae
(18.143--145). In this way the masses of scalar particles and their numbers per unit
volume in both de Sitter phases and for our contemporary epoch depend on
geometric parameters in our theory.

We can connect $\o \mu$ and $N$ from the formula (18.101):
$$
\o\mu=\frac{9.8}{N}\,. \eq198
$$
If $N\simeq60$, one gets
$$
\o\mu=0.16 \eq199
$$
and
$$
5{\o\mu}^2=0.13. \eq200
$$
Using Eq.\ (18.83) one finds:
$$
n_s(K)=1\pm0.13\,\D\ln K. \eq201
$$

We can also find $\o\mu$ using the formulae (18.113) and (18.113a). One gets
$$
\o\mu=\(\frac{m\dr{pl}}{m_{\u A}}\)^2 \a_s^4 F(\b,\g,\a_1)=
\(\frac{m\dr{pl}}{m_{\u A}}\)^4 F(\RG,\uP,A). \eq202
$$
Using Eq.\ (18.83) one also gets
$$
n_s(K)=1\pm5\(\frac{m\dr{pl}}{m_{\u A}}\)^8 F^2(\RG,\uP,A)\, \D\ln K. \eq203
$$
Using our simplified model on gets
$$
\al
\o\mu&=81.67\(\frac{m\dr{pl}}{m_{\u A}}\)^4
\frac{7g(\z,\mu)+\sqrt{49g^2(\z,\mu)+384h(\z,\mu)}}
{(\mu^2+4)^2(1+\z^2)}\\
&\times
\frac{|\z|\((5\z^2+4)\ln 2-2\mu^2(1+\z^2)\)}
{\(98g^2(\z,\mu)+(\mu^2+4)\sqrt{49g^2(\z,\mu)+384h(\z,\mu)}+32H(\z,\mu)\)}
\eal
\eq204
$$
and respectively
$$
\al
n_s(K)&=1\pm408.35\,
\frac{\(7g(\z,\mu)+\sqrt{49g^2(\z,\mu)+384h(\z,\mu)}\)^2}
{(\mu^2+4)^4(1+\z^2)^2}\\
&\times
\frac{|\z|^2\((5\z^2+4)\ln 2-2\mu^2(1+\z^2)\)^2\,\D\ln K}
{\(98g^2(\z,\mu)+(\mu^2+4)\sqrt{49g^2(\z,\mu)+384h(\z,\mu)}+32H(\z,\mu)\)^2}\,.
\eal
\eq205
$$

It is easy to see that we can get $n_s(K)$ close to~1 (a flat power spectral
function) using parameters $\mu$ and~$\z$ in such a way that
$$
(5\z^2+4)\ln2 - 2\mu^2(1+\z^2)
$$
is close to zero. One can express $\cos\f$ (see Eq.~(18.5)) in
terms of parameters of the theory and get
$$
\al
\cos\f&=-\(\frac{m_{\u A}}{\a_s^2 m\dr{pl}}\)^{n/2}
\frac{P_1^{n/4}}
{3\cdot2^{(n+8)/4}(n+2)^{n/4}\(\RG\)^{n/4}}\\
&\times\frac{40A\(\dfrac{m_{\u A}}{\a_s^2 m\dr{pl}}\)^n m_{\u A}^2
P_1^{n/2}
+9B\a_s^4 2^{n/2}(n+2)^{n/2}\(\RG\)^{n/2}}
{5A\(\dfrac{m_{\u A}}{\a_s^2 m\dr{pl}}\)^n m_{\u A}^2
P_1^{n/2}
-3\a_s^4 B 2^{n/2}(n+2)^{n/2}\(\RG\)^{n/2}}\,,
\eal
\eq206
$$
where
$$
P_1=-n\uP + \sqrt{n^2\uP^2+4(n^2-4)A\RG}. \eq207
$$
For the \ct\ $a$ (see Eq.~(18.6)) one gets
$$
a=\a_s^2\(\frac{\sqrt B}{m_{\u A}}\)\(\frac{\a_s^2 m\dr{pl}}{m_{\u A}}\)^{n/2}
\frac{2^{n/4}(n+2)^{n/4}\(\RG\)^{n/4}}
{\sqrt A\(-n\uP + \sqrt{n^2\uP^2+4(n^2-4)A\RG}\)^{n/4}}
\eq208
$$
(see Refs [129], [130]).

}\vfill\eject

{\def\eq#1 {\eqno(19.\text{\rm#1})}
\def\rf#1 {\text{\rm(19.#1)}}
\let\al\aligned
\let\eal\endaligned
\let\ealn\eqalignno
\let\dsl\displaylines
\def\cL{\Cal L}
\let\o\overline
\let\ul\underline
\let\stpr\overrightarrow
\let\a\alpha
\let\b\beta
\let\d\delta
\let\e\varepsilon
\let\f\varphi
\let\D\Delta
\let\g\gamma
\let\G\varGamma
\let\k\kappa
\let\om\omega
\let\P\varPsi
\let\F\varPhi
\let\la\lambda
\let\La\varLambda
\let\s\sigma
\let\z\zeta
\let\pa\partial
\let\t\widetilde
\let\u\tilde
\def\sgn{\operatorname{sgn}}
\def\Sh{\operatorname{sinh}}
\def\Ch{\operatorname{cosh}}
\def\ctgh{\operatorname{ctgh}}
\def\tgh{\operatorname{tgh}}
\def\ar{\operatorname{arccos}}
\def\ars{\operatorname{arsinh}}
\def\ns{\operatorname{ns}}
\def\sc{\operatorname{sc}}
\def\sh{\operatorname{sinh}}
\def\sn{\operatorname{sn}}
\def\cn{\operatorname{cn}}
\def\dn{\operatorname{dn}}
\def\sd{\operatorname{sd}}
\def\cd{\operatorname{cd}}
\def\nd{\operatorname{nd}}
\def\arctg{\operatorname{arctg}}
\def\ctg{\operatorname{ctg}}
\def\tg{\operatorname{tg}}
\def\Li{\operatorname{Li}}
\def\Re{\operatorname{Re}}
\def\uP{\ul{\t P}}
\def\RG{\t R(\t\G)}
\def\wiel{2\mu^3+7\mu^2+5\mu+20}
\def\lnz{\ln\(|\z|+\sqrt{\z^2+1}\)+2\z^2+1}
\def\mpl{m\dr{pl}}
\def\br#1#2{\overset\text{\rm #1}\to{#2}}
\def\ddt{\frac d{dt}}
\def\({\left(}\def\){\right)}
\def\[{\left[}\def\]{\right]}
\def\dr#1{_{\text{\rm#1}}}
\def\gi#1{^{\text{\rm#1}}}
\def\ap{approximation}
\def\ce{contemporary epoch}
\def\ct{constant}
\def\co{cosmological}
\def\dc{dependen}
\def\ef{effective}
\def\ex{exponential}
\def\gr{gravitational}
\def\il{inflation}
\def\lg{lagrangian}
\def\pt{parameter}
\def\pc{particle}
\def\q{quint\-essence}
\def\sy{symmetr}
\def\tr{transform}
\def\U{Universe}
\def\wrt{with respect to }
\def\KK{Kaluza--Klein}
\def\JT{Jordan--Thiry}
\def\nos{nonsymmetric}
\def\up#1{\uppercase{#1}}
\def\eu{\expandafter\up}
\def\dint{-\kern-11pt\intop}
\def\tin#1{\hbox to\parindent{\hfil #1)\enspace}}

\null
\vskip36pt plus4pt minus4pt
\centerline{{\four\bf 19. Properties of a Quintessence}}
\vskip24pt plus2pt minus2pt
\write0{19. Properties of a Quintessence \the\pageno}

In this section we consider several consequences of the \eu\nos\ \KK\ (\JT)
Theory. We consider a value of a mass of \q\ \pc, several interesting
relations among energy scales, radiation density in the second de Sitter phase.
We find a spatial dependence of a \q\ field (and an \ef\ \gr\ \ct~$G\dr{eff}$)
in a case of spherical-statical \sy y, cylindrical-statical \sy y,
flat-static \sy y. We find a time \dc ces of a \q\ field (with no spatial
\dc ce). We get a solution for a \q\ field (a~travelling wave) and
two-dimensional wave  solution applying those solutions for~$G\dr{eff}$. We
propose some kind of statistical approach to our results. We calculate a
speed of sound in a \q.

Let us consider a value of mass of a \q\ \pc\ (a~scalar \pc) (18.183)
obtained from \eu\nos\ \KK\ (\JT) Theory:
$$
m_0=\sqrt{\frac n2}\(\frac n{n+2}\)^{n/4} |\g|^{1/2}
\(\frac{|\g|}\b \)^{n/4}. \eq1
$$

The value of this mass has been obtained by this
\pc\ during the second de Sitter phase. Moreover during our contemporary
epoch it is the same. The \pt s $n$, $\g$ and~$\b$ are defined in Sec.~14
(Eq.~(14.33)). During the second de Sitter phase a \co\ \ct\ has been calculated
and one finds
$$
\la_{c0}=6H_1^2=\frac{2|\g|^{(n+2)/2}n^{n/2}}{\b^{n/2}(n+2)^{(n+2)/2}}\,.
\eq2
$$
The value of a \co\ \ct\ remains the same during our contemporary epoch.
$H_1$~is a Hubble \ct\ during the second de Sitter phase (see Eq.~(14.79)). 

Thus for our contemporary epoch one gets
$$
m_0=\frac12 \sqrt{n(n+2)}\,\sqrt{\la_{c0}}\,. \eq3
$$
In this way we can connect the value of a \co\ \ct\ of our contemporary epoch
to the value of a mass of a \q\ \pc. From recent observational data we get
$$
\la_{c0}=\La=10^{-52}\frac1{\text{\rm m}^2}\,. \eq4
$$
Moreover in order to get a correct dimension for a mass we should add a
factor with~$\hbar$ and~$c$ (now we abandon the system of units with
$\hbar=c=1$).

One gets
$$
m_0=\frac12 \frac{\hbar}c \,\sqrt{n(n+2)}\,\sqrt{\la_{c0}}\,. \eq3a
$$
Using the value of $\la_{c0}$ one finally gets
$$
m_0\simeq \sqrt{n(n+2)}\cdot 0.17\cdot 10^{-39}\,\text{g} \eq5
$$
or
$$
m_0\simeq \sqrt{n(n+2)}\cdot 0.95\cdot 10^{-5}\,\text{eV}. \eq5a
$$
For example, if we take $n=14(=\dim G2)$, one finally gets
$$
m_0\simeq 14.2\cdot 10^{-5}\,\text{eV}. \eq6
$$

This value is bigger than that considered by different authors. Moreover, 
still sufficiently small. The \pc\ interacts only gravitationally and because
of this it is undetectable by using known experimental methods.
Taking a density of dark energy as $0.7$ of a critical density,
$$
\rho_c=1.88 h^2 \cdot 10^{-29}\frac{\text{g}}{\text{cm}^3}\,, \eq7
$$
one gets a number of \q\ \pc s per unit volume
$$
\o n=\frac{h^2}{\sqrt{n(n+2)}}\cdot1.31 \cdot10^{10}\, \frac1{\text{cm}^3} \eq8
$$
where $h$ is a dimensionless Hubble \ct\ $0.7<h<1$. Taking $n=14$ and $h=0.7$
one finally gets
$$
\o n=4\cdot10^8\,\frac1{\text{cm}^3} \eq8a
$$
which is many orders of magnitude smaller than Loschmidt number. Thus a gas
of \q\ \pc s is not so dense from the point of view of our earth conditions.
However, if this number of \pc s per unit volume is considered in a container of
size 200\,Mpc, the gas can be considered as extremely dense.

In order to settle---is this gas dense or not---we should calculate a mean
scattering length. The scattering cross-section for a \q\ \pc
$$
\sigma=\frac1{\la_{c0}}=10^{52}\,\text{m}^2. \eqno(*)
$$
A mean scattering length
$$
l=\frac1{\sigma n} \eqno(*{*})
$$
where $n$ is a number of \q\ \pc s per unit volume (Eq.~\rf8a ).

One gets
$$
l=10^{-60}\,\text{m}. \eqno(*{*}*)
$$
It means that a gas of \q\ \pc s is extremely dense (if we apply the Knudsen
criterion---a gas is dense if $l\ll L$, where $L$ is the size of the
container) even in the Solar System.

Let us consider the Eq.~(16.170) which connects several scales of
energy and gives an account of the smallness of \gr\ interactions in our
theory. We rewrite this equation in the form
$$
\(\frac{m\dr{pl}}{m\dr{EW}}\)\(\frac{m_{\u A}}{m\dr{EW}}\)^{n_1}=
\(\frac{n|\g|}{(n+2)\b}\)^{(n+2)(n_1+2)/2} 
\eq9
$$
where $m\dr{EW}$ is an electro-weak interactions energy scale. Taking
$$
\al
&m_{\u A}=m\dr{EW}\\
&n_1=2\ (M=S^2)\\
&n=14\ (H=G2)
\eal
$$
one gets
$$
\frac{m\dr{pl}}{m\dr{EW}}=\(\frac78\,\k\)^{24} \eq10
$$
where
$$
\k=\frac{|\g|}{\b} \eq11
$$
and eventually one gets
$$
\k=\(\frac{m\dr{pl}}{m\dr{EW}}\)^{1/24}\cdot\frac87\,. \eq12
$$
Taking $m\dr{EW}\simeq80\,$GeV and $m\dr{pl}\simeq2.4\cdot10^{18}\,$GeV one
gets 
$$
\k\simeq6.19 \eq13
$$
which is very reasonable for it establishes a relation between two \co\ terms
$\g$ and~$\b$ as of the same order. Simultaneously this is a consistency
condition for our model with energy scales (the 6-dimensional Planck's mass
is equal to~$m\dr{EW}$). In this way we can calculate a mass of a \q\ \pc\
for our \ce\ from Eq.~\rf1
$$
m_0=2\cdot10^{15}\,\text{eV}\cdot|\uP|^{1/2}. \eq14
$$
This gives us an estimation for $|\uP|$ and $\RG$:
$$
\RG=\frac1\k \(\frac{m\dr{EW}}{m\dr{pl}\a\dr{em}}\)^2|\uP| \eq15
$$
where $\a\dr{em}=\frac1{137}$ is a fine coupling \ct. Using Eq.~\rf13 \ and
values of $m\dr{pl}$ and $m\dr{EW}$ one gets
$$
\RG=24.75\cdot 10^{-31}|\uP|. \eq16
$$
From Eq.\ \rf14 \ and Eq.\ \rf6 \ one gets
$$
|\uP|=5\cdot10^{-39} \eq17
$$
and
$$
\RG\cong 1.2\cdot 10^{-68}. \eq18
$$
Moreover in our simplified theory we have
$$
\RG=\frac{2(\wiel)}{(\mu^2+4)^2} \eq19
$$
and $\mu$ should be very close to the root of the polynomial
$$
W(\mu)=\wiel. \eq20
$$
From \rf18 \ and \rf19 \ one gets
$$
\wiel\cong1.3\cdot10^{-66}. \eq21
$$
Thus $\mu$ is very close to the 70-digit approximation of the root of the
polynomial \rf20 \ ($W(\o \mu)=0.1\cdot10^{-67}$).
Due to this we can control the \co\ terms.

In the case of $|\uP|$ we have the formula (8.64) and one gets
$$
\ealn{
&\uP(\z_0+\e)\cong \uP(\z_0)+\frac{d\uP}{d\z}(\z_0)\e &\rf22 \cr
&\uP(\z_0=\pm1.38\ldots)=0 &\rf23 \cr
&\Bigl|\frac{d\uP}{d\z}(|\z_0|=1.38\ldots)\Bigr|=25. &\rf24 
}
$$
Thus one gets from \rf17 \ and \rf23--24 
$$
\e\simeq 2\cdot10^{-40}. \eq25
$$
Thus we need an \ap\ of $\z_0$ up 40-digit arithmetics. We see that \co\
terms coming from the \eu\nos\ \KK\ (\JT) Theory are very small, but not
zero, and that they are easily controllable by $\mu$ and~$\z$ \pt s.

Let us consider a self-interaction potential for a \q\ field for our \ce\
(which is the same as for the second de Sitter phase). One gets from \co\
terms $\la_{c0}(\P)$
$$
\ealn{
\la_{c0}(\P)&=-\frac12\(\b e^{2\P}-|\g|\)e^{n\P} &\rf26 \cr
U(q_0)&=-\frac{|\g|}{2(n+2)}\(\frac{n|\g|}{(n+2)\b}\)^\frac n2
\exp\(\frac{nm\dr{pl}}{2\sqrt{2\pi|\o M|}}q_0\)\cr
&\qquad{}\times\(n\exp\(\frac{m\dr{pl}}{\sqrt{2\pi|\o M|}}q_0\)-(n+2)\) &\rf27 \cr
}
$$
or
$$
U(q_0)=-\frac14\la_{c0}
\exp\(\frac{nm\dr{pl}}{2\sqrt{2\pi|\o M|}}q_0\)
\(n\exp\(\frac{m\dr{pl}}{\sqrt{2\pi|\o M|}}q_0\)-(n+2)\) 
\eq27a
$$
where $\la_{c0}$ is a \co\ \ct\ for our \ce\ and
$$
\P=\P_0+\frac{m\dr{pl}}{2\sqrt{2\pi|\o M|}}q_0=\P_0+\o\b q_0=\P_0+\f. \eq28
$$
For $\la_{c0}$ is very small ($10^{-52}\frac1{\text{m}^2}$), this interaction
is very small. For $n=14(=\dim H=\dim G2)$ one gets
$$
U(q_0)=-\frac12\la_{c0}
\exp\(\frac{7m\dr{pl}}{\sqrt{2\pi|\o M|}}q_0\)
\(7\exp\(\frac{m\dr{pl}}{\sqrt{2\pi|\o M|}}q_0\)-8\) .
\eq29
$$

Even in the potential $U(q_0)$ we have exponential terms, the strength of the
interactions is negligible for small value of~$q_0$. The interesting point in
our theory is the \ef\ \gr\ \ct\ (depending on a scalar field~$\P$). Let us
describe it by a \q\ field~$q_0$. One gets
$$
G\dr{eff}=G_0e^{-(n+2)\P}. \eq30
$$
After some calculations one finds
$$
G\dr{eff}=G_0\(\frac{(n+2)\b}{n|\g|}\)^{n+2}
\exp\(-\frac{n+2}{2} \frac{m\dr{pl}}{\sqrt{2\pi|\o M|}}q_0\) \eq31
$$
or
$$
G\dr{eff}=\frac{G_0}{\la_{c0}^2}\cdot \frac{4(n+2)^2\b^2}{n^2}
\exp\(-\frac{n+2}{2} \frac{m\dr{pl}}{\sqrt{2\pi|\o M|}}q_0\). \eq31a
$$
It is easy to see that the r\^ole of a Newton's \ct~$G_N$ is played by
$$
G_N=\frac{G_0}{\la_{c0}^2}\(\frac{2(n+2)\b}{n}\)^2. \eq32
$$
This is the reason that we put a \ct\ $G_0$ in the formula \rf30 .

One gets
$$
G_0=G_N\la_{c0}^2\(\frac{n}{2(n+2)\b}\)^2. \eq33
$$
For 
$$
\b=\a^2_sm\dr{pl}^2\RG=\a^2_s\frac{\RG}{l\dr{pl}^2} \eq34
$$
one gets
$$
G_N=\frac{G_0\a_s^4}{(\la_{c0}l\dr{pl}^2)^2}\(\frac{2(n+2)\RG}{n}\)^2 \eq35
$$
or
$$
G_0=G_N\(\frac{n}{2(n+2)\RG}\)^2\(\frac{\la_{c0}l\dr{pl}^2}{\a_s^2}\)^2. \eq36
$$

Let us connect to $G_0$ a new Planck's mass $m^0\dr{pl}$. In terms of this
mass and ordinary $m\dr{pl}$ mass one gets
$$
m^0\dr{pl}=m\dr{pl}\,\frac{2(n+2)\RG}{n}\,\frac{\a^2_s}{\la_{c0}l^2\dr{pl}}
\eq37
$$
or
$$
m\dr{pl}=m^0\dr{pl}\,\frac{n}{2(n+2)\RG}\,\frac{\la_{c0}l^2\dr{pl}}{\a^2_s}\,.
\eq38
$$

If we take our simplified model with $\RG$ given by Eq.~\rf19
$$
\a_s^2=\a\dr{em}=\tfrac1{137},\ n=14,
$$
we get
$$
\mpl^0=\mpl\,\frac{32(\wiel)}{7(\mu^2+4)^2}\,\frac{\a\dr{em}}{\la_{c0}
l^2\dr{pl}} \eq39
$$
or
$$
G_0=G_N\(\frac{7(\mu^2+4)^2}{32(\wiel)}\)^2\(\frac{\la_{c0}l^2\dr{pl}}
{\a\dr{em}}\)^2. \eq40
$$
Moreover we can connect unknown \ct~$G_0$ ($\mpl^0$) with an energy scale of
electro-weak interactions $m\dr{EW}$ in a way suggested and
developed here (Sec.~16).

It is enough to use the formula \rf10 \ and we get
$$
\mpl^0=m\dr{EW}\(\frac{7\k}8\)^{24}
\frac{32(\wiel)}{7(\mu^2+4)^2}\,\frac{\a\dr{em}}{\la_{c0}l^2\dr{pl}}. \eq41
$$
If we take for $\la_{c0}$ the known value of the \co\ \ct\ (Eq.~\rf4 ) for
$l\dr{pl}=4\cdot 10^{-32}\,\text{cm}$ and $\k=6.19$ we get
$$
\mpl^0=8.12\cdot10^{132}\cdot\frac{\wiel}{(\mu^2+4)^2}\,m\dr{EW}\,. \eq42
$$
Moreover we see that $\mu$ must be very close to the root of the polynomial
$\wiel$ (see (16.199)) and we get
$$
\mpl^0=2\cdot10^{130}(\wiel)m\dr{EW}\,. \eq43
$$
If we take 70-digit \ap\ of the root of the polynomial \rf20 \ and considered
value of $W(\mu)$ (Eq.~\rf18 ) obtained here, we get
$$
\mpl^0=4.6\cdot 10^{64}m\dr{EW} \eq44
$$
or
$$
\mpl^0=4.6\cdot 10^{64}\,\frac{m\dr{EW}}{\mpl}\,\mpl=
1.5\cdot 10^{48}\mpl = 3.2\cdot 10^{43}\text{g}=1.6\cdot10^{10}M_\odot
\eq45
$$
where $M_\odot=1.99\cdot10^{33}\,$g is a Solar mass. (This is a mass of our
Galaxy.)

Moreover the present critical density of matter in the \U\ is
$$
\rho_{c,0}=2.775h^{-1}\times 10^{11}\,\frac{M_\odot}{(h^{-1}\,\text{Mpc})^3}\,.
\eq46
$$
So we can find a volume containing $\mpl^0$. One gets
$$
V_0=\frac{\mpl^0}{\rho_{c,0}}=5\cdot10^{-3}h^2(\text{Mpc})^3. \eq47
$$
Thus the size of this volume is of order
$$
L_0\simeq \root 3 \of{h^2}\,0.17\,\text{Mpc}. \eq48
$$

Taking under consideration the fact that our calculations are quite rough
(for example $\k$ could be bigger a little, or $W(\t \mu)$ a little bigger),
we come to the conclusion that $\mpl^0$ is of order of the mass of our visible
\U\ (a~volume of size of $10^4$\,Mpc, see the second point of Ref.~[129]). In this way it seems natural to suppose that
$$
\mpl^0=M_U \eq49
$$
where $M_U$ is the total mass of the visible \U. However, this intriguing
conjecture demands many investigations.

Moreover it is in a real spirit of Dirac's large number hypothesis
and can give a link between cosmology and fundamental interactions theory.
Thus if we suppose \rf49 \ we get
$$
G_0=G_U=\frac1{M_U^2}\,. \eq50
$$
Let us come back to the equation for an \ef\ \gr\ \ct\ (Eq.~\rf32 ):
$$
G\dr{eff}=G_N \exp\(-\frac{n+2}2\,\frac{\mpl}{\sqrt{2\pi |\o M|}}\,q_0\).
\eq51
$$
$q_0$---a \q\ scalar field---possesses a mass and because of this it has a
finite range with Youkawa type behaviour
$$
q_0=\frac{\a}{R}\exp\(-\frac R{r_0}\) \eq52
$$
where 
$$
r_0=\frac2{\sqrt{n(n+2)}\sqrt{\la_{c0}}}\quad\hbox{and $\a$ is a positive
\ct} \eq53
$$
or taking for $n=14$ and for $\la_{c0}$ Eq.~\rf4 ,
$$
r_0=3\cdot 10^{23}\,\text{m}=10\,\text{Mpc}. \eq54
$$
The formula \rf52 \ is valid for $R\simeq r_0$. Thus we cannot observe scalar
\gr\ radiation from closed binary sources and the quadrupole radiation
formula is satisfied for a \gr\ field.

Moreover for $R<r_0$ we should take under consideration a full
self-interaction potential of a field~$q_0$ (Eq.~\rf29 ). Moreover, because
of the \ct\ $\la_{c0}$ in front of the formula \rf29 \ this is negligible.
Because of this we can repeat some considerations from Sections 10 and~14
coming back to the field~$\P$ and consider different sources of mass
for this field due to interaction with the matter. In this way in both cases
$$
G\dr{eff}=G_N\,. \eq55
$$
However, we can expect some small effects on short distances.

Let us notice that in the previous \ap\ we consider a weak field, it means
$|q_0|\ll 1$. This is different from small field. A small field considered
below is such that $q_0$ is small in a usual sense. It is negative. Moreover
it can be big in the sense of the absolute value.

Let us calculate a radiation density in the moment when the second de Sitter
phase starts. It means the radiation released after the phase transition. One
gets
$$
\rho_r=\la_{c1}(\P_1)-\la_{c0}(\P_0)=
\la_{c0}(\P_0)\(\frac{H_0^2}{H_1^2}-1\). \eq56
$$
Using Eqs (18.103a) and (18.111a) one gets
$$
\al
\rho_r&=\la_{c0}(\P_0)\cr
&\times \(\frac{n^2|\uP|^2+|\uP|\sqrt{n^2|\uP|^2+4(n^2-4)A\RG}
+4A\RG(n+2)}{2^{(n+2)/2}|\uP|^{(n+2)/2}n^{n/2}}\right.\cr
&\qquad{}\times\left.\vphantom{\frac{\sqrt{|\uP|^2}}{|\uP|^{(n+2)/2}}}
\(n|\uP|+\sqrt{n^2|\uP|^2+4(n^2-4)A\RG}\)^{(n-2)/2}
-1\),
\eal
\eq57
$$
$$
\la_{c0}(\P_0)=6H_1^2=2\(\frac{m_{\u A}}{\mpl}\)^n
\frac{m_{\u A}^2}{\a_s^{2(n+1)}} \cdot
\frac{|\uP|^{(n+2)/2}n^{n/2}}{(n+2)^{(n+2)/2}\(\RG\)^{n/2}}.
\eq58
$$
The density $\rho_r$ is of course an \ef\ radiation density, because we
adsorbe into $\rho_r$ a factor with an \ef\ \gr\ \ct.

In our simplified model we get
$$
\al
\rho_r&=\La
\(\frac{98g^2(\z,\mu)+(\mu^2+4)\sqrt{49g^2(\z,\mu)+384h(\z,\mu)}
+32H(\z,\mu)}
{g^8(\z,\mu)(\mu^2+4)^2\cdot 2^{11}\cdot7^{14}}\right.\cr
&\qquad\qquad{}\times
\(7g(\z,\mu)+\sqrt{49g^2(\z,\mu)+384h(\z,\mu)}\)^6\cr
&\qquad\qquad{}\times\left.\vphantom{\frac{\sqrt{49g^2(\z,\mu)}}{(\mu^2+4)^2}}
\(\ln\(|\z|+\sqrt{\z^2+1}\)+2\z^2+1\)^6-1\) ,
\eal
\eq59
$$
$$
\al
\La&=\(\frac{m_{\u A}}{\mpl}\)^{14}\cdot\frac{m_{\u A}^2}{\a_s^{36}}
\cdot\frac{7^{14}}{2^{24}}\cr 
&\times\frac{g^8(\z,\mu)(\mu^2+4)^8}
{(\wiel)^7\(\ln\(|\z|+\sqrt{\z^2+1}\)+2\z^2+1\)^8}\,.
\eal
\eq60
$$
In the formulas \rf59--60 \ we should put $m_{\u A}=m\dr{EW}$ and
$\a_s^2=\a\dr{em}=\frac1{137}$. Let us notice that $\La$ is our contemporary
\co\ \ct\ given by \rf4 . It gives a scale for $\rho_r$.

The functions $g(\z,\mu)$, $h(\z,\mu)$, $H(\z,\mu)$ are given by Eqs (18.143),
(18.144a),\break (18.144b), (18.145). The numerical factor in front of~$\La$ can
be calculated  and we get
$$
\La=1.04\cdot 10^{-178}\,m\dr{EW}^2a(\z,\mu) \eq61
$$
where
$$
a(\z,\mu)=\frac{g^8(\z,\mu)(\mu^2+4)^8}{(\wiel)^7\(\lnz\)^8}\,. \eq62
$$
The calculated radiation density evolves in time according to the theory
developed in Sec.~14.

Let us write an equation for the scalar field $\P$ in terms of the scalar
field $\f$ (see Eq.\ (19.28)). One easily gets
$$
g^{\a\b}({\t\nabla}_\a \pa_\b \f)-\frac{n(n+2)}{16\pi M}\,\la_{c0}
m\dr{pl}^2e^{n\f}(e^{2\f}-1)=0 \eq63a
$$
or in terms of $q_0$
$$
g^{\a\b}({\t\nabla}_\a \pa_\b q_0)+\t\e\o \a(\exp(2\o\b q_0)-1)=0.
\eq63b
$$

In order to find some influence of $q_0$ (\q) field on the value of the \ef\
\gr\ \ct\ we consider a field equation for the scalar $q_0$ field in empty
space. One gets
$$
\(\frac{\pa^2q_0}{\pa t^2}-{\stpr\nabla}{}^2 q_0\)+
\t \e \o \a \exp(n\o\b q_0)\(\exp(2\o\b q_0)-1\)=0 \eq63c
$$
where
$$
\ealn{
\o\a&=\frac{\la_{c0}n(n+2)\mpl^2}{16\pi\cdot\o M} &\rf64 \cr
\o\b&=\frac{\mpl}{2\sqrt{2\pi|\o M|}} &\rf65 \cr
\t\e&=\sgn\o M, \ \t\e{}^2=1.&\rf66
}
$$

Let us consider a static, spherically \sy ic case. In the spherical
coordinates one gets
$$
0=\frac1{r^2} \frac{d}{dr}\(r^2\frac{dq_0}{dr}\) - \t\e\o\a
\exp(n\o\b q_0)\(\exp(2\o\b q_0)-1\) \eq67
$$
where $q_0=q_0(r)$ is a function of $r$ only. In order to treat this equation
it is easier to come back to the old variable $\f=\o\b q_0$ (see Eq.~\rf28 ).
One gets
$$
\frac1{r^2} \frac{d}{dr}\(r^2\frac{d\f}{dr}\) - \frac14\t\e\o\la_{c0}
\exp(n\f)\(\exp(2\f)-1\)=0. \eq68
$$
We change the in\dc t variable $r$ into $\tau$
$$
\frac1{\tau^2} \frac{d}{d\tau}\(\tau^2\frac{d\f}{d\tau}\) - \t\e
\exp(n\f)\(\exp(2\f)-1\)=0 \eq69
$$
where
$$
r=\frac2{\sqrt{\o\la_{c0}}}\,\tau \eq70
$$
and
$$
\o\la_{c0}=\frac{n(n+2)\la_{c0}}{8\pi\o M}\,\mpl^2\,. \eq71
$$

We consider Eq.\ \rf69 \ in two regions:
\item{1)} for small fields $\f$,
\item{2)} for large fields $\f$.

In the first region we get
$$
\frac1{\tau^2} \frac{d}{d\tau}\(\tau^2\frac{d\f}{d\tau}\) + \t\e
e^{n\f}=0. \eq72
$$
In the second region we get
$$
\frac1{\tau^2} \frac{d}{d\tau}\(\tau^2\frac{d\f}{d\tau}\) - \t\e
e^{(n+2)\f}=0. \eq73
$$
Let us notice that both equations have similar nature and can be reduced to
the equation
$$
\frac1{x^2} \frac{d}{dx}\(x^2\frac{dy}{dx}\) +\e \t\e e^y=0 \eq74
$$
where in the first region $\e=1$,
$$
\ealn{
y&=n\f,&\rf75 \cr
x&=\sqrt n\,\tau, &\rf76
}
$$
and in the second region $\e=-1$,
$$
\ealn{
y&=(n+2)\f,&\rf77 \cr
x&=\sqrt {n+2}\,\tau, &\rf78
}
$$
We can transform \rf74 \ into
$$
x\,\frac{d^2y}{dx^2}+2\,\frac{dy}{dx}+\e\t\e xe^y=0 \eq79
$$
which is the celebrated Emden-Fowler equation known in the theory of gaseous
spheres (see~[131]).
Let us notice that the first region (small fields) means large distances and
the second region (large fields) means small distances. 

In this way we should consider Eq.~\rf79 \ in the region of small and
large~$x$. In the case of $\e\t\e=1$ the equation \rf74 \ has an exact
solution 
$$
y=\ln\(\frac2{x^2}\). \eq80
$$
Let us apply this to both regions (remembering that $\t\e$ in both cases has
a different sign).

One gets in the first region
$$
q_0=-\frac{2\sqrt{2\pi|\o M|}}{\mpl n} \ln\(\frac{r}{\frac{2\sqrt2}
{\sqrt{n\o\la_{c0}}}}\) \eq81
$$
and
$$
G\dr{eff}=G_N\(\frac{r}{\frac{2\sqrt2}
{\sqrt{n\o\la_{c0}}}}\)^{(n+2)/n}. \eq82
$$
In the second region
$$
q_0=-\frac{2\sqrt{2\pi|\o M|}}{\mpl (n+2)} \ln\(\frac{r}{\frac{2\sqrt2}
{\sqrt{(n+2)\o\la_{c0}}}}\) \eq83
$$
and
$$
G\dr{eff}=G_N\(\frac{r}{\frac{2\sqrt2}
{\sqrt{(n+2)\o\la_{c0}}}}\). \eq84
$$
In this way we get an interesting prediction for the behaviour of the
strength of \gr\ interactions.
In this very special solution $G\dr{eff}$ is going to zero if $r\to0$ and to
infinity if $r\to\infty$. 

Let us come to Eq.~\rf79 \ supposing $\e\t\e=1$. Thus we get
$$
x\,\frac{d^2y}{dx^2}+2\,\frac{dy}{dx}+ xe^y=0. \eq85
$$
Using an exact solution \rf80 \ we write
$$
y=y_1+\t y \eq86
$$
and consider Eq.~\rf85 \ for large $x$.

In this way we get an approximate solution (given by Chandrasekhar~[131])
$$
y=\ln\(\frac2{\eta^2}\)+\frac A{\sqrt\eta}\cos\(\frac{\sqrt7}2 \ln\eta\)
-2\ln\o\d, \quad |A|\ll 1, \eq87
$$
where
$\eta=\frac x{\,\o\d\,}$, $A$ and $\o\d$ are integration \ct s, $\o\d>0$. In this
way we get in the first region
$$
G\dr{eff}=\o G\dr{eff}\cdot\exp\(-\frac{\o A}{\sqrt \eta}\cos\(\frac{\sqrt7}
2 \ln\eta\)\) \eq88a
$$
where $\o A$ is a \ct\ ($|\o A|\ll 1$) and
$$
\eta=\frac{\sqrt{n\o\la_{c0}}\,\o\d}{2\sqrt2}\,r, \eq88b
$$
$\o G\dr{eff}$ is given by the formula \rf82 .

In the second region
$$
G\dr{eff}=\o{\o G}\dr{eff}\cdot\exp\(-\frac{\o{\o A}}{\sqrt \eta}\cos\(\frac{\sqrt7}
2 \ln\eta\)\)\eq88c
$$
where $\o{\o A}$ is a \ct\ ($|\o{\o A}|\ll1$), $\o{\o G}\dr{eff}$ is given by
the formula \rf84 \ and
$$
\eta=\frac{\sqrt{(n+2)\o\la_{c0}}\,\o\d}{2\sqrt2}\,r. \eq88d
$$
In this way we have very interesting non-Newtonian behaviour of $G\dr{eff}$
for large distances. Let us notice that the length scale is completely
arbitrary, because it is given by an integration \ct~$\o\d$. 

Let us consider Eq.\ \rf63 \ in Cartesian coordinates supposing flat \sy y
for a \q\ field $q_0=q_0(z,t)$ (nonstatic). One gets
$$
\(\frac{\pa^2q_0}{\pa t^2}-\frac{\pa^2q_0}{\pa z^2}\)
-\t \e \o \a \exp(n\o\b q_0)\(\exp(2\o\b q_0)-1\)=0. \eq89
$$
Let us change \dc t and in\dc t variables to $\xi,\eta,\f$:
$$
\ealn{
z&=\frac2{\sqrt{\o\la_{c0}}}\,\xi &\rf90 \cr
t&=\frac2{\sqrt{\o\la_{c0}}}\,\eta &\rf91 \cr
\f&=\o\b q_0. &\rf92 
}
$$
One gets
$$
\(\frac{\pa^2\f}{\pa \eta^2}-\frac{\pa^2\f}{\pa \xi^2}\)
-\t\e e^{n\f}\(e^{2\f}-1\)=0. \eq93
$$
Eq.\ \rf93 \ is an equation for flat scalar (\q) waves in our theory.
Let us consider it for large and small field~$\f$ (as before).

In this way one gets the equation
$$
\(\frac{\pa^2 y}{\pa T^2}-\frac{\pa^2y}{\pa x^2}\)
+\e\t\e e^y=0, \eq94
$$
where in the region of small field $\f$ we have $\e=1$ and
$$
\ealn{
y&=n\f &\rf95 \cr
x&=\sqrt n\, \xi &\rf96 \cr
T&=\sqrt n\, \eta &\rf97
}
$$
and in the region of large field $\f$, $\e=-1$ and
$$
\ealn{
y&=(n+2)\f &\rf98 \cr
x&=\sqrt {n+2}\, \xi &\rf99 \cr
T&=\sqrt {n+2}\, \eta. &\rf100
}
$$

Eq.\ \rf94 \ is the famous Liouville equation which can be transformed via a
B\"ack\-lund \tr ation into a two-dimensional wave equation and afterwards
solved exactly. The general solution depends on two arbitrary functions $f$
and~$g$ of one variable, sufficiently regular. It is possible to consider
several problems for this equation: Cauchy initial problem, Darboux problem
and Goursat problem.

The general solution of \rf94 \ looks like ($\e\t\e=-1$)
$$
y(T,x)=\ln\[\frac{2g'(x-T)f'(x+T)}{\bigl(g(x-T)+f(x+T)\bigr)^2}\]
\eq101
$$
where $g'$ and $f'$ are derivatives of $g$ and $f$.

Thus one gets in the first region (small field)
$$
q_0(t,z)=\frac{2\sqrt{2\pi \o M}}{\mpl n}\cdot
\ln\[\frac{2g'\bigl(\frac{z-t}a\bigr)f'\bigl(\frac{z+t}a\bigr)}
{\(g\bigl(\frac{z-t}a\bigr)+f\bigl(\frac{z+t}a\bigr)\)^2}\], \eq102
$$
$$
a=\frac{2\sqrt2}{\sqrt{n\o\la_{c0}}}\,, \eq103
$$
and
$$
G\dr{eff}=G_N\(\frac{\(g\bigl(\frac{z-t}a\bigr)+f\bigl(\frac{z+t}a\bigr)\)^2}
{2g'\bigl(\frac{z-t}a\bigr)f'\bigl(\frac{z+t}a\bigr)}\)^{(n+2)/n}. \eq104
$$
In the second region (large field) 
$$
q_0(t,z)=\frac{2\sqrt{2\pi \o M}}{\mpl (n+2)}\cdot
\ln\[\frac{2g'\bigl(\frac{z-t}b\bigr)f'\bigl(\frac{z+t}b\bigr)}
{\(g\bigl(\frac{z-t}b\bigr)+f\bigl(\frac{z+t}b\bigr)\)^2}\], \eq105
$$
$$
b=\frac{2\sqrt2}{\sqrt{(n+2)\o\la_{c0}}}\,, \eq106
$$
and
$$
G\dr{eff}=G_N\cdot\frac{\(g\bigl(\frac{z-t}b\bigr)+f\bigl(\frac{z+t}b\bigr)\)^2}
{2g'\bigl(\frac{z-t}b\bigr)f'\bigl(\frac{z+t}b\bigr)}. \eq107
$$

In this way we get a spatio-temporal pattern of changing the \ef\ \gr\ \ct\
for small and large field regions. In both cases we have
$\e\t\e=-1$. However, in the small field region we have $\e=1$ and because of
this $\t\e=-1$. In the case of large field region $\e=-1$ and $\t\e=1$. In
order to be in line with our assumptions we should consider in the first case
such functions $f$ and~$g$ that the expression in~\rf105 \ is small and for
the second case vice versa.

Let us consider Eq.\ \rf63 \ in cylindrical coordinates supposing cylindrical
\sy y for the field~$q_0$, $q_0=q_0(\rho)$. One gets
$$
\frac1\rho\(\rho\,\frac{dq_0}{d\rho}\)-\e \o\a\exp(n\o\b q_0)\(\exp(2\o\b
q_0)-1\)=0. \eq108
$$
This equation can be \tr ed into
$$
\frac1\tau\,\frac{d}{d\tau}\(\tau\,\frac{d\f}{d\tau}\)-\t\e \exp(n\f)\(e^{2\f}
-1\)=0, \eq109
$$
$$
\rho=\frac2{\sqrt{\o\la_{c0}}}\,\tau. \eq110
$$
As usual we consider Eq.\ \rf109 \ in two regions for small and large fields.

In the first region
$$
\frac1\tau\,\frac{d}{d\tau}\(\tau\,\frac{d\f}{d\tau}\)+\t\e e^{n\f}=0. \eq111
$$
In the second region we get
$$
\frac1\tau\,\frac{d}{d\tau}\(\tau\,\frac{d\f}{d\tau}\)-\t\e e^{(n+2)\f}=0.
\eq112
$$
Both equations can be reduced to the equation
$$
\frac1x\,\frac{d}{dx}\(x\,\frac{dy}{dx}\)+\e\t\e e^y=0, \eq113
$$
where in the first region $\e=1$ and
$$
\ealn{
y&=n\f &\rf114 \cr
x&=\sqrt n\,\tau &\rf115 
}
$$
and in the second region $\e=-1$ and
$$
\ealn{
y&=(n+2)\f &\rf116 \cr
x&=\sqrt{n+2}\,\tau. &\rf117
}
$$
We can \tr\ \rf113 \ into
$$
x\,\frac{d^2y}{dx^2}+\frac{dy}{dx}+\e\t\e xe^y=0 \eq118
$$
which is the equation considered in~[132] for $\e\t\e=1$.

Following H. Lemke we write down a solution to Eq.\ \rf118 \ in a compact
form in three cases (concerning an integration \ct\ introduced by H.~Lemke).
We adopt his solutions to our problem.

\item{1)}$C=\k^2>0$, $\k$---arbitrary positive number:
$$
y(x)=-2\ln\(\frac x{\,\o a\,}\(\(\frac{\o a}{x}\)^\k +\(\frac{x}{\,\o a\,}\)^\k \)\)
+\ln\(4\k^2\o a{}^2\) \eq119
$$
where $\o a>0$ is an arbitrary \ct.

\item{2)}$C=-\om^2<0$, $\om$ is an arbitrary positive number:
$$
y(x)=-2\ln\(2x\sin(\om \ln x + \d)\)+\ln(4\om^2) \eq120
$$
where $\d$ is an arbitrary \ct.

\item{3)}$C=0$:
$$
y(x)=-2\ln\(\frac x{\,\o a\,}\ln\(\frac x{\,\o a\,}\)\)-2\ln\o a \eq121
$$
where $\o a>0$ is an arbitrary \ct.

Using these solutions we write down a spatial \dc ce of $G\dr{eff}$ in the
case of small and large fields. 
In the case of small fields one gets
$$
\dsl{
\tin1 \hfill G\dr{eff}=G_N\(4\k^2\o a{}^2\)^{(n+2)/n}
\frac{\(1+r^{2\k}\)^{2(n+2)/n}}{r^{2(\k-1)(n+2)/n}} \hfill \rf122
}
$$
where
$$
r=\frac1{2\o a}\sqrt{n\o\la_{c0}}\,\rho. \eq123
$$
$$
\dsl{
\tin2 \hfill G\dr{eff}=\frac{G_N(\o\d)^{2(n+2)/n}}{(\om^2)^{(n+2)/n}}
\, r^{2(n+2)/n}\bigl(\sin(\om\ln r)\bigr)^{2(n+2)/n} \hfill \rf124 
}
$$
where
$$
\ealn{
&r=\frac1{2\o\d}\sqrt{n\o\la_{c0}}\,\rho &\rf125 \cr
&\om\ln\o\d=-\d &\rf126
}
$$
$$
\dsl{
\tin 3\hfill G\dr{eff}=G_N\o a^{2(n+2)/n}r^{2(n+2)/n}(\ln r)^{2(n+2)/n} \hfill
\rf127 
}
$$
and $r$ is given by Eq.\ \rf123 .

In the case of large field one gets:
$$
\dsl{
\tin1 \hfill G\dr{eff}=G_N(4\k^2\o a^2)\frac{(1+r^{2\k})^2}{r^{2(\k-1)}} \hfill
\rf128 
}
$$
where
$$
r=\frac1{2\o a}\sqrt{(n+2)\o\la_{c0}}\,\rho. \eq129
$$
$$
\dsl{
\tin2 \hfill G\dr{eff}=G_N\(\frac{\o\d}{\om}\)^2 r^2 \bigl(\sin(\om\ln r)\bigr)^2
\hfill \rf130
}
$$
where
$$
r=\frac1{2\o\d}\sqrt{(n+2)\o\la_{c0}}\,\rho , \quad \ln \o\d=-\frac\d\om. \eq131
$$
$$
\dsl{
\tin3 \hfill G\dr{eff}=G_N\o a^2r^2(\ln r)^2 \hfill \rf132
}
$$
and $r$ is given by Eq.\ \rf129 . 

Let us notice that in that spatial \dc ce
for large and small field we have $\k$ and $\o a$ ($\o\d$, $\om$) as
integration \ct s. In this way integration \ct s induce a power law of this
\dc ce and also a scale. For sufficiently big~$n$ ($n>14$) there is no
significant difference between both cases. It means the \q\ field behaves
everywhere as for large field case (in these solutions of course). It is
evident that the spatial \dc ce (in cylindrical \sy y case) of~$G\dr{eff}$
goes to some kind of the fifth force. However, we have to do not with a
universal law of Nature but rather with some kind of initial conditions. Thus
the real \dc ce of $G\dr{eff}$ on spatial coordinates can be obtained after
averaging~$G\dr{eff}$ \wrt initial conditions. Let us describe it using
Eq.~\rf128 .

Let us write $G\dr{eff}$ in a form where $\k$ and $\o a$ are explicitly
visible:
$$
\ealn{
G\dr{eff}&=G_N\(4\k^2\o a^2\)\cdot
\frac{\(1+\(\frac x{\,\o a\,}\)^{2\k}\)^2}{\(\frac x{\,\o a\,}\)^{2(\k-1)}},
&\rf133 \cr
x&=\frac12 \sqrt{(n+2)\o\la_{c0}}\,\rho. &\rf134 
}
$$
The \ct s $\o a$ and $\k$ are chosen randomly for they are \dc t on initial
conditions. Let $\mu$ be a measure defined on $(0,+\infty)^2$, positive and
normalized to~1, i.e.
$$
\intop_0^{+\infty} \,\intop_0^{+\infty} d\mu(\o a,\k)=1. \eq135
$$
This measure gives an account how frequently we have to do with some initial
conditions (i.e.\ with integration \ct s $\o a$,~$\k$). In this way an
experimental \dc ce of~$G\dr{eff}$ on~$x$ should be obtained from
$$
E(G\dr{eff})=4G_N \intop_0^{+\infty}\, \intop_0^{+\infty} 
\frac{\k^2\o a^2\(1+\(\frac x{\,\o a\,}\)^{2\k}\)^2}
{\(\frac x{\,\o a\,}\)^{2(\k-1)}}\,d\mu. \eq136
$$
It means it is an expectation value of $G\dr{eff}$ \wrt the measure~$\mu$.
$\mu$~need not be absolutely continuous \wrt the Lebesgue measure on ${\Bbb
R}^2$. The formula \rf136 \ could give an account on some serious problems
with comparisons of several measurements of~$G$ (a~Newton \ct). Maybe we have
to do with different initial conditions for \q\ field. These initial
conditions appear with different probabilities according to the measure
$d\mu(\o a,\k)$ going to an expectation value $E(G\dr{eff})$. The second
central moment of $d\mu(\o a,\k)$ (if it exists) can express a deviation from
the law described by $E(G\dr{eff})$. 

In the simplest case we can suppose a continuous Gaussian distribution
for~$\k$ only:
$$
d\mu(\k)=\frac1{\sqrt{2\pi}\,\s}\,e^{-(\k-\k_0)^2/(2\s^2)}\,d\k.
$$
In this case we have
$$
\o G\dr{eff}(x)=E(G\dr{eff}(x))=
\frac{G_N\o a^2}{\sqrt{2\pi}\,\s} \intop_{-\infty}^{+\infty}
d\k\,e^{-(\k-\k_0)^2/(2\s^2)}\k^2\,\frac{\(1+\(\frac x{\,\o a\,}\)^{2\k}\)^2}
{\(\frac x{\,\o a\,}\)^{2(\k-1)}}\,. \eq136a
$$
One gets after some algebra
$$
\al
\o G\dr{eff}(x)&=G_N \k_0 \exp\(\o a-\frac{\k_0^2}{2\s^2}\)\(\frac x{\,\o a\,}\)
^2\cr
&\times\Biggl[4\s^4\exp\frac{\(2\ln\(\frac x{\,\o a\,}\)\s^2+\k_0^2\)^2}{2\s^2}
\ln\(\frac x{\,\o a\,}\)\cr
&\qquad{} +(2\s^2+\k_0^4)
\exp\frac{\(2\ln\(\frac x{\,\o a\,}\)\s^2+\k_0^2\)^2}{2\s^2}\cr
&\qquad{}-4\s^4\exp\frac{(\s+\k_0)^2}2 \ln^2\(\frac x{\,\o a\,}\)+
2(\s^2+\k_0^2)\exp\frac{(\s+\k_0)^2}2 \Biggr].
\eal
\eq136b
$$

In the case of the solution \rf107 \ parametrized by two functions
of one variable we can consider them as random variables parametrized by~$z$
and~$t$. Moreover in this case we should consider a measure~$\mu$ on an
infinite-dimensional space of functions $f$ and~$g$, supposing that solution
\rf107 \ is generalized to this space. Afterwards we can use as~$\mu$ the
Wiener or Gaussian measure in $L^2$ space. If we use for Gaussian
distribution the normalized distribution $N(0,1)$, the formula \rf136b \
simplifies to
$$
\al
\o G\dr{eff}(x)&=e^{\bar a}\sqrt e \, G_N\cr
&\times\[4\(\frac x{\,\o a\,}\)^{2(1+\ln(x/\bar a))}\ln\(\frac x{\,\o a\,}\)
+\(\frac x{\,\o a\,}\)^{2(1+\ln(x/\bar a))} -4\ln^2\(\frac x{\,\o a\,}\)+2\]. 
\eal
\eq136c
$$

Let us consider three our cases \rf119 , \rf120 \ and \rf121 \ for small and
large fields cases.

The small field case is such that
$$
e^\f<1, \qquad\text{i.e. }\f<0. \eq137
$$
The large field case is if
$$
e^\f>1, \qquad\text{i.e. }\f>0. \eq138
$$
For \rf119 \ we have for the small case
$$
f(r)>1,
$$
where
$$
f(r)=r^{1-\k}+r^{\k+1}, \qquad r=\(\frac x{\,\o a\,}\). \eq139
$$

Let us consider two cases
$$
0<\k<1 \qquad\text{and}\qquad \k>1.
$$
In the first case $f(r)\nearrow$ in $[0,+\infty)$ and we have simply
$r>r_0$ where $r_0$ satisfies the equation
$$
r_0^{2\k+1}+r_0-1=0. \eq140
$$
In the second case $\k>1$,
$$
\lim_{r\to0} f(r)=+\infty \qquad\text{and}\qquad \lim_{r\to\infty}
f(r)=+\infty  .
$$
The function $f(r)$ has a minimum at
$$
r_1=\(\frac{\k-1}{\k+1}\)^{1/\k}. \eq141
$$
Let us calculate $f(r_1)$.
$$
f(r_1)=\(\frac{\k+1}{\k-1}\)^{\k-1/\k}+\(\frac{\k-1}{\k+1}\)^{\k+1/\k}. \eq142
$$
It is easy to see that if $\k>1$, then $\k-1/\k>0$. This means that
$$
f(r_1)>1 \eq143
$$
and therefore $f(r)>1$ for all $r>0$. Thus simultaneously we get a solution
for large field only if $\k<1$, i.e.
$$
r<r_0. \eq144
$$
If $\k=1$, we always have $f(r)\ge1$ (i.e.\ only a small field).

Let us consider \rf120 . In this case the small field condition reads
$$
h(r)=r\sin(\om\ln r)>1, \quad \text{where }r=\frac x {\,\o \d\,}\,. \eq145
$$
First of all we need $h(r)>0$. Let us observe that $|h(r)|\le r$ for
every~$r>0$. Next we see that the roots of $h(r)$ are the numbers
$$
r_{0,k}=e^{k\pi/\om}, \quad k=0,\pm1,\pm2,\ldots \eq146a
$$
and
$$
h(r)>0 \quad\text{if}\quad 
r_{0,2k}<r<r_{0,2k+1},\quad k=0,\pm1,\pm2,\ldots \eq146b
$$
The maximum of the function $h(r)$ in the interval
$(r_{0,2k},r_{0,2k+1})$ is greater than
$$
h(r_{0,2k+1/2})=e^{2k\pi/\om}\cdot e^{\pi/(2\om)}, \eq147
$$
but smaller than $r_{0,2k+1}$. Thus the maxima are smaller than~1 for
negative $k$ and greater than~1 if $k=0,1,2,\ldots$. It
means that in each interval $(r_{0,2k},r_{0,2k+1})$, $k=0,1,2,\ldots$, there
exist two numbers $r_{3,2k}$ and $r_{2,2k}$ such that $r_{3,2k}<r_{2,2k}$,
$$
h(r_{3,2k})=h(r_{2,2k})=1 \eq148
$$
and
$$
h(r)>1 \quad\text{if}\quad r_{3,2k}<r<r_{2,2k}\,. \eq149
$$

The condition for large fields, $0<h(r)<1$, is satisfied if
$$
r_{0,2k}<r<r_{0,2k+1}, \quad k=-1,-2,\ldots \eq150
$$
or
$$
r_{0,2k}<r<r_{3,2k}, \quad \text{or}\quad r_{2,2k}<r<r_{0,2k+1},
\quad k=0,1,2,\ldots \eq151
$$

In the third case, i.e.\ Eq.\ \rf121 , one has for the small field case
$$
r \ln r>1 \eq152
$$
where $r=\frac x{\,\o a\,}$. Let $r_4\ln r_4=1$, $r_4>1$. We have
$r>r_4=1.7632\ldots$. In the large field case 
$$
0<r<r_4=1.7632\ldots. \eq153
$$

Thus we have in general large fields on large distances.

In this way we have solutions for large and small distances. One can try to
connect them to get a solution for all distances. However in this case it is
necessary to be very careful, for our solutions depend on some integration
\ct s which can be different for both asymptotic regions.

Let us consider Eq.\ \rf63 \ in two special cases:

\item{I} $q_0=q(z)$ --- static and depending only on~$z$;
\item{II} $q_0=q_0(t)$ --- non-static and spatially \ct.

Let us consider also these cases for small and large fields~$\f$. In all of
these cases we come to the following equation
$$
\frac{d^2y}{dx^2} + \e_1\e\t\e e^y=0 \eq154
$$
where $\e_1=1$ for case I and $\e_1=-1$ for case II. 

Eq.\ \rf154 \ can easily be reduced to the integral
$$
x-x_0=\frac{\e_3}{\sqrt2} \int \frac{dy}{\sqrt{\e_2\om^2-\eta e^y}} \eq155
$$
where $\eta=\e_1\e\t\e$, $\eta^2=1$, $C=2\e_2\om^2$, $\om\ge0$, $\e_2^2=1$ is
an integration \ct, $\e_3^2=1$ and $x_0$ also is an integration \ct.

After some calculation we get the following solutions:
$$
\dsl{
\text{A.}\hfill y(x)=2\[\ln \om-\ln\left|\sinh\(
\frac{\sqrt2\,(x-x_0)\om}2 \)\right|\], \quad \eta=-1,\ \e_2=1\hfill \rf156 \cr
\text{B.}\hfill y(x)=2\[\ln \om-\ln\left|\sinh\(
\frac{\sqrt2\,(x-x_0)\om}2 \)\right|\], \quad \eta=1,\ \e_2=1\hfill \rf157
}
$$
where
$$
\frac{\sqrt2\,(x-x_0)\om}2 >\ln(1+\sqrt2)\eq158 
$$
or
$$
\frac{\sqrt2\,(x-x_0)\om}2 <\ln(\sqrt2-1).\eq159
$$
$$
\dsl{
\text{C.}\hfill y(x)=2\[\ln\om-\ln\left|
\cos\(\frac{\sqrt2\,(x-x_0)\om}2\)\right|\], 
\quad \eta=-1,\ \e_2=-1. \hfill \rf160
}
$$

Let us apply these solutions to our problems. First of all let us consider a
static configuration with $z$ \dc ce only. In this case $\e_1=1$ and $\eta=\e
\t\e$. For the small field case one gets, $\e=1$, $\eta=\t\e$,
$$
G\dr{eff}=\frac{G_N}{\om^{2(n+2)/n}}\left|\sinh p\right|^{(n+2)/n} \eq161
$$
where
$$
p=\frac{\om(z-z_0)}4 \,\sqrt{n\o\la_{c0}}\,. \eq162
$$
For the large field case we get, $\e=-1$, $\eta=-\t\e$,
$$
G\dr{eff}=\frac{G_N}{\om^2}\left|\cos p\right| \eq163
$$
where
$$
p=\frac{\om(z-z_0)}4 \,\sqrt{(n+2)\o\la_{c0}}\,. \eq164
$$
In this case $\t\e=1$ for $\eta=-1$.

In a nonstatic configuration $\e_1=-1$ and $\eta=-\e\t\e$. For the small
field case ($\e=1$), $\eta=-\t\e$,
$$
G\dr{eff}=\frac{G_N}{\om^{2(n+2)/n}}\left|\sinh q\right|^{(n+2)/n} \eq165
$$
where
$$
q=\frac{\e_3\om(t-t_0)}4 \sqrt{n\o\la_{c0}} \eq166
$$
and analogically for the large field case ($\e=-1$), $\eta=\t\e$,
$$
G\dr{eff}=\frac{G_N}{\om^2}\left|\cos q\right|, \eq167
$$
$$
q=\frac{\om(t-t_0)}4 \,\sqrt{(n+2)\o\la_{c0}}\,. \eq168
$$
In this case $\t\e=-1$ for $\eta=-1$.

Let us notice that in a static configuration for small field we have two
possibilities for $\eta=-1$ (no condition on~$p$) and $\eta=1$ (conditions
\rf158--159 ). Thus without conditions we have $\t\e=-1$ and with conditions
$\t\e=1$. In a non-static configuration for small field we have vice versa
$\t\e=1$ without conditions and $\t\e=-1$ with conditions \rf158--159 .

Let us come back to the Eq.\ \rf89 \ and consider it in a travelling wave
scheme. In this way we have
$$
q_0(z,t)=\t q(z-vt) \eq169
$$
where $v$ is a velocity of the travelling wave (a soliton), $|v|<1$. Let us
consider this equation in both regimes (for small and large fields). In this
way we come to the expression
$$
(1-v^2)\,\frac{d^2\chi}{d\xi^2}-\e\t\e e^\chi=0 \eq170
$$
where $\chi$ is a shape function of a soliton. Changing an independent
variable from $\xi$ to~$\la$ one gets
$$
\frac{d^2\chi}{d\la^2}-\e_1\e\t\e e^\chi=0 \eq171
$$
where $\e_1=-1$, i.e.\ we get Eq.\ \rf154 \ with $\eta=-\e\t\e$,
$$
\la=\frac \xi {\sqrt{1-v^2}},\quad \xi=\sqrt{1-v^2}\,\la. \eq172
$$

In this way we adopt our solutions A, B, C in both regimes: small and large
field (changing $\chi$ into~$\la$). For small field we get ($\e=1$,
$\eta=-\t\e$) 
$$
q_0(z,t)=\frac2{\o \b n}\[\ln\om - \ln\left|\sinh p\right|\] \eq173
$$
where
$$
p=\frac{\om\sqrt{n\o\la_{c0}}}{4\sqrt{1-v^2}}\,(z-vt). \eq174
$$
For large field we get ($\e=-1$, $\eta=\t\e$) 
$$
q_0(z,t)=\frac2{\o \b (n+2)}\[\ln\om - \ln\left|\cos p\right|\] \eq175
$$
where
$$
p=\frac{\om\sqrt{(n+2)\o\la_{c0}}}{4\sqrt{1-v^2}}\,(z-vt). \eq176
$$
For \rf173 \ we have $\eta=-\t\e$ and because of this $\t\e=1$ without any
conditions and if $\t\e=-1$ we have conditions \rf158--159 . In the case of
the formula \rf174 \ $\eta=-1$ and $\t\e=-1$.

We can write down formulas for $G\dr{eff}$ in the soliton case
$$
G\dr{eff}=\frac{G_N}{\om^{2(n+2)/n}}\left|\sinh p\right|^{(n+2)/n} \eq177
$$
and $p$ is given by the formula \rf174 \ (with or without conditions
\rf158--159 ). This is of course a small field case.

In the large field case
$$
G\dr{eff}=\frac{G_N}{\om^2}\left|\cos p\right|, \eq178
$$
and $p$ is given by the formula \rf176 . In this case $\t\e=-1$. (Let us
notice that this is a case of SO(3) group in our theory.)

Let us notice that conditions \rf158--159 \ can be considered as conditions
for small field in $z$ or $t$ domains. Let us notice that in our solutions
concerning a behaviour of an \ef\ \gr\ \ct\ we get completely arbitrary
length or time scale (given by integration \ct s). In this way a spatial or
time \dc ce of $G\dr{eff}$ can be (except the solution \rf80 \ and
simultaneously the approximate solution in the case of spherical \sy y) such
that $G\dr{eff}$ can be really \ct\ on distances (or times) accessible in
experiments. Only a statistical approach mentioned here can give a light on
this \dc ce to compare it with an experiment.

In order to connect our results to the ordinary \gr\ physics we consider
again Eq.~\rf67 \ in small field regime for initial conditions $\f(0)=0$ and
$\frac{d\f}{dx}(0)=0$. The first condition means that we want to have
$G\dr{eff}(0)=G_N$ and the second that the \q\ field does not grow quickly.
The problem cannot be solved analytically. Moreover R.~Emden in the first
point of Ref.~[131] did it for us. We quote here his results adopted to our
notation ($\e=+1$ and $\t\e=-1$).

\def\nh{\cr\noalign{\hrule}}
$$\vbox{\offinterlineskip
\halign{\strut\vrule\quad \hfil$#$\quad &\vrule\quad \hfil$#$\quad 
&\vrule\quad $#$\hfil \quad&\vrule\quad \hfil$#$\quad \hfil\vrule\cr
\noalign{\hrule}
\quad x\hfil&\quad -y\hfil&\hfil e^{y}\quad &\quad G\dr{eff}/G_N\nh
0.00&0.00000&1.00000&1.00000\nh
0.25&0.01037&0.98969&0.98823\nh
0.50&0.04113&0.95971&0.95409\nh
0.75&0.09113&0.91290&0.90109\nh
1.00&0.15903&0.85296&0.83380\nh
1.25&0.24225&0.78486&0.75816\nh
1.50&0.33847&0.71285&0.67920\nh
1.75&0.44488&0.64090&0.60143\nh
2.00&0.55967&0.57140&0.52749\nh
2.50&0.80584&0.44671&0.39813\nh
3.00&1.06226&0.34537&0.29670\nh
3.50&1.31937&0.26730&0.22138\nh
4.00&1.57071&0.20790&0.16611\nh
%}}
%$$
%$$\vbox{\offinterlineskip
%\halign{\strut\vrule\quad \hfil$#$\quad &\vrule\quad \hfil$#$\quad 
%&\vrule\quad $#$\hfil \quad&\vrule\quad \hfil$#$\quad \hfil\vrule\cr
%\noalign{\hrule}
%\quad x\hfil&\quad -y\hfil&\hfil e^{y}\quad &\quad G\dr{eff}/G_N\nh
4.50&1.81246&0.16325&0.12601\nh
5.00&2.04264&0.12968&0.09686\nh
6.00&2.46598&0.08493&0.05971\nh
7.00&2.84160&0.05833&0.03887\nh
8.00&3.17489&0.04180&0.02656\nh
9.00&3.47128&0.03108&0.01893\nh
10.00&3.73646&0.02384&0.01398\nh
100&8.59506&0.000175&5.0854\cdot10^{-5}\nh
1000&13.09847&0.000002&3.0683\cdot10^{-7}\nh
}}
$$
where
$$
\ealn{
&x=\frac12 \sqrt{n\o\la_{c0}}\,r \cong \(\frac r{10\,\text{Mpc}}\),&\rf179 \cr
&\frac{G\dr{eff}}{G_N}=\(e^{y}\)^{(n+2)/n}=\(e^{y}\)^{8/7}. &\rf180
}
$$
We take $n=14$.

It is easy to see that for large $n$
$$
\frac{G\dr{eff}}{G_N}=e^{y}. \eq181
$$

It is easy to see that on a distance of $1\,$Mpc $G\dr{eff}$ does not differ
from~$G_N$. Even on a distance of $10\,$Mpc it is about 10\% smaller. Thus in
the Solar System Newtonian \gr\ physics does not change. Even on the level of
a galaxy this change is minimal and cannot be observed. Moreover, there is an
important conclusion: on distances about $200\,$Mpc the strength of \gr\
interactions is about $10^{-5}$ times this on short distances measured in the
Solar System (and for $10^3$\,Mpc of $10^{-7}$). It is hard to tell how it influences a mass of a cluster of
galaxies if we realize that from any observational data only a product $GM$
has been obtained (not~$M$).

From the other side on distances of $100\,$Mpc the strength of \gr\
interactions is very weak (not only because of the distance). Thus if we
consider clusters of galaxies as substrat \pc s in cosmology then they do not
interact.

Let us consider Eq.\ \rf63  \ in Cartesian coordinates for two-dimensional
static case (i.e. $\frac{\pa}{\pa z}=0$, $\frac{\pa}{\pa t}=0$). One gets
$$
\(\frac{\pa^2q_0}{\pa x^2}+\frac{\pa^2q_0}{\pa y^2}\)
-\t \e \o \a \exp(n\o\b q_0)\(\exp(2\o\b q_0)-1\)=0 \eq182
$$
(where $\o\a$, $\o\b$ are given by formulas \rf64 \ and \rf65 ).

As usual, we come to the formula
$$
\(\frac{\pa^2\f}{\pa x_1^2}+\frac{\pa^2\f}{\pa x_2^2}\) - \t\e e^{n\f}
(e^{2\f}-1)=0 \eq183
$$
where
$$
\left.\matrix x\\y\endmatrix\right\} = \frac2{\sqrt{\o \la_{c0}}}\,x_i,
\qquad i=1,2. \eq183a
$$
We consider Eq.\ \rf183 \ for small and large fields and we get
$$
\(\frac{\pa^2\chi}{\pa z_1^2}+\frac{\pa^2\chi}{\pa z_2^2}\) - \e\t\e e^{\chi}
=0 \eq184
$$
where as usual for a small field $\e=1$ and
$$
\ealn{
\chi&=n\f, &\rf185 \cr
z_i&=\sqrt n\,x_i &\rf186
}
$$
and for a large field $\e=-1$ and
$$
\ealn{
\chi&=(n+2)\f, &\rf187 \cr
z_i&=\sqrt {n+2}\,x_i\,. &\rf188
}
$$
Thus we come to the equation known as Liouville equation
$$
\D \chi=e^\chi \eq184a
$$
if $\e\t\e=1$.

This equation can be explicitly solved. First of all we change in\dc t
variables into
$$
\ealn{
Z&=\frac1{\sqrt2}(z_1+iz_2) &\rf189 \cr
\chi(Z)&=-\ln\(\tfrac12(1-|g|^2)\)+\tfrac12 \ln\left|\frac{dg}{dZ}\right|
&\rf190
}
$$
where $g$ is an arbitrary analytic function on a complex plane $Z$.

In this way we get for the small field case
$$
G\dr{eff}=G_N\(\frac{1-|g(Z)|^2}2\)^{(n+2)/n}
\(\left|\frac{dg}{dZ}(Z)\right|\)^{-(n+2)/(2n)} \eq191
$$
where
$$
Z=\sqrt{\frac{\o\la_{c0}}{2n}}\cdot(x+iy). \eq192
$$

In the large field case
$$
G\dr{eff}=G_N\(\frac{1-|g(Z)|^2}2\)
\(\left|\frac{dg}{dZ}(Z)\right|\)^{-1/2} \eq193
$$
where
$$
Z=\sqrt{\frac{\o\la_{c0}}{2(n+2)}}\cdot(x+iy). \eq194
$$
Eqs \rf191 \ and \rf193 \ can have very interesting behaviour for
$\frac{dg}{dZ}$ could have some singularities. The physical interpretation of
these singularities can be very interesting.

Similarly as for Eqs \rf104 \ and \rf107 \ we can consider an expectation
value of $G\dr{eff}$ \wrt some kind of normalized measure for a space of
analytic functions on the complex plane (respectively chosen).

It seems that an assumption all the energy of a \q\ is stored as \q\ \pc s is
unrealistic. Let us suppose that only a fraction of this energy is stored as
\pc s. Let this fraction be $\eta$,
$$
0<\eta<1. \eq195
$$
$\eta$ can be a function of time and even of a space-point (locally).

Let us suppose that a gas of \q\ \pc s is a perfect gas governed by the
Clapeyron equation. Thus we have
$$
\frac{p_1}{\rho_1}=\frac{K_BT}{m_0}=\frac T{T_0}, \quad
T_0=\frac{m_0}{K_B}\simeq 0.11\,{}^\circ\text{K} \eq196
$$
(if we take $m_0\simeq 10^{-5}\,$eV). $T$ is the temperature of a gas.

One gets
$$
\ealn{
\rho_1&=\eta\rho_Q=\eta\rho &\rf197 \cr
p&=p_Q+p_1=p_Q+\eta\,\frac T{T_0}\, \rho=
-\rho(1-\eta)+\rho\,\frac T{T_0}\,\eta. &\rf198
}
$$
Eventually one gets
$$
\ealn{
p&=\rho\(\eta\(1+\frac T{T_0}\)-1\) &\rf199 \cr
p&=w\rho &\rf200 \cr
w&=\eta\(1+\frac T{T_0}\)-1. &\rf201 
}
$$

Now we can calculate an isothermic speed of sound in a \q.
$$
C_1^2=\frac p\rho=w=\eta\(1+\frac T{T_0}\)-1. \eq202
$$
For
$$
0<C_1^2<1 \eq203
$$
we get
$$
\frac{T_0}{T+T_0}<\eta<\min\[\frac{2T_0}{T_0+T},1\]. \eq204
$$
Let us remind to the reader that an isothermic sound is appropriate for low
frequency of acoustic waves (we have to do with this sound in astrophysics).
In general we have to do with so called adiabatic sound. In order to
calculate a speed of an adiabatic sound in a \q\ we should find an analogue
for a Poisson adiabate for our equation of state. One gets supposing that an
internal energy of a \q\ is an energy of one-atomic gas of \q\ \pc s
$$
dU+p\,dV=0 \eq205
$$
or
$$
\frac32\,dT=\frac{d\rho}{\rho}\(\eta-1+\frac T{T_0}\,\eta\) \eq206
$$
and finally
$$
\ealn{
&\frac p{\rho^\k}=\text{const.} &\rf207 \cr
&\k=\frac{2\eta+3}3 &\rf208
}
$$
and a speed of an adiabatic sound simply reads
$$
C_2=\sqrt{\frac{\k p}{\rho}}= \sqrt{\k\(\eta\(1+\frac T{T_0}\)-1\)}\,. \eq209
$$
It is interesting to ask what kind of a polythrope $\k$ represents. Let us
remind to the reader that in general
$$
\k=\frac{c_p-c}{c_v-c} \eq210
$$
where $c$ is a specific heat of a polythrope in mind.

One gets in our case
$$
c=\frac{3(\eta-1)}{2\eta}\,\o R \eq211
$$
where $\o R$ is a universal gas \ct.

Thus we found both speeds of sound in a \q\ for low frequency and for
high frequency of sound.
The measurement of both speeds can help us to find $\eta$ and $T$. One
gets 
$$
\eta=\frac32\(\(\frac{C_2}{C_1}\)^2-1\) \eq212
$$
and
$$
T=\frac{T_0}{3(C_2^2-C_1^2)}\(C_1^2(2C_1^2+5)-3C_2^2\). \eq213
$$
Let us notice that a gas of \q\ \pc s is quite cold.
Moreover these \pc s are highly relativistic. For the temperature
$T\simeq T_0=0.11\,{}^\circ$K one sees that a speed of a \q\ \pc \ is about
0.91 of the speed of light. Moreover we can consider lower temperatures.
It is interesting to notice that the speed of sound is of the same order.

Thus if we want to have both speeds smaller than a speed of light,
$$
0<C_i^2<1, \quad i=1,2, \eq214
$$
we get
$$
\eta>\frac{1+\sqrt{13+12t}}{1+t}=f(t) \eq215
$$
where
$$
t=\frac T{T_0}\,. \eq216
$$
The condition
$$
t=\frac T{T_0}>13 \eq217
$$
guarantees that $f(t)<1$. Moreover for $t=14$ one gets
$$
\eta\ge 0.96357. \eq218
$$
This seems quite interesting, however maybe too much. Let us calculate a mean
scattering length of \q\ \pc s for such a big $\eta=0.96357$
$$
l\dr{scattering}=\frac1{\eta\s n}=1.0378\cdot l. \eq219
$$
We have still to do with a very dense gas of \q\ \pc s.

It seems that it would be reasonable to repeat these calculations using a
different equation of state for \q\ \pc s gas. In particular we can consider
a gas of \q\ \pc s as a massless boson gas (with spin zero). In this case the
equation of state looks:
$$
p=p_Q+p_1=-(1-\eta)\rho+\frac{4\eta\s\rho}3\,T^4 \eq220
$$
and an adiabate equation
$$
dU=\o c_v\,dT+p\,dV=0 \eq221
$$
where
$$
\o c_v=16\s T^3 , \eq222
$$
$\s$ is the Stefan-Boltzmann \ct.

One easily integrates
$$
\frac p{\rho^{\bar\k}}=\text{const.} \eq223
$$
where
$$
\o\k=\frac{3+\eta}3\,. \eq224
$$
We have as before two kinds of sound: an isothermic sound
$$
\o C_1=\sqrt{\frac p\rho}=\sqrt{\eta-1+\frac{4\eta\s}3\,T^4} \eq225
$$
and an adiabatic sound
$$
\o C_2=\sqrt{\o\k\(\eta-1+\frac{4\eta\s}3\,T^4\)}=\sqrt{\frac{\o\k p}\rho}\,. 
\eq226
$$

From $0<\o C_i^2<1$, $i=1,2$, one gets
$$
\frac1{1+\(T/{\,\o T_0\,}\)^4} < \eta < \min\[1,\frac2{1+\(T/{\,\o
T_0\,}\)^4} \] \eq227
$$
and
$$
\eta>\frac{-3t^4-1+\sqrt{9t^8+18t^4+13}}{2(1+t^4)}=\o f(t) \eq228
$$
with the condition
$$
t=\frac T{\,\o T_0\,}>\frac1{\sqrt2}\simeq 0.707. \eq229
$$
This condition guarantees that
$$
\o f(t)<1 \eq230
$$
where $T_0^4=\frac3{4\s}$ or
$$
\o T_0=\frac1{K_B}\root4\of{\frac{360}\pi}\simeq\frac{3.27}{K_B}\,, \eq231
$$
$K_B$ is the Boltzmann \ct.

From the measurements of $\o C_1$ and $\o C_2$ we can obtain as before $\eta$ and
$T$. One gets:
$$
\eta=3\(\(\frac{\o C_2}{\,\o C_1\,}\)^2-1\) \eq232
$$
and
$$
T=\frac{\o T_0}{\root4\of3}\(\frac{4\o C_1^2+\o C_1^4-3\o C_2^2}{\o C_2^2-\o
C_1^2}\)^{1/4}. \eq233
$$
Recently many papers have appeared concerning the speed of sound in a \q\
(see Ref.~[133]). In some of them the authors propose to measure this speed.

There are interesting propositions to include primordial \gr\ waves to
fluctuations of a \q\ and vice versa. 

We consider a mass of a \q\ \pc, a speed of sound in a \q\ and
several solutions to \q\ equations coming to the interesting behaviour of an
\ef\ \gr\ \ct\ (see~[135]). Some of issues of our theory are considered in
self-interacting Brans-Dike theory [134].

The fraction $\eta$ can be connected with the part of an energy density of
\q\ field $Q$ in such a way that it is a fraction of fluctuation energy density
around an equilibrium~$Q_0$. In this way the $q_0$ field which can evolve due
to an evolution of~$Q$ and due to fluctuations of primordial \gr\ waves will
be a source of a gas of thermalized \pc s. It seems that an approach with a
boson equation of state is more appropriate to consider. 

It is interesting to consider a fraction of a dark energy stored as boson
particles as a dark matter in the \U.

\medskip
Now we can also calculate a mass of a skewon \pc\ in a linear \ap\ from
Sec.~17. Let us consider Eq.~(17.77) for a field $F_{\mu \rho \gamma }$ given
by Eq.~(17.78). The \ct\ in the formula (17.77) is a square of mass of a
skewon \pc\ (in a linear \ap\ of course). This is of course our $\la_{c0}$.
Thus we have
$$
m\dr{skewon}=\sqrt{\la_{c0}}\simeq 10^{-5}\,\text{eV}. \eq234
$$

Using Eqs \rf3 \ and \rf234 \ we get a relation between a skewon mass and a
\q\ mass particle
$$
m_0=\frac12\sqrt{n(n+2)}\,m\dr{skewon}\,. \eq234a
$$

Let us consider a primordial spectrum of \gr\ waves in our approach. This
spectrum is flat for our Hubble \ct\ is really \ct
$$
P\dr{grav}=\frac2{M^2\dr{pl}}\(\frac{H_0}{2\pi}\)^2 \eq235
$$
(see Ref.\ [110]). Thus
$$
n\dr{gr}=1. \eq236
$$

Now we start to examine \q\ fluctuations caused by fluctuations of a metric.
In order to do it we follow Ref.~[110] to perturb a spatial part of a metric
$$
ds^2=R^2(\tau)\[-d\tau^2+\(\d_{ij}+2\t E{}^T_{ij}\)dx^i\,dx^j\], \eq237
$$
$\tau$ is a conformal time.

If we expand Einstein equations for \rf237 \ up to linear terms, we get
$$
\frac{d^2E^T_{ij\vec k}}{d\tau^2}+2RH_K\frac{dE^T_{ij\vec k}}{d\tau}
+k^2E^T_{ij\vec k}=0, \quad K=0,1, \eq238
$$
where $\t E{}^T_{ij}$ is a spatial perturbation of the metric, $R(\tau)$ is a
scale factor depending on a conformal time, $E^T_{ij\vec k}$ is a $\vec k$
Fourier component $\t E{}^T_{ij}$
$$
\al
&\t E{}^T_{ij}=\frac1{(2\pi)^{3/2}}\int E^T_{ij\vec k}e^{-i\vec k\vec r}\,d^3\vec
k\\
&\vec k=(k_1,k_2,k_3), \quad |\vec k|^2=k^2, \quad \vec r=(x,y,z).
\eal
\eq239
$$
The important quantity is an amplitude of \gr\ wave corresponding to
$E^T_{ij}$, i.e.
$$
h_{ij\vec k}=RE^T_{ij\vec k}\,. \eq240
$$
$H_K$ is a Hubble \ct. We take $K=1$. The relation between a conformal time $\tau$ and
ordinary time~$t$ is given by
$$
\ealn{
R(\tau)&=-\frac1{H_1\tau}, \quad -\infty<\tau<0, &\rf241 \cr
R(t)&=R_1e^{H_1t}. &\rf242
}
$$
For $h_{ij\vec k}$ one gets
$$
\frac{d^2h_{ij\vec k}}{d\tau^2}-\frac2{\tau^2}\,\frac{dh_{ij\vec k}}{d\tau}
+k^2h_{ij\vec k}=0. \eq243
$$
This equation can be easily solved. For further investigations we need only
$$
\ealn{
h_{\vec k}&=h^i_{i\vec k} &\rf244 \cr
h_{\vec k}&=A_{\vec k}\(\tau|k| \cos(\tau|k|)-\sin(\tau|k|)\)+
B_{\vec k}\(\tau|k| \sin(\tau|k|)+\cos(\tau|k|)\) &\rf245
}
$$
and especially
$$
\frac{dh_{\vec k}}{d\tau}=A_{\vec k}\tau\sin(\tau|k|)+B_{\vec k}\tau
\cos(\tau|k|) \eq246
$$
where $A_{\vec k}$ and $B_{\vec k}$ are \ct s.

We need also a \q\ field time \dc ce (e.g.\ in a slow roll \ap). We use our
solution from Sec.~14 (see Eq.~(14.416)), making some simplifications and
changing $t$ into~$\tau$:
$$
\P=\ln\(\sqrt{\frac{n+2}n}\sqrt{\frac\b{|\g|}}\)+\ln\(\frac{1-C(-\tau)^{\t k}}
{1+C(-\tau)^{\u\k}}\). \eq247
$$
Moreover what we really need is a derivative of $\P$.
$$
\P'=\frac{d\P}{d\tau}=\frac{2C{\t\k}(-\tau)^{{\u\k}-1}}{1-C^2(-\tau)^{2{\u\k}}} \eq248
$$
where
$$
\ealn{
{\t\k}&=\frac{\o B\sqrt{2n}}{H_1} &\rf248a \cr
C&=e^{\o B\sqrt{2n}t_0}\(R_1H_1\)^{\o B\sqrt{2n}/H_1}=
e^{\o B\sqrt{2n}t_0}\(R_1H_1\)^{\u\k}. &\rf248b
}
$$

Now we proceed to a \q\ fluctuation equation (see Refs\ [111,133]):
$$
\d \ddot Q_{\vec k}+3H_1\d \dot Q_{\vec k}
+\(c^2_sk^2+R^2U''(Q)\)\d Q_{\vec k}= \frac1{2\o\b} h'_k\cdot \P'. \eq249
$$
One gets
$$
\al
\d \ddot Q_{\vec k}&+3H_1\d \dot Q_{\vec k}
+\(c^2_sk^2+R^2U''(Q)\)\d Q_{\vec k}\cr
&= \frac{C\t\k}{\o\b}\(A_{\vec k}\frac{(-\tau)^{{\u\k}-1}\sin(\tau|k|)}{1-C^2(-\tau)^{2{\u\k}}}
+B_{\vec k}\frac{(-\tau)^{{\u\k}-1}\cos(\tau|k|)}{1-C^2(-\tau)^{2{\u\k}}}\)\cr
&=\frac{C\t\k}{\o\b}\, C_{\vec k}\,\frac{\sin(\tau|k|+\d_{\vec
k})}{1-C^2(-\tau)^{2{\u\k}}}\,(-\tau)^{2{\u\k}}.
\eal
\eq250
$$
$\o\b$ is a normalization \ct\ in a definition of $Q=\frac\P{\o\b}$, $\frac1{\o\b{}^2}=\frac{8\pi |\o M|}{\mpl^2}$ (see
Eqs~(14.139) and \rf28 ).

Let us consider $\d Q_{\vec k}$ as a Fourier component of a $q_0$ field
subject to the following initial conditions:
$$
\frac{dq_{0\vec k}(0)}{d\tau}=0=q_{0\vec k}(0). \eq251
$$
In this case one gets
$$
\frac{d^2q_{0\vec k}}{d\tau^2}+3H_1\frac{dq_{0\vec k}}{d\tau}
+\(c^2_sk^2+\frac{m_0^2}{H_1^2\tau^2}\)q_{0\vec k}=
\frac{(-\tau)^{{\u\k}-1}C{\t\k}}{\o\b\(1{-}C^2(-\tau)^{2{\u\k}}\)}C_{\vec k}\sin\(\tau|k|
{+}\d_{\vec k}\), \eq252
$$
$c_s$ is  a speed of sound in a \q, $C_{\vec k}$ and $\d_{\vec k}$ are \ct s.
Let us notice that $\tau=0$ corresponds to $t\to\infty$. Thus in some sense
we have asymptotic conditions for \gr\ waves. 

For further convenience it is good to change $\tau$ into $-\tau$. In this way
we get
$$
\frac{d^2q_{0\vec k}}{d\tau^2}-3H_1\frac{dq_{0\vec k}}{d\tau}
+\(c^2_sk^2+\frac{m_0^2}{H_1^2\tau^2}\)q_{0\vec k}=
-\frac{C{\t\k}\tau^{{\u\k}-1}}{\o\b\(1{-}C^2\tau^{2{\u\k}}\)}C_{\vec k}\sin\(\tau|k|
{-}\d_{\vec k}\), \eq253
$$
$$
\tau\in(0,+\infty), \quad \tau=\frac1{R_1H_1}e^{-H_1t}. \eq254
$$

The fluctuations (the field $q_0$) of a \q\ $Q$ are driven by \gr\ waves and
a \q\ field~$Q$. It is easy to see that we are interested in solutions for
small~$\tau$ (large~$t$). In this case one gets
$$
\frac{d^2q_{0\vec k}}{d\tau^2}-3H_1\frac{dq_{0\vec k}}{d\tau}
+\frac{m_0^2}{H_1^2\tau^2}q_{0\vec k}=-\frac{C{\t\k}\tau^{{\u\k}-1}}{\o\b}\,
C_{\vec k}\sin(\tau|k|-\d_{\vec k}). \eq255
$$
Let us calculate a \ct\ ${\t\k}$. One gets
$$
\al
{\t\k}&=\frac{\o B\sqrt{2n}}{H_1}=\(\frac n{n+2}\)^{n/4}\(\frac{|\g|}\b\)^{n/4}
\frac{\sqrt{\o M}}{\mpl}\cdot4(n+2)|\g|^{1/2}\\
&=8\sqrt{3\pi n}(n+2)\sqrt{\o M}\(\frac n{n+2}\)^{n/4}
\(\frac{m_{\u A}}{\mpl}\)^{(n+2)/2}\,\frac{|\uP|^{1/2}}{\a_s^{(n+1)}}
\(\frac{|\uP|}{\RG}\)^{n/2}.
\eal
\eq256
$$
If we use our simplified model with $n=14$, $\a_s^2=\a\dr{em}$, $m_{\u A}=
m\dr{EW}$, we get
$$
{\t\k}\cong10^{92}. \eq257
$$
Thus this is really a large number.

Moreover for small $\tau$ we have
$$
q_{o\vec k}(\tau)\simeq0,\quad \frac{dq_{0\vec k}}{d\tau}\simeq0. \eq258
$$
In this way we arrive to an equation
$$
\frac{d^2q_{0\vec k}}{d\tau^2}=-\frac{C\t\k}{\o\b}\(B_{\vec k}k\tau^{\u\k}+A_{\vec k}\tau
^{{\u\k}-1}\) \eq259
$$
and finally
$$
q_{0\vec k}(\tau)=-\frac{B_{\vec k}|k|C{\t\k}}{\o\b({\t\k}+2)({\t\k}+1)}\,\tau^{{\u\k}+2}
-\frac{A_{\vec k}C{\t\k}}{\o\b({\t\k}+1){\t\k}}\,\tau^{{\u\k}+1} \eq260
$$
or (taking under consideration that ${\t\k}$ is very large)
$$
\ealn{
q_{0\vec k}(\tau)&=-\frac{C}{\o\b\t\k} \(B_{\vec k}k\tau^{\u\k}+A_{\vec k}\tau^{{\u\k}-1}\)
& \rf261 \cr
q_{0\vec k}(\tau)&=-\frac{C}{\o\b\t\k} \biggl(A_{\vec k}(R_1H_1)^{-\o B\sqrt{2n}/H_1}
e^{-\o B\sqrt{2n}t}\cr
&\qquad\qquad{}+kB_{\vec k}(R_1H_1)^{-((\o B\sqrt{2n}/H_1)-1)}
e^{-\o B\sqrt{2n}t}\cdot e^{H_1t}\biggr) &\rf262 \cr
q_0(t,\vec r)&=\int q_{0\vec k}(t)e^{i\vec k\vec
r}\,d\vec k. &\rf263
}
$$
One gets
$$
\al
q_0(t,\vec r)&=-\frac{C}{\o\b\t\k} \biggl(g(\vec r)\(\frac{H_1}{R_1}\)^{\o B\sqrt{2n}/H_1}
e^{-\o B\sqrt{2n}t}\\
&\qquad\qquad{}+f(\vec r)\(\frac{H_1}{R_1}\)^{(\o B\sqrt{2n}/H_1)-1}
e^{-2\o B\sqrt{2n}t}\cdot e^{H_1t}\biggr) 
\eal \eq264
$$
where
$$
\ealn{
f(\vec r)&=\int A_{\vec k}e^{i\vec k\vec r}\,d^3\vec k
&\rf265 \cr
g(\vec r)&=\int |k|B_{\vec k}e^{i\vec k\vec r}\,d^3\vec k.
&\rf266
}
$$

$g(\vec r)$ and $f(\vec r)$ characterize a spatial \dc ce of \gr\ waves
background. For large $t$ we have to do with $g(\vec r)$, i.e.
$$
q_0(t,\vec r)=-\frac{C}{\o\b\t\k} g(\vec r)(R_1H_1)^{-((\o B\sqrt{2n}/H_1)-1)}
\cdot e^{-\o B\sqrt{2n}t}e^{H_1t}. \eq267
$$
In this way one gets from an energy-momentum tensor for $q_0$
$$
T_{\mu\nu}=\pa_\mu q_0\cdot \pa_\nu q_0-\tfrac12\eta_{\mu\nu}
\(\pa^\a q_0\cdot\pa_\a q_0+m_0^2q_0^2\) \eq268
$$
$$
\al
\rho_{q_0}&=T_{44}=\dot q^2_0-\frac12\(\dot q^2_0-
|\stpr\nabla q_0|^2-m_0^2q_0^2\)\cr
&=\frac12\,q_0^2\(\(H_1-\o B\sqrt{2n}\)^2+
|\stpr\nabla\ln g(r)|^2+m_0^2\)\cr
&=\frac{\rho\dr{gr}2n\o B{}^2 R_1^2\(\(H_1-\o B\sqrt{2n}\)^2+
|\stpr\nabla\ln g(r)|^2+m_0^2\)}{\o\b{}^2 e^{2\o B\sqrt{2n}t}}\,
e^{2H_1t}e^{\o B\sqrt{2n}t_0}
\eal
\eq269
$$
where
$$
\rho\dr{gr}=\tfrac12 g^2 \eq270
$$
is an energy density of primordial \gr\ waves.

For ${\t\k}=\frac{\o B\sqrt{2n}}{H_1}$ is a large number,
$$
\o B\sqrt{2n} \gg H_1 \eq271
$$
and $\rho_{q_0}$ is going to zero if $t\to\infty$. This density will be
frozen (no \dc ce on time) at an end of the second de Sitter phase $t=
t\dr{end}\gi{II}$, $t_0=t\dr{initial}\gi{II}$. In this way one gets
$$
\frac{\rho_{q_0}}{\rho\dr{gr}}=
\frac{2n\o B{}^2 R_1^2\(\(H_1-\o B\sqrt{2n}\)^2+
|\stpr\nabla\ln g(r)|^2+m_0^2\)}{\o\b{}^2\exp\(\frac{2\o B\sqrt{2n}N_1}{H_1}\)}\,
e^{2N_1} \eq272
$$
where $N_1$ is an amount of \il\ during the second de Sitter phase.
If $g(\vec r)=\text{const.}$, $\rho_{q_0}$ is isotropic and homogeneous.

The function $g(\vec r)$ can be written in a form
$$
g(\vec r)=g_0+\d g(\vec r) \eq273
$$
where $\d g(\vec r)$ is a small deviation and $g_0=\text{const.}$ One gets
$$
\stpr\nabla\(\ln g(\vec r)\)\simeq \frac1{g_0}\stpr\nabla\d g(\vec r). \eq274
$$
In this way one writes
$$
\rho_{q_0}=\rho_{q_0}^0+\d \rho_{q_0} \eq275
$$
where
$$
\al
\rho_{q_0}^0&=\frac{2n\o B{}^2 R_1^2\rho\dr{gr}\(\(H_1-\o B\sqrt{2n}\)^2+
m_0^2\)}{\o\b{}^2\exp\(\frac{2\o B\sqrt{2n}N_1}{H_1}\)}\,
e^{2N_1}\\
\d\rho_{q_0}&=|\stpr\nabla g(\vec r)|^2 \,
\frac{2n\o B{}^2 R_1^2\rho\dr{gr}}{\o\b{}^2\exp\(\frac{2\o B\sqrt{2n}N_1}{H_1}\)}\,
e^{2N_1}
\eal
\eq276
$$

According to our ideas the energy $\rho_{q_0}$ should be stored as a gas of
\q\ \pc s. The measurement of a low frequency speed of sound and a high
frequency speed of sound can give us $\eta$ and its spatial variation. In
this way we get also a square length of a gradient of $\d g(\vec r)$. If we
suppose that $\rho\dr{gr}$ is stored as a gas of gravitons (massless \pc s
with two polarization states) then one can write
$$
p\dr{gr}=\rho\dr{gr} \(\frac{T\dr{gr}}{\,\o T_0\,}\)^4 \eq277
$$
(see Eq.\ \rf231 ).

Probably we should also consider a gas of skewons. However, this is a
different story. Taking into account \rf235  \ and \rf236 \ we should expect
$\d \rho_{q_0}$ as extremely small and we can neglect it in our theory. 
Moreover, $\rho\dr{gr}$ is very important. Let us consider Eq.~\rf253 \ for
these values of $\tau$ for which the variable~$t$ is close to
$t\dr{end}\gi{I}$ (the end of the first de Sitter phase or beginning of the
second de Sitter phase). In this way one writes
$$
\tau=\frac1{H_1R_1}\,e^{-H_1t\dr{end}\gi I}+\xi=\tau_0+\xi,
\quad 0<\xi\ll 1. \eq278
$$
Taking into account that $\xi$ is very small we arrive to the following equation
$$
\frac{d^2q_{0\vec k}}{d\xi^2}  - 3H_1\,\frac{dq_{0\vec k}}{d\xi}
+\(c_s^2k^2+\frac{m_0^2}{H_1^2\tau_0^2}\)q_{0\vec k}
=-\frac{C_{\vec k}\tau_0}{2\xi\o\b}\sin\(\xi|k|+\o\d_{\vec k}\) \eq279
$$
where
$$
\o\d_{\vec k}=\tau_0|k|-\d_{\vec k}\,. \eq280
$$
We have also initial conditions
$$
\frac{dq_{0\vec k}}{d\xi}(0)=q_{0\vec k}(0)=0. \eq281
$$
Eq.\ \rf279 \ can be solved by a Laplace \tr\ method in the case of
$\o\d_{\vec k}=0$. 

Let
$$
c_s^2k^2+\frac{m_0^2}{H_1^2\tau_0^2}=b^2. \eq282
$$
One gets
$$
s^2\o q_{0\vec k}(s)-3H_1s\o q_{0\vec k}(s)+b^2\o q_{0\vec k}(s)=
-\frac{C_{\vec k}\tau_0}{2\o\b} \arctg\(\frac{|k|}s\) \eq283
$$
$$
\o q_{0\vec k}(s)=-\frac{C_{\vec k}\tau_0}{2\o\b}\,
\frac{\arctg\bigl(\frac{|k|}s\bigr)}{s^2-3H_1s+b^2} \eq284
$$
and
$$
q_{0\vec k}(\xi)=-\frac{C_{\vec k}\tau_0}{2\o\b}\intop_0^\infty
\frac{\arctg\bigl(\frac{|k|}s\bigr)e^{s\xi}}{s^2-3H_1s+b^2}\,ds. \eq285
$$
However, we cannot get any compact form of this solution. Moreover for we are
looking for a solution around zero, for $\o\d_{\vec k}=0$ we get
$$
-\frac{C_{\vec k}\tau_0}{2\xi}\sin(\xi|k|)\simeq
-\frac{C_{\vec k}\tau_0|k|}{2}\,. \eq286
$$

In this case the Laplace \tr\ method is successful:
$$
\ealn{
q_{0\vec k}(\xi)&=-\frac{C_{\vec k}\tau_0|k|}{2\o\b}\intop_0^\infty
\frac{e^{s\xi}}{s(s^2-3H_1s+b^2)}\,ds &\rf287 \cr
q_{0\vec k}(\xi)&=-\frac{C_{\vec k}\tau_0|k|}{2\o\b}
\(\frac1{b^2}+\frac{e^{s_1\xi}}{s_1(s_1-s_2)}+\frac{e^{s_2\xi}}{s_2(s_2-s_1)}
\) &\rf287a
}
$$
where $s_1$ and $s_2$ are roots of the polynomial
$$
s^2-3H_1s+b^2=0. \eq288
$$
Moreover $\xi$ is small. Thus we write
$$
e^{s_i\xi}\cong 1+s_i\xi, \quad i=1,2, \eq289
$$
and get
$$
q_{0\vec k}(\xi)=-\frac{C_{\vec k}\tau_0|k|}{2b^2\o\b}\(1+\frac{3H_1}{\sqrt\D}\)
\eq290
$$
if
$$
\D=9H_1^2-4b^2=9H_1^2-4\Bigl(c_s^2k^2+\frac{m_0^2}{H_1^2\tau_0^2}\Bigr)>0. \eq291
$$

If $\D<0$, one gets
$$
q_{0\vec k}(\xi)=-\frac{C_{\vec k}\tau_0|k|}{2b^2\o\b}\(1+\exp(\tfrac32 H_1\xi)
\(\frac{3H_1}{\om}\sin(\om\xi)-\cos(\om\xi)\)\) \eq292
$$
\vskip-2pt
\noindent where
$$
4\om^2=-\D=4\Bigl(c_s^2k^2+\frac{m_0^2}{H_1^2\tau_0^2}\Bigr)-9H_1^2>0. \eq293
$$
For small $\xi$ one gets
$$
q_{0\vec k}(\xi)=-\frac{3 C_{\vec k}\tau_0|k|H_1}{4b^2\o\b}\,\xi, \eq294
$$
\vskip-2pt
\noindent the \dc ce on $t$ is as follows:
$$
\xi=\frac{-1}{R_1H_1}\bigl(e^{-H_1t}-e^{-H_1t\dr{end}\gi{I}}\bigr). \eq295
$$
Moreover $t$ is close to \smash{$t\dr{end}\gi{I}$} and we get
$$
\xi\simeq \frac{e^{-H_1t\dr{end}\gi{I}}}{R_1}(t-t\dr{end}\gi{I}) \eq296
$$
for $t$ very close to \smash{$t\dr{end}\gi{I}$}.

In this way we get fluctuations of a \q\ in the moment of the first order
phase transition of the first de Sitter phase to the second de Sitter phase. 
 
Finally one gets
$$
\al
q_{0\vec k}(t)&=-\frac
{C_{\vec k}|k|e^{-H_1t\gi{I}\dr{end}}}
{2H_1R_1\(c_s^2|k|^2+R_1^2m_0^2e^{2H_1t\gi{I}\dr{end}}\)\o\b}\cr
&\times\(1+\frac{3H_1}{\sqrt{9H_1^2-4\bigl(c_s^2|k|^2+R_1^2m_0^2e^{2H_1t\gi{I}\dr{end}}\bigr)}}\)
\eal
\eq297
$$
for
$$
|k|<\frac1{2c_s}\sqrt{9H_1^2-4R_1^2e^{2H_1t\gi{I}\dr{end}}m_0^2}=k_0 \eq298
$$
with a condition
$$
3H_1>2R_1e^{H_1t\gi{I}\dr{end}}m_0 \eq299
$$
and
$$
q_{0\vec k}(t)=-\frac
{3 C_{\vec k}e^{-2H_1t\gi{I}\dr{end}}(t-t\gi{I}\dr{end})|k|}
{4R_1^2\(c_s^2k^2+R_1^2m_0^2e^{2H_1t\gi{I}\dr{end}}\)\o\b}\, .
\eq300
$$
for $k>k_0$.

One finds
$$
\al
q_0(\vec r,t)&=-\frac{e^{-H_1t\gi{I}\dr{end}}}{2\o\b R_1}
\( \frac1{H_1}\intop_{|k|<k_0}
d^3\vec k\, e^{i\vec k\vec r}\frac
{C_{\vec k}|k|}{\(c_s^2k^2+R_1^2m_0^2e^{2H_1t\gi{I}\dr{end}}\)}\right.\cr
&\quad{}\times\(1+\frac{3H_1}{\sqrt{9H_1^2-4\(c_s^2k^2+R_1^2m_0^2e^{2H_1t\gi{I}\dr{end}}\)}}\)
\\
&\left.+\frac{3(t-t\gi{I}\dr{end})}{2R_1}\,e^{-H_1t\gi{I}\dr{end}}
\intop_{|k|>k_0}d^3\vec k\, e^{i\vec k\vec r}\frac
{C_{\vec k}|k|}{\(c_s^2k^2+R_1^2m_0^2e^{2H_1t\gi{I}\dr{end}}\)}\)
\eal
\eq301
$$

Equation \rf301 \ gives us space-time fluctuations of a \q\ field after a
phase transition from the first to the second de Sitter phase. Let us notice
that \rf297 \ is singular for $k\to k_0$.

In an approach developed here we consider Eq.~\rf250 \ with zero initial
conditions in various \ap s. Moreover there is a different approach. In this
approach we write a solution to Eq.~\rf250
$$
q_{0\vec k}(t)=\t q_{0\vec k}(t)+\o q_{0\vec k}(t) \eq302
$$
where $\t q_{0\vec k}(t)$ is a solution of a homogeneous equation (a general
integral) and $\o q_{0\vec k}(t)$ is a special solution to inhomogeneous
equation. This different approach consists in taking $\t q_{0\vec
k}(t)\equiv0$ and considering only $\o q_{0\vec k}(t)$. In this case we get
nonzero initial conditions for \q\ fluctuations. However, this approach is not
unambigous. Our approach is unambigous. The zero initial conditions are in
some sense distinguished.

Let us calculate an energy density of this scalar field. In order to do this
let us write $q_0(\vec r,t)$ in a form
$$
q_0(\vec r,t)=A(\vec r)+B(\vec r)(t\gi{I}\dr{end}-t) \eq303
$$
where
$$
\ealn{
A(\vec r)&=-\frac{e^{-H_1t\gi{I}\dr{end}}}{2\o\b R_1H_1}
\intop_{|k|<k_0}
d^3\vec k\, e^{i\vec k\vec r}\frac
{C_{\vec k}|k|}{\(c_s^2k^2+R_1^2m_0^2e^{2H_1t\gi{I}\dr{end}}\)}\cr
&\quad{}\times\(1+\frac{3H_1}{\sqrt{9H_1^2-4\(c_s^2k^2+R_1^2m_0^2e^{2H_1t\gi{I}\dr{end}}\)}}\)
&\rf304 \cr
B(\vec r)&=-\frac{3 e^{-2H_1t\gi{I}\dr{end}}}{4\o\b R_1^2}
\intop_{|k|>k_0}d^3\vec k\, e^{i\vec k\vec r}\frac
{C_{\vec k}|k|}{\(c_s^2k^2+R_1^2m_0^2e^{2H_1t\gi{I}\dr{end}}\)}\,.
&\rf305
}
$$
One gets from the Eq.\ \rf268
$$
\rho_{q_0}=T_{44}=\frac12\(C(\vec r)+D(\vec r)(t\gi{I}\dr{end}-t)
+E(\vec r)(t\gi{I}\dr{end}-t)^2\) \eq306
$$
where
$$
\ealn{
C(\vec r)&=B^2(\vec r)+m_0^2A^2(\vec r)+\bigl|\stpr\nabla(A(\vec r))\bigr|^2 
&\rf307 \cr
D(\vec r)&=2\(\stpr\nabla (A(\vec r))\cdot \stpr\nabla (B(\vec r))
+m_0^2A(\vec r)B(\vec r)\) &\rf308 \cr
E(\vec r)&=\(\bigl|\stpr\nabla(B(\vec r))\bigr|^2+m_0^2B^2(\vec r)\).
&\rf309
}
$$
All these formulae are satisfied for $t$ close to $t\gi{I}\dr{end}$. In
general $\rho_{q_0}$ is anisotropic and inhomogeneous. For
$t\gi{I}\dr{end}-t$ is very small we get
$$
\rho_{q_0}=\frac12\(C(\vec r)+D(\vec r)(t\gi{I}\dr{end}-t)\). \eq310
$$

It is interesting to calculate a time $t(\vec r)$ for which $\rho_{q_0}$ is
zero ($t(\vec r)$ depends on a space point). One gets (neglecting gradients
of $A(\vec r)$ and $B(\vec r)$)
$$
t(\vec r)=t\gi{I}\dr{end}+\frac1{2m_0^2}\(\o f(\vec r)+
\frac{m_0^2}{\o f(\vec r)}\) \eq311
$$
where
$$
\o f(\vec r)=\frac{3H_1}{2R_1}\cdot
\frac{e^{-H_1t\gi{I}\dr{end}}
\intop_{|k|>k_0}d^3\vec k\, e^{i\vec k\vec r}\frac
{C_{\vec k}|k|}{(c_s^2k^2+R_1^2m_0^2e^{2H_1t\gi{I}\dr{end}})}}
{\intop_{|k|<k_0}
d^3\vec k\, e^{i\vec k\vec r}\frac
{C_{\vec k}|k|}{(c_s^2k^2+R_1^2m_0^2e^{2H_1t\gi{I}\dr{end}})}
\biggl(1+\frac{3H_1}{\sqrt{9H_1^2-4(c_s^2k^2+R_1^2m_0^2e^{2H_1t\gi{I}\dr{end}})}}\biggr)}\,.
\eq312
$$

The conclusion from these calculations is that after a very short time an
energy density of \q\ fluctuations is going to zero (almost zero) and
probably is frozen on a very low level. A time of this process depends on a
space point. Moreover, after sufficiently long time (which is really very
short) we get a homogeneity and isotropy, as far as our rough calculations
advise us. Let us come back to Eq.~\rf253 . Our solutions obtained here are
for large $t$ and $t$ close to $t\gi{I}\dr{end}$. One can try to match them.
It seems reasonable to suppose that $\rho^0_{q_0}$ (see Eq.~\rf276 ) is equal
to this very low level energy density, mentioned above.

The \q\ field $q_0(\vec r,t)$ considered above can be a source of
fluctuations of an \ef\ \gr\ \ct\ $G\dr{eff}$.
$$
G\dr{eff}=G_N \exp\(-(n+2)\o\b q_0(\vec r,t)\) \eq313
$$
(see Eq.\ \rf28 ). If we use Eq.\ \rf267 , we get
$$
G\dr{eff}=G_N \exp\Biggl(-\frac
{\sqrt2\,(n+2)\sqrt{\rho_{q_0}}\,\o\b}
{\sqrt{\bigl((H_1-\o B\sqrt{2n})^2+\bigl|\stpr\nabla \ln g(\vec r)\bigr|^2
+m_0^2\bigr)}}\Biggr) \eq314
$$
and we connect contemporary fluctuations of an \ef\ \gr\ \ct\ to an energy
density of \q\ fluctuations. Taking a \q\ field from \rf303 \ we get
$$
G\dr{eff}=G_N \exp\(-(n+2)\o\b\(A(\vec r)+B(\vec r)(t\gi{I}\dr{end}-t)\)\).
\eq315
$$
In this way an \ef\ \gr\ \ct\ depends on a speed of a sound in a \q.

In order to get some comparison let us consider a different \ap \ of a \q\
evolution given by Eqs (14.402) and (14.405),
$$
Q(t)=Q_0+\frac{\f_0}{\o \b}\sin(\om_0t)  \eq316
$$
where
$$
\om_0^2=\frac{M\dr{pl}}{2\o M}\,(n+2)\(\frac n{n+2}\)^{n/2}
|\g|\(\frac{|\g|}\b\)^{n/2} \eq317
$$
or
$$
\al
\om_0&=\frac{m_{\u A}\sqrt{\mpl}}{2\a_s^{(n+1)/2}\sqrt{2H}\sqrt{|\o M|}}
\(\frac{m_{\u A}}{\mpl}\)^{n/2}(n+2)^{1/2}\\
&\times\(\frac n{n+2}\)^{n/4}
|\t P|^{1/2}\(\frac{|\t P|}{\RG}\)^{n/4}.
\eal
\eq318
$$
In terms of a conformal time $\tau$ one gets
$$
\ealn{
Q(\tau)&=Q_0-\frac{\f_0}{\o\b}\sin\(\frac{\om_0}{H_1}\ln(-\tau R_1H_1)\)
&\rf319 \cr
\frac{dQ}{d\tau}&=-\frac{\f_0\om_0}{\o\b H_1}\cdot\frac1\tau\cos\(
\frac{\om_0}{H_1}\ln(-\tau R_1H_1)\) &\rf320
}
$$
Let us notice that $\f_0$ is an arbitrary \ct.

Using Eq.\ \rf246 \ and Eq.\ \rf249 \ one gets
$$
\al
\d\ddot Q_{\vec k}&-3H_1\d \dot Q_{\vec k}
+\(c_s^2|k|^2+\frac{m_0^2}{H_1^2\tau^2}\)\d Q_{\vec k}\\
&=\frac{\f_0\om_0}{2H_1\o\b}\(A_{\vec k}\sin(\tau|k|)-B_{\vec k}\cos(\tau|k|)\)
\cos\(\frac{\om_0}{H_1}\ln(\tau R_1H_1)\) 
\eal
\eq321
$$
where we change $\tau \to -\tau$, $\tau\in(0,+\infty)$.

Let us consider $\d Q_{\vec k}$ as \q\ field $q_{0\vec k}$ (as before) and
let us consider equation \rf321 \ for
$$
\tau=\tau_0+\xi=\frac1{H_1R_1}\,e^{-H_1 t\gi{I}\dr{end}}+\xi,
\qquad 0<\xi\ll 1 \eq322
$$
(also as before).

We can consider Eq.\ \rf321 \ only for $\tau$ given by \rf322 , for our \ap\
works only here.

Changing a variable from $\tau$ to $\xi$ and doing some simplifications we get
$$
\al
\frac{d^2q_{0\vec k}}{d\xi^2}&-3H_1\,\frac{dq_{0\vec k}}{d\xi}
+\(c_s^2|k|^2+\frac{m_0^2}{H_1^2\tau_0^2}\)q_{0\vec k}\\
&=-\frac{\f_0\om_0}{2H_1\o \b}\,C_{\vec k}\cos(\om_0t\gi{I}\dr{end})
\sin(|k|\xi+\o\d_{\vec k})
\eal
\eq323
$$
where $\o\d_{\vec k}=\tau_0|k|-\d_{\vec k}$ (as before).

For a very small $\xi$ we get
$$
\sin(|k|\xi+\d_{\vec k})\simeq \sin\d_{\vec k} \eq324
$$
and we get Eq.\ \rf279 \ for $\o\d_{\vec k}=0$ and for small $\xi$, $C_{\vec
k}$ is expressed in terms of $A_{\vec k}$ and $B_{\vec k}$,
$$
C_{\vec k}=\sqrt{A_{\vec k}^2+B_{\vec k}^2}\,. \eq325
$$
In this way using a Laplace \tr\ method for this equation with initial conditions
\rf281 \ one gets
$$
q_{0\vec k}(\xi)=-F\(\frac1{b^2}+\frac{e^{s_1\xi}}{s_1(s_1-s_2)}
+\frac{e^{s_2\xi}}{s_2(s_2-s_1)}\) \eq326
$$
where $s_1$ and $s_2$ are roots of the equation \rf288 \ and
$$
F=\frac{\f_0\om_0C_{\vec k}\sin\o\d_{\vec k}\cos(\om_0t\gi{I}\dr{end})}
{2H_1\o\b}\,. \eq327
$$
In this way we get similar solutions as before
$$
\al
q_{0\vec k}(t)&=-\frac{C_{\vec k}\sin\o\d_{\vec k}\cos(\om_0t\gi{I}\dr{end})
\f_0\om_0}
{2\o\b H_1\(c_s^2|k|^2+R_1^2m_0^2e^{2H_1t\gi{I}\dr{end}}\)}\\
&\times\(1+\frac{3H_1}
{\sqrt{9H_1^2-4\bigl(c_s^2|k|^2+R_1^2m_0^2e^{2H_1t\gi{I}\dr{end}}\bigr)}}\)
\eal
\eq328
$$
for $|k|$ satisfying condition \rf298 \ and
$$
q_{0\vec k}(t)=-\frac
{3\f_0\om_0C_{\vec k}\sin\o\d_{\vec k}\cos(\om_0t\gi{I}\dr{end})
e^{-H_1t\gi{I}\dr{end}}}
{4\o\b R_1H_1\(c_s^2|k|^2 +R_1^2m_0^2e^{2H_1t\gi{I}\dr{end}}\)}\(t-t\gi{I}\dr{end}\)
\eq329
$$
for $k>k_0$ (see Eq.\ \rf298 ).

In this way one gets
$$
\al
q_0(\vec r,t)&=-\frac{\f_0\om_0\cos(\om_0t\gi{I}\dr{end})}{2\o\b H_1}
\( \intop_{|k|<k_0}
d^3\vec k\, e^{i\vec k\vec r}\frac
{C_{\vec k}\sin\o\d_{\vec k}}
{\(c_s^2|k|^2+R_1^2m_0^2e^{2H_1t\gi{I}\dr{end}}\)}\right.\\
&\quad{}\times\(1+\frac{3H_1}
{\sqrt{9H_1^2-4\bigl(c_s^2|k|^2+R_1^2m_0^2e^{2H_1t\gi{I}\dr{end}}\bigr)}}\)
\\
&\left.+\frac{(t-t\gi{I}\dr{end})}{2R_1}\,e^{-H_1t\gi{I}\dr{end}}
\intop_{|k|>k_0}d^3\vec k\, e^{i\vec k\vec r}\frac
{C_{\vec k}\sin\o\d_{\vec k}}
{\(c_s^2|k|^2+R_1^2m_0^2e^{2H_1t\gi{I}\dr{end}}\)}\)
\eal
\eq330
$$

Thus we get similar \dc ce on time as before (see Eq.\ \rf303 ). Now with the
functions
$$
\ealn{
\o A(\vec r)&=-\frac{\f_0\om_0\cos(\om_0t\gi{I}\dr{end})}{2\o\b H_1}
\intop_{|k|<k_0}
d^3\vec k\, e^{i\vec k\vec r}\frac
{C_{\vec k}\sin\o\d_{\vec k}}
{\(c_s^2|k|^2+R_1^2m_0^2e^{2H_1t\gi{I}\dr{end}}\)}\cr
&\quad{}\times\(1+\frac{3H_1}
{\sqrt{9H_1^2-4\bigl(c_s^2|k|^2+R_1^2m_0^2e^{2H_1t\gi{I}\dr{end}}\bigr)}}\)
&\rf331 \cr
\o B(\vec r)&=\frac
{3\f_0\om_0\cos(\om_0t\gi{I}\dr{end})}{4\o\b R_1H_1}
\,e^{-H_1t\gi{I}\dr{end}}
\intop_{|k|>k_0}d^3\vec k\, e^{i\vec k\vec r}\frac
{C_{\vec k}\sin\o\d_{\vec k}}
{\(c_s^2|k|^2+R_1^2m_0^2e^{2H_1t\gi{I}\dr{end}}\)}
\qquad\qquad&\rf332
}
$$
and
$$
\al
\o f(\vec r)&=\frac
{\displaystyle
3e^{-H_1t\gi{I}\dr{end}}
\int_{|k|>k_0}d^3\vec k\, e^{i\vec k\vec r}\frac
{C_{\vec k}\sin\o\d_{\vec k}}
{\bigl(c_s^2|k|^2+R_1^2m_0^2e^{2H_1t\gi{I}\dr{end}}\bigr)}}
{\displaystyle
2R_1\int_{|k|<k_0}
d^3\vec k\, e^{i\vec k\vec r}\frac
{C_{\vec k}\sin\o\d_{\vec k}}
{\bigl(c_s^2|k|^2+R_1^2m_0^2e^{2H_1t\gi{I}\dr{end}}\bigr)}}\\
&\times
\(1+\frac{3H_1}
{\sqrt{9H_1^2-4\bigl(c_s^2|k|^2+R_1^2m_0^2e^{2H_1t\gi{I}\dr{end}}\bigr)}}\)
\eal
\eq333
$$
in the place of $A(\vec r)$, $B(\vec r)$, $f(\vec r)$ (see Eqs
\rf306$\div$312 \ and a discussion below which is applicable here). 

The general conclusion from our recent calculations is such that a behaviour 
of \q\ fluctuations does not depend on an evolution law of a \q\ for time
close to $t\gi{I}\dr{end}$.

Let us consider Eq.\ (14.402) for $\o M<0$. In this case one gets
$$
Q=Q_0+\f_0\sh(\om_0t) \eq334
$$
where $\om_0$ is given by Eq.\ \rf 318 . It is easy to see that we can repeat
all the considerations given above, changing $\sin$ into $\Sh$ and $\cos$
into $\Ch$. In this way one gets
$$
\al
q_{0\vec k}(t)&=-\frac
{C_{\vec k}\sin\o\d_{\vec k}\Ch(\om_0t\gi{I}\dr{end})\f_0\om_0}
{2\o\b H_1\(c_s^2|k|^2 +R_1^2m_0^2e^{2H_1t\gi{I}\dr{end}}\)}\\
&\times
\(1+\frac{3H_1}
{\sqrt{9H_1^2-4\bigl(c_s^2|k|^2+R_1^2m_0^2e^{2H_1t\gi{I}\dr{end}}\bigr)}}\)
\eal
\eq335
$$
for $|k|$ satisfying condition \rf 298 \ and
$$
q_{o\vec k}(t)=-\frac
{3\f_0\om_0C_{\vec k}\sin\o\d_{\vec k}\Ch(\om_0t\gi{I}\dr{end})e^{-H_1t\gi{I}\dr{end}}}
{4\o\b R_1H_1\(c_s^2|k|^2 +R_1^2m_0^2e^{2H_1t\gi{I}\dr{end}}\)}
(t-t\gi{I}\dr{end}) \eq336
$$
for $k>k_0$ (see Eq.\ \rf 298 ).

One finds
$$
\al
q_0(\vec r,t)&=-\frac{\f_0\om_0\Ch(\om_0t\gi{I}\dr{end})}{2\o\b H_1}
\( \intop_{|k|<k_0}
d^3\vec k\, e^{i\vec k\vec r}\frac
{C_{\vec k}\sin\o\d_{\vec k}}
{\(c_s^2|k|^2+R_1^2m_0^2e^{2H_1t\gi{I}\dr{end}}\)}\right.\\
&\quad{}\times\(1+\frac{3H_1}
{\sqrt{9H_1^2-4\bigl(c_s^2|k|^2+R_1^2m_0^2e^{2H_1t\gi{I}\dr{end}}\bigr)}}\)
\\
&\left.+\frac{(t-t\gi{I}\dr{end})}{2R_1}\,e^{-H_1t\gi{I}\dr{end}}
\intop_{|k|>k_0}d^3\vec k\, e^{i\vec k\vec r}\frac
{C_{\vec k}\sin\o\d_{\vec k}}
{\(c_s^2|k|^2+R_1^2m_0^2e^{2H_1t\gi{I}\dr{end}}\)}\)
\eal
\eq337
$$
and consequently
$$
\ealn{
\o{\o A}(\vec r)&=\frac{\Ch(\om_0t\gi{I}\dr{end})}
{\cos(\om_0t\gi{I}\dr{end})}\,\o A(\vec r) &\rf338 \cr
\o{\o B}(\vec r)&=\frac{\Ch(\om_0t\gi{I}\dr{end})}
{\cos(\om_0t\gi{I}\dr{end})}\,\o B(\vec r) &\rf339 \cr
\o{\o f}(\vec r)&=\o f(\vec r) &\rf340
}
$$
with the same conclusions as for a case with $\o M>0$. In this case an amount
of \il\ reads
$$
\al
N_1&=\D tH_1=\frac{H_1}{\om_0}\left|\ars\(\frac1{\f_0n}\)\right|
=\frac{H_1}{\om_0}\ln\(\frac{1+\sqrt{\f_0^2n^2+1}}{\f_0n}\)\\
&=\frac{M\dr{pl}}{\sqrt{6|\o M|}\,(n+2)}
\ln\(\frac{1+\sqrt{1+\f_0^2n^2}}{\f_0n}\)
\eal
\eq341
$$

\let\th\theta
\def\<{\left\langle}
\def\>{\right\rangle}

Let us give simple applications of our \il\ theory (which is effectively
similar to a hybrid \il) to CMB anisotropy problem.
In order to do this we remind to the reader some notions of CMB anisotropy
theory: 
$$
\frac{\D T(\vec e)}T = \sum_{l,m}a_{lm}Y_{l,m}(\vec e) \eq342
$$
where $Y_{lm}(\vec e)=Y_{lm}(\th,\f)$ are spherical harmonics and $a_{lm}$
are multipole coefficients of~CMB. $\frac{\D T(\vec e)}T$ means fluctuations
of a temperature of CMB measured in a direction of~$\vec e$ parametrized by
two angles $\th$ and~$\f$. A~direction of a photon is $\vec n=-\vec e$.

After averaging one gets
$$
\<a_{lm}a^*_{l'm'}\>=\d_{ll'}\d_{mm'}C_l. \eq343
$$
For a two point correlation function one gets
$$
\<\frac{\D T}T(\vec e_1)\cdot \frac{\D T}T(\vec e_2)\>
_{\vec e_1\cdot\vec e_2=\cos\th}=
\frac1{4\pi}\sum_l(2l+1)C_lP_l(\cos\th) \eq344
$$
(see Refs [110] for more details) where $P_l(\cos\th)$ is a Legendre
polynomial. Thus in order to compare a theory with observations it is
necessary to calculate $C_l$ in some region of~$l$ (i.e., $2<l<l_{\max}$).
Usually we subtract effects of relative movement and $C_0=C_1=0$.

$C_l$ can have different origins coming from several types of perturbations
(i.e.\ scalar, vector, tensor) (see Refs [110] for details). In our theory we
have scalar perturbations due to \il\ and tensor perturbations due to \gr\
waves. In our three models of Higgs' field evolution we get a power spectrum
of primordial scalar fluctuations (see Eqs (18.87), (18.69), (18.10)) and a
power spectrum for tensorial fluctuations (\gr\ waves fluctuations)
Eq.~\rf235 . The tensor fluctuations are scale invariant 
$$
n_T=n\dr{gr}-1=0.
$$
In the case of scalar fluctuations we calculate $n_s$ and $\frac{dn_s}{d\ln
K}$ (see Eqs (18.14), (18.15), (18.72), (18.74), (18.90), (18.91)) and we
find conditions for $n_s=1$. Thus we get (under some conditions) a
Harrison-Zel$'$dovich spectrum. Moreover $\frac{dn_s}{d\ln K}$ can be in
general nonzero and it is possible to use it to compare with observations in
order to falsificate some models. In general we get in three cases
$$
P^s_R(K)=P^s_0K^{n_s-1}. \eq345
$$
Thus via a standard procedure (see Refs [110]) one gets
$$
C^s_l=\frac{\(P_0^s\)^2}9 \cdot \frac
{\G(3-n_s)\G(l-\frac12+\frac{n_s}2)}
{2^{3-n_s}\G^2(2-\frac{n_s}2)\G(l+\frac52-\frac{n_s}2)} \eq346
$$
where $n_s\simeq1$.

If we take $n_s=1$ we get 
$$
l(l+1)C^s_l=\text{const.} \eq347
$$
which is in an agreement with observational data. In the case of tensorial
(\gr\ waves) perturbations one gets
$$
P^T(K)= P\gi{gr}_0 K^{n\dr{gr}-1}=P\gi{gr}_0K^{n_T} \eq348
$$
and by a standard procedure
$$
C^T_l=C\gi{gr}_l= \(P\gi{gr}_0\)^2\cdot \frac{(l+2)!}{(l-2)!}
\cdot \frac{\G(6-n_T)\G(l-2+\frac{n_T}2)}
{2^{6-n_T}\G^2(\frac72-n_T)\G(l+4-\frac{n_T}2)}\,. \eq349
$$
For $n_T=0$ or $n\dr{gr}=1$ one gets
$$
C\gi{gr}_l=\(P\gi{gr}_0\)^2\cdot\frac8{15\pi}\cdot
\frac{l^2(l+1)^2}{(l+3)(l-2)} \eq350
$$
(see Refs [110]). The singularity for $l=2$ is not realistic, there is only
an amplification for $l=2$.

Let us notice that our power flat spectrum (scale-invariant) is a good
starting point for every structure formation theory in the \U. Using our full
$P^s_R(K)$ spectras we can do more precise calculations for~$C^s_l$.
Especially in order to compare with future very precise \co\ observations
(Planck satellite) we should take under consideration $\frac{dn_s}{d\ln K}$
and deviations from scale invariant spectrum, which has been reported from
WMAP data. Some further developments of our $P_R(K)$ spectras are beyond the
scope of this work for they strongly depend on details of \co\ models
involved (see Refs~[110]). In particular we can consider $C^s_l$ and
$C\gi{gr}_l$ only for $l\lesssim100$. For $l\gtrsim100$, $\frac{\D T}T$ is
dominated by acoustic oscillation (see Refs~[110]). $C\gi{gr}_l$ 
are very small for $l\ge60$ (they decay on sub-horizon
scales). Finally we write
$$
C^s_l l(l+1) = \text{const.} \cong
\<\(\frac{\D T}T(\th_l)\)^2\>, \eq351
$$
$$
\th_l=\frac\pi l \eq352
$$
and
$$
C\gi{gr}_l l(l+1)=\(P\gi{gr}_0\)^2 \cdot \frac8{15\pi}\cdot\frac{l(l+1)}
{(l+3)(l-2)}\,. \eq353
$$
The behaviour of $C_l$ for $l>100$ is beyond the scope of this work (now).

The interesting point in future calculations is to consider an influence of
\q\ fluctuations. Those fluctuations are of scalar nature. For they depend on
\gr\ waves fluctuations their influence on $C^s_l$ is negligible.

We can consider some applications of a \q\ in astrophysics. First of all we
can develop a formalism concerning a macroscopic flow of a \q\ gas (i.e., a
relativistic hydrodynamics of a \q). For a \q\ field has an influence on an
\ef\ \gr\ \ct, such a flow can disturb \gr\ interactions between galaxies or
even stars. In this approach we should add some assumptions, e.g.\ that a \q\
energy density is an energy density of \q\ field in a statistical field theory.
However, this is beyond a scope of this work.

Secondly, it is interesting to consider a Bose-Einstein condensation of \q\
\pc s. They are massive scalar \pc s. Under some assumptions (beyond this
theory) we can find a wave function of a condensat. This wave function can be
considered a field of a \q\ and afterwards enters the formula for
$G\dr{eff}$. In this way Einstein-Bose condensation can influence effective
\gr\ interactions between galaxies and stars. This is also beyond a scope of
the work.

As a typical astrophysical application we can consider a problem of
$N$~bodies with the \q\ field. In this problem $N$~points interact \gr ly
with an \ef\ \gr\ \ct\ depending on a distance. The simplest case is of
course a two bodies problem. However, this problem cannot be solved
analytically. It can be considered numerically using some codes for the
$N$~bodies problem. This is also beyond a scope of our work. Moreover, some
numerical solutions can be applied for galaxies movement in a cluster of
galaxies. 

In all of those problems we should consider \q\ fluctuations developed here
as a source of \q\ gas and Bose-Einstein condensat. In the third problem we
should also consider an interaction of a \q\ field with mass points.

It is interesting to consider a continuous distribution of a matter (a dust)
under \gr\ interaction and a \q. In this way we consider hydrodynamic
equation coupled to a \q\ field in a Newtonian physics limit (i.e.\
selfgravitating system with $G\dr{eff}$ depending on a \q).

Let us notice the following fact. In our theory we have two natural
candidates for a dark matter. They are a skewon and a \q\ particle. Both
particles are massive with a mass of the same order ($10^{-5}$\,eV, see Eqs
(19.234--234a)). They are weakly interacting with an ordinary matter, they
are a part of gravity. Skewon is a quant of a quantized (in a linear
approximation) skew-\sy ic part of the metric. In contradiction to a graviton
it obtains a mass due to a \co\ \ct. A~\q\ \pc\ is a quant of a quantized
scalar field $\P$ in a linear approximation. The scalar field $\P$ plays many
roles in the theory. It influences an \ef\ \gr\ \ct\ and \co\ terms.

Simultaneously we explain \co\ \ct\ via our \q\ scenario.

Both \pc s interact \gr ly---they are part of gravity. A~dark matter problem
appears on many levels in the \U. On the level of galaxies, clusters of
galaxies and on the level of \co\ models. It seems, it is quite universal.
This universality is similar to the universality of \gr\ interactions. Some
researchers claim that because of this it is natural to change a \gr\ theory
in order to give a pure \gr\ explanation of the effect of a dark matter.
Moreover, in our proposal we have to do with a deformation of General
Relativity (due to \nos\ metric and a scalar field from the \eu\nos\ \JT\
Theory). The theory satisfies a Bohr correspondence principle. The additional
\gr\ degrees of freedom are universal as a gravity itself. Thus both massive
\pc s corresponding to these degrees of freedom can be considered seriously
as a dark matter (gravitons cannot be ``a~dark matter'', they are massless, they
are a part of a radiation). \eu\co\ models with such a dark matter will be
examined elsewhere.

Let us consider a scale of length from (19.63a)
$$
L=\sqrt{\frac{n(n+2)\la_{\text{co}}}{16\pi|\ov M|}}
\simeq 10\,\text{Mpc}, \eq354
$$
where 
$\la_{\text{co}}$ is a \co\ \ct.

$L$ has been calculated for $n=14=\dim G2$ and for $|\ov M|=1$. Moreover, 
$G2$ has been considered as a
group $H$ in the Nonsymmetric Kaluza--Klein (Jordan--Thiry) Theory. 
$G2$ is important
only for Glashow--Weinberg--Salam model of unification of electroweak
interactions. If we want to unify all fundamental interactions we need a
bigger group~$H$, such that $G2\subset H$. We need of course a group $G$ such that
$$
\text{SU}(2)_L \times \text{U}(1) \times \text{SU}(3)_c \subset G. \eq355
$$
There are a lot of possibilities. One of the most promising is $G=\text{SO}(10)$.
Moreover, we need also a group $G_0$ such that $M=G/G_0$.

In our world $G_0=\text{U}(1)_{\text{em}} \times \text{SU}(3)_c$. The group $H$ for $G=\text{SO}(10)$
and $G_0=\text{U}(1)_{\text{em}}\times \text{SU}(3)_c$ should be such that
$$
\text{SO}(10)\times \bigl(\text{U}(1)_{\text{em}}\times \text{SU}(3)_c\bigr) \subset H.
\eq356
$$
The simplest choice is $H=\text{SO}(16)$. Why?

First of all $G2\subset \text{SO}(16)$ and $\text{SO}(10)\times\text{SO}(6)\subset\text{SO}(16)$.
Moreover, $\text{SO}(6)\simeq\text{SU}(4)$ and $\text{U}(1)\times \text{SU}(3) \subset\text{SU}(4)$. Thus if we
identify $\text{U}(1)$ with $\text{U}(1)_{\text{em}}$ and $\text{SU}(3)$ with $\text{SU}(3)_c$ we get
what we want. In this way
$$\dsl{\indent
M=\raise6pt\hbox{SO(10)}\bigg/
\lower6pt\hbox{$\text{U}(1)\times\text{SU}(3)$}, \quad 
S^2\subset M,\hfill\cr 
\hfill \dim\text{SO}(16)=120, \ \dim\text{SO}(10)=45, \ n_1=\dim M=36.\indent}
$$
Thus the scale of length should be rescaled and we get for $n=120$
$$
L=\frac{80.2\,\text{Mpc}}{\sqrt{|\ov M|}}\,.\eq357
$$
For a scale of time
$$
T=\frac{258.56}{\sqrt{|\ov M|}}\times10^6\,\text{yr}.\eq358
$$
Moreover, we still have $L=\xi\cdot 10$\,Mpc, $T=\xi\cdot32\times10^6$\,yr, where
$\xi$ is of order~1.

}
\vfill\eject

{\let\ealn\eqalignno
\let\eq\equiv
\let\ot\otimes
\let\ov\overline
\let\os\overrightarrow
\let\pa\partial
\let\q\quad
\def\qh#1{\quad\hbox{#1}\quad}
\let\ul\underline
\let\wt\widetilde
\let\a\alpha
\let\b\beta
\let\d\delta
\let\D\varDelta
\let\ve\varepsilon
\let\g\gamma
\let\G\varGamma
\let\l\lambda
\let\La\Lambda
\let\o\omega
\let\O\varOmega
\let\vf\varphi
\let\si\sigma
\let\t\theta
\let\z\zeta
\def\m{{\mu\nu}}
\def\cL{{\Cal L}}
\def\dg #1,#2,#3,{{{#1}_{#2}}^{\!#3}}
\def\gd #1,#2,#3,{{{#1}^{#2}}_{\!#3}}
\def\nad#1#2{\overset\text{\rm #1}\to{#2}}
\def\falg{{\mathrel{\lower5pt\hbox{${\scriptstyle\sim}$}\hskip-5pt g}}}
\def\falH{{\mathrel{\lower4pt\hbox{${\scriptstyle\sim}$}\hskip-7.5pt H}}}
\def\hor{\operatorname{hor}}
\def\div{\operatorname{div}}
\def\pp#1#2{\frac{\pa #1}{\pa #2}}
\def\eqref#1{{\rm(#1)}}
\let\dsl\displaylines
\def\em{\text{\rm em}}
\def\crt{\text{\rm crt}}

\let\bal\aligned \let\eal\endaligned
\let\bga\gathered \let\ega\endgathered

\def\up#1{\uppercase{#1}}
\def\eu{\expandafter\up}
\def\cf{coefficient}
\def\cfn{confinement}
\def\cn{connection}
\def\dv{derivative}
\def\elm{electromagnetic}
\def\e{equation}
\def\f{function}
\def\gr{gravitation}
\def\nos{nonsymmetric}
\def\NK#1{Nonsymmetric Kaluza--Klein #1Theory}
\def\so{solution}
\def\st{such that }
\def\s{symmetric}
\def\wrt{with respect to }
\def\tag#1{\eqno(#1)}

\null
\vskip36pt plus4pt minus4pt
\centerline{{\four\bf 20. Charge Confinement}}

\vskip4pt 
\centerline{{\four\bf in the \NK{}}}
\vskip24pt plus2pt minus2pt
\write0{20. Charge Confinement in the Nonsymmetric Kaluza--Klein Theory \the\pageno}

We give some elements of the \NK{} in order to remind some notions to the
reader. 

Let $\ul P$ be a principal fibre bundle with a structural group $G=U(1)$ over
a space-time $E$ with a projection $\pi$ and let us define on this bundle a
connection~$\a$. We call this bundle an {\it \elm\ bundle} and $\a$ an {\it
\elm\ connection}. We define a curvature $2$-form for the connection~$\a$
$$
\O=d\a\eq \frac12\,\pi^*(F_\m\ov\t^\mu\wedge\ov\t^\mu)
\tag{20.1}$$
where
$$
F_\m=\pa_\mu A_\nu- \pa_\nu A_\mu, \q e^*\a=A_\mu\ov\t^\mu.
\tag{20.2}$$
$A_\mu$ is a 4-potential of the \elm\ field, $e$ is a local section of~$\ul
P$, $F_\m$ is an \elm\ field strength, and $\ov\t^\mu$ is a frame
on~$E$. Bianchi identity is
$$
d\O=0,
\tag{20.3}$$
so the 4-potential exists. This is of course simply the first pair of Maxwell
\e s. On the space-time $E$ we define a \nos\ metric tensor $g_{\a\b}$ \st
$$
\bal {}&g_{\a\b}=g_{(\a\b)}+g_{[\a\b]}\\
&g_{\a\b}g^{\g\b}=g_{\b\a}g^{\b\g}=\d^\g_\a,
\eal
\tag{20.4}$$
where the order of indices is important. In such a way we suppose that
$$
g=\det g_{\a\b}\ne0.
\tag{20.5}$$
We suppose also that
$$
\wt g=\det g_{(\a\b)}\ne0,
\tag{20.6}$$
defining an inverse tensor ${\wt g}^{(\a\b)}$ for $g_{(\a\b)}$ \st 
${\wt g}^{(\a\b)}g_{(\b\mu)}=\d_\mu^\a$.
We define also on $E$ two connections ${{\ov w}^\a}_{\!\b}$ and ${{\ov W}^\a}_{\!\b}$
$$
{{\ov w}^\a}_{\!\b}={{\ov \G}^\a}_{\!\b\g}{\ov \t}^\g
\tag{20.7}$$
and
$$
{{\ov W}^\a}_{\!\b}={{\ov W}^\a}_{\!\b\g}\ov \t^\g
\tag{20.8}$$
\st
$$
{{\ov W}^\a}_{\!\b}={{\ov w}^\a}_{\!\b}-\tfrac23\,{\d^\a}_{\!\b}\ov W,
\tag{20.9}$$
where
$$
\ov W=\ov W_\g\ov \t^\g = \frac12\bigl({{\ov W}^\si}_{\!\g\si}-
{{\ov W}^\si}_{\!\si\g}\bigr)\ov \t^\g.
$$
For the connection ${{\ov w}^\a}_{\!\b}$ we suppose the following conditions
$$
\bga
\ov D g_{\a+\b-}=\ov Dg_{\a\b}-g_{\a\d}{\ov Q}^\d_{\b\g}(\ov\G){\ov\t}^\g=0\\
\gd \ov Q,\a,\b\a,(\ov\G)=0,
\ega
\tag{20.10}$$
where $\ov D$ is the exterior covariant \dv\ \wrt $\gd\ov w,\a,\b,$ and $\gd
\ov Q,\a,\b\a,(\ov \G)$ is a torsion of $\gd\ov w,\a,\b,$. Thus we have defined
on space-time all quantities present in Moffat's theory of \gr\ (NGT, see
Ref.~[34]). Let us introduce on~$\ul P$ a frame
$$
\t^A=\bigr(\pi^*({\ov\t}^\a),\l\a=\t^5\bigr), \q \l=\text{\rm const}.
\tag{20.11}$$

Now we turn to the natural \nos\ metrization of the bundle $\ul P$. We have
$$
\g=\pi^*g-\t^5\ot\t^5=\pi^*(g_{\a\b}{\ov\t}^\a\ot{\ov\t}^\b)-\t^5\ot\t^5.
\tag{20.12}$$
From the classical Kaluza--Klein theory we know that $\l=2\frac{\sqrt
G}{c^2}$. We work with such a system of units that $G=c=1$ and $\l=2$. Thus
we have
$$
\bga
\g_{AB}=\left(\matrix g_{\a\b} & \ & 0 \\ 0 && -1 \endmatrix \right),\\
\g=\g_{AB}\t^A\ot\t^B.
\ega
\tag{20.13}$$

Now we define on $\ul P$ a \cn\ $\gd w,A,B,$ \st
$$
\bga
\gd w,A,B,=\gd\G,A,BC,\t^C\\
D\g_{A+B-}=D\g_{AB}-\g_{AD}\gd Q,D,BC,(\G)\t^C=0,
\ega
\tag{20.14}$$
which is invariant \wrt an action of the group $U(1)$ on $\ul P$. $D$~is an
exterior covariant \dv\ \wrt the \cn\ $\gd w,A,B,$ and $\gd Q,D,BC,(\G)$ is
its torsion. In the fifth position of Ref.~[18] it is shown that
$$
\gd w,A,B,=\left( \matrix
\pi^*(\gd \ov w,\a,\b,)+g^{\g\a}H_{\g\b}\t^5
&\vrule height9pt depth 5pt width0.4pt &  H_{\b\g}\t^\g 
\vrule height0pt depth 5pt width0pt \\
\noalign{\hrule}
g^{\a\b}(H_{\g\b}+2F_{\b\g})\t^\g
&\vrule height10pt depth 5pt width0.4pt & 0 
\vrule height10pt depth 0pt width0pt 
\endmatrix \right)
\tag{20.15}$$
where $H_{\b\g}$ is a tensor on $E$ \st
$$
g_{\d\b}g^{\g\d}H_{\g\a}+g_{\a\d}g^{\d\g}H_{\b\g}=
2g_{\a\d}g^{\d\g}F_{\b\g}.
\tag{20.16}$$
It is possible to prove that
$$
H_{\g\b}=-H_{\b\g} \q (\hbox{if }F_\m=-F_{\nu\mu}).
\tag{20.17}$$

We define on $\ul P$ a second \cn
$$
\bga
\gd W,A,B,=\gd w,A,B,-\tfrac49\gd \d,A,B,\ov W\\
\ov W=\hor\ov W.
\ega
\tag{20.18}$$
Let us calculate a Moffat--Ricci curvature scalar for $\gd W,A,B,$. One gets
$$
R(W)=\ov R(\ov W)+\bigl(2(g^{[\m]}F_\m)^2-H^\m F_\m\bigr)
\tag{20.19}$$
where
$$
\ov R(\ov W)=g^\m\ov R_\m(\ov \G)+\tfrac23 g^{[\mu\nu]}\ov W_{[\mu,\nu]}
\tag{20.20}$$
is a Moffat--Ricci scalar for the \cn\ $\gd \ov W,\a,\b,$ and $\ov R_{\a\b}
(\ov\G)$ is a Moffat--Ricci tensor for the \cn\ $\gd\ov w,\a,\b,$. In
particular 
$$
\ov R_\m(\ov\G)=\gd\ov R,\a,\m\a,(\ov\G)+\tfrac12\gd\ov R,\a,\a\m,
(\ov\G),
\tag{20.21}$$
where $\gd R,\a,\m\rho,(\ov\G)$ are components of the ordinary curvature
tensor  for $\ov\G$. In addition
$$
H^{\mu\a}=g^{\b\mu}g^{\g\a}H_{\b\g}.
\tag{20.22}$$
Using Palatini variation principle \wrt $g_{\a\b}$, $A_\mu$, $\gd \ov
W,\a,\b,$, i.e.\ 
$$
0=\d\int_V \sqrt{\det\g_{AB}}\,R(W)\,d^5x=
2\pi\d\int_U\sqrt{-g}\Bigl(\ov R(\ov W)+\bigl(2(g^{[\m]}F_\m)^2-H^\m F_\m\bigr)
\Bigr)\,d^4x
$$
where $V=U\times U(1)$, $U\subset E$,
one gets from \eqref{20.19} fields \e s
$$\dsl{\indent\hfill
\ov R_{\a\b}(\ov W)-\tfrac12g_{\a\b}\ov R(\ov W)=8\pi\nad{em}T_{\a\b}\hfill
(20.23)\cr
\indent\hfill{\falg^{[\m]}}_{,\nu}=0 \hfill(20.24)\cr
\indent\hfill g_{\m,\si}-g_{\z\nu}\gd\ov\G,\z,\mu\si,-g_{\mu\z}\gd\ov\G,\z,\si\nu,=0
\hfill(20.25)\cr
\indent\hfill \pa_\mu(\falH^{\a\mu})=2\falg^{[\a\b]}\pa_\b(g^{[\m]}F_\m) 
\hfill(20.26)
}$$
where
$$\eqalignno{
\nad{em}T_{\a\b}&=\frac1{4\pi}\Bigl\{g_{\g\b}g^{\tau\mu}g^{\ve\g}H_{\mu\a}
H_{\tau\ve}-2g^{[\m]}F_\m F_{\a\b}\cr
&\qquad {}-\tfrac14\,g_{\a\b}\bigl(H^\m
F_\m-2(g^{[\m]}F_\m)^2\bigr)\Bigr\}\ &(20.27)\cr
\falg^{[\m]}&=\sqrt{-g}\,g^{[\m]} &(20.28)\cr
\falH^\m&=\sqrt{-g}\,g^{\b\mu}g^{\g\a}H_{\b\g}=\sqrt{-g}\,H^{\mu\a}
&(20.29)
}$$
One can prove
$$\dsl{
\indent\hfill H^\m H_\m=F^\m H_\m,\hfill(20.30)\cr
\indent\hfill g^{[\m]}F_\m=g^{[\m]}H_\m\hfill(20.31)
}$$
and
$$
g^{\si\nu}g^{\a\nu}H_{\si\a}F_\m+g^{\mu\si}g^{\nu\b}H_{\b\si}F_\m=2g^{\mu\si}
g^{\nu\b}F_\m F_{\b\si}.
$$
We have also
$$
g^{\a\b}\nad{em}T_{\a\b}=0.
\tag{20.32}$$

Equations \eqref{20.21}--\eqref{20.26} can be written in the form
$$\dsl{\indent\hfill
\ov R_{(\a\b)}(\ov\G)=8\pi \nad{em}T_{(\a\b)}\hfill(20.33)\cr
\indent\hfill \ov R_{[[\a\b],\g]}(\ov\G)=8\pi\nad{em}T_{[[\a\b],\g]} \hfill(20.34)\cr
\indent\hfill {\ov\G}_\mu=0 \hfill(20.35)\cr
\indent\hfill g_{\m,\si}-g_{\z\nu}{\ov\G}^\z_{\mu\si}-g_{\mu\z}{\ov\G}^\z_{\si\nu}=0
\hfill(20.36)\cr
\indent\hfill \pa_\mu\bigl(\falH^{\a\mu}-2\falg^{[\a\mu]}(g^{[\nu\b]}F_\m)\bigr)=0
\hfill(20.37)
}$$
where $\ov R_{\a\b}(\ov\G)$ is a Moffat--Ricci tensor for the \cn\
$\gd\ov w,\a,\b,=\gd \ov\G,\a,\b\g,{\ov\t}^\g$ and
$$
\ov\G_\mu=\gd\ov\G,\a,[\mu\a],,
\tag{20.38}$$
$_{,\g}$ means a partial \dv\ \wrt $x^\g$ (as usual).

In the theory we get a current density
$$
{\ov J}^\a=2\pa_\mu\bigl(\sqrt{-g}\,g^{[\a\mu]}(g^{[\nu\b]}F_{\nu\b})\bigr)
\tag{20.39}$$
which is conserved by its definition (in this way it is a topological
current). Equation \eqref{20.37} can be written in the form
$$\dsl{\indent\hfill 
\ov\nabla_\mu H^{\a\mu}=J^\a\hfill(20.40)\cr
\indent\hfill J^\a=2\ov\nabla_\mu\bigr(g^{[\a\mu]}(g^{[\nu\b]}F_{\nu\b})\bigr)
=2g^{[\a\b]}\ov\nabla_\mu(g^{[\nu\b]}F_{\nu\b})\hfill(20.41)
}$$
where $\ov\nabla_\mu$ is a covariant \dv\ for the connection $\gd\ov
w,\a,\b,$. 

Equation \eqref{20.16} can be solved \wrt $H_{\nu\mu}$
$$
H_{\nu\mu}=F_{\nu\mu}-{\wt g}^{(\tau\a)}F_{\a\nu}g_{[\mu\tau]}
+{\wt g}^{(\tau\a)}F_{\a\mu}g_{[\nu\tau]}.
\tag{20.42}$$
However the form of Eq.\ \eqref{20.16} is easier to handle from theoretical
point of view. Writing $H_\m$ in the form
$$
H_\m=F_\m -4\pi M_\m
\tag{20.43}$$
we get
$$
Q_\m^5=8\pi M_\m=2{\wt g}^{(\tau\a)}\bigl(F_{\a\mu}g_{[\nu\tau]}
-F_{\a\nu}g_{[\mu\tau]}\bigr),
\tag{20.44}$$
where $Q_\m^5$ is a torsion in the fifth dimension for the \cn\ $\gd w,A,B,$
on~$\ul P$ and $M_\m$ is an \elm\ polarization tensor induced by a \nos\
tensor $g_{\a\b}$ (if $g_{\a\b}=g_{(\a\b)}$, $F_\m=H_\m$). In this way $H_\m$
can be considered as an induction tensor of an \elm\ field. Moreover, the
second pair of Maxwell \e s \eqref{20.40} suggests that rather $H^\m$ should be
considered as an induction tensor.

In the theory we get an \elm\ field lagrangian
$$
\cL_{\em}=-\frac1{8\pi}\bigl(H^\m F_\m-2(g^{[\m]}F_\m)^2\bigr)
\tag{20.45}$$
which can be written in the form
$$
\cL_{\em}=-\frac1{8\pi}\bigl((g^{\mu\a}g^{\nu\b}-g^{\nu\b}{\wt g}^{(\mu\a)}
+g^{\nu\b}g^{\mu\o}{\wt g}^{(\tau\a)}g_{\o\tau})
F_{\a\b}F_\m-2(g^{[\m]}F_\m)^2\bigr) 
\tag{20.46}$$
or
$$
\cL_{\em}=-\frac1{8\pi}\bigl(F^\m F_\m - 2(g^{[\m]}F_\m)^2
+(g^{\nu\b}g^{\mu\o}{\wt g}^{(\tau\a)}g_{\o\tau}-g^{\nu\b}{\wt g}^{(\mu\a)})
F_{\a\b}F_\m\bigr)
\tag{20.47}$$
where
$$
F^\m=g^{\mu\a}g^{\nu\b}F_{\a\b}.
$$

Let us consider energy--momentum tensor of an \elm\ field in the \NK{}, i.e.\
$\nad{em}T_{\a\b}$. Using Eq.~\eqref{20.42} one gets
$$
\nad{em}T_{\a\b}=\nad{o}T_{\a\b}+\frac1{4\pi}\,t_{\a\b}
\tag{20.48}$$
where
$$
\nad{o}T_{\a\b}=\frac1{4\pi}\Bigl(\gd F,\tau,\a,F_{\tau\b}-\frac14\,g_{\a\b}
F^{\m}F_\m\Bigr)
\tag{20.49}$$
is an energy--momentum tensor of an \elm\ field in N.G.T.,
$$
\gd F,\tau,\a,=g^{\tau\g}F_{\g\a}=-\dg F,\a,\tau,
\tag{20.50}$$
and
$$\dsl{
\ t_{\a\b}=g_{\g\b}\dg F,\nu,\tau,F_{\o\tau}g^{\ve\g}{\wt g}^{(\rho\nu)}
{\wt g}^{(\d\o)}g_{[\a\rho]}g_{[\ve\d]}
-g_{\g\b}{\wt g}^{(\rho\nu)}\bigl(F^{\mu\g}F_{\nu\mu}g_{[\a\rho]}
+F_{\mu\a}\dg F,\nu,\mu,g^{\ve\g}g_{[\a\rho]}\bigr)\hfill\cr
\hfill{}-2g^{[\m]}F_\m F_{\a\b}
+\frac14\,g_{\a\b}\Bigl(2\bigl(g^{[\m]}F_\m\bigr)^2
-\bigl(g^{\nu\d}g^{\mu\o}{\wt g}^{(\tau\ve)}g_{\o\tau}-g^{\nu\d}{\wt g}^{(\mu\ve)}
\bigr)F_{\ve\d}F_\m\Bigr)\ (20.51)
}$$
is a correction coming from the \NK{}.

Let us consider the second pair of Maxwell \e s in the \NK{}. One writes them
in the following form (using \eqref{20.42}):
$$
\ov\nabla_\mu F^{\a\mu}=\gd J,\a,p,+J^\a
\tag{20.52}$$
where $J^\a$ is a topological current and $\gd J,\a,p,$ is a polarization
current 
$$\dsl{
\ \gd J,\a,p,=4\pi \ov\nabla_\mu M^{\a\mu}=\frac{4\pi}{\sqrt{-g}}\,
\pa_\mu\bigl(\sqrt{-g}\,M^{\a\mu}\bigr)=
\ov\nabla_\mu\Bigl(g^{\a\b}g^{\mu\g}{\wt g}^{(\tau\rho)}
\bigl(F_{\rho\g}g_{[\b\tau]}-F_{\rho\b}g_{[\g\tau]}\bigr)\Bigr)\hfill\cr
\hfill{}=\frac1{\sqrt{-g}}\,\pa_\mu\Bigl(\falg^{\a\b}g^{\m}{\wt g}^{(\tau\rho)}
\bigl(F_{\rho\g}g_{[\b\tau]}-F_{\rho\b}g_{[\g\tau]}\bigr)\Bigr).
\ (20.53)
}$$

The energy momentum tensor in the form \eqref{20.48} and the second pair of
Maxwell \e s can be obtained directly from Palatini variation principle \wrt
$\gd \ov W,\a,\b,$, $g_\m$ and $A_\mu$ for
$$
R(W)=R(\ov W)+8\pi\cL_{\em}
\tag{20.54}$$
where $\cL_{\em}$ is given by Eq.~\eqref{20.47}.

Writing as usual
$$\ealn{
F_\m&=\left(\matrix
0 &\ &-B_3 &\ &B_2 &\ &-E_1\\
B_3&&0&&-B_1&&-E_2\\
-B_2&&B_1&&0&&-E_3\\
E_1&&E_2&&E_3&&0 \endmatrix \right)&(20.55)\cr
H^\m&=\left(\matrix
0 &\ &-H^3&\ &H^2&\ &-D^1\\
H^3&&0&&-H^1&&-D^2\\
-H^2&&H^1&&0&&-D^3\\
D^1&&D^2&&D^3&&0 \endmatrix \right) &(20.56)
}$$
and introducing Latin indices $a,b=1,2,3$ we get
$$\ealn{
E_a=F_{4a}, &\q D^a=H^{4a}&(20.57)\cr
\os D=(D^1,D^2,D^3), &\q \os E=(E_1,E_2,E_3) &(20.58)\cr
\os B=-(F_{23},F_{31},F_{12}), &\q \os H=-(H^{23},H^{31},H^{12}) &(20.59)
}$$
or
$$\ealn{
B_a=-\frac12\,{\ve_a}^{bc}F_{bc}, &\q F_{cm}=-\dg\ve,cm,e,B_e&(20.60)\cr
H^a=-\frac12\,\gd\ve,a,bc,H^{bc}, &\q H^{cm}=-\gd\ve,cm,e,H^e. &(20.61)
}$$
$\ve_{abc}$ is a usual 3-dimensional antisymmetric symbol. $\ve_{123}=1$ and it
is unimportant for it if its indices are in up or down position. We keep those
indices in up and down position only  for a convenience.

Using Eq.\ \eqref{20.42} one gets
$$\ealn{
D^a&=A^{ac}E_c+C^{ad}B_d &(20.62)\cr
H^a&={\ov A}^{ac}E_c+{\ov C}^{ad}B_d. &(20.63)
}$$
In formulae \eqref{20.62} and \eqref{20.63} $A^{ac}$ can be identified with
$\ve_{ac}$ (a~dielectric constant tensor) and ${\ov C}^{ab}$ with $(\mu^{-1})
_{ab}$ (an inverse of magnetic constant tensor). Remaining \cf s have more
complex interpretation. Moreover, it is possible to think about them as on
material properties of some kind generalized medium. 
A medium with nonzero $C^{ad}$ and $\ov C{}^{ad}$ is called {\it
bianisotropic}. In our case they are induced
by the \nos\ tensor $g_{\a\b}$, where
$$\eqalignno{
A^{me}&=g^{\mu e}g_{[\d\mu]}\wt g^{(4\d)}g^{m4}
-g^{44}\bigl(g_{[\d\mu]}g^{\mu e}\wt g^{(m\d)}-g^{me}\bigr)\cr
&\qquad {}+g^{\o4}g_{[\d\o]}g^{4e}\wt g^{(m\d)}-g^{m4}g^{4e}
&(20.64)\cr
C^{pe}&=\ve_{mz}{}^p\bigl(g^{z4}g^{me}+g^{\mu e}g_{[\d\mu]}g^{z4}\wt g^{(m\d)}
-g^{\o4}g_{[\d\o]}g^{me}\wt g^{(z\d)}\bigr)
&(20.65)\cr
\ov A^{pm}&=\tfrac12 \ve^p{}_{ek}\bigl(g^{4k}g^{me}-g^{mk}g^{4e}
-(g^{mk}\wt g^{(4\d)}
+g^{4k}\wt g^{(m\d)})g^{\mu e}g_{[\d\mu]}\cr
&\qquad {}-g^{\o k}g_{[\d\o]}g^{me}\wt g^{(4\d)}\bigr)
&(20.66)\cr
\ov C^{pf}&=\tfrac12\ve^p{}_{ek}\ve_{wm}{}^f\bigl(g^{wk}g^{me}-g^{wk}g^{\mu e}
g_{[\d\mu]}\wt g^{(m\d)}-g^{\o k}g_{[\d\o]}g^{me}\wt g^{(w\d)}\bigr)
&(20.67)
}$$

Let us notice that in the case of a diagonal $g_{(\a\b)}$  $F_\m=
H_\m$, also in the case of spherically symmetric $g_\m$ we have $F_\m=H_\m$,
which was extensively used in order to find an exact \so\ for field \e s
(see the fifth position of Ref.~[18]). 

Formulae \eqref{20.55}--\eqref{20.56} have a formal character. The physical
meaning of $(\os D,\os H)$ and $(\os E,\os B)$ as induction or strength
vectors of electric or magnetic fields is sound only in a stationary case.

Let us consider a \nos\ tensor for axially \s\ and stationary space-time in
cylindrical coordinates (see Ref.~[136])
$$
g_\m=\left(\matrix
-e^{2(n-l)}&\ &0&\ &ade^n&\ &de^n\\
0&&-e^{2(n-l)}&&kae^m&&ke^n\\
-ade^n&&-ake^n&&ca^2e^{2l}-r^2e^{-2l}&&ace^{2l}\\
-de^n&&-ke^n&&ace^{2l}&&ce^{2l} \endmatrix\right)
\tag{20.68}$$
where $c=1+d^2+k^2$, $x^1=r$, $x^2=z$, $x^3=\t$, $x^4=t$, and all the \f s
$n,l,a,b,d,k$ are \f s of~$r$ and~$z$ only,
$$
g=r^2e^{4(n-l)}, \q \wt g=-r^2e^{4(n-l)}(1+d^2+k^2).
\tag{20.69}$$

An \elm\ field is described by
$$\dsl{\indent\hfill 
F_\m=\left( \matrix
0&\ &0&\ &p&\ &s\\
0&&0&&q&&u\\
-p&&-q&&0&&0\\
-s&&-u&&0&&0 \endmatrix \right)\hfill(20.70)\cr
 p=B_z,\ s=-E_r,\ q=-B_r,\ u=-E_z,
}$$
and all the \f s depend on $r$ and $z$ only. 

In this case we can calculate $H_\m$ and $H^\m$. One gets
$$
H_\m=\frac1{\wt g}\left(\matrix
0&\ &r^2m_1&\ &-pr^2m_2&\ &-r^2sm_2\\
-r^2m_1&&0&&-qr^2m_2&&-r^2um_2\\
pr^2m_2&&qr^2m_2&&0&&r^2m_3\\
r^2sm_2&&r^2um_2&&-r^2m_3&&0 \endmatrix \right)
\tag{20.71}$$
where $m_1=(du-ks)e^{5n-6l}$, $m_2=(1+d^2+k^2)e^{4(n-l)}$,
$m_3=\bigl(d(p-as)+\break k(q-as)\bigr)e^{3n-2l}$,
$$
H^\m=\frac1P
\left(\matrix
0& 0& r^4m_4& ar^4m_4+r^6m_6\\
0&0&r^4m_5&-ar^4m_5+r^6m_7\\
-r^4m_4&-r^4m_5&0&0\\
-ar^4m_4-r^6m_6&ar^4m_5-r^6m_7&0&0 \endmatrix \right)
\tag{20.72}$$
where 
$$
\bal 
P=e^{12(n-l)}r^6(1+d^2+k^2),\q\\
m_4=e^{10n-8l}(1+d^2+k^2)(as-p),\q &m_5=e^{10n-8l}(1+d^2+k^2)(au-q), \\
m_6=e^{10n-12l}(s+d^2s+dku),\q &m_7=e^{10n-12l}(dks+u+k^2u).
\eal
\tag{20.73}$$

Let us come back to Eq.\ \eqref{20.26} (the second part of Maxwell \e s) and let
us consider it for $\a=4$
$$
\pa_m\falH^{4m}=2\falg^{[4b]}\pa_b \bigl(g^{[\m]}F_\m\bigr), \q m,b=1,2,3,
\tag{20.74}$$
or
$$
\div\bigr(\sqrt{-g}\,\os D\bigr)=\rho\sqrt{-g}.
\tag{20.75}$$
In this equation $\rho$ represents a density of an electric charge.

If $\os D=0$ the density of charge equals zero. The problem which we now pose
is as follows. Is it possible to have $\os D=0$ and $\os E\ne0$? This means
that we have a condition
$$
A^{ae}E_e+C^{ae}H_e=0. 
\tag{20.76}$$
This means we have a \cfn\ of a charge induced by a special properties of
``vacuum'' (i.e.\ a \gr al field described by \nos\ tensor $g_\m$).

In the case of a stationary, axially \s\ field we have conditions
$$\ealn{
as-p&=\frac{r^2(s+d^2s+dku)}{a(1+d^2+k^2)}\,e^{-4l}&(20.77)\cr
au-q&=\frac{r^2(dks+u+k^2u)}{a(1+d^2+k^2)}\,e^{-4l}&(20.78)
}$$
with $u,s\ne0$.

These conditions can be imposed on \f s $n,l,d,k,a$ to be satisfied for any
nonzero $u,s$ with some dependence on $p$ and~$q$. One gets
$$
\bga
e^{4l}a^2(1+d^2+k^2)=r^2(1+d^2)\\
p=-u\,\frac{r^2dk}{a(1+d^2+k^2)}\,e^{-4l}\\
e^{4l}a^2(1+d^2+k^2)=r^2(1+k^2)\\
q=-s\,\frac{dkr^2}{a(1+d^2+k^2)}\,e^{-4l}
\ega
\tag{20.79}$$
and finally
$$\dsl{\indent\hfill 
d=k, \q a=re^{-2l}\sqrt{\frac{1+d^2}{1+2d^2}}\,,\hfill(20.80)\cr
\indent\hfill p=-\a u,\q q=-\a s, \hfill(20.81)
}$$
where
$$
\a=\frac{rde^{2l}}{\sqrt{(1+d^2)(1+2d^2)}}
\tag{20.82}$$
or
$$\ealn{
B_z&=\a E_z &(20.83)\cr
B_r&=-\a E_r. &(20.84)
}$$

\vfill\eject

\null
\vskip36pt plus4pt minus4pt
\centerline{{\four\bf 21. Gravito--\elm\ Waves \eu\so s}}

\vskip4pt
\centerline{{\four\bf  in the \NK{}}}
\vskip24pt plus2pt minus2pt
\write0{21. Gravito--\elm\ Waves \eu\so s in the \NK{} \the\pageno}

Let us consider the following \nos\ metric in cartesian coordinates
$$
g_\m=\left(\matrix -1&\ &0&\ &r&\ &-r\\
0&&-1&&s&&-s\\
-r&&-s&&e-1&&-e\\
r&&s&&-e&&1+e\endmatrix\right)
\tag{21.1}$$
where
$$
\bal
e&=e(x,y,z-t)\\
s&=s(x,y,z-t)\\
r&=r(x,y,z-t)
\eal
$$
which describes generalized plane wave for $g_\m$ and
$$
F_\m=\left(\matrix
0&\ &0&\ &k&\ &-k\\
0&&0&&p&&-p\\
-k&&-p&&0&&0\\
k&&p&&0&&0\endmatrix \right)
\tag{21.2}$$
where
$$
\bal
p&=p(x,y,z-t)\\
q&=q(x,y,z-t)
\eal
$$
which describes generalized plane \elm\ wave. All the \f  s mentioned here
are subject of field \e s in \NK{}. Using results from Ref.~[137] one gets
the following \e s
$$
-\D e+4Q-\biggl[\Bigl(\pp rx-\pp sy\Bigr)^2+\Bigl(\pp ry+\pp sx\Bigr)^2\biggr]
=4\biggl[\Bigl(\pp Ax\Bigr)^2+\Bigl(\pp Ay\Bigr)^2\biggr]
\tag{21.3}$$
where
$$\ealn{
{}&Q=\biggl[\Bigl(\pp rx\Bigr)^2+r\,\pp{^2r}{x^2}+\frac12\,\pp sx\Bigl(\pp ry+
\pp sx\Bigr) +\frac12\,s\Bigl(\pp{^2r}{x\pa y}+\pp{^2s}{x^2}\Bigr)
+\Bigl(\pp sy\Bigr)^2\cr
&\qquad{}+s\,\pp{^2s}{y^2}+\frac12\,\pp ry\Bigl(\pp ry+\pp sx
\Bigr)+\frac12\,r\Bigl(\pp{^2r}{y^2}+\pp{^2s}{x\pa y}\Bigr)\biggr]&(21.4)\cr
&\D A(x,y,z-t)=0 &(21.5)\cr
&p=\pp Ax,\q q=-\pp Ay&(21.6)\cr
&\pp ry=\pp sx+H(x,y,z-t), \q \D H(x,y,z-t)=0 &(21.7)
}$$
where $\D=\pp{^2}{x^2}+\pp{^2}{y^2}$ is the Laplace operator in two
dimensions. Equation \eqref{21.3} is a Poisson \e\ for~$e$.
Using Eq.~\eqref{20.24} one gets
$$
\pp sy=-\pp rx \qh{or} s=-\pp Bx,\ r=\pp By.
$$
In this way we get
$$
\D B=H.
\tag{21.8}$$

$A$ and $H$ are arbitrary harmonic \f s in two dimensions with arbitrary
dependence on $(z-t)$ of $C^2$ class. $B$~is an arbitrary \so\ of Poisson \e\
and arbitrary \f\ for $(z-t)$ of $C^2$ class. 
The \f~$e$ can be obtained from the Poisson \e
$$
\D e=f
\tag{21.9}$$
where $f$ is given in terms of $B,A,H$ and takes a simple form
$$
f=-4\biggl[\Bigl(\pp Ax\Bigr)^2+\Bigl(\pp Ay\Bigr)^2\biggr]
+\biggl(\pp{^2B}{y^2}-\pp{^2B}{x^2}\biggr)^2+2\Bigl(\pp Bx\,\pp Hx+
\pp By\,\pp Hy\biggr). 
$$
The dependence on $(z-t)$ is
parametric and is given by the dependence on $(z-t)$ of $B,A,H$. This \so\
describes a generalized plane gravito-\elm\ wave.

Now we consider the following \nos\ tensor in cartesian coordinates
$$
g_\m=\left(\matrix
-a&\ &0&\ &r&\ &-r\\
0&&-a&&s&&-s\\
-r&&-s&&-b&&0\\
r&&s&&0&&b\endmatrix\right),
\tag{21.10}$$
$$
\bal
a&=a(x,y), \q r=r(x,y,z-t),\\
b&=b(x,y), \q s=s(x,y,z-t),
\eal
$$
and an \elm\ field tensor
$$\dsl{\indent\hfill 
F_\m=\left(\matrix
0&\ &0&\ &p&\ &-p\\
0&&0&&q&&-q\\
-p&&-q&&0&&0\\
p&&q&&0&&0\endmatrix\right),\hfill(21.11)\cr
\indent\hfill p=p(x,y,z-t), \q q=q(x,y,z-t).\hfill(21.12)
}$$

We put \eqref{21.10} and \eqref{21.11} into the field \e\ 
\eqref{20.23}--\eqref{20.29}. Using the results from Ref.~[138] we get the
following \e s:
$$\ealn
{&\D(\a/a)=4a\b^2-\frac2a \biggl[\Bigr(\pp \psi x\Bigr)^2
+\Bigl(\pp\psi y\Bigr)^2\biggr]&(21.13)\cr
&p=\pp \psi y, \q q=-\pp \psi x &(21.14)\cr
&\psi=\psi(x,y,z-t), \q \D\psi=0 \cr
&\b=\frac1{2a} \Bigl(\pp ry-\pp sx\Bigr)&(21.15)\cr
&\a=r^2+s^2. &(21.16)
}$$
From the remaining field \e\ we get
$$
\D\b=0,
\tag{21.17}$$
where $\D=\pp{^2}{x^2}+\pp{^2}{y^2}$,
$$
a=e^A,
\tag{21.18}$$
where $A$ is an arbitrary harmonic \f\ in two variables,
$$
\D A=0.
\tag{21.19}$$
The \f\ $b$ is given by
$$
b=G_1(z+t)G_2(z-t)
\tag{21.20}$$
where $G_1$ and $G_2$ are arbitrary \f s of $C^2$ class (of on variable). The
\f\ $\b$ should be written in the following form
$$
\b(x,y,z,t)=f_0(z-t)\b_0(x,y)
\tag{21.21}$$
where $\b_0$ is an arbitrary harmonic \f\ and $f_0$ is an arbitrary \f\ of
one variable (of $C^2$ class). Let us consider Eq.~\eqref{20.24}. One gets
$$
\pp ry+\pp sx=0.
\tag{21.22}$$
From this equation we get
$$
r=\pp \vf x, \q s=-\pp\vf y
\tag{21.23}$$
where $\vf$ is a \f\ (arbitrary) of two variables (of~$C^3$) and a \f\ of
$z-t$. 

One gets
$$
\D\vf=2e^Af_0\b_0.
\tag{21.24}$$
Let $\wt\vf$ be any arbitrary \so\ of Eq.~\eqref{21.24} (remembering that $A$,
$f_0$ and~$\b_0$ are arbitrary). In this way
$$
\a=r^2+s^2=\biggl(\pp{\wt\vf}y \biggr)^2+\biggl(\pp{\wt\vf}x\biggr)^2.
\tag{21.25}$$
Let us consider Eq.~\eqref{21.13} in the following form
$$
\D g=4e^A\b_0^2f_0^2-\frac2{e^A}\biggl[\biggl(\pp\psi x \biggr)^2
+\biggl(\pp\psi y \biggr)^2\biggr]
\tag{21.26}$$
and let $\wt g$ be any \so\ of this Poisson \e.

Thus
$$\dsl{\indent\hfill 
\frac\a a=\wt g\hfill(21.27)\cr
\indent\hfill \biggl(\pp{\wt\vf}y \biggr)^2+\biggl(\pp{\wt\vf}x \biggr)^2=\wt g
e^A.\hfill(21.28)
}$$
Eq.\ \eqref{21.28} gives us a consistency condition for an existence of
gravito-\elm\ wave.

Let us consider the following \nos\ metric
$$
g_\m=\left(\matrix
0&\ &0&\ &0&\ &1\\
0&&b&&0&&l+q\\
0&&0&&b&&m+p\\
1&&l-q&&m-p&&-v
\endmatrix\right)
\tag{21.29}$$
and an \elm\ field strength tensor
$$
F_\m=\left(\matrix
0&\ &0&\ &0&\ &0\\
0&&0&&0&&s\\
0&&0&&0&&u\\
0&&-s&&-u&&0
\endmatrix\right)
\tag{21.30}$$
in cartesian coordinates
$$
x^1=x, \q x^2=y, \q x^3=z, \q x^4=t.
$$
It is possible to consider $x^4$ as $z-t$. We suppose that
$$
\pp{}{x^1}\,g_{\m}=\pp{}{x^1}\,F_{\m}=0.
$$
Using results from Refs [139], [140] and field \e\ of the \NK{} we
get the following \e s
$$
\bal
{}&s=\pp \psi{x^3},\\
&u=\pp \psi{x^2},\\
&\biggl(\pp{^2}{(x^2)^2}+\pp{^2}{(x^3)^2}\biggr)\psi=0, \q
\psi=\psi(x^2,x^3,x^4).
\eal
\tag{21.31}$$
Supposing
$$
b=1, \q q=0
\tag{21.32}$$
we get further
$$
\bga
p=p(x^2,x^4)\\
\pp{^3}{(x^2)^3}\,p=0\\
\pp l{x^3}=\pp m{x^2}\,, \q \pp l{x^2}=-\pp m{x^3}\\
l=l(x^2,x^3,x^4), \q m=m(x^2,x^3,x^4).
\ega
\tag{21.33}$$
Thus $l$ and $m$ are harmonically conjugate.

Writing 
$$
v=1+f, \q f=f(x^2,x^3,x^4),
\tag{21.34}$$
where $f$ satisfies the equation
$$
\D f=2p^2\biggl(\frac{p_{,22}}p+\Bigl(\frac{p_{,2}}p\Bigr)^2\biggr)
-2\bigl((\psi_{,2})^2+(\psi_{,3})^2\bigr),
\tag{21.35}$$
the form of $p$ can be easily found
$$
p=\frac{(x^2)^2}2\,\vf_1(x^4)+x^2\vf_2(x^4)+\vf_3(x^4)
$$
where $\vf_i$, $s=1,2,3$, are arbitrary \f s of one variable of $C^3$ class.

This \so\ represents a gravito-\elm\ wave if we interpret $x^4$ as wave front
variable $z-t$. ``,'' means a \dv\ \wrt $x^2$ or~$x^3$, $f$~is a \so\ of a
Poisson \e\ with a parametric dependence on~$x^4$ imposed by $\psi$, $\vf_i$,
$i=1,2,3$. \eu\f s $\psi$, $l$ and $m$ are arbitrary \f s of $x^4$ variable
of $C^2$ class.

For all three \so s considered here $t_\m=M_\m=J^\mu=\gd J,\mu,p,=0$. Moreover,
$\os D\ne \os E$ for all \so s.

\vfill\eject

\null
\vskip36pt plus4pt minus4pt
\centerline{{\four\bf 22. An Influence of a Cosmological Constant}}

\vskip4pt
\centerline{{\four\bf on a \eu\so}}

\vskip4pt
\centerline{{\four\bf in the \NK{}}}
\vskip24pt plus2pt minus2pt
\write0{22. An Influence of a Cosmological Constant on a \eu\so\ in the \NK{} \the\pageno}

Let us consider Eqs.\ \eqref{20.23}--\eqref{20.29} and let us change
$\nad{em}T_{\a\b}$ into
$$
T_{\a\b}=\nad{em}T_{\a\b}-\frac{\La}{8\pi}\,g_{\a\b}.
$$
It means we introduce a cosmological constant $\La$ to the theory. 

In the \NK{} $\La=0$. Moreover, in the \eu\nos\ Nonabelian Kaluza--Klein
Theory this constant is in general non-zero. The Yang--Mills field can be
reduced to a $U(1)$ subgroup of a group $G$ (see Section~5 for more
details). We also consider \NK{} with external sources, and a cosmological
constant term can be added to the external energy momentum tensor (see
the fifth position of Ref.~[18]). 

Let us consider corrected field \e s in a static, spherically \s\ case. Using
results from the fifth position of Ref.~[18] we get the following exact \so
$$
g_\m=\left(\matrix
-\a&\ &0&\ &0&\ &\o\\
0&&-r^2&&0&&0\\
0&&0&&-r^2\sin^2\t&&0\\
-\o&&0&&0&&\g
\endmatrix\right)
\tag{22.1}$$
where
$$\ealn{
\o&=\frac{l^2}{r^2}&(22.2)\cr
\a^{-1}&=\biggl(1+\frac{Q^2}{\ov br}\,g\Bigl(\frac r{\ov b}\Bigr)+\frac{\La r^2}3
\biggr) &(22.3)\cr
\g&=\biggl(1+\frac{l^4}{{\ov b}^4}\biggr)
\biggl(1+\frac{Q^2}{\ov br}\,g\Bigl(\frac r{\ov b}\Bigr)+\frac{\La r^2}3
\biggr) &(22.4)\cr
F_{14}&=E=-\frac Q{r^2}\biggl(\frac{r^4}{r^4+{\ov b}^4}\biggr) &(22.5)\cr
\ov b&=4l^4 &(22.6)
}$$
$Q$ is an electric charge and $l$ is an integration constant of length
dimension. Let us notice that a \so\ with a cosmological constant differs
from the previous \so\ only by a term $\frac{\La r^2}3$ similarly as in
General Relativity. 
All the details concerning this \so\ can be found in the fifth position of Ref.~[18].

Let us introduce the following notation
$$\ealn{
a&=\frac{Q^2}{2l^2} \q \Bigl(a=\frac{GQ^2}{2c^2l^2}\Bigr)&(22.7)\cr
b&=\frac{2l^2\La}3 &(22.8)
}$$
In this notation one gets
$$
\a^{-1}=1+a\,\frac{g(x)}x+bx^2
\tag{22.9}$$
where
$$
g(x)=\frac1{4\sqrt2}\log\biggl(\frac{x^2+\sqrt2 x+1}{x^2-\sqrt2 x+1}
\biggr)-\frac1{2\sqrt2}\biggl[\arctan(\sqrt2 x+1)+\arctan(\sqrt2 x-1)
\biggr]
\tag{22.10}$$
(see the fifth position of Ref.~[18]),
$$
x=\frac{r}{\sqrt2 l}\,.
\tag{22.11}$$

Now it is interesting to examine an influence of the cosmological constant
$\La$ (via~$b$) on properties of the \so. The most interesting case is to
find horizons for such a \so. It means, to find zeros of the \f
$$
f(x)=\a^{-1}=1+a\,\frac{g(x)}x+bx^2=0.
\tag{22.12}$$
In the previous case $b=0$ we get that there is a critical value of $a=
a_{\crt}= 3.17\dots$ such that for $a<a_{\crt}$ there is not any horizon,
for $a=a_{\crt}$ there is one horizon, and for $a>a_{\crt}$ there are
two horizons. Let us suppose that $\La>0$ ($b>0$) and examine $a_{\crt}$
in this case. One gets
$$
\eeqalign{
&\hbox{for } b=0.001, \q && a_{\crt}\cong 3.2\cr
&\hbox{for }b=0.01, \q && a_{\crt}\cong 3.6\cr
&\hbox{for }b=0.1, \q && a_{\crt}\cong 4\cr
&\hbox{for }b=1, \q && a_{\crt}\cong 11
}$$
Thus the cosmological constant results in a higher value of $a_{\crt}$.

It is interesting to consider a negative value of $\La$ ($b<0$). In this case
we have always one more horizon (the so-called de~Sitter horizon, a
cosmological horizon). For example for $a=0.1$, $b=-0.001$ we have only one
horizon. For $a=5$, $b=-0.001$ we have three horizons, two as before for
$b>0$ and one de~Sitter horizon. For $a=3$, $b=-0.01$ we have two horizons,
one de~Sitter horizon and one double as before for $b>0$. In this case $a=3$
is a critical value for $b=-0.001$. For lower negative values of~$b$, i.e.\
$b=-0.1$, we have for $a=0.1$, $0.7$, $3$ only one (de~Sitter) horizon.

\obraz{\input linie
\newcount\nr
\newcount\nd
\newcount\nf
\newdimen\hs
\hs=0.45\hsize
\sz=\hsize
\divide\sz by3600
\jedn\sz
\def\crt{\text{\rm crt}}
\def\obrazek#1 {\global\nr=#1 \global\advance\nr by64
\ifodd\nr \hbox to\hsize\bgroup \fi \hfil \vbox \bgroup \hsize=\hs
\leftline \bgroup \hskip-0.7\hs }
\def\koniec#1 #2 {\hfil \egroup \vskip8pt 
\centerline{\hfil $a=#1$, $b=#2$\hfil }\egroup \hfil 
\ifodd\nr \else \hskip30pt \egroup \vfill \fi 
\nd=\nr \divide\nd by8 \nf=\nd \multiply\nf by8
\ifnum\nr=\nf \centerline{\nine Fig. \the\nd}\fi }
\def\kon#1;{\hfil \egroup \vskip8pt
\centerline{\hfil $#1$\hfil }\egroup \hfil
\ifodd\nr \else \hskip30pt \egroup \vfill \fi
\nd=\nr \divide\nd by8 \nf=\nd \multiply\nf by8
\ifnum\nr=\nf \centerline{\nine Fig. \the\nd}\vfill \fi }

 \input wynik-k }

Summing up, for $\La>0$ we get two horizons, one horizon or no horizon, 
$0<r_{H_1}<r_{H_2}$,
$0<r_H$ (as in the case of Reissner--Nordstr\"om \so). For $\La<0$ we have a
de~Sitter horizon
$$
\bal
0&<r_S\\
0&<r_{H_1}<r_{H_2}<r_S\\
0&<r_H<r_S
\eal
$$
and inside it one or two horizons or a case without an inside horizon.
In Figures 9, 10, 11, 12 we give plots of the \f\ $f(x)$ for several values of
parameters $a$ and $b$.

One gets a value of critical parameter $a_{\crt}$ and a critical value of
a horizon radius from the following equations
$$
\bal
0&=f(x_{\crt})=1+a_{\crt}\,\frac{g(x_{\crt})}{x_{\crt}}
+bx_{\crt}^2\\
0&=\frac{df}{dx}(x_{\crt})=2bx_{\crt}+\frac{a_{\crt}}{x_{\crt}}
\,\frac{dg}{dx}(x_{\crt})-\frac{a_{\crt}}{x_{\crt}^2}\,g(x_{\crt}).
\eal
\tag{22.13}$$
After some algebra we get
$$\dsl{\indent\hfill 
x_{\crt}^3(1+bx_{\crt}^2)+g(x_{\crt})(x_{\crt}^4+1)(
3bx_{\crt}^2+1)=0\hfill(22.14)\cr
\indent\hfill a_{\crt}=-\frac{(1+bx_{\crt}^2)x_{\crt}}{g(x_{\crt})}
=\frac{(1+x_{\crt}^4)}{x_{\crt}^2}\,(3bx_{\crt}^2+1)\hfill(22.15)
}$$
Eq.\ \eqref{22.14} has one \so\ for $b\ge0$. However in the case of $b_0<b<0$ it
has two \so s, one \so\ for $b=b_0$ and no \so s for $b<b_0$. Thus we can write
$a_{\crt}=a_{\crt}(b)$, $x_{\crt}(b)$ only for $b\ge0$ (and
$b=b_0$). In the remaining case $b_0<b<0$ we rewrite these \e s as
$$\ealn{
b&=a_{\crt}\,\frac{x^3_{\crt}+g(x_{\crt})(1+x^4_{\crt})}
{2x^3_{\crt}(x^4_{\crt}+1)}&(22.16)\cr
a_{\crt}&=-\frac{2x_{\crt}(x^4_{\crt}+1)}{x^3_{\crt}
+3g(x_{\crt})(x^4_{\crt}+1)}\,.\cr
}$$
The estimated value of $b_0$ is $-0.0302993$. It seems that in the case of de
Sitter horizon for $b<b_0=-0.0302993$ there are not any horizon except the
mentioned one.

In the case of $b<0$ we have an interesting phenomenon due to the fact that
the function~$f$ has two local extrema (one minimum and one maximum). For a
special value of $b_0$ they collapse to one inflection point. This results in
one horizon as a \so\ of a system of \e s
$$
\bal
f(x_{\crt},a_{\crt},b_0)&=0\\
\pp fx(x_{\crt},a_{\crt},b_0)&=0\\
\pp{^2f}{x^2}f(x_{\crt},a_{\crt},b_0)&=0
\eal
\tag{22.17}$$
One finds $x_{\crt}\approx2.30331$, $a_{\crt}\approx2.8446$, $b_0
=-0.0302993$.

\obraz{\vfill \input linie
\newcount\nr
\newcount\nd
\newcount\nf
\newdimen\hs
\hs=0.45\hsize
\sz=\hsize
\divide\sz by3600
\jedn\sz
\def\crt{\text{\rm crt}}
\def\obrazek#1 {\global\nr=#1 \global\advance\nr by64
\ifodd\nr \hbox to\hsize\bgroup \fi \hfil \vbox \bgroup \hsize=\hs
\leftline \bgroup \hskip-0.7\hs }
\def\koniec#1 #2 {\hfil \egroup \vskip8pt 
\centerline{\hfil $a=#1$, $b=#2$\hfil }\egroup \hfil 
\ifodd\nr \else \hskip30pt \egroup \vfill \fi 
\nd=\nr \divide\nd by8 \nf=\nd \multiply\nf by8
\ifnum\nr=\nf \centerline{\nine Fig. \the\nd}\fi }
\def\kon#1;{\hfil \egroup \vskip8pt
\centerline{\hfil $#1$\hfil }\egroup \hfil
\ifodd\nr \else \hskip30pt \egroup \vfill \fi
\nd=\nr \divide\nd by8 \nf=\nd \multiply\nf by8
\ifnum\nr=\nf \centerline{\nine Fig. \the\nd}\vfill \fi }

\obrazek33
\grub0.2pt
\MT   0.000  100.000
\LT1200.000  100.000
\MT 1140.000 115.000 \LT 1200.000 100.000 \LT 1140.000 85.000
\cput(1200.000,25.000,b)
\MT 300.000   80.000
\LT 300.000  120.000
\cput(300.000,20.000,2)
\MT 500.000   80.000
\LT 500.000  120.000
\cput(500.000,20.000,4)
\MT 700.000   80.000
\LT 700.000  120.000
\cput(700.000,20.000,6)
\MT 900.000   80.000
\LT 900.000  120.000
\cput(900.000,20.000,8)
\MT1100.000   80.000
\LT1100.000  120.000
\cput(1100.000,20.000,10)
\MT 100.000    0.000
\LT 100.000  900.000
\MT 85.000 840.000 \LT 100.000 900.000 \LT 115.000 840.000
\rput(130.000,870.000,x_{\crt})
\MT  80.000  300.000
\LT 120.000  300.000
\lput(60.000,280.000,0.5)
\MT  80.000  500.000
\LT 120.000  500.000
\lput(60.000,480.000,1.0)
\MT  80.000  700.000
\LT 120.000  700.000
\lput(60.000,680.000,1.5)
\grub0.6pt
\MT  100.064  769.959
\LT  100.064  769.959
\LT  100.127  768.446
\LT  100.191  766.955
\LT  100.256  765.485
\LT  100.385  762.606
\LT  100.515  759.805
\LT  100.646  757.077
\LT  100.778  754.418
\LT  100.911  751.824
\LT  101.044  749.292
\LT  101.179  746.818
\LT  101.314  744.401
\LT  101.451  742.037
\LT  101.588  739.724
\LT  101.726  737.460
\LT  101.865  735.242
\LT  102.005  733.069
\LT  102.146  730.938
\LT  102.359  727.817
\LT  102.574  724.783
\LT  102.791  721.830
\LT  103.010  718.954
\LT  103.231  716.151
\LT  103.454  713.416
\LT  103.679  710.746
\LT  103.906  708.138
\LT  104.135  705.589
\LT  104.366  703.096
\LT  104.599  700.657
\LT  104.835  698.269
\LT  105.072  695.929
\LT  105.312  693.636
\LT  105.634  690.648
\LT  105.960  687.735
\LT  106.290  684.893
\LT  106.624  682.119
\LT  106.962  679.408
\LT  107.303  676.758
\LT  107.649  674.166
\LT  107.998  671.629
\LT  108.440  668.532
\LT  108.888  665.513
\LT  109.342  662.567
\LT  109.803  659.690
\LT  110.269  656.880
\LT  110.743  654.133
\LT  111.222  651.445
\LT  111.806  648.295
\LT  112.400  645.222
\LT  113.003  642.221
\LT  113.616  639.291
\LT  114.238  636.426
\LT  114.871  633.624
\LT  115.621  630.429
\LT  116.386  627.312
\LT  117.164  624.266
\LT  117.957  621.289
\LT  118.765  618.377
\LT  119.706  615.124
\LT  120.667  611.947
\LT  121.647  608.843
\LT  122.648  605.807
\LT  123.670  602.835
\LT  124.714  599.924
\LT  125.913  596.720
\LT  127.141  593.586
\LT  128.397  590.517
\LT  129.682  587.512
\LT  131.146  584.243
\LT  132.648  581.042
\LT  134.188  577.908
\LT  135.769  574.836
\LT  137.391  571.823
\LT  139.055  568.866
\LT  140.763  565.962
\LT  142.515  563.109
\LT  144.314  560.304
\LT  146.346  557.272
\LT  148.438  554.293
\LT  150.591  551.363
\LT  152.806  548.482
\LT  155.087  545.645
\LT  157.434  542.852
\LT  159.628  540.349
\LT  161.881  537.878
\LT  164.195  535.438
\LT  166.572  533.027
\LT  169.013  530.646
\LT  171.522  528.292
\LT  174.099  525.963
\LT  176.747  523.660
\LT  179.469  521.381
\LT  181.983  519.350
\LT  184.561  517.336
\LT  187.204  515.341
\LT  189.914  513.361
\LT  192.693  511.398
\LT  195.544  509.450
\LT  198.469  507.517
\LT  201.469  505.598
\LT  204.547  503.694
\LT  207.352  502.012
\LT  210.222  500.340
\LT  213.159  498.677
\LT  216.167  497.025
\LT  219.245  495.381
\LT  222.398  493.746
\LT  225.626  492.119
\LT  228.933  490.501
\LT  231.892  489.092
\LT  234.914  487.688
\LT  238.002  486.290
\LT  241.156  484.897
\LT  244.378  483.509
\LT  247.671  482.127
\LT  251.037  480.750
\LT  254.477  479.377
\LT  257.993  478.008
\LT  261.588  476.644
\LT  264.734  475.478
\LT  267.940  474.315
\LT  271.209  473.155
\LT  274.541  471.998
\LT  277.940  470.843
\LT  281.405  469.691
\LT  284.939  468.541
\LT  288.544  467.393
\LT  292.221  466.248
\LT  295.972  465.105
\LT  299.799  463.964
\LT  303.705  462.825
\LT  307.021  461.877
\LT  310.393  460.930
\LT  313.824  459.985
\LT  317.315  459.040
\LT  320.866  458.097
\LT  324.480  457.154
\LT  328.157  456.213
\LT  331.899  455.273
\LT  335.708  454.333
\LT  339.584  453.394
\LT  343.531  452.456
\LT  347.549  451.518
\LT  351.640  450.581
\LT  355.805  449.645
\LT  360.047  448.708
\LT  364.367  447.773
\LT  368.768  446.838
\LT  373.251  445.903
\LT  376.898  445.156
\LT  380.599  444.408
\LT  384.357  443.661
\LT  388.172  442.914
\LT  392.045  442.166
\LT  395.977  441.419
\LT  399.971  440.672
\LT  404.026  439.925
\LT  408.144  439.177
\LT  412.327  438.430
\LT  416.575  437.683
\LT  420.890  436.935
\LT  425.275  436.187
\LT  429.729  435.439
\LT  434.254  434.691
\LT  438.852  433.943
\LT  443.525  433.194
\LT  448.274  432.445
\LT  453.101  431.695
\LT  456.773  431.133
\LT  460.490  430.571
\LT  464.254  430.008
\LT  468.064  429.445
\LT  471.923  428.881
\LT  475.829  428.318
\LT  479.785  427.754
\LT  483.791  427.190
\LT  487.848  426.625
\LT  491.957  426.060
\LT  496.118  425.495
\LT  500.333  424.930
\LT  504.603  424.364
\LT  508.928  423.797
\LT  513.309  423.231
\LT  517.748  422.663
\LT  522.245  422.096
\LT  526.802  421.528
\LT  531.419  420.959
\LT  536.097  420.390
\LT  540.839  419.821
\LT  545.643  419.251
\LT  550.513  418.680
\LT  555.449  418.109
\LT  560.452  417.538
\LT  565.523  416.965
\LT  570.665  416.392
\LT  575.877  415.819
\LT  581.161  415.245
\LT  586.520  414.671
\LT  591.953  414.096
\LT  597.463  413.519
\LT  603.050  412.943
\LT  608.717  412.366
\LT  614.465  411.788
\LT  620.295  411.209
\LT  626.209  410.629
\LT  632.209  410.050
\LT  638.297  409.468
\LT  642.404  409.081
\LT  646.552  408.693
\LT  650.740  408.304
\LT  654.969  407.915
\LT  659.240  407.526
\LT  663.553  407.136
\LT  667.909  406.746
\LT  672.309  406.356
\LT  676.752  405.965
\LT  681.240  405.574
\LT  685.773  405.182
\LT  690.352  404.790
\LT  694.977  404.398
\LT  699.649  404.005
\LT  704.369  403.611
\LT  709.137  403.218
\LT  713.954  402.824
\LT  718.821  402.429
\LT  723.738  402.034
\LT  728.706  401.638
\LT  733.726  401.242
\LT  738.798  400.846
\LT  743.924  400.449
\LT  749.103  400.051
\LT  754.338  399.654
\LT  759.627  399.255
\LT  764.973  398.856
\LT  770.377  398.456
\LT  775.838  398.057
\LT  781.358  397.656
\LT  786.938  397.255
\LT  792.578  396.853
\LT  798.279  396.451
\LT  804.043  396.049
\LT  809.870  395.645
\LT  815.762  395.242
\LT  821.718  394.837
\LT  827.741  394.432
\LT  833.830  394.027
\LT  839.988  393.620
\LT  846.215  393.214
\LT  852.512  392.806
\LT  858.880  392.398
\LT  865.321  391.990
\LT  871.835  391.581
\LT  878.424  391.171
\LT  885.089  390.760
\LT  891.831  390.349
\LT  898.650  389.937
\LT  902.090  389.731
\LT  905.549  389.525
\LT  909.029  389.318
\LT  912.529  389.112
\LT  916.050  388.905
\LT  919.591  388.698
\LT  923.153  388.491
\LT  926.736  388.283
\LT  930.340  388.076
\LT  933.966  387.868
\LT  937.613  387.660
\LT  941.281  387.452
\LT  944.972  387.244
\LT  948.684  387.035
\LT  952.419  386.827
\LT  956.176  386.618
\LT  959.955  386.409
\LT  963.758  386.200
\LT  967.583  385.990
\LT  971.431  385.781
\LT  975.303  385.571
\LT  979.198  385.361
\LT  983.116  385.151
\LT  987.059  384.941
\LT  991.026  384.730
\LT  995.017  384.520
\LT  999.032  384.309
\LT 1003.072  384.098
\LT 1007.137  383.886
\LT 1011.227  383.675
\LT 1015.343  383.463
\LT 1019.484  383.251
\LT 1023.650  383.039
\LT 1027.843  382.827
\LT 1032.062  382.615
\LT 1036.307  382.402
\LT 1040.579  382.189
\LT 1044.877  381.976
\LT 1049.203  381.762
\LT 1053.556  381.549
\LT 1057.937  381.335
\LT 1062.345  381.121
\LT 1066.781  380.907
\LT 1071.246  380.692
\LT 1075.739  380.478
\LT 1080.261  380.263
\LT 1084.811  380.048
\LT 1089.392  379.833
\LT 1094.001  379.617
\LT 1098.640  379.402
\LT 1103.310  379.186
\kon x_{\crt}(b),\ b>0;
\obrazek34
\grub0.2pt
\MT   0.000  100.000
\LT1200.000  100.000
\MT 1140.000 115.000 \LT 1200.000 100.000 \LT 1140.000 85.000
\cput(1200.000,25.000,b)
\MT 300.000   80.000
\LT 300.000  120.000
\cput(300.000,20.000,2)
\MT 500.000   80.000
\LT 500.000  120.000
\cput(500.000,20.000,4)
\MT 700.000   80.000
\LT 700.000  120.000
\cput(700.000,20.000,6)
\MT 900.000   80.000
\LT 900.000  120.000
\cput(900.000,20.000,8)
\MT1100.000   80.000
\LT1100.000  120.000
\cput(1100.000,20.000,10)
\MT 100.000    0.000
\LT 100.000  900.000
\MT 85.000 840.000 \LT 100.000 900.000 \LT 115.000 840.000
\rput(130.000,870.000,a)
\MT  80.000  300.000
\LT 120.000  300.000
\lput(60.000,280.000,10)
\MT  80.000  500.000
\LT 120.000  500.000
\lput(60.000,480.000,20)
\MT  80.000  700.000
\LT 120.000  700.000
\lput(60.000,680.000,30)
\grub0.6pt
\MT  100.064  163.573
\LT  100.064  163.573
\LT  101.246  165.601
\LT  102.502  167.615
\LT  103.830  169.633
\LT  105.232  171.665
\LT  106.624  173.604
\LT  108.086  175.568
\LT  109.618  177.562
\LT  111.126  179.469
\LT  112.700  181.411
\LT  114.343  183.389
\LT  115.947  185.279
\LT  117.616  187.207
\LT  119.351  189.174
\LT  121.032  191.048
\LT  122.775  192.959
\LT  124.582  194.911
\LT  126.455  196.906
\LT  128.256  198.797
\LT  130.118  200.728
\LT  132.042  202.700
\LT  134.032  204.715
\LT  135.929  206.615
\LT  137.886  208.555
\LT  139.904  210.536
\LT  141.985  212.560
\LT  143.950  214.454
\LT  145.972  216.386
\LT  148.053  218.358
\LT  150.195  220.371
\LT  152.399  222.427
\LT  154.667  224.526
\LT  156.787  226.474
\LT  158.964  228.461
\LT  161.199  230.488
\LT  163.495  232.557
\LT  165.852  234.667
\LT  168.274  236.822
\LT  170.510  238.800
\LT  172.802  240.815
\LT  175.150  242.870
\LT  177.556  244.964
\LT  180.022  247.100
\LT  182.551  249.278
\LT  185.143  251.499
\LT  187.800  253.766
\LT  190.219  255.819
\LT  192.693  257.910
\LT  195.224  260.040
\LT  197.812  262.209
\LT  200.460  264.419
\LT  203.169  266.671
\LT  205.942  268.966
\LT  208.778  271.305
\LT  211.315  273.388
\LT  213.905  275.508
\LT  216.547  277.664
\LT  219.245  279.857
\LT  222.000  282.089
\LT  224.812  284.360
\LT  227.683  286.672
\LT  230.616  289.024
\LT  233.611  291.420
\LT  236.670  293.859
\LT  239.796  296.342
\LT  242.528  298.507
\LT  245.312  300.707
\LT  248.148  302.942
\LT  251.037  305.213
\LT  253.981  307.521
\LT  256.981  309.867
\LT  260.038  312.251
\LT  263.153  314.675
\LT  266.329  317.140
\LT  269.567  319.646
\LT  272.867  322.194
\LT  276.232  324.786
\LT  279.087  326.980
\LT  281.989  329.205
\LT  284.939  331.463
\LT  287.938  333.753
\LT  290.987  336.078
\LT  294.087  338.436
\LT  297.239  340.829
\LT  300.445  343.258
\LT  303.705  345.724
\LT  307.021  348.226
\LT  310.393  350.767
\LT  313.824  353.346
\LT  317.315  355.966
\LT  320.866  358.625
\LT  324.480  361.327
\LT  328.157  364.070
\LT  331.899  366.857
\LT  335.708  369.688
\LT  339.584  372.565
\LT  342.736  374.899
\LT  345.933  377.264
\LT  349.176  379.659
\LT  352.467  382.086
\LT  355.805  384.544
\LT  359.193  387.035
\LT  362.630  389.559
\LT  366.118  392.116
\LT  369.658  394.708
\LT  373.251  397.335
\LT  376.898  399.997
\LT  380.599  402.696
\LT  384.357  405.432
\LT  388.172  408.205
\LT  392.045  411.017
\LT  395.977  413.868
\LT  399.971  416.759
\LT  404.026  419.691
\LT  408.144  422.664
\LT  412.327  425.679
\LT  416.575  428.738
\LT  419.805  431.061
\LT  423.074  433.410
\LT  426.381  435.783
\LT  429.729  438.183
\LT  433.116  440.609
\LT  436.544  443.062
\LT  440.013  445.542
\LT  443.525  448.050
\LT  447.080  450.585
\LT  450.678  453.150
\LT  454.320  455.743
\LT  458.007  458.365
\LT  461.740  461.018
\LT  465.519  463.700
\LT  469.345  466.414
\LT  473.219  469.159
\LT  477.142  471.935
\LT  481.115  474.744
\LT  485.138  477.586
\LT  489.212  480.461
\LT  493.338  483.371
\LT  497.517  486.314
\LT  501.751  489.293
\LT  506.038  492.308
\LT  510.382  495.359
\LT  514.783  498.446
\LT  519.241  501.572
\LT  523.758  504.735
\LT  528.334  507.937
\LT  532.971  511.179
\LT  537.671  514.461
\LT  542.433  517.784
\LT  547.259  521.148
\LT  552.151  524.555
\LT  557.109  528.005
\LT  562.135  531.498
\LT  567.229  535.037
\LT  572.394  538.620
\LT  577.630  542.250
\LT  582.939  545.927
\LT  588.322  549.652
\LT  591.953  552.163
\LT  595.617  554.695
\LT  599.316  557.249
\LT  603.050  559.826
\LT  606.819  562.426
\LT  610.624  565.049
\LT  614.465  567.695
\LT  618.343  570.365
\LT  622.257  573.059
\LT  626.209  575.777
\LT  630.200  578.520
\LT  634.229  581.287
\LT  638.297  584.080
\LT  642.404  586.898
\LT  646.552  589.742
\LT  650.740  592.612
\LT  654.969  595.508
\LT  659.240  598.432
\LT  663.553  601.382
\LT  667.909  604.360
\LT  672.309  607.366
\LT  676.752  610.400
\LT  681.240  613.463
\LT  685.773  616.555
\LT  690.352  619.676
\LT  694.977  622.827
\LT  699.649  626.008
\LT  704.369  629.220
\LT  709.137  632.462
\LT  713.954  635.737
\LT  718.821  639.043
\LT  723.738  642.381
\LT  728.706  645.752
\LT  733.726  649.156
\LT  738.798  652.594
\LT  743.924  656.065
\LT  749.103  659.572
\LT  754.338  663.113
\LT  759.627  666.690
\LT  764.973  670.303
\LT  770.377  673.953
\LT  775.838  677.639
\LT  781.358  681.363
\LT  786.938  685.126
\LT  792.578  688.927
\LT  798.279  692.767
\LT  804.043  696.647
\LT  809.870  700.567
\LT  815.762  704.528
\LT  821.718  708.531
\LT  827.741  712.576
\LT  830.777  714.615
\LT  833.830  716.664
\LT  836.901  718.724
\LT  839.988  720.795
\LT  843.093  722.877
\LT  846.215  724.971
\LT  849.355  727.075
\LT  852.512  729.191
\LT  855.687  731.318
\LT  858.880  733.456
\LT  862.092  735.606
\LT  865.321  737.768
\LT  868.569  739.941
\LT  871.835  742.126
\LT  875.120  744.323
\LT  878.424  746.532
\LT  881.747  748.753
\LT  885.089  750.986
\LT  888.450  753.231
\LT  891.831  755.489
\LT  895.231  757.759
\LT  898.650  760.042
\LT  902.090  762.337
\LT  905.549  764.645
\LT  909.029  766.966
\LT  912.529  769.299
\LT  916.050  771.646
\LT  919.591  774.006
\LT  923.153  776.379
\LT  926.736  778.765
\LT  930.340  781.165
\LT  933.966  783.578
\LT  937.613  786.005
\LT  941.281  788.446
\LT  944.972  790.901
\LT  948.684  793.369
\LT  952.419  795.852
\LT  956.176  798.349
\LT  959.955  800.860
\LT  963.758  803.386
\LT  967.583  805.926
\LT  971.431  808.481
\LT  975.303  811.050
\LT  979.198  813.634
\LT  983.116  816.234
\LT  987.059  818.849
\LT  991.026  821.478
\LT  995.017  824.124
\LT  999.032  826.784
\LT 1003.072  829.461
\LT 1007.137  832.153
\LT 1011.227  834.861
\LT 1015.343  837.585
\LT 1019.484  840.325
\LT 1023.650  843.082
\LT 1027.843  845.855
\LT 1032.062  848.644
\LT 1036.307  851.450
\LT 1040.579  854.273
\LT 1044.877  857.113
\LT 1049.203  859.970
\LT 1053.556  862.845
\LT 1057.937  865.737
\LT 1062.345  868.646
\LT 1066.781  871.573
\LT 1071.246  874.518
\LT 1075.739  877.481
\LT 1080.261  880.462
\LT 1084.811  883.462
\LT 1089.392  886.480
\LT 1094.001  889.516
\LT 1098.640  892.572
\LT 1103.310  895.646
\kon a_{\crt}(b),\ b>0;
 
\obrazek39
\grub0.2pt
\MT   0.000  100.000
\LT1200.000  100.000
\MT 1140.000 115.000 \LT 1200.000 100.000 \LT 1140.000 85.000
\cput(1200.000,25.000,b)
\MT 940.000   80.000
\LT 940.000  120.000
\MT 780.000   80.000
\LT 780.000  120.000
\cput(780.000,20.000,-0.010)
\MT 620.000   80.000
\LT 620.000  120.000
\MT 460.000   80.000
\LT 460.000  120.000
\cput(460.000,20.000,-0.020)
\MT 300.000   80.000
\LT 300.000  120.000
\MT 140.000   80.000
\LT 140.000  120.000
\cput(140.000,20.000,-0.030)
\MT1100.000    0.000
\LT1100.000  900.000
\MT 1085.000 840.000 \LT 1100.000 900.000 \LT 1115.000 840.000
\rput(1130.000,870.000,x_{\crt})
\MT1080.000  250.000
\LT1120.000  250.000
\rput(1140.000,230.000,2)
\MT1080.000  400.000
\LT1120.000  400.000
\rput(1140.000,380.000,4)
\MT1080.000  550.000
\LT1120.000  550.000
\rput(1140.000,530.000,6)
\MT1080.000  700.000
\LT1120.000  700.000
\rput(1140.000,680.000,8)
\grub0.6pt
\MT 1097.970  225.935
\LT 1097.970  225.935
\LT 1077.706  226.227
\LT 1057.515  226.524
\LT 1037.397  226.826
\LT 1017.352  227.133
\LT  997.381  227.445
\LT  977.482  227.761
\LT  957.657  228.083
\LT  937.906  228.412
\LT  918.228  228.746
\LT  898.625  229.085
\LT  879.096  229.432
\LT  859.641  229.784
\LT  840.260  230.143
\LT  820.955  230.509
\LT  801.725  230.883
\LT  782.570  231.264
\LT  763.491  231.655
\LT  744.488  232.053
\LT  725.561  232.459
\LT  706.710  232.876
\LT  687.937  233.301
\LT  669.241  233.738
\LT  650.623  234.185
\LT  632.083  234.644
\LT  613.622  235.115
\LT  597.075  235.551
\LT  580.592  235.996
\LT  564.174  236.453
\LT  547.821  236.924
\LT  531.534  237.408
\LT  515.314  237.905
\LT  499.160  238.416
\LT  483.074  238.945
\LT  467.055  239.490
\LT  451.106  240.054
\LT  435.225  240.637
\LT  419.415  241.242
\LT  403.675  241.870
\LT  388.008  242.524
\LT  374.142  243.128
\LT  360.334  243.756
\LT  346.586  244.411
\LT  332.897  245.094
\LT  319.269  245.809
\LT  305.704  246.561
\LT  292.201  247.351
\LT  278.763  248.188
\LT  267.059  248.963
\LT  255.407  249.782
\LT  243.808  250.653
\LT  233.909  251.446
\LT  224.052  252.291
\LT  214.237  253.195
\LT  206.091  254.002
\LT  197.977  254.864
\LT  189.895  255.796
\LT  183.454  256.599
\LT  177.036  257.465
\LT  172.238  258.164
\LT  167.454  258.913
\LT  162.684  259.722
\LT  157.929  260.608
\LT  154.768  261.249
\LT  151.615  261.943
\LT  148.469  262.698
\LT  146.899  263.105
\LT  145.332  263.534
\LT  143.766  263.991
\LT  142.203  264.478
\LT  140.643  265.004
\LT  139.085  265.577
\LT  137.530  266.211
\LT  135.978  266.928
\LT  134.429  267.764
\LT  132.885  268.801
\LT  131.345  270.298
\LT  130.501  272.022
\LT  130.425  272.642
\LT  130.425  272.853
\LT  130.501  273.483
\LT  131.338  275.310
\LT  132.855  276.990
\LT  134.368  278.212
\LT  135.878  279.233
\LT  137.386  280.134
\LT  138.892  280.954
\LT  140.396  281.714
\LT  141.898  282.427
\LT  143.399  283.102
\LT  144.899  283.746
\LT  146.397  284.363
\LT  149.391  285.535
\LT  152.381  286.637
\LT  155.367  287.685
\LT  159.840  289.175
\LT  164.308  290.589
\LT  168.770  291.942
\LT  174.713  293.672
\LT  180.649  295.334
\LT  188.062  297.335
\LT  196.949  299.652
\LT  207.308  302.266
\LT  219.139  305.169
\LT  232.446  308.358
\LT  250.191  312.531
\LT  278.319  319.053
\LT  302.060  324.545
\LT  321.401  329.051
\LT  337.811  332.913
\LT  352.771  336.478
\LT  367.774  340.102
\LT  381.319  343.424
\LT  394.906  346.810
\LT  407.023  349.880
\LT  419.179  353.011
\LT  429.850  355.806
\LT  440.554  358.655
\LT  451.295  361.565
\LT  462.073  364.537
\LT  471.342  367.138
\LT  480.642  369.792
\LT  489.974  372.502
\LT  499.339  375.271
\LT  508.739  378.102
\LT  516.600  380.512
\LT  524.486  382.970
\LT  532.400  385.478
\LT  540.341  388.040
\LT  548.312  390.657
\LT  556.312  393.333
\LT  564.343  396.070
\LT  572.407  398.872
\LT  580.504  401.741
\LT  588.636  404.683
\LT  595.167  407.090
\LT  601.722  409.549
\LT  608.301  412.059
\LT  614.906  414.625
\LT  621.536  417.248
\LT  628.193  419.932
\LT  634.878  422.680
\LT  641.592  425.493
\LT  646.646  427.648
\LT  653.412  430.585
\LT  658.507  432.837
\LT  663.620  435.134
\LT  668.751  437.477
\LT  673.901  439.867
\LT  679.071  442.306
\LT  684.260  444.797
\LT  689.470  447.343
\LT  694.701  449.944
\LT  699.954  452.603
\LT  705.229  455.324
\LT  710.527  458.108
\LT  715.848  460.959
\LT  721.194  463.879
\LT  726.564  466.873
\LT  731.960  469.943
\LT  737.383  473.093
\LT  741.013  475.239
\LT  744.656  477.424
\LT  748.311  479.649
\LT  751.979  481.915
\LT  755.660  484.223
\LT  759.355  486.576
\LT  763.064  488.975
\LT  766.786  491.421
\LT  770.523  493.917
\LT  774.275  496.463
\LT  778.042  499.062
\LT  781.824  501.717
\LT  785.623  504.429
\LT  789.437  507.200
\LT  793.268  510.034
\LT  797.116  512.931
\LT  800.982  515.897
\LT  804.865  518.932
\LT  806.814  520.476
\LT  810.725  523.623
\LT  812.688  525.226
\LT  814.656  526.848
\LT  816.629  528.491
\LT  818.606  530.154
\LT  820.589  531.839
\LT  822.577  533.546
\LT  824.570  535.276
\LT  826.568  537.028
\LT  828.572  538.804
\LT  830.581  540.604
\LT  832.596  542.429
\LT  834.616  544.279
\LT  836.642  546.154
\LT  838.674  548.057
\LT  840.712  549.986
\LT  842.755  551.945
\LT  844.805  553.931
\LT  846.861  555.948
\LT  848.923  557.994
\LT  850.992  560.072
\LT  853.067  562.181
\LT  855.148  564.324
\LT  857.236  566.500
\LT  859.331  568.711
\LT  861.433  570.959
\LT  863.542  573.242
\LT  865.658  575.565
\LT  867.782  577.925
\LT  869.913  580.327
\LT  872.051  582.769
\LT  874.197  585.255
\LT  876.351  587.785
\LT  878.513  590.361
\LT  880.683  592.983
\LT  882.862  595.655
\LT  885.049  598.376
\LT  887.244  601.150
\LT  889.449  603.978
\LT  891.662  606.861
\LT  893.885  609.802
\LT  896.117  612.803
\LT  898.359  615.866
\LT  900.610  618.993
\LT  902.871  622.187
\LT  905.143  625.451
\LT  907.425  628.786
\LT  909.718  632.197
\LT  912.022  635.686
\LT  914.338  639.255
\LT  916.664  642.910
\LT  919.003  646.653
\LT  921.354  650.488
\LT  923.717  654.419
\LT  926.093  658.451
\LT  928.482  662.588
\LT  930.885  666.835
\LT  933.301  671.198
\LT  935.732  675.683
\LT  938.177  680.295
\LT  940.638  685.041
\LT  943.114  689.927
\LT  945.606  694.962
\LT  948.115  700.154
\LT  950.641  705.511
\LT  953.184  711.043
\LT  955.746  716.761
\LT  958.327  722.677
\LT  960.927  728.800
\LT  963.548  735.147
\LT  966.190  741.730
\LT  968.854  748.567
\LT  971.541  755.674
\LT  974.251  763.073
\LT  976.986  770.783
\LT  979.747  778.829
\LT  982.535  787.238
\LT  985.352  796.040
\LT  988.198  805.267
\LT  991.075  814.959
\LT  993.986  825.157
\LT  996.931  835.910
\LT  999.912  847.274
\kon x_{\crt}(b),\ b<0;
\obrazek40
\grub0.2pt
\MT   0.000  100.000
\LT1200.000  100.000
\MT 1140.000 115.000 \LT 1200.000 100.000 \LT 1140.000 85.000
\cput(1200.000,25.000,b)
\MT 940.000   80.000
\LT 940.000  120.000
\MT 780.000   80.000
\LT 780.000  120.000
\cput(780.000,20.000,-0.010)
\MT 620.000   80.000
\LT 620.000  120.000
\MT 460.000   80.000
\LT 460.000  120.000
\cput(460.000,20.000,-0.020)
\MT 300.000   80.000
\LT 300.000  120.000
\MT 140.000   80.000
\LT 140.000  120.000
\cput(140.000,20.000,-0.030)
\MT1100.000    0.000
\LT1100.000  150.000
\MT1100.000  220.000
\LT1100.000  900.000
\MT 1085.000 840.000 \LT 1100.000 900.000 \LT 1115.000 840.000
\rput(1130.000,870.000,a_{\crt})
\MT1080.000  250.000
\LT1120.000  250.000
\rput(1140.000,230.000,3)
\MT1080.000  400.000
\LT1120.000  400.000
\rput(1140.000,380.000,4)
\MT1080.000  550.000
\LT1120.000  550.000
\rput(1140.000,530.000,5)
\MT1080.000  700.000
\LT1120.000  700.000
\rput(1140.000,680.000,6)
\grub0.6pt
\MT 1097.970  275.864
\LT 1097.970  275.864
\LT 1077.706  275.013
\LT 1057.515  274.160
\LT 1037.397  273.306
\LT 1017.352  272.452
\LT  997.381  271.596
\LT  977.482  270.739
\LT  957.657  269.881
\LT  937.906  269.022
\LT  918.228  268.162
\LT  898.625  267.300
\LT  879.096  266.437
\LT  859.641  265.572
\LT  840.260  264.706
\LT  820.955  263.838
\LT  801.725  262.968
\LT  782.570  262.097
\LT  763.491  261.224
\LT  744.488  260.349
\LT  725.561  259.472
\LT  706.710  258.593
\LT  687.937  257.712
\LT  669.241  256.828
\LT  650.623  255.942
\LT  632.083  255.054
\LT  613.622  254.163
\LT  595.240  253.269
\LT  578.764  252.461
\LT  562.354  251.652
\LT  546.008  250.840
\LT  529.729  250.025
\LT  513.516  249.207
\LT  497.369  248.386
\LT  481.290  247.562
\LT  465.280  246.734
\LT  449.338  245.904
\LT  433.465  245.069
\LT  417.662  244.230
\LT  401.931  243.387
\LT  386.271  242.540
\LT  370.684  241.688
\LT  355.172  240.831
\LT  339.734  239.969
\LT  324.373  239.101
\LT  309.089  238.226
\LT  293.886  237.345
\LT  280.440  236.556
\LT  267.059  235.760
\LT  253.747  234.957
\LT  240.504  234.147
\LT  227.333  233.329
\LT  214.237  232.502
\LT  202.842  231.769
\LT  191.509  231.028
\LT  180.243  230.276
\LT  169.047  229.512
\LT  159.512  228.846
\LT  150.041  228.167
\LT  142.203  227.588
\LT  134.429  226.992
\LT  131.345  226.745
\LT  130.501  226.676
\LT  130.425  226.670
\LT  130.425  226.670
\LT  130.501  226.676
\LT  131.338  226.748
\LT  137.386  227.297
\LT  146.397  228.162
\LT  156.859  229.221
\LT  168.770  230.487
\LT  182.132  231.978
\LT  195.468  233.537
\LT  208.787  235.163
\LT  222.097  236.855
\LT  235.403  238.614
\LT  248.712  240.443
\LT  262.028  242.342
\LT  275.355  244.313
\LT  288.698  246.359
\LT  300.574  248.243
\LT  312.468  250.190
\LT  324.381  252.203
\LT  336.317  254.285
\LT  348.278  256.437
\LT  358.767  258.382
\LT  369.277  260.385
\LT  379.812  262.450
\LT  390.372  264.578
\LT  400.960  266.774
\LT  411.577  269.038
\LT  422.224  271.376
\LT  431.377  273.440
\LT  440.554  275.562
\LT  449.758  277.744
\LT  458.989  279.990
\LT  468.249  282.302
\LT  477.539  284.682
\LT  485.304  286.721
\LT  493.092  288.811
\LT  500.904  290.955
\LT  508.739  293.156
\LT  516.600  295.415
\LT  524.486  297.734
\LT  532.400  300.118
\LT  540.341  302.567
\LT  548.312  305.086
\LT  556.312  307.677
\LT  562.734  309.804
\LT  569.178  311.982
\LT  575.642  314.211
\LT  582.127  316.495
\LT  588.636  318.836
\LT  595.167  321.235
\LT  601.722  323.695
\LT  608.301  326.219
\LT  614.906  328.809
\LT  621.536  331.469
\LT  628.193  334.201
\LT  633.204  336.300
\LT  638.232  338.443
\LT  643.275  340.631
\LT  648.335  342.867
\LT  653.412  345.151
\LT  658.507  347.486
\LT  663.620  349.874
\LT  668.751  352.317
\LT  673.901  354.816
\LT  679.071  357.375
\LT  684.260  359.994
\LT  689.470  362.678
\LT  694.701  365.428
\LT  699.954  368.247
\LT  705.229  371.138
\LT  710.527  374.105
\LT  715.848  377.151
\LT  721.194  380.279
\LT  724.771  382.412
\LT  728.360  384.585
\LT  731.960  386.798
\LT  735.572  389.053
\LT  739.196  391.352
\LT  742.833  393.695
\LT  746.482  396.085
\LT  750.143  398.523
\LT  753.818  401.011
\LT  757.506  403.549
\LT  761.207  406.141
\LT  764.923  408.788
\LT  768.653  411.492
\LT  772.397  414.255
\LT  776.157  417.080
\LT  779.931  419.968
\LT  783.721  422.923
\LT  787.528  425.946
\LT  791.350  429.041
\LT  795.190  432.211
\LT  797.116  433.825
\LT  799.047  435.459
\LT  800.982  437.113
\LT  802.921  438.787
\LT  804.865  440.483
\LT  806.814  442.201
\LT  808.767  443.940
\LT  810.725  445.702
\LT  812.688  447.487
\LT  814.656  449.296
\LT  816.629  451.129
\LT  818.606  452.987
\LT  820.589  454.869
\LT  822.577  456.778
\LT  824.570  458.713
\LT  826.568  460.676
\LT  828.572  462.666
\LT  830.581  464.684
\LT  832.596  466.732
\LT  834.616  468.810
\LT  836.642  470.918
\LT  838.674  473.058
\LT  840.712  475.230
\LT  842.755  477.435
\LT  844.805  479.674
\LT  846.861  481.948
\LT  848.923  484.257
\LT  850.992  486.604
\LT  853.067  488.988
\LT  855.148  491.411
\LT  857.236  493.875
\LT  859.331  496.379
\LT  861.433  498.926
\LT  863.542  501.516
\LT  865.658  504.151
\LT  867.782  506.833
\LT  869.913  509.562
\LT  872.051  512.340
\LT  874.197  515.169
\LT  876.351  518.050
\LT  878.513  520.985
\LT  880.683  523.976
\LT  882.862  527.025
\LT  885.049  530.133
\LT  887.244  533.303
\LT  889.449  536.536
\LT  891.662  539.836
\LT  893.885  543.203
\LT  896.117  546.642
\LT  898.359  550.154
\LT  900.610  553.742
\LT  902.871  557.410
\LT  905.143  561.159
\LT  907.425  564.994
\LT  909.718  568.918
\LT  912.022  572.934
\LT  914.338  577.047
\LT  916.664  581.260
\LT  919.003  585.577
\LT  921.354  590.004
\LT  923.717  594.545
\LT  926.093  599.205
\LT  928.482  603.990
\LT  930.885  608.906
\LT  933.301  613.959
\LT  935.732  619.156
\LT  938.177  624.504
\LT  940.638  630.011
\LT  943.114  635.686
\LT  945.606  641.536
\LT  948.115  647.573
\LT  950.641  653.807
\LT  953.184  660.248
\LT  955.746  666.910
\LT  958.327  673.806
\LT  960.927  680.950
\LT  963.548  688.358
\LT  966.190  696.049
\LT  968.854  704.040
\LT  971.541  712.354
\LT  974.251  721.013
\LT  976.986  730.044
\LT  979.747  739.474
\LT  982.535  749.336
\LT  985.352  759.666
\LT  988.198  770.502
\LT  991.075  781.892
\LT  993.986  793.884
\LT  996.931  806.537
\LT  999.912  819.918
\kon a_{\crt}(b),\ b<0;
\obrazek45
\grub0.2pt
\MT   0.000  425.000
\LT1400.000  425.000
\MT 160.000  425.000
\LT 160.000  435.000
\MT 220.000  425.000
\LT 220.000  435.000
\MT 280.000  425.000
\LT 280.000  435.000
\MT 340.000  425.000
\LT 340.000  435.000
\cput(340.000,365.000,2)
\MT 400.000  425.000
\LT 400.000  435.000
\MT 460.000  425.000
\LT 460.000  435.000
\MT 520.000  425.000
\LT 520.000  435.000
\MT 580.000  425.000
\LT 580.000  435.000
\cput(580.000,365.000,4)
\MT 640.000  425.000
\LT 640.000  435.000
\MT 700.000  425.000
\LT 700.000  435.000
\MT 760.000  425.000
\LT 760.000  435.000
\MT 820.000  425.000
\LT 820.000  435.000
\cput(820.000,365.000,6)
\MT 880.000  425.000
\LT 880.000  435.000
\MT 940.000  425.000
\LT 940.000  435.000
\MT1000.000  425.000
\LT1000.000  435.000
\MT1060.000  425.000
\LT1060.000  435.000
\cput(1060.000,365.000,8)
\MT1120.000  425.000
\LT1120.000  435.000
\MT1180.000  425.000
\LT1180.000  435.000
\MT1240.000  425.000
\LT1240.000  435.000
\MT1300.000  425.000
\LT1300.000  435.000
\cput(1300.000,365.000,10)
\MT 100.000    0.000
\LT 100.000  850.000
\MT  92.000   50.000
\LT 108.000   50.000
\MT  92.000   87.500
\LT 108.000   87.500
\MT  92.000  125.000
\LT 108.000  125.000
\MT  92.000  162.500
\LT 108.000  162.500
\MT  92.000  200.000
\LT 108.000  200.000
\MT  92.000  237.500
\LT 108.000  237.500
\MT  92.000  275.000
\LT 108.000  275.000
\MT  92.000  312.500
\LT 108.000  312.500
\MT  92.000  350.000
\LT 108.000  350.000
\MT  92.000  387.500
\LT 108.000  387.500
\MT  92.000  425.000
\LT 108.000  425.000
\MT  92.000  462.500
\LT 108.000  462.500
\MT  92.000  500.000
\LT 108.000  500.000
\MT  92.000  537.500
\LT 108.000  537.500
\MT  92.000  575.000
\LT 108.000  575.000
\MT  92.000  612.500
\LT 108.000  612.500
\MT  92.000  650.000
\LT 108.000  650.000
\MT  92.000  687.500
\LT 108.000  687.500
\MT  92.000  725.000
\LT 108.000  725.000
\MT  92.000  762.500
\LT 108.000  762.500
\MT  92.000  800.000
\LT 108.000  800.000
\MT  84.000   50.000
\LT 116.000   50.000
\lput(80.000,31.250, -1.0)
\MT  84.000  237.500
\LT 116.000  237.500
\lput(80.000,218.750, -0.5)
\MT  84.000  425.000
\LT 116.000  425.000
\MT  84.000  612.500
\LT 116.000  612.500
\lput(80.000,593.750,  0.5)
\MT  84.000  800.000
\LT 116.000  800.000
\lput(80.000,781.250,  1.0)
\grub0.6pt
\MT  100.000  800.000
\LT  101.000  799.973
\LT  102.000  799.893
\LT  103.000  799.760
\LT  104.000  799.572
\LT  105.000  799.332
\LT  106.000  799.038
\LT  107.000  798.691
\LT  108.000  798.290
\LT  109.000  797.836
\LT  110.000  797.328
\LT  111.000  796.767
\LT  112.000  796.152
\LT  113.000  795.484
\LT  114.000  794.763
\LT  115.000  793.988
\LT  116.000  793.160
\LT  117.000  792.279
\LT  118.000  791.344
\LT  119.000  790.356
\LT  120.000  789.315
\LT  121.000  788.220
\LT  122.000  787.073
\LT  123.000  785.872
\LT  124.000  784.618
\LT  125.000  783.312
\LT  126.000  781.953
\LT  127.000  780.541
\LT  128.000  779.076
\LT  129.000  777.559
\LT  130.000  775.990
\LT  131.000  774.368
\LT  132.000  772.695
\LT  133.000  770.970
\LT  134.000  769.193
\LT  135.000  767.365
\LT  136.000  765.486
\LT  137.000  763.556
\LT  138.000  761.576
\LT  139.000  759.546
\LT  140.000  757.466
\LT  141.000  755.336
\LT  142.000  753.158
\LT  143.000  750.931
\LT  144.000  748.656
\LT  145.000  746.334
\LT  146.000  743.965
\LT  147.000  741.549
\LT  148.000  739.087
\LT  149.000  736.581
\LT  150.000  734.029
\LT  151.000  731.434
\LT  152.000  728.796
\LT  153.000  726.116
\LT  154.000  723.394
\LT  155.000  720.632
\LT  156.000  717.829
\LT  157.000  714.989
\LT  158.000  712.110
\LT  159.000  709.195
\LT  160.000  706.244
\LT  161.000  703.258
\LT  162.000  700.239
\LT  163.000  697.188
\LT  164.000  694.106
\LT  165.000  690.993
\LT  166.000  687.853
\LT  167.000  684.685
\LT  168.000  681.491
\LT  169.000  678.273
\LT  170.000  675.032
\LT  171.000  671.769
\LT  173.000  665.184
\LT  175.000  658.531
\LT  177.000  651.822
\LT  180.000  641.682
\LT  187.000  617.890
\LT  190.000  607.744
\LT  192.000  601.028
\LT  194.000  594.366
\LT  196.000  587.768
\LT  197.000  584.497
\LT  198.000  581.247
\LT  199.000  578.018
\LT  200.000  574.813
\LT  201.000  571.632
\LT  202.000  568.477
\LT  203.000  565.348
\LT  204.000  562.247
\LT  205.000  559.175
\LT  206.000  556.133
\LT  207.000  553.122
\LT  208.000  550.143
\LT  209.000  547.196
\LT  210.000  544.283
\LT  211.000  541.405
\LT  212.000  538.562
\LT  213.000  535.754
\LT  214.000  532.984
\LT  215.000  530.250
\LT  216.000  527.554
\LT  217.000  524.896
\LT  218.000  522.277
\LT  219.000  519.697
\LT  220.000  517.156
\LT  221.000  514.655
\LT  222.000  512.194
\LT  223.000  509.773
\LT  224.000  507.393
\LT  225.000  505.053
\LT  226.000  502.754
\LT  227.000  500.496
\LT  228.000  498.279
\LT  229.000  496.103
\LT  230.000  493.967
\LT  231.000  491.873
\LT  232.000  489.819
\LT  233.000  487.806
\LT  234.000  485.833
\LT  235.000  483.901
\LT  236.000  482.008
\LT  237.000  480.156
\LT  238.000  478.343
\LT  239.000  476.570
\LT  240.000  474.835
\LT  241.000  473.140
\LT  242.000  471.483
\LT  243.000  469.864
\LT  244.000  468.283
\LT  245.000  466.739
\LT  246.000  465.232
\LT  247.000  463.762
\LT  248.000  462.328
\LT  249.000  460.930
\LT  250.000  459.567
\LT  251.000  458.240
\LT  252.000  456.946
\LT  253.000  455.687
\LT  254.000  454.461
\LT  255.000  453.268
\LT  256.000  452.108
\LT  257.000  450.980
\LT  258.000  449.883
\LT  259.000  448.818
\LT  260.000  447.784
\LT  261.000  446.779
\LT  262.000  445.805
\LT  263.000  444.859
\LT  264.000  443.943
\LT  265.000  443.054
\LT  266.000  442.194
\LT  267.000  441.360
\LT  268.000  440.554
\LT  269.000  439.774
\LT  270.000  439.019
\LT  271.000  438.291
\LT  272.000  437.587
\LT  273.000  436.907
\LT  274.000  436.252
\LT  275.000  435.620
\LT  276.000  435.011
\LT  277.000  434.424
\LT  278.000  433.860
\LT  279.000  433.318
\LT  280.000  432.797
\LT  281.000  432.297
\LT  282.000  431.817
\LT  284.000  430.917
\LT  286.000  430.094
\LT  288.000  429.345
\LT  290.000  428.666
\LT  292.000  428.055
\LT  294.000  427.508
\LT  296.000  427.023
\LT  298.000  426.597
\LT  300.000  426.227
\LT  302.000  425.911
\LT  304.000  425.646
\LT  306.000  425.430
\LT  308.000  425.260
\LT  310.000  425.134
\LT  312.000  425.051
\LT  315.000  425.000
\LT  318.000  425.033
\LT  321.000  425.143
\LT  324.000  425.324
\LT  327.000  425.571
\LT  330.000  425.878
\LT  333.000  426.241
\LT  336.000  426.655
\LT  340.000  427.278
\LT  344.000  427.973
\LT  348.000  428.734
\LT  353.000  429.762
\LT  358.000  430.863
\LT  364.000  432.264
\LT  371.000  433.981
\LT  379.000  436.021
\LT  394.000  439.956
\LT  409.000  443.890
\LT  419.000  446.447
\LT  428.000  448.676
\LT  436.000  450.585
\LT  443.000  452.193
\LT  450.000  453.736
\LT  457.000  455.210
\LT  464.000  456.612
\LT  471.000  457.939
\LT  478.000  459.189
\LT  485.000  460.359
\LT  491.000  461.297
\LT  497.000  462.175
\LT  503.000  462.992
\LT  509.000  463.747
\LT  515.000  464.442
\LT  521.000  465.074
\LT  527.000  465.644
\LT  533.000  466.153
\LT  539.000  466.599
\LT  545.000  466.984
\LT  551.000  467.308
\LT  557.000  467.570
\LT  563.000  467.771
\LT  569.000  467.911
\LT  575.000  467.991
\LT  582.000  468.009
\LT  589.000  467.946
\LT  596.000  467.802
\LT  603.000  467.579
\LT  610.000  467.276
\LT  617.000  466.896
\LT  624.000  466.438
\LT  631.000  465.904
\LT  638.000  465.293
\LT  645.000  464.608
\LT  652.000  463.848
\LT  659.000  463.014
\LT  666.000  462.108
\LT  673.000  461.129
\LT  680.000  460.079
\LT  687.000  458.958
\LT  694.000  457.767
\LT  701.000  456.506
\LT  708.000  455.177
\LT  715.000  453.780
\LT  722.000  452.316
\LT  729.000  450.785
\LT  736.000  449.188
\LT  743.000  447.525
\LT  750.000  445.798
\LT  757.000  444.007
\LT  764.000  442.152
\LT  771.000  440.234
\LT  778.000  438.253
\LT  785.000  436.210
\LT  793.000  433.801
\LT  801.000  431.312
\LT  809.000  428.744
\LT  817.000  426.098
\LT  825.000  423.376
\LT  833.000  420.576
\LT  841.000  417.701
\LT  849.000  414.750
\LT  857.000  411.725
\LT  865.000  408.626
\LT  873.000  405.454
\LT  881.000  402.208
\LT  889.000  398.891
\LT  897.000  395.502
\LT  905.000  392.041
\LT  913.000  388.511
\LT  921.000  384.910
\LT  929.000  381.239
\LT  937.000  377.499
\LT  945.000  373.691
\LT  953.000  369.814
\LT  961.000  365.870
\LT  969.000  361.859
\LT  977.000  357.780
\LT  985.000  353.635
\LT  993.000  349.424
\LT 1001.000  345.147
\LT 1009.000  340.805
\LT 1017.000  336.397
\LT 1025.000  331.926
\LT 1033.000  327.389
\LT 1042.000  322.210
\LT 1051.000  316.950
\LT 1060.000  311.609
\LT 1069.000  306.190
\LT 1078.000  300.691
\LT 1087.000  295.113
\LT 1096.000  289.457
\LT 1105.000  283.723
\LT 1114.000  277.911
\LT 1123.000  272.022
\LT 1132.000  266.056
\LT 1141.000  260.014
\LT 1150.000  253.896
\LT 1159.000  247.701
\LT 1168.000  241.431
\LT 1177.000  235.086
\LT 1186.000  228.666
\LT 1195.000  222.171
\LT 1204.000  215.602
\LT 1213.000  208.958
\LT 1222.000  202.241
\LT 1231.000  195.450
\LT 1240.000  188.586
\LT 1249.000  181.649
\LT 1258.000  174.639
\LT 1267.000  167.557
\LT 1276.000  160.402
\LT 1285.000  153.175
\LT 1294.000  145.876
\LT 1300.000  140.970
\koniec     3.0349 -0.0145067
\obrazek46
\grub0.2pt
\MT   0.000  425.000
\LT1400.000  425.000
\MT 160.000  425.000
\LT 160.000  435.000
\MT 220.000  425.000
\LT 220.000  435.000
\MT 280.000  425.000
\LT 280.000  435.000
\MT 340.000  425.000
\LT 340.000  435.000
\cput(340.000,365.000,2)
\MT 400.000  425.000
\LT 400.000  435.000
\MT 460.000  425.000
\LT 460.000  435.000
\MT 520.000  425.000
\LT 520.000  435.000
\MT 580.000  425.000
\LT 580.000  435.000
\cput(580.000,365.000,4)
\MT 640.000  425.000
\LT 640.000  435.000
\MT 700.000  425.000
\LT 700.000  435.000
\MT 760.000  425.000
\LT 760.000  435.000
\MT 820.000  425.000
\LT 820.000  435.000
\cput(820.000,365.000,6)
\MT 880.000  425.000
\LT 880.000  435.000
\MT 940.000  425.000
\LT 940.000  435.000
\MT1000.000  425.000
\LT1000.000  435.000
\MT1060.000  425.000
\LT1060.000  435.000
\cput(1060.000,365.000,8)
\MT1120.000  425.000
\LT1120.000  435.000
\MT1180.000  425.000
\LT1180.000  435.000
\MT1240.000  425.000
\LT1240.000  435.000
\MT1300.000  425.000
\LT1300.000  435.000
\cput(1300.000,365.000,10)
\MT 100.000    0.000
\LT 100.000  850.000
\MT  92.000   50.000
\LT 108.000   50.000
\MT  92.000   87.500
\LT 108.000   87.500
\MT  92.000  125.000
\LT 108.000  125.000
\MT  92.000  162.500
\LT 108.000  162.500
\MT  92.000  200.000
\LT 108.000  200.000
\MT  92.000  237.500
\LT 108.000  237.500
\MT  92.000  275.000
\LT 108.000  275.000
\MT  92.000  312.500
\LT 108.000  312.500
\MT  92.000  350.000
\LT 108.000  350.000
\MT  92.000  387.500
\LT 108.000  387.500
\MT  92.000  425.000
\LT 108.000  425.000
\MT  92.000  462.500
\LT 108.000  462.500
\MT  92.000  500.000
\LT 108.000  500.000
\MT  92.000  537.500
\LT 108.000  537.500
\MT  92.000  575.000
\LT 108.000  575.000
\MT  92.000  612.500
\LT 108.000  612.500
\MT  92.000  650.000
\LT 108.000  650.000
\MT  92.000  687.500
\LT 108.000  687.500
\MT  92.000  725.000
\LT 108.000  725.000
\MT  92.000  762.500
\LT 108.000  762.500
\MT  92.000  800.000
\LT 108.000  800.000
\MT  84.000   50.000
\LT 116.000   50.000
\lput(80.000,31.250, -1.0)
\MT  84.000  237.500
\LT 116.000  237.500
\lput(80.000,218.750, -0.5)
\MT  84.000  425.000
\LT 116.000  425.000
\MT  84.000  612.500
\LT 116.000  612.500
\lput(80.000,593.750,  0.5)
\MT  84.000  800.000
\LT 116.000  800.000
\lput(80.000,781.250,  1.0)
\grub0.6pt
\MT  100.000  800.000
\LT  101.000  799.970
\LT  102.000  799.879
\LT  103.000  799.728
\LT  104.000  799.517
\LT  105.000  799.245
\LT  106.000  798.913
\LT  107.000  798.521
\LT  108.000  798.068
\LT  109.000  797.555
\LT  110.000  796.982
\LT  111.000  796.348
\LT  112.000  795.654
\LT  113.000  794.899
\LT  114.000  794.084
\LT  115.000  793.209
\LT  116.000  792.274
\LT  117.000  791.278
\LT  118.000  790.222
\LT  119.000  789.106
\LT  120.000  787.930
\LT  121.000  786.694
\LT  122.000  785.398
\LT  123.000  784.042
\LT  124.000  782.626
\LT  125.000  781.150
\LT  126.000  779.614
\LT  127.000  778.019
\LT  128.000  776.365
\LT  129.000  774.651
\LT  130.000  772.879
\LT  131.000  771.047
\LT  132.000  769.157
\LT  133.000  767.208
\LT  134.000  765.202
\LT  135.000  763.137
\LT  136.000  761.014
\LT  137.000  758.835
\LT  138.000  756.598
\LT  139.000  754.305
\LT  140.000  751.955
\LT  141.000  749.550
\LT  142.000  747.089
\LT  143.000  744.574
\LT  144.000  742.004
\LT  145.000  739.381
\LT  146.000  736.705
\LT  147.000  733.976
\LT  148.000  731.196
\LT  149.000  728.364
\LT  150.000  725.482
\LT  151.000  722.551
\LT  152.000  719.571
\LT  153.000  716.543
\LT  154.000  713.469
\LT  155.000  710.349
\LT  156.000  707.183
\LT  157.000  703.975
\LT  158.000  700.723
\LT  159.000  697.430
\LT  160.000  694.097
\LT  161.000  690.724
\LT  162.000  687.314
\LT  163.000  683.868
\LT  164.000  680.386
\LT  165.000  676.871
\LT  166.000  673.323
\LT  167.000  669.745
\LT  168.000  666.137
\LT  169.000  662.502
\LT  170.000  658.841
\LT  171.000  655.155
\LT  172.000  651.447
\LT  174.000  643.969
\LT  176.000  636.421
\LT  178.000  628.816
\LT  181.000  617.337
\LT  187.000  594.296
\LT  190.000  582.835
\LT  192.000  575.249
\LT  194.000  567.724
\LT  195.000  563.988
\LT  196.000  560.271
\LT  197.000  556.577
\LT  198.000  552.905
\LT  199.000  549.259
\LT  200.000  545.638
\LT  201.000  542.045
\LT  202.000  538.480
\LT  203.000  534.946
\LT  204.000  531.444
\LT  205.000  527.974
\LT  206.000  524.538
\LT  207.000  521.136
\LT  208.000  517.771
\LT  209.000  514.443
\LT  210.000  511.153
\LT  211.000  507.902
\LT  212.000  504.690
\LT  213.000  501.519
\LT  214.000  498.389
\LT  215.000  495.301
\LT  216.000  492.256
\LT  217.000  489.254
\LT  218.000  486.295
\LT  219.000  483.381
\LT  220.000  480.511
\LT  221.000  477.686
\LT  222.000  474.906
\LT  223.000  472.171
\LT  224.000  469.483
\LT  225.000  466.840
\LT  226.000  464.243
\LT  227.000  461.693
\LT  228.000  459.188
\LT  229.000  456.730
\LT  230.000  454.318
\LT  231.000  451.952
\LT  232.000  449.632
\LT  233.000  447.358
\LT  234.000  445.130
\LT  235.000  442.947
\LT  236.000  440.809
\LT  237.000  438.717
\LT  238.000  436.669
\LT  239.000  434.666
\LT  240.000  432.707
\LT  241.000  430.792
\LT  242.000  428.920
\LT  243.000  427.092
\LT  244.000  425.306
\LT  245.000  423.562
\LT  246.000  421.860
\LT  247.000  420.200
\LT  248.000  418.580
\LT  249.000  417.001
\LT  250.000  415.461
\LT  251.000  413.961
\LT  252.000  412.500
\LT  253.000  411.078
\LT  254.000  409.693
\LT  255.000  408.346
\LT  256.000  407.035
\LT  257.000  405.761
\LT  258.000  404.523
\LT  259.000  403.319
\LT  260.000  402.151
\LT  261.000  401.016
\LT  262.000  399.915
\LT  263.000  398.847
\LT  264.000  397.812
\LT  265.000  396.809
\LT  266.000  395.837
\LT  267.000  394.895
\LT  268.000  393.984
\LT  269.000  393.103
\LT  270.000  392.251
\LT  271.000  391.428
\LT  272.000  390.633
\LT  273.000  389.865
\LT  274.000  389.125
\LT  275.000  388.411
\LT  276.000  387.723
\LT  277.000  387.061
\LT  278.000  386.423
\LT  279.000  385.811
\LT  280.000  385.222
\LT  281.000  384.657
\LT  282.000  384.115
\LT  283.000  383.596
\LT  284.000  383.099
\LT  285.000  382.624
\LT  286.000  382.170
\LT  287.000  381.736
\LT  289.000  380.930
\LT  291.000  380.202
\LT  293.000  379.548
\LT  295.000  378.966
\LT  297.000  378.452
\LT  299.000  378.002
\LT  301.000  377.615
\LT  303.000  377.287
\LT  305.000  377.016
\LT  307.000  376.798
\LT  309.000  376.632
\LT  311.000  376.514
\LT  313.000  376.443
\LT  315.000  376.416
\LT  317.000  376.430
\LT  320.000  376.526
\LT  323.000  376.705
\LT  326.000  376.960
\LT  329.000  377.284
\LT  332.000  377.674
\LT  335.000  378.122
\LT  338.000  378.625
\LT  341.000  379.177
\LT  345.000  379.982
\LT  349.000  380.858
\LT  353.000  381.794
\LT  358.000  383.038
\LT  363.000  384.351
\LT  369.000  385.997
\LT  376.000  387.992
\LT  386.000  390.928
\LT  407.000  397.166
\LT  417.000  400.070
\LT  425.000  402.329
\LT  433.000  404.516
\LT  440.000  406.361
\LT  447.000  408.136
\LT  454.000  409.836
\LT  460.000  411.229
\LT  466.000  412.560
\LT  472.000  413.829
\LT  478.000  415.033
\LT  484.000  416.172
\LT  490.000  417.243
\LT  496.000  418.246
\LT  502.000  419.180
\LT  508.000  420.045
\LT  514.000  420.841
\LT  520.000  421.567
\LT  526.000  422.223
\LT  532.000  422.809
\LT  538.000  423.325
\LT  544.000  423.771
\LT  550.000  424.148
\LT  556.000  424.456
\LT  562.000  424.694
\LT  568.000  424.864
\LT  574.000  424.966
\LT  580.000  425.000
\LT  586.000  424.967
\LT  592.000  424.866
\LT  598.000  424.700
\LT  604.000  424.467
\LT  610.000  424.169
\LT  616.000  423.806
\LT  622.000  423.379
\LT  628.000  422.888
\LT  634.000  422.333
\LT  640.000  421.716
\LT  646.000  421.037
\LT  652.000  420.296
\LT  658.000  419.494
\LT  664.000  418.631
\LT  671.000  417.549
\LT  678.000  416.386
\LT  685.000  415.142
\LT  692.000  413.819
\LT  699.000  412.418
\LT  706.000  410.939
\LT  713.000  409.383
\LT  720.000  407.750
\LT  727.000  406.043
\LT  734.000  404.260
\LT  741.000  402.403
\LT  748.000  400.473
\LT  755.000  398.470
\LT  762.000  396.395
\LT  769.000  394.249
\LT  776.000  392.032
\LT  783.000  389.744
\LT  790.000  387.387
\LT  797.000  384.962
\LT  804.000  382.468
\LT  811.000  379.906
\LT  818.000  377.276
\LT  825.000  374.580
\LT  832.000  371.818
\LT  839.000  368.990
\LT  846.000  366.097
\LT  853.000  363.139
\LT  860.000  360.117
\LT  867.000  357.032
\LT  874.000  353.883
\LT  881.000  350.671
\LT  888.000  347.396
\LT  895.000  344.060
\LT  902.000  340.662
\LT  910.000  336.703
\LT  918.000  332.665
\LT  926.000  328.549
\LT  934.000  324.354
\LT  942.000  320.081
\LT  950.000  315.731
\LT  958.000  311.304
\LT  966.000  306.801
\LT  974.000  302.223
\LT  982.000  297.569
\LT  990.000  292.840
\LT  998.000  288.037
\LT 1006.000  283.160
\LT 1014.000  278.209
\LT 1022.000  273.185
\LT 1030.000  268.088
\LT 1038.000  262.919
\LT 1046.000  257.678
\LT 1054.000  252.365
\LT 1062.000  246.981
\LT 1070.000  241.526
\LT 1078.000  236.000
\LT 1086.000  230.405
\LT 1094.000  224.739
\LT 1102.000  219.003
\LT 1110.000  213.198
\LT 1118.000  207.324
\LT 1126.000  201.381
\LT 1134.000  195.370
\LT 1142.000  189.290
\LT 1150.000  183.143
\LT 1158.000  176.927
\LT 1166.000  170.644
\LT 1174.000  164.295
\LT 1182.000  157.878
\LT 1190.000  151.394
\LT 1198.000  144.844
\LT 1206.000  138.227
\LT 1214.000  131.544
\LT 1222.000  124.796
\LT 1230.000  117.982
\LT 1238.000  111.102
\LT 1246.000  104.157
\LT 1254.000   97.147
\LT 1262.000   90.072
\LT 1270.000   82.933
\LT 1278.000   75.729
\LT 1286.000   68.460
\LT 1294.000   61.128
\LT 1300.000   55.586
\koniec     3.4281 -0.0163862
\obrazek47
\grub0.2pt
\MT   0.000  500.000
\LT1400.000  500.000
\MT 160.000  500.000
\LT 160.000  510.000
\MT 220.000  500.000
\LT 220.000  510.000
\MT 280.000  500.000
\LT 280.000  510.000
\MT 340.000  500.000
\LT 340.000  510.000
\cput(340.000,440.000,2)
\MT 400.000  500.000
\LT 400.000  510.000
\MT 460.000  500.000
\LT 460.000  510.000
\MT 520.000  500.000
\LT 520.000  510.000
\MT 580.000  500.000
\LT 580.000  510.000
\cput(580.000,440.000,4)
\MT 640.000  500.000
\LT 640.000  510.000
\MT 700.000  500.000
\LT 700.000  510.000
\MT 760.000  500.000
\LT 760.000  510.000
\MT 820.000  500.000
\LT 820.000  510.000
\cput(820.000,440.000,6)
\MT 880.000  500.000
\LT 880.000  510.000
\MT 940.000  500.000
\LT 940.000  510.000
\MT1000.000  500.000
\LT1000.000  510.000
\MT1060.000  500.000
\LT1060.000  510.000
\cput(1060.000,440.000,8)
\MT1120.000  500.000
\LT1120.000  510.000
\MT1180.000  500.000
\LT1180.000  510.000
\MT1240.000  500.000
\LT1240.000  510.000
\MT1300.000  500.000
\LT1300.000  510.000
\cput(1300.000,440.000,10)
\MT 100.000    0.000
\LT 100.000  850.000
\MT  92.000   50.000
\LT 108.000   50.000
\MT  92.000   80.000
\LT 108.000   80.000
\MT  92.000  110.000
\LT 108.000  110.000
\MT  92.000  140.000
\LT 108.000  140.000
\MT  92.000  170.000
\LT 108.000  170.000
\MT  92.000  200.000
\LT 108.000  200.000
\MT  92.000  230.000
\LT 108.000  230.000
\MT  92.000  260.000
\LT 108.000  260.000
\MT  92.000  290.000
\LT 108.000  290.000
\MT  92.000  320.000
\LT 108.000  320.000
\MT  92.000  350.000
\LT 108.000  350.000
\MT  92.000  380.000
\LT 108.000  380.000
\MT  92.000  410.000
\LT 108.000  410.000
\MT  92.000  440.000
\LT 108.000  440.000
\MT  92.000  470.000
\LT 108.000  470.000
\MT  92.000  500.000
\LT 108.000  500.000
\MT  92.000  530.000
\LT 108.000  530.000
\MT  92.000  560.000
\LT 108.000  560.000
\MT  92.000  590.000
\LT 108.000  590.000
\MT  92.000  620.000
\LT 108.000  620.000
\MT  92.000  650.000
\LT 108.000  650.000
\MT  92.000  680.000
\LT 108.000  680.000
\MT  92.000  710.000
\LT 108.000  710.000
\MT  92.000  740.000
\LT 108.000  740.000
\MT  92.000  770.000
\LT 108.000  770.000
\MT  92.000  800.000
\LT 108.000  800.000
\MT  84.000   50.000
\LT 116.000   50.000
\lput(80.000,35.000, -1.5)
\MT  84.000  200.000
\LT 116.000  200.000
\lput(80.000,185.000, -1.0)
\MT  84.000  350.000
\LT 116.000  350.000
\lput(80.000,335.000, -0.5)
\MT  84.000  500.000
\LT 116.000  500.000
\MT  84.000  650.000
\LT 116.000  650.000
\lput(80.000,635.000,  0.5)
\MT  84.000  800.000
\LT 116.000  800.000
\lput(80.000,785.000,  1.0)
\grub0.6pt
\MT  100.000  800.000
\LT  101.000  799.979
\LT  102.000  799.916
\LT  103.000  799.811
\LT  104.000  799.663
\LT  105.000  799.474
\LT  106.000  799.242
\LT  107.000  798.969
\LT  108.000  798.653
\LT  109.000  798.296
\LT  110.000  797.896
\LT  111.000  797.454
\LT  112.000  796.970
\LT  113.000  796.444
\LT  114.000  795.876
\LT  115.000  795.266
\LT  116.000  794.614
\LT  117.000  793.920
\LT  118.000  793.184
\LT  119.000  792.406
\LT  120.000  791.586
\LT  121.000  790.724
\LT  122.000  789.820
\LT  123.000  788.875
\LT  124.000  787.888
\LT  125.000  786.859
\LT  126.000  785.789
\LT  127.000  784.677
\LT  128.000  783.523
\LT  129.000  782.329
\LT  130.000  781.093
\LT  131.000  779.816
\LT  132.000  778.498
\LT  133.000  777.140
\LT  134.000  775.741
\LT  135.000  774.301
\LT  136.000  772.822
\LT  137.000  771.302
\LT  138.000  769.743
\LT  139.000  768.144
\LT  140.000  766.506
\LT  141.000  764.829
\LT  142.000  763.113
\LT  143.000  761.360
\LT  144.000  759.568
\LT  145.000  757.739
\LT  146.000  755.873
\LT  147.000  753.970
\LT  148.000  752.032
\LT  149.000  750.057
\LT  150.000  748.048
\LT  151.000  746.004
\LT  152.000  743.926
\LT  153.000  741.814
\LT  154.000  739.671
\LT  155.000  737.495
\LT  156.000  735.287
\LT  157.000  733.049
\LT  158.000  730.782
\LT  159.000  728.485
\LT  160.000  726.160
\LT  161.000  723.808
\LT  162.000  721.429
\LT  163.000  719.025
\LT  164.000  716.596
\LT  165.000  714.144
\LT  166.000  711.669
\LT  167.000  709.172
\LT  169.000  704.119
\LT  171.000  698.992
\LT  173.000  693.802
\LT  175.000  688.556
\LT  178.000  680.608
\LT  182.000  669.911
\LT  188.000  653.820
\LT  191.000  645.829
\LT  193.000  640.545
\LT  195.000  635.305
\LT  197.000  630.120
\LT  199.000  624.998
\LT  201.000  619.947
\LT  202.000  617.450
\LT  203.000  614.975
\LT  204.000  612.520
\LT  205.000  610.088
\LT  206.000  607.680
\LT  207.000  605.295
\LT  208.000  602.935
\LT  209.000  600.600
\LT  210.000  598.291
\LT  211.000  596.009
\LT  212.000  593.754
\LT  213.000  591.527
\LT  214.000  589.328
\LT  215.000  587.158
\LT  216.000  585.017
\LT  217.000  582.906
\LT  218.000  580.825
\LT  219.000  578.774
\LT  220.000  576.753
\LT  221.000  574.764
\LT  222.000  572.805
\LT  223.000  570.878
\LT  224.000  568.982
\LT  225.000  567.117
\LT  226.000  565.284
\LT  227.000  563.483
\LT  228.000  561.714
\LT  229.000  559.976
\LT  230.000  558.270
\LT  231.000  556.595
\LT  232.000  554.953
\LT  233.000  553.341
\LT  234.000  551.761
\LT  235.000  550.213
\LT  236.000  548.695
\LT  237.000  547.209
\LT  238.000  545.753
\LT  239.000  544.327
\LT  240.000  542.932
\LT  241.000  541.567
\LT  242.000  540.232
\LT  243.000  538.927
\LT  244.000  537.650
\LT  245.000  536.403
\LT  246.000  535.184
\LT  247.000  533.994
\LT  248.000  532.832
\LT  249.000  531.697
\LT  250.000  530.590
\LT  251.000  529.510
\LT  252.000  528.457
\LT  253.000  527.429
\LT  254.000  526.428
\LT  255.000  525.453
\LT  256.000  524.502
\LT  257.000  523.577
\LT  258.000  522.676
\LT  259.000  521.799
\LT  260.000  520.946
\LT  261.000  520.116
\LT  262.000  519.310
\LT  263.000  518.525
\LT  264.000  517.763
\LT  265.000  517.023
\LT  266.000  516.305
\LT  267.000  515.607
\LT  268.000  514.930
\LT  270.000  513.637
\LT  272.000  512.422
\LT  274.000  511.283
\LT  276.000  510.216
\LT  278.000  509.219
\LT  280.000  508.288
\LT  282.000  507.423
\LT  284.000  506.618
\LT  286.000  505.873
\LT  288.000  505.184
\LT  290.000  504.549
\LT  292.000  503.966
\LT  294.000  503.432
\LT  296.000  502.946
\LT  298.000  502.504
\LT  300.000  502.106
\LT  303.000  501.584
\LT  306.000  501.147
\LT  309.000  500.791
\LT  312.000  500.507
\LT  315.000  500.292
\LT  318.000  500.139
\LT  321.000  500.044
\LT  324.000  500.003
\LT  327.000  500.010
\LT  331.000  500.089
\LT  335.000  500.238
\LT  339.000  500.450
\LT  344.000  500.790
\LT  349.000  501.203
\LT  355.000  501.775
\LT  362.000  502.525
\LT  370.000  503.461
\LT  383.000  505.079
\LT  401.000  507.344
\LT  411.000  508.543
\LT  420.000  509.556
\LT  428.000  510.391
\LT  436.000  511.156
\LT  443.000  511.761
\LT  450.000  512.302
\LT  457.000  512.776
\LT  464.000  513.180
\LT  471.000  513.512
\LT  478.000  513.770
\LT  485.000  513.952
\LT  492.000  514.057
\LT  499.000  514.085
\LT  506.000  514.034
\LT  513.000  513.904
\LT  520.000  513.695
\LT  527.000  513.406
\LT  534.000  513.038
\LT  541.000  512.591
\LT  548.000  512.065
\LT  555.000  511.460
\LT  562.000  510.776
\LT  569.000  510.014
\LT  576.000  509.175
\LT  583.000  508.258
\LT  590.000  507.265
\LT  597.000  506.195
\LT  604.000  505.050
\LT  611.000  503.829
\LT  618.000  502.534
\LT  625.000  501.166
\LT  632.000  499.723
\LT  639.000  498.208
\LT  646.000  496.621
\LT  653.000  494.963
\LT  660.000  493.233
\LT  667.000  491.433
\LT  674.000  489.563
\LT  681.000  487.623
\LT  688.000  485.615
\LT  695.000  483.539
\LT  702.000  481.395
\LT  709.000  479.185
\LT  716.000  476.907
\LT  723.000  474.564
\LT  730.000  472.155
\LT  737.000  469.681
\LT  744.000  467.142
\LT  751.000  464.539
\LT  758.000  461.873
\LT  765.000  459.144
\LT  772.000  456.351
\LT  779.000  453.497
\LT  786.000  450.581
\LT  794.000  447.173
\LT  802.000  443.685
\LT  810.000  440.118
\LT  818.000  436.473
\LT  826.000  432.749
\LT  834.000  428.949
\LT  842.000  425.071
\LT  850.000  421.117
\LT  858.000  417.088
\LT  866.000  412.982
\LT  874.000  408.802
\LT  882.000  404.548
\LT  890.000  400.220
\LT  898.000  395.818
\LT  906.000  391.342
\LT  914.000  386.794
\LT  922.000  382.174
\LT  930.000  377.482
\LT  938.000  372.718
\LT  946.000  367.883
\LT  954.000  362.977
\LT  962.000  358.000
\LT  970.000  352.954
\LT  978.000  347.837
\LT  986.000  342.651
\LT  994.000  337.395
\LT 1002.000  332.071
\LT 1010.000  326.678
\LT 1018.000  321.217
\LT 1026.000  315.687
\LT 1034.000  310.090
\LT 1042.000  304.425
\LT 1050.000  298.693
\LT 1058.000  292.893
\LT 1066.000  287.027
\LT 1074.000  281.094
\LT 1082.000  275.095
\LT 1090.000  269.030
\LT 1098.000  262.899
\LT 1106.000  256.702
\LT 1114.000  250.440
\LT 1122.000  244.112
\LT 1130.000  237.719
\LT 1138.000  231.261
\LT 1146.000  224.739
\LT 1154.000  218.152
\LT 1162.000  211.500
\LT 1170.000  204.785
\LT 1179.000  197.153
\LT 1188.000  189.440
\LT 1197.000  181.647
\LT 1206.000  173.774
\LT 1215.000  165.820
\LT 1224.000  157.787
\LT 1233.000  149.673
\LT 1242.000  141.481
\LT 1251.000  133.208
\LT 1260.000  124.857
\LT 1269.000  116.427
\LT 1278.000  107.917
\LT 1287.000   99.330
\LT 1296.000   90.663
\LT 1300.000   86.786
\koniec     2.9677 -0.0207742
\obrazek48
\grub0.2pt
\MT   0.000  550.000
\LT1400.000  550.000
\MT 160.000  550.000
\LT 160.000  560.000
\MT 220.000  550.000
\LT 220.000  560.000
\MT 280.000  550.000
\LT 280.000  560.000
\MT 340.000  550.000
\LT 340.000  560.000
\cput(340.000,490.000,2)
\MT 400.000  550.000
\LT 400.000  560.000
\MT 460.000  550.000
\LT 460.000  560.000
\MT 520.000  550.000
\LT 520.000  560.000
\MT 580.000  550.000
\LT 580.000  560.000
\cput(580.000,490.000,4)
\MT 640.000  550.000
\LT 640.000  560.000
\MT 700.000  550.000
\LT 700.000  560.000
\MT 760.000  550.000
\LT 760.000  560.000
\MT 820.000  550.000
\LT 820.000  560.000
\cput(820.000,490.000,6)
\MT 880.000  550.000
\LT 880.000  560.000
\MT 940.000  550.000
\LT 940.000  560.000
\MT1000.000  550.000
\LT1000.000  560.000
\MT1060.000  550.000
\LT1060.000  560.000
\cput(1060.000,490.000,8)
\MT1120.000  550.000
\LT1120.000  560.000
\MT1180.000  550.000
\LT1180.000  560.000
\MT1240.000  550.000
\LT1240.000  560.000
\MT1300.000  550.000
\LT1300.000  560.000
\cput(1300.000,490.000,10)
\MT 100.000    0.000
\LT 100.000  850.000
\MT  92.000   50.000
\LT 108.000   50.000
\MT  92.000  100.000
\LT 108.000  100.000
\MT  92.000  150.000
\LT 108.000  150.000
\MT  92.000  200.000
\LT 108.000  200.000
\MT  92.000  250.000
\LT 108.000  250.000
\MT  92.000  300.000
\LT 108.000  300.000
\MT  92.000  350.000
\LT 108.000  350.000
\MT  92.000  400.000
\LT 108.000  400.000
\MT  92.000  450.000
\LT 108.000  450.000
\MT  92.000  500.000
\LT 108.000  500.000
\MT  92.000  550.000
\LT 108.000  550.000
\MT  92.000  600.000
\LT 108.000  600.000
\MT  92.000  650.000
\LT 108.000  650.000
\MT  92.000  700.000
\LT 108.000  700.000
\MT  92.000  750.000
\LT 108.000  750.000
\MT  92.000  800.000
\LT 108.000  800.000
\MT  84.000   50.000
\LT 116.000   50.000
\lput(80.000,25.000, -2.0)
\MT  84.000  300.000
\LT 116.000  300.000
\lput(80.000,275.000, -1.0)
\MT  84.000  550.000
\LT 116.000  550.000
\MT  84.000  800.000
\LT 116.000  800.000
\lput(80.000,775.000,  1.0)
\grub0.6pt
\MT  100.000  800.000
\LT  101.000  799.983
\LT  102.000  799.932
\LT  103.000  799.847
\LT  104.000  799.728
\LT  105.000  799.575
\LT  106.000  799.388
\LT  107.000  799.168
\LT  108.000  798.913
\LT  109.000  798.624
\LT  110.000  798.301
\LT  111.000  797.945
\LT  112.000  797.554
\LT  113.000  797.129
\LT  114.000  796.671
\LT  115.000  796.178
\LT  116.000  795.652
\LT  117.000  795.091
\LT  118.000  794.497
\LT  119.000  793.869
\LT  120.000  793.207
\LT  121.000  792.511
\LT  122.000  791.782
\LT  123.000  791.018
\LT  124.000  790.222
\LT  125.000  789.391
\LT  126.000  788.527
\LT  127.000  787.629
\LT  128.000  786.698
\LT  129.000  785.733
\LT  130.000  784.736
\LT  131.000  783.705
\LT  132.000  782.641
\LT  133.000  781.544
\LT  134.000  780.414
\LT  135.000  779.252
\LT  136.000  778.057
\LT  137.000  776.830
\LT  138.000  775.571
\LT  139.000  774.280
\LT  140.000  772.958
\LT  141.000  771.604
\LT  142.000  770.218
\LT  143.000  768.802
\LT  144.000  767.356
\LT  145.000  765.879
\LT  146.000  764.372
\LT  147.000  762.835
\LT  148.000  761.269
\LT  149.000  759.675
\LT  150.000  758.052
\LT  151.000  756.401
\LT  152.000  754.723
\LT  153.000  753.018
\LT  154.000  751.286
\LT  155.000  749.528
\LT  156.000  747.745
\LT  157.000  745.937
\LT  158.000  744.105
\LT  159.000  742.249
\LT  160.000  740.371
\LT  161.000  738.470
\LT  162.000  736.548
\LT  164.000  732.642
\LT  166.000  728.660
\LT  168.000  724.607
\LT  170.000  720.490
\LT  172.000  716.318
\LT  174.000  712.096
\LT  177.000  705.689
\LT  180.000  699.216
\LT  189.000  679.677
\LT  192.000  673.214
\LT  195.000  666.822
\LT  197.000  662.611
\LT  199.000  658.449
\LT  201.000  654.342
\LT  203.000  650.297
\LT  205.000  646.319
\LT  207.000  642.414
\LT  209.000  638.586
\LT  210.000  636.702
\LT  211.000  634.839
\LT  212.000  632.998
\LT  213.000  631.179
\LT  214.000  629.381
\LT  215.000  627.607
\LT  216.000  625.855
\LT  217.000  624.127
\LT  218.000  622.422
\LT  219.000  620.741
\LT  220.000  619.084
\LT  221.000  617.451
\LT  222.000  615.843
\LT  223.000  614.259
\LT  224.000  612.700
\LT  225.000  611.166
\LT  226.000  609.657
\LT  227.000  608.173
\LT  228.000  606.713
\LT  229.000  605.279
\LT  230.000  603.869
\LT  231.000  602.485
\LT  232.000  601.125
\LT  233.000  599.790
\LT  234.000  598.480
\LT  235.000  597.195
\LT  236.000  595.934
\LT  237.000  594.697
\LT  238.000  593.485
\LT  239.000  592.296
\LT  240.000  591.131
\LT  241.000  589.991
\LT  242.000  588.873
\LT  243.000  587.779
\LT  244.000  586.708
\LT  245.000  585.659
\LT  246.000  584.633
\LT  247.000  583.630
\LT  248.000  582.648
\LT  249.000  581.688
\LT  250.000  580.750
\LT  251.000  579.833
\LT  252.000  578.937
\LT  253.000  578.062
\LT  254.000  577.207
\LT  256.000  575.557
\LT  258.000  573.984
\LT  260.000  572.488
\LT  262.000  571.065
\LT  264.000  569.712
\LT  266.000  568.428
\LT  268.000  567.209
\LT  270.000  566.055
\LT  272.000  564.961
\LT  274.000  563.926
\LT  276.000  562.948
\LT  278.000  562.023
\LT  280.000  561.151
\LT  282.000  560.329
\LT  284.000  559.555
\LT  286.000  558.826
\LT  288.000  558.141
\LT  290.000  557.497
\LT  293.000  556.606
\LT  296.000  555.799
\LT  299.000  555.069
\LT  302.000  554.412
\LT  305.000  553.821
\LT  308.000  553.291
\LT  311.000  552.819
\LT  314.000  552.400
\LT  317.000  552.028
\LT  321.000  551.601
\LT  325.000  551.244
\LT  329.000  550.949
\LT  334.000  550.655
\LT  339.000  550.432
\LT  345.000  550.242
\LT  352.000  550.103
\LT  360.000  550.023
\LT  373.000  549.990
\LT  391.000  549.970
\LT  402.000  549.891
\LT  411.000  549.755
\LT  419.000  549.569
\LT  427.000  549.310
\LT  434.000  549.020
\LT  441.000  548.665
\LT  448.000  548.243
\LT  455.000  547.750
\LT  462.000  547.184
\LT  469.000  546.543
\LT  476.000  545.825
\LT  483.000  545.030
\LT  490.000  544.156
\LT  497.000  543.202
\LT  504.000  542.168
\LT  511.000  541.054
\LT  518.000  539.858
\LT  525.000  538.582
\LT  532.000  537.224
\LT  539.000  535.786
\LT  546.000  534.266
\LT  553.000  532.666
\LT  560.000  530.985
\LT  567.000  529.224
\LT  574.000  527.384
\LT  581.000  525.464
\LT  588.000  523.464
\LT  595.000  521.386
\LT  602.000  519.230
\LT  609.000  516.996
\LT  616.000  514.685
\LT  623.000  512.297
\LT  630.000  509.833
\LT  637.000  507.292
\LT  644.000  504.676
\LT  651.000  501.985
\LT  658.000  499.220
\LT  665.000  496.381
\LT  672.000  493.468
\LT  679.000  490.481
\LT  686.000  487.422
\LT  693.000  484.291
\LT  700.000  481.088
\LT  707.000  477.814
\LT  714.000  474.469
\LT  721.000  471.053
\LT  728.000  467.567
\LT  735.000  464.012
\LT  742.000  460.387
\LT  749.000  456.693
\LT  756.000  452.930
\LT  763.000  449.099
\LT  770.000  445.201
\LT  777.000  441.234
\LT  784.000  437.201
\LT  791.000  433.101
\LT  798.000  428.934
\LT  805.000  424.701
\LT  812.000  420.402
\LT  819.000  416.038
\LT  826.000  411.608
\LT  833.000  407.114
\LT  840.000  402.554
\LT  847.000  397.931
\LT  854.000  393.243
\LT  861.000  388.491
\LT  868.000  383.675
\LT  875.000  378.796
\LT  882.000  373.854
\LT  889.000  368.849
\LT  896.000  363.781
\LT  903.000  358.651
\LT  910.000  353.458
\LT  917.000  348.204
\LT  924.000  342.887
\LT  931.000  337.509
\LT  938.000  332.070
\LT  946.000  325.778
\LT  954.000  319.407
\LT  962.000  312.956
\LT  970.000  306.427
\LT  978.000  299.818
\LT  986.000  293.131
\LT  994.000  286.365
\LT 1002.000  279.521
\LT 1010.000  272.599
\LT 1018.000  265.600
\LT 1026.000  258.523
\LT 1034.000  251.369
\LT 1042.000  244.137
\LT 1050.000  236.829
\LT 1058.000  229.444
\LT 1066.000  221.983
\LT 1074.000  214.446
\LT 1082.000  206.832
\LT 1090.000  199.142
\LT 1098.000  191.377
\LT 1106.000  183.536
\LT 1114.000  175.620
\LT 1122.000  167.628
\LT 1130.000  159.562
\LT 1138.000  151.420
\LT 1146.000  143.204
\LT 1154.000  134.913
\LT 1162.000  126.547
\LT 1170.000  118.107
\LT 1178.000  109.593
\LT 1186.000  101.005
\LT 1194.000   92.343
\LT 1202.000   83.607
\LT 1210.000   74.797
\LT 1218.000   65.913
\LT 1226.000   56.957
\LT 1234.000   47.926
\LT 1242.000   38.823
\LT 1250.000   29.646
\LT 1258.000   20.396
\LT 1266.000   11.073
\LT 1274.000    1.678
\LT 1282.000   -7.791
\LT 1290.000  -17.332
\LT 1298.000  -26.945
\LT 1300.000  -29.360
\koniec     2.8446 -0.0302993
 }

The points $x_{\crt}\approx2.30331$, $a_{\crt}\approx2.8446$ correspond
to the minimum of the \f\ $b(x)$ defined by the first \e\ \eqref{22.16}.

In Figures 13, 14 we give plots for $x_{\crt}(b)$ and $a_{\crt}(b)$ for $b>0$
and for $b<0$. In the case of $b<0$ we have two branches $x_{\crt}$ and
$a_{\crt}$. In the same figure we give also plots of the function $f$ for
some interesting values of $a$ and $b$ ($b<0$).

}
\vfill\eject

{\def\ap{approximation}
\def\cfn{confinement}
\def\ce{contemporary epoch}
\def\cn{connection}
\def\ct{constant}
\def\cd{coordinate}
\def\co{cosmological}
\def\dc{dependen}
\def\di{dimension}
\def\ef{effective}
\def\elm{electromagnetic}
\def\e{equation}
\def\ex{exponential}
\def\gr{gravitational}
\def\il{inflation}
\def\iv{invariant}
\def\lg{lagrangian}
\def\nos{nonsymmetric}
\def\pt{parameter}
\def\pc{particle}
\def\pt{potential}
\def\q{quint\-essence}
\def\SS{Solar System}
\def\so{solution}
\def\spt{space-time}
\def\sy{symmetr}
\def\s{symmetric}
\def\tr{transform}
\def\Un{Universe}
\def\U{\text{U}}
\def\wrt{with respect to }
\def\KK{Kaluza--Klein}
\def\JT{Jordan--Thiry}
\def\nos{nonsymmetric}
\def\NK#1{Nonsymmetric Kaluza--Klein #1Theory}
\def\up#1{\uppercase{#1}}
\def\eu{\expandafter\up}
\let\P\varPsi
\def\gd #1,#2,#3,{{{#1}^{#2}}_{\!#3}}
\def\(#1){{(#1)}}
\def\[#1]{{[#1]}}
\let\G\varGamma
\let\Ps\varPsi
\let\la\lambda
\let\b\beta
\let\a\alpha
\def\U{\text{U}}
\def\Pl{Planck}

\null
\vskip36pt plus4pt minus4pt
\centerline{{\four\bf Conclusions}}
\vskip24pt plus2pt minus2pt
\write0{Conclusions \the\pageno}

The \nos\ \KK\ (\JT) Theory has been developed. It unifies the gauge invariance
principle with the coordinate invariance principle but in more than
four-dimensional space-time. In particular in the case of electromagnetic and
\gr\ interactions in 5-dimensional theory.

A general nonabelian Yang--Mills fields have been unified with gravity in
$(n+4)$-\di al space-time ($n$---a \di\ of gauge group).
The theory uses a non\s\ metric defined on a metrized (in a non\s\
way) principal fibre bundle over a \spt\ with a structural group $\U(1)$ in an
\elm\ case and in general case nonabelian semi-simple compact group~$G$. The
\cn\ on \spt\ and on a metrized principal fibre bundle is compatible with
this metric. This \cn\ is similar to a \cn\ from Einstein's Unified Field
Theory, however we use its higher \di al analogue. This \cn\ is
right-\iv\ \wrt an action of the group~$G$ (a~gauge group). 

In the \elm\ case the metric and the \cn\ are bi-\iv\ \wrt the group~$\U(1)$.
The theory has been developed to
include a scalar field leading to an effective \gr\ constant and \spt\
dependent cosmological terms. It is possible to extend the theory to include
Higgs' fields and spontaneous symmetry breaking of a gauge group to get
massive vector boson fields. Some exact \so\ of the fields equations has been
found and a proposition  to solve a \cfn\ of colour in QCD has
been posed.

The theory is fully relativistic and unifies \elm\ field, gauge fields,
Higgs' field and scalar forces with NGT (Non\s\ Gravitation Theory) 
in a nontrivial way. 
By `in a nontrivial way' we mean that we get from the theory
something more than NGT, ordinary Kaluza--Klein (\JT) Theory, classical
electrodynamics, Yang--Mills' field theory with Higgs' field and spontaneous
symmetry breaking. These new features are some kind of ``interference
effects'' between all of them. 
This theory unifies two important approaches in higher-dimensional
philosophy: Kaluza--Klein principle and a \di al reduction principle.

The beautiful theories such as Kaluza--Klein theory (a~Kaluza miracle) and
its descendents should pass the following test if they are treated as real
unified theories. They should incorporate chiral fermions. Since the
fundamental scale in the theory is a Planck's mass, fermions should be
massless up to the moment of spontaneous symmetry breaking. Thus they should
be zero modes. In our approach they can obtain masses on a \di al reduction
scale. Thus they are zero modes in $(4+n_1)$-\di al case. In this way
$(n_1+4)$-\di al fermions are not chiral (according to very well known
Witten's argument on an index of a Dirac operator). Moreover, they are
not zero modes after a \di al reduction, i.e., in 4-\di al case. It means we
can get chiral fermions under some assumptions.

We expect some nonrelativistic effects leading
to nonnewtonian gravity. 

In this work we consider the \eu\nos\ \KK\ (\JT) Theory with spontaneous \sy
y breaking and Higgs' mechanism with a scalar field~$\P$ (see Introduction
for a short description). We get \il\ from this theory with a \q\ and several
testable \co\ predictions. 
We find a dynamical model for a \co\ \ct.
In the models of the \Un\ we get several phase
transitions of the second and of the first order. The real source of a \q\ is
a scalar field~$\P$. Due to its very unusual selfinteraction potential
(coming from higher dimensions) it can proceed to the very unexpectable
features. Of course this is not the end of the story. This is really a
beginning. 

Especially we should develop a super\sy ic (super\gr) extension of our
theory. In order to do this it is necessary to use a general formalism of
supermanifolds [141] and super-Lie algebra theory [142,~143]. We should develop
Einstein-Kaufmann connections on supermanifolds and super-Yang-Mills fields
based on super-Lie group. From physical point of view the formalism by P.~Nath
and R.~Arnowitt [144] will be very helpful. On the other side we can use also
a theory of symplectic structure on homogeneous spaces [145].

In that way we want to connect our approach to classical approach with
super\sy y and supergravity [146] with a geometric formulation [147]
geometrizing fermion fields.  Perhaps our approach is one of the phases  of
M-theory (see [148]) as a low energy theory. 

There are some further prospects:

\item{1.} To find conditions for \cfn\ (of colour) in a nonabelian version in
the theory.
\item{2.} To find gravito--Yang--Mills waves.
\item{3.} To find spherical and cylindrical waves in the theory.

\def\gr{gravitation}
Finally, we give some remarks. There are some misunderstandings connecting
Kaluza--Klein Theory, Einstein's Unified Field Theory, \eu\nos\ \eu\gr\
Theory (NGT), \eu\nos\ Kaluza--Klein Theory (NKKT), \eu\nos\ Jordan--Thiry
Theory (NJTT).

\def\elm{electromagneti}
{\bf1.} First of all we comment a \ct\ $\la=\frac{2\sqrt{G_N}}{c^2}$. The \ct\ $\la$
appeared as a free parameter in this theory. Moreover in order to get
Einstein \e s with \elm c sources known from GR it is fixed and it is not free
any more. Why is there not a Planck's length?
I explain it shortly. The Kaluza theory is classical for a paper published by him
is classical as a classical paper in the scientific literature. It is also classical
for this theory is not quantum. For this we cannot get here a Planck's \ct.
This is simply for we need a Planck's \ct\ in order to construct the Planck's
length. \Pl's \ct\ is absent in Kaluza theory for this theory is classical
(non-quantum). The \Pl's length appeared in the further development done by
O.~Klein. O.~Klein considered a Klein--Gordon \e\ in 5-dimensional extension.
The \Pl's \ct\ is present in Klein--Gordon \e.
This \e\ can be considered as an \e\ for a classical scalar field. In
Kaluza--Klein theory \Pl's length appears as a scale of length.

\def\tr{transformation}
{\bf2.}
The classical Kaluza theory has been abandoned by 1950's. Moreover,
due to some mathematical investigations a deep structure has been discovered
behind the theory. Let me describe it shortly. First of all it happens that
behind Maxwell theory of \elm sm there is a principal fibre bundle over a
\spt\ with a structural group $\U(1)$ and a \cn\ defined on this bundle is an
\elm c field. Gauge \tr, four-\pt, the first pair of Maxwell \e\ obtained a
clear geometrical meaning in terms of a fibre bundle approach.

It happens also that a classical Kaluza theory is a theory of metrized (in a
natural way) \elm c fibre bundle (see Ref.~[51]). 

This is a true unification of the two fundamental principles of invariance in
physics: a gauge invariance principle and a \cd\  invariance principle, as we
mention in Section~1.

In Section 2 of Ref.~[16] a classical KKT in this setting has been described
(see also the last two lines of page 576 with a fixing of the \ct~$\la$).

Moreover this paper is devoted to the KKT with torsion in such a way that we
put in the place of GR the Einstein--Cartan theory obtaining new features the
so-called ``interference effects'' between gravity and \elm sm going to some
effects which are small, moreover in principle measurable in experiment.

{\bf3.}
Let us consider Einstein Unified Field Theory. A.~Einstein started this
theory in 1920's. In 1950 he came back to this theory describing it in
Appendix~II of the fifth edition of his famous book {\it The Meaning of
Relativity} (see Ref.~[3]).

It is worth to mention that there are many versions of this theory. The
oldest Einstein--Thomas theory and after that Einstein--Strauss theory,
Einstein--Kaufmann theory. There are also two approaches, weak and strong
field \e s. The Einstein Unified Field Theory can be also considered as a
real theory and Hermitian theory. A slight deviation is the so-called
Bonnor's Unified Field Theory. In all of these approaches there are two
fundamental notions: \nos\ affine \cn\
$\gd\G,\la,\mu\nu,\ne\gd\G,\la,\nu\mu,$ and the \nos\ metric $g_{\mu\nu}
\ne g_{\nu\mu}$. \eu\cn\ and metric can be real or Hermitian. In this theory
there is also a second \cn\ $\gd W,\la,\mu\nu,\ne\gd W,\la,\nu\mu,$. \eu\cn\ 
$\gd\G,\la,\mu\nu,$ is a so-called constrained \cn, $\gd W,\la,\mu\nu,$ is called
unconstrained. All of these approaches have no free parameters. Some
parameters which appear in \so s of field \e s are {\it integration \ct s}.

What was an aim to construct such theories? The aim was to find a unified
theory of gravity and \elm sm in such a way that GR and Maxwell theory appear
as some limit of the theory. This approach ended with fiasco. It was
impossible to obtain a Lorentz force.
It was impossible to obtain a Coulomb law too. 

One can find all references to all versions of Einstein Unified Field Theory
in Refs [4], [5], [6] and we will not quote them here. Moreover, it
is worth to mention that A.~Einstein considered this theory as a theory of an
extended \gr. Moreover, there is a reference of A.~Einstein's idea to treat
this theory as a theory of an extended gravity only. A.~Einstein published a
paper on it in {\it Scientific American} (the only one Einstein's paper in
this journal). 

Geometrical--mathematical properties of Einstein Unified Field Theory have
been described in a book by Vaclav Hlavat\'y (see Ref.~[43]).

In those times A.~Einstein started a program of geometrization of physics. Some
notions of this program have been described in Ref.~[149].

There is also an approach to this theory going in a different direction. It
has been summarized in the book by A.~H.~Klotz (see Ref.~[150].

{\bf 4.}
Let us comment NGT (\eu\nos\ \eu\gr al Theory) by J.~W.~Moffat. J.~W.~Moffat
reinterpreted Einstein Unified Field Theory as a theory of a pure \gr al
field (see Ref.~[34]). He introduced material sources to the formalism.
He and his co-workers developed this idea getting many interesting
results which are in principle testable by astronomical observations in the 
\SS\ and beyond. He was using both real and Hermitian theory. Simultaneously
he developed a formalism with two \cn s $\gd\G,\la,\mu\nu,$ and $\gd
W,\la,\mu\nu,$. I~refer to some of these papers.

{\bf5.}
Let us comment the \NK{and the \eu\nos\break Jordan--Thiry }. I~posed and
developed these theories using the \nos\ metrization of an \elm c fibre
bundle using differential forms formalism as in my paper (see Ref.~[16]).

Early results concerning the \NK{} have been published (see Refs [18] (the
first item), [25], [30]).

The final result of the theory with some developments has been published in
the fifth position of Ref.~[18]. The paper contains also an extension to the \eu\nos\ Jordan--Thiry
Theory with a scalar field $\Ps$ (or~$\rho$). In order to get a pure \NK{} it
is enough to put ${\Ps=0}$ (or~${\rho=1}$). All new features as some
``interference effects'' between \elm c fields and \gr\ have been quoted in
Introduction. The theory has no free parameters except integration \ct s in
\so s.

It is possible to get an extension of the theory to the non-Abelian case. In
this case we have one free parameter. Moreover, this parameter can be fixed
by a \co\ \ct. The final version of this theory can be found in
Ref.~[18] (the fourth and fifth items).  In Ref.~[18] (the second item) 
one can find also an extension to the case with
Higgs' field and spontaneous symmetry breaking. In the last case there are
three free parameters which can be fixed by a \co\ \ct\ and scales of
masses.

I do not refer in my work to the paper Ref.~[151], for the authors are using completely different approach (it is
better to say {\it three approaches}). This approach is far away from
investigations in my work. Moreover, in future both approaches can meet and we
will shake hands. The only one point which is now common is a starting point,
a classical Kaluza Theory. We do not refer to Ref.~[152].

This paper deals with some problems
in NGT. However, NGT considered by them has only a little touch with NGT
considered here. They introduced a mass for skew-\s\ field
$B_{\mu\nu}=-B_{\nu\mu}$ (in our notation it is $g_\[\mu\nu]$). Moreover, the
$g_\[\mu\nu]$ can obtain a mass in a linear \ap\ of \eu\nos\ Non-Abelian
Kaluza--Klein Theory due to a \co\ \ct\ and it is not necessary to
introduce a mass term. It seems that this is a completely different
approach (see Ref.~[152]). For a cure of NGT by a \co\ \ct\ see also
Ref.~[153].

Let us notice the following fact. Einstein's Unified Field Theory has been
abandoned for it has been proved using EIH (Einstein--Infeld--Hoffman) method
that there is not a Lorentz force term and Coulomb like law.

These are
disadvantages of Einstein Unified Field Theory but not NGT. This works now
for our advantage, for we do not see any term like Lorentz force and
Coulomb-like law in \gr al physics (I~do not mean a Newton \gr al law which
can be obtained in Einstein Unified Field Theory). 
Someone said: ``{\it it is clever to use advantages, moreover, more clever is
to use disadvantages''} and this is a case. Moreover, in the \NK{} we get
Lorentz force term from $(N+4)$-dimensional (5-dimensional in an \elm c case) 
geodetic \e s (see Ref.~[18] and Section 15). 

All additional notions in the \NK{\JT\ } have been described in Section~5. We
get from $(N+4)$-\di al theory ($N=n+n_1$) four-\di al \e s due to an
invariance of a \nos\ metric and a \cn\ \wrt the right invariance action of
the group (in the \elm c case this is a biinvariance of the group $\U(1)$).

Let us notice also the following fact. \eu\e s obtained in the \NK{\JT\ } are
different from these in pure NGT. Due to this we can obtain nonsingular \so s
of field \e s in the \elm c case. These \so s possess a nonsingular metric
$g_\(\a\b)$ and nonsingular electric field. The asymptotic behaviour is as in
the case of Reissner--Nordstr\"om \so\ (see Ref.~[18], the fifth item, and
Section~22). This is impossible to get in pure NGT.

\vfill\eject
}

{\def\eq#1 {\eqno(\text{\rm A.#1})}
\def\rf#1 {\text{\rm(A.#1)}}
\let\al\aligned
\let\eal\endaligned
\let\ealn\eqalignno
\let\dsl\displaylines
\def\cL{\Cal L}
\let\o\overline
\let\ul\underline
\let\stpr\overrightarrow
\let\a\alpha
\let\b\beta
\let\d\delta
\let\e\varepsilon
\let\G\varGamma
\let\om\omega
\let\F\varPhi
\let\la\lambda
\let\z\zeta
\let\t\widetilde
\let\u\tilde
\let\wh\widehat
\def\Ad{\operatorname{Ad}\nolimits}
\def\br#1#2{\overset\text{\rm #1}\to{#2}}
\def\brgn#1{\br{gauge}{\nabla_{#1}}}
\def\brg#1{\br{gauge}{\nabla^{#1}}}
\def\({\left(}\def\){\right)}
\def\[{{\textstyle[}}
\def\]{{\textstyle]}}
\def\dr#1{_{\text{\rm#1}}}
\def\cn{connection}
\def\co{cosmological}
\def\lg{lagrangian}
\def\sy{symmetr}
\def\wrt{with respect to }
\def\KK{Kaluza--Klein}
\def\JT{Jordan--Thiry}
\def\nos{nonsymmetric}
\def\up#1{\uppercase{#1}}
\def\eu{\expandafter\up}
\def\gtn#1{{\t g}^{(#1)}}
\def\gtk#1{g^{[#1]}}
\def\ink#1{_{[#1]}}
\def\dg#1,#2,{_{#1}{}^{#2}}
\def\gd#1,#2,{^{#1}{}_{\!#2}}
\def\ba{\bar a}

\def\bd{\bar d}
\def\be{\bar e}
\def\Bf{\bar f}
\def\bk{\bar k}
\def\bm{\bar m}
\def\bp{\bar p}

\def\bv{\bar v}
\def\bw{\bar w}
\def\bz{\bar z}
\def\hi{\hat\imath}\def\hj{\hat\jmath}

\null
\vskip36pt plus4pt minus4pt
\centerline{{\four\bf Appendix A}}
\vskip24pt plus2pt minus2pt
\write0{Appendix A \the\pageno}

In this Appendix we give some useful formulae in the \eu\nos\ \KK\ (\JT)
Theory. First of all we can solve explicitly Eq.~(5.22) getting
$$\ealn{
L^n{}_{\om\mu}&=H^n{}_{\om\mu}+\mu h^{na}k_{ad}H^d{}_{\om\mu}
+\bigl(H^n{}_{\a\om}\gtn{\a\d}g\ink{\d\mu}-H^n{}_{\a\mu}\gtn{\a\d}g\ink{\d\om}
\bigr)\cr
&-2\mu h^{na}k_{ad}\gtn{\d\tau}\gtn{\a\b}H^d{}_{\d\a}g\ink{\tau\om}g\ink{\b\mu}
-2\mu h^{na}k_{ad}\gtn{\d\b}\gtn{\a\tau}H^d{}_{\b[\om}g_{\mu]\tau}g\ink{\d\a}\cr
&+2\mu^2h^{na}h^{bc}k_{ac}k_{bd}\gtn{\a\b}
H\gd d,\a\lower2pt\hbox{$\[$}\om,g_{[\mu\lower2pt\hbox{$\]$}\b]}
&\text{(A.1)}}
$$
where
$$
l_{ab}=h_{ab}+\mu k_{ab}
$$
as usual.

We can raise indices $L^n_{\om\nu}$ and we get
$$
L^{n\tau\e}=g^{\om\tau}g^{\nu\e}L^n{}_{\om\nu}. \eq2
$$
Now we write $H^a{}_{\mu\nu}$ and $L^{a\mu\nu}$ in terms of $\stpr E{}^a=
(E_{\bar a}{}^a)=(E_1^a,E_2^a,E_3^a)$, $\o a=1,2,3$, and $\stpr B{}^a=
(B_{\bar a}{}^a)=(B_1^a,B_2^a,B_3^a)$, $L^{n\mu\nu}$ in terms of $\stpr D{}^a=
(D^{\bar aa})=(D^{1a},D^{2a},D^{3a})$ and $\stpr H{}^a=
(H^{\bar aa})=(H^{1a},H^{2a},H^{3a})$, i.e.
$$
\ealn{
H_{\mu\nu}^a&=\(\matrix
0 & -B_3^a & B_2^a & -E_1^a\\
B_3^a & 0 & -B_1^a & -E_2^a\\
-B_2^a & B_1^a & 0 & -E_3^a\\
E_1^a & E_2^a & E_3^a & 0 \endmatrix\) &\text{(A.3)}\cr
\noalign{\vskip3pt}
L^{a\mu\nu}&=\(\matrix
0 & -H^{3a} & H^{2a} & -D^{1a} \\
H^{3a} & 0 & -H^{1a} & -D^{2a} \\
-H^{2a} & H^{1a} & 0 & -D^{3a} \\
D^{1a} & D^{2a} & D^{3a} & 0 \endmatrix\) &\text{(A.4)}}
$$

In this way
$$\dsl{
E_{\bar a}^a=H_{4\bar a}^a, \quad D^{\bar aa}=L^{a4\bar a}\cr
\hphantom{(A.6)}\hfill \stpr B{}^a=-(H_{23}^a,H_{31}^a,H_{12}^a) \hfill \text{(A.5)}\cr
\stpr H{}^a=-(L^{a23},L^{a31},L^{a12})}
$$
or
$$
\dsl{
\hphantom{(A.6)}\hfill B_{\bar a}^a=-\tfrac12\, \e\dg\bar a,\bar b\bar c,H_{bc}^a, \quad H_{\bar c\bar m}^a
=-\e\dg\bar c\bar m,\bar e,B_{\bar e}^a \hfill \text{(A.6)}\cr
\hphantom{(A.6)}\hfill H^{\bar a a}=-\tfrac12\,\e\gd\bar a,\bar b\bar c,L^{a\bar b\bar c}, \quad
L^{a\bar c\bar m}=-\e\gd\bar c\bar m,\bar e,H^{\bar ea} \hfill \text{(A.7)}}
$$
$\e_{\bar a\bar b\bar c}$, $\o a,\o b,\o c=1,2,3$, is a usual 3-dimensional
anti\sy ic symbol, $\e_{123}=1$ and it is unimportant for it if its indices
are in up or down position. We keep these indices in up or down position only
for a convenience.

One gets
$$
D^{n\bar e}=\o B{}^n{}_f{}^{\bar d\bar e} B^f{}_{\bar d}
+\o A{}^n{}_f{}^{\bar v\bar e}E^f{}_{\bar v} \eq8
$$
where
$$\ealn{
\o B^n{}_d{}^{\bp\be}&=\e\dg\bm\bz,\bp,\bigl(g^{\bz4}g^{\bm\be}\d\gd n,d,
+\mu k\gd
n,d,g^{\bz4}g^{\bm\be}+g^{\mu\be}g\ink{\d\mu}g^{\bz4}\gtn{\bm\d}\d\gd n,d,\cr
&-g^{\om4}g\ink{\d\om}g^{\bm\be}\gtn{\bz\d}\d\gd n,d,
+\mu k\gd n,d,g^{\mu\be}\gtn{\a\tau}g\ink{\mu\tau}g\ink{\d\a}g^{\bz4}\gtn{\d\bm}\cr
&+\mu k\gd n,d, g^{\om4}\gtn{\a\tau}g\ink{\mu\tau}g\ink{\d\a}g^{\bm\be}\gtn{\d\bz}
-\mu^2 k\gd n,c,k\gd c,d,g^{\mu\tau}g\ink{\mu\b}g^{\bz4}\gtn{\bm\b}\cr
&-\mu^2 k\gd n,c,k\gd c,d,g^{\om4}g\ink{\om\b}g^{\bm\be}\gtn{\bz\b}
-2\mu k\gd n,d,g^{\om4}g^{\mu\tau}g\ink{\tau\om}g\ink{\b\mu}\gtn{\bz\tau}
\gtn{\bm\b}\bigr)
&\text{(A.9)}\cr
\o A^n{}\dg d,\bm\be,&=g^{44}g^{\bm\be}\d\gd n,d,
-\mu k\gd n,d,g^{\bm4}g^{4\be}+\mu k\gd n,d,g^{44}g^{\bm\be}\cr
&+g^{\mu\be}g\ink{\d\mu}\gtn{4\d}g^{\bm 4}\d\gd n,d,
-g^{\mu\be}g\ink{\d\mu}g^{44}\gtn{\bm\d}\d\gd n,d,\cr
&-g^{\om4}g\ink{\d\om}\gtn{4\d}g^{\bm\be}\d\gd n,d,
+g^{\om4}g\ink{\d\om}g^{4\be}\gtn{\bm\d}\d\gd n,d,\cr
&-\mu k\gd n,d,g^{\mu\be}\gtn{\a\tau}g\ink{\mu\tau}g\ink{\d\a}g^{\u m4}\gtn{\d\mu}
+\mu k\gd n,d,g^{\mu\tau}\gtn{\a\tau}g\ink{\mu\tau}g\ink{\d\a}g^{44}\gtn{\d\bm}\cr
&+\mu k\gd n,d,g^{\om4}\gtn{\a\tau}g\ink{\mu\tau}g\ink{\d\a}g^{\bm\be}
\gtn{\d\mu}
-\mu k\gd n,d,g^{\om4}\gtn{\a\tau}g\ink{\mu\tau}g\ink{\d\a}g^{4\be}\gtn{\d\bm}\cr
&-\mu^2k\gd n,c,k\gd c,d,g^{\mu\be}g\ink{\mu\b}g^{\bm4}\gtn{\mu\b}
-\mu^2k\gd n,c,k\gd c,d,g^{\mu\be}g\ink{\mu\b}g^{44}\gtn{\bm\b}\cr
&-\mu^2k\gd n,c,k\gd c,d,g^{\om4}g\ink{\om\b}g^{\bm\be}\gtn{4\b}
+\mu^2k\gd n,c,k\gd c,d,g^{\om4}g\ink{\om\b}g^{4\be}\gtn{\bm\b}\cr
&+2\mu k\gd n,d,g^{\om4}g^{\mu\be}g\ink{\tau\om}g\ink{\b\mu}
\gtn{\bm\tau}\gtn{4\b}\cr
&-2\mu k\gd n,d,g^{\om4}g^{\mu\be}g\ink{\tau\om}g\ink{\b\mu}
\gtn{4\tau}\gtn{\bm\b}-\d\gd n,d,g^{\bm4}g^{4\be}
&\text{(A.10)}}
$$
$$
H^{n\bd}=\o C{}^n{}_f{}^{\bd\bp}B^f{}_{\bar p}+\o D{}^n{}_f{}^{\bd\bv}E^f{}_{\bv} 
\eqno\text{(A.11)}
$$
where
$$\ealn{
\o C{}^n{}_d{}^{\bp\Bf}&=\tfrac12\,\e\gd\bp,\be\bk,\e\dg \bw\bm,\Bf,
\bigl(g^{\bw\bk}g^{\bm\be}\d\gd n,d,
+\mu k\gd n,d,g^{\bw\bk}g^{\bm\be}
-g^{\bw\bk}g^{\mu\be}g\ink{\d\mu}\gtn{\bm\d}\d\gd n,d,\cr
&-g^{\om\bk}g\ink{\d\om}g^{\bm\be}\gtn{\bw\d}\d\gd n,d,
-2\mu k\gd n,d,g^{\om\bk}g^{\mu\be}g\ink{\tau\om}g\ink{\b\mu}\gtn{\bw\b}
\gtn{\bm\tau}\cr
&+\mu k\gd n,d,g^{\om\bk}\gtn{\a\tau}g\ink{\om\tau}g\ink{\d\a}g^{\bm\be}\gtn{\d\bw}
-\mu^2 k\gd n,c,k\gd c,d,g^{\om\bk}g\ink{\om\b}g^{\bm\be}\gtn{\bw\b}\bigr)
& \text{(A.12)}\cr
\o D{}^n{}_d{}^{\bp\bm}&=\tfrac12\,\e\gd \bp,\be\bk,\bigl(
g^{4\bk}g^{\bm\be}\d\gd n,d,-g^{\bm\bk}g^{4\be}\d\gd n,d,-\mu k\gd n,d,
g^{\bm\bk}g^{4\be}\cr
&+\mu k\gd n,d,g^{\bw\bk}g^{4\bk}+g^{\bm\bk}g^{\mu\be}g\ink{\d\mu}\gtn{4\d}
\d\gd n,d,-g^{4\bk}g^{\mu\be}\gtn{\bm\d}g\ink{\d\mu}\d\gd n,d,\cr
&-g^{\om\bk}g\ink{\d\om}g^{\bm\be}\gtn{4\d}\d\gd n,d,
+2\mu k\gd n,d,g^{\om\bk}g^{\mu\be}g\ink{\tau\om}g\ink{\b\mu}
\gtn{\bm\b}\gtn{4\tau}\cr
&-2\mu k\gd n,d,g^{\om\bk}g^{\mu\be}g\ink{\tau\om}g\ink{\b\mu}
\gtn{4\b}\gtn{\bm\tau}
+\mu k\gd n,d,g^{\mu\be}g\ink{\d\a}g\ink{\mu\tau}\gtn{\a\tau}g^{4\bk}\gtn{\d
\bm}\cr
\noalign{\eject}
&-\mu k\gd n,d,g^{\mu\be}g\ink{\d\a}g\ink{\mu\tau}\gtn{\a\tau}g^{\bm\bk}
\gtn{\d4}
+\mu k\gd n,d,g^{\om\bk}\gtn{\a\tau}g\ink{\om\tau}g\ink{\d\a}g^{\bm\be}
\gtn{4\d}\cr
&+\mu^2 k\gd n,c,k\gd c,d,g^{\mu\be}g\ink{\mu\b}g^{\bm\bk}\gtn{4\b}
-\mu^2k\gd n,c,k\gd c,d,g^{\mu\be}g\ink{\mu\b}g^{4\bk}\gtn{\bm\b}\cr
&-\mu^2k\gd n,c,k\gd c,d,g^{\om\bk}g\ink{\om\b}g^{\bm\be}\gtn{4\b}
+\mu^2k\gd n,c,k\gd c,d,g^{\om\bk}g\ink{\om\b}g^{4\be}\gtn{\bm\b}\bigr)
& \text{(A.13)}}
$$
The confinement condition in this theory means
$$
D^{\ba a}=0 \eq14
$$
with $E^{\ba a}\ne0$ and can be satisfied by special arrangement of the \nos\
tensor $g_{\mu\nu}$. This generalizes a notion of a charge confinement from
Section~20 and can be considered as a color confinement in the case of
$G=\SU(3)_c$ (QCD).

The condition (6.4) can be explicitly solved.
%up to the second order of $\xi$ and the first of~$\z$. 
One gets
$$
\ealn{
L^n{}_{\om\u m} &= \brgn\om \F\gd n,\u m,+\xi k\gd n,d,\brgn\om \F\gd d,\u m,
-\Bigl(\z\brgn\om \F\gd n,\u a,h^{o\u a\u d}k_{o\u d\u m}+\gtn{\a\mu}
\brgn\a\F\gd n,\u m,g\ink{\mu\om}\Bigr)\cr
&-2\xi\z k\gd n,d,\brgn\om \F\gd d,\u d,\gtn{\d\a}g\ink{\a\om}h^{o\u d\u a}
k\gd o,\u a\u m,\cr
&+\xi k\gd n,d,\Bigl(\z^2 h^{\u d\u a}\brgn\om \F\gd d,\u a,
k\gd o,\u d\u b,k\gd o,\u m\u c,h^{o\u c\u b}
+\brgn\b \F\gd d,\u m,\gtn{\d\b}g\ink{\d\a}g\ink{\om\mu}\gtn{\a\mu}\Bigr)\cr
&-\xi^2k^{nb}k_{bd}\Bigl(\z\brgn\om \F\gd d,\u a,h^{o\u a\u b}k\gd o,\u m\u b,
+\gtn{\a\b}\brgn\a \F\gd d,\u m,g\ink{\om\b}\Bigr),
&\text{(A.15)}}
$$
where
$$
k^{nb}=h^{na}h^{bp}k_{ap}.  \eqno\text{(A.16)}
$$
The condition (6.5) can be explicitly solved. One gets
$$
\ealn{
L\gd n,\u w\u m,&=H\gd n,\u w\u m,+\mu k\gd n,d,H\gd d,\u w\u m,
+\z\bigl(h^{o\u a\u d}H\gd n,\u a\u w, k\gd o,\u d\u m,-h^{o\u a\u d}
H\gd n,\u a\u m,k\gd o,\u a\u w,\bigr)\cr
&-2\mu \z^2h^{o\u d\u c}h^{o\u a\u b}H\gd d,\u d\u a,k\gd o,\u c\u w,
k\gd o,\u b\u m,
-2\mu\z k\gd n,d, h^{o\u a\u p}h^{o\u d\u b}H\gd d,\u b[\u w,k\gd o,\u m]\u p,
k_{\u d\u a}\cr
&+2\mu^2\z k^{nb}k_{bd}H\gd d,\u a[\u w,k\gd o,\u m]\u p,h^{o\u p\u a}.
&\text{(A.17)}}
$$

Using the above formulae we can express the Yang--Mills \lg\ in the \eu\nos\
Nonabelian \KK\ Theory and the \lg\ for Higgs' field.

The Yang--Mills \lg\ reads
$$\ealn{\hskip-20pt
\cL\dr{YM}&=\frac1{8\pi}\,\biggl(
h_{nk}H^{k\om\mu}H\gd n,\om\mu,-2h_{cd}H^cH^d+2h_{nk}H^{k\om\mu}
H\gd n,\d\om,g\ink{\a\mu}\gtn{\a\d}\cr
&+\mu\Bigl[2k_{nk}H^{k\om\mu}H\gd n,\d\om,\gtn{\d\a}g\ink{\a\mu}
-2k_{kd}H^{k\om\mu}H\gd d,\d\a,\gtn{\d\b}\gtn{\a\rho}g\ink{\b\om}g\ink{\rho\mu}\cr
&-k_{kd}H^{k\om\mu}H\gd d,\eta\om,\gtn{\eta\b}\gtn{\a\rho}
g\ink{\mu\a}g\ink{\b\rho}
+k_{kd}H^{k\om\mu}H\gd d,\eta\mu,\gtn{\eta\d}\gtn{\a\rho}g\ink{\d\rho}
g\ink{\om\d}\Bigr]\cr
&+\mu^2\Bigl[k_{nk}k\gd n,d,H^{k\om\mu}H\gd d,\eta\mu,\gtn{\rho\b}\gtn{\eta\a}
g\ink{\om\b}g\ink{\a\rho}\cr
&-2k_{nk}k\gd n,d,H^{k\om\mu}H\gd d,\d\a,\gtn{\d\eta}\gtn{\a\rho}
g\ink{\eta\om}g\ink{\rho\mu}\cr
&-k_{nk}k\gd n,d,H^{k\om\mu}H\gd d,\eta\om,\gtn{\rho\a}\gtn{\eta\b}
g\ink{\mu\a}g\ink{\b\rho}
+k\dg k,b,k_{bd}H^{k\om\mu}H\gd d,\a\om,\gtn{\a\b}g\ink{\mu\a}\cr
\noalign{\eject}
&+k\dg k,b,k_{bd}H^{k\om\mu}H\gd d,\a\om,\gtn{\a\b}g\ink{\mu\a}
-k\dg k,b,k_{bd}H^{k\om\mu}H\gd d,\a\mu,\gtn{\a\b}g\ink{\om\b}\cr
&+k\gd p,n,k_{pk}H^{k\om\mu}H\gd n,\om\mu,\Bigr]\cr
&+\mu^3 \Bigl[k_{nk}k^{nb}k_{bd}H^{k\om\mu}H\gd d,\a\om,\gtn{\a\b}
g\ink{\mu\b}\cr
&-k_{nk}k^{nb}k_{bd}H^{k\om\mu}H\gd d,\a\mu,H^{k\om\mu}
\gtn{\a\b}g\ink{\om\b}\Bigr]\biggr)
&\text{(A.18)}}
$$

The kinetic term for Higgs' field is given by
$$\ealn{
\cL\dr{kin}(\nabla\F)&=\frac{1}{V_2}\int_M {\sqrt{|\t g|}}\,d^{n_1}x\,
\Bigl[l_{nk}g^{\om\mu}g^{\u m\u p}\brgn \mu \F\gd k,\u p,
\Bigl\{\brgn\om \F\gd n,\u m,+\xi k\gd n,d,\brgn\om \F\gd d,\u m,\cr
&-\z\brgn\om \F\gd d,\u a,h^{o\u a\u q}k\gd o,\u q\u m,
-\brgn\a \F\gd a,\u m,\gtn{\a\eta}g\ink{\eta\om}\cr
&-2\xi\z\brgn\d \F\gd d,\u a,k\gd n,d,\gtn{\d\a}g\ink{\a\om}
h^{o\u d\u q}k\gd o,\u q\u m,\cr
&-\xi\bigl(\z^2 k\gd n,d,\brgn\om \F\gd d,\u a,h^{o\u b\u q}h^{o\u a\u w}
k\gd o,\u q\u m,k\gd o,\u w\u b,
+k\gd n,d,\brgn\b \F\gd d,\u m,\gtn{\a\nu}\gtn{\b\rho}g\ink{\nu\om}
g\ink{\rho\a}\bigr)\cr
&+\xi^2\bigl(\z k^{nb}k_{bd}\brgn\om \F\gd d,\u a,h^{o\u a\u q}
k\gd o,\u q\u m,+\brgn\a \F\gd d,\u m,\gtn{\a\b}g\ink{\b\om}\bigr)\Bigr\}
\Bigr]
&\text{(A.19)}}
$$

In the case of $g_{\mu\nu}=\eta_{\mu\nu}$ (a Minkowski tensor)
one gets
$$\ealn{
\cL\dr{kin}(\nabla\F)&=\frac{1}{V_2}\int_M {\sqrt{|\t g|}}\,d^{n_1}x\,
\Bigl[l_{nk}g^{\u m\u p}\brg\om \F\gd k,\u p,
\Bigl\{\brgn\om \F\gd n,\u m,
+\xi k\gd n,d,\brgn\om \F\gd d,\u m,\cr
&-\z\brgn\om \F\gd d,\u a,k\gd o\u a,\u m,
-\xi\z^2 k\gd n,d,k\gd o\u b,\u m, k\gd o\u a,\u b,\brgn\om\F\gd d,\u a,\cr
&+\xi^2\z k^{nb}k_{bd}k\gd o\u a,\u m,\brgn\om \F\gd d,\u a,\Bigr\}\Bigr],
&\text{(A.20)}}
$$
where $\brg\om \F\gd k,\u p,=\eta^{\om\mu}\brgn\mu \F\gd k,\u p,$, 
$k\gd o\u a,\u b,=h^{o\u a\u c}k_{o\u c\u b}$.

A mass matrix for broken
gauge bosons can be calculated.
$$
\ealn{
{}M_{ij}^2(\F\dr{crt}^k)&=\frac{\a_S^2}{\hbar c}\,\frac1{V_2}
\int_M \sqrt{|\t g|}\,d^{n_1}x\cr
&\Bigl\{l_{np}g^{\u m\u p}\br{$(k)$}B\gd p,\u pi,
\Bigl(\br{$(k)$}B\gd d,\u mj,
+\xi k\gd n,d,\br{$(k)$}B\gd d,\u mj,
-\z \br{$(k)$}B\gd d,\u aj,k\gd o\u a,\u m,\cr
&-\xi\z^2k\gd n,d,k\gd o\u b,\u m,k\gd o\u a,\u b,\br{$(k)$}B\gd d,\u aj,
+\xi^2\z k^{nb}k_{bd}k\gd o\u a,\u m,\br{$(k)$}B\gd d,\u aj,\Bigr)\Bigr\}
&\text{(A.21)}}
$$
$k=0,1$, where 
$$
\br{$(k)$}B\gd b,\u ni,=\bigl[\d\gd\u m,\u n, C\gd b,ms,\a\gd s,i,
+\d\gd b,m,f\gd \u m,\u ni,\bigr][\F^k\dr{crt}]\gd m,\u m,. \eq22
$$

In the case of symmetric theory ($l_{ab}=h_{ab}$, $g_{\u a\u b}=h\gd
o,\u ab,$) one gets
$$
M^2_{ij}=\frac{\a^2_S}{\hbar c}\,\frac1{V_2} \int_M\sqrt{|\t g|}\,d^{n_1}x
\,\bigl\{h_{bn}h^{o\u m\u p} B\gd b,\u pi,B\gd n,\u mj,\bigr\}. \eq23
$$

The Higgs' potential is given by
$$\ealn{
V(\F)&=\frac1{V_2}\int_M\sqrt{|\t g|}\,d^{n_1}x\biggl\{g^{\u w\u p}g^{\u m\u q}
\Bigl[h_{nk}H\gd n,\u w\u m,+2\z h_{nk}H\gd n,\u d\u w,k\gd o\u d,\u m,\cr
&+\mu\z\Bigl(2k_{nk}H\gd n,\u d\u w,k\gd o\u d,\u m,
+\z\bigl(-2k_{kd}H\gd d,\u d\u a,k\gd o\u d,\u w,k\gd o\u a,\u m,\cr
&-k_{kd}H\gd d,\u l\u w,k\gd o,\u m\u a,k^{o\u l\u a}+k_{kd}H\gd d,\u l\u m,
k^{o\u l\u a}k\gd o,\u w\u a,\bigr)\Bigr)\cr
&+\mu^2\z\Bigl(k\dg n,b,k_{bd}H\gd d,\u a\u w,k\gd o,\u m,{}^{\u a}
-k\dg k,b,k_{bd}H\gd d,\u a\u m,k\gd o,\u w,{}^{\u a}
+\z\bigl(k_{nk}k\gd n,d,H\gd d,\u l\u m,k\gd o,\u w,{}^{\u r}k\gd \u l,\u r,\cr
&-2k_{nk}k\gd n,d,H\gd d,\u d\u a,k\gd o\u d,\u w,k\gd o\u a,\u m,
-k_{nk}k\gd n,d,H\gd d,\u l\u w,k\gd o,\u m,{}^{\u r}k\gd \u d,\u r,
\bigr)\Bigr)\cr
&+\mu^3\z \bigl(k_{nk}k^{nb}k_{bd}H\gd d,\u a\u w,k\gd o,\u m,{}^{\u a}
-k_{nk}k^{nb}k_{bd}H\gd d,\u a\u m,k\gd o,\u w,{}^{\u a}\bigr)\Bigr]
\cdot H\gd k,\u p\u q,\cr
&-2h_{cd}\bigl(H\gd c,\u p\u q,\gtk{\u p\u q}\bigr)
\bigl(H\gd d,\u a\u b,\gtk{\u a\u b}\bigr)\biggr\}
& \text{(A.24)}}
$$
or
$$\ealn{
V(\F)&=\frac1{V_2}\int_M\sqrt{|\t g|}\,d^{n_1}x
\Bigl(P\dg kl,[\u p\u q][\u a\u b],H\gd k,\u p\u q,H\gd l,\u a\u b,
-2h_{kl}\bigl(H\gd k,\u p\u q,\gtk{\u p\u q}\bigr)\bigl(H\gd l,\u a\u b,
\gtk{\u a\u b}\bigr)\Bigr)\cr
&=\frac1{V_2}\int_M\sqrt{|\t g|}\,d^{n_1}x \, Q\dg sk,[\u c\u d][\u p\u q],
H\gd s,\u c\u d,H\gd k,\u p\u q, & \text{(A.25)}\cr
&Q\dg sk,[\u c\u d][\u p\u q],=Q\dg ks,[\u p\u q][\u c\u d],=
-Q\dg sk,[\u d\u c][\u p\u q],=-Q\dg sk,[\u c\u d][\u q\u p],=
Q\dg sk,[\u d\u c][\u q\u p],}
$$
$$\ealn{
P\dg sk,[\u c\u d][\u p\u q],
&=g^{[\u c\[\u p}g^{\u d]\u q\]}h_{sk}-2\z h_{sk}k\gd o[\u d,|\u e|,g^{\u c
][\u p}g^{|\u e|\u q]}\cr
&+\mu\z\Bigl(-2k_{sk}k\gd o[\u d,|\u e|,g^{\u c][\u p}g^{|\u e|\u q]}
+\z\bigl(2k_{sk}k\gd o[\u c,|\u e|,k\gd o\u d],\u f,g^{\u e[\u p}
g^{|\u f|\u q]}\cr
&-k_{sk}k\gd o,\u e\u a,k^{o[\u d|\u a|}g^{\u c][\u p}g^{|\u e|\u q]}
-k_{sk}k^{o[\u c|\u a|}g^{\u d][\u q}g^{|\u e|\u p]}k\gd o,\u e\u a,\bigr)\Bigr)\cr
&+\mu^2\z\Bigl(-k_{bs}k\dg k,b,k\gd o[\u d,|\u a|,g^{\u c][\u p}
g^{|\u a|\u q]}
-k_{bs}k\dg k,b,k\gd o,\u a,{}^{[\u c}g^{|\u a|\[\u p}g^{\u d]\u q\]}\cr
&+\z\bigl(k\gd n,s,k_{nk}k\gd o,\u a,{}^{\u r}k\gd o[\u c,|\u r|,
g^{\u a\[\u p}g^{\u d]\u q\]}
-2k\gd n,s,k_{nk}k\gd o[\u c,|\u e|,k\gd o\u d],\u f,g^{\u e[\u p}g^{|\u f|
\u q]}\bigr)\Bigr)\cr
&+\mu^3 \z\bigl(-k_{bs}k^{nb}k_{nk}k\gd o,\u e,{}^{[\u d}g^{\u c][\u p}
g^{|\u e|\u q]}
-k_{bs}k^{nb}k_{ns}k\gd o,\u e,{}^{[\u c}g^{|\u e|\[\u p}g^{\u d]\u q\]}\bigr)\cr
&+\mu^2g^{[\u c\[\u p}g^{\u d]\u q\]}k\gd p,s,k_{pk}
& \text{(A.26)}\cr
Q\dg sk,[\u c\u d][\u p\u q],&=P\dg sk,[\u c\u d][\u p\u q],
-2h_{sk}\gtk{\u c\u d}\gtk{\u p\u q}}
$$

One can decompose the Higgs field into independent components according to
Ref.~[80] and rules from Chapter~8 in the following way
$$
\F^c_{\u b}=\t\F{}^c_{\u b}=\sum_{(\ul n{}_i,\ul n{}'_j)}
\(\F_{\ul m{}_j}^{(\ul n{}_i,\ul n{}'_j)}\)_{\u b}^c \eq27
$$
where
$$
\ealn{
\Ad_G\bigr|_{G_0}&=\sum_i\oplus \ul n{}_i\oplus\Ad_{G_0} &\text{(A.28)}\cr
\Ad_H\bigr|_{G_0\oplus G}&=\sum_j\oplus\bigl(\ul n{}'_j\otimes \ul m{}_j
\bigr). & \text{(A.29)}}
$$
$\ul n{}_i$ are irreducible representations of $G_0$ and $\ul m{}_j$ are
irreducible representations of~$G$. In the sum \rf27 \ $\ul n{}_i$ and $\ul
n{}'_j$ are identical representations of~$G_0$ from \rf28 \ and \rf29 \
decomposition. In this way a constraint is satisfied identically and we
can put $\t\F{}_{\u b}^c$ given by \rf27 \ into (A.16--17,18--25). Thus the
analysis of a mass spectrum in the theory can be simplified (a~little).

For $k=0$, $\F^0\dr{crt}$, $H\gd k,\u p\u q,=0$ one gets the following
matrix for Higgs' bosons
$$\ealn{
m\gd 2\u h,f,{}\gd \u e,a,&=\frac{-1}{V_2}\int_M\biggl\{
\frac{8\a^2_S}{\hbar c}\,Q\dg sk,[\u e\u a][\u h\u q],C\gd s,ac,C\gd k,ef,
(\F^0\dr{crt})\gd c,\u a,(\F^0\dr{crt})\gd e,\u q,\cr
&-\frac{2\a_S}{\sqrt{\hbar c}}\,Q\dg as,[\u p\u q][\u h\u a],f\gd \u e,\u p\u q,
C\gd s,ef,(\F^0\dr{crt})\gd e,\u a,
+\frac{4\a_S}{\sqrt{\hbar c}}\,Q\dg sf,[\u e\u a][\u p\u q],f\gd \u h,\u p\u q,
C\gd s,ea,(\F^0\dr{crt})\gd a,\u a,\cr
&+Q\dg af,[\u c\u d][\u p\u q],f\gd e,\u c\u d,f\gd \u h,\u p\u q,
\biggr\}\sqrt{|\u g|}\,d^{n_1}x
& \text{(A.30)}\cr}
$$

For $k=1$, $\F^1\dr{crt}$, $H\gd k,\u p\u q,\ne0$ and $\F^1\dr{crt}$ (if
exists) satisfies the following equation:
$$
\frac{2\a_S}{\sqrt{\hbar c}}\,Q\dg sk,[\u e\u a][\u p\u q],C\gd s,ac,
(\F^1\dr{crt})\gd c,\u a,=Q\dg ak,[\u c\u d][\u p\u q],f\gd \u e,\u c\u d,
\eq31
$$
and a supplementary condition
$$
\F_{\u b}^c f_{\hi\u d}^{\u b}-\mu_{\hi}^a \F_{\u a}^b C_{ab}^c=0. \eq32
$$

A mass matrix for Higgs' bosons looks like
$$
m\gd 2\u h,f,{}\gd \u e,a,=\frac{-1}{V_2}\int_M
\biggl(\frac{4\a_S}{\sqrt{\hbar c}}\,Q\dg sk,[\u e\u h][\u p\u q],
H\gd k,\u p\u q,(\F^1\dr{crt})C\gd s,af,\biggr)\sqrt{|\u g|}\,d^{n_1}x.
\eq33
$$

$$\ealn{\hskip-10pt 
H_{\u m\u n}^b(\F\dr{crt}^k)&=\a_S\frac1{\sqrt{\hbar c}}\,C_{cd}^b
(\F\dr{crt}^k)_{\u n}^c(\F\dr{crt}^k)_{\u m}^d
-\frac1{\a_S}\sqrt{\hbar c}\,\mu_{\hi}^b f_{\u n\u m}^{\hi}
-(\F\dr{crt}^k)_{\u c}^b f_{\u n\u m}^{\u c}. & \text{(A.34)}\cr
H\gd b,\u m\u n,(\F\dr{crt}^0)&=0 & \text{(A.35)}\cr}
$$

Finally, let us give a formula for a \lg\ of an electromagnetic field in the
NKK. 
$$
\cL\dr{em}=\frac1{8\pi}\Bigl(2\bigl(\gtk{\mu\nu}F_{\mu\nu}\bigr)^2-
\bigl(g^{\mu\a}g^{\nu\b}-g^{\nu\b}\gtn{\mu\a}+g^{\nu\b}g^{\mu\om}
\gtn{\tau\a}g_{\om\tau}\bigr)F_{\a\b}F_{\mu\nu}\Bigr) \eq36
$$

In order to calculate \co\ terms in the theory it is necessary to know
Einstein--Kaufmann \cn s on a group~$G$ and on a homogeneous manifold
$M=G/G_0$. Let us give those \cn s. On a group~$G$ a right invariant
Einstein--Kaufmann \cn\ reads:
$$
\dsl{
\indent \G_{wm}^n=-\tfrac12\,C_{wm}^n +\tfrac12\bigl(K\dg wm,n,-2\mu^2k\dg [m,a,
K_{w]ab}k^{nb}\bigr)\hfill\cr
\hfill{}+h^{me}\Bigl\{\mu K\dg e(w,a,k_{m)a}+\mu^2k\dg c,b,\bigl[k\dg (m,c, k\dg w,c, K_{w)ab}
k\dg e,a,-K_{eab}k\dg(w,a,k\dg m),c,\bigr]\Bigr\}\indent 
\text{(A.37)}}
$$
where
$$
K_{abc}=-\mu \bigl(\t\nabla_a k_{bc}-\t\nabla_b k_{ca}+\t\nabla_c k_{ab}
\bigr). \eq38
$$
$\t\nabla_a$ means a Riemannian covariant derivative on a Lie-semisimple
group $G$ \wrt a biinvariant Killing tensor $h_{ab}$.

One gets
$$
\t\nabla k_{bc}=-\tfrac12\bigr(C_{bc}^f k_{fc}+C_{ca}^f k_{bf}\bigr) \eq39
$$
and
$$
K_{abc}=\mu\bigl(C_{ba}^f k_{fc}+C_{ac}^fk_{fb}-C_{bc}^f k_{fa}\bigr). \eq40
$$
Let us remind that $k_{ab}$ is a right-invariant anti\sy ic tensor on~$G$.

If we write a \cn\ $\G$ on~$G$ in the form
$$
\G_{wm}^n = -\tfrac12\, C_{wm}^n +u\gd n,wn,, \eq41
$$
one gets
$$
R_{bd}=\t R_{bd}+\t\nabla_a u\gd a,bd,-\t\nabla_du\gd a,ba,
+\tfrac12\bigl(\t\nabla_b u\gd a,ad,-\t\nabla_d u\gd a,ab,\bigr) \eq42
$$
where
$$
\dsl{
\hphantom{(A.39)}\hfill \t\nabla_a u\gd c,ed,=-\tfrac12 \bigl(C_{ea}^fu\gd c,fd,
+C_{da}^f u\gd c,ef,-C_{fa}^c u\gd f,ed,\bigr)\hfill \text{(A.43)}\cr
\hphantom{(A.39)}
u\gd n,wm,=\tfrac12\,\mu\bigl(L\dg wm,n,-2\mu k\dg [m,a,L_{w]ab}
k^{nb}\bigr)\hfill\cr
\hfill{}-\mu^2 h^{ne}\bigl(L\dg e(w,a, k_{m)a}+\mu^2 k\dg c,b,
\bigl[k\dg (m,c,L_{w)ab}k\dg e,a,-L_{eb}k\dg(w,a,
k\dg m), c,\bigr]\bigr) \indent \text{(A.44)}}
$$
where
$$
L_{abc}=C_{ba}^f k_{fc}+C_{ac}^fk_{fb}-C_{bc}^fk_{fa}. \eq45
$$

Finally
$$
R_{bd}=\t R_{bd}-\tfrac12\,C_{db}^f(u\gd a,af,+u\gd a,fa,)
-\tfrac14\,C_{fa}^a(2u\gd f,ba,+u\gd f,ab,)+\tfrac14\,(C_{fb}^au\gd f,ad,
-C_{bd}^fu\gd a,fd,), \eq46
$$
$$
\t R_{bd}=-\tfrac14\,h_{bd}. \eq47
$$
For example, for $G=\text{SO}(3)$,
$$
\t R_{bd}(\text{SO}(3))=\tfrac12\,\d_{bd}. \eq48
$$

If 
$$\ealn{
{}&k_{ab}=C_{ab}^f V_f  &\text{(A.49)}\cr
&\wh \nabla_k V_f=0 &\text{(A.50)}}
$$
one gets
$$
\ealn{\hskip-20pt
u\gd n,wm,&=\frac\mu2\,C\gd sn,w,C\gd p,ms,V_p-\mu^3C\gd p,as,V_p
C^{rnb}V_rC^q{}\dg[m,a,V_qC\gd s,bw],\cr
&{}+\mu^4C^f{}\dg c,b,V_f \bigl[C^p{}\dg(m,c,V_pC\gd s,bw),
C\gd q,as,V_qC^{rna}V_r \cr
&\quad{}- C^s{}\dg b,n,C\gd p,as,V_p C^q{}\dg (w,a,
V_q C\gd r,m),V_r\bigr]. & \text{(A.51)}}
$$

Let us remind to the reader that
$$
\t\nabla_kV_f=-\tfrac12\,C\gd f,fk,V_e. \eq52
$$
$u\gd n,wm,$ can be calculated explicitly in a general form. One gets
$$
\ealn{
u\gd n,wm,&=\tfrac12\,\mu\(C\gd f,mw,k\dg f,n,+C^f{}\dg w,n,
k_{fm}-C^f{}\dg m,n,k_{fw}\)\cr
&-\tfrac12\,\mu^2\Bigl[C\gd f,bw,k_{fa}(k^{na}k\dg m,b,-k^{nb}k\dg
m,a,)+C\gd f,mb,k_{fa}(k^{na}k\dg w,b,-k^{nb}k\dg w,a,)\cr
&-2k^{nb}k\dg m,a,C\gd f,ab,k_{fw}-\bigl(k\dg f,a,k_{mc}C^f{}\dg
w,n,+2k_{fm}C^{anf}k_{wa}\cr
& - C^f{}\dg w, a, k\dg f,a,k_{ma}
+C^f{}\dg m,n,k\dg f, a,k_{wa}-C^f{}\dg m,a,k\dg f,a,
k_{wa}\bigr)\Bigr]\cr
&+\tfrac12\,\mu^4\Bigl[3k\dg c,b,C\gd f,ab,k\dg f,n,k\dg w,a,
k\dg m,a,+C\gd f,bw,k_{fa}k\dg m,c,(k\dg c,a,k^{nb}
-k\dg c,b,k^{na})\cr
&+C\gd f,bm,k_{fa}k\dg w,a,(k\dg c,a,k^{nb}-k\dg c,b,k^{na})
+C\gd f,ab,k\dg c,b,k\dg w,c,k\gd a, m,k\gd n, f,\cr
&+C\gd fn,b,k_{fa}k\dg m,c,(k\dg c,a,k\dg w,b,-k\dg c,b,
k\dg w,a,)\cr
&+C\gd an,f,k\dg b,f,(k\dg c,b,k_{ma}
k\dg w,c,-k\dg w,b,k\dg m,c,k_{ac})\Bigr] & \text{(A.53)}}
$$
$$
R_G=l^{ab}R_{ab}. \eq54
$$
In the case of the Einstein--Kaufmann \cn\ on a $M=G/G_0$ manifold one gets 
$$\ealn{
\kern-20pt \wh{\o\G}{}\gd \u n,\u w\u m,&=\left\{\matrix \t n\\\t w\t m\endmatrix\right\}
+\frac12\bigl(K\dg\u w\u m,\u n,-2\t g\dg[\u m,\u a],K_{\u w]\u a\u b}
\t g_[{}\gd \u n\u b,],\bigr)\cr
&+h^{0\u n\u e}\left\{K_{\u e}{}\gd \u a,(\u w, \t g_{|\u m|\u a)}
+\t g{}_{[\u c}{}\gd \u b,], \bigl[\t g\dg [(|\u m|,\u c,{}_]K_{\u w)\u a\u b}
\t g_{[\u e}{}\gd \u a,],-K_{\u c\u a\u b}\t g_{[(\u w}{}\gd \u a,],
\t g_{[\u m)}{}\gd\u c,],\bigr]\right\} &\text{(A.55)}}
$$
where
$$
K_{\u a\u b\u c}=-\t\nabla_{\u a}\t g_{[\u b\u c]}-\t\nabla_{\u b}
\t g\ink{\u c\u a}+\t\nabla_{\u c}\t g\ink{\u a\u b}=
\z\bigl(-\t\nabla_{\u a}k_{\u b\u c}^0-\t\nabla_{\u b}k_{\u c\u a}^0
+\t\nabla_{\u c}k_{\u a\u b}^0\bigr) \eq56
$$
where $\t\nabla$ means a covariant derivative \wrt the Riemannian \cn\ on a
manifold $M=G/G$ with a left-invariant metric tensor $h_{\u a\u b}^0$, 
$\left\{\matrix \u a\\\u b\u c\endmatrix\right\}$ mean Christoffel symbols
built from $h_{\u a\u b}^0$. In this way a \co\ term reads
$$
\ul P=\frac1{V_1}\int_M \sqrt{|\t g|}\, \wh{\o R}(\wh{\o \G})\,d^{n_1}x
\eq57
$$
where
$$
\dsl{
\hphantom{(A.52)}\hfill \wh{\o R}(\wh{\o \G})=g^{\u a\u b}\wh{\o R}_{\u a\u b}
(\wh{\o \G}) \hfill \text{(A.58)}\cr
\hphantom{(A.52)}\hfill 
\wh{\o R}_{\u b\u d}=\t R_{\u b\u d}+\t\nabla_{\u a}u\gd \u a,\u b\u d,
-\t\nabla_{\u d}u\gd\u a,\u b\u a,+\tfrac12\,\bigl(\t\nabla_{\u b}
u\gd\u a,\u a\u d,-\t\nabla_{\u d}u\gd\u a,\u a\u b,\bigr)
\hfill \text{(A.59)}}
$$
where $\t R_{\u b\u d}$ is a Ricci tensor for a Riemannian \cn\ on~$M$ formed
from $g_{(\u a\u b)}=h_{\u a\u b}^0$.

In order to facilitate some calculations of a particle spectrum in the theory
let us write
$$
H^b{}_{\u m\u n}=\sum_i \bigl(H^{(\ul n{}_i,\ul n{}'_j)}_{\ul m{}_j}
\bigr)^b{}_{\u m\u n} \eq60
$$
where
$$
\ealn{
\bigl(H^{(\ul n{}_i,\ul n{}'_j)}_{\ul m{}_j}\bigr)^b{}_{\u m\u n}
&=\a_S\,\frac1{\sqrt{\hbar c}}(\ul n{}'_j\otimes \ul m{}_j)^b{}_{ef}
\bigl(\Phi^{(\ul n{}_i,\ul n{}'_j)}_{\ul m{}_j}\bigr)\gd e,\u m,
\bigl(\Phi^{(\ul n{}_i,\ul n{}'_j)}_{\ul m{}_j}\bigr)\gd f,\u n,\cr
&-\frac1{\a_S}\sqrt{\hbar c}\Bigl(\bigl(\ul n{}_i\bigr)\gd b,\u m\u n,
+\bigl(\Ad_{G_0}\bigr)\gd b,\u m\u n,\Bigr)\cr
&-\bigl(\ul n{}_i\bigr)\gd\u e,\u m\u n,
\bigl(\Phi^{(\ul n{}_i,\ul n{}'_j)}_{\ul m{}_j}\bigr)\gd b,\u e,
-\bigl(\Ad_{G_0}\bigr)\gd \u e,\u m\u n,
\bigl(\Phi^{(\ul n{}_i,\ul n{}'_j)}_{\ul m{}_j}\bigr)\gd b,\u e,.&
\text{(A.61)}}
$$
In this formalism the supplementary condition \rf32 \ is satisfied identically.

Let us notice that we can use all the formulae for Higgs' field potential,
kinetic energy of Higgs' field, a mass matrix for broken gauge bosons and a
mass matrix for Higgs' bosons derived here for hierarchy problem of symmetry
breaking as described in Section~9.

Let us consider some geometrical notion in the theory. Using Eq.~(A.1) and
the results from the fifth position of Ref.~[18] one gets
$$\ealn{
Q\gd n,\om\mu,&=-2\Bigl(\mu h^{na}k_{ad}H^d{}_{\om\mu}
+\bigl(H^n{}_{\a\om}\gtn{\a\d}g\ink{\d\mu}-H^n{}_{\a\mu}\gtn{\a\d}g\ink{\d\om}
\bigr)\cr
&-2\mu h^{na}k_{ad}\gtn{\d\tau}\gtn{\a\b}H^d{}_{\d\a}g\ink{\tau\om}g\ink{\b\mu}
-2\mu h^{na}k_{ad}\gtn{\d\b}\gtn{\a\tau}H^d{}_{\b[\om}g_{\mu]\tau}g\ink{\d\a}\cr
&+2\mu^2h^{na}h^{bc}k_{ac}k_{bd}\gtn{\a\b}
H\gd d,\a\lower2pt\hbox{$\[$}\om,g_{[\mu\lower2pt\hbox{$\]$}\b]}\Bigr)
&\text{(A.62)}}
$$
where $Q\gd n,\om\mu,$ is a tensor of a torsion in additional dimensions
connected to the Yang--Mills' field. In the same way we can write for the
Higgs' field, using Eq.~(A.17),
$$\ealn{
Q\gd n,\u w\u m,&=-2\Bigl(
\mu k\gd n,d,H\gd d,\u w\u m,
+\z\bigl(h^{o\u a\u d}H\gd n,\u a\u w, k\gd o,\u d\u m,-h^{o\u a\u d}
H\gd n,\u a\u m,k\gd o,\u a\u w,\bigr)\cr
&-2\mu \z^2h^{o\u d\u c}h^{o\u a\u b}H\gd d,\u d\u a,k\gd o,\u c\u w,
k\gd o,\u b\u m,
-2\mu\z k\gd n,d, h^{o\u a\u p}h^{o\u d\u b}H\gd d,\u b[\u w,k\gd o,\u m]\u p,
k_{\u d\u a}\cr
&+2\mu^2\z k^{nb}k_{bd}H\gd d,\u a[\u w,k\gd o,\u m]\u p,h^{o\u p\u a}\Bigr),
&\text{(A.63)}}
$$
where $Q\gd n,\u w\u m,$ is a tensor of a torsion over a manifold $M=G/G_0$
for Higgs' field. And finally (using (A.15))
$$
\ealn{Q\gd n,\om\u m,&=-\Bigl(
\xi k\gd n,d,\brgn\om \F\gd d,\u m,
-\Bigl(\z\brgn\om \F\gd n,\u a,h^{o\u a\u d}k_{o\u d\u m}+\gtn{\a\mu}
\brgn\a\F\gd n,\u m,g\ink{\mu\om}\Bigr)\cr
&-2\xi\z k\gd n,d,\brgn\om \F\gd d,\u d,\gtn{\d\a}g\ink{\a\om}h^{o\u d\u a}
k\gd o,\u a\u m,\cr
&+\xi k\gd n,d,\Bigl(\z^2 h^{\u d\u a}\brgn\om \F\gd d,\u a,
k\gd o,\u d\u b,k\gd o,\u m\u c,h^{o\u c\u b}
+\brgn\b \F\gd d,\u m,\gtn{\d\b}g\ink{\d\a}g\ink{\om\mu}\gtn{\a\mu}\Bigr)\cr
&-\xi^2k^{nb}k_{bd}\Bigl(\z\brgn\om \F\gd d,\u a,h^{o\u a\u b}k\gd o,\u m\u b,
+\gtn{\a\b}\brgn\a \F\gd d,\u m,g\ink{\om\b}\Bigr)\Bigr),
&\text{(A.64)}}
$$
where $Q\gd n,\om\u m,=-Q\gd n,\u m\om,$ is a tensor of torsion over $E$ and
$M$. 

}
\vfill\eject

\font\ninecyr=wncyr9
\font\ninecyit=wncyi9
\def\cyracc{\def\cydot{{\kern0pt}}%
	\def\cprime{\char"7E }\def\Cprime{\char"5E }%
	\def\u##1{\if \i##1\accent"24 i%
		\else \accent"24 ##1\fi }%
}

\def\cyr{\ninecyr\cyracc}
\def\cyrit{\ninecyit\cyracc}

\font\ntt=cmtt9
\font\nsl=cmsl9
\font\nineeuf=eufm9 at9\trpt
\def\fg{\hbox{\nineeuf g}}
\overfullrule0pt

\write0{References \the\pageno}
\null
\vskip2cc
\maxref={[100]}
\uref{
%\advance\baselineskip by6pt plus.7pt minus.7pt
\item{[1]} \sp{1}{Kaluza} T., {\it Zum Unit\"atsproblem der 
Physik}\/, Sitzgsber. Preuss. Akad. Wiss., p. 966, (1921). 

\item{[2]} \sp{1}{Klein} O. Z., Phys. {\bf 37}, p. 895 (1926). 
\item{}\sp{1}{Klein} O., {\it On the theory
of charged fields}\/, in New theories in Physics. Conference
organized in collaboration with International Union of
Physics and the Polish Co-operation Committee, Warsaw, May
30th-June 3rd, 1938, published in Paris, p. 77, (1939). 

\item{[3]} \sp{1}{Einstein} A., {\it The meaning of 
relativity}\/, Appendix II,
Fifth Edition, revised, p. 127 (Methuen and Co.: London,
U.K.), (1951). 

\item{}
\sp{1}{Jakubowicz} A. and \sp{1}{Klekowska} J., {\it The necessary and
sufficient condition for the existence of the unique
connection of the two-dimensional generalized Riemannian
space}\/, Tensor N.S {\bf 20} p. 72, (1969). 

\item{}
\sp{1}{Chung} K. T. and \sp{1}{Lee} Y. J., {\it Curvature Tensors and Unified
Field Equations in SEX$_{n}$}\/, Int. Journal of Theor. Phys. {\bf 27},
p. 1083 (1988). 

\item{}
\sp{1}{Chung} K. T. and \sp{1}{Hwang} H. J., {\it Three and 
Five-Dimensional Considerations of the Geometry of Einstein's
$^{*}\fg$-Unified Field Theory}\/, Int. Journal of theor. 
Phys. {\bf 27}, p. 1105, (1988). 

\item{}
\sp{1}{Shavokhina} N. S., {\it Nonsymmetric Metric in Nonlinear Field
Theory}\/, Preprint of the JINR, P2-86-685, Dubna 1986. 

\item{[4]} \sp{1}{Kaufmann} B., {\it Mathematical structure of the nonsymmetric
field theory}\/, Helv. Phys. Acta. Suppl. (1956), p. 227;

\item{}
\sp{1}{Chung} K. T., {\it Some Reccurence Relations for Einstein's
Connection for 2-Dimensional Unified Theory of Relativity}\/,
Acta Mathematica Hungarica {\bf 41} (1--2) p. 47 (1983). 

\item{[5]} \sp{1}{Einstein} A. and \sp{1}{Kaufmann} B., Ann. of Math., 
{\bf 59} (1954), p. 230; 

\item{}
\sp{1}{Kaufmann} B., Ann. of Math., {\bf 46} (1945), p. 578. 

\item{[6]} \sp{1}{Einstein} A., {\it A generalization of the relativistic theory
of  gravitation}\/, Ann.  of  Math., {\bf 46}, p. 578, (1945). 

\item{}
\sp{1}{Einstein} A. and \sp{1}{Strauss} E. G., {\it A generalization of the
relativistic theory of gravitiation II}\/, Ann.  of
Math., {\bf 47}, p. 731, (1946).

\item{[7]} \sp{1}{Kerner} R., {\it Generalization of Kaluza--Klein theory 
for an arbitrary nonabelian gauge group}\/, Ann. Inst. H. Poincare
IX, p. 143, (1968). 

\item{[8]} \sp{1}{Cho} Y. M., {\it Higher dimensional unification of gravitation and
gauge theories}\/, Journal of Math. Phys., {\bf 16}, p. 2029, (1975). 

\item{}
\sp{1}{Cho} Y. M. and \sp{1}{Freund} P. G. O., {\it Nonabelian gauge fields as
Nambu--Goldstone fields}\/, Phys. Rev. {\bf D12}, p. 1711, (1975). 

\item{[9]} \sp{1}{Kopczy{\'n}ski} W., {\it A fibre bundle description of coupled
gravitational and gauge fields}\/, in Differential
Geometrical Methods in Mathematical Physics (Springer,
Verlag, Berlin, Heidelberg, New York), p. 462, (1980). 

\item{[10]} \sp{1}{Kalinowski} M. W., {\it Vanishing of the cosmological constant in
nonabelian Klein--Kaluza theories}\/, Int. Journal of
Theor. Phys., {\bf 22}, p. 385, (1983). 

\item{[11]} \sp{1}{Thirring} W., {\it Five dimensional theories and $CP$ 
violation}\/, Acta Phys. Austriaca Suppl. IX (Wien, New York,
Springer), p. 256, (1972).

\item{[12]} \sp{1}{Kalinowski} M. W., {\it $PC$ nonconservation and dipole electric
moment of fermion in the Kaluza--Klein Theory}\/, Acta Phys. 
Austriaca {\bf 53}, p. 229, (1981). 

\item{[13]} \sp{1}{Kalinowski} M. W., {\it Spinor  fields in nonabelian
Klein--Kaluza theories}\/, Int. Journal  of  Theor. 
Physics, {\bf 23}, p. 131, (1984). 

\item{[14]} \sp{1}{Kalinowski} M. W., {\it $3/2$ spinor field in the 
Klein--Kaluza theory}\/, Acta Phys. Austriaca {\bf 55}, p. 197, (1983). 

\item{[15]} \sp{1}{Kalinowski} M. W., {\it Rarita--Schwinger in nonabelian
Klein--Kaluza theories}\/, Journal of Phys. {\bf A} (Math and 
Gen), 15, p. 2441. 144, (1982). 

\item{[16]} \sp{1}{Kalinowski} M. W., {\it Gauge fields with torsion}\/, Int. Journal
of Theor. Physics, {\bf 20}, p. 563, (1981). 

\item{[17]} \sp{1}{Einstein} A., {\it Zur Elektrodynamik bewegter
K\"oper}\/, Ann. Phys. (Germany), {\bf 17}, p. 891, (1905). 

\item{[18]} \sp{1}{Kalinowski} M. W., {\it The nonsymmetric Kaluza--Klein 
Theory}\/. Journal of Math. Phys., {\bf 24}, p. 1835, (1983). 

\item{} \sp{1}{Kalinowski} M. W., {\it Nonsymmetric Fields theory and
its Applications}\/, World Scientific, Singapore, New Jersey, London,
Hong Kong, 1990.

\item{} \sp{1}{Kalinowski} M. W., {\it Can we get Confinement from
extra dimensions}\/, in Physics of Elementary Interactions World Scientific, 
Singapore, New Jersey, London, Hong Kong 1991 (ed. Z. Ajduk, S.
Pokorski, A. K. Wr{\'o}blewski).

\item{} \sp{1}{Kalinowski} M. W., {\it Nonsymmetric Kaluza--Klein
(Jordan--Thiry) in a general Nonabelian Case}\/, Int. Journal
of Theor. Phys. {\bf 30} p.~281 (1991).

\item{} \sp{1}{Kalinowski} M. W., {\it Nonsymmetric Kaluza--Klein
(Jordan--Thiry) Theory in the Electromagnetic Case}\/, Int. Journal
of Theor. Phys. {\bf 31} p.~611 (1992).

\item{} \sp{1}{Kalinowski} M. W., {\it Geodetic equations in
nonsymetrics Jordan--Thiry theory}\/, I and II, Bull. Soc. Sci. 
Lettres \L\'od\'z 45 Ser. Rech. Deform. 
{\bf 19} p.~89 (1995), {\bf 19}, p.~117 (1995).

\item{} \sp{1}{Kalinowski} M. W., {\it Deformation of Riemannian Geometry and
Kaluza--Klein Theory}\/, Bull. Soc. Sci. Lettres \L\'od\'z {\bf XXV} p.~113
(1998). 

\item{[19]} \sp{1}{Kalinowski} M. W., {\it On the nonsymmetric
Jordan--Thiry theory}\/, Canadian Journal of Physics, {\bf 61}, p. 844, (1983). 

\item{[20]} \sp{1}{Jordan} P., {\it Schwerkraft und Weltal}\/, (Vieweg Verlag:
Braunschweig, 1955). 

\item{[21]} \sp{1}{Thiry} Y., {\it \'Etude mat\'ematique de equations d'une theorie
unitare  a  quinze  variables  de  champ}\/,
(th\'ese), (Gautiers-Villars, 1951).

\item{[22]} \sp{1}{Lichnerowicz} A., {\it Theorie relativistes de la gravitation et
de l'electromagnetisme}\/, (Masson: Paris), (1955). 

\item{[23]} \sp{1}{Kalinowski} M. W., {\it The  nonsymmetric-nonabelian
Kaluza--Klein Theory}\/, Journal of Physics A (Math. and
Gen.), {\bf 16}, p. 1669, (1983). 

\item{[24]} \sp{1}{Kalinowski} M. W., {\it The nonsymmetric-nonabelian 
Jordan--Thiry Theory}\/, II Nuovo Cimento LXXXA, p. 47, (1984).

\item{[25]} \sp{1}{Kalinowski} M. W., {\it Material sources in the nonsymmetric
Kaluza--Klein theory}\/, Journal of Math. Phys., {\bf 25}, 
p. 1045, (1984). 

\item{[26]} \sp{1}{Kalinowski} M. W., {\it Spontaneous symmetry breaking and Higgs'
mechanism in the nonsymmetric Kaluza--Klein theory}\/, Ann. 
of Phys., {\bf 148}\/, (New York), p. 241, (1983). 

\item{[27]} \sp{1}{Kalinowski} M. W., {\it Spontaneous symmetry breaking and Higgs'
mechanism in the nonsymmetric Jordan--Thiry theory}\/,
Fortschritte der Physik {\bf 34}, p. 361, (1986). 

\item{[28]} \sp{1}{Kalinowski} M. W. and \sp{1}{Mann} R. B., 
{\it Linear approximation in the nonsymmetric Kaluza--Klein theory}\/, 
Clasical and Quantum Gravity, {\bf 1}, p. 157, (1984). 

\item{[29]} \sp{1}{Kalinowski} M. W. and \sp{1}{Mann} R. B., 
{\it Linear approximation in the nonsymmetric Jordan--Thiry theory}\/, 
Nuovo Cimento {\bf 91B}, p. 67, (1986). 

\item{[30]} \sp{1}{Kalinowski} M. W. and \sp{1}{Kunstatter} G.,
{\it Spherically Symmetric solution in the nonsymmetric 
Kaluza--Klein theory}\/, Journal of Math. Phys., {\bf 25}, p. 117, (1984). 

\item{[31]} \sp{1}{Mann} R. B., {\it Exact solution of an algebraically extended
Kaluza--Klein theory}\/, Journal of Math.  Phys., {\bf 26}, 
p. 2308, (1985). 

\item{[32]} \sp{1}{Kalinowski} M. W., {\it $PC$ nonconservation and a dipole electric
moment of fermion in the nonsymmetric Kaluza--Klein
theory}\/, Int. Journal of Theor. Physics, {\bf 26}, p. 21, (1987). 

\item{[33]} \sp{1}{Kalinowski} M. W., {\it Spinor  field  in  the
nonsymmetric-nonabelian Kaluza--Klein theory}\/, Int. Journal
of Theor. Physics, {\bf 26}, p. 565, (1987). 

\item{[34]} \sp{1}{Moffat} J. W., {\it Generalized theory of gravitation and its
physical consequences}\/, in Proceeding of the VII
international School of Gravitation and Cosmology. Erice,
Scilly ed. by V. de Sabbata (World Scientific Publishing
Co.: Singapore), p. 127, (1982). 

\item{[35]} \sp{1}{Kunstatter} G., \sp{1}{Moffat} J. W. and 
\sp{1}{Malzan} J., {\it Geometrical
interpretation of a generalized theory of gravitation}\/,
Journal of Math. Phys., {\bf 24}, p. 886, (1983). 

\item{[36]} \sp{1}{Hilbert} D., Gott. Nachr., {\bf 12}, (1916). 

\item{[37]} \sp{1}{Lev{i-C}ivita}, Atti. R. Accad Naz. 
Lincei Mam. Cl. Sci. Fiz. Mat. Nat. {\bf 26}, p. 311, (1917); 
\sp{1}{Thiry} Y., J. Math. Pure Appl., {\bf 30}, p. 275, (1951). 

\item{[38]} \sp{1}{Lichnerowicz} A., {\it Sur certains problems 
globaux relatifs au
systeme des equations d'Einstein}\/, (Herman: Paris), 1939. 

\item{[39]} \sp{1}{Einstein} A. and \sp{1}{Pauli} W., Ann. Math.,
{\bf 44}, p. 131, (1943). 

\item{}
\sp{1}{Einstein} A., Rev. Univ. Nac. Tucuman, {\bf 2}, p. 11, (1941). 

\item{[40]} \sp{1}{Werder} R., {\it Absense of stationary solutions to 
Einstein--Yang--Mill's equations}\/, Phys. Rev., {\bf D25}, p. 2515, (1982). 

\item{}
\sp{1}{Bartnik} R. and \sp{1}{McKinnon} J., {\it Particlelike Solutions of the
Einstein--Yang--Mill's Equations}\/, Phys. Rev. Lett., {\bf 61}, p. 141,
(1988).

\item{[41]} \sp{1}{Kunstatter} G., {\it Finite Energy Electric Monopoles in an
Extended theory of gravitations}\/, Journal of Math. Phys., {\bf 25},
p. 2691, (1984). 

\item{[42]} \sp{1}{Roseveare} N. T., {\it Mercury's Perihelion: From Le 
Verier to Einstein}\/, Claredon Press Oxford, (1982). 

\item{[43]} \sp{1}{Hlavaty} V., {\it Geometry of Einstein's Unified Field 
Theory}\/, Nordhoff-Verlag, Groningen 1957.

\item{}
\sp{1}{Tonnelat} M. A., {\it Einstein's Unified Field Theory}\/, Gordon
and Breach Science Publishers, New York, London, Paris (1966). 

\item{[44]} \sp{1}{Hill} H. A., \sp{1}{Bos} R. J. and \sp{1}{Goode}
P. R., {\it Preliminary determination
of the quadrupole moment of the sun from Rotational
splitting of global oscillations and its relevance to test
of general relativity}\/, Physical Reviev Lett. {\bf 33},
p. 709, (1983). 

\item{}
\sp{1}{Hill} H. A., {\it Rotational splitting of Global Solar
Oscillations and its Relevance to test of General
Relativity}\/, Int. Journal of Theor. Physics {\bf 23}, p. 689,
(1984).

\item{}
\sp{1}{Gough} D. O., {\it Inertial rotation and gravitational quadrupole
moment of the Sun}\/, Nature {\bf 298}, p. 334, (1982). 

\item{[45]} \sp{1}{Moffat} J. W., {\it Consequences of a New experimental
Determination of the Quadrupole Moment of the Sun for
Gravitational Theory}\/, Physical Rev.  Lett. {\bf 50}, p. 709,
(1983). 

\item{}
\sp{1}{Campbell} L. and \sp{1}{Moffat} J. B., {\it Quadrupole moment of the Sun
and Planetary Orbits}\/, Astrophysical Journal {\bf 275}, p. L77,
(1983). 

\item{[46]} \sp{1}{Moffat} J. W., {\it The orbit of Icarus as a test of a Theory of
gravitation}\/, University of Toronto preprint, May (1982). 

\item{}
\sp{1}{Campbell} L., \sp{1}{McDow} J. C., \sp{1}{Moffat} J. W. and
\sp{1}{Vincent} D., {\it The sun's quadrupole moment and the perihelion precesion of
Mercury}\/, Nature {\bf 305}, p. 508, (1983).

\item{[47]} \sp{1}{Moffat} J. W., {\it Generalized theory of
gravitation}\/, Foundation
of Physics {\bf 14}, p. 1217, (1984).

\item{}
\sp{1}{Moffat} J. W., {\it Test of a theory of gravitation using the
Data from the Binary Pulsar $1913+16$}\/, University of Toronto
Report, August 1981. 

\item{}
\sp{1}{Krisher} T. P., {\it Dipole gravitational radiation in the 
non-symmetric theory of Moffat}\/, Phys. Rev. {\bf D32}, p. 329, (1985). 

\item{}
\sp{1}{Will} M. C., {\it Violation of weak equivalence principle in
theories of gravity with nonsymetric metric}\/, Phys. Rev. 
Left. {\bf 62}, p. 369, (1989).

\item{[48]} \sp{1}{Moffat} J. W., {\it Experimental Consequences of the Nonsymmetric
Gravitation theory including the Apsidal motion of
Binaries}\/, Lecture given at the conference on General
Relativity and Relativistic Astrophysics, University of
Dalhouse, Halifax, Nova Scotia, April 1985. 

\item{[49]} \sp{1}{Mcdow} J. C., {\it Testing the nonsymmetric theory of
gravitation}\/, PhD thesis, University of Toronto, 1983.

\item{}
\sp{1}{Hoffman} J. A., \sp{1}{Masshal} H. L. and \sp{1}{Lewin} W. G. H.,
Nature {\bf 271}, p. 630, (1978).

\item{[50]} \sp{1}{Bergman} P. G., {\it Comments on the scalar-tensor
theory}\/, Int. Journal of Theor. Physics, {\bf 1}, p. 52, (1968).

\item{[51]} \sp{1}{Trautman} A., {\it Fibre bundles associated with
space-time}\/, Rep. of Math. Phys. {\bf 1}, p. 29, (1970).

\item{[52]} \sp{1}{Utiyama} R., {\it Invariant Theoretical Interpretation of
Interaction}\/, Phys. Rev. {\bf 101}, p. 1597, (1956).

\item{[53]} \sp{1}{Stacey} F. D., \sp{1}{Tuck} G. J., \sp{1}{Moore} G. J.,
\sp{1}{Holding} S. C., \sp{1}{Goldwin} B. D. and \sp{1}{Zhou} R., {\it Geophysics 
and the law of gravity}\/, Rev. of Modern Physics {\bf 59}, p. 157,
(1987).

\item{}
\sp{1}{Ander} M. E., \sp{1}{Goldman} T., \sp{1}{Hughs} R. J. and
\sp{1}{Nieto} M. M.,  {\it Possible
Resolution of the Brookhaven and Washington E\"otv\"os
Experiments}\/, Phys. Rev. Lett. {\bf 60}, p. 1225, (1988). 

\item{}
\sp{1}{Eckhardt} D. H., \sp{1}{Jekeli} C., \sp{1}{Lazarewicz} A. R., 
\sp{1}{Romaides} A. J. and \sp{1}{Sands} R. W., {\it Tower Gravity Experiments: 
Evidence for Non-Newtonian Gravity}\/, Phys. Rev. Lett. {\bf 60},
p. 2567, (1988). 

\item{}
\sp{1}{Moore} G. I., \sp{1}{Stacey} F. D., \sp{1}{Tuck} G. J., 
\sp{1}{Goodwin} B. D., \sp{1}{Linthorne} N. P., \sp{1}{Barton} M.~A., 
\sp{1}{Reid} D. M. and \sp{1}{Agnew} G. D.,
{\it Determination of the gravitational constant at an
effective mass separation of 22 m}\/, Phys. Rev. {\bf D38} p. 1023,
(1988).

\item{[54]} \sp{1}{Fischbach} E., \sp{1}{Sudarsky} D., \sp{1}{Szafer} A.,
\sp{1}{Tolmadge} C. and \sp{1}{Arnson} S. H., {\it Reanalysis of the E\"otv\"os 
experiment}\/, Phys. Rev. Lett. {\bf 56}, p. 3, (1985).

\item{[55]} \sp{1}{Thieberg} P., {\it Search for a Substance-Dependent Force with a
New Differential Accelerometer}\/, Phys. Rev. Letters {\bf 58},
p. 1066, (1987).

\item{[56]} \sp{1}{Wesson} P. S., {\it Does gravity change with a time?}\/, 
Physics Today {\bf 33}, p. 32, (1980).

\item{[57]} \sp{1}{Gillies} G. T. and \sp{1}{Ritter} R. C., {\it Experiments on Variation of the
Gravitational Constant using Precision Rotations}\/, in
Precision Measurements and Fundamental Constants II ed. by
B. N. Taylor and W. D. Phillips, Natl. Bur. Stand. (US) Spec. 
Publ. 617, p. 629. (1984).

\item{[58]} \sp{1}{Rayski} J., {\it Unified theory and modern
physics}\/, Acta Physica Polonica Vol. XXVIII, p. 89, (1965).

\item{[59]} \sp{1}{Kobayashi} S. and \sp{1}{Nomizu} K., {\it Foundation  of  differential
geometry}\/, Vol. I and Vol. II, New York, (1963). 

\item{}
\sp{1}{Kobayashi} S., {\it Transformation Groups in differential
geometry}\/, Springer Verlag, Berlin (1972). 

\item{[60]} \sp{1}{Lichnerowicz} A., {\it Th\'eori\'e globale des connexions et de
group d'holonomie}\/, Ed. Cremonese, Rome, (1955). 

\item{[61]} \sp{1}{Hermann} R., {\it Yang--Mills, Kaluza--Klein 
and the Einstein Program}\/, Math. Sci. Press, Brookline, Mass. 1978. 

\item{}
\sp{1}{Coquereaux} R. and \sp{1}{Jadczyk} A., {\it Riemannian geometry, Fibre
bundle, Kaluza--Klein theory and all that $\ldots $}\/, World
Scientific Publishing, Singapore 1988. 

\item{[62]} \sp{1}{Zalewski} K., {\it Lecture on rotational group}\/,
PWN, Warsaw 1987 (in Polish).

\item{}
\sp{1}{Barut} A. O., \sp{1}{R{\c a}czka} R., {\it Theory of group representations and
applications}\/, PWN, Warsaw, (1980). 

\item{[63]} \sp{1}{Moffat} J. W., {\it Static spherically symmetric solution for the
field of a charged particle in a new theory of gravity}\/,
Phys. Rev. {\bf D19}, p. 3562, (1978). 

\item{[64]} \sp{1}{Kalinowski} M. W., {\it Comment on the nonsymmetric
Kaluza--Klein theory with material\break sources}\/, Zeitschrift f\"ur Phisik 
C (Particles and Fields) {\bf 33}, p. 76, (1986).

\item{[65]} \sp{1}{Moffat} J. W., {\it New theory of 
gravitation}\/, Phys. Rev. D {\bf 19}, p. 3557, (1979).

\item{[66]} \sp{1}{Moffat} J. W., {\it Gauge invariance and string interactions in a
generalized theory of gravitation}\/, Phys. Rev. D {\bf 23}, p. 2870,
(1981).

\item{[67]} \sp{1}{Moffat} J. W. and \sp{1}{Woolgar} E., 
{\it The Apsidal Motion of the Binary
Star in the Nonsymmetric Gravitational Theory}\/, University
of Toronto Report, 1984. 

\item{}
\sp{1}{Moffat} J. W., {\it The Orbital motion of DI Hercules as a test
of the Theory of Gravitation}\/, University of Toronto
Report, 1984. 

\item{[68]} \sp{1}{de\ Groot} S. R. and \sp{1}{Suttorp} R. G., 
{\it Foundations  of Electrodynamics}\/, North Holland 
Publishing Company, Amsterdam, 1972. 

\item{[69]} \sp{1}{Pleba{\'n}ski} J., {\it Nonlinear 
Electrodynamics}\/, Nordita Copenhagen, 1970. 

\item{[70]} \sp{1}{Kalinowski} M. W., {\it On a generalization of  the
Einstein--Cartan theory and Kaluza--Klein theory}\/, Lett in
Math. Phys. {\bf 5}, p. 489, (1981).

\item{}
\sp{1}{Kalinowski} M. W., {\it Torsion and the Kaluza--Klein 
theory}\/, Acta Phys. Austriaca {\bf 27}, p. 45, 1958. 

\item{[71]} \sp{1}{Wang} H. C., {\it On Invariant connection on principal fibre
bundle}\/, Nagoye Math. Journal {\bf 13}, p. 19, (1958). 

\item{[72]} \sp{1}{Harnad} J., \sp{1}{Shnider} S. and \sp{1}{Vinet} L., 
{\it Group actions on Principal Bundles and Invariance conditions for Gauge
Fields}\/, Journal of Math. Phys., {\bf 21}, p. 2719, (1980). 

\item{[73]} \sp{1}{Harnad} J., \sp{1}{Shnider} S. and \sp{1}{Tafel} J., {\it Group actions on
Principal Bundles and Dimensional Reduction}\/, Lett. in
Math. Phys. {\bf 4}, p. 107, (1980). 

\item{[74]} \sp{1}{Forgacs} P. and \sp{1}{Manton} N. S., {\it Space-time Symmetries in gauge
Theories}\/, Comm. Math. Phys. {\bf 72}, p. 15, (1980). 

\item{[75]} \sp{1}{Manton} N. S., {\it A New Six-Dimensional Approach to the
Weinberg--Salam Model}\/, Nucl. Phys. B {\bf 158}, p. 141, (1979). 

\item{} \sp{1}{Paw{\l}owski} M. and \sp{1}{R{\c a}czka} R., {\it A
Higgs' free model for Fundamental Interactions, Part I:
Formulation of the Model}\/, Preprint SINS -- 1P/VIII/1995
ILAS/EP--3--1995.

\item{} \sp{1}{Paw{\l}owski} M. and \sp{1}{R{\c a}czka} R., {\it A
Higgs' free model for Fundamental Interactions, Part II:
Predictions for Electroweak observables}\/, Preprint SINS -- 2P/VIII/1995
ILAS/EP--4--1995.

\item{[76]} \sp{1}{Trautman} A., {\it Geometrical Aspect of Gauge 
Configurations}\/, in new Developments in Mathematical Physics, Acta Phys. 
Austriaca Suppl. XXIII, p. 401, (1981), Springer Verlag
(1981). 

\item{[77]} \sp{1}{Witten} E., {\it Some Exact Multipseudo-Particle Solutions of
Clasical Yang--Mill's Theory}\/, Phys. Rev. Lett. {\bf 38}, p. 121,
(1977). 

\item{} \sp{1}{Hooft} G., {\it Magnetic monopoles in unified
gauge theones}\/, Nuclear Phys. B: 79 p.~276 (1976).

\item{[78]} \sp{1}{Maclenburg} W., {\it Geometrical unification of gauge and
Higgs' fields}\/, Triesre 1979 preprint JCTP JC/79/131. 

\item{[79]} \sp{1}{Maclenburg} W., {\it Towards a unifield theory for gauge and
Higgs' fields}\/, Trieste 1981 preprint JTCP JC/81/8. 

\item{[80]} \sp{1}{Chapline} G. and \sp{1}{Manton} N. S., {\it The Geometrical Significance of
Certain Higgs' Potentials: an Approach to grand
Unification}\/, Nucl. Phys. B {\bf 184}, p. 391, (1981). 

\item{[81]} \sp{1}{Sato} K., {\it Cosmological Consequences on a First Order Phase
Transition of a Vacuum}\/, Proceedings of the 4th Kyoto
summer Institute ed. M. Konuma and T. Maskawe, World
Scientific Publishing Co., Singapore (1981). 

\item{[82]} \sp{1}{Linde} A., {\it Coleman--Weinberg theory and inflationary
universe scenerio}\/, Phys. Lett. {\bf 114B}, p. 431, (1982). 

\item{[83]} \sp{1}{Linde} A. D., {\it A new Inflationary Universe Scenario: a
Possible Solution of the Horizon, Flatness, Homogenity,
Isotropy and Primordial Monopoles Problems}\/, Phys. Lett. 
{\bf 1088}, p. 389, (1982). 

\item{[84]} \sp{1}{Sher} M. A., {\it Effects of phase transitions of the evolution
of the early universe}\/. Phys. Rev. D {\bf 22}, p. 2989, (1980). 

\item{[85]} \sp{1}{Guth} A. H., {\it Inflantionary universe: A possible solution to
the horizon and flatness problems}\/, Phys. Rev. D {\bf 23},
p. 347, (1981). 

\item{[86]} \sp{1}{Helgason} S., {\it Differential Geometry, Lie groups and
Symmetric Spaces}\/, Academic Press, New York 1987. 

\item{}
\sp{1}{Helgason} S., {\it Groups and Geometric Analysis}\/, Academic
Press Inc., London, 1984. 

\item{}
\sp{1}{Wolf} J. A. and \sp{1}{Gray} A., {\it Homogeneous spaces defined by Lie
group automorphisms}\/ I, II Journal of Differential Geometry
{\bf 2} p. 77, (1968), {\bf 2} p. 115, (1968). 

\item{}
\sp{1}{Gray} A., {\it Nearly K\"ahlerian manifolds}\/, Journal of
Differential Geometry, {\bf 4}, p. 283, (1970). 

\item{}
{\cyr \sp{1}{{X}tukar{\cprime}} V. L.}, {\cyrit Invariantnye svyaznosti 
i metriki na odnorodnyh prostranstvah, sootvet\cydot stvu{yu}wih global\cprime nym 
tro{\u\i}kam}\/,
{\cyr Matematiqeskie zametki T. 26, str. 449 (1979)}.

\item{[87]} \sp{1}{Cheeger} J. and \sp{1}{Ebin} D. G., {\it Comparision Theorem in Riemannian
Geometry}\/, North Holland/American Elsevier, New York,
Amsterdam, Oxford (1975). 

\item{[88]} \sp{1}{Coleman} S. and \sp{1}{Weinberg} E., {\it Radiative Corrections as the
Origin of the Spontaneous Symetry Breaking}\/, Phys. Rev. 
D {\bf 7}, p. 1888, (1973).

\item{[89]} \sp{1}{Weinberg} S., {\it Gauge hierarchies}\/, Phys. 
Lett. {\bf 82B} p. 387. 

\item{[90]} \sp{1}{Moffat} J. W., {\it Phase Transition in the Early Universe as a
consequence of a Generalized Theory of Gravitation}\/,
invited talk presented at the Conference of Differential
Geometric Methods in Theoretical Physics, Trieste, Italy
June 30-July 3, (1981). 

\item{[91]} \sp{1}{Moffat} J. W. and \sp{1}{Vincent} D., {\it The Early Universe in a
Generalised Theory of Gravitation}\/, Can. Journal of Phys. 
{\bf 60}, p. 659, (1982). 

\item{[92]} \sp{1}{Kalinowski} M. W., {\it The minimal coupling scheme for Dirac's
field in the nonsymmetric theory of gravitation}\/, Journal
of Modern Physics {\bf A1}, p. 185, (1981). 

\item{[93]} \sp{1}{Langacker} P., {\it Grand Unified Theories and Proton 
Decay}\/, Phys. Rep. {\bf 72}, p. 185, (1981). 

\item{[94]} \sp{1}{Mann} R. B. and \sp{1}{Moffat} J. W., {\it Post-Newtonian approximation of new
theory of gravity}\/, Can J. Phys. {\bf 59}, p. 1592, (1981). 

\item{[95]} \sp{1}{Moffat} J. W., {\it Test Particle Motion in the Nonsymmetric
Gravitation Theory}\/, Phys. Rev. {\bf D35} p. 3733, (1987).

\item{[96]} \sp{1}{Moffat} J. W., \sp{1}{Savaria} R. and
\sp{1}{Woolgar} E., {\it E\"otv\"os Test of the
Nonsymmetric Gravitation Theory and the Fifth Force}\/,
University of Toronto Report, 1986. 

\item{}
\sp{1}{Moffat} J. W. and \sp{1}{Woolgar} E., {\it Motion of masive bodies: Testing
the nonsymmetric gravitation theory}\/, Phys. Rev. {\bf D37},
p. 918, (1988). 

\item{}
\sp{1}{Krisher} T. P., Astrophys. J. (Lett) {\bf 320} p. 147, (1987). 

\item{}
\sp{1}{Moffat} J. W., {\it Cosmions in the Nonsymmetric Gravitational
Theory}\/, Phys. Rev. {\bf D39}, p. 474, (1989). 

\item{}
\sp{1}{Mann} R. B., \sp{1}{Moffat} J. W. and \sp{1}{Palmer} J. H., 
{\it Comment on ``Violation
of the Weak Equivalence Principle in Theories of Gravity
with a Nonsymmetric Metric"}\/. MATPHYS--TH 89/01, U of T
Report, Toronto 1989.

\item{}
\sp{1}{Krisher} T. P., {\it Possible test at Jupiter of the nonsymmetric
gravitational theory}\/, Phys. Rev. {\bf D40} p. 1373, (1989).

\item{[97]} \sp{1}{Einstein} A., {\it The Bianchi identities in the generalized
theory of gravitation}\/, Can. J. Math. {\bf 2}, p. 120, (1950). 

\item{[98]} \sp{1}{Kunstatter} G. and \sp{1}{Moffat} J. W., {\it Conservation laws in a
generalized theory of gravitation}\/, Phys. Rev. {\bf D19},
p. 1084, (1979). 

\item{[99]} \sp{1}{Rudjobing} M., Ann Astrophys. {\bf 22}, p. 111, (1959). 

\item{}
\sp{1}{Martynov} D. Ya. and \sp{1}{Khaliulin} Kh. F., Astrophysics Space Sc. {\bf 71},
p. 147, (1980). 

\item{}
\sp{1}{Popper} D. M., Ap. J. {\bf 188}, p. 559, (1974). 

\item{}
\sp{1}{Popper} D. M., Ap. J. {\bf 254}, p. 209, (1982). 

\item{}
\sp{1}{Guinan} E. F. and \sp{1}{Maloney} F. P., 163-rd AAS Meeting, Las Vegas
(Abstract);

\item{}
\sp{1}{Khaliulin} Kh. F., Sov. Astron. {\bf 27}, p. 43, (1983). 

\item{}
\sp{1}{Koch} R. H., Ap. J. {\bf 183}, p. 275, (1973).

\item{}
\sp{1}{Koch} R. H., Astron. J. {\bf 82}, p. 653, (1977). 

\item{[100]} \sp{1}{Will} C. M., {\it Theory and Experiment in Gravitational
Physics}\/, Cambridge University Press, 1981.

\item{[101]} \sp{1}{Cornish} N. J., \sp{1}{Moffat} J. M., {\it
Non-singular gravity without black holes}\/, Journal of Math. Phys.
{\bf 35}, p.~6628 (1994).

\item{[102]} \sp{1}{Cornish} N. J., \sp{1}{Moffat} J. M., {\it
A non-singular theory of gravity~?}\/, Phys. Lett. {\bf B 336} p.~337.

\item{[103]} \sp{1}{Cornish} N. J., {\it Analysis of the Nonsingular
Wyman--Schwarzschild Metric}\/, Modern Physics Letters {\bf A9},
p.~3629 (1994).

\def\ii#1 {\item{[#1]}}

\ii104 \sp{1}{Bahcall} N. A., \sp{1}{Ostriker} J. P., \sp{1}{Perlmutter} S., 
\sp{1}{Steinhardt} P. J.,
{\it The cosmic triangle: revealing the state of the Universe}\/,
Science {\bf284}, p.~1481 (1999).

\item{} \sp{1}{Wang} L., \sp{1}{Caldwell} R. R., \sp{1}{Ostriker} J. P., 
\sp{1}{Steinhardt} P. J.,
{\it Cosmic concordance and quintessence}\/, 
arXiv: astro-ph/9901388v2.

\item{} \sp{1}{Wiltshire} D. L., 
{\it Supernovae Ia, evolution and quintessence}\/, 
arXiv: astro-ph/0010443.

\item{} \sp{1}{Primack} J. R.,
{\it Cosmological parameters},
Nucl. Phys.~B (Proc. Suppl.) {\bf87}, p.~3 (2000).

\item{} \sp{1}{Bennett} C. L. et al., 
{\it First Year Wilkinson Microwave Anisotropy Probe (WMAP) Observations:
Determination of Cosmological Parameters}\/,
arXiv: astro-ph/0302209v2.

\ii105 \sp{1}{Caldwell} R. R., \sp{1}{Dave} R., \sp{1}{Steinhardt} P. J.,
{\it Cosmological imprint of an energy component with general equation
of state}\/,
Phys. Rev. Lett. {\bf80}, p.~1582 (1998).

\item{} \sp{1}{Armeendariz-Picon} C., \sp{1}{Mukhanov} V., 
\sp{1}{Steinhardt} P. J.,
{\it Dynamical solution to the problem of a small cosmological constant and
late-time cosmic acceleration}\/,
Phys. Rev. Lett. {\bf85}, p.~4438 (2000).

\item{} \sp{1}{Maor} J., \sp{1}{Brustein} R., \sp{1}{Steinhardt} P. J.,
{\it Limitations in using luminosity distance to determine the equation of
state of the Universe}\/,
Phys. Rev. Lett. {\bf86}, p.~6 (2001).

\ii106 \sp{1}{Axenides} M., \sp{1}{Floratos} E. G., \sp{1}{Perivolaropoulos} L.,
{\it Some dynamical effects of the cosmological constant}\/,
Modern Physics Letters {\bf A15}, p.~1541 (2000).

\ii107 \sp{1}{Vishwakarma} R. G.,
{\it Consequences on variable $\Lambda$-models from distant type Ia
supernovae and compact radio sources}\/,
Class Quantum Grav. {\bf18}, p.~1159 (2001).

\ii108 \sp{1}{Di{\'a}z-Rivera} L. M., \sp{1}{Pimentel} L. O.,
{\it Cosmological models with dynamical $\Lambda$ in scalar-tensor 
theories}\/,
Phys. Rev. {\bf D66}, p.~123501-1 (1999).

\item{} \sp{1}{Gonz{\'a}lez-Di{\'a}z} P. F.,
{\it Cosmological models from quintessence}\/,
Phys. Rev. {\bf D62}, p.~023513-1 (2000).

\item{} \sp{1}{Frampton} P. H.,
{\it Quintessence model and cosmic microwave background}\/,
astro-ph/0008412.

\item{} \sp{1}{Dimopoulos} K.,
{\it Towards a model of quintessential inflation}\/,
Nucl. Phys.~B (Proc. Suppl.) {\bf95}, p.~70 (2001).

\item{} \sp{1}{Charters} T. C., \sp{1}{Mimoso} J. P., \sp{1}{Nunes} A.,
{\it Slow-roll inflation without fine-tun\-ning}\/,
Phys. Lett. {\bf B472}, p.~21 (2000).

\item{} \sp{1}{Brax} Ph., \sp{1}{Martin} J.,
{\it Quintessence and supergravity}\/,
Phys. Lett. {\bf B468}, p.~40 (1999).

\item{} \sp{1}{Sahni} V.,
{\it The cosmological constant problem and quintessence}\/,
Class. Quantum Grav. {\bf19}, p.~3435 (2002).

\ii109 \sp{1}{Liddle} A. R.,
{\it The early Universe}\/,
arXiv: astro-ph/9612093.

\item{} \sp{1}{Gong} J. O., \sp{1}{Stewart} E. D.,
{\it The power spectrum for multicomponent inflation to second-order
corrections in the slow-roll expansion}\/,
Phys. Lett. {\bf B538}, p.~213 (2002).

\item{} \sp{1}{Sasaki} M., \sp{1}{Stewart} E. D.,
{\it A general analytic formula for the spectral index of the density
perturbations produced during inflation}\/,
Progress of Theoretical Physics {\bf95}, p.~71 (1996).

\item{} \sp{1}{Nakamura} T. T., \sp{1}{Stewart} E. D.,
{\it The spectrum of cosmological perturbations produced by a
multicomponent inflation to second order in the slow-roll approximation}\/,
Phys. Lett. {\bf B381}, p.~413 (1996).

\item{} \sp{1}{Turok} N.,
{\it A critical review of inflation}\/,
Class. Quantum Grav. {\bf19}, p.~3449 (2002).

\ii110 \sp{1}{Liddle} A. R., \sp{1}{Lyth} D. H., 
{\it Cosmological Inflation and Large-Scale Structure}\/,
Cambridge Univ. Press, Cambridge 2000.

\item{} \sp1{Peebles} P. J. E.,
{\it Principles of Physical Cosmology\/},
Princeton Univ. Press, Princeton, 1993.

\item{} \sp1{Tsagas} Ch. G.,
{\it Cosmological Perturbations\/},
arXiv: astro-ph/0201405v1.

\item{} \sp1{Durrer} R.,
{\it The Theory of CMB Anisotropies\/},
arXiv: astro-ph/0109522v1.

\item{} \sp1{Bertschinger} E.,
{\it Cosmological Perturbation Theory and Structure Formation\/},
arXiv: astro-ph/0101009v1.

\item{} \sp1{Brandenberger} R. H.,
{\it Lectures on the Theory of Cosmological Perturbations\/},
arXiv: hep-th/0306071v1.

\item{} \sp1{Challinor} A.,
{\it Anisotropies in the Cosmic Microwave Background\/},
arXiv: astro-ph/0403344v1.

\ii111 \sp{1}{Steinhardt} P. J., \sp{1}{Wang} L., \sp{1}{Zlatev} I.,
{\it Cosmological tracking solutions}\/,
Phys. Rev. {\bf D59}, p.~123504-1 (1999).

\item{} \sp{1}{Steinhardt} P. J.,
{\it Quintessential Cosmology and Cosmic Acceleration}\/,
Web page\hfil\break {\ntt http://feynman.princeton.edu/\~{}steinh}

\ii112 \sp1{Painlev{\'e}} P.,
{\it Sur les \'equations diff\'erentielles du second ordre et d'ordre
sup\'erieur dont l'int\'e\-grale g\'en\'erale est uniforme\/},
Acta Mathematica {\bf25}, p.~1 (1902).

\def\h{\hskip2pt plus0.2pt minus0.2pt}
\ii113 \sp{1}{Steigman} G., 
%{\it Primordial\h alchemy:\h from\h the\h big-bang\h to\h the\h present\h Universe},
{\it Primordial alchemy: from the big-bang to the present Universe},
arXiv: astro-ph/0208186.

\item{} \sp{1}{Steigman} G.,
{\it The baryon density though the (cosmological) ages}\/.

\item{} \sp{1}{Scott} J., \sp{1}{Bechtold} J., \sp{1}{Dobrzycki} A., 
\sp{1}{Kul-Karni} V.~P.,
{\it A uniform analysis of the LY-$\alpha$ forest of $Z=0-5$}\/,
arXiv: astro-ph/0004155.

\item{} \sp{1}{Hui} L., \sp{1}{Haiman} Z., \sp{1}{Zaldarriaga} M., 
\sp{1}{Alexander} T.,
{\it The cosmic baryon fraction and the extragalactic ionizing background}\/,
arXiv: astro-ph/0104442.

\item{} \sp{1}{Balbi} A., \sp{1}{Ade} P., \sp{1}{Bock} J., \sp{1}{Borrill} J., 
\sp{1}{Boscaleri}~A., \sp{1}{De\ Bernardis}~P.,
{\it Constraints on cosmological parameters from Maxima-$1$}\/,
arXiv: astro-ph/0005124.

\item{} \sp{1}{Jeff} A. H. et al.,
{\it Cosmology from Maxima-$1$, BOOMERANG \& COBE/DMR CMB observations}\/,
arXiv: astro-ph/0007333.

\ii114 \sp{1}{Serna} A., \sp{1}{Alimi} J. M.,
{\it Scalar-tensor cosmological models}\/,
arXiv: astro-ph/9510139.

\item{} \sp{1}{Serna} A., \sp{1}{Alimi} J. M.,
{\it Constraints on the scalar-tensor theories of gravitation from
primordial nucleosynthesis}\/,
arXiv: astro-ph/9510140.

\item{} \sp{1}{Navarro} A., \sp{1}{Serna} A., \sp{1}{Alimi} J. M.,
{\it Search for scalar-tensor gravity theories with a non-monotonic time
evolution of the speed-up factor}\/,
Class. Quantum Grav. {\bf19}, p.~4361 (2002).

\item{} \sp{1}{Serna} A., \sp{1}{Alimi} J. M., \sp{1}{Navarro} A., 
{\it Convergence of scalar-tensor theories toward General Relativity and
primordial nucleosynthesis}\/,
arXiv: gr-qc/0201049.

\ii115 \sp1{Kalinowski} M. W.,
{\it Dynamics of Higgs' field and a quintessence
in the nonsymmetric Kaluza--Klein (Jordan--Thiry) theory}, 
arXiv: hep-th/0306241. 

\item{} \sp1{Kalinowski} M. W.,
{\it Cosmological models
in the nonsymmetric Kaluza--Klein theory}, 
arXiv: astro-ph/0310745v3. 

\ii116 \sp{1}{K{\"a}lbermann} G., 
{\it Communication through an extra dimension}\/,
Int.~J. of Modern Phys.~{\bf A15}, p.~3197 (2000).

\item{} \sp{1}{Visser} M.,
{\it An exotic class of Kaluza-Klein model}\/,
Phys. Lett. {\bf B159}, p.~22 (1985).

\item{} \sp{1}{Li} Li-Xin, \sp{1}{Gott} III J. R.,
{\it Inflation in Kaluza-Klein theory: Relation between the fine-structure
constant and the cosmological constant}\/,
Phys. Rev. {\bf D58}, p.~103513-1 (1998).

\item{} \sp{1}{Squires} E. J.,
{\it Dimensional reduction caused by a cosmological constant}\/,
Phys. Lett. {\bf B167}, p.~286 (1986).

\item{} \sp{1}{Caldwell} R., \sp{1}{Langlois} D.,
{\it Shortcuts in the fifth dimension}\/,
Phys. Lett. {\bf B511}, p.~129 (2001).

\item{} \sp1{Loup} F., \sp1{Santos} P. A., \sp1{Martins\ da\ Silva\ Santos}
D., 
{\it Can Geodesics in Extra Dimensions Solve Cosmic Light Speed Limit?\/},
Gen. Rel. and Grav. {\bf 35}, p.~1849 (2003).

\item{} \sp1{Loup} F., \sp1{Santos} P. A., \sp1{Martins\ da\ Silva\ Santos}
D., 
{\it The Dynamics of Chung-Freese Braneworld\/},
Gen. Rel. and Grav. {\bf 35}, p.~2035 (2003).

\ii117 \sp{1}{Dick} R.,
{\it Brane worlds}\/,
Class. Quantum Grav. {\bf18}, p.~R1 (2001).

\item{} \sp{1}{Brax} Ph., \sp{1}{Falkowski} A., \sp{1}{Lalak} Z.,
{\it Non-BPS branes of supersymmetric brane worlds}\/,
Phys. Lett. {\bf B521}, p.~105 (2001).

\ii118 \sp{1}{Randall} L., \sp{1}{Sundrum} R.,
{\it An alternative to compactification}\/,
arXiv: hep-th/9906064.

\ii119 \sp{1}{Arkami-Hamed} N., \sp{1}{Dimopoulos} S., \sp{1}{Dvali} G.,
{\it The hierarchy problem and new dimensions at a milimeter}\/,
Phys. Lett. {\bf B429}, p.~263 (1998).

\ii120 \sp1{Kalinowski} M. W.,
{\it A Warp Factor in the nonsymmetric Kaluza--Klein (Jordan--Thiry) theory},
arXiv: hep-th/0306157v3.

\item{} \sp1{Kalinowski} M. W.,
{\it Hierarchy of symmetry breaking in the nonsymmetric Kaluza--Klein\break
(Jordan--Thiry) theory}, 
arXiv: hep-th/0307135v2.

\item{} \sp1{Kalinowski} M. W.,
{\it Higgs' Fields in the Nonsymmetric Field Theory}, Warsaw 1995.

\item{[121]} \sp{1}{Damour} T., \sp{1}{Deser} S. and
\sp{1}{MacCarthy} J., {\it Nonsymmetric gravity theories: Inconnstences 
and a cure}\/, Phys. Rev. {\bf D47} p.~1541 (1993).

\item{[122]} \sp{1}{Misner} C. W., \sp{1}{Thorne} K. S. and
\sp{1}{Wheeler} J. A., {\it Gravitation}\/, San Francisco:
Freeman (1973).

\item{} \sp{1}{Hulse} R. A. and \sp{1}{Taylor} J. H., Astrophys.
J. {\bf 195} p.~L51 (1975).

\item{[123]} \sp{1}{Moffat} J. W., {\it Nonsymmetric 
grvitational theory}\/, J. Math. Phys. {\bf 36}, p. 3722 (1995).

\item{[124]} \sp{1}{Cones} A., {\it Noncommutative 
Geometry}\/, Academic
Press Inc., London UK, San Diego California USA, 1994.

\item{} \sp{1}{Martin} C. P., \sp{1}{Garcia-Bondia} J. M.
and \sp{1}{Varily} J. C., 
{\it The Standard Model as a noncommutative geometry: 
the low-energy regime}\/, 
Physics Reports {\bf 294}, p. 363 (1998).

\item{} \sp{1}{Jochun} B. and \sp{1}{Sch{\"u}cker} T., 
{\it Yang--Mills--Higgs versus Connes--Lott}\/,
Comm. in Math. Physics {\bf 178} p. 1 (1996).

\item{} \sp{1}{Jochun} B. and \sp{1}{Sch{\" u}cker} T., 
{\it A Left-Right Symmetric Model {\`a} la Connes--Lott}\/, 
Lett. in Math. Physic {\bf 32}, p. 153 (1994).

\item{} \sp{1}{Coquereaux} R., \sp{1}{H{\" a}ubling} R., 
 \sp{1}{Papadopoulos} N. A. and \sp{1}{Scheck} F., 
{\it Generalized gauge transformations and hidden 
symmetry in the standard model}\/, Int. Journal of 
Modern Phys. {\bf 7}, p. 2809 (1992).

\item{} \sp{1}{Dubois-Violette} M., \sp{1}{Kerner} R. 
and \sp{1}{Madore} J., 
{\it Noncommutative differential
geometry of matrix algebras}\/, J. Math. Phys. {\bf 31}, p.
316 (1990).

\item{} \sp{1}{Madore} J., {\it An Introduction to Noncommutative 
Differential Geometry and Its Physical Application}\/,
Cambridge University Press, Cambridge 1995.

\ii 125 \sp{1}{Ashtekar} A.,
{\it New variables for classical and quantum gravity}\/,
Phys. Rev. Lett. {\bf 57}, p.~2244 (1986).

\item{} \sp{1}{Ashtekar} A.,
{\it New Hamiltonian formulation of general relativity}\/,
Phys. Rev. {\bf D36}, p.~1587 (1987).

\item{} \sp{1}{Ashtekar} A., \sp{1}{Lewandowski} J., \sp{1}{Marolf} D.,
\sp{1}{Mour{\~a}o} J. and~J., \sp{1}{Thiemann} T.,
{\it Quantization of diffeomorphism invariant theories of connections with
local degrees of freedom}\/,
J.~Math. Phys. {\bf36}, p.~6456 (1995).

\item{} \sp{1}{Ashtekar} A., \sp{1}{Lewandowski} J. and \sp{1}{Sahlmann}~H.,
{\it Polymer and Fork representations for scalar field}\/,
Class. Quantum Gravity {\bf20}, L1--11 (2003).

\item{} \sp{1}{Rovelli} C.,
{\it Notes for a brief history of quantum gravity}\/,
arXiv: gr-qc/0006061.

\item{} \sp1{Gaul} M., \sp1{Rovelli} C.,
{\it Loop quantum gravity and the meaning of diffeomorphism invariance}\/,
Lect. Notes Phys. {\bf 541}, p.~277 (2000).

\item{} \sp1{Thiemann} T.,
{\it Lectures on Loop Quantum Gravity\/},
arXiv: gr-qc/0210094v1.

\item{} \sp1{Thiemann} T.,
{\it Quantum Spin Dynamics $($QSD$):$ {\rm VII.}~Symplectic Structures and
Continuum Lattice Formulations of Gauge Field Theories\/},
arXiv: hep-th/0005232v1, Class. Quant. Grav. {\bf18}, p.~3293 (2001).

\item{} \sp1{Ashtekar} A.,
{\it A Generalized Wick Transform for Gravity}\/,
arXiv: gr-gc/9511083v2, Phys. Rev. {\bf D53}, p.~2865 (1996).

\item{} \sp1{Ashtekar} A., \sp1{Lewandowski} J.,
{\it Background Independent Quantum Gravity: A Status Report\/},
arXiv: gr-gc/0404018v1.

\ii126 \sp1{Donoghue} J. F.,
{\it The Quantum Theory of General Relativity at Low Energies}\/,
arXiv: gr-gc/9607039v1.

\item{} \sp1{Donoghue} J. F.,
{\it General relativity as an effective field theory: The leading quantum
corrections\/},
Phys. Rev. {\bf D50}, p.~3874 (1994).

\item{} \sp1{Donoghue} J. F.,
{\it Introduction to the Effective Field Theory. Description of Gravity\/},
arXiv: gr-gc/9512024v1.

\item{} \sp1{Donoghue} J. F., \sp1{Torma} T.,
{\it On the power counting of loop diagrams in general relativity\/},
arXiv: hep-th/9602121v1.

\item{} \sp1{Bjerrum-Bohr} N. E. J., \sp1{Donoghue} J. F., \sp1{Holstein} B. R.,
{\it Quantum corrections to the Schwarzschild and Kerr metrics\/},
Phys. Rev. {\bf D68}, p.~084005 (2003).

\item{} \sp1{Donoghue} J. F.,
{\it Leading Quantum Correction to the Newtonian Potential\/},
Phys. Rev. Lett. {\bf72}, p.~2996 (1994).

\item{} \sp1{Donoghue} J. F., \sp1{Torma} T.,
{\it Infrared behaviour of graviton-graviton scattering\/}
arXiv: hep-ph/9901156v1.

\item{} \sp1{Donoghue} J. F., \sp1{Holstein} B. R., \sp1{Garbrecht} B.,
\sp1{Konstandin} T.,
{\it Quantum Corrections to Reissner-Nordstr\"om and Kerr-Newman metrics\/},
Phys. Lett. {\bf B529}, p.~132 (2002).

\item{} \sp1{Bjerrum-Bohr} N. E. J., \sp1{Donoghue} J. F., \sp1{Holstein} B. R.,
{\it Quantum gravitational corrections to the nonrelativistic scattering
potential of two masses\/},
Phys. Rev. {\bf D67}, p.~084033 (2003).

\ii127 \sp1{Colombeau} J. F.,
{\it Differential Calculus and Holomorphy\/},
North-Holland Publishing Comp., Amsterdam 1982.

\item{} \sp1{Colombeau} J. F.,
{\it New Generalized Functions and Multiplication of Distribution\/},
North-Holland Publishing Comp., Amsterdam 1984.

\item{} \sp1{Colombeau} J. F.,
{\it Elementary Introduction to New Generalized Functions\/},
North-Holland Publishing Comp., Amsterdam 1985.

\item{} \sp1{Colombeau} J. F.,
{\it Multiplication of Distributions\/},
Springer-Verlag, Berlin 1992 (Lecture Notes in Mathematics 1532).

\item{} \sp1{Charpentier} E.,
{\it Sur l'\'elimination des ``infinis'' en th\'eorie quantique des champs:
la r\'egu\-la\-ri\-sa\-tion zeta \`a l'\'epreuve de l'interpr\'etation de Colombeau
ou vice verse\/},
Dissertationes Mathematicae CCCLXXXIII, Warszawa 1999.

\ii128 {\ntt www.cmbfast.org}

\item{}{\ntt www.cmbeasy.org}

\item{} \sp1{Dorau} M., 
{\it CMBEASY: an Object Oriented Code for the Cosmic Microwave Background\/},
arXiv: astro-ph/0302138v1.

\ii129 \sp1{Kalinowski} M. W., 
{\it On some cosmological consequences of the Nonsymmetric
Kaluza--Klein (Jordan--Thiry) Theory\/}, 
arXiv: hep-th/0307289v2.

\item{} \sp1{Gott} III J. R. et al., 
{\it A Map of the Universe\/},
arXiv: astro-ph/0310571v1.

\ii130 \sp1{Kalinowski} M. W., 
{\it A short note on a total amount of inflation}\/, 
arXiv: hep-th/0307299v2.

\ii131 \sp1{Emden} R.,
{\it Gaskugeln: Anwendungen der mechanischer W\"arme\/},
B.~G.~Teubner, Leipzig und Berlin 1907.

\item{} \sp1{Chandrasekhar} S.,
{\it Introduction to the Study of Stellar Structure\/},
New York, Dover 1939.

\ii132 \sp1{Lemke} H.,
{\it \"Uber die Differentialgleichungen, welche den Gleichgewichtszustand
eines gas\-f\"or\-migen Himmelsk\"orpers bestimmen, dessen Teile gegeneinander
nach dem Newtonschen Gesetze gra\-vi\-tieren\/},
Journal f\"ur Mathematik, Bd.~142, Heft~2, S.~118 (1913).

\ii133 \sp1{Steinhardt} P. J.,
{\it The quintessential introduction to dark energy\/},
Lecture given at {\nsl The search of dark matter and dark energy in the
Universe\/}, The Royal Society, January 22--23, 2003.

\item{} \sp1{DeDeo} S., \sp1{Caldwell} R. R., \sp1{Steinhardt} P. J.,
{\it Effects of the sound speed of quintessence on the microwave background and large
scale structure\/},
arXiv: astro-ph/0301284v2.

\item{} \sp1{Erickson} J. K., \sp1{Caldwell} R. R., \sp1{Steinhardt} P. J.,
\sp1{Armendriz-Picon} C., \sp1{Mukhanov}~V., 
{\it Measuring the speed of sound of quintessence\/},
arXiv: astro-ph/0112438v1.

\item{} \sp1{Rahul} D., \sp1{Caldwell} R. R., \sp1{Steinhardt} P. J.,
{\it Sensitivity of the cosmic microwave background anisotropy to initial
conditions in quintessence cosmology\/},
arXiv: astro-ph/0206372v2.

\item{} \sp1{Caldwell} R. R., \sp1{Steinhardt} P. J.,
{\it The imprint of gravitational waves in models dominated by a dynamical cosmic
scalar field\/},
arXiv: astro-ph/9710062v1.

\ii134 \sp1{Chakraborty} S., \sp1{Chakraborty} N. C., \sp1{Debnath} U.,
{\it A quintessence problem in self-inter\-act\-ing Brans-Dicke theory\/},
arXiv: gr-gc/0305103v1.

\item{} \sp1{Chakraborty} S., \sp1{Chakraborty} N. C., \sp1{Debnath} U.,
{\it Quintessence problem and Brans-Dicke theory\/},
arXiv: gr-qc/0306040v1.

\ii135 \sp1{Kalinowski} M. W.,
{\it Comment on a quintessence particle mass in the Kaluza--Klein Theory
and properties of its field\/},
arXiv: astro-ph/0310300v6.

\ii136
\sp1{Antoci} S.,
{\it Stationary Axially Symmetric Solutions of the Hermitian Theory of
Relativity\/{\rm:} Generalization of the Papapetrou Class of Vacuum Fields}\/,
Lettere al Nuovo Cimento {\bf 36}, p.~16 (1983).

\ii137
\sp1{Lal} K. B., \sp1{Ali} Najaf,
{\it The wave solutions of the field equations of Einstein's, Bonnor's and
Schr\"odinger's non-symmetric unified field theories in generalized
Takeno-space-time}\/, 
Tensor (N.S.) {\bf21}, p.~241 (1970).

\ii138
\sp1{Lal} K. B., \sp1{Khare} D. C.,
{\it On solutions of Einstein's unified field equations of gravitation and
electromagnetism in $V_2\times V_2$ space-time}\/,
Tensor (N.S.) {\bf20}, p.~335 (1969).

\ii139
\sp1{Vaidya} P. C.,
{\it Unified gravitational and electromagnetic waves}\/,
Progr. Theoret. Phys. {\bf25}, p.~305 (1961).

\ii140
\sp1{Joshi} R. L., \sp1{Husain} S. I.,
{\it Total radiation in Einstein's unified field theory}\/,
Tensor (N.S.) {\bf 15}, p.~66 (1964).

\ii141 \sp1{De\ Witt} B.,
{\it Supermanifolds\/}, second edition,
Cambridge University Press, Cambridge 1992.

\ii142 \sp1{Kac} V. G.,
{\it A sketch of Lie superalgebra\/},
Comm. in Math. Phys. {\bf53}, p.~31 (1977).

\ii143 \sp1{Kac} V. G.,
{\it Lie superalgebras\/},
Advances in Math. {\bf26}, p.~8 (1977).

\ii144 \sp1{Nath} P., \sp1{Arnowitt} R., \sp1{Chamseddine} A. H.,
{\it Applied $N=1$ Supergravity\/},
World Scientific, London, Singapore, Hongkong 1984.

\ii145 \sp1{Zwart} Ph. B., \sp1{Boothby} W. M.,
{\it On compact homogeneous symplectic manifolds\/}, 
Ann. Inst. Fourier Grenoble {\bf30}, p.~129 (1980).

\item{} \sp1{Tralle} A., \sp1{Oprea} J.,
{\it Symplectic manifolds with non-K\"ahler structure},
Springer-Verlag, Berlin-Heidelberg 1997.

\ii146 \sp1{Wess} J., \sp1{Bagger} J.,
{\it Supersymmetry and Supergravity}\/,
second edition revised and expanded,
Princeton University Press, Princeton, New Jersey 1992.

\ii147 \sp1{Bin{\'e}truy} P., \sp1{Girardi} G., \sp1{Grimm} R.,
{\it Supergravity couplings: A geometric formulation}\/,
Phys. Reports {\bf343}, p.~255 (2001).

\ii148 \sp1{Polchinski} J.,
{\it String Theory\/}, vol. I+II,
Cambridge University Press, Cambridge, Mass. 1998.

\ii149
\sp1{Kalinowski} M. W., 
{\it The program of geometrization of physics. Some philosophical remarks}, 
Synthese {\bf77}, p.~129 (1988).

\ii150
\sp1{Klotz} A. H., 
{\it Macrophysics and Geometry: From Einstein's Unified Field Theory to 
Cosmology}, 
Cambridge Univ.\ Press, Cambridge, 1982. 

\ii151
\sp1{Overduin} J. M., \sp1{Wesson} P. S.,
{\it Kaluza--Klein Gravity}, 
Phys.\ Rep.\ {\bf283}, p.~303 (1997).

\ii152
\sp1{Janssen} T., \sp1{Prokopec} T.,
{\it Instabilities in the Nonsymmetric Theory of Gravitation},
Classical and Quantum Gravity {\bf23}, p.~4967 (2006).

\ii153
\sp1{Damour} T., \sp1{Deser} S., \sp1{McCarthy} J.,
{\it Nonsymmetric gravity theories: Inconsistencies and a cure},
Phys.\ Rev.~D {\bf47}, p.~1541 (1993).

}
\vfill \eject

\null
\vskip36pt plus4pt minus4pt
\centerline{{\four\bf Acknowledgement}}
\vskip24pt plus2pt minus2pt
\write0{Acknowledgement \the\pageno}

We would like to thank University of \L\'od\'z, Poland, for a partial support
by Badania W\l asne U\L\ Nr.\ 505/485 on early stages of development of this
work. We thank also Higher Vocational State School in Che\l m for a
continuous support during the academic year 2002/2003 under own scientific
programme conducted in the Institute for Mathematics and Computer Science.

I would like to thank Professor B.~Lesyng for the opportunity to carry out
computations using Mathematica~7 in the Centre of Excellence
BioExploratorium, Faculty of Physics, University of Warsaw. 

\vfill \eject

\closeout0
$\hbox{}$
\null
\vglue3cc plus4pt minus4pt
\ab{In this book we construct the Nonsymmetric Jordan--Thiry Theory
unifying\break N.G.T., the Yang--Mills' field, the Higgs' fields and scalar
forces in a geometric manner. In this way we get masses from higher
dimensions. We discuss spontaneous symmetry breaking, the Higgs'
mechanism and a mass generation in the theory. The scalar field
${\varPsi}$ (as
in the classical Jordan--Thiry Theory) is connected to the effective
gravitational constant. This field is massive and has Yukawa-type
behaviour. We discuss the relation between ${\sym R}_{+}$ invariance and 
${\bold U}(1)_{\rF}$
from G.U.T. within Einstein $\lambda $-transformation, and fermion number
conservation. In this way we connect $\ov{W}_{\mu}$-field from N.G.T. with a gauge
field $\widetilde{A}^{\rF}_{\mu}$ from G.U.T. We derive the equation of motion for a test
particle from conservation laws in the hydrodynamic limit. We consider
a truncation procedure for a tower of massive $\rho _{k}$ (or ${\varPsi} _{k})$ scalar 
fields using Friedrichs' theory and an approximation procedure for the
lagrangian involving Higgs' field up to the second order with respect
to $h_{\mu,\nu} = g_{\mu,\nu} - \eta _{\mu,\nu}$ and $\zeta
k^{0}_{\tilde{a}\tilde{b}}$. The geodetic equations on the
Jordan--Thiry manifold are considered with an emphasis to terms
involving Higgs' field. We consider also field equations in
linear approximation and an influence of electromagnetic and
$\ov{W}_{\mu}$ fields on field equations.

We consider a dynamics of Higgs' field in the framework of cosmological
models involving the scalar field~$\varPsi$. The field~$\varPsi$ plays here a
r\^ole of a quintessence field. We consider phase transition in cosmological
models of the second and the first order. Due to a tower of massive scalar
fields~$\varPsi_k$ derived from the theory we are able to get a warp factor
known from some modern approaches. Using a warp factor we build a
time-machine in the theory.

We consider a mass of a quintessence particle, various properties of a
quintessence field. We calculate a speed of sound in a quintessence and
fluctuations of a quintessence caused by primordial metric fluctuations.

The \nos\ \KK\ (\JT) Theory has been developed here. It unifies the gauge invariance
principle with the coordinate invariance principle but in more than
four-dimensional space-time. In particular in the case of electromagnetic and
\gr\ interactions in 5-dimensional theory.

A general nonabelian Yang--Mills fields have been unified with gravity in
$(n+4)$-\di al space-time ($n$---a \di\ of gauge group).
The theory uses a non\s\ metric defined on a metrized (in a non\s\
way) principal fibre bundle over a \spt\ with a structural group $\U(1)$ in an
\elm\ case and in general case nonabelian semi-simple compact group~$G$. The
\cn\ on \spt\ and on a metrized principal fibre bundle is compatible with
this metric. This \cn\ is similar to a \cn\ from Einstein's Unified Field
Theory, however we use its higher \di al analogue. This \cn\ is
right-\iv\ \wrt an action of the group~$G$ (a~gauge group). 

In the \elm\ case the metric and the \cn\ are bi-\iv\ \wrt the group~$\U(1)$.
The theory has been developed to
include a scalar field leading to an effective \gr\ constant and \spt\
dependent cosmological terms. It is possible to extend the theory to include
Higgs' fields and spontaneous symmetry breaking of a gauge group to get
massive vector boson fields. 

The theory is fully relativistic and unifies \elm\ field, gauge fields,
Higgs' field and scalar forces with NGT (Non\s\ Gravitation Theory) 
in a nontrivial way. 
By `in a nontrivial way' we mean that we get from the theory
something more than NGT, ordinary Kaluza--Klein (\JT) Theory, classical
electrodynamics, Yang--Mills' field theory with Higgs' field and spontaneous
symmetry breaking. These new features are some kind of ``interference
effects'' between all of them. 
This theory unifies two important approaches in higher-dimensional
philosophy: Kaluza--Klein principle and a \di al reduction principle.

The beautiful theories such as Kaluza--Klein theory (a~Kaluza miracle) and
its descendents should pass the following test if they are treated as real
unified theories. They should incorporate chiral fermions. Since the
fundamental scale in the theory is a Planck's mass, fermions should be
massless up to the moment of spontaneous symmetry breaking. Thus they should
be zero modes. In our approach they can obtain masses on a \di al reduction
scale. Thus they are zero modes in $(4+n_1)$-\di al case. In this way
$(n_1+4)$-\di al fermions are not chiral (according to very well known
Witten's argument on an index of a Dirac operator). Moreover, they are
not zero modes after a \di al reduction, i.e., in 4-\di al case. It means we
can get chiral fermions under some assumptions.

We expect some nonrelativistic effects leading
to nonnewtonian gravity. 

The real motivation of this book is to find a geometric unifications of
fundamental interactions of Nature and to find applications in Modern
Cosmology including inflation, dark matter and dark energy using a paradigm
of physics, which unifies two fundamental concepts of invariance in physics:
coordinate invariance principle and gauge invariance principle. The first is
known in General Relativity and in vaiable alternative theories of gravitation. The
second is known in Electrodynamics and Yang--Mills field theory. Yang--Mills
field theory governs Standard Model, i.e.: Glashow--Salam--Weiberg (GSW) model of
electro-weak interactions and the theory of strong interactions (QCD) using 
$\text{SU}(2)_L\times \U(1)$ and SU(3)$_c$ groups as gauge groups. It governs
also any Grand Unified Theorems.

The book is addressed to graduate students in theoretical physics,
mathematical physics, cosmology, particle physics and field theory. It is
also addressed to scientists working in these domains and to any reader who
is interested in these rapidly developing domains.

We find exact solutions of the field equations of the Nonsymmetric
Kaluza--Klein Theory in an electromagnetic case, i.e.\
gravito--electromagnetic waves and spherically symmetric stationary solution
with a cosmological constant. We consider also a charge confinement in this
theory with a generalization to color confinement in nonabelian theory (e.g.\
QCD).
}
\vfill\eject

\vfill \eject
\def\sym#1{{\Bbb #1}}
\def\rF{\text{F}}
\def\ov#1{#1\llap{$\overline{\phantom{\text{\rm#1}}}$}}

\centerline{\bf A brief information of a text}
\centerline{\hfil}

In this book we construct the Nonsymmetric Jordan--Thiry Theory
unifying N.G.T. (Nonsymmetric Gravitation Theory),
the Yang--Mills' field, the Higgs' fields and scalar
forces in a geometric manner. In this way we get masses from higher
dimensions. The Nonsymmetric Jordan--Thiry Theory unifies a coordinate
invariance principle and a gauge invariance principle.
We discuss spontaneous symmetry breaking, the Higgs'
mechanism and a mass generation in the theory. The scalar field
${\varPsi}$ (as
in the classical Jordan--Thiry Theory) is connected to the effective
gravitational constant. This field is massive and has Yukawa-type
behaviour. We discuss the relation between ${\sym R}_{+}$ invariance and
${\bold U}(1)_{\rF}$
from G.U.T. (Grand Unified Theory) within Einstein $\lambda $-transformation, and fermion number
conservation. In this way we connect $\ov{W}_{\mu}$-field from N.G.T. with a gauge
field $\widetilde{A}^{\rF}_{\mu}$ from G.U.T. We derive the equation of motion for a test
particle from conservation laws in the hydrodynamic limit. We consider
a truncation procedure for a tower of massive $\rho _{k}$ (or ${\varPsi} _{k})$ scalar
fields using Friedrichs' theory and an approximation procedure for the
lagrangian involving Higgs' field up to the second order with respect
to $h_{\mu\nu} = g_{\mu\nu} - \eta _{\mu\nu}$ and $\zeta
k^{0}_{\tilde{a}\tilde{b}}$. The geodetic equations on the
Jordan--Thiry manifold are considered with an emphasis to terms
involving Higgs' field. We consider also field equations in
linear approximation and an influence of electromagnetic and
$\ov{W}_{\mu}$ fields on field equations.

We consider a dynamics of Higgs' field in the framework of cosmological
models involving the scalar field~$\varPsi$. The field~$\varPsi$ plays here a
r\^ole of a quintessence field. We consider phase transition in cosmological
models of the second and the first order. Due to a tower of massive scalar
fields~$\varPsi_k$ derived from the theory we are able to get a warp factor
known from some modern approaches. Using a warp factor we build a
time-machine in the theory.

We consider a mass of a quintessence particle, various properties of a
quintessence field. We calculate a speed of sound in a quintessence and
fluctuations of a quintessence caused by primordial metric fluctuations.

\end

\end